\documentclass[11pt]{book}

\usepackage{bures}         
\usepackage{a4wide}
\usepackage{graphics}

\usepackage{amstex}    
\usepackage{amsthm}
\usepackage{amsbsy}                   
\usepackage{amstext}          
\usepackage{ctagsplt}               
\usepackage{amsfonts}           
\usepackage{amssymb}
\usepackage{amscd}

\DeclareFontFamily{U}{rsf}{}
\DeclareFontShape{U}{rsf}{m}{n}{
  <5> <6> rsfs5 <7> <8> <9> rsfs7 <10-> rsfs10}{}
\DeclareMathAlphabet\Scr{U}{rsf}{m}{n}

\begin{document}


\newcommand{\HRule}{\rule{\linewidth}{1mm}}


\newfont{\bol}{cmbxti10 at 11 pt}

\setlength{\parskip}{\medskipamount}

\def\thefootnote{\fnsymbol{footnote}}

\makeatletter

\renewcommand{\section}{\@startsection%
		{section}{1}{0cm}{2\baselineskip}%
		{0.5\baselineskip}%
		{\centering\Large\bf}}

\renewcommand{\subsection}{\@startsection%
		{subsection}{2}{-0.5mm}{\baselineskip}%
			{0.1mm}%
			{\raggedright\large\bf}}

\makeatother

\newcommand{\paragraphe}[1]{\smallskip{\it #1.~---~}}

\renewcommand{\thechapter}{\arabic{chapter}}
\renewcommand{\thesection}{\S\arabic{section}.}
\renewcommand{\thesubsection}{\arabic{section}.\arabic{subsection}.}
\renewcommand{\theequation}{\arabic{equation}}
\renewcommand{\thefigure}{\arabic{figure}}

\newenvironment{remarque}{\smallskip{\it Remarque.}~---~}{\smallskip}
\newenvironment{remarques}{\smallskip{\it Remarques.}~---~}{\smallskip}
\newenvironment{exemples}{\smallskip{\it Exemples.}~---~}{\smallskip}
\newenvironment{exemple}{\smallskip{\it Exemple.}~---~}{\smallskip}
\newenvironment{notations}{\smallskip{\it Notations.}~---~}{\smallskip}
\newenvironment{thm}[1]{\smallskip{\sc Th\'eor\`eme{\rm#1}%
			.}~---~\em}{\rm\smallskip}
\newenvironment{propos}[1]{\smallskip{\sc Proposition{\rm#1}%
			.}~---~\em}{\rm\smallskip}
\newenvironment{cor}{\smallskip{\sc Corollaire.}~---~\em}{\rm\smallskip}
\newenvironment{lemme}[1]{\smallskip{\sc Lemme{\rm#1}%
			.}~---~\em}{\rm}


\newcommand{\ie}{{\em i.e.}~}
\newcommand{\cad}{\mbox{c.-\`a-d. }}

\def\<<{\leavevmode
  \raise0.28ex\hbox{$\scriptscriptstyle\langle\!\langle$}\nobreak
  \hskip -.6pt plus.3pt minus.2pt\,}
\def\>>{\,\nobreak\hskip -.6pt plus.3pt minus.2pt
  \raise0.28ex\hbox{$\scriptscriptstyle\rangle\!\rangle$}}

\def\dateUSenglish{\def\today{\ifcase\month\or
 January\or February\or March\or April\or May\or June\or
 July\or August\or September\or October\or November\or December\fi
 \space\number\day, \number\year}}
\def\datefrench{\def\today{\ifnum\day=1\relax 1\raisebox{.6ex}{\small er}\else
  \number\day\fi \space\ifcase\month\or
  janvier\or f\'evrier\or mars\or avril\or mai\or juin\or
  juillet\or ao\^ut\or septembre\or octobre\or novembre\or d\'ecembre\fi
  \space\number\year}}
\datefrench

\newcommand{\qq}{\begin{eqnarray}}
\newcommand{\qqq}{\end{eqnarray}}
\newcommand{\non}{\nonumber}

\newcommand{\NK}{{{\bf K}}}
\newcommand{\NR}{{{\bf R}}}
\newcommand{\NA}{{{\bf A}}}
\newcommand{\NP}{{{\bf P}}}
\newcommand{\NC}{{{\bf C}}}
\newcommand{\NT}{{{\bf T}}}
\newcommand{\NZ}{{{\bf Z}}}
\newcommand{\NH}{{{\bf H}}}
\newcommand{\NM}{{{\bf M}}}
\newcommand{\NN}{{{\bf N}}}
\newcommand{\NS}{{{\bf S}}}
\newcommand{\NW}{{{\bf W}}}
\newcommand{\NV}{{{\bf V}}}
\newcommand{\NL}{{\bf L}}

\renewcommand{\mathcal}[1]{{\Scr #1}}
\newcommand{\CA}{{\mathcal{A}}}
\newcommand{\CB}{{\mathcal{B}}}
\newcommand{\CC}{{\mathcal{C}}}
\newcommand{\CalD}{{\mathcal{D}}}
\newcommand{\CE}{{\mathcal{E}}}
\newcommand{\CF}{{\mathcal{F}}}
\newcommand{\CG}{{\mathcal{G}}}
\newcommand{\CH}{{\mathcal{H}}}
\newcommand{\CI}{{\mathcal{I}}}
\newcommand{\CJ}{{\mathcal{J}}}
\newcommand{\CK}{{\mathcal{K}}}
\newcommand{\CL}{{\mathcal{L}}}
\newcommand{\CM}{{\mathcal{M}}}
\newcommand{\CN}{{\mathcal{N}}}
\newcommand{\CO}{{\mathcal{O}}}
\newcommand{\CP}{{\mathcal{P}}}
\newcommand{\CQ}{{\mathcal{Q}}}
\newcommand{\CR}{{\mathcal{R}}}
\newcommand{\CS}{{\mathcal{S}}}
\newcommand{\CT}{{\mathcal{T}}}
\newcommand{\CU}{{\mathcal{U}}}
\newcommand{\CV}{{\mathcal{V}}}
\newcommand{\CW}{{\mathcal{W}}}
\newcommand{\CX}{{\mathcal{X}}}
\newcommand{\CY}{{\mathcal{Y}}}
\newcommand{\CZ}{{\mathcal{Z}}}

\newcommand{\N}{{\mathbb{N}}}
\newcommand{\Z}{{\mathbb{Z}}}
\newcommand{\R}{{\mathbb{R}}}
\newcommand{\C}{{\mathbb{C}}}
\newcommand{\Q}{{\mathbb{Q}}}
\newcommand{\J}{{\mathbb{J}}}
\newcommand{\PP}{{\mathbb{P}}}
\newcommand{\poincare}{{\mathbb{H}}}

\newcommand{\Slr}{{\rm SL}_2\R}     
\newcommand{\Slc}{{\rm SL}_2}    
\newcommand{\Su}{{\rm SU}_2}       
\newcommand{\Gc}{G^{\,\C}}
\newcommand{\dC}{{}^{\,\C}}
\newcommand{\eC}{{}^{\C}}
\newcommand{\lieg}{{\mathfrak{g}}}      
\newcommand{\lieh}{{\mathfrak{h}}}
\newcommand{\liet}{{\mathfrak{t}}}     
\newcommand{\lietc}{{\mathfrak{t}^\C}}
\newcommand{\liehc}{{\mathfrak{h}^\C}}
\newcommand{\lien}{{\mathfrak{n}}}
\newcommand{\liegc}{{\mathfrak{g}^\C}}        
\newcommand{\slc}{{\mathfrak{sl}_2}}     
\newcommand{\sun}{{\mathfrak{su}(n)}}         
\newcommand{\glV}{{\mathfrak{gl}\,(V)}}        
\newcommand{\sudeux}{{\mathfrak{su}_2}}    
\newcommand{\conf}{{\mathfrak{C}}}         
\newcommand{\confproj}{{\rm P}\Slc}    

\newcommand{\bigphase}{{\boldsymbol{P}}}
\newcommand{\phase}{{\mathcal{P}}}

\renewcommand{\bar}[1]{\overline{#1}}

\newcommand{\XX}{\mathfrak{X}}
\newcommand{\Nr}{{\boldsymbol{r}}}
\newcommand{\Nl}{{\boldsymbol{\ell}}}
\newcommand{\Bpsi}{{\boldsymbol{\Psi}}}
\newcommand{\Bgamma}{\gamma}
\newcommand{\Bphi}{{\boldsymbol{\Phi}}}
\newcommand{\BW}{{\mathfrak{W}}}
\newcommand{\BK}{{\mathfrak{K}}}
\newcommand{\Bnabla}{{\boldsymbol{\nabla}}}
\newcommand{\conbar}{\overline{\boldsymbol{\partial}}}
\newcommand{\con}{\boldsymbol{\partial}}

\newcommand{\bigt}{\bigotimes}          
\newcommand{\bigp}{\bigoplus}
\newcommand{\de}{\overline{\partial}}
\newcommand{\da}{\partial}
\newcommand{\ee}{{\rm e}}               
\renewcommand{\det}{\mbox{det}\,}    
\newcommand{\Vtau}{\underline{\tau}}
\newcommand{\Vbar}{\overline{V}}
\newcommand{\vbar}{\overline{v}}
\newcommand{\chir}{\chi\sous{R}}          
\newcommand{\chiR}{\chi\sous{R'}}      
\newcommand{\gr}{g\sous{R}}
\newcommand{\eqr}{\lbrack R \rbrack}   
\newcommand{\eqR}{\lbrack R' \rbrack} 
\newcommand{\aij}{{\alpha;\,i,j}}        
\newcommand{\bkl}{{\beta;\,k,l}}        
\newcommand{\tr}{{\rm tr}}             
\newcommand{\Tr}{{\rm Tr}}            
\newcommand{\pun}{^{\phantom{-}}}    
\newcommand{\STOP}{S\sos{\rm top}}  
\newcommand{\Ssigma}{S\sos{\sigma}}    
\newcommand{\SCP}{S\sous{\C P^1}}         
\newcommand{\sur}[1]{^{{#1}}}           
\newcommand{\supr}[1]{^{^{#1}}}
\newcommand{\un}[1]{\underline{#1}}
\newcommand{\sous}[1]{_{\mbox{\tiny $#1$}}}
\newcommand{\sos}[1]{_{#1}}
\newcommand{\grl}{\mathop{\otimes}\limits_\ell g(\xi_\ell)\sous{R_\ell}}  
\newcommand{\hrl}{\mathop{\otimes}\limits_\ell h(\xi_\ell)\sous{R_\ell}}  
\newcommand{\hseul}{h(\xi_\ell)\sous{R_\ell}}
\newcommand{\hinv}{\mathop{\otimes}\limits_\ell h(\xi_\ell)\sous{R_\ell}^{-1}}  
\newcommand{\hAR}[1]{\sur{#1}\hspace{-0.15cm}A^{01}}    
\newcommand{\hAL}[1]{\sur{#1}\hspace{-0.15cm}A^{10}}   
\newcommand{\hA}[1]{\sur{#1}\hspace{-0.15cm}A}        
\newcommand{\AR}{A^{01}}                 
\newcommand{\AL}{A^{10}}                
\newcommand{\barT}{\overline{T}}
\newcommand{\barJ}{\overline{J}}           
\newcommand{\Jbar}{\overline{\CJ}}
\newcommand{\CTbar}{\overline{\CT}}      
\newcommand{\Cas}{C_{12}}          
\newcommand{\deltamu}{\delta\mu^{z}_{\overline{z}}}
\newcommand{\deltanu}{\delta\nu^{z}_{z}}
\newcommand{\posiJ}{{_{1+i\J}\over^2}}
\newcommand{\negaJ}{{_{1-i\J}\over^2}}

\newcommand{\esprep}{V_{\hspace{-0.04cm}\underline{\lambda}}}
\newcommand{\baresprep}{\overline{V}_{\hspace{-0.04cm}\underline{\lambda}}}
\newcommand{\espmeca}{V_{\hspace{-0.04cm}\underline{\ell}}}
\newcommand{\esprepspin}{V_{\hspace{-0.04cm}\underline{j}}}
\newcommand{\ChSi}{\BW_{g,\underline{R}}}
\newcommand{\chsi}{\CW(\Sigma,\underline{\xi},\underline{R})}
\newcommand{\chsiSph}{\CW(\C P^1,\underline{\xi},\underline{R})}
\newcommand{\chsispin}{\CW(\C P^1,\underline{\xi},\underline{j})}
\newcommand{\moduleR}{\CM_{g,N}}
\newcommand{\vv}{\hspace{-0.06cm}}
\newcommand{\espR}{V_{\hspace{-0.04cm}\lambda}}
\newcommand{\bibendum}{V\sous{\hspace{-0.04cm}\underline{R}}}

\newcommand{\partialbaruj}{\partial_{\hspace{0.02cm}\overline{u}^j}}
\newcommand{\partialbartau}{\partial_{\hspace{0.02cm}\overline{\tau}}}
\newcommand{\partialbarz}{\partial_{\hspace{0.02cm}\overline{z}}}
\newcommand{\partialbarw}{\partial_{\hspace{0.02cm}\overline{w}}}
\newcommand{\partialzbar}{\partial_{\hspace{0.02cm}\overline{z}_\ell}}
\newcommand{\KZmubar}{\Bnabla_{\hspace{-0.08cm}\overline{\delta\mu}}}
\newcommand{\KZmu}{\Bnabla_{\hspace{-0.1cm}\delta\mu}}
\newcommand{\KZz}{\Bnabla_{\hspace{-0.1cm}z_\ell}}
\newcommand{\KZzbar}{\Bnabla_{\hspace{-0.08cm}\overline{z}_\ell}}
\newcommand{\KZtau}{\Bnabla_{\hspace{-0.1cm}\tau}}
\newcommand{\KZtaubar}{\Bnabla_{\hspace{-0.08cm}\overline{\tau}}}

\newcommand{\exponen}{\ee^{\frac{ik}{2\pi}\int\tr\,A^{10}\wedge\,A^{01}}}
\newcommand{\exponenB}{\ee^{\frac{ik}{2\pi}\int\tr\,B^{10}\wedge\,B^{01}}}

\newcommand{\lsca}{{\boldsymbol{\langle\!\!\langle}}}
\newcommand{\rsca}{{\boldsymbol{\rangle\!\!\rangle}}}
\newcommand{\scal}{{\boldsymbol{|\!\!||\!\!|}}}
\newcommand{\sca}{{\boldsymbol{|\!\!|}}}
\newcommand{\Blsca}{{\boldsymbol{(\!\!(}}}
\newcommand{\Brsca}{{\boldsymbol{)\!\!)}}}

\newcommand{\BA}{{\bf A}}
\newcommand{\BF}{{\bf F}}

\newcommand{\widegamma}{\widetilde{\mbox{\boldmath$\Gamma$}}}
\newcommand{\Bv}{\boldsymbol{v}}
\newcommand{\partialA}{\overline{\partial}_{\hspace{-0.6mm}A^{01}}}
\newcommand{\partialh}{\overline{\partial}_{\hspace{-0.6mm}h^{-1}\de h}}
\newcommand{\partialAu}{\overline{\partial}_{\hspace{-0.6mm}A^{01}_u}}
\newcommand{\deltapsi}{{_{\delta\Bpsi}\over^{\delta A^{01}}}}
\newcommand{\expoS}{\ee^{k\,S(h,\,A^{01})}}
\newcommand{\Ahun}{\hAR{h^{-1}}}
\newcommand{\partialpha}{\partial_{n^\alpha}}
\newcommand{\expordonne}{\overset{\leftarrow}{\text{P}}\ee}
\newcommand{\expordonner}{\overset{\rightarrow}{\text{P}}\ee}
\newcommand{\Dbar}{\overline{D}}

\newcommand{\deltaW}{{_{\delta}\over^{\delta A^{01}(w)}}}
\newcommand{\deltaZ}{{_{\delta}\over^{\delta A^{01}(z)}}}
\newcommand{\deltaZpsi}{{_{\delta\Bpsi}\over^{\delta A^{01}(z)}}}
\newcommand{\deltaw}{{_{\delta}\over^{\delta A_{\overline{w}}}}}
\newcommand{\deltaz}{{_{\delta}\over^{\delta A_{\overline{z}}}}}
\newcommand{\deltazpsi}{{_{\delta\Bpsi}\over^{\delta A_{\overline{z}}}}}

\newcommand{\beltrami}{\mu^{z}_{\overline{z}}}
\newcommand{\fbeltrami}{\beltrami\,\partial_z\otimes d\overline{z}}
\newcommand{\Green}{{\bf G}}
\newcommand{\partialn}{\partial_{n^\gamma}}
\newcommand{\quotient}[2]{{_{#1}\over^{#2}}}
\newcommand{\afaire}{\marginpar{$\clubsuit$}}
\newcommand{\id}{\boldsymbol{id}}
\newcommand{\imtau}{\tau_2}
\newcommand{\comp}{{G}'_{a\pm}}
\newcommand{\green}{\boldsymbol{g}}
\newcommand{\baseA}{\mathfrak{A}}
\newcommand{\baseB}{\mathfrak{B}}
\newcommand{\noyau}{P}
\newcommand{\sumprime}{\sideset{}{'}{\sum}}
\newcommand{\Bnablaz}{\Bnabla_{\hspace{-0.1cm}z_\ell}}
\newcommand{\Bnablazbar}{\Bnabla_{\hspace{-0.08cm}\overline{z}_\ell}}
\newcommand{\singulier}{\,{}^\bullet_\bullet\,}

\newcommand{\nombre}{\addtocounter{equation}{1}}
\newcommand{\taga}[1]{\tag{\ref{#1}.a}}
\newcommand{\tagb}[1]{\tag{\ref{#1}.b}}
\newcommand{\refa}[1]{\ref{#1}.a}
\newcommand{\refb}[1]{\ref{#1}.b}

\newcommand{\widex}{{\boldsymbol{x}}}
\newcommand{\Axb}{A^{01}_{\widex,b}}
\newcommand{\Ad}{{\rm Ad}}
\newcommand{\ad}{{\rm ad}}
\newcommand{\zozo}{\stackrel{\otimes}{,}}

\renewcommand{\leq}{\leqslant}
\renewcommand{\geq}{\geqslant}
\newcommand{\wi}{\widetilde}
\newcommand{\wih}{\widehat}

\newcommand{\grm}{\mathop{\otimes}\limits_m g(\xi_m)\sous{R_m}}
\newcommand{\sumini}{\mbox{\small$\mathop{\sum}\limits_\ell$}}



\thispagestyle{empty}
\renewcommand{\thepage}{}
\begin{setlength}{\parindent}{0mm}
\begin{setlength}{\parskip}{0mm}
\marginpar{{\vspace{-2cm}\hspace{-6.125in}\large N${}^{\rm o}$ d'ordre : 5344}}

\begin{center}
{\LARGE\bf TH\`ESE}\\[5mm]
\normalsize pr\'esent\'ee par\\[5mm]
{\LARGE\bf Pascal~Tran-Ngoc-Bich}\\[10mm]
pour obtenir le titre de\\[5mm]
\Large Docteur en Sciences de l'Universit\'e d'Orsay (Paris 11)\\[5mm]
\normalsize Sp\'ecialit\'e~: Physique Th\'eorique\\[10mm]
\HRule\\
\LARGE\bf R\'esolution exacte de certains mod\`eles sigma\\
en th\'eorie quantique des champs
\end{center}
\HRule
\vspace{5mm}
\begin{center}
{\large Soutenue le 6 juillet 1998}
\end{center}
\vspace{10mm}
{\large
\begin{center}
	{\Large Jury}\\[5mm]
\begin{tabular}{ll}
Olivier Babelon & \\
Denis Bernard & Rapporteur\\
Pierre Bin\'etruy & Pr\'esident\\
Michel Dubois-Violette & \\
Benjamin Enriquez & Rapporteur\\
Krzysztof Gaw\c{e}dzki & Directeur de th\`ese
\end{tabular}
\end{center}
}
\vspace*{\stretch{3}}
\end{setlength}
\end{setlength}
\thispagestyle{empty}
\cleardoublepage
\thispagestyle{empty}
\phantom{blabla}


\setcounter{page}{1}
\renewcommand{\thepage}{\roman{page}}
\begin{setlength}{\parindent}{0mm}
\begin{setlength}{\parskip}{0mm}
\vspace*{\stretch{1}}
\begin{center}
{\sc PASCAL TRAN-NGOC-BICH}\\[50mm]
\Large\bf R\'esolution exacte de certains mod\`eles sigma\\
en th\'eorie quantique des champs\\
\end{center}
\vspace*{\stretch{20}}
\begin{center}
{\textsc{UNIVERSIT\'E d'ORSAY}}
\end{center}
\end{setlength}
\end{setlength}

\clearpage
\vspace*{\stretch{1}}
\begin{flushleft}
\begin{minipage}[t]{10cm}
\begin{verse}
\dots

\protect\hspace{2.7cm}Elle a l'aspect charmant\\
\protect\hspace{2.7cm}D'une adorable rousse

Ses cheveux sont d'or on dirait\\
Un bel \'eclair qui durerait\\
Ou ces flammes qui se pavanent\\
Dans les roses-th\'e qui se fanent

Mais riez riez de moi\\
Hommes de partout surtout gens d'ici\\
Car il y a tant de choses que je n'ose vous dire\\
Tant de choses que vous ne me laisseriez pas dire\\
Ayez piti\'e de moi\\[8mm]

Fragment d'un Calligramme \<<La jolie rousse\>>\\
{\sc Guillaume Appolinaire}
\end{verse}
\end{minipage}
\end{flushleft}
\vspace*{\stretch{2}}


\cleardoublepage


\vspace*{\stretch{1}}
\begin{center}
{\bf\Large Remerciements}
\end{center}

\begin{quote}
\setlength{\parindent}{1em}

\protect\hspace{\parindent}%
C'est un r\'eel plaisir de saisir l'occasion qui m'est
donn\'ee d'exprimer toute ma gratitude envers Krzysztof Gaw\c{e}dzki 
pour avoir accept\'e de me faire partager son activit\'e
scientifique. Au cours de ces ann\'ees, Krzysztof a toujours
r\'epondu avec gentillesse \`a toutes mes questions, m\^eme
les plus stupides. Sa confiance
et son soutien permanents, malgr\'e mon
absence prolong\'ee, m'ont \'et\'e d'un grand secours.
Je le remercie sinc\`erement pour tout cela.
\medskip

\indent Alors que j'\'etais encore \'etudiant en ma\^\i trise, Pascal Degiovanni
a guid\'e mes premiers pas en physique th\'eorique. Je lui dois
en partie d'\^etre ici maintenant.
\medskip

\indent Je remercie Denis Bernard et Benjamin Enriquez d'avoir accept\'e
la t\^ache de rapporteur, Olivier Babelon et Michel
Dubois-Violette d'avoir accept\'e de faire partie du jury
et Pierre Bin\'etruy d'en \^etre le pr\'esident.
\medskip

\indent Je remercie l'I.H.\'E.S. pour son hospitalit\'e et
les membres du personnel pour leur accueil
irr\'eprochable, tout particuli\`erement Madame
Vergne qui a r\'ealis\'e les fi\-gures de cette th\`ese.
Je remercie le laboratoire de physique de l'\'E.N.S. Lyon de 
m'avoir offert l'opportunit\'e d'exposer mon travail.
\bigskip

\indent Je voudrais d\'edier cette th\`ese \`a ma famille, mes parents
et mon fr\`ere, qui m'ont toujours soutenu alors
qu'ils sont les premiers \`a supporter mon sale caract\`ere.
\medskip

\indent Beaucoup de temps a pass\'e depuis le d\'ebut de cette th\`ese,
j'ai eu le privil\`ege de lier de nombreuses amiti\'es, 
parfois courtes parfois intenses,
mais toutes me tiennent \`a coeur. 
Un chaleureux merci \`a Magali et I avec qui j'ai form\'e un trio
musical des plus folklorique. Merci \`a
Christine, \'Emeline, Estelle,
Ida, Sylvie, Anne, Pascale mais aussi les Seb, les Vincent, Marc,  Bruno, les compagnons
de gal\`ere Richard, Michel et Vincent et, toutes et tous les
autres auxquels je pense.
Estelle a accept\'e spontan\'ement de parcourir ma th\`ese
\`a la recherche des fautes d'orthographe. Je l'en remercie.
Peut-\^etre de mani\`ere inhabituelle, je remercierais aussi mes 
\'el\`eves de m'avoir communiqu\'e leur bonne humeur.
\enlargethispage{6cm}

\end{quote}

\vspace*{\stretch{2}}

\cleardoublepage


\tableofcontents



\chapter*{Introduction\markboth{\sc Introduction}{\sc Introduction}}

\addcontentsline{toc}{chapter}{\protect\numberline{}Introduction}

\renewcommand{\thepage}{\arabic{page}}
\setcounter{page}{1}


\protect\hspace{\parindent}%
Dans cette th\`ese, on s'attache \`a l'\'etude de certaines th\'eories
des champs bidimensionnelles connues sous le nom de mod\`eles 
de Wess-Zumino-Novikov-Witten (WZNW). On essaye de r\'esoudre 
explicitement ces mod\`eles au niveau quantique. Il s'agit 
de trouver les fonctions de Green qui donnent
les valeurs moyennes des op\'erateurs de ces th\'eories
sur le vide. Un mod\`ele de WZNW est un mod\`ele sigma 
dont les champs classiques sont des applications de 
l'espace-temps bidimensionnel (une surface de Riemann
$\Sigma$ dans le cas euclidien) dans un groupe de Lie $G$,
l'espace cible. C'est une g\'en\'eralisation de la m\'ecanique 
quantique d'une particule sur le groupe $G$, quantifiant 
le mouvement g\'eod\'esique sur $G$. La dynamique de cette particule
peut \^{e}tre compl\`etement r\'esolue gr\^{a}ce 
au large groupe des sym\'etries que forme le groupe $G\times G$. 
Witten~\cite{witten:wzw} a montr\'e que cette 
propri\'et\'e peut \^{e}tre reproduite dans la th\'eorie 
des champs bidimensionnelle si on ajoute \`a l'action standard 
du mod\`ele sigma un terme de nature topologique~:
l'action de Wess-Zumino. En plus, on obtient de cette
mani\`ere une th\'eorie des champs invariante conforme.
Outre la grande richesse math\'ematique du mod\`ele de WZNW, 
l'un des faits les plus importants est la possibilit\'e 
de reproduire, par construction quotient, d'autres th\'eories 
conformes des champs. Ces th\'eories d\'ecrivent les 
diff\'erentes classes d'universalit\'e de comportement 
critique en m\'ecanique statistique des syst\`emes 
bidimensionnels. Elles sont aussi employ\'ees en th\'eorie des cordes 
comme des vides classiques autour desquels on quantifie la corde 
de fa\c{c}on perturbative.

Au niveau classique, le groupe des sym\'etries du mod\`ele de WZNW 
est donn\'e par deux copies du groupe des lacets $LG\times LG$, auxquels
s'ajoutent les sym\'etries conformes. Pour quantifier le mod\`ele
de WZNW on peut utiliser soit la th\'eorie des repr\'esentations 
des groupes des lacets soit une quantification \`a la Feynman.
Dans cette derni\`ere approche, il est bien de coupler
le mod\`ele aux champs de jauge externes $A$. La sym\'etrie
$LG\times LG$ s'\'etend alors \`a la sym\'etrie chirale de
jauge. Cette sym\'etrie de jauge permet d'\'ecrire des expressions 
pour les fonctions de Green qui factorisent leur
d\'ependance en $A$ en composantes chirales $A^{10}$ et $A^{01}$
du champ de jauge. Ainsi fait, l'\'etude du mod\`ele de WZNW se r\'eduit 
\`a l'\'etude d'une th\'eorie chirale 
(ou holomorphe dans l'espace-temps euclidien). 

La d\'ependance holomorphe des fonctions de Green \`a composante
chirale fix\'ee co\"{\i}ncide avec celle des \'etats quantiques d'une 
th\'eorie topologique de jauge tridimensionnelle connue sous 
le nom de th\'eorie de Chern-Simons (CS)~\cite{elitzur,witten:jones}. 
On montre formellement que les fonctions de Green du mod\`ele de WZNW 
sont d\'etermin\'ees par le produit scalaire (\`a la Bargmann) 
des \'etats quantiques de CS permettant de mettre ensemble 
les deux parties chirales de la th\'eorie. Les calculs sont men\'es 
\`a un niveau de rigueur ne satisfaisant pas les crit\`eres 
des math\'ematiciens. Le processus de quantification \`a 
la Feynman utilise l'int\'egrale fonctionnelle qu'on ne sait 
pas d\'efinir~; initialement, le produit scalaire des \'etats CS
est aussi exprim\'e par une telle int\'egrale. 
Ayant men\'e les calculs le plus loin possible, on utilise 
les r\'esultats pour red\'efinir {\it a posteriori}
les objets de la th\'eorie. On doit alors v\'erifier que 
ceux-ci ont un sens math\'ematique --- par exemple, on regarde 
la convergence d'int\'egrales --- et qu'ils satisfont aux divers 
crit\`eres physiques de la th\'eorie qu'on a pu observer 
sur les expressions formelles.

En pratique, on doit donc comprendre la structure de
l'espace des \'etats de CS et calculer le
produit scalaire. Les difficult\'es rencontr\'ees dans la mise
en {\oe}uvre de ce programme sont principalement contenues dans
l'espace des orbites $\CN=\CA^{01}/\CG^\C$ des champs
de jauge chiraux $A^{01}$ modulo les transformations de jauge 
chirales. L'espace $\CN$ peut aussi \^{e}tre interpr\'et\'e comme
l'espace des modules des $\Gc$-fibr\'es principaux holomorphes 
au-dessus de la surface de Riemann $\Sigma$. 
On d\'efinit un \'etat de CS sur un ouvert 
dense dans $\CA^{01}$ compos\'e des orbites du groupe $\CG^\C$ 
et on essaye de d\'eterminer sous quelles conditions on peut 
\'etendre (holomorphiquement) cet \'etat \`a tout $\CA^{01}$.
On obtient de cette mani\`ere des r\`egles de fusion 
(g\'en\'eralis\'ees) qui d\'eterminent compl\`etement l'espace
des \'etats de CS. 

Le calcul du produit scalaire est r\'eduit,
par un changement de variables, \`a une int\'egrale sur
$\CG^\C$, et en une int\'egrale sur l'espace des modules $\CN$. 
L'espace $\CN$ n'est explicitement d\'ecrit que dans un petit nombre 
de cas, ce qui laisse pr\'esager de nombreuses difficult\'es. En genre 
z\'ero et un, les grandes \'etapes ont \'et\'es men\'ees 
\`a bien~\cite{gaw97:unitarity,gaw91:scalar}. Par contre, 
la convergence du produit scalaire n'a \'et\'e montr\'ee que dans 
un petit nombre de cas~\cite{gaw96:elliptic,gaw89:quadrature}.
On conjecture~\cite{gaw89:construc} que l'int\'egrale donnant 
le produit scalaire \`a la Bargmann converge si et seulement 
si on la calcule sur les \'etats de CS qui satisfont
les r\`egles de fusion. Nous parcourons dans la th\`ese ces 
th\`emes en restant pr\`es des travaux originaux.

L'un des principaux objectifs est l'\'etude 
de l'espace des \'etats de CS quand on fait varier 
la structure complexe $\J$ de $\Sigma$ et les coordonn\'ees 
des points o\`u on ins\`ere des champs. On note $\un\xi$ 
la s\'equence de ces points. L'objet math\'ematique
qui contient l'information cherch\'ee est la connexion de
Knizhnik-Zamolodchikov-Bernard (KZB). En genre z\'ero, il
s'agit d'un ensemble d'\'equations diff\'erentielles satisfaites
par les parties chirales des fonctions de Green du mod\`ele de WZNW, 
obtenues pour la premi\`ere fois par Knizhnik et Zamolodchikov~\cite{kz}.  
La g\'en\'eralisation en genre sup\'erieur est le fait 
de Bernard~\cite{bernard:kzbanyg,bernard:kzb}.
On propose une construction g\'en\'erale 
de la connexion de KZB dans un esprit proche de la philosophie 
d\'evelopp\'ee par Axelrod, Della Pietra et Witten~\cite{axelrod}.
Cette fa\c{c}on de proc\'eder permet de retrouver les r\'esultats
connus en genre z\'ero et un. En termes plus abstraits,
l'ensemble des espaces des \'etats devrait former un fibr\'e vectoriel
holomorphe, appel\'e fibr\'e de Friedan-Shenker~\cite{friedan}, 
au-dessus de l'espace des modules des surfaces de Riemann avec 
points marqu\'es --- les $\un\xi$. Le produit
scalaire \`a la Bargmann permet d'impl\'ementer une structure 
hermitienne sur ce fibr\'e. La connexion de KZB est la connexion 
m\'etrique (ou unitaire) sur le fibr\'e de Friedan-Shenker, \ie la connexion 
qui respecte les structures holomorphe et hermitienne. 
Nous utilisons les expressions formelles donnant la structure
hermitienne pour d\'eriver la connexion de KZB. On repr\'esente
la connexion soit dans le langage de l'espace $\CA^{01}$
des champs chiraux de jauge soit dans le langage de l'espace 
des modules $\CN$, cette  derni\`ere repr\'esentation
 permettant d'obtenir des formules 
explicites pour une surface de Riemann $\Sigma$ de genre z\'ero
ou un. Un probl\`eme important est de montrer, que les expressions 
finales donnent bien la connexion avec les propri\'et\'es voulues 
(par exemple que la connexion est projectivement plate). 
L'unitarit\'e de la connexion, qui a \'et\'e obtenue 
en utilisant cette propri\'et\'e au niveau formel, reste le probl\`eme 
ouvert sauf dans des cas sp\'eciaux.

La partie chirale (holomorphe) de la th\'eorie WZNW, li\'ee 
\`a la th\'eorie tridimensionnelle de CS peut \^{e}tre
aussi vue comme une quantification d'une th\'eorie chirale
de jauge bidimensionnelle. C'est une th\'eorie propos\'ee 
par Hitchin~\cite{hitchin} comme une usine \`a syst\`emes
int\'egrables. Un syst\`eme de Hitchin a pour
espace des configurations l'espace des modules $\CN$
et pour espace des phases le fibr\'e cotangent
(holomorphe) $T^*\CN$.
L'approche de Hitchin peut \^{e}tre aussi g\'en\'eralis\'ee aux 
cas des surfaces de Riemann avec points marqu\'es. On essaye en 
ce qui nous concerne de trouver explicitement un ensemble complet 
de Hamiltoniens en involution pour le syst\`eme de Hitchin. En genre 
z\'ero et un ce n'est pas difficile. La discussion du cas de genre 
deux, pour $G^\C={\rm SL}_2$, est, certainement, la partie la plus  
originale de cette th\`ese. On utilise la description de l'espace 
des modules d\'etermin\'ee par Narasimhan et Ramanan~\cite{nar-ram}. 
Les premi\`eres \'etapes vers la d\'etermination des Hamiltoniens 
sont le fruit d'un travail de van Geemen et Previato~\cite{VGP}. 
Notre principal r\'esultat est la propri\'et\'e d'auto-dualit\'e 
du syst\`eme de Hitchin en genre deux, \cad  l'invariance de ses 
Hamiltoniens par \'echange des positions et moments de l'espace 
des phases. Elle a \'et\'e d\'emontr\'ee en utilisant une
formule explicite pour les valeurs des Hamiltoniens de Hitchin
en points g\'en\'eriques de $T^*\CN$ --- van Geemen et Previato
ont trouv\'e seulement l'expression sur des quartiques (de Kummer)
dans les fibres de $T^*\CN$. L'auto-dualit\'e des Hamiltoniens
nous a permis de compl\'eter le travail~\cite{VGP} 
et de d\'emontrer la conjecture des ses auteurs sur 
la forme des Hamiltoniens de Hitchin. Nous avons aussi d\'ecouvert le lien 
entre le mod\`ele de Hitchin ${\rm SL}_2$ en genre deux et les syst\`emes
int\'egrables de Neumann. La technique de la matrice de Lax d\'evelopp\'ee 
pour ces derniers nous a permis de trouver les variables
d'action-angle pour notre mod\`ele. Ces r\'esultats ont \'et\'e
expos\'es dans des articles originaux (soumis aux journaux 
scientifiques) qui sont joints \`a la th\`ese.

La quantification (g\'eom\'etrique) des syst\`emes de Hitchin 
reproduit essentiellement la partie chirale de la th\'eorie WZNW.
On fait varier la structure complexe sur $\Sigma$, variation
mesur\'ee par une diff\'erentielle de Beltrami. On peut alors 
coupler les Hamiltoniens quadratiques de Hitchin avec les diff\'erentielles
de Beltrami $\delta\mu$. On a encore des Hamiltoniens
en involution sur l'espace des phases. L'espace des \'etats quantiques
devient l'espace des sections holomorphes d'une puissance d'un 
fibr\'e vectoriel holomorphe (du fibr\'e en droites d\'eterminant 
pour les cas sans insertions des points) au-dessus de $\CN$.
Il co\"{\i}ncide avec l'espace des \'etats de CS. La quantification 
des Hamiltoniens quadratiques donne alors la connexion de KZB. 
Nous d\'ecrivons une telle construction en genre z\'ero, un et deux, 
dans le dernier cas uniquement pour $G={\rm SU}_2$.

\newpage
\vspace*{\stretch{1}}
\begin{gather*}
\mbox{{\sc Sch\'ema g\'en\'eral}}\\[1cm]
\begin{CD}
\boxed{\mbox{Mod\`ele de WZNW classique}}\\
@V\mbox{Quantification \`a la Feynman\quad}VV\\
\boxed{\mbox{Mod\`ele de WZNW quantique}} @<<<
	\mbox{Th\'eorie conforme}\atop{
	\mbox{des champs}}\\
@V\mbox{Factorisation holomorphe\quad}VV\\
\boxed{\mbox{\'Etats de Chern-Simons}}  @<<<    
\mbox{R\`egles de fusion}\\
@V\mbox{Variation de $\J$ et $\un\xi$\quad}VV\\
\boxed{\mbox{Fibr\'e de Friedan-Shenker}}       
@<<< \mbox{Principe de factorisation}\atop
	{\mbox{Formule de Verlinde}}\\
@V\mbox{Structure hermitienne\quad}VV\\
\boxed{\mbox{Produit scalaire \`a la Bargmann}} 
@<<<    \mbox{Convergence}\atop{\mbox{
	$\stackrel{?}{\Longleftrightarrow}$ R\`egles de fusion}}\\
@V\mbox{Connexion m\'etrique\quad}VV\\
\boxed{\mbox{Connexion de KZB}} @<<<    \mbox{Unitarit\'e ?}\atop
		{\mbox{Ansatz de Bethe}}\\
@A\mbox{Quantification g\'eom\'etrique\quad}AA\\
\boxed{\mbox{Hamiltoniens coupl\'es \`a $\delta\mu$}}\\
@A\mbox{Variation de $\J$\quad}AA\\
\boxed{\mbox{Syst\`emes de Hitchin}}    
@<<< \mbox{Syst\`emes int\'egrables}
\end{CD}
\end{gather*}
\vspace*{\stretch{2}}
\newpage
\vspace*{\stretch{1}}
\begin{center}

{\sc Plan de la th\`ese}

\end{center}

\begin{description}

\item[Chapitre 1] le premier chapitre traite sommairement l'action 
	gouvernant le mouvement g\'eod\'esique d'une particule sur 
	un groupe de Lie et sa quantification~;
	
\item[Chapitre 2]  on aborde divers aspects
	du mod\`ele de WZNW. On commence par la construction du 
	mod\`ele classique et la description du formalisme canonique sur
	l'espace des phases. Ensuite, on quantifie le mod\`ele
	et on donne les diverses fa\c{c}ons de repr\'esenter 
	le groupe des sym\'etries. On en d\'eduit
	la forme g\'en\'erale de l'espace des \'etats~;

\item[Chapitre 3] 
	on explique comment les \'etats de Chern-Simons 
	rentrent en jeu. On \'etudie l'espace et
	le produit scalaire des \'etats de Chern-Simons 
	en genre z\'ero et un. On dit quelques mots sur
	le genre sup\'erieur pour le groupe $G={\rm SU}_2$~;

\item[Chapitre 4] on construit la connexion de KZB.
	On en d\'eduit des expressions explicites en genre z\'ero et un.
	On \'etudie \'egalement les syst\`emes de Hitchin en genre
	z\'ero et un~;

\item[Chapitre 5] ce chapitre reproduit les articles 
	\'ecrits avec K.~Gaw\c{e}dzki~: \<<Self-duality of
	the ${\rm SL}_2$ Hitchin integrable system at genus $2$\>>,
	 \`a para\^\i tre dans Communications
	in Mathematical Physics, et
	\<<Hitchin systems at low genera\>>. Le d\'ebut est un r\'esum\'e
	en fran\c{c}ais.

\end{description}

\vskip 2cm

Un petit commentaire sur les notations. Lorsqu'on se r\'ef\`ere
\`a la section 1.{\bf 1.1}, il s'agit de la sous-section {\bf 1} 
de la section {\bf 1} du chapitre 1~!
Quand on se r\'ef\`ere \`a l'\'equation (1.1), il s'agit de l'\'equation
1 du chapitre 1, et quand on se r\'ef\`ere \`a l'\'equation (1) il s'agit
de la premi\`ere \'equation du chapitre courant. On utilise \`a l'envi les
abr\'eviations CS, KZB et WZNW pour respectivement
Chern-Simons, Knizhnik-Zamolodchikov-Bernard et Wess-Zumino-Novikov-Witten.

\vspace*{\stretch{2}}



\chapter{M\'ecanique quantique d'un mod\`ele sigma}

\medskip
\section{\'El\'ements de la th\'eorie des repr\'esentations}

Soit $V$ un espace vectoriel complexe non r\'eduit \`a 
z\'ero. Une {\bol repr\'esentation} $R$ d'un groupe de Lie $G$ dans $V$ est un
homomorphisme  $\Phi\sous{R}$ de $G$ dans ${\rm Aut}\,V$, groupe des
isomorphismes de $V$ dans lui-m\^eme. L'espace $V$ est appel\'e l'espace
de repr\'esentation de $R$. Pour simplifier,
si $g\in G$ et $v\in V$, on note $\gr v\equiv\Phi_{_R}(g)(v)$. 
Si $V$ est de dimension finie $n$, on dit que la repr\'esentation
est de dimension finie $n$.
Si $V=\C^n$, la repr\'esentation est dite matricielle.
Si $V= L^2(G,d_\ell g)$ o\`u $d_\ell g$ est une mesure invariante 
\`a gauche, la repr\'esentation
r\'eguli\`ere \`a gauche, not\'ee $\CL$, est d\'efinie par~:
$(g\sous{\CL} f)(x)=f(g^{-1}x)$. On d\'efinit aussi une repr\'esentation
r\'eguli\`ere \`a droite, not\'ee $\CR$ : $V=L^2(G,d_r g)$ et
$(g\sous{\CR} f)(x)=f(xg)$.

Une repr\'esentation est dite fid\`ele si $\Phi\sous{R}$ est injective.
Un sous-espace invariant est un sous-espace vectoriel $U$ 
tel que $\gr(U)\subset U$, pour tout $g\in G$.
Une repr\'esentation est dite {\bol irr\'eductible} si $V$ et $0$ sont ses seuls 
sous-espaces invariants.
Si $V$ est muni d'un produit scalaire hermitien, pour lequel tout $\gr$
est unitaire, \cad $\gr^\dagger\, \gr=\gr\,\gr^\dagger={\rm id}$,
la repr\'esentation est dite  {\bol unitaire}. 
Deux repr\'esentations $R$ et $R'$ de $G$, respectivement dans $V$ et
$V'$, sont dites {\bol \'equivalentes} s'il existe une application 
lin\'eaire inversible $E:V\rightarrow V'$ telle
que $g_{_{R'}}E=E\,\gr$, pour tout $g\in G$. 
L'application $E$ est appel\'ee un op\'erateur d'entrelacement.
Cette notion d'\'equivalence induit naturellement une
relation d'\'equivalence sur l'ensemble des repr\'esentations.
Suivant le {\bol lemme de Schur}, si
$R$ est une repr\'esentation irr\'eductible de $G$
dans l'espace $V$, un endomorphisme $L$ de $V$ tel que, 
pour tout $g \in G$, $\gr L=L\,\gr$,
est scalaire, \ie il existe un $\lambda\in\C$ tel que $L=\lambda\,{\rm
id}$.

On peut effectuer sur les repr\'esentations les op\'erations
usuelles de dualit\'e, conjugaison, somme directe ou produit tensoriel. 
Soient $R$ et $R'$ deux repr\'esentations de $G$
agissant res\-pectivement dans $V$ et $V'$~:

---~la repr\'esentation duale $R^*$ (ou contragradiante)
de $R$ est donn\'ee par les endomorphismes transpos\'es
$(g^{-1}_{_R})^{\rm t}$ agissant dans l'espace dual $V^*$~;

---~la repr\'esentation conjugu\'ee $\overline{R}$
est donn\'ee par les applications $\gr$ vues
comme endomorphismes de l'espace conjugu\'e
complexe $\overline{V}$. Ce dernier est l'espace $V$ muni de
la multiplication scalaire par les nombres complexes conjugu\'es.
On notera $\overline{v}$ un \'el\'ement $v$ de $V$ mais 
vu comme \'el\'ement de $\Vbar$.
Pour $R$ unitaire, les repr\'esentations $R^*$ et $\overline{R}$
sont \'equivalentes par l'isomorphisme naturel entre $V^*$
et $\overline{V}$ induit par le produit scalaire sur $V$~;

---~la somme (resp. le produit) de deux
repr\'esentations, not\'ee $R\oplus R'$
(resp. $R\otimes R'$) agit dans l'espace $V\oplus V'$ 
(resp. $V\otimes V'$) comme suit : 
\[
g_{_{R\oplus R'}}=\gr\oplus g_{_{R'}}\quad({\rm resp.}\ 
g_{_{R\otimes R'}}=\gr\otimes g_{_{R'}}).
\]

Dans la suite nous consid\'ererons des repr\'esentations
o\`u $V$ est muni d'une structure d'espace
vectoriel topologique. On supposera alors que, pour tout $v\in V$,
l'application $G\ni g\mapsto \gr v\in V$ est continue.

\subsection{Groupes compacts}

On suppose ici que $G$ est un groupe compact.
Soit $dg$ la mesure de Haar sur $G$ (bi-invariante) normalis\'ee 
par $\int_G\,dg=1$. 
Si l'espace de repr\'esentation $V$ est de dimension finie ---~ les
repr\'esentations irr\'eductibles d'un groupe compact
le sont automatiquement~--- on peut 
munir $V$ d'un produit scalaire hermitien rendant $R$ unitaire. 
Pour cela, il suffit de prendre
\[
(u,v) =\int_G \langle\,\gr u\mid\gr v\,\rangle\,dg
\]
o\`u $\langle.\,,.\rangle$ est un produit hermitien
quelconque sur $V$.

Soit $R_\alpha$ une repr\'esentation unitaire et irr\'eductible
de $G$ dans l'espace $V_\alpha$.  On se donne 
une base orthonorm\'ee $\lbrace e_j^\alpha\rbrace$ de $V_\alpha$.
On note $g_{\alpha;\,i,j}$
le $(i,j)$-\`eme \'el\'ement matriciel,
soit $(e_i^\alpha,g_{_{R_\alpha}}\, e_j^\alpha)$.
On conna\^\i t le produit scalaire des fonctions $g_{\alpha;\,i,j}$ gr\^ace aux 
{\bol relations d'orthogonalit\'e de Schur}.
Si $R_\alpha$ et $R_\beta$ sont deux repr\'esentations
irr\'eductibles in\'equivalentes, alors 
\[
\int_G\overline{g_{\alpha;\,i,j}}\,\,g_{\beta;\,k,l}\,\,dg
=0.
\]
Pour une seule repr\'esentation,
\[
\int_G\overline{g_{\alpha;\,i,j}}\,\, g_{\alpha;\,k,l}\,\,dg=
\frac{\delta_{i,k}\,\delta_{j,l}}%
{d_\alpha}
\]
o\`u $d_\alpha$ est la dimension de $V_\alpha$.
Autrement dit, les fonctions $g_{\alpha;\,i,j}$
sont orthogonales pour le produit scalaire standard sur $L^2(G,dg)$.

\label{thm:peter}
Soit $\lbrace R_\alpha\rbrace$ un ensemble maximal de
repr\'esentations irr\'eductibles,
unitaires et deux \`a deux  in\'equivalentes. 
D'apr\`es le {\bol th\'eor\`eme de Peter-Weyl}, la famille $\lbrace {d_\alpha^{1/2}}%
g_{\alpha;\,i,j}\rbrace$ forme une base orthonorm\'ee de $L^2(G,dg)$.
La repr\'esentation $R_\alpha$ d\'efinit une application lin\'eaire
$\rho_\alpha:\,\Vbar_{\vv\alpha}\otimes V_\alpha\rightarrow L^2(G,dg)$ 
par la formule
\[
\rho_\alpha(\overline{v}\otimes v')= d_\alpha^{1/2}\,(v,
g_{_{R_\alpha}} v').
\]
L'image de $\rho_\alpha$ est engendr\'ee
par les \'el\'ements
matriciels $g_{\alpha;\,i,j}$. D'apr\`es les relations d'orthogo\-nalit\'e
de Schur, $\rho_\alpha$ pr\'eserve le produit
scalaire. Le th\'eor\`eme de Peter-Weyl
implique que $\oplus_\alpha\rho_\alpha$
est un isomorphisme unitaire, soit
\qq
\mathop{\bigp}\limits_\alpha \Vbar_{\vv\alpha}\otimes
V_\alpha\,\cong\,L^2(G,dg).
\label{decc}
\qqq
Un calcul rapide donne
\[
\begin{cases}
g_{_\CL}\,\rho_\alpha(\vbar\otimes v')&=
		\rho_\alpha(\,\overline{g}_{_{R_\alpha}}\vbar\otimes v'),\\
g_{_\CR}\,\rho_\alpha(\vbar\otimes v')&=
		\rho_\alpha(\vbar\otimes g_{_{R_\alpha} }v').
\end{cases}
\]
Ainsi, l'application $\oplus_\alpha\rho_\alpha$ 
entrelace la repr\'esentation
r\'eguli\`ere gauche (resp. droite) dans $L^2(G,dg)$ avec
l'action naturelle de $G$ sur les facteurs de gauche~(\footnote{L'action 
\`a gauche de $g\in G$ sur $\oplus_\alpha\,(\vbar_\alpha\otimes v'_\alpha)$
conduit \`a~: \(\mathop{\oplus}\limits_\alpha\,(\,%
\overline{g}_{_{R_\alpha}}\vbar_\alpha\otimes v'_\alpha).\)})
(resp. droite) dans
$\bigoplus_\alpha \Vbar_{\vv\alpha}\otimes V_\alpha$.

Soit $f$ un \'el\'ement de $L^2(G,dg)$. 
D'apr\`es le th\'eor\`eme pr\'ec\'edent,
$f$ se d\'ecompose de mani\`ere unique 
sur $\lbrace g_{\alpha;\,i,j}%
\rbrace$ :
\begin{equation}
f(g)=\sum_\alpha\sum_{i,j} N_\aij\, 
{d_\alpha^{1/2}}\,g_{\alpha;\,i,j}
\label{dec}
\end{equation}
o\`u
\[
N_\aij=d_\alpha^{1/2}\int_G\overline{h_{\alpha;\,i,j}}\, 
f(h)\, dh
\]
et la somme dans~(\ref{dec}) converge dans $L^2(G,dg)$.
Posons
\[
f_{_{R_\alpha}}=d_\alpha^{1/2}\int_G f(h^{-1})\, h_{_{R_\alpha}}\,dh.
\]
On a donc $N_\aij=(e^\alpha_j,f_{_{R_\alpha}}e^\alpha_i)$.

L'isomorphisme~(\footnote{%
L'inverse de $\iota$ est l'application $\iota^{-1} :%
{\rm End}(V_\alpha)\ni\phi%
\rightarrow \mathop{\sum}\limits_i \overline{e^\alpha_i}%
\otimes\phi(e^\alpha_i)\:\in%
\Vbar_{\vv\alpha}\otimes V_\alpha$.})
$\,\iota:\,\Vbar_{\vv\alpha}\otimes V_\alpha
\rightarrow {\rm End}(V_{\alpha})\,$ tel que
\[
\iota(\vbar\otimes v')(w)=(v,w)\,v'
\]
identifie canoniquement les espaces
$\Vbar_{\vv\alpha} \otimes V_\alpha$ et ${\rm End}(V_\alpha)$, identification admise
une fois pour toutes.
On peut composer l'application
 $$L^2(G,dg)\ni f\longmapsto f_{_{R_\alpha}}\in
{\rm End}(V_{\alpha})
\,\cong\,\Vbar_{\vv\alpha}\otimes V_\alpha$$ 
avec $\rho_\alpha$.
D'une part, 
$\Vbar_{\vv\alpha}\otimes V_\alpha \stackrel{\rho_\alpha}{\longrightarrow}%
L^2(G,dg)\rightarrow {\rm End}(V_\alpha)\,\cong\,\Vbar_{\vv
\alpha}\otimes V_\alpha$
donne l'identit\'e de 
$\Vbar_{\vv\alpha}\otimes V_\alpha$. D'autre part,
$L^2(G,dg)\rightarrow {\rm End}(V_\alpha)\,\cong\,\Vbar_{\vv\alpha}\otimes%
V_\alpha\stackrel{\rho_\alpha}{\longrightarrow}L^2(G,dg)$
projette sur le composant $\Vbar_{\vv\alpha}\otimes V_\alpha$ de
la d\'ecomposition~(\ref{decc}). La formule~(\ref{dec}) 
peut \^etre mise sous forme de transform\'ee de Fourier~:
\[
f(g)=\sum_\alpha d_\alpha^{1/2}\,{\rm tr}\, f_{_{R_\alpha}} g_{_{R_\alpha}}
\]
exprimant la d\'ecomposition (\ref{decc}) 
des repr\'esentations r\'eguli\`eres 
de $G$ en repr\'esentations irr\'eductibles. L'unitarit\'{e}
de cette d\'ecomposition conduit \`a la formule
de Parseval-Plancherel 
\[
\|f\|^2_{_{L^2(G)}}=\int_G|f(g)|^2\,dg=
\sum_\alpha \|f_{_{R_\alpha}}\|_{_{\rm HS}}^2
\]
o\`u la norme de Hilbert-Schmidt d'une application lin\'eaire
$A$ est d\'efinie par $\Vert A\Vert^2_{_{\rm HS}}={\rm tr}\,A^\dagger A$.

Enfin, chaque repr\'esentation unitaire de dimension finie est 
\'equivalente \`a une somme de repr\'esentations irr\'eductibles%
~(\footnote{L'assertion reste vraie pour des repr\'esentations 
de dimension infinie dans la cat\'egorie hilbertienne.})
--- le compl\'ement orthogonal de chaque
 sous-espace invariant reste invariant.

Le {\bol caract\`ere} d'une repr\'esentation $R$ de dimension finie  
est la fonction $\chir:G\rightarrow \C$ telle que
\[
\chir(g)={\rm tr}\,g_{_{R}}.
\]
Pour les repr\'esentations irr\'eductibles $R_\alpha$,
$\chi\sous{R_\alpha}(g)=\sum_ig_{\alpha;i,i}.$
Les caract\`eres r\'ev\`elent des propri\'et\'es int\'eressantes.
La plus importante est l'invariance sur les classes
de conjugaison, soit 
\[
\chir(hgh^{-1})=\chir(g).
\]
On dit alors que le caract\`ere de $R$ est une fonction de classe.
Au niveau des caract\`eres, l'\'equivalence entre repr\'esentations duale
et conjugu\'ee se traduit par :
\[
\chi\sous{R^*}(g)=\overline{\chir(g^{-1})}=\chi_{_{\overline{R}}}(g^{-1}).
\]
Si $R$ et $R'$ sont 
deux repr\'esentations quelconques de $G$~:
\qq
\chi_{_{R\oplus R'}} & = & \chir+\chi_{_{R'}}\,,\non\\
\chi_{_{R\otimes R'}} & = & \chir\,\chi_{_{R'}}\,\non.
\qqq
Ces deux  propri\'et\'es justifient  en partie
le paragraphe suivant consacr\'e \`a l'anneau des repr\'esentations.

Les relations d'orthogonalit\'e\'{e} de Schur impliquent
des relations d'orthogonalit\'e de 
caract\`eres. Si $R_\alpha$ et $R_\beta$ sont deux repr\'esentations 
irr\'eductibles in\'equivalentes
de $G$, alors
\[
\int_G\overline{\chi\sous{R_\alpha}(g)}\,\chi\sous{R_\beta}(g)\,dg=0
\]
et pour une seule repr\'esentation 
\[
\int_G|\chi\sous{R_\alpha}(g)|^2\,dg=1.
\]
Les caract\`eres $\chi\sous{R_\alpha}$ forment une base orthonormale
dans le sous-espace des fonctions de classe de $L^2(G,dg)$. 

La notion de caract\`ere est importante, car elle permet de
d\'eterminer les repr\'esentations \`a \'equivalence pr\`es. 
En effet, si $R$ et $R'$ sont \'equivalentes alors $\chir=\chiR$.
L'inverse est aussi vrai~: si $R=\oplus_\beta R_{\beta}$ alors
$\int_G\overline{\chi_{_{R_\alpha}}(g)}\, \chi_{_{R}}(g)\,dg$
donne la multiplicit\'e de la repr\'esentation irr\'eductible 
$R_\alpha$ dans la somme $\oplus_\beta R_{\beta}$. 
On peut ainsi reconstruire $R$ de $\chi\sous{R}$ \`a \'equivalence pr\`es.

\subsection{Anneau des repr\'esentations $\NK(G)$}

Il est naturel de consid\'erer l'ensemble $\NC(G)$
des classes d'\'equivalence
de repr\'esentations de dimensions finies de $G$ et d'essayer 
de lui adjoindre une structure d'anneau~; la construction 
suivante reste valable pour un groupe non-compact. 

Notons $\eqr$ la classe d'\'equivalence de la repr\'esentation $R$.
On d\'efinit les op\'erations produit et somme sur les classes
d'\'equivalence comme prolongement des op\'erations sur les 
repr\'esentations~:
\[
\left\{\begin{array}{ll}
\eqr+\eqR=[R\oplus R']=\eqR+\eqr\\
\eqr\cdot \eqR=[R\otimes R']=\eqR\cdot\eqr.
\end{array}\right.
\]
\`A ce stade, $\NC(G)$ est un semi-groupe.
Pour obtenir un anneau, il nous faut construire l'inverse pour l'addition. 
Suivant Grothendieck, on d\'efinit une
relation d'\'equivalence sur $\NC(G)\times\NC(G)$ par
\[
([R_1],[R_2])\sim([R'_1],[R'_2])\,\Longleftrightarrow\,
[R_1]+[R'_2]=[R'_1]+[R_2].
\]
La transitivit\'e est une 
cons\'equence de la r\`egle d'annulation pour la loi $+$
sur $\NC(G)$~: 
\[
[R]+[Q]=[R']+[Q]\,\Longrightarrow\,[R]=[R'].
\]

Soit $\NK(G)$ l'ensemble des classes d'\'equivalence de
$\NC(G)\times\NC(G)$ vis \`a vis de $\sim$. 
Les op\'erations somme et produit dans $\NC(G)$ descendent \`a
$\NK(G)$.
On v\'erifie ais\'ement que
$ -([R_1],[R_2])_\sim=%
([R_2],[R_1])_\sim$.
Pour simplifier, on notera $[R_1]-[R_2]$ la
classe d'\'equivalence $([R_1],[R_2])_\sim$.
Un \'el\'ement de $\NK(G)$ est appel\'e
une {\bol repr\'esentation virtuelle} de $G$.
Le triplet $(\NK(G),+,\hspace{0.025cm}\cdot\hspace{0.03cm})$
est alors un anneau commutatif, l'{\bol anneau des repr\'esentations}.

On a d\'efini pr\'ec\'edemment les notions de repr\'esentations
duale et conjugu\'ee~;
ces derni\`eres s'\'etendent \`a $\NK(G)$.
Il est clair que ces deux op\'erations sont des involutions de $\NK(G)$.
De plus, pour les groupes compacts,
la dualit\'e et la conjugaison dans $\NK(G)$
co\"\i ncident.

Pour un groupe de Lie compact, simple, connexe et simplement connexe
il existe une correspondance bijective entre poids dominants
et repr\'esentations irr\'eductibles (cf. section {\bf 2}). On note $\rho_i$ la classe
d'\'equivalence de la repr\'esentation
irr\'eductible unitaire correspondante au poids fondamental $\lambda_i$.
On montre alors~\cite[p. 164]{adams}

\begin{propos}{}%
L'anneau des repr\'esentations est isomorphe \`a
l'ensemble des polyn\^omes \`a coefficients dans $\Z$ sur les classes 
d'\'equivalence de repr\'esentations de plus hauts poids fondamentaux, 
\ie
\[
\NK(G)\,\cong\,\Z\,[\,\rho_1,\ldots,\rho_r\,].
\]
\end{propos}

\medskip
\section{Alg\`ebres de Lie simples}

Soit $G$ un groupe de Lie compact, simple, connexe et simplement connexe. 
On note $\lieg$ l'alg\`ebre de Lie de $G$ et $\liegc$ 
l'alg\`ebre de Lie complexifi\'ee de $\lieg$, \cad $\liegc=\lieg\oplus i\lieg$. 
Le groupe de Lie complexifi\'e $\Gc$ est le groupe connexe, simplement
connexe (unique \`a isomorphisme
pr\`es) ayant pour alg\`ebre de Lie $\liegc$. 
L'alg\`ebre $\lieg$ est fix\'ee par l'involution anti-lin\'eaire 
$X\mapsto X^\dagger$ de $\liegc$. Nous utilisons la convention des 
physiciens pour laquelle l'application exponentielle qui envoie $\lieg$ dans 
le groupe $G$~ est $\lieg\ni X\mapsto \exp iX$.  Le groupe $G$ est 
un sous-groupe 
compact maximal de $\Gc$ fix\'e par $gg^\dagger=1$, \ie $(\exp iX)^\dagger
=\exp -iX$. Une excellente
r\'ef\'erence pour tout ce qui suit est~\cite{knapp}.
  
On d\'ecompose $\liegc$ relativement \`a une {\bol sous-alg\`ebre 
de Cartan} $\liet$, soit une sous-alg\`ebre ab\'elienne maximale dans $\lieg$,
\[
\liegc=\lietc\oplus\bigoplus_{\alpha\in\Delta}\C e_\alpha
\]
o\`u $e_\alpha$ est un vecteur propre dans $\liegc$ 
pour l'action adjointe de $\liet$~:
\[
[X,e_\alpha]=\alpha(X)\,e_\alpha,\quad \forall X\in\liet.
\]
Par construction, les $\alpha$ sont des \'el\'ements de $\liet^*$, appel\'es
{\bol racines} de $\liegc$. On note $\Delta$ l'ensemble 
des racines. Il est possible de choisir les vecteurs $e_\alpha$ de telle 
sorte que $e_\alpha^\dagger=e_{-\alpha}$ et le commutateur
\[
[e_\alpha,e_{-\alpha}]\equiv\alpha^\vee\,,
\]
qui appartient \`a l'alg\`ebre de Cartan $\liet$, satisfait 
\[
\alpha(\alpha^\vee)=2\,.
\]
On appelle $\alpha^\vee$ la {\bol coracine} associ\'ee \`a 
la racine $\alpha$~; on note $\Delta^\vee$ l'ensemble des coracines.
Par contre, si $\alpha,\beta\in\Delta$, $\alpha\neq-\beta$, on a
\[
[e_\alpha, e_\beta]=N_{\alpha\beta}\, e_{\alpha+\beta}
\]
et $N_{\alpha\beta}=0$ si $\alpha+\beta\notin\Delta$.

La {\bol forme de Killing} de $\lieg$ est la forme bilin\'eaire
non d\'eg\'en\'er\'ee $K(X,Y)\equiv
\tr\,\ad_X\circ\ad_Y$ sur $\lieg\times\lieg$, o\`u $\tr$
est la trace lin\'eaire. On pr\'ef\`ere renormaliser la
forme de Killing en posant $\tr\, XY=K(X,Y)/I_\phi$.
On fixe le coefficient $I_\phi$ plus loin.
La forme de Killing est invariante dans l'action adjointe,
\ie $\tr\,[X,Y]\,Z+\tr\, Y\,[X,Z]=0$, mais aussi par tout automorphisme
de l'alg\`ebre. Elle est unique \`a normalisation
pr\`es et s'\'etend naturellement \`a $\liegc$. Avec nos hypoth\`eses sur $G$,
sa restriction $\tr\mathop{|}_{\liet\times \liet}$ est non 
d\'eg\'en\'er\'ee, donc
$\forall \lambda\in\liet^*$, il existe un unique $X_\lambda\in\liet$ tel que 
$\lambda(X)=\tr\,X_\lambda\,X$ pour tout $X\in\liet$ 
et $\tr\,\lambda\,\mu\equiv\tr\,X_\lambda\,X_\mu
=\lambda(X_\mu)=\mu(X_\lambda)$, pour $\lambda$ et $\mu$ dans $\liet^*$.
On a 
\[
\lambda(\alpha^\vee)=\tr\,[e_\alpha,e_{-\alpha}]\,X_\lambda
	=\tr\,e_{-\alpha}\,[X_\lambda,e_\alpha]
	=\tr\,\lambda\alpha\ \tr\,e_{-\alpha}\,e_\alpha.
\]
Pour $\lambda=\alpha$, on obtient $2=\tr\,\alpha^2\,
\tr\,e_{-\alpha}\,e_\alpha$. 
Il suit donc que 
\[
\alpha^\vee=\frac{2}{\tr\, \alpha^2}\, X_\alpha\,,\quad\ 
\tr\, (\alpha^\vee)^2=\frac{4}{\tr\alpha^2}\,,\quad\ 
X_\alpha=\frac{2}{\tr\,(\alpha^\vee)^2}\,\alpha^\vee\,.
\]
On peut toujours choisir
$(h^j)_{j=1,\cdots, r}$ une base orthonorm\'ee
de la sous-alg\`ebre de Cartan $\lietc$, avec $(h^j)^\dagger=h^j$.
On montre que $\tr\,e_\alpha h^j=0=\tr\,e_\alpha e_\beta$, 
$\alpha+\beta\neq 0$.
L'ensemble des $e_\alpha$ et des $h^j$ forme une {\bol base
de Cartan-Weyl} pour $\liegc$. Une base de la forme compacte $\lieg$ est
donn\'ee par $h^j,i(e_\alpha-e_{-\alpha}),e_\alpha+e_{-\alpha}$.
Dans les applications, on privil\'egie souvent l'usage d'une 
{\bol base orthogonale} $(t^a)$ avec
\[
\tr\,t^at^b=\delta^{ab}/2,\qquad [t^a,t^b]=if^{abc}t^c
\]
(sommation sur $c$ !). Les $f^{abc}$ sont les constantes de structure 
de l'alg\`ebre de Lie.
Ces conventions sont en accord avec la base utilis\'ee pour l'alg\`ebre 
$\mathfrak{sl}_2$, \cad $t^a=\sigma^a/2$ o\`u les $\sigma^a$ sont les
matrices de Pauli. En ce qui concerne la base
de Cartan-Weyl de $\mathfrak{sl}_2$~:
$e_{\pm \alpha}=\sigma^\pm$ et $h=\sigma^3/\sqrt{2}$. On a deux racines
$\pm\alpha$ avec $\alpha(h)=\sqrt{2}$.

Les alg\`ebres $\mathfrak{n}_{\pm}=\oplus_{\alpha>0}\C e_{\pm \alpha}$ sont 
deux sous-alg\`ebres nilpotentes maximales dans $\liegc$. On a
$\liegc=\mathfrak{n}_+\oplus\lietc\oplus\mathfrak{n}_-$. 
On appelle traditionnellement sous-alg\`ebres de Borel 
les sous-alg\`ebre $\mathfrak{b}_\pm=\lietc\oplus\mathfrak{n}_\pm$.
Ici, on va utiliser aussi la sous-alg\`ebre 
$\mathfrak{b}=i\liet\oplus\mathfrak{n}_+$.
La {\bol d\'ecomposition d'Iwasawa} certifie que
$\liegc=\lieg\oplus \mathfrak{b}$. Au niveau des groupes de Lie, cela 
donne une d\'ecomposition de $\Gc$ par rapport au groupe $B$ d'alg\`ebre $\mathfrak{b}$.
Ainsi, tout \'el\'ement $h$ de $\Gc$ se d\'ecompose de mani\`ere unique en
\[
h=\ee^{\sum_{\alpha>0}v_\alpha e_\alpha}\,\ee^{\varphi/2}\,g
\]
avec $g$ dans le groupe compact $G$, \cad $gg^\dagger=1$,
$v_\alpha\in\C$ et $\varphi\in\liet$. Pour rappel, $\ee^{\varphi/2}$ n'est
pas dans le groupe compact, avec nos conventions sur l'application exponentielle.

L'ensemble des racines engendre $\liet^*$ mais ce 
n'est pas une base de $\liet^*$.
Une sous-famille $(\alpha_1,\dots,\alpha_r)$ de $\Delta$ est 
appel\'ee une base de $\Delta$ si $B$ est une base de $\liet^*$ 
et toute racine $\alpha$ de $\liegc$ est 
une combinaison lin\'eaire des $\alpha_i\in B$, \`a 
coefficients entiers tous $\geq0$ (pour $\alpha\in\Delta_+\subset\Delta$)
ou tous $\leq0$ (pour $\alpha\in\Delta_-\subset\Delta$). 
Le {\bol rang} de $\liegc$, not\'e $r$, est \'egal \`a la dimension de $\liet$.
Les $\alpha_i,\ i=1,\cdots,r,$ sont appel\'ees les {\bol racines simples}.
Les racines de $\liegc$ appartenant \`a $\Delta_+$
sont les racines positives, et celles appartenant \`a $\Delta_-$ les
racines n\'egatives. 

Parmi toutes les racines positives, il existe 
une racine privil\'egi\'ee, la {\bol racine la plus grande} $\phi$. C'est
l'unique racine telle que $\phi+\alpha\not\in\Delta$, pour tout $\alpha\in
\Delta_+$. On normalise alors la forme de Killing $\tr$ par $\tr\,\phi^2=2$.
Ceci fixe compl\`etement le coefficient $I_\phi$. Si on d\'ecompose
$\phi$ et $\phi^\vee$, la coracine de $\phi$, sur une base 
avec $\phi=\sum_ik_i\alpha_i$ et $\phi^\vee=\sum_ik^\vee_i\alpha_i^\vee$
--- les coefficients $k_i$ ($k_i^\vee$) sont les labels de Kac (duaux) ---
le nombre de Coxeter et le {\bol nombre de Coxeter dual} sont respectivement
\[
g=1+\sum_ik_i,\qquad g^\vee=1+\sum_ik^\vee_i.
\]
\`A la fin de la section, on montre que la normalisation choisie
pour la forme de Killing conduit
\`a $I_\phi=2 g^\vee$. En particulier, on obtient~: dans la base orthogonale,
\[
f^{abc}f^{bcd}=g^\vee\delta^{ad}\,.
\]

La table suivante pr\'esente la {\bol classification de Cartan} des
alg\`ebres simples. L'indice $r$ indique le rang de l'alg\`ebre, $d$ la dimension et 
$g^\vee$ le nombre de Coxeter dual~:

\begin{center}
\begin{tabular}{ccccc}
\hline
\hline
\hspace{0.1cm}
$\liegc$        & $\lieg$       & $d$      & $r$          & $g^\vee$\\ \hline
$A_r$   & $\mathfrak{su}(r+1)$  & $r^2+2r$ & $1,2,\cdots$ & $r+1$\\ 
$B_r$   & $\mathfrak{so}(2r+1)$ & $2r^2+r$ & $2,3,\cdots$ & $2r-1$\\
$C_r$   & $\mathfrak{sp}(2r)$   & $2r^2+r$ & $3,4,\cdots$ & $r+1$\\
$D_r$   & $\mathfrak{so}(2r)$   & $2r^2-r$ & $4,5,\cdots$ & $2r-2$\\
\hline
\hline
\end{tabular}
\qquad
\begin{tabular}{ccc}
\hline
\hline
$\liegc$ & $d$ & $g^\vee$\\ \hline
$E_6$    & $78$  & $12$\\
$E_7$    & $133$  & $18$\\
$E_8$    & $248$  & $30$\\
$F_4$    & $52$   & $9$ \\
$G_2$    & $14$   & $4$ \\\hline
\hline
\end{tabular}
\end{center}

Le {\bol groupe de Weyl}\, $W$ est le sous-groupe de $\hbox{GL}(\liet^*)$ 
engendr\'e par les r\'eflexions par rapport aux hyperplans 
$L_\alpha=\{\beta\in\liet^*\ |\ \tr\,\beta\alpha=0\}$,
$r_\alpha : \liet^*\ni\mu\longmapsto\mu-\mu(\alpha^\vee)\,\alpha\in\liet^*$.
On appelle {\bol chambre de Weyl} une des composantes 
connexes de $\liet^*\setminus\bigcup_{\alpha\in\Delta}L_\alpha$.
Le groupe de Weyl satisfait les propri\'et\'es suivantes~:

--~$W$ est fini, $W$ pr\'eserve la forme de Killing~; 

--~l'action de $W$ sur l'ensemble des chambres de Weyl est simplement 
transitive, \ie il existe un unique \'el\'ement $w\in W$ tel que
$w(C)=C'$, si $C$ et $C'$ sont deux chambres de Weyl~;

--~si $B$ est une base de $\Delta$, $W$ est engendr\'e par les 
$r_{\alpha_i}$ ($\alpha_i\in B$), appel\'ees r\'eflexions simples~;

--~l'adh\'erence $\overline{C}$ d'un chambre de Weyl
$C$ est un domaine fondamental pour $W$~:
toute orbite de $W$ dans $\liet^*$ intersecte $\overline{C}$ 
en un seul point.

\noindent Soit $T$ un tore maximal dans $G$, \cad un sous-groupe de Cartan
d'alg\`ebre de Lie $\lieg$. On peut identifier $W$ et $N(T)/T$ o\`u $N(T)$ est
le normaliseur de $T$ dans $G$, \ie $N(T)=\{g\in G\ \big|\ gTg^{-1}=T\}$.
Le quotient $N(T)/T$ est juste l'ensemble des automorphismes de $T$ induits
par les automorphismes int\'erieurs de $G$. Ainsi, si $w\in N(T)/T$, 
il existe $g\in N(T)\subset G$ tel que 
$w(t)=gtg^{-1}$, pour tout $t\in T$. L'\'el\'ement $w$ induit 
un automorphisme lin\'eaire de $\liet$ et donc de $\liet^*$.
Dans ce langage, les r\'eflexions simples $r_{\alpha_i}$
sont donn\'ees par $\exp [\pi i\,(e_{\alpha_i}+e_{-\alpha_i})/2]\in G$.

La {\bol matrice de Cartan} $A$ est la matrice ayant pour \'el\'ement 
$(i,j)$~:
$a_{ij}\equiv 2\,\tr\,\alpha_i\,\alpha_j\,/\,\tr\,\alpha_j^2$.
La matrice $A$ v\'erifie les propri\'et\'es suivantes
\begin{gather}
\label{matricecartan1}
\begin{cases}
	a_{ii}=2, & \text{si $i,j=1,\ldots,r$}\\
	a_{ij}=0\Leftrightarrow a_{ji}=0,& \text{si $i,j=1,\ldots,r$}\\
	a_{ij}\in\Z_-,& \text{si $i,j=1,\ldots,r,\,i\not=j$}
\end{cases}\\
\label{matricecartan2}
\mbox{$A$ est d\'efinie positive},
\end{gather}
\cad tous les mineurs principaux de
$A$ sont strictement positifs. En particulier, $A$ est une matrice
de rang $r$.
Cette matrice caract\'erise compl\`etement l'alg\`ebre $\liegc$.
Ainsi la classification des alg\`ebres de Lie simples se r\'eduit \`a la
classification des matrices de Cartan.

Le sous-groupe additif de $\liet^*$ engendr\'e par $\Delta$
est un r\'eseau, appel\'e r\'eseau 
des racines et not\'e $Q$.
Les coracines engendrent \'egalement un r\'eseau dans $\liet$,
appel\'e r\'eseau des coracines et not\'e $Q^\vee$.
Un {\bol poids} $\lambda$ est un \'el\'ement de $\liet^*$ tel que
$\lambda(q^\vee)\in\Z$, pour tout $q^\vee\in Q^\vee$.
L'ensemble des poids forme un r\'eseau $P\subset\liet^*$, appel\'e 
r\'eseau des poids. Par construction, $P=(Q^\vee)^*$ est le r\'eseau  
dual \`a $Q^\vee$ et $Q\subset P\subset \liet^*$.
Le r\'eseau dual \`a $Q$, not\'e $P^\vee$, est le r\'eseau des 
{\bol copoids}~; $Q^\vee\subset P^\vee\subset\liet$. 
Si $\hbox{exp}:\lieg\rightarrow G$ est l'application exponentielle,
$Q^\vee=\{X\in\liet\ |\ \hbox{exp}(2\pi iX)=1\}$ et
$P^\vee=\{X\in\liet\ |\ \hbox{exp}(2\pi iX)\in Z(G)\}$,
o\`u $Z(G)$ est le centre de $G$.
$Z(G)\cong P^\vee\,/\,Q^\vee\cong P\,/\,Q$ et $|Z(G)|=\det A$. 
Les poids tels que $\lambda_i(\alpha_i^\vee)=\delta_{ij}$ sont les
poids {\bol fondamentaux}. Les poids fondamentaux forment une base de $P$~:
tout poids $\lambda$ s'\'ecrit de mani\`ere unique 
$\lambda=\sum_i n_i\lambda_i$, avec $n_i\in\Z$. On note $P_+$
l'ensemble des {\bol poids dominants}, pour lesquels $n_i\geq 0$.

Une repr\'esentation (de dimension finie) de $\liegc$ dans un espace vectoriel
($V$ de dimension finie) est un homomorphisme d'alg\`ebres
de Lie $\liegc\rightarrow \glV$. On utilise \`a l'envi le
terme {\bol module} pour parler de repr\'esentation. 
Un module $M$ est un espace vectoriel plus une action de
l'alg\`ebre. Clairement, il s'agit juste de deux mani\`eres de
consid\'erer le m\^eme objet. 
Comme on s'y attend, on \'echange repr\'esentations d'alg\`ebres
et de groupes par int\'egration ou d\'erivation.
Un {\bol module} $M_\lambda$ est dit {\bol de plus haut poids} s'il existe 
$\lambda\in\liet^*$ et un vecteur $v_\lambda\in M_\lambda$, tels que
\[
e_\alpha v_\lambda=0\quad \forall \alpha\in\Delta_+,\qquad Xv_\lambda=%
\lambda(X)\,v_\lambda\quad \forall X\in\liet,~\quad{\rm et}\quad
M_\lambda=\CU(\liegc)v_\lambda
\]
o\`u $\CU\,(\liegc)$ est l'alg\`ebre enveloppante de $\liegc$.
On dit que $\lambda$ est le plus haut poids de la repr\'esentation ---
m\^eme si rien ne nous dit que $\lambda$ est un poids dans
le sens donn\'e auparavant --- et
$v_\lambda$ est un vecteur de plus haut poids. 
Tout vecteur de $M_\lambda$ appartient \`a $\CU(\mathfrak{n}_-)\,v_\lambda$.
Une repr\'esentation est dite unitaire s'il existe
un produit scalaire sur $M$ tel que la conjugaison hermitienne
correspondante envoie $X\in\liegc$ vers $X^\dagger$.

Parmi toutes les repr\'esentations de plus haut poids $\lambda$
fix\'e, il existe une repr\'esentation plus grande que toutes
les autres, le {\bol module de Verma} $V_\lambda$ \`a partir duquel
on obtient n'importe quel autre module de plus haut poids $\lambda$ 
en quotientant par un sous-espace invariant, et une plus petite, la repr\'esentation
irr\'eductible $H_\lambda$, obtenue en quotientant par le plus grand sous-espace propre
invariant. Le module de Verma admet une unique forme hermitienne, la
{\bol forme de Shapovalov}, telle que $X^\dagger=X$ et $\langle v_\lambda,
v_\lambda\rangle= 1$. Le plus grand sous-espace propre invariant est alors 
le sous-espace des vecteurs orthogonaux \`a tous les autres. La forme de Shapovalov
descend donc sur $H_\lambda$ pour donner une forme hermitienne non-d\'eg\'en\'er\'ee
sur $H_\lambda$.

Les repr\'esentations unitaires de plus haut poids sont celles pour lesquelles
la forme de Shapovalov sur $H_\lambda$ est positive. C'est le cas
si, et seulement si, $\lambda \in P_+$.
Ainsi, le {\bol th\'eor\`eme du plus haut poids} dit que
les repr\'esentations de $\liegc$ 
irr\'eductibles, unitaires de plus haut poids sont \'enum\'er\'ees
par les plus hauts poids $\lambda\in P_+$. Elles sont toutes de 
dimension finie, et s'int\`egrent en des repr\'esentations irr\'eductibles,
unitaires de $G$. R\'eciproquement, toute repr\'esentation irr\'eductible,
unitaire de $G$ se d\'erive en une repr\'esentation de $\liegc$,
irr\'eductible, unitaire, de plus haut poids. Ainsi, on peut 
\'egalement \'enum\'erer les classes d'\'equivalence de repr\'esentations
de $G$ irr\'eductibles et unitaires par les poids dominants.
En plus, l'action de $\liet$ sur $M_\lambda$ est diagonale, 
on peut donc d\'ecomposer l'espace de repr\'esentation en une
somme directe 
\[
M_\lambda=\bigoplus_{\mu\in\liet^*}M_\lambda(\mu),\qquad
M_\lambda(\mu)=\big\{v\in M_\lambda\ \big|\ Xv=\mu(X)\,v, \forall X\in\liet\big\},
\]
\'eventuellement vide. Si $M_\lambda(\mu)\neq\emptyset$, $\mu$ est un poids de
l'alg\`ebre de Lie et $\mu=\lambda-\sum_\ell
\alpha_\ell$, o\`u $\alpha_\ell\in\Delta_+$.

Le {\bol Casimir quadratique} est l'\'el\'ement de
l'alg\`ebre enveloppante
\[
C=\quotient{1}{2}
	\sum_{i,j}(\tr\,e_i e_j)^{-1}\,
	\,e_i e_j
\]
o\`u $(e_i)$ est une base quelconque de $\liegc$, dont le choix ne
change pas $C$. Mieux encore, le Casimir est
dans le centre de l'alg\`ebre enveloppante.
Dans la base orthogonale et dans la base de Cartan-Weyl,
\[
C\,=\,t^at^a\,=\,\sum_{\alpha>0}\frac{\tr\,\alpha^2}{2}
	\,(e_\alpha e_{-\alpha}+e_{-\alpha}e_\alpha)
	+\quotient{1}{2}\sum_{j=1}^rh^jh^j.
\]
Le Casimir quadratique   
dans une repr\'esentation $R$ de plus haut poids 
$\lambda$ est obtenu  en rempla\c{c}ant les
$e_i$ par les $(e_i)_{_{R}}$ dans $C$.
Suivant le lemme de Schur, ce Casimir agit scalairement. On note
$C_\lambda$ ce scalaire. On trouve ais\'ement
que $C_\lambda=\frac{1}{2}\,\tr\,\lambda\,(\lambda+2\rho)$, o\`u
$\rho$ est le vecteur de Weyl~: $\rho\equiv\frac{1}{2}%
\sum_{\alpha>0}\alpha=\sum_i\lambda_i$.
En particulier, la racine la plus grande $\phi$ est le plus
haut poids de la repr\'esentation adjointe.
On constate que le Casimir $C_\phi$ dans la repr\'esentation adjointe 
est $\frac{1}{2}\,\tr\,\phi^2+g^\vee-1$.
D'autre part, en utilisant la base orthogonale, on trouve $C_\phi=
I_\phi/2$. Au total,
\[
I_\phi=2g^\vee+\tr\,\phi^2-2.
\]
En convenant, comme on le fera toujours, que $\tr\,\phi^2=2$ on voit
que $I_\phi=2g^\vee$ et que le nombre de Coxeter dual est aussi 
le Casimir quadratique de la repr\'esentation adjointe.

Soit $R$ une repr\'esentation de $G$ de dimension 
finie. Le caract\`ere de $G$ est une fonction de classe.
Or, tout \'el\'ement de $G$ est conjugu\'e \`a un \'el\'ement 
du tore maximal $T$, donc
le caract\`ere est compl\`etement d\'etermin\'e par ses valeurs sur $T$. 
On peut
poser
\[
\chi\sous{R}(u)=\tr_{_R}\,\ee^{iu}
\]
pour $u\in\liet$ et $\tr_{_R}$ repr\'esente la trace dans 
la repr\'esentation $R$. Le caract\`ere dans une repr\'esentation 
irr\'eductible unitaire de plus haut poids
$\lambda$ est donn\'e par
\[
\chi_\lambda(u)=\sum_\mu\dim M_\lambda(\mu)\,\ee^{i\mu(u)},
\]
ou plus explicitement par la {\bol formule des caract\`eres de Weyl}~:
\[
\chi_\lambda(u)=\Pi(u)^{-1}\sum_{w\in W}(-1)^{\ell(w)}
	\ee^{iw\,(\lambda+\rho)(u)}
\]
o\`u $\ell(w)$ est le nombre de racines positives transform\'ees par $w$
en racines n\'egatives et $\Pi(u)$ est le {\bol d\'enominateur de Weyl}
\[
\Pi(u)=\sum_{w\in W}(-1)^{\ell(w)}\ee^{i w\,\rho(u)}
	=\ee^{i\rho(u)}\prod_{\alpha>0}(1-\ee^{-i\alpha(u)}).
\]
On obtient la {\bol formule de la dimension de Weyl} donnant la dimension de
$M_\lambda$ en \'evaluant $\chi_\lambda$ en $0$,
\[
d_\lambda=\prod_{\alpha>0}\frac{\tr\,(\lambda+\rho)\alpha}{\tr\,\rho\alpha}.
\]

\medskip
\section{Mouvement g\'eod\'esique d'une particule sur un groupe}

Le {\bol mouvement g\'eod\'esique} d'une particule sur une surface riemannienne
$M$ munie d'une m\'etrique $\gamma=\gamma_{\mu\nu}dx^\mu dx^\nu$ est
habituellement gouvern\'e par l'action donn\'ee par la
longueur g\'eod\'esique~:
\[
S_{l.g.}(x(\cdot))=\int ds=\int_0^T \Big(\gamma_{\mu\nu}(x)
\frac{dx^\mu}{d\tau}\frac{dx^\nu}{d\tau}\Big)^{1/2} d\tau
\]
o\`u $x:[0,T]\ni\tau\rightarrow x(\tau)\in M$ est la trajectoire
de la particule. L'\'equation $\delta S_{l.g.}(x(\cdot))=0$
d'une g\'eod\'esique est
\[
\frac{d^2x^\mu}{d\tau^2}+\Gamma^{\mu}_{\nu\lambda}(x)\,
	\frac{dx^\nu}{d\tau}\,\frac{dx^\lambda}{d\tau}=0
\]
o\`u $\Gamma$ est le symbole de Christoffel. La longueur
g\'eod\'esique est
invariante par reparam\'etrisation --- c'est un objet purement 
g\'eom\'etrique.  La pr\'esence de la racine carr\'ee pr\'efigure
des difficult\'es techniques. C'est pourquoi
on pr\'ef\`ere utiliser une autre action classiquement \'equivalente~:
\qq
\label{geodesique}
S(x(\cdot))={_1\over^2}\,\int_0^T\,\gamma_{\mu\nu}(x)\,
\frac{dx^\mu}{d\tau}\frac{dx^\nu}{d\tau}\,d\tau.
\qqq
La nouvelle action n'est plus invariante par reparam\'etrisation. 
L'{\bol action de Polyakov} 
\[
S_{{\rm P}}(x(\cdot),h(\cdot))=\quotient{1}{2}\int\gamma_{\mu\nu}(x)\,
\frac{dx^\mu}{d\tau}\frac{dx^\nu}{d\tau}\,h^{-1/2}d\tau
	+\quotient{1}{2}\int h^{1/2}d\tau
\]
permet de retrouver la longueur g\'eod\'esique en minimisant $h(\cdot)$. 
Quand on fixe la
\<<jauge\>> \`a $h\equiv 1$, on trouve l'action~\eqref{geodesique} 
(\`a une constante pr\`es) mais on brise
l'invariance par reparam\'etrisation.

On s'int\'eresse au mouvement d'une particule sur un groupe
de Lie $G$ compact, simple. La forme de Killing fournit une m\'etrique 
riemannienne sur $G$. 
La m\'ecanique de la particule $\tau\mapsto g(\tau)$
est gouvern\'ee par l'action
\[
S(g(.))=-{_k\over^4}\int\tr\,(g^{-1}\partial_\tau g)^2\,d\tau
\]
o\`u $k$ est une constante de couplage strictement positive.
L'action se transforme simplement dans la multiplication point par point
\[
S(g_1g_2)=S(g_1)+S(g_2)+{_k\over^2}\int\tr\,(g^{-1}_1\partial_\tau g_1)\,(g_2\partial_\tau
g^{-1}_2)\,d\tau
\]
La variation de l'action suivant la transformation $g\mapsto g+\delta g$,
fix\'ee sur le bord, est
\[
\delta S(g)={_k\over^2}\int\tr\left((g^{-1}\delta g)\,\partial_\tau%
	(g^{-1}\partial_\tau g)\right)\,d\tau.
\]
Par cons\'equent, l'\'equation classique du mouvement $\delta S(g)=0$
conduit \`a
\qq
\label{Class}
\partial_\tau(g^{-1}\partial_\tau g)=0.
\qqq
L'\'equation~(\ref{Class}) est invariante par les 
transformations
\[
g\rightarrow g_1g\,g_2^{-1}
\]
o\`u $g_1$ et $g_2$ 
sont deux \'el\'ements quelconques fix\'es de $G$.   
Comme tout \'el\'ement de $G$ est conjugu\'e \`a un \'el\'ement du
sous-groupe de Cartan, on peut utiliser cette invariance chirale de sorte 
que l'espace des solutions soit param\'etris\'e par
\qq
\label{Para}
g(\tau)=g\sous{\CL}\ee^{i\tau v}\,g^{-1}\sous{\CR}
\qqq
o\`u $(g\sous{\CL},g\sous{\CR})\in G\times G$ et $v\in\liet$. 

L'ensemble des configurations instantan\'ees du syst\`eme 
forme l'espace des configurations $G$. 
L'espace cin\'ematique est le fibr\'e tangent \`a $G$, not\'e $TG$.
On peut identifier $TG$ \`a $G\times\lieg$, ainsi on
param\'etrise $TG$ par les donn\'ees de Cauchy
$(g(0),\,\frac{1}{i}\,(g^{-1}\partial_\tau g)(0))$,
soit $(g\sous{\CL}\, g\sous{\CR}^{-1},\,g\sous{\CR}
\,v\,g\sous{\CR}^{-1})$.
L'espace des phases est le fibr\'e cotangent \`a $G$, not\'e $T^*G$.
Il peut \^etre aussi identifi\'e \`a $G\times\lieg$ 
o\`u l'action d'un covecteur sur un vecteur tangent devient
\[
\label{Covecteur}
\left\langle\, (g,X)\,|\,(g,Y)\,\right\rangle \equiv 
\langle\, X,Y\,\rangle 
=\tr\,XY.
\]
Soit $g^{-1}dg$ la $1$-forme sur $G$ \`a valeurs dans $i\lieg$
d\'efinie par
\[
	\langle\,(g,X)\,|\,g^{-1}dg\,\rangle=i\,X.
\]
Le covecteur $(g,Y)$ co\"\i ncide alors avec $\frac{1}{i}\,\tr\,Y\,g^{-1}dg\,(g)$.
Le passage de $TG$ \`a $T^*G$ se fait par la transformation de Legendre
$\Lambda$, \cad l'application $\Lambda : TG\ni(g,X)\longmapsto 
(g,\Lambda_g(X))\in T^*G$, avec
\[
\langle\, W,\Lambda_g(X)\,\rangle\equiv \left.\quotient{d}{d\epsilon}\right 
|_{\epsilon=0}L(g,X+\epsilon W),
\]
$L$ \'etant le Lagrangien du syst\`eme~: $L(g,X)=\frac{k}{4}\,\tr\,X^2$.
On obtient $\langle\, W ,\Lambda_g(X)\,\rangle=\frac{k}{2}\,\tr\,XW$, d'o\`u
\[
\Lambda_g(X)={_k\over^2}\,X.
\]
Le Hamiltonien du syst\`eme est d\'efini par $H(g,\Lambda_g(X))
=-L(g,X)+\tr\, X\Lambda_g(X)$ d'o\`u
\[
H(g,Y)\,=\,{_1\over^k}\,\tr\, Y^2\,.
\]

L'espace des phases est un espace symplectique, \cad une vari\'et\'e 
\'equip\'ee d'une $2$-forme ferm\'ee non
d\'eg\'en\'er\'ee. Consid\'erons la $1$-forme sur $T^*G$ 
\[
	\omega(g,Y)={_1\over^i}\,\tr\,(Y\,g^{-1}dg)\circ
	T_{_{(g,Y)}}\pi
\]
o\`u $T\pi$ est l'application tangente \`a la projection canonique
$\pi:T^*G\rightarrow G$ --- omise dans la suite.
La $2$-forme $\Omega=d\omega$ est la forme symplectique canonique sur $T^*G$.

Nous pouvons param\'etriser l'espace de phase $T^*G$ par 
la transformation de Legendre des donn\'ees de Cauchy :
\[
(g,Y)\,=\,(g\sous{\CL}g^{-1}\sous{\CR},\,{_k\over^2}\,g\sous{\CR}\,
v\,g^{-1}\sous{\CR})\,.
\]
Avec cette param\'etrisation
\[
	\omega\,=\,\frac{_k}{^{2i}}\,\tr\,v\left(g\sous{\CL}^{-1}
	dg\sous{\CL}-g\sous{\CR}^{-1}
	dg\sous{\CR}\right).
\]
La forme symplectique $\Omega$ se scinde alors en deux composantes 
\[
	\Omega=d\omega=\Omega\sous{\CL}-\Omega\sous{\CR}
\]
o\`u
\[
\Omega\sous{\CL}=\,\frac{_{ki}}{^{2}}\,
	\tr\left(v\,(g\sous{\CL}^{-1}dg\sous{\CL})^{\wedge 2}
	-dv\wedge g^{-1}\sous{\CL}dg\sous{\CL}\right)
\]
et la m\^eme expression pour $\Omega\sous{\CR}$
mais en rempla\c{c}ant $g\sous{\CL}$ par $g\sous{\CR}$.
Le Hamiltonien devient
\[
H(g\sous{\CL},\, g\sous{\CR},\, v)\ =\ {_k\over^4}\,\tr\, v^2\,.
\]

\medskip
\section{Quantification}

Une quantification naturelle consiste \`a prendre,
comme {\bol espace des \'etats quantiques}, l'espace
de Hilbert des fonctions de carr\'e sommable $\NH=L^2(G,dg)$ --- pour
une vari\'et\'e riemannienne quelconque, on aurait pris
$\NH=L^2(M,\Omega_\gamma)$, $\Omega_\gamma$ \'etant le volume
riemannien. On normalise la mesure de Haar par $\int_G dg=1$.
D'apr\`es le th\'eor\`eme de Peter-Weyl,
\[
\NH\cong\mathop{\bigp}\limits_{\alpha} 
	\overline{V}_{\hspace{-0.06cm}\lambda_\alpha}\otimes 
	V_{\lambda_\alpha}
\]
o\`u les $R_{\alpha}$ forment un ensemble
maximal de repr\'esentations irr\'eductibles, unitaires de $G$ 
agissant dans les espaces vectoriels (de dimension finie) 
$V_{\lambda_\alpha}$, $\lambda_\alpha$ \'etant le plus haut poids 
de la repr\'esentation $R_{\alpha}$.

Soit $(t^a)$, comme auparavant, une base orthogonale
de l'alg\`ebre de Lie $\lieg$ de $G$ telle que
$\tr\, t^at^b=\delta^{ab}/2$ et $[t^a,t^b]=if^{abc}\,t^c$. 
Pour d\'ecrire l'action infinit\'esimale de l'alg\`ebre, on introduit
deux op\'erateurs $J^a\equiv\frac{1}{i}\,d\CL(t^a)$
et $\overline{J}^a\equiv\frac{1}{i}\,d\CR(t^a)$, \cad
\[
(J^af)(g)={_1\over^i}\,\left.\quotient{d}{d\epsilon}\right|_{\epsilon=0}%
	f(\ee^{-i\epsilon\,t^a}\,g),\qquad
(\overline{J}^af)(g)={_1\over^i}\,\left.\quotient{d}{d\epsilon}
	\right|_{\epsilon=0} f(g\,\ee^{i\epsilon\,t^a}).
\]
On v\'erifie que ces op\'erateurs satisfont les relations 
de commutation suivantes 
\[
[J^a,J^b]=if^{abc}\,J^c,
\qquad[\overline{J}^a,\overline{J}^b]=if^{abc}\,\overline{J}^c.
\]

Pour d\'efinir le Hamiltonien du syst\`eme, on introduit
l'op\'erateur de Laplace-Beltrami. Si $M$ est
une vari\'et\'e riemannienne, on pose
\[
(\Psi,-\Delta_{{\rm LB}}\Phi)
	=\int \gamma^{\mu\nu}\partial_\mu \Psi\,\partial_\nu \Phi\  
	\Omega_\gamma
\]
o\`u $(.,.)$ est le produit scalaire $L^2$.
Pour $M=G$, l'op\'erateur de Laplace-Beltrami, not\'e $\Delta_G$,
est un op\'erateur sym\'etrique, auto-adjoint, positif.
Restreint \`a $\overline{V}_{\hspace{-0.06cm}\lambda_\alpha}
\otimes V_{\lambda_\alpha}$, il agit comme 
$-C_{\lambda_\alpha}\,{\bf 1}$, $C_{\lambda_\alpha}$ \'etant le Casimir 
quadratique de la 
repr\'esentation $R_{\alpha}$. Le {\bol Hamiltonien} 
est alors $\CH=-\frac{1}{2}\,\Delta_{{\rm LB}}$,
\cad
\[
\CH={_2\over^k}\,J^a\,J^a={_2\over^k}\,\overline{J}^a\,
	\overline{J}^a=-{_2\over^k}\,\Delta_G.
\]

Comme toujours, on peut aussi utiliser une quantification \`a la Feynman.
Il est bien connu qu'en m\'ecanique quantique les deux approches
sont \'equivalentes. Consid\'erons le noyau de l'op\'erateur 
$\ee^{-\tau\,\CH}$, \ie
\[
\big(\ee^{-\tau\,\CH}\Psi\big)(g_1)
	=\int \big(\ee^{-\tau\,\CH}\big)(g_1,g_2)\,\Psi(g_2)\,Dg_2
\]
pour $\Psi\in\CH$.
On peut exprimer ce noyau par l'int\'egrale fonctionnelle
sur tous les chemins $\tau\,\rightarrow\, g(\tau)$
fix\'es sur le bord~:
\[
\big(\ee^{-(t'-t)\,\CH}\big)(g_1,g_2)=
	\mathop{\int}\limits_{\scriptstyle
			 g(t)=g_1\atop\scriptstyle
			 g(t')=g_2}
\ee^{-S(g)}\,Dg
\]
o\`u la mesure $Dg$ est le produit formel
des mesures de Haar $\prod_{0\leq\tau\leq T}dg(\tau)$ sur le groupe $G$
et l'int\'egration sur le temps dans $S(g)$ est limit\'ee \`a
l'intervalle $[t,t']$.

Une cons\'equence de cette repr\'esentation est
la {\bol formule de Feynman-Kac} (version euclidienne)~\cite{glim} :
 pour $0\leq t_1\leq\dots
\leq t_N\leq T$,
\begin{multline*}
\int \overline{\Psi_2(g(T))}\,
	\mathop{\otimes}\limits_\ell g(t_\ell)\sous{R_\ell}\,
	\Psi_1(g(0))\,\ee^{-S(g)}\,Dg\\
=\big\langle \Psi_2\,\big|\,
	 \ee^{-(T-t_N)\,\CH}\,g\sous{R_N}\,%
	\ee^{-(t_N-t_{N-1})\,\CH}\,\ldots\,\ee^{-(t_2-t_1)\,\CH}\,%
	g\sous{R_1}\,\ee^{-t_1\,\CH}\,
	\Psi_1\big\rangle
\end{multline*}
o\`u les deux c\^{o}t\'es appartiennent \`a $\bigotimes_\ell
{\rm End}\, V_{\lambda_\ell}\cong\bigotimes_\ell(
\bar V_{\lambda_\ell}\otimes V_{\lambda_\ell})$. A droite,
$g\sous{R_\ell}$ doit \^{e}tre interpr\'et\'{e} comme la matrice
des op\'erateurs (born\'es) de multiplication dans $\NH$ par 
les \'el\'ements matriciels de $g\sous{R_\ell}$. L'int\'egrale
sur les chemins \`a droite peut \^{e}tre d\'efinie rigoureusement
comme celle d'un {\bol mouvement brownien} sur $G$.
En int\'egrant sur les chemins p\'eriodiques $[0,T]\ni\tau\,
\rightarrow\, G$, $g(0)=g(T)$, on obtient une autre version
de la formule de Feynman-Kac~:
\begin{align*}
\Gamma\,\equiv\,\int\limits_{g\ \mbox{\tiny p\'eriodiques}}&
\mathop{\otimes}\limits_\ell g(t_\ell)\sous{R_{\ell}}
\,\,{\rm e}^{-S(g)}\,\, Dg\\
&=
\tr_{_\NH}\,\big( \ee^{-(T-t_N)\,\CH}\,g\sous{R_N}\,%
\ee^{-(t_N-t_{N-1})\,\CH}\,\ldots\,\ee^{-(t_2-t_1)\,\CH}\,%
g\sous{R_1}\,\ee^{-t_1\,\CH}\big)
\end{align*}
o\`u la trace $\tr_{_\NH}$ est prise sur $\NH$.
Nous allons appeler $\Gamma$ la {\bol fonction de Green}.
Bien que sous-entendues, les fonctions de Green d\'ependent
de $\underline{t}$ et de $\underline{R}$.

\medskip
\section{M\'ecanique quantique quotient}

La construction quotient --- en anglais \<<coset construction\>> --- 
prend ses racines dans la th\'eorie conforme des champs. 
On peut faire agir sur l'espace $\NH=L^2(G,dg)$ le groupe $G\times G
\equiv G_{\CL}\times G_{\CR}$, produit des actions \`a gauche et \`a
droite. Soient $H$ un sous-groupe de $G\times G$ et
$L^2(G,dg)_H$ le sous-espace de $\NH$ form\'e des fonctions 
invariantes par $H$. Si $H$ est un sous-groupe de $G_{\CL}$ ou 
de $G_{\CR}$ ou un produit de tels sous-groupes, on obtient 
des fonctions invariantes sur les espaces quotients correspondants. 
Ici, on choisit un sous-groupe dans la partie diagonale $G_{{\rm diag}}
\subset G_\CL\times G_\CR$, \cad $H\subset G_
{{\rm diag}}$; on obtient alors des 
fonctions invariantes par l'action adjointe de $H$, \cad des 
fonctions d\'efinies sur $G/\Ad(H)$.
On peut d\'ecomposer les repr\'esentations irr\'eductibles
de $G$ par rapport \`a $H$~:
\[
V_{{\lambda_G}}\cong \bigoplus_{{\lambda_H}} M^{\lambda_G}
	_{\lambda_H}\otimes V_{{\lambda_H}}
\]
o\`u $\lambda_H$ fait r\'ef\'erence au plus haut poids d'une repr\'esentation
irr\'eductible de $H$. Les espaces interm\'ediaires $M^{\lambda_G}_{\lambda_H}$
sont canoniquement isomorphes \`a
${\rm Hom}(V_{{\lambda_H}},V_{{\lambda_G}})^H$. Il suit
\[
L^2(G,dg)\cong \bigoplus_{\lambda_G,\lambda_H,\lambda'_H} 
	\overline{M}^{\lambda_G}_{\lambda_H}\otimes\overline{V}_{{\lambda_H}}
	\otimes M^{\lambda_G}_{\lambda'_H}\otimes
	V_{\lambda'_H}.
\]
Or, d'apr\`es le lemme de Schur, l'espace des vecteurs de $V_{{\lambda_H}}\otimes
\overline{V}_{\lambda'_H}$ invariants par l'action diagonale de $H$ est
simplement $\delta_{\lambda_H,\lambda'_H}\,\C$, donc 
\[
L^2(G,dg)_H\cong \bigoplus_{\lambda_G,\lambda_H} 
\overline{M}^{\lambda_G}_{\lambda_H}
	\otimes M^{\lambda_G}_{\lambda_H}.
\]
Soit $\lieh^\perp$ le compl\'ement orthogonal de l'alg\`ebre de Lie $\lieh$ de $H$. 
On d\'ecompose la base $(t^a)$ en une base $(t^{\rm a})$ 
de $\lieh$ et une base $(t^\alpha)$ de $\lieh^\perp$. 
Les Laplaciens $\Delta_G$, $\Delta_H=
-J^{\rm a}J^{\rm a}$ et $\overline{\Delta}_H=-\overline{J}^{\rm a}
\overline{J}^{\rm a}$ 
commutent avec l'action (infinit\'esimale) \`a gauche et 
\`a droite de $\lieh$. On peut donc prendre 
\[
-\quotient{2}{k}\,(\Delta_G-\Delta_H)
	=-\quotient{2}{k}\,(\Delta_G-\overline{\Delta}_H)\ \equiv\ \CH
\]
comme Hamiltonien sur $L^2(G,dg)_H$. Notons que le Hamiltonien trouv\'e
n'est pas l'op\'erateur de Laplace-Beltrami pour la m\'etrique
projet\'ee sur $G/\Ad(H)$. 
Les fonctions de Green de la m\'ecanique quantique correspondante,
qu'on va appeler pour des raisons historiques \<<$G/H$\>>, 
peuvent \^{e}tre d\'efinies comme
\[
\Gamma_{G/H}\,\equiv\,
\,\Tr\,\big( \ee^{-(T-t_N)\,\CH}\, f_N\,%
\ee^{-(t_N-t_{N-1})\,\CH}\,\ldots\,\ee^{-(t_2-t_1)\,\CH}\,%
f_1\,\ee^{-t_1\,\CH}\big)
\]
o\`u $f_\ell\in L^2(G,dg)_H$. Elles sont \`a valeurs num\'eriques.
Dans le cas particulier o\`u $H=G$, l'espace $L^2(G,dg)_G$ est compos\'e
de fonctions de classes et les caract\`eres
des repr\'esentations irr\'eductible forment sa base orthonormale.
Dans ce cas, le Hamiltonien $\CH$ s'annule et les fonctions
de Green $\Gamma_{G/G}$ divergent.

On peut r\'eobtenir le syst\`eme quotient \<<$G/H$\>>
en couplant le mouvement 
g\'eod\'esique sur le groupe $G$ \`a un champ de jauge $A_\pm$ 
\`a valeurs dans $\liehc$. L'action coupl\'ee est
\[
S(g,A_+,A_-)\equiv S(g)+{_k\over^2}\int \tr\,\left( A_+g^{-1}\partial_\tau g
+g\partial_\tau g^{-1}A_-+gA_+g^{-1}A_--A_+A_-\right)d\tau
\]
o\`u, plus g\'en\'eralement, on peut prendre $g$ \`a valeurs dans $\Gc$.
La nouvelle action est laiss\'ee invariante par la transformation de jauge 
\[
^hg  \equiv  hgh^{-1}, \qquad
^hA_\pm  \equiv  hA_\pm h^{-1}+h\partial_\tau h^{-1}
\]
o\`u $\tau\mapsto h(\tau)$ est \`a valeurs dans $H$.
Si on d\'ecouple les actions \`a gauche et \`a droite, on trouve
\qq
S(h_1gh_2,\hA{h_2}_+,\hA{h_1}_-) & = & S(g,A_+,A_-)-S(h_1^{-1}h_2,A_+,A_-),
\non\\
 & = & S(g,A_+,A_-)+S(h_1h_2^{-1},^{h_2}\!\!A_+,^{h_1}\!\!A_-).\non
\qqq
Les fonctions de Green $\Gamma(A_+,A_-)$ obtenues en rempla\c{c}ant 
l'action $S$ par 
l'action coupl\'ee changent dans une transformation de jauge comme
\[
\Gamma(\hA{h_2\,}_+,\hA{h_1}_-)=\ee^{S(h^{-1}_1h_2,A_+,A_-)}\,
\mathop{\otimes}\limits_\ell h_{1}(\tau_\ell)\sous{R_\ell}\, \Gamma(A_+,A_-)
	\,\mathop{\otimes}
\limits_\ell h_{2}(\tau_\ell)\sous{R_\ell}^{-1}.
\]
Afin de d\'ecoupler les actions de gauche et de droite, 
on peut changer un tout petit
peu l'action en posant $\widetilde{S}(g,A_+,A_-)=S(g,A_+,A_-)
+\frac{k}{2}\int\tr\,A_+ A_-d\tau$. 
Les fonctions de Green $\widetilde{\Gamma}$ qui 
en r\'esultent changent donc comme~(\footnote{
Pour l'action modifi\'ee,
\qq
\widetilde{S}(h_1gh_2,\hA{h_2}_+,\hA{h_1}_-) & = &
	\widetilde{S}(g,A_+,A_-)-\widetilde{S}(h_1^{-1},A_-)
	-\widetilde{S}(h_2,A_+),\non\\
 & = & \widetilde{S}(g,A_+,A_-)+
\widetilde{S}(h_1h_2^{-1},^{h_2}\!\!A_+,^{h_1}\!\!A_-).\non
\qqq
})
\[
\widetilde{\Gamma}(\hA{h_2\,}_+,\hA{h_1}_-)=\ee^{S(h^{-1}_1,A_-)
	+S(h_2,A_+)}\,
\mathop{\otimes}\limits_\ell h_{1}(\tau_\ell)\sous{R_\ell}\, 
	\widetilde{\Gamma}(A_+,A_-)
	\,\mathop{\otimes}
\limits_\ell h_{2}(\tau_\ell)\sous{R_\ell}^{-1}.
\]
Pour les fonctions de Green normalis\'ees
$\langle\,\otimes_\ell g(\tau_\ell)\sous{R_\ell}\,\rangle
_{A_+,A_-}\equiv Z(A_+,A_-)^{-1}\,\Gamma(A_+,A_-)$,
o\`u $Z$ est la fonction de partition pour l'action jaug\'ee
$Z(A_+,A_-)=\int \ee^{-S(g,A_+,A_-)}Dg$,
on obtient :
\[
\langle\,\mathop{\otimes}\limits_\ell g(\tau_\ell)\sous{R_\ell}
	\,\rangle _{_{A_+,A_-}}
=\mathop{\otimes}\limits_\ell h_{1}(\tau_\ell)\sous{R_\ell}\,
\langle\,\mathop{\otimes}\limits_\ell g(\tau_\ell)\sous{R_\ell}
	\,\rangle _{\hA{h_2\ }_+,\hA{h_1\ }_-}\,\mathop{\otimes}
\limits_\ell h_{2}(\tau_\ell)\sous{R_\ell}^{-1}.
\]
Que l'on prenne $S$ ou $\widetilde{S}$, cela ne change rien \`a l'affaire.

Les fonctions de Green $\Gamma_{G/H}$ du mod\`ele \<<$G/H$\>> sont obtenues 
en int\'egrant sur des chemins p\'eriodiques $g(\tau)$
et sur les champs de jauge $A_-=-A_+^\dagger$~:
\[
\Gamma_{G/H}\ =\ \int\prod\limits_\ell f_\ell(g)
\,\,{\rm e}^{-S(g,\, A_+,A_-)}\,\, Dg\,\, DA_+\, DA_-\,.
\]
Notons que l'int\'egrale sur $g$ peut \^{e}tre exprim\'ee
par la fonction de Green $\Gamma(A_+,A_-)$.
Les champs de jauge apparaissent quadratiquement dans l'action
et on peut d'abord int\'egrer sur eux comme
l'int\'egration est gaussienne. On reste alors avec l'int\'egrale
sur $g$ avec l'action effective
\[
S_{\rm eff}(g)=-\quotient{k}{4}
	\int \tr\,\big(
	(g^{-1}\partial_\tau g)^2
	+2\,(g^{-1}\partial_\tau g)\,
	(1-E\,\Ad_gE)^{-1}E\,\Ad_g(g^{-1}\partial_\tau g)
	\big)\,d\tau
\]
o\`u $E$ est le projecteur orthogonal de $\lieg$ sur $\lieh$.
Comme pour le mouvement g\'eod\'esique sur $G$ quantifi\'e,
les fonctions de Green de la m\'ecanique quantique \<<$G/H$\>>
peuvent \^{e}tre calcul\'ees par l'analyse harmonique sur $G$
(et sur $H$).



\chapter{Mod\`ele de Wess-Zumino-Novikov-Witten}

\medskip
\section{Action de WZNW}

\subsection{Mod\`ele sigma bidimensionnel}

Au cours du pr\'ec\'edent chapitre, on s'est int\'eress\'e 
au mouvement g\'eod\'esique d'une parti\-cule sur un groupe compact $G$. 
Maintenant, on \'etudie une action analogue d\'efinie pour des applications
$\Sigma\ni\xi\mapsto g(\xi)\in G$, $\Sigma$ \'etant une 
vari\'et\'e r\'eelle $\CC^\infty$, bidimensionnelle, compacte,
munie d'une orientation et d'une m\'etrique riemannienne 
$\gamma=\gamma_{\mu\nu}\,dx^\mu\,dx^\nu$.
On obtient (la version euclidienne d') un {\bol mod\`ele sigma 
bidimensionnel} d'espace cible
un groupe compact $G$ d\'efini par l'action
\[
k\,\Ssigma(g)=-{_{k}\over^{8\pi}}\,\int_\Sigma\tr\,\gamma(g^{-1}dg,g^{-1}dg)\,
\Omega_{\gamma}
\]
o\`u $\Omega_{\gamma}$ est le volume riemannien et $k$
est une constante de couplage positive. Si on se donne 
un syst\`eme de coordonn\'ees locales $(x,y)$ compatible avec
l'orientation donn\'ee, l'action devient
\begin{equation}\label{sigmaun}
k\,\Ssigma(g)=-{_{k}\over^{8\pi}}\,\int_\Sigma\tr\,\gamma^{\mu\nu}\,%
	g^{-1}\partial_\mu g\,g^{-1}\partial_\nu g\,\sqrt{\gamma}%
	\,d^2\sigma
\end{equation}
o\`u $d^{2}\sigma$ est la mesure associ\'ee \`a la forme $dx\wedge dy$,
$\partial_\mu=\partial/\partial x^\mu$,
$(\gamma^{\mu\nu})$ est l'inverse de la matrice $(\gamma_{\mu\nu})$
et $\sqrt{\smash[b]{\gamma}}=\sqrt{\smash[b]{{\rm det}\,(\gamma_{\mu\nu})}}$.

Deux m\'etriques riemanniennes $\gamma$
et $\gamma'$ sont conform\'ement \'equivalentes s'il existe une 
fonction $\sigma\in\CC^\infty(\Sigma,\R)$, telle que 
$\gamma=\ee^\sigma\gamma'$, \ie si
$\gamma$ et $\gamma'$ sont dans la m\^eme orbite
des transformations de Weyl $\gamma\mapsto\ee^\sigma\gamma$.
Produire une {\bol classe
conforme} de m\'etriques est \'equiva\-lent \`a la donn\'ee
d'une {\bol structure complexe} sur $\Sigma$, \cad
d'une classe d'\'equivalence d'atlas analytiques. La vari\'et\'e $\Sigma$
munie d'une structure complexe devient une surface de Riemann.
Les coordonn\'ees locales r\'eelles $(x,y)$, induites par les coordonn\'ees
complexes $z=x+iy$ de la structure complexe correspondante \`a la classe 
conforme de $\gamma$, sont des coordonn\'ees 
isothermes pour $\gamma$, soit $\gamma=\rho\,(dx^2+dy^2)$ o\`u 
$\rho$ est une fonction $\CC^\infty$ \`a valeurs r\'eelles sur
la carte correspondante. Les formes $dx\wedge dy$ d\'efinissent l'orientation 
de $\Sigma$. On dit que la structure complexe sur $\Sigma$
correspondante \`a une classe conforme de m\'etriques est compatible
avec les m\'etriques de la classe.

Une {\bol structure presque complexe} sur une vari\'et\'e bidimensionnelle 
r\'eelle $\Sigma$ est la donn\'ee d'un automorphisme $\J$
du fibr\'e tangent $T\Sigma$ tel que $\J^2=-1$, o\`u $1$ est l'identit\'e
de $T\Sigma$. L'application $\J$ s'\'etend en un automorphisme 
du fibr\'e complexifi\'e $T^\C\Sigma=T\Sigma\otimes\C$~ (not\'e 
encore par $\J$). Comme tel, $\J$ a deux valeurs propres 
$\pm i$. On d\'ecompose $T^\C\Sigma$ en $T^{1,0}\Sigma\oplus T^{0,1}\Sigma$
avec \label{decompo}
\[
T^{1,0}\Sigma=\{\pi_{1,0}X,\ X\in T^\C\Sigma\},\quad
T^{0,1}\Sigma=\{\pi_{0,1}X,\ X\in T^\C\Sigma\}
\]
o\`u $\pi_{1,0}=\frac{1}{2}\,(1+i\J),\ \pi_{0,1}=
\frac{1}{2}\,(1-i\J)$ sont les projecteurs sur les valeurs
propres de $\J$. De m\^eme, 
$T^{\ast}\eC\Sigma=T^{\ast\,1,0}\Sigma\oplus T^{\ast\,0,1}\Sigma$
o\`u $T^{\ast\,1,0}\Sigma=\pi_{1,0}^*\,T^{\ast\,\C}\Sigma$ 
et $T^{\ast\,0,1}\Sigma=\pi_{0,1}^*\,T^{\ast\,\C}\Sigma$, 
pour les projecteurs transpos\'es $\pi_{1,0}^*$ et $\pi_{0,1}^*$.
Les deux op\'erateurs 
$\partial=\pi^\ast_{1,0}\circ d$ et $\de=\pi^\ast_{0,1}\circ d$
satisfont $d=\partial+\de$.
On dit qu'une structure presque complexe est 
int\'egrable si le commutateur de champs vectoriels
de type $(1,0)$ (\cad \`a valeurs dans $T^{1,0}\Sigma$)
est aussi de type $(1,0)$. De fa\c{c}on g\'en\'erale, une structure complexe 
induit une structure presque complexe int\'egrable sur la vari\'et\'e
diff\'erentielle sous-jacente telle que
\[
\partial=dz\,{_\partial\over^{\partial z}}\quad {\rm et}\quad 
\de=d\overline{z}\,{_\partial\over^{\partial\overline{z}}}\ .
\]
Le {\bol th\'eor\`eme de Newlander et Nirenberg} montre 
que toutes les structures
presque complexes int\'egrables sont de ce type. La dimension $2$ est
sp\'eciale puisque toute structure presque complexe
est int\'egrable. On dit qu'une m\'etrique riemannienne $\gamma$ 
est compatible avec une structure presque complexe si pour tous 
vecteurs $X$ et $Y$, on a $\gamma(\J X,\J Y)=\gamma(X,Y)$. 
Cette notion co\"\i ncide avec celle d\'efinie plus haut
pour une structure complexe.
On vient donc de voir que les notions de structure complexe 
ou presque complexe
et de classe conforme de m\'etriques co\"\i ncident pour une vari\'et\'e 
diff\'erentielle $\CC^\infty$ bidimensionnelle orient\'ee. 
On note  $\overline{\Sigma}$ la vari\'et\'e
diff\'erentielle $\Sigma$ munie de la structure (presque) complexe
oppos\'ee $-\J$. La surface $\overline{\Sigma}$ h\'erite de l'orientation
oppos\'ee \`a celle de $\Sigma$.
Plus tard, on fera varier la structure complexe. Concr\`etement, 
on utilisera des variations infinit\'esimales 
$\delta\J$ de $\J$. On reliera alors $\delta\J$ \`a 
la diff\'erentielle de Beltrami, qui appara\^\i t 
naturellement dans le contexte des classes conformes de m\'etriques.

Maintenant, $\Sigma$ est une surface de Riemann 
compacte dont la structure complexe est compatible avec la m\'etrique 
riemannienne $\gamma$. La m\'etrique est de la forme 
\[
\gamma=2\,\gamma_{z\overline{z}}\,dz\,d\overline{z}.
\]
L'orientation est donn\'ee par $i\,dz\wedge d\overline{z}$. 
Le volume riemannien est $\Omega_{\gamma}=
i\,\gamma_{z\overline{z}}\,dz\wedge d\overline{z}$. 
La forme de courbure est la $2$-forme (imaginaire) 
$R_{\gamma}=\de\partial\ln\,\gamma_{z\overline{z}}$.
On a $i\,R_{\gamma}=K_{\gamma}\,\Omega_{\gamma}$
o\`u $K_{\gamma}$ est la courbure gaussienne ou courbure scalaire.
Le th\'eor\`eme de Gauss-Bonnet nous apprend que
\[
\int_\Sigma R_{\gamma}=4\pi i\,(g-1)=-2\pi i\,\chi(\Sigma)
\]
o\`u $g$ est le genre de la surface de $\Sigma$ et $\chi(\Sigma)=2-2g$
est la caract\'eristique d'Euler.

Avec l'\'equation~(\ref{sigmaun}), on a soulign\'e la d\'ependance
du mod\`ele vis-\`a-vis de la m\'etrique. 
Dor\'enavant, on pr\'ef\`ere rendre la structure complexe explicite
\[
k\,\Ssigma(g)=-{_{k}\over^{8\pi}}\,\int_\Sigma\tr\,g^{-1}dg\wedge \ast g^{-1}dg
\]
o\`u $\ast$ est l'op\'erateur de dualit\'e de Hodge qui,
en dimension deux, co\"\i ncide avec l'op\'erateur $\J$ transpos\'e. 
En fonction de $\partial$ et $\de$, l'action est
\[
k\,\Ssigma(g)=-{_{ik}\over^{4\pi}}\,\int_\Sigma\tr\,(g^{-1}%
\partial g)\wedge(g^{-1}\de g).
\]

\subsection{Action de Wess-Zumino}

En l'\'etat, il est difficile de quantifier ce mod\`ele.
N\'eanmoins, Witten~\cite{witten:wzw} 
a montr\'e qu'on peut modifier le mod\`ele en rajoutant un terme 
topologique \`a l'action pour obtenir un mod\`ele facilement 
quantifiable, tout en pr\'eservant ses sym\'etries classiques. 
Pr\'ecis\'ement, on rajoute l'{\bol action topologique de Wess-Zumino}~:
\begin{equation}\label{topo}
k\,\STOP(g)={_{ik}\over^{4\pi}}\,\int_\Sigma\tr\,(g^*\omega)
\end{equation}
o\`u $\omega$ est une $2$-forme telle que 
$d\omega=-\frac{1}{3}\,(g^{-1}dg)^{\wedge 3}\equiv\beta$. 
La $3$-forme $\beta$ sur $G$ est ferm\'ee mais elle n'est pas exacte, ainsi
$\beta\in H^3(G,\R)$, et $\omega$ n'existe que localement.
Si $B$ est une vari\'et\'e tridimensionnelle de bord
$\partial B=\Sigma$, on utilise le th\'eor\`eme de Stokes pour r\'e\'ecrire 
$\STOP$
\[
k\,\STOP(g)=-{_{ik}\over^{12\pi}}\,%
\int_B\tr\,(\widetilde{g}^{-1}d\widetilde{g})^{\wedge 3}
\]
o\`u $\widetilde{g}:B\rightarrow G$ est une extension de $g$ \`a $B$.
Bien entendu, cette extension peut ne pas exister (si le groupe
$G$ n'est pas simplement connexe) et a priori le r\'esultat
d\'epend du choix de $\widetilde{g}$. 

Pour simplifier, on va supposer que le groupe $G$ est connexe, 
simplement connexe et simple. En particulier, $\pi_1(G)$ est trivial
et $H_1(G,\Z)\cong\pi_1(G)\,/\,[\pi_1(G),\pi_1(G)]=0$. 
D'apr\`es l'isomorphisme de Hurewicz~\cite{rotman}, 
$H_2(G,\Z)\cong\pi_2(G)$. Or $\pi_2(G)=0$ d\`es que
$G$ est compact~\cite[p. 142]{pressley}, donc $H_2(G,\Z)=0$. Ainsi,
il est toujours possible de trouver une extension $\widetilde{g}$ de $g$.
 
Soient $\widetilde{g}':B'\rightarrow G$ et $\widetilde{g}'':B\rightarrow G$
deux extensions de $g$. Soit $B$ la $3$-vari\'et\'e obtenue
en collant, suivant leur bord $\Sigma$, $B'$ et $-B''$, 
la vari\'et\'e $B''$ avec l'orientation oppos\'ee. On a 
\[
-{_{ik}\over^{12\pi}}\,\int_{B'}\tr\,(\widetilde{g}'{}^{-1}d\widetilde{g}')^{\wedge 3}%
+{_{ik}\over^{12\pi}}\,\int_{B''}\tr\,(\widetilde{g}^{-1}d\widetilde{g})^{\wedge 3}=
-{_{ik}\over^{12\pi}}\,%
\int_B\tr\,(\widetilde{g}^{-1}d\widetilde{g})^{\wedge 3}
\]
o\`u $\widetilde{g}$ co\"\i ncide avec $\widetilde{g}'$~(resp.~$\widetilde{g}''$) 
sur $B'$~(resp.~$B''$). Comme $\partial B=\emptyset$, l'ambigu\"\i t\'e
dans la d\'efinition du terme topologique r\'eside dans les p\'eriodes
de la forme $\beta$.
On montre que $H_3(G,\Z)\cong\Z$ est engendr\'e par la classe d'homologie 
de $\widetilde{g}_{\phi}$~: le rel\`evement de l'homomorphisme
d'alg\`ebres de Lie $g_{\phi}:\sudeux\rightarrow\lieg$,
$\phi$ \'etant la racine la plus grande de $\liegc$,
\[
\sigma^\pm\mapsto e_{\pm\phi},\qquad\sigma^3\mapsto\phi^\vee
\]
en un homomorphisme de groupe $\widetilde{g}_\phi:\Su\rightarrow G$.
Il suit
\[
-{_{ik}\over^{12\pi}}\,%
\int_{\Su}\tr\,(\widetilde{g}_{\phi}^{-1}d\widetilde{g}_{\phi})^{\wedge 3}=%
-{_{ik}\over^{12\pi}}\,%
\int_{\Su}\tr\,(g^{-1}dg)^{\wedge 3}=%
-{_{ik}\over^{12\pi}}\,24\pi^2=-2\pi ik.
\]
Au passage, on montre que $-\frac{1}{8\pi^2}\,\beta$ est dans l'image
de l'application $H^3(G,\Z)\rightarrow H^3(G,\R)$.
Le terme topologique est donc d\'efini modulo $2\pi ik\Z$.
Les quantit\'es $\ee^{-k\,\STOP(g)}$, que nous appellerons 
{\bol amplitudes} ou, par analogie avec la physique 
statistique, {\bol facteurs de Boltzmann}, 
sont donc correctement d\'efinies d\`es que $k\in\N$.

Dans la suite, on sous-entend souvent le produit 
altern\'e $\wedge$. L'action compl\`ete 
\[
k\,S(g)=-{_{ik}\over^{4\pi}}\,\int_\Sigma\tr\,(g^{-1}\partial
g)\,(g^{-1}\de g)-{_{ik}\over^{12\pi}}\,%
\int_B\tr\,(\widetilde{g}^{-1}d\widetilde{g})^3\quad(\hbox{mod}\ 2\pi i)
\]
d\'efinit le {\bol mod\`ele de Wess-Zumino-Novikov-Witten}~(WZNW) de
niveau $k$, une d\'efinition correcte des facteurs de Boltzmann
imposant la \<<quantification\>> du niveau $k$~: $k\in\N$.
Il est aussi possible de construire une action de WZNW dans
des situations topologiques plus compliqu\'ees~\cite{gaw88:topo}
comme par exemple pour des groupes
non simplement connexes~\cite{gaw88:spectra,gaw88:spectrazero}.

\subsection{Sym\'etries classiques}

Dor\'enavant, on admet les champs prenant leurs valeurs dans le groupe
complexifi\'e $\Gc$. L'analyse effectu\'ee dans la section pr\'ec\'edente
reste valable. Afin de trouver les points stationnaires de l'action de WZNW,
on s'int\'eresse aux variations de l'action
dans la transformation $g\rightarrow g+\delta g$. Un calcul 
rapide conduit \`a
\begin{align*}
\delta\Ssigma(g) & =-{_{i}\over^{4\pi}}\,\int_\Sigma\tr\,(g^{-1}\delta g)\,%
(\de(g^{-1}\partial g)-\partial(g^{-1}\de g)),\\
\delta\STOP(g) & =-{_{i}\over^{4\pi}}\,\int_\Sigma\tr\,(g^{-1}\delta g)\,%
(g^{-1}dg)^2.
\end{align*}
Par suite,
\begin{equation}
\label{varS}
\delta S(g)={_{i}\over^{2\pi}}\,\int_\Sigma\tr\,(g^{-1}\delta g)\,%
\partial(g^{-1}\de g).
\end{equation}
L'\'equation classique (euclidienne) du mouvement satisfaite par 
les points stationnaires est donc 
\[
\partial(g^{-1}\de g)=0
\]
ou encore $\de(g\partial g^{-1})=0$. Localement, les solutions
se d\'ecomposent en un produit d'une fonction holomorphe et
d'une fonction antiholomorphe, toutes deux \`a valeurs dans $\Gc$.
Les solutions classiques sont invariantes par 
les transformations suivantes

---~{\bol sym\'etrie conforme},
$g\rightarrow g\circ f$, \ $g\rightarrow (g\circ \overline{f})^{-1}$,
o\`u $f$ est une transformation locale holomorphe de $\Sigma$~;

---~{\bol sym\'etrie chirale}, $g\rightarrow g_{1}gg_{2}^\dagger$,
o\`u $g_{1}$ et $g_{2}$ sont deux transformations locales
holomorphes de $\Sigma$ dans $\Gc$.

La propri\'et\'e fondamentale de l'action de WZNW est son comportement
dans la multiplication point par point des champs, donn\'e
par la {\bol formule de Polyakov-Wiegmann}~\cite{polyakov}~:
\begin{equation}
\label{PW}
S(g_{1}g_{2})=S(g_{1})+S(g_{2})-\Gamma\sous{\Sigma}(g_{1}%
,g_{2})\quad(\hbox{mod}\ 2\pi i)
\end{equation}
o\`u
\[
\Gamma\sous{\Sigma}(g_{1},g_{2})\,\equiv\,
{_{i}\over^{2\pi}}\,\int_\Sigma\tr\,(g_{1}^{-1}\de g_{1})%
\,(g_{2}\partial g_{2}^{-1}).
\]

\medskip
\section{Structure symplectique sur l'espace des phases}

Suivant un sch\'ema tr\`es g\'en\'eral~\cite{kijowski}~--- on peut
\'egalement consulter les r\'ef\'erences biblio\-graphiques
dans~\cite{gaw91:qg}~--- on explique comment obtenir un {\bol formalisme 
canonique} pour une th\'eorie des champs bidimensionnelle.
Notre th\'eorie est gouvern\'ee par une action du type
\[
S(\phi)=\int_\Sigma\CL(x^\mu,\phi^a,\partial_\nu \phi^a)\,d{\bf x}
\]
o\`u $\CL$ est la densit\'e de Lagrangien
d\'efinie sur l'espace $\bigphase$, 
l'ensemble des $(x^\mu,\phi^a,\xi^a_\nu)$. Ici, $x^\mu$
($\mu=0,1$) est un syst\`eme de coordonn\'ees locales sur 
l'espace-temps $\Sigma$. G\'eom\'etriquement, $\bigphase$
est l'espace des 1-jets de sections d'un fibr\'e $\CB$
sur $\Sigma$ de coordonn\'ees $(\phi^a)$ 
le long des fibres. Ainsi,  on peut voir $\bigphase$  comme un
fibr\'e sur $\CB$ ou sur $\Sigma$. Le  moment conjugu\'e
\`a $\xi_\nu$ est $\pi^{\nu a}={\partial\CL}/{\partial\xi^a_\nu}$.
Consid\'erons la $2$-forme suivante sur $\bigphase$ 
\[
\alpha=\CL\,dx^{0}\wedge dx^{1}+\pi^{0\,a}\,(d\phi^a-\xi^a_{0}\,dx^{0})
	\wedge dx^{1}+\pi^{1\,a}\,dx^{0}\wedge(d\phi^a-\xi^a_{1}\,dx^{1})
\]
et $\omega=d\alpha$, la $3$-forme ferm\'ee sur $\bigphase$. 
Un  \'etat classique $\Phi$ est une section de $\bigphase$
sur $\Sigma$ telle que $\Phi^*(i(X)\,\omega)=0$,  o\`u $i(X)$ est l'op\'erateur
de contraction avec $X$, pour tout champ de vecteur $\boldsymbol{X}$ tangent 
\`a $\bigphase$ et d\'efini sur l'image de $\Phi$. Cette \'egalit\'e code 
les \'equations d'Euler-Lagrange
pour $\phi^a(x)$ ainsi que les relations 
$\xi^a_\nu=\partial_{\nu}\phi^a$. L'espace des \'etats classiques,
not\'e $\phase$, correspond donc bien aux solutions du probl\`eme 
variationnel $\delta\int\CL$. Dans des cas int\'eressants,
on peut param\'etriser les solutions par des donn\'ees de Cauchy,
c.-\`a-d. par les restrictions de $\Phi$ \`a une hypersurface 
$\Sigma_0$ dans $\Sigma$. L'espace $\phase$ h\'erite alors d'une structure
de vari\'et\'e (de dimension infinie). Ensuite, l'espace tangent $T_\Phi \phase$
\`a $\phase$ est l'espace des champs de vecteurs
$X$ tangents \`a $\bigphase$, d\'efinis sur l'image de $\Phi$
et satisfaisant aux contraintes provenant de la lin\'earisation
des \'equations classiques autour de $\Phi$.
La $2$-forme $\Omega$ sur $\bigphase$, donn\'ee par la formule 
\begin{equation}
\label{Formesym}
\Omega_{\Phi}(X,X')\equiv\int_{\Sigma_0} {\Phi}^*(i(X\wedge X')\,\omega),
\end{equation}
est ferm\'ee. Cette forme ne d\'epend que de la classe d'homologie de
l'hypersurface $\Sigma_0$. Si $\Omega$ n'est pas d\'eg\'en\'er\'ee, 
elle d\'efinit une forme symplectique sur $\phase$, appel\'e alors
espace des phases de la th\'eorie~---~si $\Omega$ est d\'eg\'en\'er\'ee, 
on r\'eduit d'abord  par d\'eg\'en\'erescence l'espace des solutions classiques
pour obtenir l'espace des phases.
\`A toute fonction $\varphi$ sur l'espace des phases on associe un 
champ de vecteurs hamiltonien $X_\varphi$
d\'efini par la relation $d\varphi=i(X_\varphi)\,\Omega$. 
L'espace des phases est alors une 
vari\'et\'e de Poisson, dont le crochet de Poisson est donn\'e par
\[
\{\varphi,\varphi'\}\equiv X_\varphi(\varphi').
\]

Appliquons cette construction au 
mod\`ele de WZNW en prenant pour $\Sigma$ un cylindre
muni d'une m\'etrique euclidienne ou minkowskienne.
Soit $y^0,y^1$ un syst\`eme de coordonn\'ees locales sur $\Sigma$,
avec $y^1$ d\'efinie modulo $2\pi$ et $y^0$ appartenant
\`a un intervalle donn\'e. On munit $\Sigma$ de la m\'etrique
euclidienne $\gamma$ plate, soit 
$\gamma=(dy^0)^2+(dy^1)^2$. L'action euclidienne de WZNW est 
\[
k\,S(g)=-{_{k}\over^{8\pi}}\,\int_\Sigma
	\tr\,(g^{-1}\partial_\mu g)\,(g^{-1}\partial^\mu g)\,dy^0\,dy^1
	-{_{ik}\over^{12\pi}}\,\int_\Sigma d^{-1}\,\tr\,(g^{-1}dg)^{\wedge 3}
\]
o\`u, pour la partie topologique, on a utilis\'e l'\'equation~(\ref{topo}), 
apr\`es avoir remarqu\'e que localement $\omega=d^{-1}\beta$. 
Pour obtenir le mod\`ele minkowskien, 
on transforme $\Sigma$ par (l'inverse de) la rotation de Wick~:
$y^0\rightarrow -i\,y^0\equiv x^0$ et $x^1=y^1$.
La surface est alors munie de la m\'etrique habituelle
$\gamma\sous{\rm Mink.}=(dx^0)^2-(dx^1)^2$. 
L'action minkowskienne est \'egale \`a
$i$ fois l'action euclidienne (\'ecrite apr\`es rotation), \cad
\[
k\,S\sous{\rm Mink.}(g)=-{_k\over^{8\pi}}\,\int_\Sigma\tr\,(g^{-1}\partial_\mu g)
\,(g^{-1}\partial^\mu g)\,
dx^0\,dx^1+{_k\over^{12\pi}}\,\int_\Sigma d^{-1}\,\tr\,(g^{-1}dg)^{\wedge 3}.
\]
Cette action v\'erifie la contrainte de r\'ealit\'e, \cad
$k\,S\sous{\rm Mink.}(g)\in\R$ pour $g$ \`a valeurs dans 
le groupe compact $G$, contrainte n\'ecessaire quand on consid\`ere
un syst\`eme physiquement coh\'erent~\cite[p. 300]{weinberg}. 

Commen\c{c}ons par l'\'etude du cas minkowskien.
La partie \<<sigma\>> de l'action produit le Lagrangien
\[
\CL_{\sigma}(x^{\mu},g,\xi_\mu)=-{_k\over^{8\pi}}\,\tr\,\xi_{\mu}\,\xi^{\mu},
\quad \xi_\mu\,\xi^\mu=(\xi_0)^2-(\xi_1)^2.
\]
On peut lui appliquer la construction pr\'ec\'edente. 
Soit $\bigphase$ l'espace form\'e des triplets $(x^\mu,g,\xi_\mu)$, o\`u chaque
$\xi_\mu$ prend ses valeurs dans l'alg\`ebre de Lie $\lieg$.
Les moments associ\'es \`a $\xi_0^a$ et $\xi^a_1$ sont donn\'es par
\[
\pi^{0\,a}=-{_k\over^{8\pi}}\,\xi^{0\,a} ,
\quad\pi^{1\,a}=-{_k\over^{8\pi}}\,\xi^{1\,a}.
\]
La $2$-forme canonique $\alpha_{\sigma}$ d\'eduite du Lagrangien 
$\CL_{\sigma}$ est 
\[
\alpha_{\sigma}={_k\over^{8\pi}}\,\tr\,\xi_{\mu}\,\xi^{\mu}\,dx^{0}\wedge dx^{1}
-{_k\over^{4\pi}}\,\tr\,\xi^{0}\,(g^{-1}dg)\wedge dx^{1}
-{_k\over^{4\pi}}\,\tr\,\xi^{1}\,dx^{0}\wedge (g^{-1}dg).
\]
Une mani\`ere simple de prendre en compte le terme de Wess-Zumino 
consiste \`a rajouter \`a $\omega_{\sigma}=d\alpha_{\sigma}$ la $3$-forme 
$\frac{k}{12\pi}\,\tr\,(g^{-1}dg)^{\wedge 3}$. On obtient
\begin{multline*}
\omega= {_k\over^{4\pi}}
	\,\tr\,\xi_\mu\,d\xi^\mu\wedge dx^0\wedge dx^1
	-{_k\over^{4\pi}}\,\tr\,d\xi^0\wedge (g^{-1}dg)\wedge dx^1
	-{_k\over^{4\pi}}\,\tr\,d\xi^1\wedge dx^0\wedge (g^{-1}dg)\\
+{_k\over^{4\pi}}\,\tr\,\xi^0\,(g^{-1}dg)^{\wedge 2}\wedge dx^1
	-{_k\over^{4\pi}}\,\tr\,\xi^1\,(g^{-1}dg)^{\wedge 2}\wedge dx^0
	+{_k\over^{12\pi}}\,\tr\,(g^{-1}dg)^{\wedge 3}.
\end{multline*}
En coordonn\'ees $x^{\pm}=x^{1}\pm x^{0}$, les \'etats classiques
correspondent aux solutions des \'equations
\[
\xi_\mu=g^{-1}\partial_\mu g\quad{\rm et}\quad
	\partial_{x^+}(g^{-1}\partial_{x^-} g)=0.
\]
Les solutions g\'en\'erales sur le cylindre se factorisent localement en 
\[
g(x^0,x^1)=g\sous{{\CL}}(x^+)\,g\sous{{\CR}}(x^-)^{-1}.
\]
Globalement, la condition de p\'eriodicit\'e~: $g(x^0,x^1+2\pi)=g(x^0,x^1)$
se traduit par
\[
g\sous{{\CL},{\CR}}(x^{\pm}+2\pi)=g\sous{{\CL},{\CR}}(x^{\pm})\,\mu
\]
o\`u $\mu\in G$ est la monodromie. Si $\phase\sous{{\CL},{\CR}}\equiv
\{g\sous{{\CL},{\CR}}:\R\rightarrow G\ \big|\ g\sous{{\CL},{\CR}}(x+2\pi)=g\sous{{\CL},{\CR}}(x)\,
\mu\sous{{\CL},{\CR}}\}$, l'espace des phases devient $\phase=\Delta/G$,  
o\`u $\Delta$ est la sous-vari\'et\'e de $\phase\sous{{\CL}}\times \phase\sous{{\CR}}$ 
pour laquelle les monodromies de gauche et de droite sont \'egales, soit
$\mu\sous{{\CL}}=\mu\sous{{\CR}}$~(\footnote{
L'action de $g\in G$ sur $\phase\sous{{\CL}}\times \phase\sous{{\CR}}$ est simplement 
$(g\sous{{\CL}},g\sous{{\CR}})\rightarrow(g\sous{{\CL}}g,g\sous{{\CR}}g)$.}). 
Si on param\'etrise l'espace des phases
$\phase$ par les conditions initiales $g(0,x)$ et $g^{-1}\partial_0g(0,x)$,
la $2$-forme d\'eduite de $\omega$ par la formule~(\ref{Formesym})  est
\[
\Omega={_k\over^{4\pi}}\,\int_0^{2\pi}\tr\,\left(-d\xi_0\wedge(g^{-1}dg)
	+(\xi_0+g^{-1}\partial_{x^1}g)\,(g^{-1}dg)^{\wedge 2}\right)
	\Big|_{(0,x)}\,dx.
\]
On v\'erifie que $\Omega$ n'est pas d\'eg\'en\'er\'ee. Les crochets de Poisson
sur $\phase$ sont alors
\begin{align*}
\{g(x)_{1},g(y)_{2}\}&=0\phantom{{_8\over^k}},\\
\{g(x)_{1},\xi_0(y)_{2}\}&=-{_{8\pi}\over^k}\,g(x)_{1}\,\Cas
\,\delta(x-y),\\
\{\xi_0(x)_{1},\xi_0(y)_{2}\}&={_{8\pi}\over^k}\,[\Cas,\xi_0(x)_{1}%
	+(g^{-1}\partial_xg)(x)_{1}]\,\delta(x-y) 
\end{align*}
o\`u on a utilis\'e les notations habituelles 
$F_{1}\equiv F\otimes 1$, $F_{2}\equiv 1\otimes F$ 
et $\Cas=t^a\otimes t^a$ (somme sur $a$).\label{tenseurr}

Si on r\'e\'ecrit $\Omega$ en s\'eparant les parties
gauche et droite $g\sous{{\CL}}$ et $g\sous{{\CR}}$, on obtient
\[
\Omega=\Omega\sous{{\CL}}+\Omega\sous{{\CR}}
\]
o\`u
\[
\Omega\sous{{\CL}}={_k\over^{4\pi}}\,\int_0^{2\pi}\tr\,\Big[
	(g\sous{{\CL}}^{-1}dg\sous{{\CL}})\wedge
	\partial_x(g\sous{{\CL}}^{-1}dg\sous{{\CL}})
	+(g\sous{{\CL}}^{-1}dg\sous{{\CL}})(0)
	\wedge(d\mu\sous{{\CL}})\mu\sous{{\CL}}^{-1}\Big]
	 -{_k\over^{4\pi}}\,\rho(\mu\sous{{\CL}})
\]
et $\Omega\sous{{\CR}}$ est donn\'ee par la m\^eme 
formule mais avec $g\sous{{\CR}}$ 
et $\mu\sous{{\CR}}$ \`a la place de $g\sous{{\CL}}$ et $\mu\sous{{\CL}}$.
Quand \`a $\rho$, c'est une $2$-forme sur $G$ ne d\'ependant que de 
$\mu\sous{{\CL},{\CR}}$.
La pr\'esence de $\rho$ vient de ce que l'on ne conna\^\i t $\Omega$ que sur 
la diagonale $\Delta$ dans $\phase\sous{{\CL}}\times \phase\sous{{\CR}}$. 
Il semble naturel de choisir $\Omega\sous{{\CL}}$ pour obtenir une forme
symplectique sur $\phase\sous{{\CL}}$. H\'elas, comme
\[
d\Omega\sous{{\CL}}={_k\over^{2\pi}}\,\tr\,(\mu\sous{{\CL}}^{-1}
d\mu\sous{{\CL}})%
^{\wedge 3}-{_k\over^{4\pi}}\,d\rho(\mu\sous{{\CL}}),
\]
$d\Omega\sous{{\CL}}$ n'est pas ferm\'ee~---~
$\tr\,(\mu\sous{{\CL}}^{-1}d\mu\sous{{\CL}})^{\wedge 3}$ 
est une forme ferm\'ee
mais elle n'est pas exacte sur $G$. 
Il existe plusieurs mani\`eres de contourner cette
difficult\'e. Par exemple~\cite{gaw91:quantum,gaw93:lattice}, 
on peut choisir un $\rho$ tel que $d\Omega\sous{{\CL}}$
soit nulle sur un sous-espace dense de $G$. On montre alors que
les solutions locales de notre probl\`eme sont en correspondance bijective
avec une paire de solutions $r^{\pm}$ de l'\'equation de Yang-Baxter classique.
Les crochets de Poisson sur $\phase\sous{{\CL}}$ sont alors donn\'es par
\begin{align*}
\{g\sous{{\CL}}(x)_{1},g\sous{{\CL}}(y)_{2}\}&
=g\sous{{\CL}}(x)_{1}\,g\sous{{\CL}}(y)_2\,r^\pm,\\
\{g\sous{{\CL}}(x)_{1},(\mu\sous{{\CL}})_2\}&=g\sous{{\CL}}(x)_{1}\,
	(\mu\sous{{\CL}})_2\,r^- -g\sous{{\CL}}(x)_{1}\,r^+\,
	(\mu\sous{{\CL}})_2,\\
	\{(\mu\sous{{\CL}})_1,(\mu\sous{{\CL}})_2\}&=r^+\,
	(\mu\sous{{\CL}})_1\,(\mu\sous{{\CL}})_2
	-(\mu\sous{{\CL}})_1\,r^+\,(\mu\sous{{\CL}})_2-(\mu\sous{{\CL}})_2
	\,r^-\,(\mu\sous{{\CL}})_1 +(\mu\sous{{\CL}})_1\,
	(\mu\sous{{\CL}})_2\,r^-
\end{align*}
o\`u, dans la premi\`ere \'equation, on choisit le signe suivant que
$x<y$ ou $x>y$, $|x-y|<2\pi$.

Dans le cas euclidien, on peut r\'ep\'eter les constructions
pr\'ec\'edentes \`a condition de consi\-d\'erer des solutions classiques
\`a valeurs dans $\Gc$. Localement, ces solutions sont de la forme
$g(y_0,y_1)=g\sous{{\CL}}(w)\, 
g\sous{{\CR}}(\overline{w})^{-1}$, o\`u $w=y^0+iy^1$
et $g\sous{{\CL}}(w)$ (resp. $g\sous{{\CR}}(\overline{w})$) est une fonction 
holomorphe (resp. antiholomorphe) \`a valeurs dans $\Gc$. Pour $\Sigma$ 
un cylindre euclidien, on obtient l'espace des phases $\phase\sous{E}$,
param\'{e}tris\'{e} par les donn\'ees de Cauchy
$g(0,y)$ et $\xi_0(0,y)=(g^{-1}\partial_{y^0}g)(0,y)$, chacune prenant
ses valeurs dans $\Gc$ et dans $\liegc$. De fa\c{c}on naturelle, 
$\phase\sous{E}$ devient une vari\'et\'e complexe 
(de dimension infinie). La formule
\[
\Omega\sous{E}={_k\over^{4\pi}}\,\int_0^{2\pi}\tr\,
	\left(-i\, d\xi_0\wedge(g^{-1}dg)
	+(i\xi_0+g^{-1}\partial_{y^1}g)\,(g^{-1}dg)^{\wedge 2}\right)
	\left|_{(0,y)} \right.\,dy
\]
d\'efinit une 2-forme symplectique holomorphe sur $\phase\sous{E}$.

\begin{figure}[h]
\makebox[5cm][l]{\hspace{2.5cm}
\includegraphics{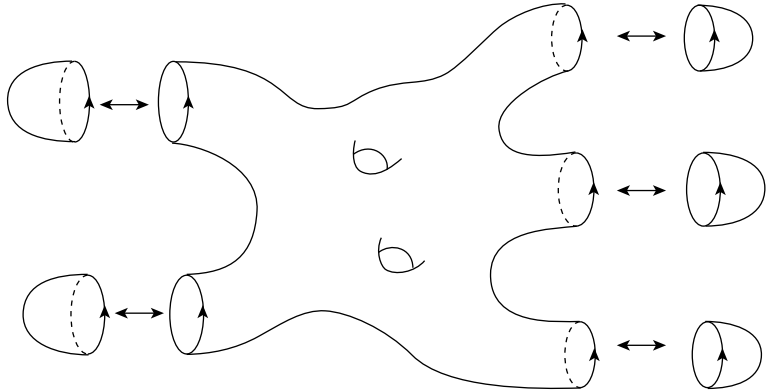}}
\caption{Surface de Riemann \`a bords}
\label{Rb}
\end{figure}

\label{riemman}
Consid\'erons maintenant une surface de Riemann $\Sigma$
\`a bord $\partial\Sigma$, form\'e de
composantes connexes not\'ees $(\partial\Sigma)_i$, $i\in I$ (cf. figure~\ref{Rb}).
Nous supposons que les composantes du bord sont param\'etris\'ees
par des applications analytiques r\'eelles 
$p_{i}:S^1\rightarrow(\partial\Sigma)_i$. On distingue les bords 
\<<sortants\>> et \<<entrants\>> suivant que la param\'etrisation
correspondante respecte ou non l'orientation de $(\partial\Sigma)_i$.
On scinde l'ensemble $I$ en deux sous-ensembles $I_\pm$, o\`u 
$I_-$ \'enum\`ere les bords entrants et $I_+$ les bords sortants.
Les applications $p_i$ pour $i\in I_+$ ($i\in I_-$) s'\'etendent 
en des applications holomorphes sur les petits anneaux 
$\{\,z\ \big|\ 1-\epsilon<|z|\leq 1\,\}$ 
(ou $\{\,z\ \big|\ 1\leq|z|<1+\epsilon\,\}$),
\cad les cylindres (euclidiens) obtenus des anneaux par le changement 
de variables $z=\ee^w=\ee^{y^0+iy^1}$. On garde la m\^{e}me
notation pour d\'esigner les $p_i$ et les extensions.

\setlength{\unitlength}{1mm}
Consid\'erons la vari\'et\'e complexe symplectique suivante
\begin{eqnarray*}
\phase\sous{\Sigma}=
\Big(\begin{picture}(7,10)(-5,-1)
\thicklines
\put(-2.5,-2.5){\line(1,1){5}}
\put(-2.5,2.5){\line(1,-1){5}}
\put(-3,-7){\makebox(8,5)[bl]{$\scriptstyle i\in I_-$}}
\end{picture}
\vspace{-3mm}
-\phase\sous{E}\Big)\times
\Big(\begin{picture}(7,10)(-5,-1)
\thicklines
\put(-2.5,-2.5){\line(1,1){5}}
\put(-2.5,2.5){\line(1,-1){5}}
\put(-3,-7){\makebox(8,5)[bl]{$\scriptstyle i\in I_+$}}
\end{picture}
\ \phase\sous{E}\Big)
\end{eqnarray*}
o\`u $-\phase\sous{E}$ est l'espace $\phase\sous{E}$ muni de la forme 
symplectique $-\Omega\sous{E}$.
Chaque application $g:\Sigma\rightarrow\Gc$, solution de
l'\'equation classique $\partial(g^{-1}\bar\partial g)=0$,
d\'efinit un \'el\'ement $\widetilde{g}\in \phase\sous{\Sigma}$
dont la composante d'indice $i$ correspond \`a l'\'etat
classique $g\circ p_i$ sur un cylindre. On peut montrer que l'ensemble
des $\widetilde{g}$ forme une sous-vari\'et\'e lagrangienne complexe
$\CA_\Sigma$ de la vari\'et\'e symplectique complexe 
$\phase\sous{\Sigma}$. Notons la sym\'etrie de $\CA_\Sigma$ :

\begin{center}
{si $g_{1,2}:\Sigma\rightarrow\Gc$ sont holomorphes
alors $\, g_1\, \CA_\Sigma\, g_2^\dagger\,=\,\CA_\Sigma\,.$}
\end{center}

\noindent Nous d\'ecrirons plus tard la version 
quantique de cette construction.

\vspace{2cm}

\newpage

\medskip
\section{Mod\`ele de WZNW sur une surface \`a bord}

On se propose d'\'etendre la d\'efinition de l'action de WZNW
\`a des champs $g$ d\'efinis sur une surface de Riemann $\Sigma$
\`a bord. On verra que les amplitudes ne sont plus des nombres complexes
mais les \'el\'ements d'un fibr\'e. On commence par consid\'erer le cas 
o\`u $\Sigma$ est le disque ferm\'e
$D=\{z\in\C\ \big|\ |z|\leq1\}$. 

\subsection{$\boldsymbol{\Sigma=D}$}

On \'etend les champs $g:D\rightarrow\Gc$ \`a $\C P^1=D\cup D'$, o\`u
$D'=\{z\in\C\ \big|\ |z|\geq1\}\cup\{\infty\}$.
Une extension de $g$ est enti\`erement d\'etermin\'ee par la donn\'ee
d'un champ $g':D'\rightarrow\Gc$ tel que 
$g'\mathop{|}_{S^1}=g\mathop{|}_{S^1}$.
Une extension est donc une application $g\,\sharp\,g':\C P^1\rightarrow\Gc$ 
d\'efinie par
\[
g\,\sharp\,g'=\begin{cases}
			g, & \text{sur $D$}\\
			g', & \text{sur $D'$}.
		\end{cases}
\]

On sait alors correctement d\'efinir $\ee^{-k\,\SCP(g\,\sharp\,g')}$,
n\'eanmoins cette quantit\'e d\'epend de l'exten\-sion choisie.
Consid\'erons une autre extension, soit $g\,\sharp\,(g'\mu')$ o\`u
$\mu'\in\CC^\infty_1(D',\Gc)$, l'ensemble des applications
$\CC^\infty$ de $D'$ dans $\Gc$ qui se prolongent contin\^ument
\`a $\C P^1$ par $1$. D'apr\`es la formule de Polyakov-Wiegmann,
\[
\SCP(g\,\sharp\,(g'\mu'))=\SCP(g\,\sharp\,g')+\SCP%
(1\,\sharp\,\mu')-\Gamma\sous{\C P^1}(g\,\sharp\,g',1\,\sharp\,\mu')%
\quad(\hbox{mod}\ 2\pi i)
\]
et $\Gamma\sous{\C P^1}(g\,\sharp\,g',1\,\sharp\,\mu')=\Gamma\sous{D'}(%
g',\mu')$. 

On utilise cette r\`egle pour d\'efinir un fibr\'e holomorphe en droites
$\CL^k$ au-dessus du groupe de lacets $L\Gc\,\equiv\,\CC^\infty(S^1,\Gc)%
\,\cong\,\CC^\infty(D',\Gc)\,/\,\CC^\infty_1(D',\Gc)$.
\[
\CL^k\,\equiv\,\left(\CC^\infty(D',\Gc)\times\C\right)/\sim'
\]
o\`u
\[
(g',z')\,\sim'\,(g'\mu',z'\,\ee^{-k\,\SCP(1\,\sharp\,\mu')+k\,\Gamma\sous{D'}
(g',\mu')})
\]
pour $\mu'\in\CC^\infty_1(D',\Gc)$. On v\'erifie que $\sim'$ est bien une
relation d'\'equivalence. La projection est simplement
$\pi':\CL^k\ni[(g',z')]\sous{\sim'}\rightarrow g'\mathop{|}_{S^1}\in L\Gc$.
Le fibr\'e $\CL^k$ est la $k$-i\`eme puissance du fibr\'e $\CL$,
obtenu en prenant $k=1$.
On peut reformuler la relation d'\'equivalence $\sim'$ de 3 autres 
fa\c{c}ons~:
\begin{align*}
(g',z')\, & \sim'\,(g'\mu',z'\,\ee^{-k\,\SCP(1\,\sharp\,\mu')+k\,\Gamma\sous{D'}
		  (g',\mu')})\,=\,%
		  (g'\mu',z'\,\ee^{-k\,\SCP(1\,\sharp\,\mu',\,0\,\sharp\,%
		  (g'{}^{-1}\de g'))}),\\
       \, & \sim'\,(\mu'g',z'\,\ee^{-k\,\SCP(1\,\sharp\,\mu')+k\,\Gamma\sous{D'}
		  (\mu',g')})\,=\,%
		  (\mu'g',z'\,\ee^{-k\,\SCP(1\,\sharp\,\mu',\,0\,\sharp\,%
		  (g'\da g'{}^{-1}))}).
\end{align*}

Soit $g:D\rightarrow\Gc$. On note $\gamma=g\mathop{|}_{S^1}\in L\Gc$ 
et on d\'efinit
\[
\ee^{-k\,S\sous{D}(g)}\,\equiv\,[(g',\ee^{-k\,\SCP(g\,\sharp\,g')})]
	\sous{\sim'}%
\,\in\,\pi'{}^{-1}(\gamma)=(\CL^k)_\gamma
\]
o\`u $g'$ est un \'el\'ement de $\CC^\infty(D',\Gc)$ co\"\i ncidant avec
$\gamma$ sur $S^1$. Ainsi l'amplitude (ou facteur de Boltzman)
devient un \'el\'ement du fibr\'e $\CL^k$.

On peut d\'ecrire le fibr\'e dual \`a $\CL^k$ par
\[
\CL^{-k}\,\equiv\,\left(\CC^\infty(D,\Gc)\times\C\right)/\sim
\]
o\`u
\[
(g,z)\,\sim\,(g\mu,z\,\ee^{-k\,\SCP(\mu\,\sharp\,1)+k\,\Gamma\sous{D}%
(g,\mu)})
\]
pour $\mu\in\CC^\infty_1(D,\Gc)$ et 
$\pi:\CL^{-k}\ni[(g,z)]_{\sim}\rightarrow 
g\mathop{|}_{S^1}\in L\Gc$. L\`a encore, on peut 
d\'efinir $\CL^{-k}$ de plusieurs mani\`eres \'equivalentes~:
\begin{align*}
(g,z)\,&\sim\,(g\mu,z\,\ee^{-k\,\SCP(\mu\,\sharp\,1)+k\,\Gamma\sous{D}%
		  (g,\mu)})\,=\,%
		  (g\mu,z\,\ee^{-k\,\SCP(\mu\,\sharp\,1,\,0\,\sharp\,%
		  (g^{-1}\de g))}),\\
      \,&\,\sim\,(\mu g,z\,\ee^{-k\,\SCP(\mu\,\sharp\,1)+k\,\Gamma\sous{D}%
		  (\mu,g)})\,=\,%
		  (\mu g,z\,\ee^{-k\,\SCP(\mu\,\sharp\,1,\,0\,\sharp\,%
		  (g\da g^{-1}))}).
\end{align*}
La dualit\'e entre $\CL^k$ et $\CL^{-k}$ est simplement
\[
\langle\,[(g',z')]_{\sim'}\,,\,[(g,z)]_\sim\,\rangle\,\equiv\,%
z'z\,\ee^{k\,\SCP(g\,\sharp\,g')}.
\]
L'amplitude du champ $g':D'\rightarrow\Gc$ est
\[
\ee^{-k\,S\sous{D'}(g')}\,\equiv\,[(g,\ee^{-k\,\SCP(g\,\sharp\,g')})]\sous{\sim}
\,\in\,\pi^{-1}(\gamma')=(\CL^{-k})_{\gamma'}
\]
si $\gamma'=g|_{S^1}\in L\Gc$.

On peut utiliser $z\,\ee^{-k\,S\sous{D}(g)}$ (resp.  $z'\,\ee^{-k\,S\sous{D'}(g')}$)
pour repr\'esenter les \'el\'ements de $\CL^k$ (resp. $\CL^{-k}$). Dans ce 
langage, la dualit\'e est donn\'ee par
\[
\langle\, z\,\ee^{-k\,S\sous{D}(g)}\,,\,z'\,\ee^{-k\,S\sous{D'}(g')}\,\rangle\,=\,%
zz'\,\ee^{-k\,\SCP(g\,\sharp\,g')}.
\]

\subsection{Surface de Riemann g\'en\'erale\label{generale}}

On consid\`ere maintenant une surface de Riemann $\Sigma$
\`a bord, comme \`a la fin de la section~{\bf 2}.
On peut compl\'eter $\Sigma$ pour former une surface de Riemann compacte 
$\widetilde{\Sigma}$. \`A cet effet, on \<<colle\>> 
contin\^ument une copie $D_i$ de $D$ suivant $(\partial\Sigma)_i$, $i\in I_-$,
et une copie $D_i'$ de $D'$ si $i\in I_+$.

Soient $g:\Sigma\rightarrow\Gc$ et 
$\widetilde{g}:\widetilde{\Sigma}\rightarrow\Gc$ une 
extension de $g$. L'amplitude $\ee^{-k\,S\sous{\Sigma}(g)}$ est 
l'unique \'el\'ement
de $(\bigotimes_{i\in I_-}\CL^{-k}_{g_{i}})\,\otimes\,%
(\bigotimes_{i\in I_+}\CL^{k}_{g_{i}})$~--- $g_{i}=g\circ p_{i}$~--- 
tel que
\[
\langle\,\ee^{-k\,S\sous{\Sigma}(g)}\,,\,%
(\bigotimes_{i\in I_-}\ee^{-k\,S\sous{D}(\widetilde{g}|_{D_i})})%
\,\otimes\,(\bigotimes_{i\in I_+}\ee^{-k\,S\sous{D'}(\widetilde{g}|_{D'_i})})%
\,\rangle\,=\,\ee^{-k\,S\sous{\widetilde{\Sigma}}(\widetilde{g})}.
\]

Les {\bol amplitudes quantiques} du mod\`ele de WZNW sont donn\'ees par 
les int\'egrales fonctionnelles formelles
\begin{equation}
\label{opera}
\NA\sous{\Sigma}((\gamma_{i})_{i\in I})%
\,\equiv\,\mathop{\int}_{\scriptstyle%
			 g:\Sigma\rightarrow G\atop\scriptstyle%
			 g_{i}=\gamma_{i}}%
\,\ee^{-k\,S\sous{\Sigma}(g)}\,Dg\,=\,\mathop{\int}_{\raggedleft\scriptstyle%
			 g:\Sigma\rightarrow G\atop\raggedleft\scriptstyle%
			 g_{i}=1}%
\,\ee^{-k\,S\sous{\Sigma}(\gamma g)}\,Dg\,\in\,%
(\bigotimes_{i\in I_-}\CL^{-k}_{\gamma_{i}})\,\otimes\,%
(\bigotimes_{i\in I_+}\CL^{k}_{\gamma_{i}})
\end{equation}
o\`u $\gamma_{i}\in LG$ 
et $\gamma:\Sigma\rightarrow G$ telle que
$\gamma\circ p_{i}=\gamma_{i}$, sont fix\'ees. 
Par invariance de la mesure de Haar,
la d\'efinition pr\'ec\'edente ne d\'epend pas des valeurs 
de $\gamma$ \`a l'int\'erieur de $\Sigma$, m\^eme
si $\gamma$ prend ses valeurs dans $\Gc$ ---~%
ceci \`a condition que l'int\'egrande soit holomorphe.
Ainsi, les int\'egrales peuvent \^etre aussi d\'efinies pour
$\gamma_i\in L\Gc$ et on peut assimiler les amplitudes quantiques aux
sections holomorphes (formelles) d'un fibr\'e~:
\[
\NA\sous{\Sigma}\,\in\,\big(\bigotimes_{i\in I_-}H^0(\CL^{-k})\big)
\,\otimes\,\big(\bigotimes_{i\in I_+}H^0(\CL^{k})\big).
\]
Il appara\^\i t que $H^0(\CL^k)$ est l'espace naturel des
\'etats quantiques du mod\`ele de WZNW.
Les amplitudes $\NA\sous{\Sigma}$
quantifient les sous-vari\'et\'es lagrangiennes $\CA\sous{\Sigma}$
de la section {\bf 2}.

\subsection{Sym\'etrie de Kac-Moody}

On d\'efinit un produit sur 
$(\CL^k)^\times=\CL^k\,\setminus\,\{\mbox{section z\'ero}\}$~:
\[
[(g'_{1},z'_{1})]_{\sim'}\bullet[(g'_{2},z'_{2})]_{\sim'}%
\,\equiv\,[(g'_{1}g'_{2},\,z'_{1}z'_{2}\,
\ee^{k\,\Gamma\sous{D'}(g'_{1},g'_{2})})]_{\sim'}.
\]
On v\'erifie que l'op\'eration $\bullet$ est bien d\'efinie et
induit une structure de groupe sur $(\CL^k)^\times$. 
Si on utilise $z\,\ee^{-k\,S\sous{D}(g)}$ pour repr\'esenter
les \'el\'ements de $((\CL^k)^\times,\bullet)$, l'\'el\'ement neutre est
$\ee^{-k\,S\sous{D}(1)}$, l'inverse est 
\[
(z\,\ee^{-k\,S\sous{D}(g)})^{-1}=%
z^{-1}\,\ee^{-k\,\Gamma\sous{D}(g,g^{-1})}%
\,\ee^{-k\,S\sous{D}(g^{-1})}
\]
et l'op\'eration $\bullet$ devient
\[
(z_{1}\,\ee^{-k\,S\sous{D}(g_{2})})\bullet(z_{2}\,
\ee^{-k\,S\sous{D}(g_{2})})=%
z_{1}z_{2}\,\ee^{-k\,\Gamma\sous{D}(g_{1},g_{2})}\,
\ee^{-k\,S\sous{D}(g_{1}g_{2})}.
\]
En particulier, on obtient une g\'en\'eralisation de la 
formule de Polyakov-Wiegmann pour une surface de Riemann \`a bord
\[
\ee^{-k\,S\sous{\Sigma}(g_{1}g_{2})}=
	\ee^{k\,\Gamma\sous{\Sigma}(g_{1},g_{2})}\ %
\ee^{-k\,S\sous{\Sigma}(g_{1})}\bullet\ee^{-k\,S\sous{\Sigma}(g_{2})}
\]

Muni de l'op\'eration $\bullet$, $(\CL^k)^\times$ est
 le {\bol groupe de Kac-Moody $\widehat{LG}\dC_k$ de niveau $k$}~\cite{pressley}
qui est l'extension centrale du groupe des lacets $L\Gc$, 
\ie il existe une suite exacte d'homomorphis\-mes
\[
1\rightarrow\C^\times\stackrel{i_k}{\rightarrow}
\widehat{LG}\dC_k\rightarrow L\Gc\rightarrow 1.
\]
La premi\`ere fl\`eche envoie $z\in\C^\times$ sur
$z\,\ee^{-k\,S\sous{D}(1)}$ et la seconde est la projection $\pi'$
induite par la structure de fibr\'e.
\`A partir de l'extension centrale satisfaite par
$\widehat{LG}\dC_1$, on obtient une extension centrale de
$\widehat{LG}\dC_k$ en divisant $\widehat{LG}\dC_1$ par $i_1(\{
{}^k\hspace{-0.13cm}\sqrt{1}\})$.
On peut \'egalement relever l'involution
$g\mapsto g^\dagger$ sur le fibr\'e $\CL^k$. On pose --- et on v\'erifie
que c'est bien d\'efini ---~: $[(g',z')]\sous{\sim'}^\dagger
=[(g'{}^\dagger,\overline{z}')]\sous{\sim'}$ soit encore
\[
\big(z\,\ee^{-k\,S\sous{D}(g)}\big)^\dagger
	=\overline{z}\,\ee^{-k\,S\sous{D}(g^\dagger)}.
\]
On v\'erifie aussi que $(\widehat{h}_1\bullet\widehat{h}_2)^\dagger=
\widehat{h}_2^\dagger\bullet\widehat{h}_1^\dagger$.
On dit qu'un \'el\'ement $\widehat h$ de $\widehat{LG}\dC_k$ est
{\bol unitaire} si $\widehat h^\dagger=\widehat h^{-1}$~; comme pour
le groupe $L\Gc$. Explicitement, cela donne pour
$h=z\,{\rm e}^{-k\, S_D(g)}$
\[
gg^\dagger=1,\qquad |z|^2=\ee^{k\,\Gamma\sous{D}(g,g^{-1})}.
\]
On note $\widehat{LG}_k$ le sous-groupe de $\widehat{LG}\dC_k$ form\'e
des \'el\'ements unitaires. 
Le fibr\'e en droite $\CL^k\rightarrow LG$ est muni d'une 
{\bol structure hermitienne} donn\'ee par
\[
\big|z\,\ee^{-k\,S\sous{D}(g)}\big|^2
	=|z|^2\,\ee^{-k\,\Gamma\sous{D}(g,g^{-1})}.
\]
On peut donc dire que $\widehat{LG}$ est l'ensemble des vecteurs
de $\CL^k\rightarrow LG$ de longueur $1$.
Cette fois, $\widehat{LG}_k$ est une extension
centrale de $LG$~:
\[
1\rightarrow S^1\rightarrow
\widehat{LG}_k\rightarrow LG\rightarrow 1.
\]

Le groupe $\widehat{LG}\dC_k$ agit \`a gauche et \`a droite sur 
$H^0(\CL^k)$ par 
\[
(\Nl(\widehat h)\psi)(\gamma)=\widehat h\bullet
\psi(h^{-1}\gamma)\,,\qquad
(\Nr(\widehat h)\psi)(\gamma)=\psi(\gamma h)\bullet\widehat h
\]
o\`u $\widehat h\in \widehat{LG}\dC_k$ et
$\psi\in H^0(\CL^k)$. Les amplitudes
$\NA\sous{\Sigma}$ poss\`edent (formellement) les
propri\'et\'es d'invariance
\[
\Nl(\ee^{-k\,S\sous{\Sigma}(g_{1})})
	\,\Nr(\ee^{-k\,S\sous{\Sigma}(g_{2}^\dagger)})\, \NA\sous{\Sigma}
	\,=\,\NA\sous{\Sigma}
\]
pour des applications holomorphes $g_{1,2}:\,\Sigma\rightarrow \Gc$.
C'est la version quantique des propri\'et\'es de sym\'etrie des sous-vari\'et\'es
lagrangiennes $\CA_\Sigma$ de la section $\bf 2$.
En effet, si $g_{1,2\,i}=g_{1,2}\circ p_{i}$,
\begin{setlength}{\multlinegap}{20pt}
\begin{multline*}
\ee^{-k\,S\sous{\Sigma}(g_{1})}\bullet
	\NA\sous{\Sigma}((\gamma_{i})_{i\in I})
\bullet\ee^{-k\,S\sous{\Sigma}(g_{2}^\dagger)}\\
=\,\mathop{\int}_{\scriptstyle%
			 g:\Sigma\rightarrow G\atop\scriptstyle%
			 g_{i}=\gamma_{i}}%
\,\ee^{-k\,S\sous{\Sigma}(g_{1})}\bullet
\ee^{-k\,S\sous{\Sigma}(g)}\bullet\ee^{-kS\sous{\Sigma}(g_{2}^\dagger)}
\,Dg\,
=\,\mathop{\int}_{\scriptstyle
	 g:\Sigma\rightarrow G\atop\scriptstyle
	 g_i=\gamma_i}
\,\ee^{-k\,S\sous{\Sigma}(g_1gg_2^\dagger)}\,Dg\\
=\,\mathop{\int}_{\raggedleft\scriptstyle%
	 g:\Sigma\rightarrow G\atop\raggedleft\scriptstyle%
	 g_{i}=g_{1i}\,\gamma_{i}\,g_{2i}^\dagger}%
	\ee^{-k\,S\sous{\Sigma}(g)}\, Dg
\ =\  
\NA\sous{\Sigma}((g_{1i}\,\gamma_{i}\,g_{2i}^\dagger)_{i\in I }).
\end{multline*}
\end{setlength}
\noindent Cette invariance est une r\'eminiscence 
au niveau quantique de la sym\'etrie chirale
not\'ee pr\'ec\'edemment au niveau classique. 
La sym\'etrie classique $L\Gc\times L\Gc$ se 
traduit par une
action projective de $L\Gc\times L\Gc$ sur l'espace des \'etats.

Au niveau infinit\'esimal, on d\'ecrit 
cette sym\'etrie  par les g\'en\'erateurs
\begin{align*}
J^{a}_{n}&\equiv{_1\over^i}\,
	\left.{_d\over^{d\epsilon}}\right|_{\epsilon=0}
	\Nl\big(\ee^{-k\,S\sous{D}(\ee^{i\epsilon\,t^a\,z^n})}\big),\quad
	 \text{si $n\geq 0$},\\
J^{a}_{n}&\equiv{_1\over^i}\,
	\left.{_d\over^{d\epsilon}}\right|_{\epsilon=0}
	\Nl\big(\ee^{-k\,S\sous{D}
	(\ee^{i\epsilon\,t^a\,{\overline z}^{-n}})}\big),\quad
	 \text{si $n\leq 0$}
\end{align*}
ainsi que les $\overline{J}^{a}_{n}$ obtenus de la m\^eme mani\`ere mais en rempla\c{c}ant 
$\Nl$ par $\Nr$. On montre alors~\cite{gaw88:topo} que
\[
[J^{a}_{n},J^{b}_{m}]=if^{abc}\,
	J^{c}_{n+m}+{_{kn}\over^2}\,\delta^{ab}\,
	\delta_{n+m,0}
\]
ainsi que pour les $\overline{J}^{a}_{n}$. Les $J^a_n$
et les $\overline{J}^{a}_{n}$ commutent entre eux.
Ces relations de commutation
d\'efinissent l'alg\`ebre de Kac-Moody affine 
$\widehat{L\lieg}\eC$ (cf. section {\bf 7.2}). 
L'espace $H^0(\CL^k)$ est donc accompagn\'e d'une repr\'esentation de 
$\widehat{L\lieg}\eC\oplus\widehat{L\lieg}\eC$.
Dans le restant de cette th\`ese, on abandonne ce point de
vue pour ne plus consid\'erer que des surfaces de Riemann compactes ---
sans bord --- le lecteur int\'eress\'e pourra consulter les 
articles~\cite{gaw92:laced,gaw88:topo,gaw90:wzw} (cf. aussi la section 
{\bf 5.4}).

\medskip
\section{Fonctions de Green et leurs sym\'etries}

\paragraphe{Strat\'egie g\'en\'erale}%
On se propose de quantifier le mod\`ele de WZNW. 
Pour cela, on utilise une quantification \`a la Feynman. 
Les fonctions de Green de la th\'eorie sont donn\'ees par des int\'egrales 
fonctionnelles formelles. On ne donne pas un sens math\'ematiquement 
rigoureux \`a de tels objets et les manipulations effectu\'ees sur les int\'egrales
fonctionnelles restent formelles. En particulier, les fonctions de Green n\'ecessitent sans
doute d'\^etre renormalis\'ees. N\'eanmoins, le mod\`ele poss\'edant 
des propri\'et\'es de sym\'etrie
tr\`es riches, on esp\`ere trouver des expressions explicites
pour les fonctions de Green, \cad arriver \`a la \<<quadrature des th\'eories
conformes\>>~\cite{gaw89:quadrature}. 
Une fois ce travail effectu\'e, il s'agira de remonter le chemin en 
sens inverse, soit utiliser les r\'esultats pour d\'efinir les fonctions
de Green et ensuite v\'erifier qu'elles r\'epondent 
aux divers crit\`eres requis par notre mod\`ele.

\subsection{Couplage \`a un champ de jauge}

On d\'esire jauger la sym\'etrie $G\times G$ du mod\`ele
WZNW. Soit $A$ une $1$-forme sur $\Sigma$
\`a valeurs dans $\liegc$~; $A=\AL+\AR$, si on s\'epare
les composantes $(1,0)$  et $(0,1)$.
En physique, on dit que $A$ est un champ de jauge (complexe).
Localement, $\AL=A_{z}\,dz$ et $\AR=A_{\overline{z}}\,d\overline{z}$.
On peut alors d\'efinir~\cite{gaw:book} l'{\bol action de WZNW coupl\'ee
\`a un champ de jauge} :
\[
S(g,A)=S(g)+{_{i}\over^{2\pi}}\,\int_\Sigma\tr\,[A^{10}\,(g^{-1}\de g)+%
(g\partial g^{-1})\,A^{01}+gA^{10}g^{-1}A^{01}-A^{10}A^{01}].
\]
G\'eom\'etriquement, $g$ devrait \^etre vu comme une section
du fibr\'e associ\'e \`a la 
repr\'esentation adjointe du $\Gc$-fibr\'e principal trivial 
$P=\Sigma\times\Gc$. Dans ce langage, le champ $A$ est une connexion sur $P$.
Comme le fibr\'e est
trivial, utiliser le langage g\'eom\'etrique n'est pas tr\`es int\'eressant.
On d\'ecouple les transformations de jauge chirales de gauche et droite. 
Si $h_{1,2}:\Sigma\rightarrow\Gc$,
\begin{alignat*}{2}
g &\mapsto h_{2}g, &\quad
	&\begin{cases}
		\AL\mapsto\AL\\
		\AR\mapsto\hAR{h_{2}}\equiv%
		      h_{2}\AR h_{2}^{-1}+h_{2}\de h_{2}^{-1}
	\end{cases},\\
g &\mapsto gh_{1}^{-1},&\quad%
	&\begin{cases}
		\AL\mapsto\hAL{h_{1}}\equiv%
	    h_{1}\AL h_{1}^{-1}+h_{1}\partial h_{1}^{-1}\\
		\AR\mapsto\AR
	\end{cases}.
\end{alignat*}
Les relations suivantes traduisent le comportement de l'action jaug\'ee
dans les transformations chirales
\begin{equation}
S(h_{2}gh_{1}^{-1},\hAL{h_{1}}+\hAR{h_{2}})=%
\begin{cases}
	S(g,A)-S(h_{2}^{-1}h_{1},A)\\
	S(g,A)+S(h_{2}h_{1}^{-1},\hAL{h_{1}}+\hAL{h_{2}})
\end{cases}       \quad(\hbox{mod}\ 2\pi i).
\label{pwjauge}
\end{equation}
La deuxi\`eme \'egalit\'e se d\'eduit imm\'ediatement de la premi\`ere.
La premi\`ere est une cons\'equence de la formule de Polyakov-Wiegmann.
On constante que l'action de WZNW n'est pas pr\'eserv\'ee dans une 
transformation de jauge chirale. Par contre, si $h_{1}=h=h_{2}$, on
obtient $S(hgh^{-1},\hA{h})=S(g,A)$. Le mod\`ele jaug\'e est donc invariant
par cette transformation non chirale.

Un champ de jauge unitaire est d\'efini par les relations
$A=-A^\dagger$, ou encore $\AL=-(\AR)^\dagger$. Ces conditions
sont pr\'eserv\'ees par les transformations de jauge chirales avec 
$h_{1}=(h^\dagger)^{-1}$, $h_{2}=h$, \cad
\begin{align*}
\AL & \mapsto\hAL{h^{\dagger-1}}\,\equiv\,%
  (h^\dagger)^{-1}\AL h^\dagger+(h^\dagger)^{-1}\partial h^\dagger\\
\AR & \mapsto\hAR{h}\,\equiv\,h\AR h^{-1}+h\de h^{-1}.
\end{align*}
Les \'equations~(\ref{pwjauge}) deviennent
\[
S(hgh^\dagger,\hAL{h^{\dagger-1}}+\hAR{h})=%
\begin{cases}
	S(g,A)-S(h^{-1}(h^\dagger)^{-1},A)\\
	S(g,A)+S(hh^\dagger,\hAL{h^{\dagger-1}}+\hAR{h})
\end{cases}       \quad(\hbox{mod}\ 2\pi i).
\]
En particulier, pour $hh^\dagger=1$, \ie pour une transformation
de jauge unitaire (\`a valeurs dans le groupe compact $G\subset\Gc$), on a
$S(hgh^\dagger,\hA{h})=S(g,A)$.

\subsection{Fonctions de Green}

Soit $\xi_{1},\ldots,\xi_{N}$ une s\'equence $\un{\xi}$ de $N$ points 
distincts de $\Sigma$~(\footnote{Quel que soit le type 
d'insertion effectu\'e, on suppose
que les points o\`u on ins\`ere les champs sont deux \`a deux distincts.}). 
\`A chaque point $\xi_\ell$, on associe une coordonn\'ee locale
$z_\ell$.
Soit $R_{1},\ldots,R_{N}$ une s\'equence
$\un{R}$ de $N$ repr\'esentations irr\'eductibles
de $G$, chacune agissant dans un espace de Hilbert not\'e $V_{\lambda_\ell}$.
On note $\esprep=\otimes_\ell V_{\lambda_\ell}$ o\`u $\un{\lambda}$ est
la s\'equence de plus hauts poids.
Ces repr\'esentations s'\'etendent naturellement en des repr\'esentations
holomorphes de $\Gc$.
Les {\bol fonctions de Green} (euclidiennes) sont formellement d\'efinies 
par l'int\'egrale fonctionnelle suivante
\[
\mbox{\boldmath $\Gamma$}(\un{\xi},
	\un{R},A)\,\equiv\,\int\grl\,\ee^{-k\,S(g,A)}\,Dg%
\ \in\ \mathop{\otimes}\limits_{\ell}{\rm End}(V_{\lambda_\ell})
\,\cong\,\baresprep\otimes\esprep
\]
o\`u $Dg$ est le produit formel $\prod_{\xi\in\Sigma}dg(\xi)$ 
des mesures de Haar sur $G$. Bien qu'on ait \'etendu les champs au groupe
complexifi\'e $\Gc$, l'int\'egration porte uniquement sur les champs
\`a valeurs dans le groupe compact $G$. On utilise aussi les {\bol fonctions
de Green normalis\'ees}
\[
\langle\,\grl\,\rangle\sous{A}\,\equiv\,Z\sous{A}^{-1}\int\grl\,%
\ee^{-k\,S(g,A)}\,Dg\,=\,Z\sous{A}^{-1}\ %
\mbox{\boldmath $\Gamma$}(\un{\xi},\un{R},A)
\]
o\`u
\[
Z\sous{A}\,\equiv\,\int\ee^{-k\,S(g,A)}\,Dg
\]
est la {\bol fonction de partition}. On oublie l'indice $A$ en champ de jauge nul.

\subsection{Identit\'e de Ward chirale --- version globale}

Soient $h_{1,2}:\Sigma\rightarrow\Gc$ des transformations 
de jauge chirales de gauche et de
droite. D'apr\`es l'invariance formelle des mesures de Haar 
dans l'int\'egrale fonctionnelle
et suivant la formule de Polyakov-Wiegmann jaug\'ee~(\ref{pwjauge}), 
les fonctions de Green se transforment simplement~: 
\begin{align*}
\mbox{\boldmath$\Gamma$}(\un{\xi},\un{R},
\hAL{h_{1}}+\hAR{h_{2}})&=%
\int\mathop{\otimes}\limits_\ell(h_{2}gh_{1}^{-1})(\xi_\ell)%
\sous{R_\ell}\,\ee^{-k\,S(h_{2}gh_{1}^{-1},\hAL{h_{1}\ }%
+\hAR{h_{2}\,})}\,Dg\\
&=\ee^{k\,S(h_{2}^{-1}h_{1},A)}\,%
\mathop{\otimes}\limits_\ell h_{2}(\xi_\ell)\sous{R_\ell}\,%
\mbox{\boldmath$\Gamma$}(\un{\xi},\un{R},A)\,%
\mathop{\otimes}\limits_\ell h_{1}(\xi_\ell)^{-1}\sous{R_\ell}.
\end{align*}
En th\'eorie des champs, on appelle
une telle relation une {\bol identit\'e de Ward} (chirale, globale). 
Elles peuvent aussi \^{e}tre \'ecrites comme
\begin{align*}
Z\sous{\hAL{h_{1}\ }+\hAR{h_{2}\,}} & =
\ee^{k\,S(h_{2}^{-1}h_{1},A)}\,Z\sous{A},\\
\langle\,\grl\,\rangle\sous{\hAL{h_{1}\ }+\hAR{h_{2}\,}} & =
\mathop{\otimes}\limits_\ell h_{2}(\xi_\ell)\sous{R_\ell}\,%
\langle\,\grl\,\rangle\sous{A}
\mathop{\otimes}\limits_\ell h_{1}(\xi_\ell)^{-1}\sous{R_\ell},
\end{align*}
exprimant ainsi la covariance des fonctions de Green normalis\'ees
et la loi de transformation non-triviale de la fonction de partition
sous l'action des transformation de jauge chirales. 

Il est souvent plus
commode de travailler avec des lois de transformation factoris\'ees,
d\'ecouplant ainsi les actions de gauche et de droite. Pour cela, il suffit 
de modifier l'action jaug\'ee comme suit
\[
\widetilde{S}(g,A)\,\equiv\,S(g,A)+{_i\over^{2\pi}}\int\tr\,(\AL\,\AR).
\]
Pour la nouvelle action, l'\'equation~(\ref{pwjauge}) devient
\[
\widetilde{S}(h_{2}gh_{1}^{-1},\hAL{h_{1}}+\hAR{h_{2}})=%
\widetilde{S}(g,A)-S(h_{1},\AL)-S(h_{2}^{-1},\AR).
\]
Si $\widetilde{Z}\sous{A}$ et 
$\widetilde{\mbox{\boldmath$\Gamma$}}(\un{\xi},\un{R},A)$
sont les quantit\'es obtenues en rempla\c{c}ant $S$ par $\widetilde{S}$, on a
\[
\widetilde{Z}\sous{\hAL{h_{1}\ }+\hAR{h_{2}\,}}=%
\ee^{k\,S(h_{1},\AL)+k\,S(h_{2}^{-1},\AR)}\,\widetilde{Z}\sous{A}.
\]
Par cons\'equent, les fonctions de Green modifi\'ees
$\widetilde{\mbox{\boldmath$\Gamma$}}(\un{\xi},\un{R},A)%
=\widetilde{Z}\sous{A}\,\langle\,
\otimes_\ell g(\xi_\ell)\sous{R_\ell}
\,\rangle\sous{A}$ v\'erifient
l'identit\'e de Ward chirale, globale, d\'{e}coupl\'{e}e
\begin{equation}
\begin{split}
\widetilde{\mbox{\boldmath$\Gamma$}}%
(\un{\xi},\un{R},&\hAL{h_{1}}+\hAR{h_{2}})\\
&=\ee^{k\,S(h_{1},\AL)+k\,S(h_{2}^{-1},\AR)}
\mathop{\otimes}\limits_\ell h_{2}(\xi_\ell )\sous{R_\ell }\,%
\widetilde{\mbox{\boldmath$\Gamma$}}(\{\xi_\ell \},\{R_\ell \},A)\,%
\mathop{\otimes}\limits_\ell h_{1}(\xi_\ell )^{-1}\sous{R_\ell }.
\label{ward}
\end{split}
\end{equation}

\subsection{Identit\'es de Ward chirales --- version infinit\'esimale}

\`A partir des identit\'es de Ward globales, 
il est naturel de vouloir d\'eriver des identit\'es de 
Ward infinit\'esimales. On suppose que les fonctions de Green d\'ependent
de mani\`ere $\CC^\infty$~(au sens de Fr\'echet) du champ de jauge $A$. 
En particulier, toutes les d\'eriv\'ees fonctionnelles existent
au sens des distributions.
Introduisons les {\bol fonctions de Green avec insertions de courants}
\[
\langle\,%
\prod_{j=1}^M J^{a\sous{j}}_{z\sous{j}}\,%
\prod_{\overline{\jmath}=1}^{\overline{M}}%
J^{b\sous{\overline{\jmath}}}_{\overline{z}\sous{\overline{\jmath}}}%
\,\grl\,\rangle\sous{A}%
\,\equiv\,\widetilde{Z}\sous{A}^{-1}\,%
\frac{%
(-\pi\delta)^{M+\overline{M}}%
	}{%
\mathop{\prod}\limits_j%
	\delta A^{a\sous{j}}_{\overline{z}\sous{j}}\,%
\mathop{\prod}\limits_{\overline{\jmath}}%
	\delta A
^{b\sous{\overline{\jmath}}}_{z\sous{\overline{\jmath}}}%
	}%
\,\Big(\widetilde{Z}\sous{A}\,\langle\,\grl\,\rangle\sous{A}\Big).
\]
Les exposants $a\sous{j}$ et $b\sous{\overline{\jmath}}$ d\'esignent 
des composantes dans l'alg\`ebre
de Lie. Les indices $z\sous{j}$ et $z\sous{\overline{\jmath}}$ sont 
des coordonn\'ees locales
pour les points o\`u on ins\`ere les courants%
~--- on fera bien attention \`a ne pas confondre ces coordonn\'ees
avec celles des points $\xi_\ell $. On a simplement not\'e 
$\AL(\zeta)=A^{a}_{z}(\zeta)\,t^a\,dz$ et 
$\AR(\zeta)=A^{a}_{\overline{z}}(\zeta)\,
t^a\,d\overline{z}$, si $z$ est une coordonn\'ee
locale pour le point $\zeta$ o\`u on ins\`ere le courant. Posons 
$h_{1}\equiv 1$ et $h_{2}=\ee^{\delta\Lambda}$.
D'apr\`es l'identit\'e de Ward globale~(\ref{ward}),
\[
\widetilde{\mbox{\boldmath$\Gamma$}}(\AL+\hAR{\ee^{\delta\Lambda}})%
\,=\,\ee^{k\,S(\ee^{-\delta\Lambda},\AR)}\,%
\mathop{\otimes}\limits_\ell(\ee^{\delta\Lambda})(\xi_\ell )\sous{R_\ell }\,%
\widetilde{\mbox{\boldmath$\Gamma$}}(A).
\]
Pour ne pas alourdir le calcul, on sous-entend $\un{\xi},\un{R}$.
On r\'e\'ecrit cette \'equation au premier ordre en $\delta\Lambda$~---
dans ce qui suit, l'\'egalit\'e $\simeq$ signifie \<<\`a des termes
du second ordre en $\delta\Lambda$ pr\`es\>>.
Le terme de gauche donne
\[
\widetilde{\mbox{\boldmath$\Gamma$}}(\AL+\hAR{\ee^{\delta\Lambda}})%
\simeq\widetilde{\mbox{\boldmath$\Gamma$}}(A)+%
\int_\Sigma\frac{\delta\widetilde{\mbox{\boldmath$\Gamma$}}}%
{\delta A^{a}_{z}}(A)\,\delta A^{a}_{z}\,d^{2}z%
+\int_\Sigma\frac{\delta\widetilde{\mbox{\boldmath$\Gamma$}}}%
{\delta A^{a}_{\overline{z}}}(A)%
\,\delta A^{a}_{\overline{z}}\,d^{2}z.
\]
La variation du champ de jauge est~: $\delta\AL=0$ et 
$\delta\AR=\hAR{\ee^{\delta\Lambda}\!\!}-\AR$, soit
$\delta A^{a}_{\overline{z}}=if^{abc}\,\delta\Lambda^{b}\,
A^{c}_{\overline{z}}%
-\partialbarz(\delta\Lambda^{a})$. On a donc
\[
\widetilde{\mbox{\boldmath$\Gamma$}}(\AL+\hAR{\ee^{\delta\Lambda}})%
\simeq\widetilde{\mbox{\boldmath$\Gamma$}}(A)+%
\int_\Sigma\frac{-\delta}%
{\delta A^{a}_{\overline z}}\widetilde{\mbox{\boldmath$\Gamma$}}(A)%
\,\left(\partialbarz(\delta\Lambda^{a})-%
if^{abc}\,\delta\Lambda^{b}\,A^{c}_{\overline{z}}\right)\,d^{2}z.
\]
Un calcul rapide utilisant l'\'equation~(\ref{varS}) donne
$\ee^{k\,S(\ee^{-\delta\Lambda},\AR)}\simeq 1+\frac{k}{2\pi}%
\,\int_\Sigma\partial_z(\delta\Lambda^{a})\,A^{a}_{\overline z}
\,d^{2}z$. On a aussi 
$\otimes_\ell(\ee^{\delta\Lambda})(\xi_\ell )\sous{R_\ell }%
\simeq 1+\sum_\ell\delta\Lambda^{a}(\xi_\ell )\,t^a_\ell $,
o\`u $t^a_\ell =1\otimes\cdots\otimes\underbrace{dR_\ell (t^a)}_\ell%
\otimes\cdots\otimes 1$.
On aboutit finalement \`a
\begin{equation}\label{ward1}
\tag{\ref{ward1}.a}
\begin{split}
\int_\Sigma\langle\,J^{a}_{z}\,\grl\,\rangle\sous{A}
&\left(\partialbarz(\delta\Lambda^{a})-%
if^{abc}\,\delta\Lambda^{b}\,A^{c}_{\overline{z}}\right)\,d^{2}z\\
&\simeq\Big({_k\over^2}\,\int_\Sigma\partial_z(\delta\Lambda^{a})%
 \,A^{a}_{\overline z}\,d^{2}z%
+\pi\,\sum_\ell\delta\Lambda^{a}(\xi_\ell )\,t^a_\ell \Big)\,%
\langle\,\grl\,\rangle\sous{A}.
\end{split}
\end{equation}
De m\^eme, en utilisant $h_{1}=\ee^{\delta\Lambda}$ et $h_{2}\equiv 1$,
on trouve
\begin{equation}
\tag{\ref{ward1}.b}
\begin{split}
\int_\Sigma\langle\,J^{a}_{\overline z}\,\grl\,\rangle\sous{A}&
\left(\partial_{z}(\delta\Lambda^{a})-%
if^{abc}\,\delta\Lambda^{b}\,A^{c}_{z}\right)\,d^{2}z\\
&\simeq\langle\,\grl\,\rangle\sous{A}%
 \Big({_k\over^2}\,\int_\Sigma\partialbarz(\delta\Lambda^{a})%
 \,A^{a}_{z}\,d^{2}z%
-\pi\,\sum_\ell\delta\Lambda^{a}(\xi_\ell )\,t^a_\ell \Big).%
\end{split}
\addtocounter{equation}{1}
\end{equation}
Apr\`es int\'egration par parties dans les \'equations~(\ref{ward1}.a-b),
si on observe que $\delta\Lambda$ a \'et\'e choisi 
quelconque, on obtient
\begin{equation}\label{ward2}
\tag{\ref{ward2}.a}
\begin{split}
\partialbarz\langle\,J^{a}_{z}\,\grl\,\rangle\sous{A}%
	&+if^{abc}\,A^{b}_{\overline{z}}\,%
	\langle\,J^{c}_{z}\,\grl\,\rangle\sous{A}\\
&=\Big({_k\over^2}\,\partial_zA^{a}_{\overline z}%
	-\pi\,\sum_\ell\delta^{(2)}(z-z_\ell )\,t^a_\ell \Big)\,%
	\langle\,\grl\,\rangle\sous{A},
\end{split}
\end{equation}
\begin{equation}
\tag{\ref{ward2}.b}
\begin{split}
\partial_{z}\langle\,J^{a}_{\overline z}\,\grl\,\rangle\sous{A}%
&+if^{abc}\,A^{b}_{z}\,%
\langle\,J^{c}_{\overline z}\,\grl\,\rangle\sous{A}\\
&=\langle\,\grl\,\rangle\sous{A}%
\Big({_k\over^2}\,\partialbarz A^{a}_{z}%
+\pi\,\sum_\ell\delta^{(2)}(z-z_\ell )\,t^a_\ell \Big).
\end{split}
\addtocounter{equation}{1}
\end{equation}
Nous appellerons ces deux \'egalit\'es les 
{\bol identit\'es de Ward chirales locales}. 
Derri\`ere ces \'equations se cachent
des \<<armes\>> particuli\`erement redoutables.
Tout de suite, elle permettent de d\'eterminer le comportement 
\`a courtes distances des fonctions 
de Green \`a une ou deux insertions de courants.

\subsection{D\'eveloppements \`a courtes distances I}

On reprend l'\'equation~(\ref{ward2}.a) en champ de jauge 
nul~(\footnote{Il suffit que $A$ soit localement nul,
\cad qu'il s'annule autour des points d'insertion.}),
\begin{equation}\label{kzinit}
\partialbarz\langle\,J^{a}_{z}\,\grl\,\rangle=%
-\pi\,\sum_\ell\delta^{(2)}(z-z_\ell )\,t^a_\ell \,%
\langle\,\grl\,\rangle.
\end{equation}
D\'ej\`a, on constate que $\langle\,J^{a}_{z}\,\otimes
_\ell g(\xi_\ell)\sous{R_\ell}\,\rangle$
est analytique en $z$ dans tout domaine ne contenant pas les points
d'insertion. De m\^eme, l'insertion de $J^{a}_{\overline z}$ conduit \`a
une fonction de Green analytique en $\overline{z}$ sauf aux points d'insertion.
Utilisons les relations 
$\pi\,\delta^{(2)}(z-w)=\partialbarz(z-w)^{-1}$ et
$\pi\,\delta^{(2)}(z-w)=\partial_{z}(\overline{z}-\overline{w})^{-1}$,
prises au sens des distributions.
On obtient des {\bol d\'eveloppements \`a courtes distances}
\begin{align}\label{ward6}
\tag{\ref{ward6}.a}
\langle\,J^{a}_{z}\,\grl\,\rangle&=%
-\sum_\ell{_{t^a_\ell }\over^{z-z_\ell }}\,%
\langle\,\grl\,\rangle+\cdots,\\
\tag{\ref{ward6}.b}
\langle\,J^{a}_{\overline z}\,\grl\,\rangle&=%
\langle\,\grl\,\rangle%
\sum_\ell{_{t^a_\ell }\over^{\overline{z}-\overline{z}_\ell }}+\cdots.%
\addtocounter{equation}{1}
\end{align}
Dans l'\'equation~(\ref{ward6}.a), les pointill\'es~$\cdots$ symbolisent des termes 
analytiques en $z$ autour des points d'insertion et, dans 
l'\'equation~(\ref{ward6}.b),
ils repr\'esentent des termes analytiques en $\overline z$ autour des
points d'insertion. 
Notons $J(z)$ (resp. $\overline{J}(\overline{z})$) l'insertion de $J_{z}$ 
(resp. $J_{\overline z}$) \`a l'int\'erieur des fonctions de Green, en 
champ de jauge (localement) nul.
On r\'e\'ecrit alors les \'egalit\'es~(\ref{ward6}.a-b) sous la 
forme plus compacte~:
\begin{align}\label{unpoint}
\tag{\ref{unpoint}.a}
J^{a}(z)\,g(\xi_\ell)\sous{R_\ell }&=%
-{_{t^a_\ell }\over^{z-z_\ell }}\,g(\xi)\sous{R_\ell }+\cdots,\\
\tag{\ref{unpoint}.b}
\barJ^{a}(\overline{z})\,g(\xi_\ell)\sous{R_\ell }&=%
g(\xi)\sous{R_\ell }{_{t^a_\ell }\over^{\overline{z}-\overline{z}_\ell }}+\cdots
\addtocounter{equation}{1}
\end{align}
\`a l'int\'erieur des fonctions de Green. 
Dans la suite, il faudra toujours garder \`a
l'esprit qu'on s'int\'eresse au comportement 
d'insertions \`a l'int\'erieur des fonctions
de Green, m\^eme si cela n'appara\^\i t pas explicitement.
Plus g\'en\'eralement, on conviendra que dans un d\'eveloppement 
\`a courtes distances du type
$\varphi(\zeta)\,\varphi(\zeta')$ les pointill\'es 
d\'esigneront des termes r\'eguliers en 
$\zeta$ autour de $\zeta'$.

Pour obtenir les d\'eveloppements
\`a courtes distances \`a deux courants, on commence par
d\'eriver l'\'equation~(\ref{ward2}.a) par rapport \`a $A^{b}_{\overline w}$~;
pr\'ecis\'ement, appliquons l'op\'erateur 
$\widetilde{Z}\sous{A}^{-1}\,\frac{-\pi\delta}{\delta A^{b}_{\overline{w}}}\,%
\widetilde{Z}\sous{A}$ \`a l'\'equation en question,
\begin{align}
\partialbarz\langle\,J^{a}_{z}\,J^{b}_{w}%
\,\grl\,\rangle\sous{A}&-i\pi\,f^{ab\beta}\,\delta^{(2)}(z-w)\,%
\langle\,J^{\beta}_{z}\,\grl\,\rangle\sous{A}\non\\
 &  +i\,f^{a\alpha\beta}\,A^{\alpha}_{\overline z}\,%
\langle\,J^{\beta}_{z}\,J^{b}_{w}\,\grl\,\rangle\sous{A}\,=
-{_k\over^2}\,\pi\,\delta^{ab}\,\partial_z\delta^{(2)}(z-w)\,%
 \langle\,\grl\,\rangle\sous{A}\non\\
 &  +\Big({_k\over^2}\,\partial_zA^{a}_{\overline z}%
-\pi\,\sum_\ell\delta^{(2)}(z-z_\ell )\,t^a_\ell \Big)\,%
\langle\,J^{b}_{w}\,\grl\,\rangle\sous{A}.
\label{deuxpoints}
\end{align}
En $A=0$,
\begin{multline*}
\partialbarz\langle\,J^{a}_{z}\,J^{b}_{w}%
\,\grl\,\rangle-i\pi\,f^{abc}\,\delta^{(2)}(z-w)\,%
\langle\,J^{c}_{z}\,\grl\,\rangle\,=\\
-{_k\over^2}\,\pi\,\delta^{ab}\,\partial_z\delta^{(2)}(z-w)\,%
 \langle\,\grl\,\rangle%
-\pi\,\sum_\ell\delta^{(2)}(z-z_\ell )\,t^a_\ell \,%
\langle\,J^{b}_{w}\,\grl\,\rangle\sous{A}.
\end{multline*}
Au passage, la fonction $\delta^{(2)}(z-w)$ dans le second terme permet 
de  remplacer $J^{c}_{z}$ par $J^{c}_{w}$.
Cela implique que
\qq
\langle\,J^{a}_{z}\,J^{b}_{w}%
\,\grl\,\rangle\,=\,{_{k\,\delta^{ab}/2}\over^{(z-w)^2}}%
\,\langle\,\grl\,\rangle+{_{if^{abc}}\over^{z-w}}\,%
\langle\,J^{c}_{w}\,\grl\,\rangle\non\\
-\sum_\ell{_{t^a_\ell }\over^{z-z_\ell }}\,%
\langle\,J^{b}_{w}\,\grl\,\rangle+\cdots
\label{ward7bis}
\qqq
\`a des termes analytiques pr\`es en $z$ autour de $w$ et $z_\ell $.
Pour $z$ et $w$ diff\'erents des $z_\ell $,
\begin{align}\label{ward7}
\tag{\ref{ward7}.a}
J^{a}(z)\,J^{b}(w)%
&={_{k\,\delta^{ab}/2}\over^{(z-w)^2}}%
+{_{if^{abc}}\over^{z-w}}\,J^{c}(w)+\cdots,\\
\tag{\ref{ward7}.b}
\barJ^{a}(\overline z)\,\barJ^{b}(\overline w)%
&={_{k\,\delta^{ab}/2}\over^{(\overline{z}-\overline{w})^2}}%
+{_{if^{abc}}\over^{\overline{z}-\overline{w}}}\,\barJ^{c}(\overline w)+\cdots.%
\nombre
\end{align}
Le d\'eveloppement \`a courtes distances pour les courants
anti-holomorphes s'obtient de la m\^eme fa\c{c}on, en utilisant
l'\'equation~(\refb{ward2}).
Il est clair que l'\'equation~(\refa{ward7}) peut \^etre reprise
sous une forme plus sym\'etrique~:
\begin{equation}
J^{a}(z)\,J^{b}(w)={_{k\,\delta^{ab}/2}\over^{(z-w)^2}}%
+{_{if^{abc}/2}\over^{z-w}}\,(J^{c}(z)+J^{c}(w))+\cdots,\label{ward74}
\end{equation}
cette fois les termes restants sont analytiques en $z$ et $w$ diff\'erents
des $z_\ell $. Un calcul similaire nous donnerait
\[
\langle\,J^{a}_{z}\,J^{b}_{\overline w}\,\grl\,\rangle=%
-\sum_\ell{{t^a_\ell }\over^{z-z_\ell }}\,\langle\,J^{b}_{\overline w}%
\,\grl\,\rangle+\cdots.
\]
Il suit $J^{a}(z)\,{\overline J}^{b}({\overline w})=\,\cdots$\,.

\subsection{Tenseur d'\'energie-impulsion}

Bien que jusque-l\`a on n'ait pas \'ecrit explicitement la d\'ependance des fonctions de Green vis-\`a-vis de la m\'etrique $\gamma$, celle-ci est essentielle. 
Le tenseur d'\'energie-impulsion est justement l'objet qui va \<<mesurer\>> pour nous
le comportement du syst\`eme dans un changement de m\'etrique.
On munit l'espace des m\'etriques de la topologie $\CC^\infty$ et
on suppose que les fonctions de Green d\'ependent de mani\`ere $\CC^\infty$
(au sens de Fr\'echet) de la m\'etrique. 
Soient $M$ points distincts
$\varsigma_{1},\cdots,\varsigma_{M}$
 pris sur la surface de Riemann. Par d\'efinition, l'insertion de 
{\bol tenseurs d'\'energie-impulsion} dans les fonctions de Green est
\[
\langle\,%
\prod_{k=1}^M T_{\mu\sous{k}\nu\sous{k}}(\varsigma\sous{k})\,%
\,\grl\,\rangle\sous{A,\gamma}%
\,\equiv\,\widetilde{Z}\sous{A,\gamma}^{-1}\,%
\frac{%
(4\pi\delta)^{M}%
	}{%
\mathop{\prod}\limits_k%
	\delta \gamma^{\mu\sous{k}\nu\sous{k}}(\varsigma\sous{k})
	}%
\,\Big(\widetilde{Z}\sous{A,\gamma}\,\langle\,\grl\,\rangle\sous{A,\gamma}\Big)
\]
o\`u on a r\'etabli la d\'ependance explicite dans la m\'etrique.
Si $z$ et $\overline z$ sont des coordonn\'ees locales en un point d'insertion $\varsigma$,
le tenseur d'\'energie-impulsion a quatre composantes $T_{zz}$, $T_{z\overline{z}}$,
$T_{\overline{z}z}$ et $T_{\overline{z}\,\overline{z}}$.
Notons qu'\`a l'int\'erieur des fonctions de Green, 
$T_{z\overline{z}}=T_{\overline{z}z}$ de part les sym\'etries 
de la m\'etrique.


La fonction de partition et les fonctions
de Green sont contraintes par
la covariance par rapport aux diff\'eomorphismes 
$D:\Sigma\rightarrow\Sigma$ pr\'eservant l'orientation :
\begin{gather*}
Z\sous{A,\gamma}=Z\sous{D^*A,D^*\gamma}\,,\qquad\qquad\widetilde Z\sous{A,\gamma}
=\widetilde Z\sous{D^*A,D^*\gamma}\,,\\
\langle\,\mathop{\otimes}\limits_\ell g(D(\xi_\ell))\sous{R_\ell}\,
\rangle\sous{A,\gamma}\ =\ 
\langle\,\mathop{\otimes}\limits_\ell g(\xi_\ell)\sous{R_\ell}\,
\rangle\sous{D^*A,D^*\gamma}\,.
\end{gather*}
Si on change la m\'etrique par une 
{\bol transformation de Weyl}\ $\gamma\rightarrow{\rm e}^\sigma\gamma$,
la fonction de partition et les fonctions de Green changent.
Pour la fonction de partition, le changement est donn\'e 
par l'\'equation
\begin{equation}\label{weyl1}
Z\sous{A,\ee^\sigma\gamma}\,=\,\ee^{\,\frac{ic}{24\pi}\,
S\sous{\rm L}(\sigma)}\,Z\sous{A,\gamma}
\end{equation}
o\`u $S\sous{\rm L}$ est l'action de Liouville
\[
S\sous{\rm L}(\sigma)=\int_\Sigma
	({_1\over^2}\,\partial\sigma\,\de\sigma+\sigma\,R\sous{\gamma})
\]
et le nombre $c=k\,\dim G/(k+g^\vee)$ est la {\bol charge centrale}. 
Les fonctions de Green ont le comportement suivant sous
les transformations de Weyl :
\begin{equation}\label{weyl2}
\langle\,\grl\,\rangle\sous{A,\ee^\sigma\gamma}=%
\prod_{l}\ee^{-\Delta_\ell \sigma(\xi_\ell )}\,%
\langle\,\grl\,\rangle\sous{A,\gamma}
\end{equation}
o\`u  $\Delta_\ell=C_\ell/(k+g^\vee)$ est le {\bol poids conforme} 
--- $C_\ell$ est le Casimir quadratique 
dans la repr\'esentation $R_\ell$. 
Le contenu de ces \'equations, exprimant {\bol l'anomalie 
conforme}, n'est pas \'el\'ementaire. 
On reporte leur d\'emonstration \`a des chapitres ult\'erieurs. 

Si on effectue une transformation de Weyl 
$\gamma\mapsto\ee^\sigma\gamma$, avec
$\sigma$ nulle autour des points d'insertion, les 
fonctions de Green ne changent pas. 
On dira qu'une m\'etrique est {\bol localement plate}
si elle est compatible avec la structure complexe
donn\'ee sur $\Sigma$ et si elle est plate 
autour des points d'insertion, \ie de la forme
$\vert dz\vert^2$. Dans ce cas, on supprimera l'indice 
$\gamma$ dans les fonctions de Green.
Gr\^ace aux propri\'et\'es~(\ref{weyl1}) et~(\ref{weyl2}),
on pourra retrouver les fonctions de Green en  m\'etrique quelconque.
Si on reprend la d\'efinition du tenseur d'\'energie-impulsion
et si on utilise l'\'equation ~(\ref{weyl1}), on obtient
\[
4\pi\,\widetilde{Z}^{-1}\sous{A,\gamma}\left.
{_\delta\over^{\delta\sigma}}\right|_{\sigma=0}
\widetilde{Z}\sous{A,\ee^\sigma\gamma}=
-\gamma^{zz}\langle\,T_{zz}\,\rangle\sous{A,\gamma}
-2\gamma^{z\overline{z}}\langle\,T_{z\overline{z}}\,\rangle\sous{A,\gamma}
-\gamma^{\overline{z}\,\overline{z}}\langle\,T_{\overline{z}
\,\overline{z}}\,\rangle\sous{A,\gamma}={_c\over^6}\,R_{\gamma}.
\]
Par cons\'equent, comme $R_{\gamma}=0$ en m\'etrique localement plate, 
on trouve
\[
\langle\,T_{z\overline{z}}\,\rangle\sous{A}=0=
\langle\,T_{\overline{z}z}\,\rangle\sous{A}.
\]
La covariance des fonctions de Green par rapport aux diff\'eomorphismes
infinit\'esimaux implique, si on travaille un peu plus \cite{gaw97:cft}, 
que
\[
\partialbarz\langle\,T_{zz}\,\rangle\sous{A}=0=\partial_z\langle\,
T_{\overline{z}\,\overline{z}}\,\rangle\sous{A}.
\]
Les \'egalit\'es $T_{z\bar z}=0=T_{\bar z z}$, $\partialbarz T_{zz}=0
=\partial_z T_{\bar z\,\bar z}$, en m\'etrique localement 
plate, tiennent aussi dans des fonctions de Green plus g\'en\'erales 
loin des autres points d'insertion.

Une {\bol transformation PCT} envoie la surface $\Sigma$ vers la surface conjugu\'ee 
$\overline{\Sigma}$, munie de la structure complexe conjugu\'ee.
Si $(z,\overline{z})$ est un syst\`eme de coordonn\'ees complexes induites par la 
structure complexe $\J$, on notera encore par $(z,\overline{z})$ les coordonn\'ees
complexes pour $-\J$. La coordonn\'ee \<<holomorphe\>> sera donc $z$ pour $\Sigma$
et $\overline{z}$ pour $\overline{\Sigma}$, et inversement pour la coordonn\'ee
anti-holomorphe.
Au niveau de l'action classique, 
le passage de $\Sigma$ \`a $\overline{\Sigma}$ 
se traduit par~(%
\footnote{L'action de WZNW d\'efinie sur la surface conjugu\'ee 
\`a $\Sigma$ est 
\[
k\,S\sous{\overline{\Sigma}}(g)=-{_{ik}\over^{4\pi}}\,\int_\Sigma\tr\,(g^{-1}\partial
g)\,(g^{-1}\de g)+{_{ik}\over^{12\pi}}\,%
\int_B\tr\,(\widetilde{g}^{-1}d\widetilde{g})^3\quad(\hbox{mod}\ 2\pi i).
\]
La partie sigma ne voit pas de diff\'erence entre $\Sigma$ et $\overline{\Sigma}$ car
l'\'echange $z\leftrightarrow \overline{z}$ est compens\'e par le changement 
d'orientation entre $\Sigma$ et $\overline{\Sigma}$.
})
\[
\overline{S\sous{\Sigma}(g,A)}=S\sous{\Sigma}(g^\dagger,-A^\dagger),\quad%
S\sous{\overline{\Sigma}}(g^{-1},A)=S\sous{\Sigma}(g,A).
\]
Si on reproduit la m\^eme op\'eration sur les int\'egrales fonctionnelles, on 
constate que
\[
\overline{Z}{}\sous{A,\Sigma}=Z\sous{-A^\dagger,\overline{\Sigma}},\qquad
\overline{\langle\,\grl\,\rangle}{}\sous{A,\Sigma}
=\langle\,\mathop{\otimes}\limits_\ell
g(\xi_\ell )\sous{\overline{R}_\ell }\,\rangle
\sous{-A^\dagger,\overline{\Sigma}}.
\]
Ce r\'esultat repose sur deux points fondamentaux~: d'abord, on 
int\`egre uniquement sur
les champs \`a valeurs dans le groupe compact, ensuite, 
$\overline{g(\xi_\ell )}{}\sous{R_\ell }=g(\xi_\ell )
\sous{\overline{R}_\ell }$ o\`u $\overline{g(\xi_\ell )}\sous{R_\ell }$
est \'el\'ement de $\overline{{\rm End}(V_{\lambda_\ell})}
\cong{\rm End}(\overline{V}_{\hspace{-0.06cm}\lambda_\ell})$.
Si on consid\`ere des fonctions de Green o\`u on a 
\'eventuellement ins\'er\'e des courants
ou le tenseur d'\'energie-impulsion, le ph\'enom\`ene persiste. En clair,
\begin{gather*}
\begin{aligned}
\overline{\langle\,J^{a}_{z}\,\grl\,\rangle}{}\sous{A,\Sigma}
        =&-\langle\,J^{a}_{\overline{z}}\,\mathop{\otimes}\limits_\ell
        g(\xi_\ell )\sous{\overline{R}_\ell }
        \,\rangle\sous{-A^\dagger,\overline{\Sigma}},\\
\overline{\langle\,T_{z z}\,\grl\,\rangle}{}\sous{A,\Sigma}
         =&\langle\,T_{\overline{z}\,\overline{z}}\,
         \mathop{\otimes}\limits_\ell
         g(\xi_\ell )\sous{\overline{R}_\ell }\,\rangle
         \sous{-A^\dagger,\overline{\Sigma}},
\end{aligned}
\begin{aligned}
\bar{\langle\,J^a_{\bar z}\,\grl\,\rangle}{}\sous{A,\Sigma}
        =&-\langle\,J^a_z\,\mathop{\otimes}\limits_\ell
        g(\xi_\ell )\sous{\bar R_\ell}
        \,\rangle\sous{-A^\dagger,\bar \Sigma},\\
\bar{\langle\,T_{\bar z\,\bar z}\,\grl\,
        \rangle}{}\sous{A,\Sigma} =&
        \langle\,T_{zz}\,\mathop{\otimes}\limits_\ell
        g(\xi_\ell)\sous{\overline{R}_\ell}\,\rangle
        \sous{-A^\dagger,\overline{\Sigma}},
\end{aligned}\\
\overline{\langle\,T_{z\overline{z}}\,\grl\,\rangle}{}\sous{A,\Sigma}
	 =\langle\,T_{\overline{z}z}\,\mathop{\otimes}\limits_\ell
g(\xi_\ell )\sous{\overline{R}_\ell }\,\rangle\sous{-A^\dagger,
\overline{\Sigma}}.
\end{gather*}
La covariance par rapport aux diff\'eomorphismes et aux transformations
de Weyl ainsi que la sym\'etrie PCT traduisent les 
propri\'et\'es conformes de la th\'eorie quantique.

Dans la suite, on notera par $T(z)$ (resp.~$\barT(\overline{z})$) l'insertion de
$T_{zz}$ (resp.~$T_{\overline{z}\,\overline{z}}$) dans les fonctions de Green, en 
m\'etrique localement plate et en champ de jauge (localement) nul.
En fait, le tenseur d'\'energie-impulsion n'est pas ind\'ependant des courants.
Ainsi, on verra plus tard que $T$ et $\barT$ peuvent \^etre obtenus
gr\^ace \`a la {\bol construction de Sugawara}~\cite{sugawara}
\begin{align}
\label{sugawara}
\taga{sugawara}
T(z)&\,\equiv\,{_2\over^{k+g^\vee}}\,\mathop{\lim}\limits_{w\rightarrow z}%
\left(\tr\,J(w)\,J(z)-{_{k\,\dim G}\over^{4\,(w-z)^2}}\right),\\
\tagb{sugawara}
\barT(\overline{z})&\,\equiv\,{_2\over^{k+g^\vee}}\,\mathop{\lim}%
\limits_{\overline{w}\rightarrow \overline{z}}%
\left(\tr\,\barJ(\overline{w})\,\barJ(\overline{z})-%
{_{k\,\dim G}\over^{4\,(\overline{w}-\overline{z})^2}}\right).
\nombre
\end{align}
On peut interpr\'eter ce r\'esultat
comme le versant quantique de l'\'egalit\'e classique 
$\left.T_{zz}\right|^{\text{class.}}=\frac{2}{k}\,\tr\,(
\left.J_z\right|^{\text{class.}})^2$ et idem avec $\overline{z}$.
Au niveau classique, on d\'efinit et on calcule~:
\begin{gather*}
\label{SUGACOUR}
\left.J_{z}\right|^{\text{class.}}
	\equiv{_{\pi k\,\delta}\over^{\delta A_{\overline{z}}}}
	\,S(g,A)=
	-{_k\over^2}\,(\partial_zg+[A_{z},g])\,g^{-1},\\
\left.J_{\overline{z}}\right|^{\text{class.}}
	\equiv{_{\pi k\,\delta}\over^{\delta A_{z}}}
	\,S(g,A)=
	{_k\over^2}\,g^{-1}\,(\partialbarz g+[A_{\overline{z}},g]),\\
\left.T_{\mu\nu}\right|^{\text{class.}}
	\equiv-{_{4\pi k\,\delta}\over^{\delta\gamma^{\mu\nu}}}\,S(g,A),\\
\left.T_{zz}\right|^{\text{class.}}=\quotient{k}{2}\,
	\sqrt{\gamma}\,\tr\,\big(g^{-1}\,(\partial_zg+[A_{z},g])\big)^2,\\
\left.T_{\bar{z}\,\bar{z}}\right|^{\text{class.}}=\quotient{k}{2}\,
	\sqrt{\gamma}\,\tr\,\big(g^{-1}\,(\partialbarz g+[A_{z},g])\big)^2,\\
	T_{z\bar{z}}=0=T_{\bar{z}\,z}.
\end{gather*}
En m\'etrique localement plate, on a 
$\sqrt{\smash[b]{\gamma}}=1$, d'o\`u le r\'esultat.
Au niveau classique, la cons\-truction de Sugawara reste vraie en
champ de jauge non-nul.

L'introduction du tenseur d'\'energie-impulsion permet d'expliciter deux termes 
suppl\'ementaires dans le d\'eveloppement \`a courtes distances~(\refa{ward7}) 
restreint \`a $a=b$ et somm\'e sur $a$, 
\qq
\label{longdeuxpoints}
J^{a}(z)\,J^{a}(w)={_{k\,\dim G}\over^{2\,(z-w)^2}}+(k+g^\vee)\,T(w)+{_{k+g^\vee}\over^2}\,
(z-w)\,\partial_wT(w)+\cdots
\qqq
o\`u les termes restants sont analytiques en $z$ et $w$.
La d\'emonstration repose sur la sym\'etrie en $z$ et $w$. En effet, si on reprend 
ce d\'eveloppement en \'echangeant $z$ et $w$, on doit trouver le m\^eme r\'esultat. 
D\'ej\`a, le terme constant est trivialement donn\'e par la construction de Sugawara. 
On a donc $J^{a}(z)\,J^{a}(w)=\frac{k\,
\dim G}{2\,(z-w)^2}+(k+g^\vee)\,T(w)+(z-w)\,\CF(w)+\cdots$,
o\`u $\CF$ est une fonction analytique. La diff\'erence entre cette \'equation et sa sym\'etris\'ee
donne~: $0=(k+g^\vee)\,(T(w)-T(z))+(z-w)\,(\CF(z)+\CF(w))+\cdots$. On conclut 
facilement apr\`es avoir d\'evelopp\'e $T(w)$ et $\CF(w)$ autour de $z$.

\subsection{ D\'eveloppements \`a courtes distances II}%

On peut d\'eduire les d\'eveloppements 
\`a courtes distances faisant intervenir le tenseur 
\'energie-impulsion \`a partir des propri\'et\'es conformes des fonctions de Green,
n\'eanmoins, on va plut\^ot partir de la construction de Sugawara.
Avant, fixons quelques termes propres au jargon conforme~\cite{bpz}. 

Une {\bol transformation conforme locale} dans l'espace muni d'une 
m\'etrique riemannienne ou pseudo-riemannienne 
est un changement de coordonn\'ees locales qui ne change 
pas la classe conforme de la m\'etrique. Une telle transformation 
conserve les angles. La particularit\'e 
de la dimension $2$ est que toute transformation analytique 
locale du plan complexe muni de la m\'etrique $\vert dz\vert^2$ 
est conforme. L'inverse est aussi vrai pour les transformations
qui pr\'eservent l'orientation. Si on compactifie le plan passant 
au $\C P^1$, les seules transformations conformes globales 
qui pr\'eservent l'orientation sont les transformations
projectives $z\mapsto{az+b\over cz+d}$ de $\C P^1$. 
Ces derni\`eres forment le {\bol groupe conforme global} 
(ou projectif), not\'e $\confproj=\Slc/\Z_2$.

Reprenons l'\'etude pour une surface de Riemann compacte.
On dit qu'un champ $\varphi(z,\overline{z})$ est {\bol primaire}\label{primaire}, de
poids conforme $(\Delta, \overline{\Delta})$, si toute transformation
conforme locale laisse invariante la forme 
$\varphi(z,\overline{z})\,dz^\Delta d\overline{z}^{\overline{\Delta}}$. Les 
quantit\'es $\Delta$ et $\overline{\Delta}$ sont r\'eelles et ind\'ependantes.
La diff\'erence $s\equiv\Delta-\overline{\Delta}$ 
repr\'esente le {\bol spin} du champ~; on suppose
que $s$ est entier ou demi-entier. La somme $d\equiv\Delta+\overline{\Delta}$ est 
la {\bol dimension} du champ. G\'eom\'etriquement, on peut 
voir un champ primaire comme
une section de $K^\Delta\otimes \bar K^{\bar\Delta}$ ou de $K^s$, 
o\`u $K$ est le fibr\'e canonique, \cad le fibr\'e cotangent holomorphe,
au-dessus de $\Sigma$. En effet, gr\^ace \`a la m\'etrique 
riemannienne $\gamma$, on peut identifier le 
fibr\'e cotangent anti-holomorphe $\bar K$ avec le
fibr\'e en droite dual $K^{-1}$.
Un champ {\bol quasi-primaire} est un champ qui 
ne poss\`ede la propri\'et\'e pr\'ec\'edente
que pour les changements de coordonn\'ees appartenant 
\`a $\confproj$. Enfin, un champ 
{\bol secondaire} 
(ou {\bol descendant}) est un champ qui n'est ni primaire, 
ni quasi-primaire. Pour ce qui nous int\'eresse ici,
les champs $g(\xi_\ell )\sous{R_\ell }$ 
sont des champs primaires de poids conforme 
$(\Delta_\ell ,\Delta_\ell )$.
On en reporte la  d\'emonstration au chapitre 4. 
N\'eanmoins, remarquons d\'ej\`a que l'\'equation~(\ref{weyl2}) 
indique que 

---~$g(\xi_\ell)\sous{R_\ell}$
est un champ primaire de poids $(\Delta_\ell ,\Delta_\ell )$. 
En effet, si on 
fixe des coordonn\'ees locales $(z_\ell ,\overline{z}_\ell )$ 
au point $\xi_\ell $, on a
\[
\langle\,g(z'_\ell ,\overline{z}'_\ell )\sous{R_\ell }\,\rangle\sous{A}=
\langle\,g(z_\ell ,\overline{z}_\ell )\sous{R_\ell }\,
\rangle\sous{A,\left|dz'_\ell \,/\,dz_\ell \right|^2\,dz_\ell \,d\overline{z}_\ell }
=\left|{_{dz'_\ell }\over^{dz_\ell }}\right|^{-2\Delta_\ell }\,
\langle\,g(z_\ell ,\overline{z}_\ell )\sous{R_\ell }\,\rangle\sous{A}
\]
dans un changement de coordonn\'ees locales $z_\ell \rightarrow z'_\ell $~;

---~les courants $J^{a}(z)$ et $\barJ^{a}(\overline{z})$ sont des champs primaires 
		de poids conformes respectifs $(1,0)$ et $(0,1)$.
Cette assertion signifie juste que $J^{a}(z)\,dz$ (resp. 
$\overline{J}^{a}(\overline{z})\,d\overline{z}$) 
est une $(1,0)$-forme (resp. $(0,1)$-forme).
Ceci provient de la d\'efinition m\^eme des courants~;

---~les tenseurs $T(z)$ et $\barT(\overline{z})$ sont des champs quasi-primaires
		de poids conformes respectifs $(2,0)$ et $(0,2)$.
Si on effectue un changement de coordonn\'ees dans la construction de Sugawara, 
on obtient
facilement que
\begin{align*}
T(z')\,dz'{}^2&=T(z)\,dz^2-{_c\over^{12}}\,\{z';z\}\,dz^2\\
\barT(\overline{z}')\,d\overline{z}'{}^2&=\barT(\overline{z})\,d\overline{z}^2-
	{_c\over^{12}}\,\{\overline{z}';\overline{z}\}\,d\overline{z}^2
\end{align*}
o\`u
\[
\{z';z\}\equiv {_{d^3z'\,/\,dz^3}\over^{dz'\,/\,dz}}-
{_3\over^2}\,\left({_{d^2z'\,/\,dz^2}\over^{dz'\,/\,dz}}\right)^2
\]
est la {\bol d\'eriv\'ee de Schwarz} d'un changement de variable.  
La d\'eriv\'ee de Schwarz est nulle si, et seulement 
si, le changement de coordonn\'ees 
appartient \`a $\confproj$. La loi de 
transformation du tenseur d'\'energie-impulsion
d\'efinit une {\bol connexion projective} 
sur la surface $\Sigma$. 
G\'eom\'etriquement, la donn\'ee d'une connexion projective est 
\'equivalente \`a la donn\'ee d'une structure projective sur 
la surface sp\'ecifi\'ee par un atlas des coordonn\'ees
qui s'accordent mo\-dulo transformations projectives~\cite{tyurin}.

On consacre la fin de cette section \`a d\'emontrer les d\'eveloppements
\`a courtes distances faisant intervenir le tenseur d'\'energie-impulsion. 
Le proc\'ed\'e utilise la construction de Su\-gawara et les d\'eveloppements
\`a courtes distances obtenus pour les courants. Il n'est pas inutile
de dire qu'il est possible d'obtenir tous les d\'eveloppements qui
suivent uniquement \`a partir des propri\'et\'es conformes
de la th\'eorie~\cite{gaw97:cft}. Les d\'emonstrations n'en sont que plus
claires. En fait, la m\'ethode choisie ici ne permet pas de trouver
le d\'eveloppement~\eqref{ward12}. Pour cela il faudrait d'abord 
g\'en\'eraliser la construction de Sugawara en champ de jauge non-nul.
C'est possible mais un peu inutile puisqu'on peut d\'emontrer le
d\'eveloppement directement.
\begin{align}
T(z)\,g(\xi_\ell )\sous{R_\ell }&=%
	{_{\Delta_\ell }\over^{(z-z_\ell )^2}}\,%
	g(\xi_\ell )\sous{R_\ell }+%
	{_1\over^{z-z_\ell }}\,\partial_{z_\ell }%
	g(\xi_\ell )\sous{R_\ell }+\cdots,
	\label{ward9}\taga{ward9}\\
\barT({\overline z})\,g(\xi_\ell )\sous{R_\ell }&=%
	\,{_{\Delta_\ell }\over^{(\overline{z}-\overline{z}_\ell )^2}}\,%
	g(\xi_\ell )\sous{R_\ell }+%
	{_1\over^{\overline{z}-\overline{z}_\ell }}\,\partial_{
	\hspace{0.02cm}\overline{z}_\ell }%
	g(\xi_\ell )\sous{R_\ell }+\cdots,
	\tagb{ward9}\nombre\\
T(z)\,J^{a}(w)&= {_1\over^{(z-w)^2}}\,J^{a}(w)+%
	{_1\over^{z-w}}\,\partial_wJ^{a}(w)+\cdots\label{tenseur1}
	\taga{tenseur1},\\
\barT(\overline{z})\,\barJ^{a}(w)&=
	{_1\over^{(\overline{z}-\overline{w})^2}}%
	\,\barJ^{a}(\overline{w})+{_1\over^{\overline{z}-\overline{w}}}%
	\,\partialbarw\barJ^{a}(\overline{w})+\cdots,
	\tagb{tenseur1}\nombre\\
T(z)\,\barJ^{a}(w)&= \cdots,\label{mixte1}
	\taga{mixte1}\phantom{{_1\over^{(z-w)^2}}}\\
\barT(z)\,J^{a}(w)&= \cdots,\tagb{mixte1}\nombre\\
T(z)\,T(w)&={_{c/2}\over^{(z-w)^4}}+{_2\over^{(z-w)^2}}\,T(w)%
	+{_1\over^{z-w}}\partial_wT(w)+\cdots\label{ward11}
	\taga{ward11},\\
\barT(\overline{z})\,\barT(\overline{w})&=%
	\,{_{c/2}\over^{(\overline{z}-\overline{w})^4}}+
	{_2\over^{(\overline{z}-\overline{w})^2}}\,%
	\barT(\overline{w})+{_1\over^{\overline{z}-\overline{w}}}%
	\partialbarw\barT(\overline{w})+\cdots,
	\tagb{ward11}\nombre\\
T(z)\,\barT(\overline{w})&=-{_{\pi c}\over^{12}}\,\partial_z
	\partialbarz\,\delta^{(2)}(z-w)+\cdots.\label{ward12}
\end{align}

Pour conna\^\i tre le comportement des fonctions de Green o\`u on ins\`ere trois
courants, on commence par d\'eriver l'\'equation~(\ref{deuxpoints}) par rapport \`a 
$A^{c}_{\overline y}$, 
\begin{equation}\label{troispoints}
\begin{split}
\partialbarz&\langle\,J^{a}_{z}\,J^{b}_{w}%
\,J^{c}_{y}\,\grl\,\rangle\sous{A}%
-i\pi\,f^{ab\mu}\,\delta^{(2)}(z-w)\,%
\langle\,J^{\mu}_{z}\,J^{c}_{y}\,\grl\,\rangle\sous{A}\\
& -i\pi\,f^{ac\mu}\,\delta^{(2)}(z-y)\,%
\langle\,J^{\mu}_{z}\,J^{b}_{w}\,\grl\,\rangle\sous{A}%
+i\,f^{a\nu\mu}\,A^{\nu}_{\overline z}\,%
\langle\,J^{\mu}_{z}\,J^{b}_{w}\,%
J^{c}_{y}\,\grl\,\rangle\sous{A}\\
=& -{_k\over^2}\,\pi\,\delta^{ab}\,\partial_z\delta^{(2)}(z-w)\,%
 \langle\,J^{c}_{y}\,\grl\,\rangle\sous{A}%
 -{_k\over^2}\,\pi\,\delta^{ac}\,\partial_z\delta^{(2)}(z-y)\,%
 \langle\,J^{b}_{w}\,\grl\,\rangle\sous{A}\\
& +\Big({_k\over^2}\,\partial_zA^{a}_{\overline z}%
-\pi\,\sum_\ell\delta^{(2)}(z-z_\ell )\,t^a_\ell \Big)\,%
\langle\,J^{b}_{w}\,J^{c}_{y}\,\grl\,\rangle\sous{A}.
\end{split}
\end{equation}
Gr\^ace aux distributions de Dirac, on remplace $J^{\mu}_{z}$
par $J^{\mu}_{w}$, dans le seconde terme, et par
$J^{\mu}_{y}$, dans le troisi\`eme.
En champ de jauge nul, on int\`egre cette \'equation par 
rapport \`a $\overline{z}$
\begin{equation}
\begin{split}\label{nonsymtrois}
\langle\,J^{a}_{z}\,J^{b}_{w}\,J^{c}_{y}%
\,\grl\,\rangle=&\,
{_{k\,\delta^{ab}/2}\over^{(z-w)^2}}\,%
\langle\,J^{c}_{y}\grl\,\rangle+%
{_{k\,\delta^{ac}/2}\over^{(z-y)^2}}\,%
\langle\,J^{b}_{w}\grl\,\rangle\\
&+{_{if^{ab\mu}}\over^{z-w}}\,%
\langle\,J^{\mu}_{w}\,J^{c}_{y}\,\grl\,\rangle
+{_{if^{ac\mu}}\over^{z-y}}\,%
\langle\,J^{\mu}_{y}\,J^{b}_{w}\,\grl\,\rangle\\
& -\sum_\ell{_{t^a_\ell }\over^{z-z_\ell }}\,%
\langle\,J^{b}_{w}\,J^{c}_{y}\,\grl\,\rangle+\cdots,
\end{split}
\end{equation}
\`a des termes analytiques pr\`es en $z$ autour de $w$, $y$ et $z_\ell $.
Le comportement en $z_\ell $ ne nous int\'eressant pas, on ne tient pas
compte des termes en $1/(z-z_\ell )$. Par contre, on a aussi besoin du comportement 
des fonctions de Green par rapport \`a $w$ et $y$. On peut aussi calculer
les d\'eriv\'ees partielles par rapport \`a $\overline w$ et $\overline y$. 
On conna\^\i t donc le comportement des fonctions 
$J^{a}(z)\,J^{b}(w)\,J^{c}(y)$ par
rapport \`a $z$, $y$ et $w$. N\'eanmoins, on ne peut pas directement \<<sym\'etriser\>>
l'\'equation~(\ref{nonsymtrois}). Pour le moment, on peut juste dire qu'il existe deux
fonctions $\varrho^{abc}$ et $\chi^{abc}$, ne d\'ependant pas de $w$, telles que 
\begin{multline}\label{symtroistmp}
J^{a}(z)\,J^{b}(w)\,J^{c}(y)=%
	{_{k\,\delta^{ab}/2}\over^{(z-w)^2}}\,J^{c}(y)+%
	{_{k\,\delta^{ac}/2}\over^{(z-y)^2}}\,J^{b}(w)+%
	{_{1}\over^{(w-y)^2}}\,\varrho^{abc}(z,y)\\
+{_{if^{ab\mu}}\over^{z-w}}\,J^{\mu}(w)\,J^{c}(y)%
	+{_{if^{ac\mu}}\over^{z-y}}\,J^{\mu}(y)\,J^{b}(w)%
	+{_{1}\over^{w-y}}\,\chi^{abc}(z,y)+\cdots,
\end{multline}
\`a des termes analytiques pr\`es en $z$, $w$ et $y$, diff\'erents des $z_\ell $.
On commence par calculer la limite, quand $w$ tend vers $y$, de 
$(w-y)^2\,J^{a}(z)\,J^{b}(w)\,J^{c}(y)$ d'une part gr\^ace \`a l'\'equation~%
(\ref{symtroistmp}), d'autre part gr\^ace \`a l'\'egalit\'e analogue \`a 
l'\'equation~(\ref{nonsymtrois}) et donnant le comportement vis \`a vis de $w$.
Si on compare les r\'esultats, on trouve $\varrho^{abc}$. 
Pour d\'eterminer $\chi^{abc}$,
on proc\`ede de la m\^eme mani\`ere, 
mais avec $\mathop{\lim}\limits_{w\rightarrow y}\,
(w-y)\,(J^{a}(z)\,J^{b}(w)\,J^{c}(y)-\frac{1}{(w-y)^2}\,\varrho^{abc}(z,y))$.
Finalement, 
\begin{equation}
\begin{split}\label{symtrois}
J^{a}(z)\,J^{b}(w)\,J^{c}(y)=&\,
	{_{k\,\delta^{ab}/2}\over^{(z-w)^2}}\,J^{c}(y)+%
	{_{k\,\delta^{ac}/2}\over^{(z-y)^2}}\,J^{b}(w)+%
	{_{k\,\delta^{bc}/2}\over^{(w-y)^2}}\,J^{a}(z)\\
&+{_{if^{ab\mu}}\over^{z-w}}\,J^{\mu}(w)\,J^{c}(y)%
	+{_{if^{ac\mu}}\over^{z-y}}\,J^{\mu}(y)\,J^{b}(w)%
	+{_{if^{bc\mu}}\over^{w-y}}\,J^{\mu}(y)\,J^{a}(z)\\
& -{_{ik\,f^{abc}/2}\over^{(w-y)\,(z-y)^2}}+%
	{_{f^{bc\mu}\,f^{a\mu\nu}}\over^{(w-y)\,(z-y)}}\,J^{\nu}(y)%
	+\cdots,
\end{split}
\end{equation}
\`a des termes analytiques pr\`es en $z$, $w$, $y$ diff\'erents des $z_\ell $.
\`A premi\`ere vue, cette \'equation n'est pas sym\'etrique en $z$, $w$ et $y$. 
En fait, cette sym\'etrie est cach\'ee. 
Par exemple, si on \'echange les r\^oles tenus
par $(a,z)$ et $(b,w)$, on v\'erifie 
ais\'ement que la diff\'erence entre le r\'esultat obtenu et
l'\'equation~(\ref{symtrois}) est analytique dans les trois variables. 
Au d\'epart, pour sym\'etriser l'\'equation~(\ref{nonsymtrois}), 
on devait rajouter des termes
analytiques en $z$ exhibant le p\^ole d'ordre $2$ en $w=y$.
Quand on regarde de plus pr\`es l'\'equation~(\ref{symtrois}), 
on constate qu'on a
justement rajout\'e les termes \<<\'evidents\>> 
provenant du troisi\`eme p\^ole, 
mais sans les termes singuliers en $z$.
En $b=c$, on obtient  
\begin{equation*}
\begin{split}
J^{a}(z)\,\left(J^{b}(w)\,J^{b}(y)-{_{k\,\dim G}\over^{2(w-y)^2}}\right)=&\,
{_{k/2}\over^{(z-w)^2}}\,J^{a}(y)+{_{k/2}\over^{(z-y)^2}}\,J^{a}(w)\\
&+{_{if^{ab\mu}}\over^{z-w}}\,J^{\mu}(w)\,J^{b}(y)%
	+{_{if^{ab\mu}}\over^{z-y}}\,J^{\mu}(y)\,J^{b}(w)+\cdots
\end{split}
\end{equation*}
Si on utilise les propri\'et\'es de sym\'etrie des constantes de structure
de l'alg\`ebre de Lie, on constate qu'il n'y a r\'eellement qu'une fonction 
\`a deux points, soit $J^{b}(y)\,J^{c}(w)$, pond\'er\'ee par
\[
if^{abc}\,\left({_1\over^{z-w}}-{_1\over^{z-y}}\right)=
-if^{abc}\,{_1\over^{z-w}}\,%
\left({_{y-w}\over^{z-w}}+\cdots\right).
\]
Ensuite, on introduit le d\'eveloppement~(\refa{ward7}) et on prend 
la limite quand $y\rightarrow w$
\[
(k+g^\vee)\,J^{a}(z)\,T(w)={_{f^{abc}\,f^{bcd}}\over^{(z-w)^2}}\,J^{d}(w)
+{_{k}\over^{(z-w)^2}}\,J^{a}(w)+\cdots.
\]
Or $f^{abc}\,f^{bcd}=g^\vee\,\delta^{ad}$, donc 
$J^{a}(z)\,T(w)=\frac{1}{(z-w)^2}\,J^{a}(w)+\cdots$.
Cette \'equation exprime le comportement de $J^{a}(z)\,T(w)$ quand $z$ tend vers
$w$~; nous, c'est l'inverse qui nous int\'eresse.
Pour conclure, il suffit de remplacer
$J^{a}(w)$ par $J^{a}(z)+(w-z)\,\partial_zJ^{a}(z)+\cdots$.
D\'emontrer le d\'eveloppement~(\refb{tenseur1}) ne suscite aucune difficult\'e
suppl\'ementaire~; quant aux \'equations~(\ref{mixte1}.a-b), elles suivent
des \'egalit\'es 
\begin{align}
\label{troismixte1}
\taga{troismixte1}
J^{a}(z)\,J^{b}(w)\,\barJ^{c}(\overline{y})&=%
	{_{i\,f^{ab\mu}}\over^{z-w}}\,J^{\mu}(w)\,\barJ^{c}(\overline{y})%
	+{_{k\,\delta^{ab}/2}\over^{(z-w)^2}}\,\barJ^{c}(\overline{y})+\cdots,\\
\tagb{troismixte1}
\barJ^{a}(\overline{z})\,\barJ^{b}(\overline{w})\,J^{c}(y)&=%
	{_{i\,f^{ab\mu}}\over^{\overline{z}-\overline{w}}}\,\barJ^{\mu}(\overline{w})%
	\,J^{c}(y)%
	+{_{k\,\delta^{ab}/2}\over^{(\overline{z}-\overline{w})^2}}\,J^{c}(y)+\cdots,
\nombre
\end{align}
o\`u les termes r\'esiduels, dans 
l'\'equation~(\refa{troismixte1})~(resp.~(\refb{troismixte1})),
sont analytiques en $z$, $w$ et $\overline{y}$~(resp.~$\overline{z}$, 
$\overline{w}$ et $y$),
loin des points $\xi_\ell $. 

Lorsqu'on d\'esire ins\'erer deux tenseurs d'\'energie-impulsion, comme
dans les \'equations~(\ref{ward11}.a-b),
on doit d'abord d\'eriver 
l'\'equation~(\ref{troispoints}) par rapport \`a $A^{d}_{\overline{v}}$,
\begin{eqnarray*}
\lefteqn{\partialbarz\langle\,J^{a}_{z}\,J^{b}_{w}%
\,J^{c}_{y}\,J^{d}_{v}\,\grl\,\rangle\sous{A}%
-i\pi\,f^{ab\mu}\,\delta^{(2)}(z-w)\,%
\langle\,J^{\mu}_{z}\,J^{c}_{y}\,J^{d}_{v}\,%
\grl\,\rangle\sous{A}}\\
 & & -i\pi\,f^{ac\mu}\,\delta^{(2)}(z-y)\,%
\langle\,J^{\mu}_{z}\,J^{b}_{w}\,J^{d}_{v}\,%
\grl\,\rangle\sous{A}%
-i\pi\,f^{ad\mu}\,\delta^{(2)}(z-v)\,%
\langle\,J^{\mu}_{z}\,J^{b}_{w}\,J^{c}_{y}\,%
\grl\,\rangle\sous{A}\\
 & & +i\,f^{a\nu\mu}\,A^{\nu}_{\overline z}\,%
\langle\,J^{\beta}_{z}\,J^{b}_{w}\,%
J^{c}_{y}\,J^{d}_{v}\,\grl\,\rangle\sous{A}\,=%
-{_k\over^2}\,\pi\,\delta^{ab}\,\partial_z\delta^{(2)}(z-w)\,%
 \langle\,J^{c}_{y}\,J^{d}_{v}\,\grl\,\rangle\sous{A}\\
 & & -{_k\over^2}\,\pi\,\delta^{ac}\,\partial_z\delta^{(2)}(z-y)\,%
 \langle\,J^{b}_{w}\,J^{d}_{v}\,\grl\,\rangle\sous{A}%
-{_k\over^2}\,\pi\,\delta^{ad}\,\partial_z\delta^{(2)}(z-v)\,%
 \langle\,J^{b}_{w}\,J^{c}_{y}\,\grl\,\rangle\sous{A}\\
 & & +\Big({_k\over^2}\,\partial_zA^{a}_{\overline z}%
-\pi\,\sum_\ell\delta^{(2)}(z-z_\ell )\,t^a_\ell \Big)\,%
\langle\,J^{b}_{w}\,J^{c}_{y}\,J^{d}_{v}%
\,\grl\,\rangle\sous{A}.
\end{eqnarray*}
Comme pr\'ec\'edemment, on int\`egre cette \'equation, en champ 
de jauge nul, par rapport \`a $\overline z$,
\begin{setlength}{\multlinegap}{0pt}
\begin{multline*}
J^{a}(z)\,J^{b}(w)\,J^{c}(y)\,J^{d}(v)=%
	{_{k\,\delta^{ab}/2}\over^{(z-w)^2}}\,J^{c}(y)\,J^{d}(v)%
	+{_{k\,\delta^{ac}/2}\over^{(z-y)^2}}\,J^{b}(w)\,J^{d}(v)%
	+{_{k\,\delta^{ad}/2}\over^{(z-v)^2}}\,J^{b}(w)\,J^{c}(y)\\
+{_{if^{ab\mu}}\over^{z-w}}\,J^{\mu}(w)\,J^{c}(y)\,J^{d}(v)%
	+{_{if^{ac\mu}}\over^{z-y}}\,J^{\mu}(y)\,J^{b}(w)\,J^{d}(v)%
	+{_{if^{ad\mu}}\over^{z-v}}\,J^{\mu}(v)\,J^{b}(w)\, J^{c}(y)+\cdots,
\end{multline*}
\end{setlength}
\`a des termes analytiques pr\`es en $z$ autour de $w$, $y$ et $v$ et loin des points $\xi_\ell $.
Un calcul similaire \`a celui qui nous conduisait vers l'\'equation~(\ref{symtrois})
--- mais beaucoup plus long --- permet d'expliciter compl\`etement
le comportement de $J^{a}(z)\,J^{b}(w)\,J^{c}(y)\,J^{d}(v)$
par rapport \`a $z$, $w$, $y$ et $v$~:
\begin{equation}
\label{symquatre}
\begin{split}
J^{a}(z)\,J^{b}(&w)\,J^{c}(y)\,J^{d}(v)=
	{_{k\,\delta^{ab}/2}\over^{(z-w)^2}}\,J^{c}(y)\,J^{d}(v)
	+{_{k\,\delta^{ac}/2}\over^{(z-y)^2}}\,J^{b}(w)\,J^{d}(v)
	+{_{k\,\delta^{ad}/2}\over^{(z-v)^2}}\,J^{b}(w)\,J^{c}(y)\\
&+{_{if^{ab\mu}}\over^{z-w}}\,J^{\mu}(w)\,J^{c}(y)\,J^{d}(v)%
	+{_{if^{ac\mu}}\over^{z-y}}\,J^{\mu}(y)\,J^{b}(w)\,J^{d}(v)%
	+{_{if^{ad\mu}}\over^{z-v}}\,J^{\mu}(v)\,J^{b}(w)\, J^{c}(y)\\
&+{_1\over^{(w-y)^2}}\,\varrho^{abcd}_{1}(z,y,v)
	+{_1\over^{(w-v)^2}}\,\varrho^{abcd}_{2}(z,y,v)
	+{_1\over^{(y-v)^2}}\,\varrho^{abcd}_{3}(z,w,v)\\
&+{_1\over^{w-y}}\,\chi^{abcd}_{1}(z,y,v)
	+{_1\over^{w-v}}\,\chi^{abcd}_{2}(z,y,v)
	+{_1\over^{y-v}}\,\chi^{abcd}_{3}(z,w,v)+\cdots
\end{split}
\end{equation}
\`a des termes analytiques pr\`es en $z$, $w$, $y$ et $v$, diff\'erents des $z_\ell $.
Les fonctions $\varrho^{abcd}_{\bullet}$ et $\chi^{abcd}_{\bullet}$ sont donn\'ees par
\begin{align*}
\varrho^{abcd}_{1}(z,y,v)=&\,{\scriptstyle k\,\delta^{bc}/2}\,J^{a}(z)\,J^{d}(v)
		-{_{ik\,\delta^{bc}\,f^{ad\mu}/2}\over^{z-v}}\,J^{\mu}(v)
		-{_{k^2\,\delta^{ad}\,\delta^{bc}/4}\over^{(z-v)^2}},\\
\varrho^{abcd}_{2}(z,y,v)=&\,{\scriptstyle k\,\delta^{bd}/2}\,J^{a}(z)\,J^{c}(y)
		-{_{ik\,\delta^{bd}\,f^{ac\mu}/2}\over^{z-y}}\,J^{\mu}(y)
		-{_{k^2\,\delta^{ac}\,\delta^{bd}/4}\over^{(z-y)^2}},\\
\varrho^{abcd}_{3}(z,w,v)=&\,{\scriptstyle k\,\delta^{cd}/2}\,J^{b}(w)\,J^{a}(z)
		-{_{ik\,\delta^{cd}\,f^{ab\mu}/2}\over^{z-w}}\,J^{\mu}(w)
		-{_{k^2\,\delta^{ab}\,\delta^{cd}/4}\over^{(z-w)^2}},\\
\chi^{abcd}_{1}(z,y,v) =&\,{\scriptstyle if^{bc\mu}}
			\,J^{\mu}(y)\,J^{a}(z)\,J^{d}(v)
	+{_{f^{bc\mu}\,f^{a\mu\nu}}\over^{z-y}}\,J^{\nu}(y)\,J^{d}(v)
	+{_{f^{ad\mu}\,f^{bc\nu}}\over^{z-v}}\,J^{\nu}(y)\,J^{\mu}(v)\\
	&-{_{ik\,f^{abc}/2}\over^{(z-y)^2}}\,J^{d}(v)
	-{_{ik\,\delta^{ad}\,f^{bc\mu}/2}\over^{(z-v)^2}}\,J^{\mu}(y),\\ 
\chi^{abcd}_{2}(z,y,v) =&\,{\scriptstyle if^{bd\mu}}
			\,J^{\mu}(v)\,J^{a}(z)\,J^{c}(y)
	+{_{f^{bd\mu}\,f^{a\mu\nu}}\over^{z-v}}\,J^{\nu}(v)\,J^{c}(y)
	+{_{f^{ac\mu}\,f^{bd\nu}}\over^{z-y}}\,J^{\nu}(v)\,J^{\mu}(y)\\
	&-{_{ik\,f^{abd}/2}\over^{(z-v)^2}}\,J^{c}(y)
	-{_{ik\,\delta^{ac}\,f^{bd\mu}/2}\over^{(z-y)^2}}\,J^{\mu}(v),\\ 
\chi^{abcd}_{3}(z,w,v) =&\,{\scriptstyle if^{cd\mu}}
			\,J^{\mu}(v)\,J^{b}(w)\, J^{a}(z)
	+{_{f^{cd\mu}\,f^{a\mu\nu}}\over^{z-v}}\,J^{\nu}(v)\,J^{b}(w)
	+{_{f^{ab\mu}\,f^{cd\nu}}\over^{z-w}}\,J^{\nu}(v)\,J^{\mu}(w)\\
	&-{_{ik\,f^{acd}/2}\over^{(z-v)^2}}\,J^{b}(w)
	-{_{ik\,\delta^{ac}\,f^{cd\mu}/2}\over^{(z-w)^2}}\,J^{\mu}(v).
\end{align*}
L\`a encore, ce n'est pas la seule solution au probl\`eme pos\'e. Comme
pour les fonctions de Green \`a trois courants les autres solutions
s'obtiennent en sym\'etrisant l'\'equation~(\ref{symquatre}). On v\'erifie alors
que la diff\'erence entre deux solutions est analytique dans les quatres variables.
Quand on reprend cette \'equation avec $a=b$ et $c=d$, presque tous les termes 
disparaissent. On obtient alors l'\'equation~(\refa{ward11}) en
faisant tendre $y$ vers $v$, puis $z$ vers $w$. 
Au passage, notons qu'on doit utiliser le d\'eveloppement \`a courtes 
distances~(\ref{longdeuxpoints}). On proc\`ede de m\^eme
pour arriver \`a l'\'equation~(\refb{ward11}).

\subsection{\'Equation de Knizhnik-Zamolodchikov}%

Momentan\'ement, on suppose que $\Sigma$ est la sph\`ere de Riemann $\C P^1$.
Lorsqu'on int\`egre l'\'equation~(\ref{kzinit}), on obtient
\[
\langle\,J^{a}_{z}\,\grl\,\rangle=%
-\sum_\ell{_{t^a_\ell }\over^{z-z_\ell }}\,%
\langle\,\grl\,\rangle.
\]
Par rapport \`a l'\'equation~(\ref{ward6}.a), il n'y a pas 
de termes r\'esiduels parce que sur $\C P^1$ il n'y a pas
de $(1,0)$-forme globale holomorphe.
De la m\^{e}me fa\c{c}on, si on reprend l'\'equation~(\ref{ward7bis}), avec
$b=a$~:
\qq\label{kztemp}
\langle\,T(z)\,\grl\,\rangle=
{_1\over^{k+g^\vee}}\sum_{\ell,m}{_{t^a_\ell }\over^{z-z_\ell }}\,
{_{t^a_{m}}\over^{z-z_{m}}}
\langle\,\grl\,\rangle
\qqq
o\`u on a directement pris la limite $w\rightarrow z$. D'apr\`es 
l'\'equation~(\refa{ward9}), le terme apparaissant \`a gauche dans  
l'\'equation pr\'ec\'edente est analytique sauf aux points $\xi_\ell $
o\`u il pr\'esente un p\^ole d'ordre $2$. Si on compare 
les coefficients devant $1\over(z-z_\ell)^2$ et les r\'esidus
provenant des \'equations~(\refa{ward9}) et~(\ref{kztemp}),
on retrouve l'expression $\Delta_\ell=C_\ell/(k+g^\vee)$ 
et on obtient l'{\bol \'equation de Knizhnik-Zamolodchikov}~\cite{kz}~:
\qq
\Big(\partial_{z_\ell }-{_2\over^{k+g^\vee}}%
\,\sum_{m\neq\ell}{_{t^a_\ell \,t^a_{m}}\over^{z_\ell -z_{m}}}%
\Big)\,\langle\,\grl\,\rangle=0.
\qqq
On associe \`a ce syst\`eme diff\'erentiel la connexion
\[
\nabla\equiv d-{_2\over^{k+g^\vee}}\sum_{\ell}
	H_\ell (\un{\xi})\,dz_\ell ,
\]
o\`u $H_\ell\in{\rm End}(V_{\un\lambda})$ 
est le {\bol Hamiltonien de Gaudin}
\[
H_\ell (\un{\xi})=\,\sum_{m\neq \ell}
	{_{t^a_\ell \,t^a_{m}}\over^{z_\ell -z_m}},
\]
sur le fibr\'e trivial $Y_N\times V_{\un\lambda}$ 
au-dessus de $Y_N$, l'ensemble des s\'equences $\un{\xi}$ de
points deux \`a deux diff\'erents dans le plan complexe. 
Comme les Hamiltoniens de Gaudin commutent pour des $\ell$ diff\'erents,
la connexion $\nabla$ est plate. Son holonomie d\'efinit 
une repr\'esentation du groupe $P_N$ des tresses pures, \ie
le groupe fondamental de $Y_N$, dans l'espace $V_{\un\lambda}$~\cite{kassel}.
Au chapitre 4, on construit une \'equation diff\'erentielle
analogue pour $\Sigma$ de genre quelconque.

\medskip
\section{D'Euclide \`a Hilbert}

Pour le moment, nous avons
 focalis\'e notre attention sur le formalisme euclidien.
On explore maintenant un point de vue plus \<<traditionnel\>> 
de la th\'eorie des champs, car plus proche du formalisme
originel de la m\'ecanique quantique. Dans ce formalisme, dit op\'eratoriel,
les fondations de la th\'eorie reposent sur trois entit\'es 

--- l'espace de Hilbert des \'etats ;

--- une repr\'esentation du groupe de sym\'etrie ;

--- une certaine famille d'op\'erateurs. 
 
En th\'eorie des champs, on sait qu'il est possible de jongler entre ces deux approches.
Dans cette section, on extrait concr\`etement le formalisme op\'eratoriel
du formalisme euclidien. Notre discussion repose sur une condition physique 
(dite d'{\bol Osterwalder-Schrader} ~\cite{oster1,oster2}) de positivit\'e 
sur les fonctions de Green sur la sph\`ere de Riemann $\C P^1$. On calque 
la pr\'esentation  de~\cite{gaw97:cft}, mais adapt\'ee au mod\`ele de WZNW.

\subsection{Condition physique de positivit\'e}

Dans cette section, $\Sigma$ est la sph\`ere de Riemann $\C P^1$. 
Soit $\vartheta:\C P^1\ni z\mapsto\overline{z}^{-1}\in\C P^1$. 
Cette application envoie $D$ sur son compl\'ementaire $D'$ dans $\C P^1$ (et vice versa)
et laisse le cercle $S^1$ globalement invariant.
Soit $\gamma\sous{D}$ une m\'etrique 
riemannienne sur $D$ compatible avec la structure 
complexe, de la forme $|z|^{-2}|dz|^2$ au voisinage de $\partial D=S^1$ ---~on dira qu'une telle
m\'etrique est plate le long du bord.
On \<<colle\>> (r\'eguli\`erement) la m\'etrique $\vartheta^*\gamma\sous{D}$ sur $D'$ et
la m\'etrique $\gamma\sous{D}$ pour obtenir une m\'etrique r\'eguli\`ere
$\vartheta^*\gamma\sous{D}\,\sharp\,\gamma\sous{D}$ sur $\C P^1$.
D\'efinissons l'op\'eration $\Theta$ de {\bol r\'eflexion en temps} sur 
les champs (quasi-)primaires $\varphi$ de poids conformes 
$(\Delta,\overline{\Delta})$~: soit $z\in D$, on pose
\[
\Theta \varphi(z,\overline{z})\equiv (-\overline{z}^{-2})^\Delta
	(-z^{-2})^{\overline{\Delta}}\,\varphi^{\rm PCT}(\overline{z}^{-1},z^{-1}).
\]
Le champ $\varphi^{\rm PCT}$ est le champ obtenu par transformation PCT \`a 
partir de $\varphi$~:
\[
\overline{\langle\,\varphi(z,\overline{z})\,\,\dots\,\rangle}\sous{\C P^1}
	=\langle\,\varphi^{\rm PCT}(\overline{z},z)\,\,\dots\,\rangle
	\sous{\overline{\C P^1}}\,.
\]
Pour l'ensemble des champs (quasi-)primaires mis en \'evidence
dans la section pr\'ec\'edente, on a vu que
\begin{gather*}
\begin{aligned}
J^{a}(z)&\stackrel{\mbox{\tiny PCT}}{\longmapsto} -J^{a}(z),\\
T(z)& \stackrel{\mbox{\tiny PCT}}{\longmapsto} T(z),
\end{aligned}
\qquad
\begin{aligned}
\barJ^{a}(\overline{z})&
	\stackrel{\mbox{\tiny PCT}}{\longmapsto}-\barJ^{a}(\overline{z}),\\
\barT(\overline{z})&
	\stackrel{\mbox{\tiny PCT}}{\longmapsto} \barT(\overline{z}),
\end{aligned}\\
g(z,\overline{z})\sous{R}\stackrel{\mbox{\tiny PCT}}{\longmapsto}
	g(z,\overline{z})\sous{\overline{R}}
\end{gather*}
Notons que le champ $\varphi^{\rm PCT}$ est de m\^eme nature que
le champ original $\varphi$, \ie il est \'egalement (quasi-)primaire
de poids conformes $(\Delta,\overline{\Delta})$ mais pour $\overline{\C P^1}$ .
On a donc:
\begin{gather*}
\begin{aligned}
\Theta J^{a}(z)&=\overline{z}^{-2}\, J^{a}(\overline{z}^{-1}),\\
\Theta T(z)&=\overline{z}^{-4}\, T(\overline{z}^{-1}),
\end{aligned}
\qquad
\begin{aligned}
\Theta \barJ^{a}(\overline{z})&=z^{-2}\,\barJ^{a}(z^{-1}),\\
\Theta \barT(\overline{z})&=z^{-4}\,\barT(z^{-1}),
\end{aligned}\\
\Theta g(z,\overline{z})\sous{R}=|z|^{-4\Delta}\,%
	g(\overline{z}^{-1},z^{-1})\sous{\overline{R}}
\end{gather*}
et $\Theta$ agit sur les constantes par conjugaison complexe.
Consid\'erons la combinaison formelle (\'eventuellement vide)
\qq
\label{X}
X=\prod_{k}T(z\sous{k})\prod_{\overline{k}}\overline{T}(\overline{z}\sous{\overline{k}})%
\prod_jJ^{a\sous{j}}(z\sous{j})%
\prod_{\overline{\jmath}}\barJ^{a\sous{\overline{\jmath}}}%
	(\overline{z}\sous{\overline{\jmath}})%
\prod_\ell\prod_{[\ell]}%
	g(z_\ell,\overline{z}_\ell)\sous{R_\ell}^{[\ell]}.
\qqq
On a not\'e par
$g(z_\ell,\overline{z}_\ell)\sous{R_\ell}^{[\ell]}$
un \'el\'ement \<<matriciel\>>.
Dans la suite, on utilise $\vec{\ell}$ pour signifier le choix d'un $\ell$ et 
d'un \'el\'ement matriciel $[\ell]$.
Par d\'efinition, le support de $X$ est l'ensemble 
${\rm Supp}\,X\equiv\{z\sous{k},z\sous{\overline{k}},z\sous{j},
z\sous{\overline{\jmath}},z_\ell\}.$ On exige que ${\rm Supp}\, X\subset D$.
On \'etend le domaine d'action de $\Theta$ aux expressions $X$~:
\[
\Theta X=\prod_{k}\Theta T(z\sous{k})\prod_{\overline{k}}\Theta\overline{T}(\overline{z}%
\sous{\overline{k}})\prod_j\Theta J^{a\sous{j}}(z\sous{j})%
\prod_{\overline{\jmath}}\Theta\barJ^{a\sous{\overline{\jmath}}}%
	(\overline{z}\sous{\overline{\jmath}})%
\prod_{\vec{\ell}}\Theta g(z_\ell,\overline{z}_\ell)\sous{R_\ell }
	^{[\ell]}.
\]
Soient $\{\lambda_{\alpha}\}$ 
une famille de nombres complexes, $\{\gamma\sous{D,\alpha}\}$
une famille de m\'etriques sur $D$ (plates le long du bord) 
et $\{Y_{\alpha}\}$ une famille d'expressions uniquement compos\'ees \`a partir
de champs primaires. La {\bol condition physique de positivit\'e} impose l'in\'egalit\'e
\[
\sum_{\alpha\sous{1},\alpha\sous{2}}\overline{\lambda}_{\alpha\sous{2}}
\lambda_{\alpha\sous{1}}\,
Z\sous{\vartheta^*\gamma\sous{D,\alpha\sous{2}}\,\sharp\,\gamma\sous{D,\alpha\sous{1}}}\,
\langle\,(\Theta Y_{\alpha\sous{2}})\,Y_{\alpha\sous{1}}\,\rangle\sous{
\vartheta^*\gamma\sous{D,\alpha\sous{2}}\,\sharp\,\gamma\sous{D,\alpha\sous{1}}}
\,\geq\,0.
\]
En particulier,
$Z\sous{\vartheta^*\gamma\sous{D}\,\sharp\,\gamma\sous{D}}\geq 0$.
Par convention, la fonction de Green du produit vide vaut $1$. 
Cette in\'egalit\'e entra\^\i ne que, pour toute expression du type~\eqref{X},
\[
\sum_{\alpha\sous{1},\alpha\sous{2}}\overline{\lambda}_{\alpha\sous{2}}
\lambda_{\alpha\sous{1}}\,
\langle\,(\Theta X_{\alpha\sous{2}})\,X_{\alpha\sous{1}}\,\rangle
\,\geq\,0
\]
o\`u les fonctions de Green sont prises en champ de jauge 
localement nul et en m\'etrique localement
plate.  Une fois cette condition assur\'ee sur les fonctions de Green, la 
quantit\'e
\[
\sum_{\beta,\alpha}\overline{\lambda'}_{\beta}%
\lambda_{\alpha}\,\langle\,(\Theta X'_{\beta})%
\,X_{\alpha}\rangle
\]
d\'efinit une forme hermitienne
sur l'espace vectoriel $V\sous{D}$ engendr\'e par les expressions 
du type~(\ref{X}) avec ${\rm Supp}\, X\subset D$. 
L'hermiticit\'e de la forme, \cad 
$\overline{\langle\,(\Theta X')\,X\,\rangle}
=\langle\,(\Theta X)\,X'\,\rangle$,
est une cons\'equence des propri\'et\'es conformes des fonctions de 
Green. Il suffit de montrer que 
$\overline{\langle X\,\rangle}=\langle\,\Theta X\,\rangle$. Par construction,
\[
\overline{\langle\,\varphi(z,\overline{z})\,\,\dots\,\rangle}\sous{\C P^1}=
	\langle\,\varphi^{\rm PCT}
	(\overline{z},z)\,\,\dots\,\rangle\sous{\overline{\C P^1}}.
\]
On effectue le changement de coordonn\'ees
$\overline{\C P^1}\ni(z,\overline{z})\mapsto(w,\overline{w})=
(\frac{1}{\overline{z}},\frac{1}{z})\in\C P^1$. \`A gauche, la coordonn\'ee
$\overline{z}$ est la coordonn\'ee analytique, par cons\'equent $\varphi$
est un changement de coordonn\'ees analytiques. On sait alors que
\begin{align*}
\langle\,\varphi^{\rm PCT}
	(\overline{z},z)\,\,\dots\,\rangle\sous{\overline{\C P^1}}
&=\langle\,\big({_{dw}\over^{d\overline{z}}}\big)^\Delta
	\,\big({_{d\overline{w}}\over^{dz}}\big)^{\overline{\Delta}}
	\,\varphi^{\rm PCT}(w,\overline{w})
\ \dots\,\rangle\sous{\C P^1}\\
&=\langle\,(-\overline{z}^{-2})^\Delta
	(-z^{-2})^{\overline{\Delta}}\,
	\varphi^{\rm PCT}(\overline{z}^{-1},z^{-1})
	\ \dots\,\rangle\sous{\C P^1},
\end{align*}
ce qu'il fallait d\'emontrer.

Soit 
\[
{\NH}\,\equiv\,\overline{V\sous{D}/V\sous{D}^{\rm nul}}
\]
la compl\'etion de l'espace $V\sous{D}$, muni de la semi-norme provenant 
de la forme hermitienne positive d\'efinie ci-dessus.
Par construction, $V\sous{D}^{\rm nul}$ est l'ensemble des vecteurs 
de $V\sous{D}$ qui annulent la semi-norme. On note $\iota$ l'injection
canonique qui envoie $V\sous{D}$ dans $\NH$. On 
montre~\cite{lang:analysis} que l'image de $V\sous{D}$ par $\iota$, 
not\'ee $\NH_0$, est dense dans $\NH$. 
Le produit scalaire sur $\NH$ est simplement
\[
\big(\,\iota(X'),\,\iota(X)\,\big)=\langle\,(\Theta X')\,X\,\rangle.
\]
On convient de noter $\Omega$ le {\bol vecteur vide}, image du produit 
vide par $\iota$. L'espace $\NH$ est muni d'une {\bol
involution anti-unitaire}~(\footnote{
Une involution anti-unitaire est une involution anti-lin\'eaire pr\'eservant
le produit scalaire.}),
not\'ee $\CI$, qui envoie un vecteur $\iota(X)$ 
sur le vecteur $\overline{\iota(X)}$, 
image par $\iota$ de
\[
\overline{X}\,\equiv\,\prod_{k}T(\overline{z}\sous{k})\prod_{\overline{k}}
\overline{T}(z\sous{\overline{k}})%
\prod_jJ^{a\sous{j}}(\overline{z}\sous{j})%
\prod_{\overline{\jmath}}\barJ^{a\sous{\overline{\jmath}}}%
	(z\sous{\overline{\jmath}})%
\prod_{\vec{\ell}}g(\overline{z}_\ell,z_\ell)\sous{\overline{R}_\ell}
	^{[\ell]}.
\]

\subsection{Op\'erateurs}

D\'efinissons une {\bol op\'eration de dilatation} $S_{q}$
de param\`etre $q\in\C$, $0<|q|\leq 1$, sur les champs (quasi-)primaires
par $S_{q}\varphi(z,\overline{z})=q^\Delta\overline{q}^{\overline{\Delta}}
\varphi(qz,\overline{q}\overline{z})$. Par cons\'equent,
\begin{gather*}
\begin{aligned}
S_{q}T(z)&=q^2\,T(qz),\\
S_{q}J^{a}(z)&=q\,J^{a}(qz),
\end{aligned}
\qquad
\begin{aligned}
S_{q}\overline{T}(\overline{z})&=\overline{q}^2\,%
\barT(\overline{q}\,\overline{z}),\\
S_{q}\overline{J}^{a}(\overline{z})&=\overline{q}\,\barJ^{a}%
(\overline{q}\,\overline{z}),
\end{aligned}\\
S_{q}g(z,\overline{z})\sous{R}=|q|^{2\Delta}\,g(qz,%
\overline{q}\,\overline{z})\sous{R}.
\end{gather*}
Comme pour $\Theta$, on \'etend l'action de $S_{q}$ 
aux \'el\'ements de $V\sous{D}$,
\[
S_{q} X=\prod_{k}S_{q} T(z\sous{k})\prod_{\overline{k}}S_{q}%
\overline{T}(\overline{z}\sous{\overline{k}})%
\prod_jS_{q} J^{a\sous{j}}(z\sous{j})%
\prod_{\overline{\jmath}}S_{q}\barJ^{a\sous{\overline{\jmath}}}%
	(\overline{z}\sous{\overline{\jmath}})%
\prod_{\vec{\ell}}S_{q} g(z_\ell,\overline{z}_\ell)\sous{R_\ell}
	^{[\ell]}.
\]

\`A partir de $S_{q}$, on peut construire une op\'eration de dilatation 
agissant dans $\NH$~:
\[
\CS_{q}\iota(X)\,\equiv\,\iota(S_{q}X).
\]
L'op\'eration est bien d\'efinie \`a condition que $V\sous{D}^{\rm nul}$ soit
stable par $\CS_{q}$. Comme la forme sur $V\sous{D}$ est hermitienne, 
$V\sous{D}^{\rm nul}$ est l'ensemble des vecteurs $X$ de $V\sous{D}$
tels que $\langle\,(\Theta Y)\,X\,\rangle=0$, pour tout $Y\in V\sous{D}$. 
Il suffit donc de v\'erifier l'\'egalit\'e
\qq
\langle\,(\Theta X')\,S_{q}X\,\rangle=
\langle(\Theta S_{\overline q}X')\,X\,\rangle.
\label{contraction}
\qqq
Comme pour l'hermiticit\'e de la forme, cette \'egalit\'e est
une cons\'equence des propri\'et\'es conformes des fonctions de Green.
En effet, on a
\[
\langle\,(\Theta S_{\bar q}\,\varphi)\,\varphi'\,\,\dots\,\rangle
	=\langle\,\big(\quotient{-1}{q{\bar z}^2}\big)^\Delta
	\big(\quotient{-1}{\bar q z}\big)^{\bar \Delta}
	\varphi^{\rm PCT}\big(\quotient{1}{q\bar z},\quotient{1}{{\bar q}z}\big)
	\varphi'(z,\bar z)\ \dots\,\rangle.
\]
On effectue le changement de coordonn\'ees analytique qui envoie
$(z,\bar z)$ vers $(qz,\bar q\,\bar z)$ et on trouve exactement 
$\langle\,(\Theta \varphi')\,S_{q}\varphi\,\,\dots\,\rangle$. 
On fera attention que pour le premier champ, la coordonn\'ee
holomorphe est $\frac{1}{q\bar z}$ qui va donc vers $q/(q\bar z)=
\frac{1}{\bar z}$.
L'op\'erateur dilatation $\CS_{q}$ est donc bien d\'efini de domaine $\NH_0$ 
dense invariant dans $\NH$.
En fait, on peut extraire plus d'information de l'\'equation~(\ref{contraction}).
La famille $\{\CS_{q}\}$ forme un semi-groupe~: 
$\CS_{q\sous{1}}\CS_{q\sous{2}}=\CS_{q\sous{1}q\sous{2}}$. 
Ensuite, apr\`es application it\'erative de l'in\'egalit\'e de Schwarz et
d'apr\`es la propri\'et\'e de semi-groupe,
\begin{equation}\label{inegalsemi}
\begin{split}
\big|\big(\,\iota(X'),\CS_{q}\iota(X)\,\big)\big|&\leq \big\Vert\iota(X')
	\big\Vert\ \big\Vert\CS_{q}\iota(X)\big\Vert=%
	\big\Vert\iota(X')\big\Vert\ 
	\big(\,\iota(X),\CS_{q\overline{q}}\iota(X)\,\big)^{1/2}\\
&\leq\cdots\leq\big\Vert\iota(X')\big\Vert\ 
	\big\Vert\iota(X)\big\Vert^{\frac{1}{2}+\cdots+\frac{1}{2^{n-1}}}\,%
\big(\,\iota(X),\CS_{(q\overline{q}\,)^{2^{n-1}}}\iota(X)\,\big)^{1/2^n}.
\end{split}
\end{equation}
Supposons que, pour tout $\epsilon>0$, il existe une 
constante $C\sous{\epsilon}$ telle que
\[
|\langle\,(\Theta X')\,S_{t}X\,\rangle|\leq C\sous{\epsilon}\,t^{-\epsilon}
\]
quand $t\rightarrow 0$. Cette propri\'et\'e nous permet de majorer le terme de droite
de l'in\'egalit\'e~(\ref{inegalsemi}). On prend alors la limite quand $n$ tend vers
l'infini~; on trouve 
\[
\big|\big(\,\iota(X'),\CS_{q}\iota(X)\,\big)\big|
	\leq\big\Vert\iota(X')\big\Vert\ \big\Vert\iota(X)\big\Vert.
\]
Ainsi, le semi-groupe des dilatations 
contient uniquement des contractions de $\NH$. Qui plus est,
d'apr\`es l'\'equation~(\ref{contraction}), $\CS_{q}^\dagger=\CS_{\overline{q}}$.
Enfin, les op\'erateurs $\CS_{q}$ sont faiblement continus sur $\NH$, propri\'et\'e
assur\'ee par la continuit\'e faible sur $\NH_0$.
On sait alors que les \'el\'ements du semi-groupe $\{\CS_{q}\}$
poss\`edent la repr\'esentation suivante~\cite{reedsimon}
\[
\CS_{q}=q^{L_0}\,\overline{q}^{\overline{L}_0}
\]
o\`u $L_0$ et $\overline{L}_0$ sont des op\'erateurs auto-adjoints fortement commutants
tels que $L_0+\overline{L}_0\geq 0$. En plus, leurs domaines respectifs contiennent $\NH_0$ o\`u
\[
L_0\,\iota(X)=\partial_q\left|_{q=1}\right.\CS_{q}\iota(X),\qquad
\overline{L}_0\,\iota(X)=\partial_{\hspace{0.02cm}\overline{q}}
\left|_{q=1}\right.\CS_{q}\iota(X).
\]
Notons que $\CS_{q}\NH\sous{0}$ est dense dans $\NH$, pour tout
$q$.

Soit $z\in\C$, $0<|z|<1$~; on extrait des champs de la th\'eorie un ensemble
d'op\'erateurs $\CT(z)$, $\overline{\CT}(\overline{z})$, $\CJ^{a}(z)$,
$\overline{\CJ}^{a}(\overline{z})$ et $\widehat{g}(z,\overline{z})\sous{R}$ 
agissant dans l'espace $\CS_{z}\NH_0$ dense dans $\NH$~:
\begin{gather*}
\begin{aligned}
\CT(z)\,\iota(X)\,&\equiv\,\iota(T(z)X),\\
\CJ^{a}(z)\,\iota(X)\,&\equiv\,\iota(J^{a}(z)X),
\end{aligned}
\qquad
\begin{aligned}
\overline{\CT}(\overline{z})\,\iota(X)
\,&\equiv\,\iota(\barT(\overline{z})X),\\
\overline{\CJ}^{a}(\overline{z})\,\iota(X)\,&\equiv\,
\iota(\barJ^{a}(\overline{z})X),
\end{aligned}\\
\widehat{g}(z,\overline{z})\sous{R}\,\iota(X)\,\equiv\,
\iota(g(z,\overline{z})\sous{R}X).
\end{gather*}
Sauf indication contraire,
on notera toujours les op\'erateurs par des lettres calligraphiques. 
En prenant $\iota(X)$ dans $\CS_{z}\NH_0$~---
\cad ${\rm Supp}\, X\subset\{\,z' \,\,\vert\,\,
|z'|<\vert z\vert<1\,\}$, on \'evite toute ambigu\"\i t\'e.
En effet, ces op\'erateurs sont correctement
d\'efinis s'ils laissent $V\sous{D}^{\rm nul}$ globalement invariant~;
cons\'equence de l'\'equation~(\ref{contraction}).
On conviendra que $\CJ(z)\equiv\CJ^{a}(z)\,t^a$ et de m\^eme pour 
$\overline{\CJ}(\overline{z})$. Il suit de la d\'efinition pr\'ec\'edente
qu'un vecteur $X\in V\sous{D}$, donn\'e par l'\'equation~(\ref{X}) 
o\`u les points d'insertion sont tous de modules diff\'erents, v\'erifie 
\qq
\label{radialorder}
\CX\,=\,R\Big(
\prod_{k}\CT(z\sous{k})\prod_{\overline{k}}\overline{\CT}(\overline{z}\sous{\overline{k}})%
\prod_j\CJ^{a\sous{j}}(z\sous{j})%
\prod_{\overline{\jmath}}\overline{\CJ}^{a\sous{\overline{\jmath}}}%
	(\overline{z}\sous{\overline{\jmath}})%
\prod_{\vec{\ell}}\widehat{g}(z_\ell,\overline{z}_\ell)\sous{R_\ell}
	^{[\ell]}%
\Big)\,\Omega
\qqq
o\`u $R$ r\'eordonne les op\'erateurs de telle sorte qu'ils agissent 
dans l'ordre croissant en $|z|$. En physique, ce processus appara\^\i t sous
le nom de {\bol quantification radiale}. Enfin, si on conjugue les
op\'erateurs par l'involution $\CI$, on a 
\begin{gather*}
\begin{aligned}
\CI\,\CT(z)\,\CI &=\CT(\overline{z}),\\
\CI\,\CJ^{a}(z)\,\CI &=\CJ^{a}(\overline{z}),
\end{aligned}
\quad
\begin{aligned}
\CI\,\overline{\CT}(\overline{z})\,\CI &=\overline{\CT}(z),\\
\CI\,\overline{\CJ}^{a}(\overline{z})\,\CI &=\bar{\CJ}^{a}(z),
\end{aligned}\\
\CI\,\widehat{g}(z,\overline{z})\sous{R}\,\CI=%
\widehat{g}(\overline{z},z)\sous{\overline R}.
\end{gather*}

\subsection{Unitarit\'e}

D\'eveloppons en s\'erie de Laurent les op\'erateurs $\CJ(z)$,
$\Jbar(\overline{z})$, $\CT(z)$ et $\overline{\CT}(\overline{z})$~:
\begin{gather*}
\begin{aligned}
\CJ(z)&=\sum_{n\in\Z}J_n\,z^{-n-1},\\
\CT(z)&=\sum_{n\in\Z}L_nz^{-n-2},
\end{aligned}
\qquad
\begin{aligned}
\Jbar(\overline{z})&=\sum_{n\in\Z}\barJ_n\,{\overline z}^{-n-1},\\
\CTbar({\overline z})&=\sum_{n\in\Z}\overline{L}_n{\overline z}^{-n-2},
\end{aligned}
\end{gather*}
o\`u les modes sont donn\'es par les int\'egrales
\begin{gather*}
\begin{aligned}
J_n&={_1\over^{2\pi i}}\oint_{|z|=r<1}dz\,z^n\,\CJ(z),\\
L_n&={_1\over^{2\pi i}}\,\oint_{|z|=r<1} dz\,z^{n+1}\,\CT(z),
\end{aligned}
\qquad
\begin{aligned}
\barJ_n&=-{_1\over^{2\pi i}}\oint_{|z|=r<1}d\overline{z}\,{\overline z}^n\,\Jbar({\overline z}),\\
{\overline L}_n&=-{_1\over^{2\pi i}}\,\oint_{|z|=r<1} d\overline{z}\,%
\overline{z}^{n+1}\,\CTbar(\overline{z}).
\end{aligned}
\end{gather*}
On a vu que l'insertion des champs $J(z)$, $\barJ(\overline{z})$,
$T(z)$ et $\barT(\overline{z})$ produit des fonctions de Green analytiques
sauf aux points d'insertion, o\`u on obtient une singularit\'e.
La quantit\'e $\big(\,\iota(X'),\, L_n\iota(X)\,\big)$ 
ne d\'epend donc pas du contour choisi, tant que
ce dernier encercle les points d'insertion 
de $\iota(X)\in\CS_r\NH_0$ --- idem avec les autres
modes. Ainsi, contrairement aux conventions adopt\'ees jusque-l\`a, 
les modes, bien que not\'es par des lettres droites, 
sont des op\'erateurs de domaine $\NH_0$ dense dans $\NH$.

Soit $\epsilon>0$, choisi pour que, dans les calculs suivants, les contours
soient adapt\'es aux vecteurs $\iota(X)$ et $\iota(X')$~:
\begin{align*}
\big(\,\iota(X'),\, L_n\iota(X)\,\big) & 
		=  {_1\over^{2\pi i}}\,\oint_{|z|=1-\epsilon}dz\,%
		z^{n+1}\,\langle\,(\Theta X')\,T(z)X\,\rangle\\
	      & =  {_1\over^{2\pi i}}\,\oint_{|z|=1+\epsilon}dz\,%
		z^{n-3}\,\langle\,%
		\Theta(X'T({_1\over^{\overline{z}}}))\,X\,\rangle
\end{align*}
o\`u on a d\'eform\'e le contour d'int\'egration, puis on a remplac\'e
$T(z)$, pour $|z|=1+\epsilon$, par $z^{-4}\,\Theta T(\frac{1}{\overline z})$.
Il vient
\begin{equation*}
\begin{split}
\big(\,\iota(X'),\, L_n\iota(X)\,\big)
	&=\big(-{_1\over^{2\pi i}}\,\oint_{|z|=1+\epsilon}d\overline{z}\,%
	\overline{z}^{n-3}\,%
	\CT({_1\over^{\overline{z}}})\ \iota(X'),\,\iota(X)\,\big)\\
&= \big({_1\over^{2\pi i}}\,\oint_{|w|=(1+\epsilon)^{-1}}dw\,%
	w^{-n+1}\,\CT(w)\ \iota(X'),\,\iota(X)\,\big)
	=\big(\,L_{-n}\iota(X'),\,\iota(X)\,\big).
\end{split}
\end{equation*}
Par cons\'equent, l'op\'erateur $L_n$ est fermable~(\footnote{Pour qu'un 
op\'erateur $\CA$ soit fermable, il suffit de montrer~\cite{glim} 
que pour toute suite $(\CX_n)$ de vecteurs de $\NH_0$, telle que 
$\CX_n\rightarrow 0$ et $\CA\CX_n\rightarrow\CZ\in\NH$, alors $\CZ=0$.
On garde la m\^eme notation pour un op\'erateur et sa fermeture.}),
d'adjoint $L_n^\dagger=L_{-n}$. On proc\`ede de la m\^eme 
mani\`ere avec les autres modes~; les op\'erateurs $L_n$, 
$\overline{L}_n$, $J^{a}_n$ et $\barJ^{a}_n$
sont fermables et 
\qq
L_n^\dagger=L_{-n},\quad \overline{L}_n^\dagger=\overline{L}_{-n},\quad%
J^{a}_n{}^\dagger=J^{a}_{-n},\quad \barJ^{a}_n{}^\dagger=%
\barJ^{a}_{-n}.
\qqq
En pratique, on \'etend le domaine des op\'erateurs. On autorise la 
pr\'esence de modes dans les expressions donn\'ees par l'\'equation%
~(\ref{X}). Soit $\NH_1$ l'espace ainsi obtenu~; 
$\NH_0\subset\NH_1\subset\NH$ et $\NH_1$ est invariant sous l'action
des modes. On \'etend les op\'erateurs $\CT(z)$, 
$\overline{\CT}(\overline{z})$, $\CJ(z)$, $\overline{\CJ}(\overline{z})$
et $\widehat{g}(z,\overline{z})\sous{R}$ \`a $\CS_{z}\NH_1$ et les
modes \`a $\NH_1$.
Notons que ces op\'erateurs ne sont pas fermables. Il faut donc
regarder leurs domaines respectifs avec attention.

\subsection{Axiomes de Segal pour une th\'eorie conforme r\'eelle}

Dans cette section, on retourne \`a l'\'etude du mod\`ele d\'efini sur une surface 
de Riemann $\Sigma$ de bord $\partial\Sigma$ (cf. p.\pageref{riemman}). 
Notons $\theta:S^1\ni z\rightarrow z^{-1}\in S^1$~;
cette application renverse l'orientation de la composante $(\partial\Sigma)_i$ via
la param\'etrisation $p_{i}^\vee\equiv p_{i}\circ\theta$.
On peut compl\'eter $\Sigma$ 
(de mani\`ere unique) pour former une surface de Riemann compacte 
$\widetilde{\Sigma}$. \`A cet effet, on \<<colle\>> 
une copie $D_i$ de $D$ suivant $(\partial\Sigma)_i$, $i\in I_-$,
et une copie $D_i'$ de $D'$ si $i\in I_+$.
On peut voir~\cite{ahlfors} que la surface $\widetilde{\Sigma}$ h\'erite d'une structure
complexe.
Inversement, si on dispose d'une surface de Riemann compacte 
$\widetilde{\Sigma}$ avec de param\`etres locaux, \cad avec
un certain nombre de disques ferm\'es $\overline{D}$ et $\overline{D'}$ 
disjoints plong\'es holomorphiquement dans $\widetilde{\Sigma}$, 
on construit une surface $\Sigma$ \`a bord, dont les composantes connexes 
sont param\'etris\'ees par le cercle $S^1$.
Il suffit pour cela d'\<<enlever\>> l'int\'erieur des disques plong\'es.

On dit qu'une m\'etrique $\gamma$ (compatible avec la structure complexe)
est {\bol plate le long du bord} si, pour tout $i$, on a  
$p_{i}^*\gamma=|z|^{-2}\,|dz|^2$ au voisinage
de chaque composante du bord (apr\`es continuation analytique de $p_{i}$ \`a un voisinage 
de $S^1$). Soient $\gamma$ une m\'etrique sur $\Sigma$ et $\gamma\sous{D}$ une m\'etrique 
sur $D$~---~toutes deux plates le long du bord. Notons $z_{i}$ le plongement 
de la $i$-\`eme copie de $D$ (ou $D'$) dans $\widetilde{\Sigma}$.
Si $i\in I_+$, $\gamma_{i}=z_{i}^*\gamma\sous{D}$ est une 
m\'etrique riemannienne sur $D_{i}$~; si $i\in I_-$,
$(z_{i}\circ\vartheta)^*\gamma\sous{D}=\vartheta^*\gamma_{i}$ est une m\'etrique
riemannienne sur $D_{i}$.
On obtient une m\'etrique sur $\widetilde{\Sigma}$~:
\[
\widetilde{\gamma}\equiv\Big(\mathop{\sharp}
	\limits_{i\in I_-}\gamma_{i}\Big)\sharp\,\gamma\,
	\sharp\Big(\mathop{\sharp}\limits_{i\in I_+}
	\vartheta^*\gamma_{i}\Big).
\]
D'apr\`es la propri\'et\'e~(\ref{weyl1}), la quantit\'e suivante
\[
Z\sous{\gamma}\equiv 
Z\sous{\widetilde{\gamma}}\,
		\prod_{i\in I}(Z\sous{\vartheta^*\gamma_{i}\,\sharp
		\,\gamma_{i}})^{-1/2}
\]
ne d\'epend pas du choix de $\gamma\sous{D}$ (dans la classe conforme). 
De plus, toujours gr\^ace \`a l'anomalie conforme, $Z\sous{\gamma}$ 
change suivant l'\'equation~(\ref{weyl1})
dans une transformation de Weyl triviale au voisinage du bord. On appellera 
$Z\sous{\gamma}$
la fonction de partition pour une surface de
 Riemann \`a bord. Soient $\iota(X_i)\in\NH_0$, $i\in I$.
\`A tout surface $\Sigma$ \`a bord, on associe un op\'erateur (appel\'e aussi amplitude) 
$\NA\sous{\Sigma,(p_{i}),\gamma}$ d\'efini par ses \'el\'ements matriciels 
\begin{equation}
\label{operateur}
\big(\bigotimes_{i\in I_+}\iota(X_i)\,,\, 
\NA\sous{\Sigma,(p_{i}),\gamma}\,\bigotimes_{i\in I_-}\iota(X_i)\big)
=Z\sous{\gamma}\,\langle\,\prod_{i\in I_+}(\Theta X_i)\,\prod_{i\in I_-}X_i\,\rangle. 
\end{equation}
Si on consid\`ere une surface compacte, 
on trouve $\NA\sous{\Sigma,\gamma}=Z\sous{\gamma}$.
L'id\'ee de Segal est de prendre 
ces amplitudes pour fondations
de l'immeuble de la {\bol th\'eorie conforme des champs}.

\begin{figure}[h]
\makebox[5cm][l]{\hspace{2cm}
\includegraphics{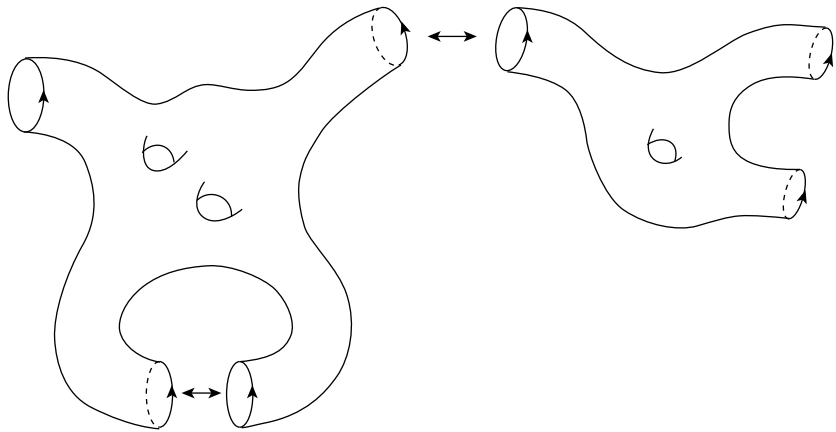}}
\caption{Unitarit\'e}
\label{segalfig}
\end{figure}

On suppose donn\'e un espace de Hilbert $\NH$ 
accompagn\'e d'une involution anti-unitaire $\CI$,
telle que, pour toute surface $\Sigma$ 
du type d\'efini ci-dessus~---~\ie $\Sigma$ est d\'ecor\'ee 
par un ensemble de param\'etrisations $p_{i}$, $i\in I_-\cup I_+$, et par une 
m\'etrique  $\gamma$ plate le long du bord~--- il existe un op\'erateur
tra\c{c}able 
		\[
		{\bf A}\sous{\Sigma,(p_{i}),\gamma}:\bigotimes_{i\in I_-}\NH\rightarrow
		\bigotimes_{i\in I_+}\NH.
		\]
On convient que le produit tensoriel vide 
donne $\C$. Ces op\'erateurs doivent v\'erifier les 
axiomes suivants 
\begin{description}
	\item[\bf A)]{\bol Localit\'e.} Si $\Sigma$ est la r\'eunion disjointe de 
		$(\Sigma_{1},(p_{1i}),\gamma_{1})$ et 
		$(\Sigma_{2},(p_{2i}),\gamma_{2})$ 
		\[
		{\bf A}\sous{\Sigma,(p_{i}),\gamma}=
		\NA\sous{\Sigma_{1},(p_{1i}),\gamma_{1}}
		\otimes\NA\sous{\Sigma_{2},(p_{2i}),\gamma_{2}} ;
		\]
	\item[\bf B)]{\bol Covariance g\'en\'erale.} Si $D:\Sigma_{1}\rightarrow\Sigma_{2}$
		est un diff\'eomorphisme holomorphe \'egal \`a l'identit\'e dans
		les param\'etrisations autour du bord, alors
		\[
		\NA\sous{\Sigma_{1},(p_{i}),\gamma}=
			\NA\sous{\Sigma_{2},(D\circ p_{i}),D^*\gamma} ;
		\]
	\item[\bf C)]{\bol Invariance PCT.} 
		\[
		\NA\sous{\overline{\Sigma},(p_{i}),\gamma}=
			\NA\sous{\Sigma,(p_{i}),\gamma}^\dagger ;
		\]
	\item[\bf D)] Soit $i_{0}\in I_+$, 
		\[
		\big(\,\iota(X')\,,\,\NA\sous{\Sigma,p_{i_{0}}^\vee,
		(p_{i'}),\gamma}
			\,\iota(X_0)\otimes\iota(X)\,\big)=
		\big(\,(\CI\,\iota(X_0))\otimes\iota(X')\,,\,
		\NA\sous{\Sigma,p_{i},\gamma}\,\iota(X)\,\big)
		\]
		o\`u $i'\neq i_{0}$, 
		$\iota(X')\in\mathop{\bigotimes}\limits_{i'\in I_+}\NH$, 
		$\iota(X)\in\mathop{\bigotimes}\limits_{i\in I_-}\NH$,
		et $\iota(X_0)$ est un 
		\'el\'ement de $\NH$ ;     
	\item[\bf E)]{\bol Unitarit\'e.} (cf. figure~\ref{segalfig})
		Si $\Sigma'$ est obtenue en identifiant 
		les bords $i_{1}\in I_-$ et $i_{2}\in I_+$, via 
		$p_{i_{2}}\circ p_{i_{1}}^{-1}$, alors
		\[
		\NA\sous{\Sigma',(p_{i'}),\gamma}=
			\tr_{i_{1},i_{2}}\,\NA\sous{\Sigma,(p_{i}),\gamma}
		\]
		o\`u $i'\neq i_{1},i_{2}$ et
		 la $\tr_{i_{1},i_{2}}$ porte sur les facteurs $i_{1}$ et 
		$i_{2}$ dans le produit tensoriel ;
	\item[\bf F)]{\bol Invariance conforme.} Si $\sigma\in
		\CC^\infty(\Sigma,\R)$ est nulle
		dans un voisinage de $\partial\Sigma$, alors
		\[
		\NA\sous{\Sigma,(p_{i}),\ee^\sigma\gamma}=
			\ee^{\,\frac{ic}{24\pi}\,S\sous{\rm L}(\sigma)}
			\NA\sous{\Sigma,(p_{i}),\gamma}.
		\]
\end{description}

Des versions des axiomes de Segal sont pr\'esent\'ees
dans un article non publi\'e~\cite{segal:cft0}
et dans les compte-rendus~\cite{segal:cft1,segal:cft2}.
Avec un langage similaire \`a celui utilis\'e dans cette
th\`ese, on trouvera dans~\cite{gaw88:cft} les
axiomes pour une th\'eorie conforme
(complexe), ainsi qu'une description de la notion
de foncteur modulaire (les blocs conformes du physicien).
Ces axiomes et ses variantes cachent une structure particuli\`erement riche
--- \'equations de Moore et Seiberg~\cite{degio},
op\'erateurs de vertex~\cite{huang,kac:vertex}, ... ---
dont on ne donnera qu'un timide aper\c{c}u.

\vspace{2cm}

\newpage

Regardons de plus pr\`es la signification des axiomes de Segal, tout particuli\`erement 
\`a la lumi\`ere de l'int\'egrale fonctionnelle.
L'existence de la famille 
d'op\'erateurs tra\c{c}ables impose des conditions de r\'egularit\'e
sur les fonctions de Green. La propri\'et\'e {\bf A)} d\'efinit les op\'erateurs
pour les surfaces non connexes. L'axiome {\bf D)} explique le comportement des op\'erateurs
quand on renverse l'orientation d'une composante du bord. Globalement, mise \`a part
l'axiome {\bf E)}, ces propri\'et\'es d\'ecoulent
des sym\'etries des fonctions de Green expliqu\'ees dans le paragraphe consacr\'e 
au tenseur d'\'energie-impulsion.
Par contre la condition {\bf E)} est nouvelle. 
En effet, l'op\'eration de recollement entre deux 
surfaces permet de relier les op\'erateurs issus 
de surfaces de topologies a priori diff\'erentes.

On a vu, dans la section~{\bf 3.2}, comment \'ecrire
l'int\'egrale fonctionnelle
pour des champs \`a valeurs fix\'ees sur le bord. 
On avait alors interpr\'et\'e
les amplitudes quantiques pour le mod\`ele de WZNW
comme les sections d'un fibr\'e en droite au-dessus du groupe
des lacets. L'amplitude
$\NA\sous{\Sigma,(p_{i}),\gamma}((g_{i})\sous{i\in I})$,
o\`u $g_i\in LG$, devient le noyau de l'op\'erateur 
$\NA\sous{\Sigma,(p_{i}),\gamma}$ dans le langage de Segal.
Formellement, la traduction de l'axiome {\bf E)} 
en terme d'int\'egrale fonctionnelle est 
\[
\NA\sous{\Sigma,(p_{i}),\gamma}((g_{i})\sous{i\in I})%
=\mathop{\int}_{\scriptstyle%
			 g:\Sigma\rightarrow G\atop\scriptstyle%
			 g\circ p_{i}=g_{i},~i\neq i_{1},i_{2}}%
		\,\ee^{-k\,S\sous{\Sigma}(g)}\,Dg
=\mathop{\int}_{LG} Dg_0\mathop{\int}_{\scriptstyle
			 g:\Sigma\rightarrow G\atop
	{                g\circ p_{i}=g_{i},~i\neq i_{1},i_{2},\atop
			 g\circ p_{i}=g_0, ~i=i_{1},i_{2}}}
\,\ee^{-k\,S\sous{\Sigma}(g)}\,Dg\,.
\]

Supposons que $\Sigma$ est la sph\`ere de Riemann et que la
m\'etrique $\gamma\sous{D}$ est localement plate.
L'espace de Hilbert de Segal $\NH_{\rm F}$ sera l'espace $\CF$ des 
sections du fibr\'e $\pi:\CL^k\rightarrow LG$.
On munit $\NH_{\rm F}$ de la norme (formelle) $L^2$ induite par la
structure hermitienne sur $\CL^k\rightarrow LG$~:
\[
\big|\CF\big|^2\sous{L^2}=\mathop{\int}\limits_{LG}\big|\CF(g_0)\big|^2
\sous{\CL^k_{g_0}}\, Dg_0\,.
\]
Expliquons le lien entre l'espace des \'etats $\NH$ construit
dans la section~{\bf 5.1} et l'espace $\NH_{\rm F}$. 
Soit $X\in V\sous{D}$. On peut
lui associer un vecteur $\iota(X)\in\NH$. D'autre part, 
on lui adjoint une section (formelle) $\CF\sous{X}$ de $\CL^k$
de $\NH_{\rm F}$ par
\[
\CF\sous{X}(g_0)\ =\ (Z\sous{\vartheta^*\gamma\sous{D}\,
\sharp\,\gamma\sous{D}})^{-1/2}
	\mathop{\int}\limits_{g\left|_{S^1}\right.=g_0}
	X\,\ee^{-k\,S\sous{D}(g)}\,Dg\in(\CL^k)\sous{g_0}\,.
\]
On peut montrer que
\begin{align*}
\vert\CF\sous{X}\vert\sous{L^2}^2=
	=(Z\sous{\vartheta^*\gamma\sous{D}\,\sharp\,\gamma\sous{D}})^{-1}
	\mathop{\int}\limits_{g\,\sharp\, g':\C P^1\rightarrow G}&
	(\Theta X)\,X\,\ee^{-k\,S\sous{\C P^1}(g\,\sharp\, g')}
	\,D(g\,\sharp\, g')\\
=& \langle\,(\Theta X)\, X\,\rangle=\big(\,\iota(X),\,\iota(X)\,\big).
\end{align*}
On a donc construit deux r\'ealisations isom\'etriques de l'espace 
des \'etats, pour l'instant toutes les deux  bas\'ees sur
l'int\'egrale fonctionnelle.

\`A partir de l'\'equation~(\ref{opera}) 
et de son interpr\'etation en termes de noyau, on peut
d\'eriver la formule~(\ref{operateur}), en utilisant
\[
Z\sous{\widetilde{\gamma}}\,\langle\,
	\prod_{i\in I_+}(\Theta X_i)\,\prod_{i\in I_-}X_i\,\rangle
	=\int\prod_{i\in I_+}(\Theta X_i)\,\prod_{i\in I_-}X_i\,
		\ee^{-k\,S\sous{\widetilde{\Sigma}}(g)}\,Dg.
\]
Le principe est d'abord de fixer des valeurs $(g_{i})$ de $g$ sur le bord de
$\partial\Sigma$, puis d'int\'egrer it\'erativement sur celles-ci.

La d\'emarche de Segal repose sur le constatation 
suivante~: on peut extraire de l'ensemble
des op\'erateurs le reste de la structure, 
\cad l'action de l'alg\`ebre de Virasoro, 
les champs primaires, etc... Pour une 
explication compl\`ete de cette remarque, je renvoie 
au cours~\cite{gaw97:cft}.

Si on s'int\'eresse au disque $D$, muni d'une m\'etrique $\gamma$ et
de bord param\'etris\'e par $p:S^1\ni z\mapsto z\in \partial D$, 
$Z^{-1}\sous{\gamma}\,\NA\sous{D,\gamma}$ est un vecteur de $\NH$
ind\'ependant de la m\'etrique. D'apr\`es l'\'equation~\eqref{operateur},
c'est le vecteur vide. Pour $D'$ muni de la m\'etrique $\vartheta^*\gamma$, on trouve
l'application lin\'eaire $\big(\Omega,.\big)$. Les propri\'et\'es {\bf C)} et {\bf D)}
entra\^\i nent que $\CI\,\Omega=\Omega$.

Les anneaux sont les surfaces derri\`ere lesquelles se cache
le semi-groupe $\CS_q$ --- leurs amplitudes forment bien un semi-groupe
vu {\bf E)}.  Soit $N_q$ l'anneau 
$|q|\leq|z|\leq 1$, $0<|q|<1$, le bord entrant \'etant param\'etris\'e par
$z\mapsto qz$ et le bord sortant par $z\mapsto z$. On utilise la m\'etrique $|dz|^2/|z|^2$.
D'apr\`es la correspondance~\eqref{operateur},
\[
\big(\iota(X'),\,\NA\sous{N_q,|dz|^2/|z|^2}\iota(X)\big)
	=Z\sous{|dz|^2/|z|^2}\,\big(
	\iota(X'),\,\CS_q\iota(X)\big)\,.
\]
Comme on peut voir que $Z\sous{|dz|^2/|z|^2}=(q\bar q)^{-c/24}$, il suit
\[
\NA\sous{N_q,|dz|^2/|z|^2}=q^{L_0-c/24}\,\bar q^{\bar L_0-c/24}.
\]
L'application $z\mapsto \ee^{iz}$ est un diff\'eomorphisme analytique entre
le cylindre $C_\tau=\{z\ {\rm mod}\ 2\pi\ \big|\linebreak 
0\leq {\rm Im}\,z\leq 2\pi\imtau\}$, le bord sortant \'etant param\'etris\'e
par $z\mapsto\frac{1}{i}\,{\rm log} z$ et le bord entrant par $z\mapsto
\tau+\frac{1}{i}\,{\rm log} z$ et l'anneau $N_q$. Ici, $\tau=\tau_1+i\imtau$,
$\tau_1\in\R$ et $q=\ee^{2\pi i \tau}$. Cette transformation induit la
m\'etrique $|dz|^2$ sur $C_\tau$. L'amplitude du cylindre est alors
\[
\NA\sous{C_\tau,|dz|^2}=\ee^{-2\pi\imtau\,H}\,\ee^{2\pi i\tau_1\,P}
\]
o\`u $H$ est le Hamiltonien quantique et $P$ est l'op\'erateur moment~(\footnote{
\label{minkow}
Anticipons un petit peu sur ce qui va suivre. L'ensemble des $L_n$ --- comme celui des
$\bar L_n$ --- forme l'alg\`ebre de Virasoro ${\rm Vir}$. On obtient ainsi
une action de ${\rm Vir}\times{\rm Vir}$ dans $\NH$ qui reproduit
les sym\'etries conformes de la th\'eorie.
L'alg\`ebre de Virasoro est une extension centrale de ${\rm Vect}\, S^1$. 
En fait, on peut identifier ${\rm Vect}\, S^1\times {\rm Vect}\, S^1$ et
l'alg\`ebre de Lie des champs de vecteurs conformes sur le cylindre
minkowskien $\{(x_0,x_1)\ \big|\ x_1\ {\rm mod}\ 2\pi\}$ avec la
m\'etrique $dx_0^2-dx_1^2$. On voit que $H=L_0+\bar L_0$ est 
le Hamiltonien quantique --- quantification de la
translation en temps $x_0$ --- et $P=L_0-\bar L_0$ est l'op\'erateur
moment --- quantification de la translation en espace $x_1$.}). 
En comparant les deux amplitudes, on trouve
\[
H=L_0+\bar L_0-\quotient{c}{12},\qquad P+L_0-\bar L_0.
\]
Les op\'erateurs devant \^etre tra\c{c}ables,
$L_0$ et $\bar L_0$ doivent avoir un spectre discret
avec des multiplicit\'es finies. En plus $L_0\,\Omega=0=\bar L_0\,\Omega$,
donc la valeur propre $0$ est isol\'ee. L'\'energie du vecteur vide
est $c/12$. En collant les bords de $C_\tau$, on obtient
la courbe elliptique $E_\tau=\C/(\Z+\tau\Z)$.  Les axiomes {\bf C)} et {\bf E)}
permettent de d\'eterminer la {\bol fonction de partition toro\"\i dale}~:
\[
Z(\tau)\,=\,\NA\sous{E_\tau,|dz|^2}\,=\,\tr\,\, q^{L_0-c/24}\,\,
	\bar q^{\bar L_0-c/24}\,.
\]
La fonction de partition $Z(\tau)$ est un invariant modulaire : 
\[
Z(\tau)\,=\, Z({_{a\tau+b}\over^{c\tau+d}})
\]
parce que les courbes elliptiques $E_\tau$ et $E_{\tau'}$ o\`u
$\tau'={a\tau+b\over c\tau+d}$ sont reli\'ees par un diff\'eomorphisme 
holomorphe $z\mapsto z/(c\tau+d)$ qui multiplie la m\'etrique
$\vert dz\vert^2$ par une constante~(\footnote{La multiplication
de la m\'etrique par une constante ne change pas $Z(\tau)$
car l'action de Liouville en $\sigma={\rm cste}.$
s'annule pour un tore.}).

Enfin, on peut trouver 
les autres g\'en\'erateurs de l'alg\`ebre de Virasoro
en consid\'erant les anneaux $N_f=D\setminus f(D)$ o\`u $f$ est 
un plongement holomorphe du disque dans son int\'erieur.

\medskip
\section{D\'eveloppements alg\'ebriques}

Dans cette section, on traduit sous forme alg\'ebrique les sym\'etries de
la th\'eorie, cod\'ees par les d\'eveloppements
\`a courtes distances de la section {\bf 4}.

Montrons que les modes de l'alg\`ebre des courants v\'erifient 
\begin{align}
[J^{a}_{n},J^{b}_{m}]&=if^{abc}\,J^{a}_{n+m}%
+{_{kn}\over^2}\,\delta^{ab}\,\delta_{n+m,0},%
\label{KM1}\taga{KM1}\\
[\overline{J}^{a}_{n},\overline{J}^{b}_{m}]&=%
if^{abc}\,\overline{J}^{a}_{n+m}%
+{_{kn}\over^2}\,\delta^{ab}\,\delta_{n+m,0},%
\tagb{KM1}\nombre\\
[J^{a}_{n},\overline{J}^{b}_{m}]&=0.\label{KM3}
\end{align}
Ces relations de commutation --- valables sur $\NH_1$ ---, 
avec $k$ remplac\'e par un \'el\'ement
abstrait $\CK$ commutant avec tous les $J^{a}_{n}$, d\'efinissent
l'{\bol alg\`ebre de Kac-Moody affine} $\widehat{L\lieg}\eC$ 
engendr\'ee par les \'el\'ements $J^{a}_{n}$ et $\CK$. 

Soient $\iota(X)$ et $\iota(X')$, deux vecteurs de $\CS_{w}\NH_1$,
\[
\big(\,\iota(X'),[J^{a}_{n},\CJ^{b}(w)]\,\iota(X)\,\big)%
=\Big({_1\over^{2\pi i}}\,\oint_{|z|=|w|+\epsilon}dz%
-{_1\over^{2\pi i}}\,\oint_{|z|=|w|-\epsilon}dz\Big)\,z^n\,%
\langle\,(\Theta X')\,J^{a}(z)\,J^{b}(w)\,X\rangle
\]
o\`u l'ordre sur les op\'erateurs est dict\'e
par la r\`egle~(\ref{radialorder}).
Ensuite, on d\'eforme le contour pour obtenir une int\'egrale sur 
$\{z\in\C\ \big|\ |z-w|=\epsilon\}$ et on utilise 
l'\'equation~(\refa{ward7})
\[
\big(\,\iota(X'),[J^{a}_{n},\CJ^{b}(w)]\,\iota(X)\,\big)=%
{_1\over^{2\pi i}}\,\oint_{|z-w|=\epsilon}dz\,z^n\,%
\langle\,(\Theta X')\, \left({_{k\,\delta^{ab}/2}\over^{(z-w)^2}}%
+{_{if^{abc}}\over^{z-w}}\,J^{c}(w)+\cdots\right)\,X\,\rangle.
\]
Enfin, on d\'eveloppe $z^n$ autour de $z=w$, soit $z^n=w^n+n(z-w)\,w^{n+1}
+\cdots$ pour obtenir
\[
\big(\,\iota(X'),[J^{a}_{n},\CJ^{b}(w)]\,\iota(X)\,\big)=%
\langle\,(\Theta X')\,\left(if^{abc}\,w^n\,J^{c}(w)+{_{kn}\over^2}\,
\delta^{ab}\,w^{n-1}\right)\,X\rangle,
\]
soit
\[
[J^{a}_{n},\,\CJ^{b}(w)]=%
if^{abc}\,w^n\,\CJ^{c}(w)+{_{kn}\over^2}\,\delta^{ab}\,w^{n-1}.
\]
Apr\`es int\'egration sur $w$, on obtient le bon commutateur~(\refa{KM1}).
La traduction de tous les autres d\'eveloppements \`a courtes distances
se fait exactement de la m\^eme mani\`ere~; je ne r\'ep\'eterai donc pas
le calcul. Ainsi, \`a partir de l'\'equation~(\refb{ward7}), on 
trouve
\[
[\barJ^{a}_{n},\,\bar\CJ^{b}(w)]=%
if^{abc}\,\bar w^n\,\bar\CJ^{c}(w)+{_{kn}\over^2}\,\delta^{ab}\,\bar w^{n-1}
\]
et $[\barJ^{a}_{n},\,\CJ^{b}(w)]=0=[J^{a}_{n},\bar\CJ^{b}(w)]$~; d'o\`u
les commutateurs~(\refb{KM1}) et~\eqref{KM3}.

\`A partir des d\'eveloppements \` a courtes distances~(\ref{unpoint}.a-b),
on obtient les relations caract\'erisant
l'alg\`ebre des {\bol champs primaires pour l'alg\`ebre de Kac-Moody affine}~:
\begin{align*}
[J^{a}_n,\,\widehat{g}(z_\ell,\overline{z}_\ell)\sous{R_\ell}]&=%
	-z_\ell^n\,t^a\sous{\ell}\,\widehat{g}(z_\ell
	,\overline{z}_\ell)\sous{R_\ell},\\
[\barJ^{a}_n,\,\widehat{g}(z_\ell,\overline{z}_\ell)\sous{R_\ell}]&=%
	\overline{z}_\ell^n\,\widehat{g}(z_\ell
	,\overline{z}_\ell)\sous{R_\ell}\,t^a\sous{\ell}.
\end{align*}
Ces relations imposent des contraintes sur le vecteur vide $\Omega$~:
\begin{gather*}
\begin{aligned}
J^{a}_{n}\,\Omega&=0,\\
J^{a}_{0}\,\widehat{g}(0)\sous{R_\ell}\,\Omega&= 
	-t^a\sous{\ell}\,\widehat{g}(0)\sous{R_\ell}\,\Omega,\\
J^{a}_{n}\,\widehat{g}(0)\sous{R_\ell}\,\Omega&=0,
\end{aligned}
\qquad
\begin{aligned}
\barJ^{a}_{n}\,\Omega&=0,\\
\barJ^{a}_{0}\,\widehat{g}(0)\sous{R_\ell}\,\Omega&=
	\widehat{g}(0)\sous{R_\ell}\,t^a\sous{\ell}\,\Omega,\\
\barJ^{a}_{n}\,\widehat{g}(0)\sous{R_\ell}\,\Omega&=0,
\end{aligned}
\qquad
\begin{aligned}
\text{si $n\geq 0$},\\
\\
\text{si $n>0$}
\end{aligned}
\end{gather*}
o\`u $\widehat{g}(0)\sous{R_\ell}\,\Omega=\lim_{z_\ell\rightarrow 0}
\widehat{g}(z_\ell,\overline{z}_\ell)\sous{R_\ell}\,\Omega$.
Pour obtenir les deux premi\`eres relations, il faut se rappeler que
l'insertion de courants est analytique. Il appara\^\i t que le sous-espace 
engendr\'e par le vecteur vide correspond \`a la repr\'esentation triviale,
tandis que le sous-espace engendr\'e par les \'el\'ements matriciels de
$\widehat{g}(0)\sous{R_\ell}\,\Omega$ correspond \`a la repr\'esentation
$(R_\ell,\bar R_\ell)$ de l'alg\`ebre de Lie $\liegc\oplus\liegc$ engendr\'ee par
les $J^a_0$ et les $\bar J^a_0$.

Maintenant, on utilise les d\'eveloppements
\`a courtes distances faisant intervenir le
tenseur d'\'energie-impulsion.
En modes, la propri\'et\'e~(\refa{sugawara}) sur le tenseur 
\'energie-impulsion devient
\qq\label{tenseurmodes}
L_n={_2\over^{k+g^\vee}}\,\sum_{m\in\Z}\tr\,:J_{n-m}\,J_m:
\qqq
o\`u on utilise l'{\bol ordre normal} (ou ordre de Wick) 
sur les modes de Laurent
\[
:J_{p}\,J_q:=\begin{cases}
			J_p\,J_q, & \text{si $p<0$}\\
			J_q\,J_p, & \text{si $p\geq 0$}.
		\end{cases}
\]
Bien entendu, on a une \'egalit\'e similaire pour $\overline{L}_n$.
On reconna\^\i t la forme habituelle de la construction de Sugawara.
Commen\c{c}ons par r\'e\'ecrire $T(z)$ sous forme int\'egrale
\[
T(z)={_2\over^{k+g^\vee}}\,\oint_{|w-z|=\epsilon}dw\,{_1\over^{w-z}}\,
\tr\,J(w)\,J(z).
\]
On d\'eforme le contour d'int\'egration afin que l'ordre sur les champs
soit compatible avec l'ordre radial sur les op\'erateurs. Il vient
\begin{equation*}
\begin{split}
L_n={_2\over^{k+g^\vee}}\,\tr\,{_1\over^{2\pi i}}\,%
	\oint_{|z|<1}dz\,z^{n+1}&\Big({_1\over^{2\pi i}}\,
	\oint_{|w|=|z|+\epsilon}dw\,{_1\over^{w-z}}\CJ(w)\,\CJ(z)\\
	&\phantom{\ \ }-{_1\over^{2\pi i}}\,\oint_{|w|=|z|-\epsilon}dw%
	\,{_1\over^{w-z}}\,\CJ(z)\,\CJ(w)\Big).
\end{split}
\end{equation*}
La suite est standard, on d\'eveloppe $1\,/\,(w-z)$ en s\'erie,
$\CJ(z)$ et $\CJ(w)$ en s\'eries de Laurent,
on obtient alors des int\'egrales connues et
\[
L_n={_2\over^{k+g^\vee}}\,\Big(\sum_{q<0}\tr\,J_q\,J_{n-q}+
\sum_{q\geq 0}\tr\,J_{n-q}\,J_q\Big).
\]
						      
Les modes satisfont \'egalement des relations de commutation
\begin{align}
[L_n,L_m]&=(n-m)\,L_{n+m}+{_c\over^{12}}\,(n^3-n)\,\delta_{n+m,0},\label{vir}
	\taga{vir}\\
[{\overline L}_n,{\overline L}_m]&=%
 (n-m)\,{\overline L}_{n+m}+{_c\over^{12}}\,(n^3-n)\,\delta_{n+m,0},
	\tagb{vir}\nombre\\
[L_n,{\overline L}_m]&=0.\label{virnul}
\end{align}
Si on repr\'esente la charge centrale $c$ par un \'el\'ement
abstrait $\CC$ commutant avec les $L_n$, 
ces relations de commutation d\'efinissent l'{\bol alg\`ebre de Virasoro}, 
not\'ee $\mbox{Vir}$, engendr\'ee par les \'el\'ements $L_n$ et $\CC$. 
L'espace des \'etats est donc muni d'une repr\'esentation unitaire
(cf. section {\bf 5.3}) de ${\rm Vir}\times {\rm Vir}$, d\'efinie
sur un sous-espace dense, et de charge centrale agissant par $c$.
Pour d\'emontrer les trois \'equations, on proc\`ede comme pour l'alg\`ebre des courants,
mais avec les d\'eveloppements \`a courtes distances~(\ref{ward11}.a-b)
et (\ref{ward12}). On trouve d'abord
\begin{align}
\label{LT}
[L_n,\CT(w)]&=\quotient{c}{12}\,(n^3-n)\,w^{n-2}
	+2(n+1)\,w^n\,\CT(w)+w^{n+1}\,\partial_w\CT(w),\taga{LT}\\
[\bar L_n,\bar\CT(\bar w)]&=\quotient{c}{12}\,(n^3-n)\,\bar w^{n-2}
	+2(n+1)\,\bar w^n\,\bar\CT(\bar w)+\bar w^{n+1}\,\partialbarw\bar\CT(\bar w),
	\tagb{LT}\nombre\\
[L_n,\bar\CT(\bar w)]&=0=[\bar L_n,\CT(w)],
\end{align}
puis les commutateurs apr\`es int\'egration sur $w$ ou $\bar w$.
On pourrait aussi utiliser 
les propri\'et\'es de l'ordre normal et la
relation~(\ref{tenseurmodes}).

\`A partir des d\'eveloppements~(\ref{ward9}.a-b), (\ref{tenseur1}.a-b)
et (\ref{mixte1}.a-b), on montre que
\begin{align}
\label{pricom}
[L_n,\widehat{g}(z_\ell,\overline{z}_\ell)\sous{R_\ell}]&=%
	\Delta_\ell\,(n+1)\,z_\ell^n\,
	\widehat{g}(z_\ell,\overline{z}_\ell)\sous{R_\ell}
	+z_\ell^{n+1}\,\partial_{z_\ell}
	\widehat{g}(z_\ell,\overline{z}_\ell)\sous{R_\ell},
	\taga{pricom}\\
[\overline{L}_n,\widehat{g}(z_\ell,\overline{z}_\ell)\sous{R_\ell}]&=%
	\Delta_\ell\,(n+1)\,\overline{z}_\ell^n
	\,\widehat{g}(z_\ell,\overline{z}_\ell)\sous{R_\ell}
	+\overline{z}_\ell^{n+1}\,\partial_{\hspace{0.02cm}
	\bar z_\ell}\widehat{g}(z_\ell,\overline{z}_\ell)\sous{R_\ell},
	\tagb{pricom}\nombre\\
\label{pricom1}
[L_n,\CJ^a(w)]&=(n+1)\,w^n\,\CJ^a(w)+w^{n+1}\,\partial_w\CJ^a(w),
	\taga{pricom1}\\
[\bar L_n,\bar \CJ^a(\bar w)]&=(n+1)\,\bar w^n\,\bar\CJ^a
	(\bar w)+\bar w^{n+1}\,\partialbarw\bar\CJ^a(\bar w),
	\tagb{pricom1}\nombre\\
[L_n,\bar\CJ^a(\bar w)]&=0=[\bar L_n,\CJ^a(w)].\label{pricom2}
\end{align}
Ces relations redisent, au niveau op\'eratoriel, que 
$\widehat{g}(z,\overline{z})\sous{R_\ell}$ et $\CJ^a(z)$, 
$\bar \CJ^a(\bar z)$ sont des {\bol champs primaires} 
(pour l'alg\`ebre de Virasoro).
On retrouve aussi que les tenseurs d'\'energie-impulsion ne sont 
pas des champs primaires --- \`a cause du terme en $c$.
Apr\`es int\'egration sur $w$ ou $\bar w$ dans les \'equations~(\ref{pricom1}.a-b),
on obtient
\begin{gather}
\begin{aligned}
[L_n,J^a_m]&=-m\,J^a_{n+m},\\
[L_n,\barJ^a_m]&=0,
\end{aligned}
\qquad
\begin{aligned}
[\bar L_n,\barJ^a_m]&=-m\,\barJ^a_{n+m},\\
[\bar L_n,J^a_m]&=0.
\end{aligned}
\end{gather}
\`A partir de ces relations, on peut montrer facilement 
que les g\'en\'erateurs du semi-groupe 
$\{\CS_{q}\}$ co\"\i ncident avec les modes z\'ero 
de l'op\'erateur d'\'energie-impulsion.

Lorsqu'on a construit $L_0$ et $\bar L_0$, on a vu que
$H=L_0+\overline{L}_0\geq 0$. Si on utilise le fait que ces deux op\'erateurs
ne sont qu'un petit bout de l'alg\`ebre de Virasoro, on peut voir~\cite{gaw97:cft}
que les deux op\'erateurs sont positifs s\'epar\'ement, 
\ie $L_0\geq 0$ et $\overline{L}_0\geq 0$. 
Ainsi, seules les {\bol repr\'esentations 
\`a \'energie positive} de l'alg\`ebre de Virasoro
interviennent. 

Enfin, les relations~(\ref{pricom}-\ref{pricom2}) 
imposent des contraintes sur le vecteur vide $\Omega$~:
\begin{gather}
\begin{aligned}
L_n\,\Omega&=0,\\
L_n\,\widehat{g}(0)\sous{R_\ell}\,\Omega&=0,\\
L_0\,\widehat{g}(0)\sous{R_\ell}\,\Omega&=\Delta_\ell\,\widehat{g}(0)\sous{R_\ell}\,\Omega,\\
L_{-1}\,\widehat{g}(0)\sous{R_\ell}\,\Omega&=\partial_{z_\ell}
	\widehat{g}(0)\sous{R_\ell}\,\Omega,\\
L_n\,\CJ^a(0)\,\Omega&=0,\\
\bar L_n\,\CJ^a(0)\,\Omega&=0,\\
L_0\,\CJ^a(0)\,\Omega&=\CJ^a(0)\,\Omega,\\
L_{-1}\,\CJ^a(0)\,\Omega&=\partial_{w} \CJ^a(0)\,\Omega,
\end{aligned}
\qquad
\begin{aligned}
\overline{L}_n\,\Omega&=0,\\
\overline{L}_n\,\widehat{g}(0)\sous{R_\ell}\,\Omega&=0,\\
\overline{L}_0\,\widehat{g}(0)\sous{R_\ell)}\,\Omega&=\Delta_\ell\,
			\widehat{g}(0)\sous{R_\ell}\,\Omega,\\
\bar L_{-1}\,\widehat{g}(0)\sous{R_\ell}\,\Omega&=\partial_{\hspace{0.02cm}
	\bar z_\ell}
	\widehat{g}(0)\sous{R_\ell}\,\Omega,\\
\overline{L}_n\,\bar\CJ^a(0)\,\Omega&=0,\\
L_n\,\bar\CJ^a(0)\,\Omega&=0,\\
\overline{L}_0\,\bar\CJ^a(0)\,\Omega&=\bar\CJ^a(0)\,\Omega,\\
\bar L_{-1}\,\bar\CJ^a(0)\,\Omega&=\partial_{\hspace{0.02cm} \bar w}
	\bar\CJ^a(0)\,\Omega,\\
\end{aligned}
\qquad
\begin{aligned}
&\text{si $n\geq-1$},\\
&\text{si $n>0$,}\\
&{}\\
&{}\\
&\text{si $n>0$},\\
&\text{si $n\geq-1$},\\
&{}\\
&{}
\end{aligned}
\label{Virasoroan}
\end{gather}
o\`u $\CJ^a(0)\,\Omega=\lim_{w\rightarrow 0}
\CJ^a(w)\,\Omega$ et idem avec $\bar\CJ^a(0)$. En particulier, le
vecteur vide est un vecteur propre pour $L_0$ et $\bar L_0$ de
plus petite valeur propre $0$. On admet qu'il n'existe qu'un seul
vecteur de cette nature. On a un peu plus, puisque $\Omega$ est tu\'e
par la sous-alg\`ebre $\mathfrak{sl}_2\times\mathfrak{sl}_2$ 
de ${\rm Vir}\times{\rm Vir}$, engendr\'ee par
les $L_0$, $L_{\pm 1}$ et les $\bar L_0$, $\bar L_{\pm 1}$. Par contre,
$\Omega$ ne l'est pas par toute l'alg\`ebre ${\rm Vir}\times {\rm Vir}$~;
on dit que la {\bol sym\'etrie conforme} est {\bol bris\'ee}. 

Les vecteurs $\widehat{g}(0)\sous{R_\ell}\,\Omega$ sont aussi des
vecteurs propres pour $L_0$ et $\bar L_0$, mais de valeurs
propres $(\Delta_\ell,\Delta_\ell)$. Comme $L_0$ et $\bar L_0$ sont
des op\'erateurs positifs, on a en particulier $\Delta_\ell>0$. 
D'apr\`es les relations~\eqref{Virasoroan}, les vecteurs
$\widehat{g}(0)\sous{R_\ell}\,\Omega$, $\CJ^a(0)\,\Omega$
et $\bar\CJ^a(0)\,\Omega$ sont annul\'es par tous les g\'en\'erateurs
$L_n$ et $\bar L_n$ avec $n>0$. On dit que ce sont des vecteurs de plus haut poids
pour ${\rm Vir}\times{\rm Vir}$. Ainsi, quand on fait tendre les
coordonn\'ees des points d'insertion vers $0$, les op\'erateurs
issus de champs primaires appliqu\'es au vecteur vide
donnent des vecteurs de plus haut poids.

\medskip
\section{Alg\`ebre de Virasoro. Alg\`ebres de Kac-Moody affines}

On vient de constater que les sym\'etries de la th\'eorie sont 
cod\'ees dans les repr\'esentations de deux alg\`ebres de dimension
infinie~: l'alg\`ebre de
Virasoro et l'alg\`ebre des courants, \ie l'alg\`ebre de Kac-Moody 
affine. La discussion qui suit est \`a mettre en parall\`ele avec 
celle du premier chapitre sur les repr\'esentations des alg\`ebres de Lie de
dimension finie.

\subsection{Alg\`ebre de Virasoro}

L'alg\`ebre de Virasoro est un proche cousin de l'alg\`ebre de Witt.
Le groupe des diff\'eomorphismes ${\rm Diff}_+\,S^1$ du cercle, pr\'eservant
l'orientation, est un groupe de 
Lie~\cite{milnor}~--- au sens de Fr\'echet~\cite{hamilton} ---
ayant pour alg\`ebre de Lie l'alg\`ebre des champs de vecteurs $\CC^\infty$ sur
le cercle~: ${\rm Vect}\,S^1$. Le crochet de Lie pour ${\rm Vect}\,S^1$
est 
\[
\big[v_1(\theta)\,\frac{d}{d\theta},v_2(\theta)\,\frac{d}{d\theta}\big]
=\big(v_1(\theta)v'_2(\theta)-v'_1(\theta)v_2(\theta)\big)\,\frac{d}{d\theta}.
\]
L'{\bol alg\`ebre de Witt} est l'alg\`ebre complexifi\'ee ${\rm Vect}^\C S^1$
engendr\'ee par les g\'en\'erateurs et relations~: 
\[
\ell_n=i\,\ee^{in\theta}\,\frac{d}{d\theta},\qquad
[\ell_n,\ell_m]=(n-m)\,\ell_{n+m}.
\]
Les extensions centrales de ${\rm Vect}\,S^1$ par $\R$ sont toutes obtenues \`a
partir de l'extension centrale universelle~\cite{gelfandfuchs} 
$\widehat{{\rm Vect}}\,S^1={\rm Vect}\,S^1\oplus\R \CZ$ munie du crochet
\[
\big[v_1\,\frac{d}{d\theta},v_2\,\frac{d}{d\theta}\big]
=\big(v_1\,v'_2-v'_1\,v_2\big)\,\frac{d}{d\theta}
	+\quotient{1}{24\pi}\int_0^{2\pi}
	d\theta\, (v_1'+v_1''')\,v_2\ \CZ,\qquad
	\big[\CZ,v_1\frac{d}{d\theta}\big]=0.
\]
Les autres cocycles sont $\R$-proportionnels au cocycle pr\'ec\'edent ---
en termes plus formels, $\dim H^2({\rm Vect}\,S^1,\R)=1$. Pour le groupe
${\rm Diff}_+\,S^1$, on a une construction similaire qui conduit 
au groupe de Virasoro-Bott, extension centrale de
${\rm Diff}_+\,S^1$ par le cocycle de Bott.
L'extension centrale complexifi\'ee est l'alg\`ebre de {\bol Virasoro}.
Ainsi, ${\rm Vir}$ est l'extension centrale universelle de l'alg\`ebre
de Witt par $\C$~:
\[
0\rightarrow \C\rightarrow {\rm Vir}\rightarrow {\rm Vect}^\C S^1\rightarrow 0
\]
o\`u la seconde fl\`eche envoie $1$ sur $\CC$ ($=i\CZ$) et la troisi\`eme envoie
$L_n$ sur $\ell_n$. On introduit la d\'ecomposition triangulaire
de l'alg\`ebre~: $\CT=\C\,\CC\oplus\C L_0$ la sous-alg\`ebre de
Cartan, $\CN_+=\oplus_{n\geq 1} \C L_n$ et
$\CN_-=\oplus_{n\geq 1}\C L_{-n}$, de sorte que ${\rm Vir}=
\CN_-\oplus\CT\oplus\CN_-$. Soit $\lambda$ un \'el\'ement
de $\CT^*$, l'alg\`ebre duale \`a $\CT$~; 
on note $\lambda(\CC)=c$, $\lambda(L_0)=\Delta$.

On est particuli\`erement int\'eress\'e par les repr\'esentations unitaires
de plus haut poids de l'alg\`ebre de Virasoro. Un module $M_{c,\Delta}$
pour l'alg\`ebre de Virasoro --- comme on l'a d\'ej\`a remarqu\'e, c'est juste
une autre fa\c{c}on de parler de repr\'esentation ---
est un {\bol module de plus haut poids}
$\lambda$, soit $(c,\Delta)$, s'il existe un vecteur $v_{c,\Delta}$
tel que
\[
\CN_+\,v_{c,\Delta}=0,\quad \CU(\CN_-)\,v_{c,\Delta}=M_{c,\Delta},
	\quad \CC\,v_{c,\Delta}=c\,v_{c,\Delta},\quad
	L_0\,v_{c,\Delta}=\Delta\,v_{c,\Delta}.
\]
On dit que $v_{c,\Delta}$ est un {\bol vecteur de plus haut poids}, 
$c$ est la {\bol charge centrale} et $\Delta$ est le {\bol poids conforme}
de la repr\'esentation. En particulier, $M_{c,\Delta}$ est 
engendr\'e par les vecteurs $L_{-n_r}L_{-n_{r-1}}\cdots L_{-n_1}\,
v_{c,\Delta}$, avec $0<n_1\leq n_2\leq\cdots\leq n_r$. Ces vecteurs ne
sont pas n\'ecessairement ind\'ependants. On dit que $\ell=\sum_in_i$ est
le {\bol niveau} du vecteur $L_{-n_r}L_{-n_{r-1}}\cdots$$L_{-n_1}\,
v_{c,\Delta}$. 
On voit qu'un vecteur de niveau $\ell$ est un vecteur propre pour
$L_0$ de valeur propre $\Delta+\ell$ et pour $\CC$ de
valeur propre $c$. En particulier, le vecteur de plus haut poids
est le vecteur ayant la plus petite valeur propre pour $L_0$.
On obtient une d\'ecomposition du module
$M_{c,\Delta}$~:
\[
M_{c,\Delta}=\bigoplus_{\ell=1}^\infty M(\ell)_{c,\Delta}
\]
o\`u $M(\ell)_{c,\Delta}$ est le sous-espace des vecteurs de niveau $\ell$~---
les vecteurs de niveaux diff\'erents sont clairement ind\'ependants~---
avec $\dim M(\ell)_{c,\Delta}\leq p(\ell)$, le nombre de partitions de $\ell$.
Un module de plus haut poids pour lequel tous les vecteurs 
$L_{-n_r}L_{-n_{r-1}}\cdots L_{-n_1}\,v_{c,\Delta}$
sont ind\'ependants est appel\'e un {\bol module de Verma} $V_{c,\Delta}$,
\cad $\dim V(\ell)_{c,\Delta}=p(\ell)$. 
Son existence est garantie pour tout couple $(c,\Delta)$ et il est unique
\`a isomorphisme pr\`es.

Parmi le vecteurs de niveau $\ell$, on isole les {\bol vecteurs singuliers},
\cad les vecteurs tu\'es par $\CN_+$.
Tout vecteur $v_s$ tu\'e par $\CN_+$ est la somme
de vecteurs singuliers. Les vecteurs singuliers de niveau $\ell$ engendrent un sous-module de 
$V_{c,\Delta}$ isomorphe \`a $V_{c,\Delta+\ell}$. 
On montre que

--- tout sous-module de $V_{c,\Delta}$ est engendr\'e par
	ses vecteurs singuliers~;

--- tout module de plus haut poids $(c,\Delta)$ est isomorphe
	\`a un quotient du module de Verma~;

--- \`a isomorphisme pr\`es, le quotient de $V_{c,\Delta}$
	par son sous-module maximal propre est l'unique
	module $H_{c,\Delta}$ irr\'eductible de plus haut poids $(c,\Delta)$.

\noindent Pour paraphraser, $H_{c,\Delta}$ est le module, de plus haut poids,
le plus petit et $V_{c,\Delta}$ est le plus grand.

Un module $M$ pour l'alg\`ebre de Virasoro est dit {\bol unitaire} s'il existe une 
forme $(.,.)$ hermitienne positive sur $M$ telle que
\[
(v,L_n\,w)=(L_{-n}\,v,w)\qquad \mbox{et}\qquad (v,\CC\,w)=(\CC\,v,w)
\]
pour tout $v,w\in M$. Dans ce cas les valeurs propres de $\CC$ et $L_0$ sont r\'eelles.
Si $c$ et $\Delta$ sont r\'eels, il existe une unique forme hermitienne $\langle.,
.\rangle$, {\bol la forme de Shapovalov}, sur le module de Verma $V_{c,\Delta}$,
telle que $\langle v,L_n\,w\rangle=\langle L_{-n}\,v,w\rangle$ pour tout
$v,w\in V_{c,\Delta}$ et telle que $\langle v_{c,\Delta},v_{c,\Delta}\rangle=1$.
Cette forme peut \^etre d\'eg\'en\'er\'ee. On introduit 
${\rm Nul}_{c,\Delta}$ le sous-espace invariant des vecteurs orthogonaux
\`a tous les autres vecteurs de $V_{c,\Delta}$. On a alors

--- si $v^{(\ell)}$ et $v^{(\ell')}$ sont des vecteurs de niveau $\ell$ et $\ell'$ diff\'erents,
	on a $\langle v^{(\ell)},v^{(\ell')}\rangle =0$~;

--- tout vecteur singulier de niveau positif appartient \`a ${\rm Nul}_{c,\Delta}$~;

--- ${\rm Nul}_{c,\Delta}$ est le sous-module maximal propre de $V_{c,\Delta}$, 
	d'o\`u $H_{c,\Delta}=V_{c,\Delta}/{\rm Nul}_{c,\Delta}$.

Il reste un probl\`eme important, celui de l'unitarit\'e de
$H_{c,\Delta}$. La question est donc de savoir si la forme hermitienne
induite sur $H_{c,\Delta}$ par la forme de Shapovalov est une forme
posi\-tive. D\'ej\`a, si $n>0$,
$\langle L_{-n}\,v_{c,\Delta},L_{-n}\,v_{c,\Delta}\rangle
	=2n\,\Delta +\frac{c}{12}\,(n^3-n)$.
On a donc la condition n\'ecessaire~: $c,\Delta\geq 0$. 
Regardons l'extension triviale, \ie $c=0$. Si $v=L_{-n}L_{-n}\,v_{0,\Delta}$
et $w=L_{-2n}\,v_{0,\Delta}$, le d\'eterminant de Gram donne
$-20n^4\,\Delta^2+32n^3\,\Delta^3$. Donc la seule possibilit\'e restante est
$\Delta=0$, \cad la repr\'esentation triviale. Ainsi, il n'existe 
pas de repr\'esentation unitaire non-triviale de l'alg\`ebre ${\rm Vect}^\C S^1$.
Pour obtenir une information plus compl\`ete, il suffit d'\'etudier
le signe de la forme de Shapovalov sur le sous-espace $H_{c,\Delta}$ des vecteurs
de niveau $\ell$. Soit $D_\ell(c,\Delta)$ le d\'eterminant de la 
matrice $m_\ell(c,\Delta)$, de taille $p(\ell)\times p(\ell)$ et d'\'el\'ements
$\langle L_{-n_r}L_{-n_{r-1}}\cdots L_{-n_1}\,v_{c,\Delta},
L_{-n'_r}L_{-n'_{r-1}}\cdots L_{-n'_1}\,v_{c,\Delta}\rangle$ avec
$\sum n_i=\sum_i n'_i=\ell$. En clair,  $H_{c,\Delta}$ est unitaire si,
 et seulement si,
la matrice $m_\ell(c,\Delta)$ est positive pour tout
$\ell$. En particulier, $D_\ell(c,\Delta)\geq 0$.
Ce d\'eterminant est donn\'e par la {\bol formule
de Kac} --- formule prouv\'ee par Feigin et Fuchs ---~:
\[
D_\ell(c,\Delta)=\kappa_\ell\mathop{\prod}_{\scriptstyle
	1\leq r,s\leq \ell\atop
	\scriptstyle 
	r,s\in \N}
	\big(\Delta-\Delta_{r,s}(m)\big)^{\,p(\ell-rs)}
\]
o\`u $m$ est la racine de l'\'equation
\[
c=c_m=1-\quotient{6}{m(m+1)},
\]
\[
\Delta_{r,s}(m)=\frac{((m+1)\,r-ms)^2-1}{4m(m+1)}, \qquad 
\kappa_\ell=\prod_{\scriptstyle
	1\leq r,s\leq \ell\atop
	\scriptstyle 
	r,s\in \N}
	\big((2r)^s\,s!\big)^{n(r,s)}
\]
avec $n(r,s)$ le nombre de partitions de $\ell$ dans lesquelles
$r$ appara\^\i t $s$ fois. 

Quand $\Delta\mapsto \infty$, $m_\ell(c,\Delta)$ devient une matrice diagonale
d'\'el\'ements positifs. Il suit de la formule de Kac que $D_\ell(c,\Delta)$
est $>0$ pour $c>1$ et $\Delta>0$. Par cons\'equent,
$m_\ell(c,\Delta)$ est non-d\'eg\'en\'er\'ee et positive pour 
$c>1$ et $\Delta>0$ et positive pour $c\geq 1$ et $\Delta\geq 0$.
Le module de Verma $V_{c,\Delta}$ est donc irr\'eductible pour 
$c>1$ et $\Delta>0$ et les repr\'esentations 
irr\'eductibles $H_{c,\Delta}$ sont unitaires si $c\geq 1$ et $\Delta\geq 0$.
Pour $c=1$, on a
\[
D_\ell(1,\Delta)=\kappa_\ell\mathop{\prod}_{\scriptstyle
	1\leq r,s\leq \ell\atop
	\scriptstyle
	r,s\in \N}
	\big(\Delta-\quotient{(r-s)^2}{4}\big)^{\,p(\ell-rs)}
\]
donc $V_{1,\Delta}$ est irr\'eductible si, et seulement si, $\Delta\neq
\frac{n^2}{4}$ pour $n\in\Z$. Ainsi, $H_{c,\Delta}$ co\"\i ncide avec
le module de Verma si, et seulement si, $c>1$ ou $c=1$ et $\Delta\neq
\frac{n^2}{4}$ pour $n\in\Z$.

Le cas le plus int\'eressant est~: $0\leq c\leq 1$ et $\Delta\geq 0$.
Dans ce contexte, une repr\'esentation $H_{c,\Delta}$,
irr\'eductible de plus haut poids $(c,\Delta)$, est unitaire si,
et seulement si, le plus haut poids appartient \`a la
s\'erie discr\`ete
\begin{gather*}
\Bigg\{
\begin{aligned}
c&=c_m,\\
\Delta&=\Delta_{r,s}(m),
\end{aligned}
\qquad
\begin{aligned}
m&=2,3,\cdots,\\
1\leq r\leq m&-1, 1\leq s\leq r.
\end{aligned}
\end{gather*}
La partie directe est un travail de Friedan, Qiu et Shenker~\cite{fqs1,fqs2}.
La r\'eciproque est donn\'ee par Goddard, Kent et Olive dans~\cite{gko1,gko2},
o\`u les auteurs construisent explicitement les repr\'esentations
irr\'eductibles unitaires de plus haut poids, par {\bol construction 
quotient}. La construction quotient peut \^etre reproduite 
par int\'egration fonctionnelle en jaugeant le mod\`ele de WZNW
par un sous-groupe de $G$~\cite{gaw89:coset} (cf. chapitre 1).

Toute repr\'esentation unitaire de plus haut poids de ${\rm Vir}$
s'int\`egre en une repr\'esentation unitaire projective 
de ${\rm Diff}_+S^1$ dans la compl\'etion de l'espace 
$H_{c,\Delta}$~\cite{goodman2}.

\subsection{Alg\`ebre de Kac-Moody affine}

Soit $G$ un groupe de Lie v\'erifiant les m\^emes hypoth\`eses que
dans la section {\bf 2} du chapitre 1. L'alg\`ebre
de Lie du groupe des lacets $LG$ --- $LG$ est
un groupe de Lie de dimension infinie~\cite{pressley} ---
est  $L^\infty\lieg=\CC^\infty(S^1,\lieg)$~(\footnote{Habituellement, on note
plut\^ot $L\lieg$ ce que l'on appelle $L^\infty\lieg$ et $L_{\rm pol}\lieg$
l'alg\`ebre des lacets polynomiaux \`a valeurs dans $\lieg$, not\'ee
ici $L\lieg$.}).
On a \'egalement
les versions complexifi\'ees~: $L\Gc$ et $L^\infty\liegc$. Soit $\widehat{L^\infty\lieg}=
L^\infty\lieg\oplus \R\CZ$ l'extension centrale de $L^\infty\lieg$ par $\R$ par
le cocycle
\[
\omega(\eta,\xi)=\quotient{1}{2\pi}\int_0^{2\pi}d\theta\,\tr\,\eta'(\theta)\,
	\xi(\theta),
\]
pour $\eta$ et $\xi$ dans $L^\infty\lieg$. Si $(t^a)$ est une base 
orthogonale de l'alg\`ebre de Lie complexifi\'ee, on note
$J^a_n=t^az^n$. Les $J^a_n$ engendrent l'{\bol
alg\`ebre de Kac-Moody affine}
\[
\widehat{L\lieg}\eC=\Big(\bigoplus_{a\,;\,n\in\Z}
	\C J^a_n\Big)\oplus \C\CK\ \subset\ \widehat{L^\infty\lieg}\eC,
\]
avec $\CK=i\CZ$. C'est une extension centrale de l'alg\`ebre $L\lieg^\C$ des
lacets polynomiaux \`a valeurs dans $\liegc$ par $\C$~:
\[
0\rightarrow \C\rightarrow \widehat{L\lieg}\eC\rightarrow L\liegc\rightarrow 0
\]
o\`u la seconde fl\`eche envoie $1$ sur $\CK$ et la troisi\`eme envoie les $J^a_n$
sur les $j^a_n=t^az^n$ qui satisfont les relations $[j^a_n,j^b_m]=if^{abc}j^c_{n+m}$.

On \'etudie plut\^ot l'alg\`ebre $\widetilde{L\lieg}\eC$, obtenue \`a partir
de l'alg\`ebre de Kac-Moody affine en ajoutant un g\'en\'erateur $\CalD$ tel que
$[\CalD,J^a_n]=n\,J^a_n$ et $[\CalD,\CK]=0$. On peut alors d\'evelopper
des techniques reproduisant les objects connus pour 
les alg\`ebres de Lie simples~\cite{kac}. On peut ainsi construire
des hyperplans affines, des chambres, etc... On restera n\'eanmoins
plus terre \`a terre.  L'alg\`ebre $\widetilde{\liet}
=\R\CalD\oplus\liet\oplus\R\CK$ sera la {\bol
sous-alg\`ebre de Cartan} de $\widetilde{L\lieg}\eC$.
On munit $\widetilde{\liet}$ du produit scalaire lorentzien~:
\[
\langle a\CalD+X+b\CK,a'\CalD+Y+b'\CK\rangle
	=\tr\,XY+ab'+a'b.
\]
L'espace dual $\widetilde{\liet}^*$ est \'egal \`a $\R\CalD^*\oplus
\liet^*\oplus\R\CK^*$, avec $\CalD^*(a\CalD+Z+b\CK)=a$,
$\CK^*(a\CalD+X+b\CK)=b$ et $\mu(a\CalD+X+b\CK)=\mu(X)$, pour $\mu\in\liet^*$
et $X\in\liet$. Les {\bol racines affines} $\widetilde{\alpha}$ de 
$\wi{L\lieg}\eC$ sont, par d\'efinition, les \'el\'ements
de $\wi{\liet}^*$ de la forme $n\CalD^*+\alpha$, avec $n\in\Z$, $\alpha\in\Delta
\cup\{0\}$ et $(n,\alpha)\neq(0,0)$. On note $e_{\wi{\alpha}}=e_\alpha z^n$, si
$n\neq 0$, et $J^j_n=h^jz^n$, si $\alpha=0$. Les  $e_{\wi{\alpha}}$ jouent 
le r\^ole des $e_\alpha$ pour $\wi{L\lieg}\eC$. On dit qu'une racine affine
est {\bol positive} si $n=0$ et $\alpha>0$, ou $n>0$ et $\alpha\in\Delta\cup\{0\}$. 
Les racines affines {\bol simples} sont les $\wi{\alpha}_i=\alpha_i$,
$i=1,\cdots r$, et 
$\wi{\alpha}_0=\CalD^*-\phi$, o\`u $\phi$ est la racine la plus grande.
La {\bol coracine} d'une racine affine $\wi{\alpha}=n\CalD^*+\alpha$ est
$\wi{\alpha}^\vee=\alpha^\vee+\frac{n}{2}\,\tr(\alpha^\vee)^2\,\CK$.
Les r\'eflexions $r_{\wi{\alpha}}$, donn\'ees par
$\wi{\liet}^*\ni\wi{\mu}\mapsto\wi{\mu}-\wi{\mu}(\wi{\alpha}^\vee)\,
\wi\alpha\in\wi{\liet}^*$, engendrent le {\bol groupe de Weyl affine}
$\wi{W}$. Le groupe $\wi{W}$ est le produit semi-direct de $W$ 
et du r\'eseau des coracines $Q^\vee$. 
Un \'el\'ement $w$ de $W$ ($q^\vee$ de $Q^\vee$) envoie
$\wi{\mu}=e\CalD^*+\mu+k\CK^*$ sur $e\CalD^*+w(\mu)+k\CK^*$ (resp.
$(e-\mu(q^\vee)-\frac{k}{2}\,\tr(q^\vee)^2\,\CalD^*+\mu+
k\,\tr\,(q^\vee\cdot)+k\CK^*$). La {\bol matrice de Cartan affine} $\wi A$ est la matrice 
$(r+1)\times(r+1)$ d'\'el\'ements $\wi{a}_{ij}=2\,\langle\wi\alpha_i,
\wi\alpha_j\rangle/\langle\wi\alpha_j,\wi\alpha_j\rangle$. La matrice $\wi A$
est d\'eg\'en\'er\'ee de rang $r$. Pour rappel, la matrice de Cartan 
$A$ n'est pas d\'eg\'en\'er\'ee. En fait, les alg\`ebres de Kac-Moody affines
s'incorporent dans un formalisme beaucoup plus g\'en\'eral. Il s'agit de 
comprendre quelles sont les alg\`ebres de Lie que l'on peut obtenir
en gardant la propri\'et\'e~\eqref{matricecartan1} des matrices de Cartan,
mais en changeant la condition~\eqref{matricecartan2}. Ce faisant on obtient
une matrice de Cartan g\'en\'eralis\'ee~; les alg\`ebres qui s'en d\'eduisent
sont appel\'ees des alg\`ebres de Kac-Moody. Les alg\`ebres de Kac-Moody affines sont
les alg\`ebres de Kac-Moody ayant une matrice de Cartan g\'en\'eralis\'ee
pour laquelle  $\det A=0$ et tous les mineurs principaux propres sont strictement
positifs --- bien entendu, on appelle matrice de Cartan affine une telle
matrice. Les alg\`ebres de Kac-Moody affines obtenues par extension 
centrale de $L\lieg\eC$ sont les
alg\`ebres de Kac-Moody affines non-tordues --- \<<untwisted\>> en anglais.
La matrice $\wi A$ caract\'erise compl\`etement l'alg\`ebre de Kac-Moody
affine. Comme pour les alg\`ebres de Lies simples, on conna\^\i t 
une classification des
alg\`ebres de Kac-Moody affines. Il y a ainsi $7$ s\'eries infinies et $9$ alg\`ebres
exceptionnelles. Les alg\`ebres non-tordues sont not\'ees g\'en\'eriquement
$X^{(1)}_r$~: quatre s\'eries infinies, $A^{(1)}_r$ ($r\geq 2$), $B^{(1)}_r$ ($r\geq 3$),
$C^{(1)}_r$ ($r\geq 2$), $D^{(1)}_r$ ($r\geq 4$) et $6$ cas exceptionnels,
$A^{(1)}_1$, $E^{(1)}_6$, $E^{(1)}_7$, $E^{(1)}_8$, $F^{(1)}_4$ et $G_2^{(1)}$.

Un \'el\'ement $\wi\lambda=e\CalD^*+\lambda+k\CK^*$ est un {\bol
poids affine} si $\wi\lambda(\wi\alpha^\vee)\in\Z$, \ie
$\lambda$ est un poids de $\liegc$ et $k\in\Z$. Les {\bol poids affines
fondamentaux} sont $\wi\lambda_0=\CK^*$ et $\wi\lambda_i
=\lambda_i+k_i^\vee\CK^*$, $i=1,\cdots r$. Rappelons que les $k_i^\vee$ sont
les labels duaux de Kac. En particulier, $\wi\lambda_i(\wi\alpha_j^\vee)
=\delta_{ij}$. Un poids affine est dit {\bol dominant} si $\wi\lambda(\wi\alpha)\geq0$,
pour toute racine affine positive ($\alpha\neq 0$). Un poids affine
est dominant si, et seulement si, $k\in\N$ et $\lambda$ est un poids
dominant de $\liegc$ avec en plus la {\bol condition
d'int\'egrabilit\'e}~:
\[
\lambda(\phi^\vee)\leq k.
\]
\`A $k$ fix\'e, il n'y a qu'un nombre fini de poids dominants
satisfaisant cette condition, qui d'ailleurs ne fait pas intervenir la
composante $\CalD^*$.

Un module $M$ pour l'alg\`ebre $\wi{L\lieg}\eC$ est un {\bol module de plus haut poids}
$\wi\lambda$ s'il existe un vecteur $v_{\wi\lambda}$ tel que
\[
e_{\wi\alpha} v_{\wi\lambda}=0,\quad \forall \wi\alpha>0,\qquad \wi Xv_{
	\wi\lambda}=\wi\lambda(\wi X)\,v_{\wi\lambda},
	\quad \forall \wi X\in\wi\liet~\quad{\rm et}\quad
M=\CU(\wi{L\lieg}\eC)v_{\wi\lambda}.
\]
On dit que $v_{\wi\lambda}$ est un {\bol vecteur de plus haut poids},
$\wi\lambda=e\CalD^*+\lambda+k\CK^*$ est le {\bol plus haut poids affine},
$-e$ est l'{\bol \'energie}, $\lambda$ est le {\bol plus haut poids},
$k$ est le {\bol niveau} de la repr\'esentation.  Un module
est dit {\bol unitaire} s'il existe une forme hermitienne 
positive sur $M$ pour laquelle
\[
(J^a_n)^\dagger=J^a_{-n},\qquad \CK^\dagger=\CK,\qquad \CalD^\dagger=\CalD.
\]
Les m\^emes d\'efinitions s'appliquent \`a l'alg\`ebre de Kac-Moody
affine $\widehat{L\lieg}\eC$, en oubliant tout ce qui concerne $\CalD$.
Si on a un module de plus haut poids pour l'alg\`ebre de Kac-Moody
affine, on peut d\'efinir une action de l'alg\`ebre de Virasoro
gr\^ace \`a la construction de Sugawara~\eqref{tenseurmodes}.
Cette action est une r\'eminiscence de l'action de ${\rm Diff}_+S^1$
sur le groupe des lacets $LG$.
Si le module pour $\widehat{L\lieg}\eC$ est unitaire, celui de l'alg\`ebre
de Virasoro est aussi unitaire, de charge centrale $c_k=k\,\dim G/(k+g^\vee)$ et
$[L_n,J^a_m]=-mJ^a_{n+m}$. On peut donc \'etendre le module de plus haut poids
pour $\widehat{L\lieg}\eC$ en un module de plus haut poids pour $\wi{L\lieg}\eC$, 
en prenant pour $\CalD$~: $-L_0+\frac{c_k}{24}$. Comme 
$L_0 v_{\wi\lambda}=\frac{C_\lambda}{k+g^\vee}\,v_{\wi\lambda}$ ---
$C_\lambda$ est le Casimir quadratique --- le vecteur de plus haut poids
affine pour le nouveau module est 
\[
\wi\lambda'= \big(-\quotient{C_\lambda}{k+g^\vee}+\quotient{c_k}{24}
	\big)\,\CalD^*+\wi\lambda.
\]

Encore une fois, parmi toutes les repr\'esentations de m\^eme plus haut poids 
de $\wi{L\lieg}\eC$ ou $\widehat{L\lieg}\eC$, on a la plus grande,
le module de Verma, et la plus petite, l'irr\'eductible. La forme
de Shapovalov est la forme hermitienne sur le module de Verma telle 
que $(J^a_n)^\dagger=J^a_{-n}$, $\CK^\dagger=\CK$,
$\CalD^\dagger=\CalD$ et 
$\langle v_{\wi\lambda},v_{\wi\lambda}\rangle=1$. Le sous-espace ${\rm Nul}$
est le plus grand sous-espace propre invariant~; la forme de Shapovalov 
descend alors en une forme hermitienne non-d\'eg\'en\'er\'ee
sur le quotient irr\'eductible du module de Verma, celle-ci \'etant positive
si, et seulement si, le plus haut poids est un poids affine dominant.

Les classes d'\'equivalence de repr\'esentations irr\'eductibles,
unitaires, de plus haut poids de $\wi{L\lieg}\eC$, ou
$\widehat{L\lieg}\eC$ sont donc en correspondance
bijective avec l'ensemble des poids affines dominants  et
donc \'enum\'er\'ees par leur niveau $k\in\N$ et
leur poids int\'egrable 
$\lambda$, \ie $\lambda(\phi^\vee)\leq k$~\cite{garland}.
Les repr\'esentations de plus haut poids ne diff\'erant que par 
leur composante $\CalD^*$ sont reli\'ees par une translation de $\CalD$~;
on peut donc toujours supposer que $\CalD$ est donn\'e par la construction de 
Sugawara. Les repr\'esentations 
unitaires de plus haut poids de $\widehat{L\lieg}\eC$
s'int\`egrent en des repr\'esentations unitaires projectives de $LG$ dans
la compl\'etion de $M$~\cite{goodman1}. 

Les {\bol caract\`eres affines} d'une repr\'esentation de $\wi{L\lieg}\eC$ sont 
\[
\chi(\wi u)=\tr\,\ee^{i\wi u}
\]
pour $\wi u\in\wi\liet$. L'espace de repr\'esentation \'etant de
dimension infinie, il faut faire un peu attention. Pour une repr\'esentation
irr\'eductible de plus haut poids, il suffit de consid\'erer
les espaces propres de $\CalD$. Les contributions des diff\'erentes
valeurs propres de $\CalD$ mises ensemble convergent \`a condition que
$\wi u=-2\pi\tau\CalD+u+b\CK$, avec $\tau=\tau_1+i\tau_2$ et $\tau_2$ positive.
Pour les repr\'esentations unitaires de plus haut poids, $\ee^{i\wi u}$ devient
un op\'erateur tra\c{c}able dans la compl\'etion de l'espace de
repr\'esentation, de sorte qu'on peut prendre la $\tr$ au
sens op\'eratoriel.
Les caract\`eres dans une repr\'esentation irr\'eductible unitaire de plus haut poids
$\wi\lambda$ sont donn\'es par la {\bol formule des caract\`eres de Weyl-Kac}~:
\[
\chi_{\wi\lambda}(\wi u)=\Pi(\wi u)^{-1}\sum_{\wi w\in\wi W}(-1)^{\ell(\wi w)}
	\ee^{i\wi w\,(\wi \lambda+\wi\rho)(\wi u)}
\]
o\`u $\wi u\in\wi\liet$, $\wi\rho=\rho+g^\vee\CK^*$,
$\ell(\wi w)$ est le nombre de racines positives transform\'ees par $\wi w$
en racines n\'egatives et $\Pi(\wi u)$ est le {\bol d\'enominateur de Weyl-Kac}
\[
\Pi(\wi u)=\sum_{\wi w\in\wi W}(-1)^{\ell(\wi w)}\ee^{i \wi w\,\wi \rho(\wi u)}
	=\ee^{i\wi\rho(\wi u)}\prod_{\wi \alpha>0}(1-\ee^{-i\wi\alpha(\wi u)}).
\]
Si $\wi\lambda=(-\frac{C_\lambda}{k+g^\vee}+\frac{c_k}{24}
)\,\CalD^*+\lambda+k\CK^*$, on utilise aussi les fonctions suivantes~:
\[
\chi_{\wi\lambda}(\wi u)=\ee^{ibk}\chi^k_\lambda(u,\tau),
	\qquad \Pi(\wi u)=q^{-d/24}\,\ee^{ibg^\vee}\Pi(\quotient{u}{2\pi},\tau)
\]
o\`u $d=\dim \lieg$ et $q=\ee^{2\pi i\tau}$. En utilisant la premi\`ere forme
du d\'enominateur et $\wi W=W\ltimes Q^\vee$, on trouve l'identit\'e
de Macdonald~:
\[
\Pi(u,\tau)=\sum_{p^\vee\in Q^\vee}
	q^{|\rho^\vee+g^\vee p^\vee|^2/(2g^\vee)}
	\sum_{w\in W}(-1)^{\ell(w)}\ee^{2\pi i\, w(\rho+g^\vee\tr\,(p^\vee\,.))u}.
\]
La somme sur $w$ fait appara\^\i tre le caract\`ere $\chi_{g^\vee\tr\,(p^\vee\,.)}(2\pi u)$
de l'alg\`ebre de Lie finie $\lieg$. Le facteur $q^{-d/24}$ appara\^\i t
via la formule \'etrange de Freudenthal-de Vries : 
$\tr(\rho^2)/(\tr(\phi^2)\, g^\vee)=d/24$.
On voit alors que le d\'enominateur de Weyl-Kac
satisfait une \'equation du type de la chaleur, \cad
\[
4\pi i\,g^\vee\,\Pi^{-1}\partial_\tau\Pi=\Pi^{-1}\Delta_u\Pi
\]
o\`u $\Delta_u$ est le Laplacien $\sum_{j=1}^r\partial^2_{u^j}$, si
$u=u^jh^j$.
En utilisant la deuxi\`eme forme du d\'enominateur, on trouve 
\[
\Pi(u,\tau)=q^{d/24}
	\prod_{\alpha\in\Delta_+}\,(\ee^{\pi iu_\alpha}-
	\ee^{-\pi iu_\alpha})\,
	\prod\limits_{\ell=1}^\infty\Big[(1- q^\ell)^{r}
	\,\prod_{\alpha\in\Delta}(1-q^\ell\ee^{2\pi iu_\alpha})
	\Big]
\]
o\`u $r=\dim\liet$, $u_\alpha=\alpha(u)$.

\medskip
\section{Espace des \'etats de WZNW}

La d\'ecomposition suivante de l'espace des \'etats est compatible
avec l'\'etude men\'ee dans la section {\bf 6}
\[
\NH=\bigoplus_{\lambda(\phi^\vee)\leq k}
	\NH^k_\lambda\otimes\NH^k_{\bar\lambda}
\]
o\`u $\NH^k_\lambda$ est la compl\'etion de l'espace de repr\'esentation
$\NV_\lambda^k$ de la repr\'esentation irr\'eductible de plus haut poids
de $\widehat{L\lieg}\eC$ de niveau $k$ et de plus haut poids $\lambda$.
On note $v^k_\lambda$ le vecteur de plus haut poids. L'espace
$\NH^k_{\bar\lambda}$ correspond \`a la repr\'esentation complexe conjugu\'ee.
Le vecteur vide est donc proportionnel au vecteur $v_0^k\otimes v_0^k$ et
les \'el\'ements matriciels de $\widehat{g}(0)\sous{R_\ell}\,\Omega$ engendrent
le sous-espace obtenu en appliquant les polyn\^omes en $J^a_0$ et $\bar J^a_0$ \`a
$v^k_\lambda\otimes v^k_{\bar\lambda}$, o\`u $\lambda$ est le plus haut poids
de la repr\'esentation $R_\ell$. 

Toutes les repr\'esentations de $G$ ne rentrent pas dans
la d\'ecomposition de l'espace des \'etats. On appelle souvent
une repr\'esentation de plus haut poids $\lambda$ telle que que 
$\lambda(\phi^\vee)\leq k$ une repr\'esentation {\bol int\'egrable}. 
Il semble ainsi que seules les repr\'esentations int\'egrables 
puissent intervenir dans le calcul des fonctions de Green (cf. chapitre 3).



\chapter{\'Etats de Chern-Simons}

\protect\hspace{\parindent}%
Notre principal objectif est la r\'esolution du mod\`ele
de WZNW quantique~; c'est \`a ce moment qu'un lien avec
la th\'eorie de Chern-Simons (CS) appara\^{\i}t.
Il se trouve que les solutions des identit\'es
de Ward chirales pour WZNW sont les \'etats quantiques
de la th\'eorie de CS. Ces derniers s'interpr\`etent
naturellement comme les sections holomorphes d'un fibr\'e
vectoriel complexe, au-dessus de l'espace des modules $\CN=\CA^{01}/\CG^\C$,
muni d'un produit scalaire \`a la Bargmann. 
La connaissance des \'etats de CS et du produit scalaire
permet alors d'exprimer les fonctions de Green du
mod\`ele de WZNW comme combinaisons sesquilin\'eaires
de facteurs holomorphes, ce que nous appelons factorisation 
holomorphe.

\medskip
\section{Espace des modules $\CN=\CA^{01}/\CG^\C$}

Le langage g\'eom\'etrique naturel des th\'eories de jauge utilise les
notions de connexion et de fibr\'es vectoriels. Au final, on travaille
dans la cat\'egorie holomorphe. On souhaite alors
r\'einterpr\'eter l'espace quotient $\CN$ des connexions modulo
les transformations de jauge comme l'espace des modules
des fibr\'es vectoriels holomorphes. Une telle
reformulation ouvre la porte vers la puissante machinerie
de la g\'eom\'etrie. On d\'emarre par des rappels 
de g\'eom\'etrie diff\'erentielle. 
La litt\'erature sur le sujet est particuli\`erement
riche~; ma petite s\'election personnelle se composerait de
\cite{kobayashi,narasimhan:real,wells}. Du point de vue de
la g\'eom\'etrie diff\'erentielle, l'article de r\'ef\'erence
sur les espaces des modules est celui d'Atiyah-Bott~\cite{atiyahbott}.
Les livres~\cite{donaldson} et surtout~\cite{koba} sont des
expos\'es remarquables sur le sujet. Les espaces des modules
ont aussi attir\'e l'attention des g\'eom\`etres alg\'ebristes,
surtout Mumford et l'\'ecole indienne du Tata Institute. 
On pourra consulter \`a ce propos les r\'ef\'erences~\cite{lepotier,seshadri}.
Bien entendu, on peut se reporter aux articles originaux et
surtout \`a ceux \'ecrits dans les ann\'ees soixantes, 
qui utilisent un langage plus accessible au commun des mortels. 

\subsection{G\'eom\'etrie diff\'erentielle}

Soit $E\rightarrow X$ un fibr\'e vectoriel complexe $\CC^\infty$ de rang $r$ 
au-dessus d'une vari\'et\'e diff\'erentielle r\'eelle  $X$. 
On note $\Gamma(E)$ l'ensemble des 
sections $\CC^\infty$ de $E$ au-dessus de $X$,
$\CE^p(E)=\Gamma(\Lambda^pT^*X\otimes E)$ l'espace des $p$-formes 
sur $X$ \`a valeurs dans $E$ et
$\CC^\infty(X)$ (resp. $\CE^p(X)$, ...) l'ensemble des fonction 
$\CC^\infty$ (resp. $p$-formes, ...) sur $X$.
Le {\bol groupe de jauge} $\CG$ de $E$ est le groupe des automorphismes
de $E$. Localement, une transformation de jauge est une application
$\CC^\infty$ de $X$ dans ${\rm GL}_r\C$. Pour un $G$-fibr\'e
principal, une transformation de jauge est localement
une application $\CC^\infty$ de $X$ dans $G$.

Une {\bol connexion} $\nabla$ sur $E\rightarrow X$ est un op\'erateur
$\C$-lin\'eaire $\nabla:\Gamma(E)\rightarrow \CE^1(E)$ tel
que $\nabla(f\xi)=(df)\,\xi+f\,\nabla\xi$,
pour tout $f\in\CC^\infty(X)$ et tout $\xi\in\Gamma(E)$.
Toute connexion s'\'etend en une diff\'erentielle ext\'erieure, 
toujours not\'ee $\nabla$, agissant sur les formes  de degr\'e quelconque sur 
$X$ \`a valeurs dans $E$~: $\nabla:\CE^*(E)
\rightarrow\CE^*(E)$ est l'unique op\'erateur co\"\i ncidant avec 
$\nabla$ sur $\Gamma(E)$ et v\'erifiant l'identit\'e de Leibniz
\[
\nabla(\omega\wedge\xi)=d\omega\wedge\xi+
(-1)^{\deg \omega}\omega\wedge\nabla\xi,
\]
si $\omega\in\CE^*(X)$ et $\xi\in\CE^*(E)$. 
\'Etant donn\'e un ouvert $U$ de $X$, il existe une $1$-forme $A$, appel\'ee
{\bol forme de connexion}, sur $U$
\`a valeurs dans $i\,\mathfrak{gl}(r,\C)$ telle que 
$\nabla\xi=(d+A\,\wedge)\xi$.
En effet, soit un rep\`ere de $E$ au-dessus de $U$, \cad une base
 $(e_{1},\cdots,e_{r})$
tel que tout \'el\'ement $\xi$ de $\CE^p(U,E)$ se d\'ecompose en 
$\xi=\xi^\mu\,e_\mu$ o\`u $\xi^\mu\in\CE^p(U)$. Soit $A$ la matrice 
$r\times r$  de $1$-formes telle que $\nabla e_{\mu}=A^\mu_\nu\,e_\mu$. 
Pour $\xi\in\CE^p(U,E)$, un petit calcul matriciel donne 
$\nabla \xi=d\xi+A\wedge\xi$.  Soient $U$ et $V$ deux ouverts de $X$. 
Dans un changement de coordonn\'ees local $g:U\cap V\rightarrow {\rm GL}_r\C$, 
\ie $e'_\mu=e_\nu\,g^\nu_\mu$ sur $U\cap V$, la forme $A$ se transforme 
comme $A\mapsto g^{-1}A\,g+g^{-1}dg$. 
C'est une d\'efinition alternative 
de la notion de connexion qui se transpose bien aux fibr\'es principaux. 
Pour un $G$-fibr\'e principal $P$, une connexion est une famille de 
$1$-formes $A$ sur $X$ \`a valeurs dans $i\lieg$ telle que 
$A\mapsto g^{-1}A\,g+g^{-1}dg$ dans un changement de coordonn\'ees locales
$g:U\cap V\rightarrow G$.

La {\bol courbure} de la connexion $\nabla$ est l'op\'erateur
$\nabla^2:\CE^0(E)\rightarrow\CE^2(E)$. La courbure mesure donc 
l'exactitude de la suite de de Rham $\CE^0(E)\rightarrow\CE^1(E)
\rightarrow\CE^2(E)\rightarrow\cdots$. 
La courbure est $\CC^\infty(X)$-lin\'eaire, donc on peut
voir $\nabla^2$ comme un \'el\'ement de $\CE^2(\text{End}\,E)$, soit
une $2$-forme \`a valeurs dans $\text{End}\,E$.
Localement, $\nabla^2\xi=F\wedge\xi$ 
o\`u $F=dA+A\wedge A$ est la {\bol forme de courbure} associ\'ee \`a la 
connexion $\nabla$. 
Pour un $G$-fibr\'e principal $P$, il
est pr\'ef\'erable d'\'ecrire $F=dA+\frac{1}{2}\,[A,A]$.

Une structure plate sur $E$ est la donn\'ee d'un recouvrement ouvert $\CU$ 
de $X$ pour lequel toutes les fonctions de transition sont constantes.
Les trois conditions suivantes sont \'equivalentes~: (1) $E$ est muni d'une 
structure plate~; (2) $E$ admet une connexion plate, \cad $F=0$~;
(3) $E$ est d\'efini par une repr\'esentation $\rho$ du 
groupe fondamental $\pi_1$ de $X$ dans ${\rm GL}_r\C$ (donn\'ee
par l'holonomie de la connexion plate). 
Cette derni\`ere condition dit que $E$ est le
fibr\'e $E_\rho=\wi X\times_\rho \C^r$ associ\'e au $\pi_1$-fibr\'e principal $\wi X\rightarrow X$
par la repr\'esentation $\rho$~: $E_\rho$ est le quotient de $\wi X\times \C^r$
par la relation d'\'equivalence $(x,v)\simeq(px,\rho(p)v)$,
si $p\in\pi_1$ --- une section de $E_\rho$ est une fonction $s$ $\CC^\infty$ telle que
$s(px)=\rho(p)s(x)$.
On dit alors que $E$ est un {\bol fibr\'e plat}. 
Le r\'esultat est \'egalement vrai pour un $G$-fibr\'e principal $P$
en rempla\c{c}ant l'assertion (3) par~: $P$ est 
d\'efini par une repr\'esentation $\rho:\pi_1\rightarrow G$, \ie $P_\rho=
\wi X\times_\rho G$.

Soit $P$ le ${\rm GL}_r\C$-fibr\'e principal associ\'e au fibr\'e vectoriel $E$. 
Soit $\widehat{P}=P\,/\,\C^*I_r$, o\`u $I_r$ est la matrice identit\'e~;
$\widehat{P}$ est un ${\rm PGL}_r\C$-fibr\'e principal, o\`u 
${\rm PGL}_r\C={\rm GL}_r\C\,/\,\C^*I_r$. Une structure projectivement 
plate sur $E$ 
est la donn\'ee d'une structure plate sur $\widehat{P}$. Une connexion est 
projectivement plate sur $E$ si la connexion induite sur $\widehat{P}$
est plate. Il y a \'equivalence entre~:
(1) $E$ est muni d'une structure projectivement plate~;
(2) $E$ admet une connexion projectivement plate.
On dit alors que $E$ est un {\bol fibr\'e projectivement plat}.
Une connexion est projectivement plate si, et seulement si, il existe
une forme  $\lambda\in\CE^2(X)$ telle que $F=\lambda I$, o\`u $I$ est 
l'identit\'e de $\text{End}\,E$.

Si $F$ est la forme de courbure d'une connexion $\nabla$ sur un fibr\'e
vectoriel complexe $E$ au-dessus d'une vari\'et\'e complexe $X$, on pose
\[
c(E,\nabla)\equiv {\rm det}\left(I_r+\quotient{i}{2\pi}\,F\right)
=1+c_1(E,\nabla)+\cdots+c_r(E,\nabla)
\]
o\`u $c_k(E,\nabla)\equiv f_k(F)$, $f_k$ \'etant le polyn\^ome sur
$\mathfrak{gl}(r,\C)$ de degr\'e $k$, ${\rm GL}_r\C$-invariant et tel
que ${\rm det}\,(t\,I_r+\frac{i}{2\pi}\,X)=t^r+t^{r-1}\,f_1(X)
+\cdots+t^0\,f_r(X)$.
Chaque $c_k(E,\nabla)$ est ind\'ependante du choix du rep\`ere o\`u on
a repr\'esent\'e $\nabla^2$~; c'est m\^eme une $2k$-forme ferm\'ee.
La {\bol $k$-i\`eme classe de Chern} $c_k(E)$ est la classe de de Rham
de $c_k(E,\nabla)$ dans $H_{{\rm dR}}^{2k}(X,\C)$ et la
forme de Chern est la classe de de Rham de $c(E,\nabla)$
dans $H_{{\rm dR}}^*(X,\C)$. En fait, $c_k(E)$ est une classe de de Rham
r\'eelle et m\^eme enti\`ere qui ne d\'epend pas de la connexion
$\nabla$. Seule la
premi\`ere classe de Chern $c_1(E)=\left[\frac{i}{2\pi}\,{\rm tr}\,F\right]$
nous int\'eresse vraiment. Entre autres propri\'et\'es, on a~:
$c_1(E)$ ne d\'epend que des classes d'isomorphismes de $E$,
la premi\`ere classe de Chern d'un fibr\'e trivial est nulle,
$c_1(E^*)=-c_1(E)$ et $c_1(\Lambda^r E)=c_1(E)$. Si $X$
est de dimension 2 et compact, le {\bol degr\'e} de $E$ est
\[
\deg E\equiv \int_X c_1(E).
\]
Comme $c_1(E)$ est une classe enti\`ere, oublier la diff\'erence
(conceptuelle) entre ${\rm deg}\,E$ et $c_1(E)$ est pleinement justifi\'e.
Notons que si $E$ est un fibr\'e de d\'eterminant $\det E\equiv
\Lambda^r E$ trivial alors $\deg E=0$.

Supposons que $X$ est une vari\'et\'e  complexe. 
On note $\CE^{p,q}(E)$ l'espace des $(p,q)$-formes \`a valeurs dans $E$.
Soit $E$ un fibr\'e vectoriel 
holomorphe au-dessus de $X$ \cad un fibr\'e
 topologiquement \'equivalent \`a un 
fibr\'e vectoriel complexe $E$ tel que la projection $\pi:E\rightarrow X$ 
est holomorphe ou encore pouvant \^{e}tre d\'efini par des fonctions
de transitions holomorphes. Une diff\'erence avec les fibr\'es vectoriels 
\<<ordinaires\>> est l'existence d'un op\'erateur privil\'egi\'e 
$\conbar:\CE^{p,q}(E)\rightarrow\CE^{p,q+1}(E)$ qui
dans un rep\`ere holomorphe (local) n'est autre que l'op\'erateur 
de Cauchy-Riemann $\de$. Une {\bol structure holomorphe} sur un 
fibr\'e vectoriel complexe sera dor\'enavant la donn\'ee 
d'un op\'erateur $\conbar$. Deux structures
holomorphes $\conbar$ et $\conbar{}'$ sont dites {\bol \'equivalentes} s'il 
existe une transformation de jauge $h$ telle que $h^{-1}\conbar h=
\conbar{}'$. Il est facile de voir que cette d\'efinition \'equivaut
\`a dire que les fibr\'es vectoriels holomorphes sous-jacents \`a
$\conbar$ et $\conbar{}'$ sont isomorphes. 

La parent\'e \'evidente entre un op\'erateur $\conbar$
et une connexion n'est pas innocente. Si $X$ est une vari\'et\'e complexe,
une connexion $\nabla$ sur un fibr\'e vectoriel complexe $E$ se d\'ecompose
naturellement en $\nabla=\nabla^{10}+\nabla^{01}$ o\`u
$\nabla^{10}:\CE^{p,q}(E)\rightarrow\CE^{p+1,q}(E)$ et 
$\nabla^{01}:\CE^{p,q}(E)\rightarrow\CE^{p,q+1}(E)$. De m\^eme,
on a $d=\da+\de$. La courbure $\nabla^2$ se scinde en trois parties~:
\begin{gather*}
\nabla^{10}\circ\nabla^{10}\in\CE^{2,0}({\rm End}\,E),\qquad 
\nabla^{10}\circ\nabla^{01}+\nabla^{01}\circ\nabla^{10}
\in \CE^{1,1}({\rm End}\,E)\\
{\rm et}\quad \nabla^{01}\circ\nabla^{01}\in\CE^{0,2}({\rm End}\,E).
\end{gather*}
Suivant le m\^eme sch\'ema, $A=A^{10}+A^{01}$ et $F=F^{20}+F^{11}+F^{02}$.
On d\'esire un crit\`ere permettant de d\'eterminer les connexions provenant
d'une structure holomorphe sur $E$, \ie pour lesquelles
$\nabla^{01}=\conbar$~; on dit qu'une telle connexion est 
{\bol int\'egrable}. Par une variante du th\'eor\`eme 
de Newlander-Nirenberg, on montre que

\begin{propos}{~\cite{atiyahbott,atiyahhit}}%
Une connexion $\nabla$ sur un fibr\'e vectoriel complexe
au-dessus de $X$ est int\'egrable si, et seulement si, $\nabla^{01}\circ
\nabla^{01}=0$. 
\end{propos}

\noindent Comme cons\'equence imm\'ediate, dans un rep\`ere holomorphe~: 
$A^{01}=0$ et $F^{02}=0$. Si $X$ est une surface de Riemann $\Sigma$, la 
condition d'int\'egrabilit\'e est toujours v\'erifi\'ee, ainsi toute 
connexion est int\'egrable. La proposition pr\'ec\'edente permet
de r\'einterpr\'eter l'espace quotient $\CA^{01}/\CG^\C$ des champs
de jauge chiraux modulo les transformations de jauge en termes plus
g\'eom\'etriques.

\subsection{Espace des modules}

Soit $E$ un fibr\'e vectoriel complexe $\CC^\infty$ au-dessus d'une surface de
Riemann $\Sigma$, de rang $r$ et de degr\'e $d$. 
Une structure complexe sur $E$ est la donn\'ee d'un op\'erateur 
$\conbar$. Le groupe des transformations de jauge 
(des automorphismes de $E$) $\CG^\C$
agit sur l'espace des structures complexes sur $E$. L'espace 
quotient $\CN$ est l'ensemble des classes d'isomorphisme 
de fibr\'es vectoriels holomorphes au-dessus de $\Sigma$ 
de rang $r$ et de degr\'e $d$. En effet, deux fibr\'es vectoriels 
holomorphes de rang $r$ et de degr\'e $d$ sur $X$ sont topologiquement 
\'equivalents et ils sont isomorphes si, et seulement 
si, leurs structures complexes respectives sont reli\'ees par 
une transformation de jauge. L'espace quotient
est commun\'ement appel\'e l'{\bol espace des modules 
de fibr\'es vectoriels holomorphes de rang $r$ et de degr\'e $d$}. 
H\'elas, pour obtenir un espace des modules sympathique on ne doit 
consid\'erer que les fibr\'es stables.
Un fibr\'e  vectoriel holomorphe $E$ 
au-dessus de $\Sigma$ est dit {\bol stable} (resp. {\bol semi-stable}) 
si, pour
tout sous-fibr\'e holomorphe propre $F$ de $E$, on a $\mu(F)<\mu(E)$
(resp. $\mu(F)\leq\mu(E)$), o\`u la 
{\bol pente} $\mu(E)$ de $E$ est la quantit\'e $\text{deg}\,E
\,/\,\text{rg}\,E$.
Si $\deg E$ et $\text{rg}\,E$ sont
premiers entre eux, il n'y a pas de diff\'erence entre stabilit\'e
et semi-stabilit\'e.

Sans la condition de stabilit\'e, l'espace quotient $\CN$ n'est
m\^eme pas Hausdorff. Mumford~\cite{mum:mod} a pu montrer qu'il existe une 
vari\'et\'e lisse $\CN_s$ de dimension $\dim \CN_s=r^2(g-1)+1$,
si $\CN_s$ n'est pas vide et $g\geq 2$, param\'etrisant 
l'ensemble des classes d'isomorphisme de fibr\'es stables de rang $r$ et 
de degr\'e $d$.
Cette vari\'et\'e admet une compactification naturelle~\cite{sesh}
not\'ee $\CN_{ss}$, 
isomorphe au quotient de l'ensemble des classes d'isomorphismes de fibr\'es 
semi-stables par la relation d'\'equivalence de Seshadri. 
Si $E$ est un fibr\'e vectoriel holomorphe semi-stable de pente $\mu$,
$E$ admet une filtration de Jordan-H\"older~\cite{sesh}
\[
\{0\}=E_{0}\subset E_1\subset\cdots\subset E_{p-1}\subset E_p=E
\]
telle que $\text{gr}_i\equiv E_i/E_{i-1}$, $i=1,\cdots,p$, soit stable et
$\mu(\text{gr}_i)=\mu$. Bien entendu, $p=1$ si $E$ est stable.
La graduation associ\'ee \`a $E$, \cad le
fibr\'e $\text{Gr}\,E\equiv\bigoplus_{i=1}^{p}\text{gr}_i$, ne d\'epend que de
la classe d'isomorphisme de $E$. On dit que deux fibr\'es semi-stables
$E$ et $E'$ sont {\bol S-\'equivalents} (ou {\bol Seshadri \'equivalents})
si $\text{Gr}\,E\cong\text{Gr}\,E'$. En particulier, si $E$ et $E'$
sont stables, ils sont S-\'equivalents si, et seulement si, ils sont
isomorphes.
Si $r$ et $d$ sont 
premiers entre eux, $\CN_s$ est d\'ej\`a compacte.
L'espace $\CN_s$ est un ouvert de $\CN_{ss}$. Les points correspondant
\`a des fibr\'es stables sont des points non-singuliers de $\CN_{ss}$.
L'espace tangent \`a $\CN_s$ est isomorphe \`a $H^1(\text{End}\,E)$.

En genre z\'ero, l'espace $\CN_s$ est vide d\`es que $r>1$ et
$\CN_{ss}$ est un point si $r$ divise $d$, vide sinon~\cite{grothendieck}
(cf. section {\bf 5} pour plus de d\'etails). 
En genre un, si $h$ est le plus grand diviseur commun de $r$ et $d$,
$\CN_{ss}$ est isomorphe au produit sym\'etrique 
$\mathfrak{S}^h\Sigma$~\cite{atiyah:vb}.
De plus, si $h=1$, $\CN_s=\CN_{ss}=\Sigma$ et, si $h>1$, $\CN_s=\emptyset$.

On suppose maintenant que $g,r\geq 2$. L'espace des modules
$\CN_s$ n'est pas vide et, except\'e dans les cas 
$r=2=g$ et $d$ pair, l'ensemble des points
singuliers de $\CN_{ss}$ est l'ensemble des points semi-stables
non-stables $\CN_{ss}\setminus\CN_s$~\cite{nar-ram}.
\`A ma connaissance, on n'a une description explicite de
$\CN_{ss}$ que dans tr\`es peu de cas.
En fait, les r\'esultats concernent
 plut\^ot l'espace des modules des fibr\'es vectoriels
holomorphes de rang $r$ et de d\'eterminant fix\'e $\CalD$ (de degr\'e $d$).
Les notations standards sont $\mathcal{U}(r,d)$ --- notre $\CN_{ss}$ 
plus haut ---
et $\mathcal{S}\mathcal{U}(r,\CalD)$. L'espace $\mathcal{S}
\mathcal{U}(r,\CalD)$ est 
la fibre de l'application ${\rm det}\,:\,\mathcal{U}(r,d)
\rightarrow J^d$, o\`u $J^d$ est la Jacobienne de $\Sigma$ param\'etrisant
les fibr\'es holomorphes en droites de degr\'e $d$, et ${\rm det}$ est
l'application qui associe \`a la classe d'isomorphisme de $E$
celle de ${\rm det}\,E$. Si $d=0$ --- d\'eterminant topologiquement
trivial ---
$\mathcal{U}(r)$ est essentiellement 
le produit $\mathcal{S}\mathcal{U}(r)\times J^0$, donc
l'\'etude de $\mathcal{U}(r)$ se ram\`ene compl\`etement \`a
celle de $\mathcal{S}\mathcal{U}(r)$. Pour simplifier, on
note indiff\'eremment tous les types
d'espaces des modules. On conna\^\i t $\CN_{ss}$
pour les configurations suivantes~: $r=2=g$ et d\'eterminant trivial~\cite{nar-ram},
surfaces hyperelliptiques, $r=2$~\cite{desale}
et surfaces non-hyperelliptiques de genre $3$, $r=2$, 
d\'eterminant trivial~\cite{nar-ram:bis}. 
D\'etaillons un petit peu le cas $r=2=g$ et d\'eterminant 
trivial puisque c'est celui qui nous int\'eressera le plus~:
$\CN_{ss}\setminus\CN_s$ est isomorphe \`a une surface
de Kummer --- ce qui est \'egalement vrai si $g\geq 3$ ---
et $\CN_{ss}$ est 
une vari\'et\'e lisse isomorphe \`a l'espace projectif 
des fonctions th\^eta de degr\'e $2$~\cite{nar-ram}, soit
l'espace projectif tridimensionnel $\mathbb{P}^3$.

Passons maintenant au cadre plus g\'en\'eral des fibr\'es 
principaux~\cite{kumar,rama}. On suppose que $G$ est groupe compact 
simple, connexe et simplement connexe. Soit $P$ un $G$-fibr\'e
principal au-dessus d'une surface de Riemann $\Sigma$. 
Sans perdre la g\'en\'eralit\'e, on peut prendre $P=\Sigma\times G$.
L'espace de connexions sur $P$ peut \^{e}tre alors identifi\'e
avec l'espace $\CA$ des $1$-formes sur $\Sigma$ \`a valeurs dans $i\lieg$.
On a d\'ej\`a d\'efini l'action du groupe de jauge sur $\CA$~:
${}^gA\equiv gAg^{-1}+g dg^{-1}$. Le groupe de jauge est 
un groupe de Lie de dimension infinie. La structure complexe sur $\Sigma$
induit une structure complexe sur $\CA$. 
\'Etant donn\'ee une structure complexe
$\J$ sur $\Sigma$, on d\'ecompose un champ $A$ (complexifi\'e)
en $A=A^{10}+A^{01}$, avec (cf. p.~\pageref{decompo})
\[
A^{10}=A\,\quotient{1+i\J}{2},\qquad
A^{01}=A\,\quotient{1-i\J}{2}.
\]
La condition d'unitarit\'e~: $A^{10}=-(A^{01})^\dagger$ permet 
de retrouver le champ original $A$. On peut donc identifier $\CA$ et
l'espace complexe $\CA^{01}$ des $(0,1)$-formes \`a valeurs dans $\liegc$.
On prolonge l'action de $\CG$ sur $\CA$ 
en une action du groupe de jauge complexifi\'e, not\'e
$\CG^\C$ ---  c'est l'ensemble des applications $\CC^\infty$ 
de $\Sigma$ dans $\Gc$. On d\'efinit : 
${}^gA^{01}=gA^{01}g^{-1}
+g\bar\partial g^{-1}$ (et ${}^gA^{10}=(g^\dagger)^{-1}A^{10}
g^\dagger+(g^\dagger)^{-1}\partial g^\dagger$).
On retrouve
l'action du groupe compact en prenant $g\,g^\dagger=1$.

Comme pour les fibr\'es vectoriels, toute connexion sur $P$ d\'etermine 
une structure holomorphe sur le $\Gc$-fibr\'e $P\dC$, 
la complexification de $P$. R\'eciproquement, tout $\Gc$-fibr\'e holomorphe,
muni d'une r\'eduction du groupe de structure \`a $G$, d\'etermine une 
$G$-connexion canonique. De plus, deux structures holomorphes sur $P\dC$
produisent des fibr\'es isomorphes si les conne\-xions sous-jacentes 
sont dans la m\^eme orbite de $\CG^\C$. Ainsi, $\CA^{01}/\CG^\C$ est 
l'ensemble des classes d'\'equivalence de $G^\C$-fibr\'es 
holomorphes. En effet, suivant~\cite{atiyahbott}, 
tout champ de jauge $A^{01}$ s'\'ecrit localement $g_\alpha^{-1}\de g_\alpha$, 
o\`u $g_\alpha:U_\alpha\rightarrow\Gc$. La quantit\'e
$g_{\alpha\beta}=g_\alpha g_\beta^{-1}$ donne un cocycle
d'un $\Gc$-fibr\'e holomorphe dont la classe modulo $(g_{\alpha\beta})
\mapsto (h_\alpha g_{\alpha\beta}h_\beta^{-1})$ d\'etermine, 
\`a isomorphisme pr\`es, le fibr\'e. D'ailleurs,
deux $A^{01}$ donnent des cocycles \'equivalents si, et seulement si,
ils sont dans la m\^eme orbite de $\CG^\C$.

Comme pr\'ec\'edemment, on peut d\'efinir des notions de stabilit\'e et
de semi-stabilit\'e. On construit alors une vari\'et\'e lisse $\CN_s$
param\'etrisant les $\Gc$-fibr\'es stables. Cette derni\`ere admet une
compactification naturelle~: l'espace des modules des classes d'\'equivalence
de Seshadri de $\Gc$-fibr\'es holomorphes. La dimension (complexe) de 
$\CN_s$ est~: 
\begin{equation}
\dim\CN_s=\begin{cases}
	0, & \text{si $g=0$}\\
	\text{rg}\,G, & \text{si $g=1$}\\
	\dim G\,(g-1), & \text {si $g>1$}
	\end{cases}.
\end{equation}
Si $G={\rm SU}_r\C$ \cad $\Gc={\rm SL}_r\C$, on obtient l'espace des 
modules $\mathcal{S}
\mathcal{U}(r)$ des fibr\'es de rang $r$ de d\'eterminant trivial. 
Le cas $r=2$ a \'et\'e \'etudi\'e par Narasimhan et 
Ramanan~\cite{nar-ram}. On utilisera ce travail au chapitre 5. 

Un r\'esultat important du \`a Narasimhan et Seshadri 
dit qu'un fibr\'e vectoriel 
de degr\'e $0$ est stable si, et seulement si, il est d\'efini par
une repr\'esentation irr\'eductible unitaire du groupe fondamental $\pi_1$
de $\Sigma$, \ie par un $\rho:\pi_1\rightarrow {\rm U}_r$. 
En particulier, un fibr\'e stable de degr\'e z\'ero
est n\'ecessairement plat.
Plus tard, la d\'emonstration de ce th\'eor\`eme fut
consid\'erablement simplifi\'ee par Donaldson~\cite{don:mod}.
La g\'en\'eralisation aux fibr\'es principaux est 
due \`a Ramanathan~\cite{rama}~:
un $\Gc$-fibr\'e $P$ est stable si, et seulement si, il provient 
d'une repr\'esentation irr\'eductible unitaire de $\pi_1$,
\cad d'une repr\'esentation irr\'eductible 
$\rho:\pi_1\rightarrow \Gc$ avec ${\rm Im}\,\rho\subset G$
--- en particulier $P$ est plat.

Pour terminer, on donne quelques d\'etails de la g\'eographie
de l'espace des modules~\cite{atiyahbott}.
Tout fibr\'e vectoriel holomorphe poss\`ede 
une filtration de Harder-Narasimham~\cite{hardernara}
\[
\{0\}=E_0\subset E_1\subset\cdots \subset E_{s-1}\subset E_s=E
\]
telle que $D_i\equiv E_i\,/\,E_{i-1}$, $i=1,\cdots,s$,
soit semi-stable et $\mu_1>\mu_2>\cdots>\mu_{s}$, o\`u $\mu_i=
\mu(D_i)$. Cette filtration est canonique, \ie unique.
Si $E$ est semi-stable, $s=1$. Notons $r_i$ (resp. $d_i$) le rang
(resp. le degr\'e) de $D_i$~; $\sum r_i=r$ et 
$\sum d_i=d$. La s\'equence $\lambda$ de paires $(r_i,d_i)$, $i=1,\cdots,s$, 
d\'efinit le {\bol type} de stabilit\'e de $E$. 

On d\'ecompose alors $\CA^{01}$ en sous-vari\'et\'es $\CA^{01}_\lambda$, 
chacune constitu\'ee de tous les fibr\'es de type $\lambda$. Comme la filtration de 
Harder-Narasimhan est canonique, les $\CA^{01}_\lambda$ sont pr\'eserv\'ees 
par l'action de $\CG^\C$ et par l\`a-m\^eme se d\'ecomposent en une union 
d'orbites. Si $\CT$ d\'esigne l'ensemble des types possibles,
\[
\CA^{01}=\bigcup_{\lambda\in\CT}\CA^{01}_\lambda.
\]
D\'efinissons un ordre partiel sur $\CT$~: 
$\lambda\preceq\mu$ si $\CP(\lambda)\subset\CP(\mu)$
o\`u $\CP(\lambda)$ est la r\'egion dans l'espace 
bidimensionnel des couples $(r,d)$, d\'elimit\'ee
par l'axe horizontal et le polyg\^one dont les sommets 
sont les points $(r_i,d_i)$ et les 
c\^ot\'es sont les segments joignant deux c\^ot\'es 
pris dans l'ordre croissant en 
$r$ (cf. figure~\ref{polygone}). 
On obtient une stratification de $\CA^{01}$~: pour tout $I\subset\CT$,
\[
\CS_{I}=\bigcup_{\mu\in I}\bigcup_{\lambda\preceq\mu}\CA^{01}_\lambda
\]
est une sous-vari\'et\'e ouverte de $\CA^{01}$. 
La codimension de $\CA^{01}_\lambda$ vaut
\[
c_\lambda=\sum_{0\leq j<i\leq r}
	\left[(r_i\,d_j-r_j\,d_i)+r_i\,r_j\,(g-1)\right]
\]
o\`u $g$ d\'esigne le genre de la surface $\Sigma$. 
La strate $\CA^{01}_{ss}$ des fibr\'es
semi-stables correspond au type minimum $\lambda_{ss}=\{(r,0)\}$ 
de $\CT$. Cette strate est l'unique strate de
codimension z\'ero et elle forme une {\bol orbite dense} dans $\CA^{01}$.
En pratique, on a besoin d'un peu plus d'information sur
cette stratification. On souhaite \'egalement conna\^\i tre les strates
de codimension $1$. On verra cela en d\'etail pour le genre z\'ero et
on esquissera le r\'esultat en genre un. On peut aussi d\'ecrire la g\'eographie
de l'espace des modules pour des fibr\'es principaux. Cela revient
essentiellement \`a se ramener au cas pr\'ec\'edent en consid\'erant
le fibr\'e $\ad\,P\dC$ associ\'e \`a $P\dC$ par la repr\'esentation adjointe
de $G$~\cite[section 10]{atiyahbott}. 
L'orbite dense dans $\CA^{01}$ est toujours la strate semi-stable.

\begin{figure}[h]
\setlength{\unitlength}{1mm}
\begin{picture}(100,70)(-30,0)
\thinlines
\put(0,0){\vector(0,1){50}}
\put(0,0){\vector(1,0){100}}
\put(-4,-4){$0$}
\put(0,0){\line(1,2){15}}
\put(15,30){\line(2,1){15}}
\put(15,30){\circle*{1}}
\put(30,37.5){\line(6,-1){6}}
\put(30,37.5){\circle*{1}}
\multiput(36.6,36.4)(0.6,-0.1){6}{\circle*{0.001}}
\put(-15,45){degr\'e}
\put(95,-5){rang}
\put(3,33){{\small $(r_1,d_1)$}}
\put(25,40){{\small $(r_2,d_2)$}}
\put(40,15){$\CP(\lambda)$}
\multiput(65,25)(0.4,-0.3){6}{\circle*{0.001}}
\put(67.4,23.2){\line(4,-3){6}}
\put(73.4,18.7){\circle*{1}}
\put(73.4,1){\line(0,-1){1}}
\put(73.4,-3){$r$}
\put(72,23){{\small $(r,d)$}}
\end{picture}
\caption{R\'egion convexe $\CP(\lambda)$.}
\label{polygone}
\end{figure}

\newpage

\medskip
\section{\'Etats de Chern-Simons}

Soit $M$ une vari\'et\'e diff\'erentielle de dimension $3$,
l'{\bol action de Chern-Simons} est
\[
S_{{\rm CS}}(B)={_i\over^{4\pi}}\int_M\tr\,(B\wedge dB+
	{_2\over^3}\,B\wedge B\wedge B)
\]
o\`u $B$ est une $1$-forme sur $M$ \`a valeurs dans $i\lieg$.
Les solutions classiques de la th\'eorie de CS sont les connexions
plates. Un objet important est l'int\'egrale fonctionnelle
\[
\int\prod_\ell \text{W}(\CC_\ell)\sous{R_\ell }\,\ee^{-k\,S_{{\rm CS}}(B)}DB
\]
o\`u $\CC_\ell$ est un n\oe ud dans $M$, soit un plongement de $S^1$ dans $M$, et
$\text{W}(\CC)\sous{R}$ est une {\bol boucle de Wilson} donn\'ee par
la trace dans la repr\'esentation $R$ de l'holonomie de la connexion $A$
le long du lacet $\CC$, \cad
\[
\text{W}(\CC)\sous{R}=\tr\sous{R}\,\text{P}\ee^{\,\int_\CC A}.
\]
L'\'etude de l'int\'egrale fonctionnelle pr\'ec\'edente a permis \`a
Witten~\cite{witten:jones} de construire (formellement) des invariants de
n\oe uds pour n'importe quelle vari\'et\'e compacte
tridimensionnelle. En particulier, cette construction produit le
{\bol polyn\^ome de Jones} en prenant $M=S^3$, $G=\Su$ et
$R_\ell $ la repr\'esentation de spin $\frac{1}{2}$.

Pour comprendre le lien entre le mod\`ele de WZNW et la th\'eorie de
CS, on utilise l'\'equation de Ward chirale locale~(2.\ref{ward2}.a)
\begin{equation*}
\begin{split}
\left(\partial_zA\sur{a}_{\overline z}+\partial_{\overline{z}}\,
	\Bigl({_{2\pi}\over^k}{_\delta\over^{\delta A\sur{b}_{\overline{z}}}}
	\Bigr)\right.
&-if^{abc}\,\Bigl({_{2\pi}\over^k}
	{_\delta\over^{\delta A\sur{b}_{\overline{z}}}}\Bigr)\,
	A\sur{c}_{\overline{z}}\\
&\left.-{_{2\pi}\over^k}\sum_\ell \delta\sur{(2)}(z-z_\ell )\,t^a_\ell \right)
	\widetilde{Z}\sous{A}\,\langle\grl\rangle\sous{A}=0.
\end{split}
\end{equation*}
On s'est content\'e de remplacer les courants par leur d\'efinition.
Introduisons deux op\'erateurs agissant sur les fonctions du champ de jauge $A$,
\[
(\BA\sur{a}_{z}\,f)(A)\equiv
	-{_{2\pi}\over^k}{_\delta\over^{\delta A\sur{a}_{\overline{z}}}}\,
	f(A),\qquad(\BA\sur{a}_{\overline{z}}\,f)(A)
	\equiv A\sur{a}_{\overline{z}}\,f(A).
\]
L'identit\'e de Ward locale prend la forme compacte suivante 
\begin{equation}
\label{platitudeW}
\left(\BF^a(z)-{_{2\pi}\over^k}\sum_\ell \delta\sur{(2)}(z-z_\ell )\,
	t^a_\ell \right)\,\widegamma(A)=0
\end{equation}
o\`u $\widegamma$ est la fonction de Green modifi\'ee introduite dans 
la section {\bf 4.1} et $\BF$ est la courbure de la \<<connexion quantique\>> 
$\BA$, d\'efinie par
\[
\BF(z)\equiv\partial_z\BA_{\overline{z}}-\partial_{\overline{z}}\BA_{z}
	+\left[\BA_{z},\BA_{\overline{z}}\right].
\]
C'est cette forme de l'identit\'e de Ward locale qui permet de relier 
les fonctions de Green du mod\`ele de WZNW aux \'etats quantiques de 
la th\'eorie de CS, dans l'image de Schr\"odinger.

On suppose maintenant que $M=\Sigma\times S^1$, o\`u $\Sigma$ est une 
surface de Riemann compacte.  La coordonn\'ee correspondante \`a $S^1$ 
joue le r\^ole du temps. Le n\oe ud $\CC_\ell $ sera le produit 
$\lbrace\xi_\ell \rbrace\times S^1$, colori\'e par une repr\'esentation 
irr\'eductible $R_\ell $ de $G$. On choisit la jauge $B_{0}=0$, 
\cad les connexions nulles dans la direction $S^1$. Les solutions 
classiques sont encore les connexions de courbure nulle. L'espace des phases 
est alors l'ensemble des connexions plates $A$ sur
le $G$-fibr\'e trivial au-dessus de $\Sigma$, muni de la structure symplectique 
h\'erit\'ee de la forme 
\[
\quotient{k}{2\pi}\int\tr\,\delta A\wedge\delta A.
\]
\'Etant donn\'ee une structure complexe 
$\J$ sur $\Sigma$, on peut donc identifier $\CA$,
l'espace des $G$-connexions, et 
l'espace $\CA^{01}$ des $(0,1)$-formes \`a valeurs dans $\liegc$.
L'introduction d'une structure complexe sur l'espace des phases nous engage 
\`a utiliser une {\bol quantification holomorphe \`a la Bargmann}
de la th\'eorie de CS. Les \'etats quantiques sont des 
fonctionnelles $\Psi$ holomorphes sur $\CA^{01}$ \`a valeurs dans $\C$. 
En plus, on impose la contrainte
\[
\BF\sur{a}(z)\,\Psi(A^{01})=0,
\]
\cad l'analogue quantique de la condition de platitude de la connexion $A$.
Si on rajoute des boucles de Wilson, les \'etats quantiques sont des 
fonctionnelles $\Psi:\CA^{01}\rightarrow\esprep$, holomorphes et satisfaisant 
la contrainte (quantique)
\begin{equation}
\label{platitudeCS}
\left(\BF^a(z)-{_{2\pi}\over^k}\sum_\ell \delta\sur{(2)}(z-z_\ell )\,
	t^a_\ell \right)\,\Psi(A^{01})=0.
\end{equation}
On note $\chsi$ l'espace des {\bol \'etats de Chern-Simons}, \ie
des \'etats quantiques de la th\'eorie de CS.

R\'einterpr\'etons la condition~\eqref{platitudeCS}. On munit $\CA^{01}$ de 
la topologie $\CC^\infty$. Les fonctions holomorphes sur $\CA^{01}$ 
sont des fonctions $\CC^\infty$ au sens de Fr\'echet de 
d\'eriv\'ees $\C$-lin\'eaires~\cite{hamilton}. Soit $\CG^\C$ 
le groupe des transformations
chirales $h:\Sigma\rightarrow \Gc$. On d\'efinit une action 
de $\CG^\C$ sur les applications holomorphes $\Psi$
\begin{equation}
\label{globalCS}
	(^h\Psi)(A^{01})=\ee^{-k\,S(h,A^{01})}\,\mathop{\otimes}\limits_\ell 
	h(\xi_\ell )\sous{R_\ell }\,\Psi(\hAR{h^{-1}}).
\end{equation}
La contrainte de platitude quantique~\eqref{platitudeCS} est la version 
locale de la condition d'invariance globale des \'etats de CS 
sous l'action de $\CG^\C$, \cad $^h\Psi=\Psi$, si $h\in\CG^\C$.
Elle est \'equivalente \`a la condition globale.

La relation~\eqref{globalCS} d\'ecrit le comportement d'un \'etat de 
CS le long des orbites de $\CG^\C$. En termes plus g\'eom\'etriques, 
un \'etat de CS est une section holomorphe du fibr\'e vectoriel 
complexe $\CV=(\CA^{01}\times\esprep)\,/\,\CG^\C$ 
au-dessus de l'espace quotient $\CN=\CA^{01}\,/\,\CG^\C$.
L'action de $\CG^\C$ sur $\CA^{01}\times\esprep$ est 
\[
(A^{01},\Bv)\sim({}^hA^{01},\ee^{k\,S(h^{-1},A^{01})}\hrl\,\Bv).
\]
C'est l\`a qu'entre en jeu l'espace des modules $\CN$. 
Apr\`es deux sections consacr\'ees \`a des rappels math\'ematiques,
on va regarder plus en d\'etail l'espace des \'etats de CS.
Comme les espaces des modules, les espaces $\chsi$ peuvent \^etre 
d\'efinis de plusieurs mani\`eres. La plupart du temps,
les constructions utilisent des objets math\'ematiques difficiles
\`a appr\'ehender pour l'humble physicien. L'int\'er\^et des math\'ematiciens
est s\^urement  apparu avec la formule de Verlinde~\cite{verlinde}
qui donne la dimension de $\chsi$. Les diff\'erentes constructions
devraient donner des espaces isomorphes m\^eme s'il 
n'y a pas de d\'emonstration compl\`ete sur le sujet~\cite{beauville2,
faltings,kumar,sorger,tsuchiya}.
Dans la section {\bf 5}, on d\'ecrira l'espace des \'etats de CS
en genre z\'ero~\cite{gaw91:zero} et dans la section {\bf 6} en genre 
un~\cite{gaw94:genus1,gaw97:unitarity}. On donnera
quelques r\'esultats en genre sup\'erieur~\cite{gaw95:higher1} dans la section {\bf 7}.

\medskip
\section{Surfaces de Riemann et courbes alg\'ebriques}

Pour \'ecrire cette section, j'ai utilis\'e la merveilleuse introduction
au sujet donn\'ee par Griffiths~\cite{griff}. Je me suis \'egalement
aid\'e de~\cite{griffiths,miranda,narasimhan}.

Soit $\Sigma$ une surface de Riemann connexe et $\widetilde{\Sigma}$ son
rev\^etement universel. D'apr\`es le th\'eor\`eme d'uniformisation 
de Riemann, on peut ranger $\Sigma$ dans l'une des trois 
cat\'egories suivantes~: elliptique (resp. parabolique,
resp. hyperbolique) si $\widetilde{\Sigma}\cong\C P^1$
(resp. $\C$, resp. $\poincare$), o\`u $\poincare$ est
le demi-plan de Poincar\'e $\poincare=\{y=y_1+iy_{2}
\in\C\ \big|\ y_{2}>0\}$. 
Au niveau \<<conforme\>>, on trouve une classification semblable~:
$\Sigma$ est elliptique (resp. parabolique, resp. hyperbolique) si, et
seulement si, elle peut \^etre munie d'une m\'etrique riemannienne $\gamma$
compatible avec la structure complexe et de courbure gaussienne
$K_{\gamma}$ \'egale \`a $1$ (resp. $0$, resp. $-1$).
Deux surfaces de Riemann sont dites isomorphes s'il existe une application
biholomorphe --- \ie holomorphe, inversible, d'inverse holomorphe~---
entre elles. Le groupe d'automorphisme de $\C P^1$ est ${\rm PSL}_2$.

Maintenant et jusqu'\`a la fin, on consid\`ere uniquement
des surfaces de Riemann $\Sigma$ compactes. 
On sait alors que $\Sigma$ est hom\'eomorphe \`a une boule 
\`a $g$ anses --- $g$ est appel\'e le {\bol genre} de la surface. Le genre est 
une quantit\'e purement topologique qui permet aussi de reproduire la 
classification pr\'ec\'edente. En genre z\'ero, toute surface est 
isomorphe \`a la sph\`ere de Riemann (cas elliptique).
En genre un, toute surface est isomorphe \` a une courbe elliptique 
$E_\tau=\C/(\Z+\tau\Z)$ (cas parabolique).
En genre $\geq 2$, toute surface est isomorphe \`a un $\poincare\,/\,
\Gamma$ o\`u $\Gamma$ est un sous-groupe discret de ${\rm PSL}_2\R$, 
op\'erant librement sur $\poincare$ et tel que l'espace 
quotient soit compact (cas hyperbolique).

Notons $\PP^n$ le plan projectif complexe de dimension $n$.
Une {\bol vari\'et\'e alg\'ebrique (projective)}
$C$ est un sous-ensemble de $\PP^n$ de la forme
\[
C=\{\xi\in\PP^n\ \big|\ F_1(\xi)=\cdots=F_m(\xi)=0\}
\]
o\`u les $F_i$ sont des polyn\^omes homog\`enes de degr\'e donn\'e
en $n+1$ variables. Si on remplace $\PP^n$ par $\C^n$ et si les $F_i$
sont des polyn\^omes \`a $n$ variables,  on obtient une {\bol vari\'et\'e
alg\'ebrique affine}. Toute vari\'et\'e alg\'ebrique $C$ d\'etermine une
unique vari\'et\'e alg\'ebrique affine $C_0=C\cap\C^n$ o\`u $\C^n$ est
plong\'e canoniquement dans $\PP^n$. Si $m=1$, $C$ est une hypersurface
alg\'ebrique de degr\'e le degr\'e de $F_1$. En plus, si $n=2$, $C$ est
une {\bol courbe alg\'ebrique plane}. Une hypersurface alg\'ebrique est {\bol
irr\'eductible} si $F_1$ est un polyn\^ome homog\`ene irr\'eductible.
Supposons donn\'ee une courbe alg\'ebrique plane irr\'eductible $C$.
Un {\bol point singulier} de $C$ est un
point o\`u toutes les d\'eriv\'ees partielles de $F$ sont nulles. Il est
remarquable que $C$ n'a qu'un nombre fini de points singuliers. Si $S$
d\'esigne l'ensemble des points singuliers et $C^*\equiv C\setminus S$, alors
on montre que $C$ et $C^*$ sont connexes et $C^*$ est une surface de Riemann
(non n\'ecessairement compacte). Une {\bol normalisation} (ou {\bol
d\'esingularisation}) de $C$ est une surface de Riemann compacte
$\widetilde{C}$ et une application holomorphe $\sigma:\widetilde{C}
\rightarrow\PP^2$ telles que~: ($i$)~$\sigma(\widetilde{C})=C$,
($ii$)~la pr\'eimage de $S$ par $\sigma$ est un
ensemble fini, ($iii$)~ $\sigma:\widetilde{C}\setminus\sigma^{-1}(S)
\rightarrow C\setminus S$ est injective. Le {\bol th\'eor\`eme de normalisation}
assure l'existence et l'unicit\'e d'une normalisation. L'unicit\'e
signifie~: si $(\widetilde{C},\sigma)$ et $(\widetilde{C}',\sigma')$ sont
deux normalisations de $C$ alors il existe un isomorphisme $\tau:\widetilde{C}
\rightarrow \widetilde{C}'$ tel que $\sigma=\sigma'\circ\tau$.

\subsection{Diviseurs}

Un diviseur $D$ sur une surface de Riemann $\Sigma$ est une application
$D:\Sigma\rightarrow\Z$ nulle sauf en un nombre fini de points. On note
\[
D=\sum_{P\in\Sigma}D(P)\,P.
\]
Muni de l'op\'eration d'addition \'evidente, l'ensemble des diviseurs
sur $\Sigma$ forme un groupe ab\'elien appel\'e groupe des
diviseurs de $\Sigma$ et not\'e ${\rm Div}\,\Sigma$. On dit qu'un diviseur
est positif (ou effectif) si $D(P)\geq 0$, pour tout 
$P\in\Sigma$ ;
on note simplement $D\geq 0$. On \'ecrit aussi $D\geq D'$, d\`es que
$D-D'\geq0$. L'application degr\'e est l'homomorphisme de groupes $\rm{deg}:
{\rm Div}\,\Sigma\rightarrow\Z$ qui envoie un diviseur $D$ sur 
${\rm deg}\,D=\sum_{P\in\Sigma}D(P)$. Le noyau de l'homomorphisme $\deg$,
not\'e ${\rm Div}^0\,\Sigma$, est le groupe des diviseurs
de degr\'e z\'ero. Plus g\'en\'eralement, ${\rm Div}^d\,\Sigma$ est
l'ensemble des diviseurs de degr\'e $d$.

Soit $\CM(\Sigma)$ l'ensemble des fonctions m\'eromorphes sur $\Sigma$. 
Soient $f\in\CM(\Sigma)$, non-identique\-ment nulle, et $P$ un point de $\Sigma$. 
Si $z$ est une coordonn\'ee locale 
dans un voisinage de $P$, centr\'ee en z\'ero, alors, dans ce 
voisinage~: $f=z^\nu\,h(z)$, o\`u $h$ est une fonction holomorphe, 
$h(0)\neq 0$ et $\nu\in\Z$. L'entier $\nu$ est appel\'e l'ordre de $f$
en $P$, not\'e $\nu_P(f)$. Si $f\equiv 0$, il est commode de poser
$\nu_P(f)=\infty$. Le diviseur associ\'e \`a $f$ est~: 
\[
(f)\equiv\sum_{P\in\Sigma}\nu_P(f)\,P.
\]
Si $f$ n'est pas constante alors ${\rm deg}\,(f)=0$, 
\ie $f$ a autant de z\'eros que de p\^oles. J'en profite pour rappeler que 
les seules fonctions holomorphes sur une surface de Riemann (compacte) sont les 
fonctions constantes. Un diviseur est principal 
s'il est le diviseur d'une fonction  m\'eromorphe non-identiquement nulle. 
Deux diviseurs $D$ et $D'$ sont dits lin\'eairement \'equivalents
si $D-D'$ est un diviseur principal~; on note alors $D\sim D'$.
Il suit imm\'ediatement des d\'efinitions que $D$ et $D'$ ont m\^eme degr\'e.
On note aussi $D$ la classe du diviseur $D$.

Soit $\CM^1(\Sigma)$ l'ensemble des 1,0-formes m\'eromorphes sur $\Sigma$.
Soit $\omega\in \CM^1(\Sigma)$,  non-identi\-que\-ment nulle. Au voisinage de $P$,
$\omega$ est de la forme $f(z)\,dz$, 
o\`u $f$ est une fonction m\'eromorphe dans ce voisinage. 
L'ordre de $\omega$ en $P$ est~: $\nu_P(\omega)=\nu_P(f)$. 
Le r\'esidu de $\omega$ en $P$ est
\[
{\rm Res}_P\,\omega\equiv\frac{1}{2\pi i}\,\int_\gamma\omega
\]
o\`u $\gamma$ est un petit cercle autour de $P$ tel que, sur le disque bord\'e
par $\gamma$, $\omega$ a au plus un p\^ole, en $P$. D'apr\`es le th\'eor\`eme
de Stokes, cette construction ne d\'epend pas de $\gamma$. Le th\'eor\`eme
des r\'esidus dit que $\sum_{P\in\Sigma}{\rm Res}_P\,\omega=0$. 
Le diviseur de $\omega$ est, par d\'efinition,
\[
(\omega)=\sum_{P\in\Sigma}\nu_P(\omega)\,P.
\]
Un diviseur est dit canonique s'il est le diviseur d'une forme
m\'eromorphe non-identiquement nulle. Deux diviseurs canoniques sont 
lin\'eairement \'equivalents, de degr\'e~: $-\chi(\Sigma)=2g-2$ (formule
de Poincar\'e-Hopf). On utilise traditionnellement $K$ pour
d\'esigner la classe d'un diviseur canonique. Ensuite, si $\Omega^1(\Sigma)$ est 
l'espace vectoriel des formes holomorphes sur $\Sigma$, on a 
$\dim \Omega^1(\Sigma)=g$.

Soit $D$ un diviseur de $\Sigma$, on pose
\begin{eqnarray*}
\CL(D) & = & \{f\in\CM(\Sigma)\ \big|\ (f)+D\geq 0\},\\
\CM^1(D) & = & \{\omega\in\CM^1(\Sigma)\ \big|\ (\omega)\geq D\}.
\end{eqnarray*}
Concr\`etement, soit $P\in\Sigma$, $f\in\CL(D)$ et $\omega\in\CM^1(D)$. 
Si $D(P)<0$ alors $f$ doit avoir un z\'ero d'ordre $\geq -D(P)$ en $P$ et
$\omega$ peut avoir un p\^ole d'ordre $\leq -D(P)$ en $P$. Si $D(P)>0$ alors
$f$ peut avoir un p\^ole d'ordre $\leq D(P)$ en $P$ et $\omega$ doit avoir
un z\'ero d'ordre $\geq D(P)$ en $P$.
Ce sont des espaces vectoriels de dimensions (finies) respectives $l(D)$ et 
$i(D)$, appel\'e indice de sp\'ecia\-lit\'e. 
Si $D$ et $D'$ sont lin\'eairement \'equivalents, on a
$\CL(D)\cong \CL(D')$ et $\CM^1(D)\cong\CM^1(D')$. La r\'eciproque
(r\'eciprocit\'e de Brill-Noether) dit que si 
$D+D'$ est un diviseur canonique, alors $\CL(D)\cong\CM^1(D')$
et $\CL(D')\cong\CM^1(D)$. En particulier, $l(K-D)=i(D)$.

Le {\bol th\'eor\`eme de Riemann-Roch} s'\'ecrit
\begin{eqnarray*}
l(D)-i(D) & = & {\rm deg}\,D+1-g,\\
{\rm ou}\ \ \ \  l(D)-l(K-D) & = & \deg D+1-g.
\end{eqnarray*}
On dispose aussi de crit\`eres d'annulation simples : si $\deg D<0$ alors 
$l(D)=0$ et si $\deg D>2g-2$ alors $i(D)=0$.

Soient $\Sigma$ et $\Sigma'$ deux surfaces de Riemann. Soient $f$
une application holomorphe non-constante de $\Sigma$ dans $\Sigma'$, $P$ un
point de $\Sigma$ et $Q$ un point de $\Sigma'$. Comme pr\'ec\'edemment,
on peut choisir une coordonn\'ee locale $z$ (resp. $w$) en $P$ (resp. $Q$) 
centr\'ee en $0$, telle que $f$ soit de la forme $w=z^\nu$, o\`u $\nu\in\N$.
On note $\nu_P(f)$ l'ordre de $f$ en $P$ --- ce dernier ne d\'epend pas
des choix effectu\'es. L'ensemble des points pour lesquels l'ordre de $f$ est
strictement sup\'erieur \`a $1$ est fini ; on appelle ces points
des {\bol points de ramification}. On d\'efinit
\[
f^{-1}(Q)\equiv\sum_{f(P)=Q}\nu_P(f)\,P.
\]
C'est un diviseur de $\Sigma$ car $f$ n'est pas constante.
Le degr\'e de $f$ est $\deg f\equiv\deg f^{-1}(Q)$.
On a bien le droit de parler du degr\'e de $f$ car, $\Sigma'$ \'etant connexe,
$Q$ ne joue qu'un r\^ole d'interm\'ediaire. Notons $n$ le degr\'e de
notre application $f$. On dit que $f$ est un {\bol rev\^etement ramifi\'e
(\'etale) \`a $n$ feuillets}~(\footnote{On rajoute parfois l'adjectif \'etale
pour faire la diff\'erence avec les rev\^etements (topologiques) du 
th\'eor\`eme d'uniformisation.}).
Le diviseur de ramification de $f$ est le diviseur
\[
R=\sum_{P\in\Sigma}(\nu_P(f)-1)\,P.
\]
La formule de Riemann-Hurwitz donne le degr\'e de $R$~:
${\rm deg}\,R=-\chi(\Sigma)+n\,\chi(\Sigma')$. 

\subsection{Genre un --- courbes elliptiques}

En genre z\'ero, toute surface de Riemann est la normalisation d'une
courbe alg\'ebrique lisse $C$ de degr\'e $3$ dans $\PP^2$ d'\'equation
affine $y^2-(x-a_1)(x-a_2)(x-a_3)=0$, 
o\`u $a_1,a_2,a_3$ sont deux \`a deux distincts.
L'\'equation projective de $C$ est  
$x_0(x_2)^2-(x_1-a_1x_0)(x_1-a_2x_0)(x_1-a_3x_0)=0$, pour
$[x_0,x_1,x_2]\in\PP^2$.
Quitte \`a composer $x$ avec un automorphisme de $\C P^1$ pour
envoyer $a_1$ sur $0$, $a_2$ sur $1$ et l'$\infty$ sur lui-m\^eme, 
on voit que toute surface de genre un 
est la normalisation d'une courbe alg\'ebrique plane 
$C_\lambda$ d'\'equation affine $y^2=x(x-1)(x-\lambda)$, o\`u $\lambda\ne 0,1$.
Deux courbes $C_\lambda$ et $C_{\lambda'}$ sont isomorphes si, et seulement 
si, il existe un automorphisme de $\C P^1$ qui envoie (globalement) 
$\{0,1,\infty,\lambda\}$ sur $\{0,1,\infty,\lambda'\}$. Les seuls cas 
possibles sont
$ \lambda'=\lambda,\ 1-\lambda,\ 1/\lambda,\ (\lambda-1)/\lambda,
\ \lambda/(\lambda-1),\ 1/(1-\lambda)$.
L'expression suivante
\[
j(\lambda)=256\,\frac{(\lambda^2-\lambda+1)^3}{\lambda^2(\lambda-1)^2}
\]
est invariante par toutes ces transformations.
L'application $j$ \'etablit donc une bijection entre les classes 
d'isomorphisme de surface de Riemann de genre un et le plan complexe~$\C$. 

Pour les courbes elliptiques~: on peut r\'ealiser explicitement $E_\tau$ 
comme la normalisation de la courbe alg\'ebrique
plane d'\'equation affine $y^2=4x^3-g_{2}x-g_{3}$, appel\'ee
forme normale de Weierstra\ss. L'application \<<normalisante\>>
 $\sigma$ envoie le r\'eseau $\Lambda_\tau=\Z+\tau\Z$ sur $[0,0,1]$ et 
$z\in E_\tau\setminus \Lambda_\tau$ sur $[1,\wp(z),\wp'(z)]\in\PP^2$, 
o\`u $\wp$ est 
la fonction de Weierstra\ss\ de p\'eriode $2\omega_1=1$ et $2\omega_2=\tau$~:
\[
\wp(z)=\frac{1}{z^2}+\sumprime \left(\frac{1}{(z+\omega)^2}
	-\frac{1}{\omega^2}\right),
\]
la somme $\sumprime$ portant sur $\omega$ dans le r\'eseau $\Lambda_\tau$
priv\'e de z\'ero.
La fonction $\wp$ satisfait bien l'\'equation 
$\wp'{}^{2}=4\,\wp^3-g_{2}\wp-g_{3}$. 
Deux courbes elliptiques $E_\tau$ et $E_{\tau'}$ sont isomorphes 
si, et seulement si, il existe $\alpha\in{\rm PSL}_2\Z$ telle que 
$\tau'=\alpha\tau$~; un \'el\'ement $\alpha$ de ${\rm PSL}_2\Z$ agit sur 
$\tau\in\poincare$ par transformation projective.
On peut \'egalement fabriquer un invariant $j$. D\'efinissons
\[
g_{2}(\tau)=60\sumprime 1/\omega^4,\qquad
g_{3}(\tau)=140\sumprime 1/\omega^6.
\]
On pose alors
\[
j(\tau)=1728\,\frac{g_{2}^3}{\Delta}
\]
o\`u $\Delta=g_{2}^3-27\,g_{3}^2$. 
La courbe elliptique $E_\tau$ n'\'etant pas singuli\`ere, la courbe 
alg\'ebrique correspondante ne l'est pas non plus et $\Delta\neq 0$. 
La fonction $j:\poincare\rightarrow\C$ est une forme modulaire de poids $0$, 
holomorphe sur $\poincare$ et ayant un p\^ole simple en $\infty$. 
Elle induit une bijection de $\poincare/{\rm PSL}_2\Z$ sur $\C$, toujours 
not\'ee $j$. Si on d\'eveloppe $j$ en puissances de $q=\ee^{2i\pi\tau}$, on a
\[
j(\tau)=\frac{1}{q}+744+196884\,q+\cdots
\]
et tous les autres coefficients sont des entiers positifs. Ceci justifie a 
posteriori le coefficient $1728$ dans $j(\tau)$.
On montre que, si une courbe elliptique $E_\tau$ est
la normalisation d'une courbe alg\'ebrique $C_\lambda$, alors 
$j(\tau)=j(\lambda)$.

\subsection{Courbes hyperelliptiques}

Une surface de Riemann $\Sigma$ compacte de genre $\geq 2$ est dite 
hyperelliptique s'il existe une application holomorphe $x$ 
de degr\'e $2$ de $\Sigma$ dans $\C P^1$. Le diviseur de ramification $R$ 
de $x$ est constitu\'e de $2g+2$ points $P_i$ distincts. 
Parfois, on dira point de Weierstra\ss\ pour point de ramification. 
Ces deux notions sont g\'en\'eralement diff\'erentes mais co\"\i ncident 
justement si $\Sigma$ est hyperelliptique. On note $a_i=x(P_i)$ --- ils sont 
deux \`a deux distincts --- et 
$x^{-1}(\infty)=\{\infty_1,\infty_2\}$. On suppose que $\infty$
n'est pas un \<<point\>> de ramification. L'involution hyperelliptique
est l'application $\iota:\Sigma\ni Q_1\mapsto Q_2\in\Sigma$, si $Q_1$ et
$Q_2$ ont la m\^eme image par $x$, \ie $\iota$ \'echange les deux feuilles. 
Cette derni\`ere induit une action sur $\CM(\Sigma)$ par
$\iota(\alpha)\equiv\iota^*\alpha$.
Si $D=(g+1)\infty_1+(g+1)\infty_2$ alors $\dim \CL(D)=g+3$. Comme $D$ 
est pr\'eserv\'e par $\iota$, on peut voir $\iota$ comme une application 
lin\'eaire sur $\C^{g+3}$ de valeurs propres $\pm 1$.
Si on d\'ecompose $\CL(D)$ en sous-espaces propres $\CL(D)^+\oplus\CL(D)^-$ 
alors $1,x,\cdots, x^{g+1}$ est une base pour $\CL(D)^+$. En particulier,
$\CL(D)^-$ est de dimension $1$. On montre alors qu'on peut choisir $y$
dans $\CL(D)^-$ tel que 
\begin{equation}
\label{hyperelliptique}
y^2=(x-a_1)\cdots(x-a_{2g+2}).
\end{equation}
Si au d\'epart $\infty$ est un point de
ramification, on ne consid\`ere qu'un produit sur $2g+1$ points.
La surface hyperelliptique $\Sigma$ de genre
$g$ est alors la  normalisation de la courbe alg\'ebrique plane
$C$ de degr\'e $2g+2$, d'\'equation affine~\eqref{hyperelliptique}
et singuli\`ere en $[0,0,1]$. L'application normalisante $\sigma:
\Sigma \rightarrow \PP^2$ envoie $P$ sur $[1,x(P),y(P)]$ et $\infty_1,\infty_2$
sur $[0,0,1]$. La projection $C\ni(x,y)\mapsto x\in\C P^1$ r\'ealise $C$
comme un rev\^etement ramifi\'e de $\C P^1$ \`a deux feuillets, 
avec $2g+2$ points de ramification.
Il est clair que sur $C$ l'involution hyperelliptique $\iota$ envoie
$(x,y)$ sur $(x,-y)$. Une base de formes holomorphes sur $\Sigma$ est donn\'ee
par 
\[
\omega_{1}=\frac{dx}{y},\ \cdots,\ \omega_{g}=\frac{x^{g-1}\,dx}{y}.
\]
La d\'emonstration passe tr\`es bien en genre z\'ero ou un, mais on pr\'ef\`ere
traiter s\'epar\'ement ces deux cas, car la courbe $C$ est alors lisse.

\subsection{Fibr\'es vectoriels holomorphes sur une surface de Riemann}

Dans cette section tous les fibr\'es
vectoriels sont des fibr\'es holomorphes sur $\Sigma$, ce que je ne 
pr\'eciserai plus. Soit $\CM^\times$ le faisceau (multiplicatif)
des germes de fonctions m\'eromorphes non-identiquement 
nulles sur $\Sigma$ et $\CO^\times$ le sous-faisceau 
des fonctions holomorphes non-nulles. On a
${\rm Div}\,\Sigma=H^0(\Sigma,\CM^\times/\CO^\times)$.
En d\'ecrivant les fibr\'es en droites par leurs
fonctions de transition, on constate que le groupe des classes 
d'isomorphismes de fibr\'es en droites est juste 
$H^1(\Sigma,\CO^\times)$~: c'est le groupe de Picard de $\Sigma$, 
not\'e ${\rm Pic}\,\Sigma$. Soit $D$ un diviseur sur $\Sigma$.
On attache \`a $D$ un fibr\'e en droites $\CO(D)$, je ne vais pas le
d\'ecrire explicitement car ce n'est pas tr\`es instructif. Par contre,
on doit retenir que le faisceau des germes de sections holomorphes
de $\CO(D)$ est canoniquement identifi\'e avec le faisceau, toujours
not\'e $\CO(D)$, d\'efini par ses sections au-dessus d'un \mbox{ouvert $U$ }
\[
\CO(D)(U)\equiv\{f\in \CM(U)\ \big|\ (f)+D|_U\geq 0\}.
\]
Ainsi, je ne ferai pas de diff\'erence entre le fibr\'e en droites
et le faisceau correspondant. De mani\`ere plus g\'en\'erale, je note
de la m\^eme fa\c{c}on un fibr\'e vectoriel $E$ et le faisceau des germes de 
sections holomorphes de $E$. La multiplication par une section 
m\'eromorphe de $\CO(D)$ permet d'identifier $\CL(D)$ et les sections
holomorphes (globales) de $\CO(D)$, \ie $H^0(\Sigma,\CO(D))$. 
On note les isomorphismes (holomorphes) suivants
\begin{gather*}
\CO(D)\otimes\CO(D')\cong\CO(D+D'),\qquad \CO(D)^{-1}\cong\CO(-D),\\
\CO(D)  \cong  \CO(D') \quad {\rm ssi} \quad D\sim D'.
\end{gather*}
Soit $L$ un fibr\'e en droites. On montre que $L$ poss\`ede toujours une
section m\'eromorphe non-nulle $s$. Si $D=(s)$, alors $L\cong\CO(D)$.
En termes cohomologiques, on peut r\'esumer ce qui pr\'ec\`ede
dans la suite exacte $0\rightarrow\CO^\times\rightarrow\CM^\times\rightarrow
\CM^\times/\CO^\times\rightarrow 0$ et la suite cohomologique qui s'en 
d\'eduit~:
$H^0(\Sigma,\CM^\times)\rightarrow {\rm Div}\,\Sigma\rightarrow{\rm Pic}
\,\Sigma$.
La premi\`ere fl\`eche associe \`a une fonction m\'eromorphe sur $\Sigma$ son
diviseur et l'application cobord $\delta$ est juste $\delta(D)=\CO(D)$.
En particulier, ${\rm Pic}\,\Sigma\cong {\rm Div}\,\Sigma/\sim$.

De la suite exacte exponentielle $0\rightarrow\Z\rightarrow 
\CO\rightarrow\CO^\times\rightarrow 0$, on extrait la suite exacte longue 
de cohomologie~: 
\[
0\rightarrow H^1(\Sigma,\Z)\rightarrow H^1(\Sigma,\CO)
\rightarrow H^1(\Sigma,\CO^\times)
\stackrel{\delta}{\rightarrow}H^2(\Sigma,\Z) \rightarrow H^2(\Sigma,\CO)=0.
\]
L'application $c_1:H^1(\Sigma,\CO^\times)\stackrel{\delta}{\rightarrow}
H^2(\Sigma,\Z)\hookrightarrow
H^2(\Sigma,\R)$ associe \`a tout fibr\'e en droites une classe de \v{C}ech,
appel\'ee premi\`ere classe de Chern. Cette terminologie est justifi\'ee
car la classe de de Rham qui repr\'esente cette classe de \v{C}ech
est exactement $c_1(L)$, comme d\'efinie aupa\-ravant. 
On v\'erifie que $\deg \CO(D)=\deg D$.
Le noyau de $c_1$ est le quotient
\[
{\rm Pic}^0\,\Sigma=H^1(\Sigma,\CO)/H^1(\Sigma,\Z),
\]
le sous-groupe de ${\rm Pic}\, \Sigma$ des fibr\'es 
de degr\'e z\'ero qui est isomorphe \`a ${\rm Div}^0\,\Sigma/\sim$.

Le fibr\'e canonique $K_\Sigma$ (ou simplement $K$)
est le fibr\'e cotangent holomorphe $T^*{}^{1,0}\Sigma$. 
Les sections m\'eromorphes de $K$ sont juste les $1$-formes m\'eromorphes. 
Le {\bol th\'eor\`eme de Riemann-Roch} pour un fibr\'e en droite $L$ dit~:
$\dim H^0(\Sigma,L)-\dim H^1(\Sigma,L)=\deg L+1-g$.
On \'ecrit toujours cette \'egalit\'e accompagn\'ee du  th\'eor\`eme de 
{\bol dualit\'e de Serre}~: $H^0(\Sigma,K\otimes L^{-1})\cong H^1(\Sigma,L)^*$.
En fait, on peut aussi \'ecrire un th\'eor\`eme de Riemann-Roch pour 
un fibr\'e vectoriel $E$ de rang~$r$. Si $h^0(\Sigma,E)\equiv\dim H^0(\Sigma,
E)$ et $h^1(\Sigma,E)\equiv\dim H^1(\Sigma,E)$, alors
\[
h^0(\Sigma,E)-h^1(\Sigma,E)=\deg E+r\,(1-g).
\]
Le th\'eor\`eme de dualit\'e de Serre reste valable~:
$H^0(\Sigma,K\otimes E^*)\cong H^1(\Sigma,E)^*$. La dualit\'e est 
simplement
\[
H^1(\Sigma,E)\times H^0(\Sigma,K\otimes E^*)
\ni(\zeta,s)  \longmapsto  \int_\Sigma \langle s
\stackrel{\wedge}{,}\zeta\rangle\in\C
\]
o\`u $s$ est vue comme une $(1,0)$-forme \`a valeurs dans $E^*$
et $\zeta$ comme une $(0,1)$-forme \`a valeurs dans $E$ par
l'isomorphisme de Dolbeaut entre $H^1(\Sigma,E)$ et $H_{{\rm Dol}}
^{0,1}(\Sigma,E)$.
\subsection{Jacobienne}

Soit $\Sigma$ une surface de Riemann de genre $\geq 1$. Soit 
$(a_1,\cdots,a_g, b_1,\cdots,b_g)$ une base symplectique canonique
de $H_1(\Sigma,\Z)$, \cad de nombres d'intersection
$(a_\alpha,a_\beta)=0=(b_\alpha,b_\beta)$ et $(a_\alpha,b_\beta)
=\delta_{\alpha,\beta}=-(b_\beta,a_\alpha)$. Soit $\omega\in\Omega^1(\Sigma)$,
on pose
\[
A_\alpha(\omega)=\int_{{a_\alpha}}\omega, \qquad
B_{\alpha}(\omega)=\int_{{b_\alpha}}\omega.
\]
Soient $\omega,\varphi\in\Omega^1(\Sigma)$, on montre les relations 
bilin\'eaires de Riemann
\[
\sum_{\alpha=1}^g\left(A_\alpha(\omega)\,B_{\alpha}(\varphi)
	-B_{\alpha}(\omega)\,A_\alpha(\varphi)
	\right)=0\qquad{\rm et }\qquad
{\rm Im}\,\sum_{\alpha=1}^g
\big(\overline{A_\alpha(\omega)}\,B_{\alpha}(\omega)\big)>0.
\]
Soit $(\omega_1,\cdots,\omega_g)$ une base de $\Omega^1(\Sigma)$.
La matrice des p\'eriodes, not\'ee $\Pi$, est la matrice 
$g\times 2g$~: $\Pi=(A,B)$, o\`u $A$ (resp. $B$) est la matrice $g\times g$
d'\'el\'ements $A_{\alpha\beta}=A_\alpha(\omega_\beta)$ (resp.
$B_{\alpha\beta}=B_\alpha(\omega_\beta)$). 
D'apr\`es les relations bilin\'eaires de Riemann, la matrice $A$
est inversible et $\tau=A^{-1}B$ est une matrice sym\'etrique de partie 
imaginaire, not\'ee $\tau_2$, d\'efinie positive.  
Dor\'enavant, on se donne une base normalis\'ee de $\Omega^1(\Sigma)$, \ie 
pour laquelle $\Pi=(I_g,\tau)$~; c'est toujours possible car $A$ est inversible.
On appelle $\tau$ la matrice des p\'eriodes normalis\'ee.
Si $\alpha=1,\cdots, g$, $e_\alpha$ d\'esigne le vecteur de $\C^g$ ayant des
z\'eros partout sauf dans sa $\alpha$-i\`eme ligne, o\`u on trouve $1$.
Par contre, si $\alpha=g+1,\cdots,2g$, $e_\alpha$ est la 
$\alpha$-i\`eme colonne de $\tau$. Comme $\tau_2$ est d\'efinie positive, 
les vecteurs $e_\alpha$ sont lin\'eairement ind\'ependants sur $\R$.
Le $\Z$-espace vectoriel $\Lambda_\tau$ engendr\'e par ces vecteurs 
forme donc un r\'eseau de $\C^g$. La Jacobienne de $\Sigma$ est 
\[
{\rm Jac}\,\Sigma\ \equiv\ \C^g/\Lambda_\tau\ =\ H^0(\Sigma,K)^*/H_1(\Sigma,\Z).
\]
Pour identifier le terme \`a l'extr\^eme gauche, il faut utiliser le
plongement de $H_1(\Sigma,\Z)$ dans $H^0(\Sigma,K)^*$ par
l'application $H_1(\Sigma,\Z)\ni\gamma\mapsto (\omega\mapsto
\int_\gamma\omega)\in H^0(\Sigma,K)^*$.

Fixons un point $P_0$ dans $\Sigma$. L'application d'Abel 
est l'application $I:\Sigma\rightarrow{\rm Jac}\,\Sigma$ qui envoie
$P\in\Sigma$ sur 
\[
I(P)=\left(\int_{P_0}^P\omega_1,\cdots,\int_{P_0}^P\omega_g\right)
\ {\rm mod}\ \Lambda_\tau.
\]
Cette application est bien d\'efinie ind\'ependamment du choix du lacet
allant de $P_0$ \`a $P$. On \'etend $I$ en une application $I:{\rm Div}\,
\Sigma\rightarrow{\rm Jac}\,\Sigma$ par~: $I(D)=\sum_in_i\,I(P_i)$, si 
$D=\sum_in_i\,P_i$. La suite 
\[
\CM(\Sigma)^\times\stackrel{(\,)}{\longrightarrow}{\rm Div}^0\,\Sigma
	\stackrel{I}{\longrightarrow}{\rm Jac}\,\Sigma\longrightarrow 0
\]
est exacte. D'une part, le th\'eor\`eme d'Abel assure l'exactitude
des deux premi\`eres fl\`eches, \ie un diviseur $D$ de degr\'e $0$ 
est principal si, et seulement si, $I(D)=0$.
D'autre part, la surjectivit\'e de $I$ est donn\'ee par
le th\'eor\`eme d'inversion de Jacobi. 
Ainsi, l'application $I$ induit un isomorphisme entre 
${\rm Jac}\, \Sigma$ et ${\rm Pic}^0\,\Sigma$, le groupe des classes
d'\'equivalence lin\'eaire des diviseurs de degr\'e z\'ero.

\medskip
\section{Th\^eta dans tous ses \'etats}

Le trait\'e de Mumford~\cite{mumford, mumford2} est incontestablement 
{\sc la} r\'ef\'erence incontournable sur les fonctions th\^eta. En ce
qui concerne les fonctions th\^eta de Jacobi, on pourra 
consulter avec bonheur les classiques~\cite{jordan,whit}. 

\subsection{Fonction th\^eta de Riemann\label{theta}}

Soit $\tau$ appartenant au demi-plan de Siegel $\,\poincare_g$, 
\cad l'ensemble des matrices complexes sym\'etriques $g\times g$ de partie
imaginaire $\tau_2$ d\'efinie positive. Bien entendu, $\poincare_1=\poincare$.
Soit ${\rm\bf Th}_k$ l'espace vectoriel des fonctions
enti\`eres sur $\C^g$ quasi-p\'eriodiques de degr\'e $k$, \cad
\[
\theta(z+p+\tau q)=\ee^{-\pi ik\,q\cdot\tau q-2\pi ik\,q\cdot z}\,\theta(z),
	\qquad p,q\in\Z^g,
\]
o\`u $q\cdot r\equiv q^t\,r$, la multiplication \`a droite \'etant la
multiplication matricielle.
Un \'el\'ement de ${\rm \bf Th}_k$ est appel\'e une {\bol fonction th\^eta de
degr\'e $k$}. Si $k=1$, ${\rm \bf Th}_1$ est un espace vectoriel de
dimension $1$ engendr\'e par la {\bol fonction th\^eta de Riemann}
\[
\vartheta(z|\tau)=\sum_{n\in\Z^g}\ee^{i\pi\,n\cdot\tau n+2\pi i\,n\cdot z}.
\]
C'est une fonction paire, holomorphe sur $\C^g\times\poincare_g$.

On suppose maintenant que $\tau$ est la matrice des p\'eriodes 
normalis\'ee d'une surface de Riemann $\Sigma$ de genre $g$. Afin de d\'eterminer
les z\'eros de $\vartheta$, on utilise la fonction interm\'ediaire~:
\[
\Sigma\ni x\mapsto\vartheta(z+\int_{x_0}^x\omega|\tau)\equiv f(x)\in\C.
\]
D'apr\`es le {\bol th\'eor\`eme d'annulation de Riemann}~: soit 
$f$ est identiquement nulle, 
soit il existe $\Delta\in\C^g$, appel\'e
{\bol vecteur des constantes de Riemann}, ne d\'ependant que du choix de la
base d'homologie et du point base $x_0$, tel que $f$ poss\`ede $g$ z\'eros
$x_1,\cdots,x_g$ avec
\begin{equation}
\label{vecrie}
z+\sum_{i=1}^{g}\int_{x_0}^{x_i}\omega=\Delta
\qquad(\text{mod}\ \Lambda_\tau).
\end{equation}
R\'eciproquement, pour tout $g$-uplet de points $(x_1,\cdots,x_g)$, si $z$
est d\'efini par l'\'equation~\eqref{vecrie} alors $f(x_i)=0$,
$\forall i=1,\cdots,g$. On obtient comme corollaire~: 
$\vartheta(z|\tau)=0$ si, et 
seulement si, il existe $x_1,\cdots,x_{g-1}\in\Sigma$ tels que
\[
z=\Delta-\sum_{i=1}^{g-1}\int_{x_0}^{x_i}\omega.
\]

Pour $k>1$, on introduit la fonction th\^eta de caract\'eristiques
$(\delta',\delta'') \in\Q^g\times\Q^g$~: 
\begin{align*}
\vartheta\left[\delta'\atop \delta''\right](z|\tau)&=
	\ee^{i\pi\,\delta'\cdot\tau\delta'+2\pi i\,\delta'\cdot
	(z+\delta'')}\,\vartheta(z+\tau\delta'+\delta''|\tau)\\
&=\sum_{n\in\Z^g}\ee^{i\pi\,(n+\delta')\cdot\tau(n+\delta')+2\pi i\,
	(n+\delta')\cdot(z+\delta'')}.
\end{align*}
La fonction th\^eta de Riemann est juste la fonction th\^eta de 
caract\'eristiques nulles. 
On v\'erifie directement les transformations suivantes
\begin{align*}
\vartheta\left[\delta'\atop\delta''\right](z+p+\tau q|\tau)&=
	\ee^{-i\pi\,q\cdot\tau q-2\pi i\,q\cdot(z+\delta'')}\,
	\ee^{2\pi i\,\delta'\cdot p}\,\vartheta\left[\delta'
	\atop\delta''\right]
	(z|\tau)\\
\vartheta\left[{\delta'+p}\atop{\delta''+q}\right](z|\tau)&=
	\ee^{2\pi i\,\delta'\cdot q}\,\vartheta\left[\delta'
	\atop\delta''\right]
	(z|\tau).
\end{align*}
L'ensemble des $\theta_{k,e}\equiv\vartheta
\left[{e/k}\atop 0\right](kz,k\tau)$,
$e\in\Z^g/k\Z^g$, \cad concr\`etement
\[
\theta_{k,e}(z)=\sum_{n\in\Z^g}\ee^{\pi ik\,(n+\frac{e}{k})\cdot
\tau(n+\frac{e}{k})
	+2\pi ik\,(n+\frac{e}{k})\cdot z},
\]
forme une base de ${\rm \bf Th}_k$. En particulier, 
$\dim {\rm \bf Th}_k=k^g$.

\subsection{Fibr\'e th\^eta}

Soit $\Lambda$ un r\'eseau de $\C^g$. Le quotient $\C^g/\Lambda$ est
une vari\'et\'e complexe, appel\'ee tore complexe.
Au passage remarquons que la raison profonde 
pour laquelle on se contente de prendre $\tau$ dans
le demi-plan de Siegel est que les seuls tores pouvant
\^etre plong\'es dans un plan projectif sont les tores $\C^g/\Lambda_\tau$ et
leurs rev\^etements finis (th\'eor\`eme de Lefschetz).
Comme $H^1(\C^g, \CO^\times)=0$~\cite[p.47]{griffiths}, tout fibr\'e 
holomorphe en 
droites au-dessus de $\C^g$ est trivial. Soit $L\rightarrow\C^g/\Lambda$
un fibr\'e en droites et $\pi:\C^g\rightarrow\C^g/\Lambda$ la
projection. Le fibr\'e pr\'eimage $\pi^*L\rightarrow\C^g$ 
est trivial, ainsi il existe une trivialisation globale $\varphi:\pi^*L
\rightarrow \C^g\times \C$. Fixons $z\in\C^g$ et $\lambda\in\C^g$. 
Comme $L_{\pi(z)}=(\pi^*L)_z=(\pi^*L)_{z+\lambda}$, 
on obtient un automorphisme de $\C$~:
\[
\C\stackrel{\varphi_z}{\longleftarrow}(\pi^*L)_z=L_{\pi(z)}
=(\pi^*L)_{z+\lambda}\stackrel{\varphi_{z+\lambda}}{\longrightarrow}\C.
\]
Un automorphisme de $\C$ est donn\'e par la multiplication par 
un nombre complexe non-nul $m_\lambda(z)$. L'ensemble des multiplicateurs
$m_\lambda$, $\lambda\in\Lambda$, est appel\'e un  
facteur d'automorphie pour $L$. Les multiplicateurs
satisfont la relation de compatibilit\'e
$m_\mu(z+\lambda)m_\lambda(z)=m_{\lambda+\mu}(z)$, $\forall z\in\C^g$.
R\'eciproquement, tout facteur d'automorphie d\'efinit un fibr\'e en droites
holomorphe sur $\C^g/\Lambda$, obtenu en quotientant $\C^g\times\C$ par 
la relation d'\'equivalence $(z,w)\sim(z+\lambda,e_\lambda(z)w)$.
Une section de ce fibr\'e en droites est une fonction holomorphe $s$ sur
$\C^g$ telle que $s(z+\lambda)=m_\lambda(z)\,s(z)$, $\forall\lambda\in\Lambda$.
De par les relations de compatibilit\'e, il suffit de conna\^\i tre les
multiplicateurs sur une base $(\lambda_1,\cdots,\lambda_{2g})$ de $\Lambda$. 
On note $m_\ell=m_{{\lambda_\ell}}$.

Soit $\Sigma$ une surface de Riemann de genre $g$ et ${\rm Jac}\,\Sigma$
sa Jacobienne. On a vu que ${\rm Jac}\,\Sigma$ est un tore complexe
model\'e sur le r\'eseau $\Lambda_\tau$ engendr\'e par les $e_\alpha$. 
Le {\bol fibr\'e th\^eta}, not\'e $L_\Theta$,
 est le fibr\'e holomorphe en droites au-dessus de ${\rm Jac}\,\Sigma$, 
d\'efini par le facteur d'automorphie~:
\[
m_{\alpha}(z)=1,\qquad 
m_{g+\alpha}(z)=\ee^{-\pi i\,\tau_{\alpha\alpha}-2\pi i\,z_\alpha}
\]
pour $\alpha=1,\cdots,g$. On aurait donc pu
d\'efinir une fonction th\^eta de degr\'e $k$ comme 
une section holomorphe de la $k$-i\`eme puissance de $L_\Theta$.
En particulier, $\dim H^0(L_\Theta^k)=k^g$. Le {\bol diviseur th\^eta},
not\'e $\Theta$, est le diviseur de la section $\vartheta$ de $L_\Theta$.
Comme corollaire du th\'eor\`eme d'annulation de Riemann, le diviseur de 
$L_\Theta$ est l'ensemble~(\footnote{Ici on parle de diviseur 
d'un fibr\'e en droites au-dessus d'une vari\'et\'e complexe et
non plus d'une surface de Riemann.})
\[
\Theta=\left\lbrace
	\Delta-\sum_{i=1}^{g-1}\int_{x_0}^{x_i}\omega\
	\text{pour}\ x_1,\cdots,x_{g-1}\in\Sigma\right\rbrace.
\]

Le fibr\'e $L_\Theta$ poss\`ede aussi une structure hermitienne~:
\[
\Vert s\Vert^2=\ee^{-2\pi\,(z-\overline{z}).\imtau^{-1}(z-\overline{z})}\,|s|^2.
\]

\subsection{Fibr\'es spin et caract\'eristiques th\^eta}

Une classe de diviseurs $D$ de $\Sigma$ telle que $2D=K$, 
o\`u $K$ est la classe canonique, est appel\'ee une
{\bol caract\'eristique} (th\^eta) de $\Sigma$. 
Notons $({\rm Pic}^0\,\Sigma)_2$ l'ensemble
des diviseurs de degr\'e z\'ero tels que $2E=0$. 
Quels que soient $D_1,D_2\in\NT$, il existe
un unique $E\in({\rm Pic}^0\,\Sigma)_2$, tel que $D_1=D_2+E$. Comme de plus
$({\rm Pic}^0\,\Sigma)_2=({\rm Jac}\,\Sigma)_2\cong(\Z/2\Z)^{2g}$, il
suit $\big|\NT\big|=2^{2g}$. 
 
Dans le langage des fibr\'es, pour parler de caract\'eristique, on utilise
la notion de fibr\'e spin.
Un {\bol fibr\'e spin} $L$ est une racine carr\'e du fibr\'e canonique, 
\ie $L^2=K$. Il n'y a pas de fibr\'e spin privil\'egi\'e. Clairement,
${\rm deg}\,L=g-1$. 

Les deux isomorphismes suivants font le lien entre  
les caract\'eristiques et
les fonctions th\^eta~:

--- l'application qui \`a $D\in\NT$ associe le lieu des points
	$I(x_1+\cdots+x_{g-1}-D)$ dans $\C^g/\Lambda_\tau$ 
	pour tout $x_1,\cdots,x_{g-1}\in\Sigma$, est
	un isomorphisme entre $\NT$ et les
	translations du diviseur $\Theta$ qui sont sym\'etriques,
	\ie les $z+\Theta$ tels que $z+\Theta=-(z+\Theta)$~;

--- l'application qui associe \`a 
	une caract\'eristique (!) $\delta=(\delta',\delta'')$
	le lieu des points tels que $\vartheta[\delta](z|\tau)=0$ 
	dans $\C^g/\Lambda_\tau$
	est un isomorphisme entre $\frac{1}{2}\Z^{2g}/\Z^{2g}$ et les
	translat\'es de $\Theta$ sym\'etri\-ques.

Si $D_0$ est la caract\'eristique qui est envoy\'ee par le premier isomorphisme
sur $\Theta$ lui-m\^eme, le vecteur des constantes de Riemann est~:
$\Delta=I(D_0+(g-1)x_0)$. Alors que $\Delta$ d\'epend \`a la fois du point
base et de la base d'homologie, la caract\'eristique $D_0$ ne d\'epend pas
du point base. Il y a donc un fibr\'e spin privil\'egi\'e \`a condition
de choisir une base d'homologie.
	
Un calcul direct montre que $\vartheta[\delta]$ est soit paire soit impaire~:
\[
\vartheta[\delta](-z|\tau)=(-1)^{4\,\delta'\delta''}\,\vartheta[\delta](z|\tau).
\]
On dira d'une caract\'eristique qu'elle est {\bol paire} 
(resp. {\bol impaire}) si $4\,\delta'\delta''$ est paire (resp. impaire)
partitionnant ainsi $\NT$ en caract\'eristiques impaires
$\NT_-$ et caract\'eristiques paires $\NT_+$. Pour les fibr\'es spin,
on d\'efinit de mani\`ere \'equivalente~: $L\in\NT$ est paire (impaire)
si $\dim H^0(L)$ l'est. Pour les dimensions, on a
\[
\big|\NT_-\big|=2^{g-1}(2^g-1),\qquad
	\big|\NT_+\big|=2^{g-1}(2^g+1).
\]

\subsection{Fonction th\^eta de Jacobi}

On suppose ici que $g=1$. Les quatre fonctions th\^eta de Jacobi 
$\vartheta_1, \vartheta_2, \vartheta_3, \vartheta_4$ sont les 
fonctions th\^eta de Riemann de caract\'eristiques respectives 
$(\frac{1}{2},\frac{1}{2})$, $(\frac{1}{2},0)$, $(0,\frac{1}{2})$ et 
$(0,0)$, soit toutes les fonctions th\^eta de caract\'eristiques 
$\delta',\delta''\in\frac{1}{2}\,\Z$. Dans cette section, nous 
utilisons exclusivement $\vartheta_1$. Sous forme 
d\'evelopp\'ee, on a
\[
\vartheta_1(z)=-i\sum_{\ell\in\Z}(-1)^\ell\,\ee^{\pi i\tau
	(\ell+\frac{1}{2})^2+\pi i z\,(2\ell+1)}.
\]
La fonction $\vartheta_1$ est analytique, impaire, de z\'eros $\Z+\tau\Z$, 
v\'erifie l'\'equation de la chaleur $\partial_z^2\vartheta_1=
4\pi i\,\partial_{\tau}\vartheta_1$ et se transforme comme~: $\vartheta_1(z+1)
=-\vartheta_1(z)$ et $\vartheta_1(z+\tau)=-\ee^{-\pi i\,(\tau+2z)}\,
\vartheta_1(z)$. On peut obtenir $\vartheta_1$ sous forme produit par 
l'interm\'ediaire du d\'enominateur de Weyl-Kac pour le 
groupe ${\rm SU}_2$~: 
\[
\vartheta_1(z)=\quotient{1}{i}\,
	\Pi(\quotient{z}{2}\sigma_3,\tau)=2q^{1/8}
	\,{\rm sin}(\pi z)\,\prod_{\ell=1}^{+\infty}
	(1-q^\ell)(1-q^\ell\ee^{2\pi iz})(1-q^\ell\ee^{-2\pi iz})
\]
pour $q=\ee^{2\pi i\tau}$. 

La fonction th\^eta de Jacobi permet de montrer quelques relations
utiles sur le d\'enominateur de Weyl-Kac pour $G$ g\'en\'eral. 
On regarde $\Pi(u,\tau)$ comme une fonction des
$u_\alpha=\alpha(u)$, $\alpha\in\Delta_+$. Pour les premi\`eres d\'eriv\'ees, on a
\begin{equation}
\label{WKone}
\Pi^{-1}\partial_{u^j}\Pi(u,\tau)=
	\sum_{\alpha>0}\alpha(h^j)\,\rho(u_\alpha),
\end{equation}
car $\Pi^{-1}\partial_{u_\alpha}\Pi(u,\tau)=\rho(u_\alpha)$
pour $\rho=\vartheta_1'/\vartheta_1$. Pour les d\'eriv\'ees
secondes, on se rappelle que $\Pi$ satisfait une \'equation du type de la chaleur,
de laquelle on peut d\'eduire
\begin{equation}
\label{WKtwo}
\begin{split}
4\pi i\,g^\vee\,\Pi^{-1}\partial_\tau\Pi(u,\tau)&=\Pi^{-1}\Delta_u\Pi(u,\tau)\\
	&=\sum_{\alpha>0}\tr(\alpha^2)\,\rho'(u_\alpha)
	+\sum_{\alpha,\beta>0}\tr(\alpha\beta)\,
	\rho(u_\alpha)\,\rho(u_\beta),\\
	&=g^\vee\,\Big(
	(d-3r)\,\eta_1+
	\sum_{\alpha>0}\frac{\vartheta_1''}{\vartheta_1}(u_\alpha)
	\Big)
\end{split}
\end{equation}
o\`u $\eta_1=-\frac{1}{6}\,\vartheta_1'''(0)/\vartheta_1'(0)$,
$d=\dim G$ et $r={\rm rang}\,G$.
La deuxi\`eme \'egalit\'e utilise l'\'equation~\eqref{WKone}. Pour obtenir
la troisi\`eme \'equation, on calcule d'abord $\vartheta''_1/\vartheta_1$
en $z$ et $\eta_1$ par passage \`a la limite en $0$, puis on compare avec
$4\pi i\,g^\vee\,\Pi^{-1}\partial_\tau\Pi(u,\tau)$ calcul\'e directement.
Au passage, on obtient
\[
-6\,\eta_1=\frac{\vartheta_1'''}{\vartheta_1'}(0)=\pi^2\,\biggl(
		24\sum_{\ell\geq 1}\frac{\ell q^\ell}{1-q^\ell}-1\biggr).
\]
$\eta_1$ est une constante standard de
la th\'eorie des fonctions elliptiques. Si $\zeta$ est, par d\'efinition,
donn\'ee par $\zeta'(z)=-\wp(z)$ alors
$2\,\eta_1=\zeta(z+1)-\zeta(z)=\zeta(1/2)$. On montre que $\eta_1=-\frac{1}{6}
\,\vartheta_1'''(0)/\vartheta_1'(0)$. Notons \'egalement les relations
suivantes~: $\zeta(z)=2\,\eta_1\,z+\rho(z)$ et $-\wp(z)=2\,\eta_1+\rho'(z)$.
Dans la litt\'erature, on trouve plut\^ot
\[
\frac{\vartheta_1'''}{\vartheta_1'}(0)=\pi^2\,\biggl(
		24\sum_{\ell\geq 1}\frac{q^\ell}{(1-q^\ell)^2}-1\biggr).
\]

On a les transformations~: $\rho(y+1)=\rho(y)$ et
$\rho(y+\tau)=\rho(y)-2\pi i$. 
Les d\'eveloppements de $\rho$ en $y=0$ nous seront aussi utiles
\[
\rho(y)=\frac{1}{y}+0+\cdots,\qquad \rho'(y)=-\frac{1}{y^2}-2\eta_1+\cdots.
\]

La fraction suivante
\[
\noyau_x(y)=\frac{\vartheta_1'(0)\,\vartheta_1(x+y)}{\vartheta_1(x)\,
\vartheta_1(y)}
\]
pr\'esente de bonnes propri\'et\'es~:
$\noyau_x(y+1)=\noyau_x(y)$ et $\noyau_x(y+\tau)=\ee^{-2\pi ix}\,\noyau_x(y)$.
De plus, $\noyau_x(y)$ est analytique sauf en $y=0$ o\`u son r\'esidu vaut un.
Pr\'ecis\'ement, on montre les d\'eveloppements en $y=0$ suivants
\[
\noyau_x(y)=\frac{1}{y}+\rho(x)+\cdots,\qquad
\noyau'_x(y)=-\frac{1}{y^2}+\quotient{1}{2}\,\frac{\vartheta_1''}
	{\vartheta_1}(x) +\eta_1+\cdots
\]

\medskip
\section{\'Etats de Chern-Simons 
sph\'eriques}

Sur la sph\`ere de Riemann, $\CA^{01}_0\equiv\{{}^h0=h^{-1}\de h, 
h\in\CG^\C\}$ forme un ouvert dense dans $\CA^{01}$ --- c'est
la strate semi-stable.
D'apr\`es l'invariance globale~\eqref{globalCS}, un \'etat $\Psi$ de CS
est enti\`erement d\'etermin\'e par sa valeur en z\'ero~:
\begin{equation}
\label{1111}
\Psi(h^{-1}\de h)=\ee^{kS(h)}\,\mathop{\otimes}\limits_\ell
	h(\xi_\ell)^{-1}\sous{R_\ell}\,\Psi(0).
\end{equation}
En reprenant cette \'equation avec $h$ constante,
on constate que $\Psi(0)$ est
un tenseur invariant sous l'action diagonale de $G$, \ie
$\Psi(0)\in\esprep^{\,G}$. On obtient ainsi un plongement
\[
\chsiSph\hookrightarrow\esprep^{\,G},
\]
en envoyant un \'etat $\Psi$ sur $\Psi(0)$. En particulier, $\chsiSph$ est
de dimension finie. On peut voir que $\Psi$ d\'efini 
sur $\CA^{01}_0$ par la relation (\ref{1111}) pour tous $\Psi(0)\in
\esprep^{\,G}$ est holomorphe (au sens de Fr\'echet). 
L'espace des \'etats de CS est donc
le sous-espace de $\esprep^{\,G}$ form\'e de telles fonctionnelles $\Psi$
s'\'etendant holomorphiquement sur tout $\CA^{01}$. 
Les conditions alg\'ebriques que l'on doit imposer sur $\Psi(0)$
pour assurer l'existence du prolongement sont 
les {\bol r\`egles de fusion}. Pour les \'etudier, on a besoin 
d'un peu plus d'information sur $\CA^{01}$. 

On peut toujours
\'ecrire localement un champ sous la forme
\[
A^{01}=\begin{cases}
		       h_1^{-1}\de h_1 , & \text{sur $D$},\\
		       h_2^{-1}\de h_2 , & \text{sur $D'$}.
		\end{cases}
\]
Le produit $h_1h_2^{-1}$ est un \'el\'ement du groupe des lacets $L\Gc$~;
c'est aussi un $1$-cocycle de fonction de transitions d'un $\Gc$-fibr\'e
holomorphe. L'orbite du champ nul correspond au fibr\'e trivial.
D'apr\`es le th\'eor\`eme de Birkhoff~\cite{pressley}, pour tout \'el\'ement 
$g\in L\Gc$, il existe $q^\vee\in Q^\vee$, le r\'eseau des coracines de $\lieg$,
et deux application $g_1$, $g_2$ holomorphes respectivement sur $D$ et $D'$, telles
que, sur le cercle $S^1$, on ait
\[
g(z)=g_1(z)\,\ee^{q^\vee\,\ln z}\,g_2(z).
\]
Il y a unicit\'e de $q^\vee$ modulo l'action du groupe de Weyl. Les lacets
pour lesquels $q^\vee=0$ forment un ouvert dense
et la strate de codimension $1$ correspond \`a $q^\vee=\phi^\vee$, la
coracine de la racine la plus grande $\phi$.
L'\'enonc\'e g\'eom\'etrique~\cite{grothendieck}
 du th\'eor\`eme de Birkhoff est~:
tout fibr\'e vectoriel holomorphe sur $\C P^1$ est isomorphe \`a $L^{a_1}
\oplus\cdots\oplus L^{a_n}$ --- $L$ \'etant un fibr\'e holomorphe ---
  la s\'equence d'entiers $(a_1,\cdots,
a_n)$ \'etant unique \`a permutation pr\`es.
On continue la discussion uniquement pour le groupe $G={\rm SU}_2$.
Tout $\Slc$-fibr\'e principal holomorphe
au-dessus de $\C P^1$ est une somme directe 
$L\oplus L^{-1}$, o\`u $L$ est un fibr\'e en droites holomorphe.
En d'autres termes, si $n$ d\'esigne le degr\'e de $L$, les fonctions de transition
d'un $\Slc$-fibr\'e principal holomorphe au-dessus de $\C P^1$ sont de la forme
\[
\begin{pmatrix}
	z^n & 0\\
	0   & z^{-n}
\end{pmatrix}.
\]
Les orbites dans $\CA^{01}/\CG^\C$ peuvent
donc \^etre \'enum\'er\'ees par les entiers $|n|$. On les note
$\CO_{|n|}$. L'orbite dense est $\CO_0$. L'ensemble des $\CO_n$ forme
une stratification de $\CA^{01}$ car
\[
\CU_{n_o}=\bigcup_{|n|\leq n_0}\CO_{|n|}
\]
est un sous-ensemble ouvert de $\CA^{01}$ et 
$\CO_{n_0}=\CU_{n_0+1}\setminus\CU_{n_0}$ 
est une sous-vari\'et\'e ferm\'ee de codimension $2n_0-1$ dans 
$\CU_{n_0}$. D'apr\`es le th\'eor\`eme de Hartogs~\cite{griffiths},
si $\Psi$ est holomorphe sur $\CU_{n_0}$, elle s'\'etend
en une application holomorphe sur $\CU_{n_0+1}$ d\`es
que la codimension de $\CU_{n_0+1}\setminus\CU_{n_0}$ est
sup\'erieure ou \'egale \`a deux. C'est toujours le cas
sauf si $n_0=1$. Par induction, il suffit donc de montrer  que
$\Psi$ est holomorphe sur $\CU_1=\CO_0\cup\CO_1$ pour obtenir une application
holomorphe sur tout $\CA^{01}$.

Dans~\cite{gaw91:zero}, il est construit
 une famille analytique de champs \`a un param\`etre
$\C\ni t\longmapsto A^{01}_t\in\CU_1$
intersectant transversalement $\CO_1$ \`a $t=0$.
Pr\'ecis\'ement, $A^{01}_t$ est donn\'e par
\[
A^{01}_t=\begin{cases}
       \qquad0, & \text{sur $D$},\\
		\begin{pmatrix}
		1 & 0\\
		-tz^{-1} &1
		\end{pmatrix}
		g_0^{-1}\de g_0
		\begin{pmatrix}
		1 & 0\\
		tz^{-1} &1      
		\end{pmatrix}, & \text{sur $D'$}
		\end{cases}
\]
o\`u $g_0$ est une fonction $\CC^\infty$ \`a valeurs 
dans ${\rm SL}_2$ qui, autour de l'\'equateur, vaut
\[
g_0=\begin{pmatrix}
	z^{-1}&0\\
	0&z
	\end{pmatrix}.
\]
Si $t=0$, $A^{01}_0$ est dans la strate de codimension un. 
Par contre, si $t\neq 0$, 
le champ $A^{01}_t$ est dans l'orbite dense car il peut se mettre sous
la forme $h_t^{-1}\de h_t$, avec
\[
h_t=\begin{cases}
 \qquad      \begin{pmatrix}
		1&t^{-1}z\\
		0 & 1
	\end{pmatrix}, & \text{sur $D$},\\
		\begin{pmatrix}
		0 & t^{-1}\\
		-t & z^{-1}
		\end{pmatrix}
		g_0
		\begin{pmatrix}
		1 & 0\\
		tz^{-1} &1      
		\end{pmatrix}, & \text{sur $D'$},
		\end{cases}
\]
fonction $\CC^\infty$ sur $\C P^1$ \`a valeurs dans ${\rm SL}_2$.
Une fonctionnelle $\Psi$ s'\'etend
holomorphiquement sur $\CU_1$ si, et seulement si,
\[
t\longmapsto \Psi(A^{01}_t)
\]
est holomorphe en $t=0$. Pour expliciter compl\`etement cette 
propri\'et\'e, on utilise la repr\'esentation 
(coh\'erente~\cite{perelomov}) de spin $j$ de ${\rm SL}_2$, 
\cad les polyn\^omes $P$ de degr\'e $\leq 2j$, de variable 
$u$, sur lesquels ${\rm SL}_2$ agit par
\[
\begin{pmatrix}
	a&b\\
	c&d
\end{pmatrix}^{-1}_j P(v)
	=(cv+d)^{2j}\,P\Big(\frac{av+b}{cv+d}\Big).
\]
Les g\'en\'erateurs (infinit\'esimaux) de ${\rm SL}_2$ sont 
\begin{equation}
\begin{split}
t^1_j &= \quotient{1}{2}\,(v^2-1)\,\partial_v-jv,\\
t^2_j &=\quotient{i}{2}\,(v^2+1)\,\partial_v-ijv,\\
t^3_j &= -v\,\partial_v+j.
\end{split}
\end{equation}
Le produit scalaire rendant l'action de ${\rm SU}_2$ unitaire est donn\'e par
\[
||P||^2=\quotient{2j+1}{\pi}\int_\C\quotient{|P(v)|^2}
	{(1+|v|^2)^{2j+2}}\,d^2v.
\]
Les tenseurs invariants dans l'action diagonale de ${\rm SL}_2$ sont alors les
polyn\^omes en $\underline{v}=(v_\ell)$, de degr\'e $\leq j_\ell$ en $v_\ell$,
et tels que
\[
P(\underline{v})=\prod_\ell(cv_\ell+d)^{2j_\ell}\,P
	\Big(\Big(\frac{av_\ell+b}{cv_\ell+d}\Big)\Big).
\]
Notons que $P$ est un polyn\^ome homog\`ene de degr\'e 
$|\underline{j}|=\sum_\ell j_\ell$, invariant par translation.

Commen\c{c}ons par supposer que tous les points d'insertion sont
dans $D$.
D'apr\`es l'invariance globale des \'etats $\Psi$, on a
\[
\Psi(A^{01}_t)=\ee^{kS(h_t)}
	\,P(\underline{v}+\quotient{1}{t}\,\underline{\xi})
	=t^{-|\underline{j}|}\,\ee^{kS(h_t)}\,
	P(\underline{\xi}+t\underline{v})
\]
o\`u $P$ repr\'esente $\Psi(0)$.
Partant de l'\'equation~(2.\ref{varS}), on peut montrer que
$\ee^{kS(h_t)}\sim t^k$ quand $t\mapsto 0$.
L'application $t\mapsto \Psi(A^{01}_t)$ est donc
holomorphe en z\'ero si, et seulement si,
$P(\underline{\xi}+t\underline{v})$ s'annule \`a l'ordre
$|\underline{j}|-k-1$ en $t$. L'espace $\chsiSph$ est invariant
par transformation de M\"obius --- le groupe d'automorphismes
de $\C P^1$ --- donc, les points d'insertion
\'etant deux \`a deux distincts, la restriction de 
d\'epart $\xi_\ell\in D$ n'est pas importante. 
La condition d'annulation est \'egalement invariante par transformation de 
M\"obius. Par contre le raisonnement ne marche que si les points d'insertion sont
diff\'erents de l'infini.
Ainsi, si aucun des points d'insertion n'est \`a l'infini, 
\begin{equation}
\chsispin=\Big\{P\in\esprepspin^{\,{\rm SU}_2}\ \big|\ 
	D^{\underline{n}}P(\underline{\xi})=0, \ |\underline{n}|
	\leq |\underline{j}|- k -1\Big\}
\end{equation}
o\`u $D^{\underline{n}}=\prod_\ell\partial_{v_\ell}^{n_\ell}$, 
$\underline{n}=(n_\ell)$ et $|\underline{n}|=\sum_\ell n_\ell$.

Si un des points d'insertion est
\`a l'infini, par exemple $\xi_1=\infty$, on d\'eduit le
r\'esultat de la configuration pr\'ec\'edente. Supposons
que les autres points d'insertion sont non-nuls. On peut se ramener
\`a l'\'etude pour la s\'equence $\un\zeta$, avec $\zeta_1=0$ et
$\zeta_\ell=-\xi_\ell^{-1}\neq \infty$, puisque $\CW(\C P^1,\underline{\xi},\underline{j})=
\CW(\C P^1,\underline{\zeta},\underline{j})$. On sait que
$t^{-|\underline{j}|}\,\ee^{kS(h_t)}\,
	P(\underline{\zeta}+t\underline{v})$
est r\'egulier en z\'ero si, et seulement si,
\[
D^{\underline{n}}P(\underline{\zeta})=0 \qquad 
	\mbox{pour} \qquad |\underline{n}|\leq |\underline{j}|- k -1.
\]
Pour obtenir un crit\`ere portant sur les d\'eriv\'ees en
$(0,\xi_2,\cdots, \xi_N)$, on utilise 
\begin{equation}
\label{babaaurhum}
\begin{split}
D^{\underline{n}}P\big|_{\underline{\zeta}}&=
	D^{\underline{n}}\prod_\ell v_\ell^{j_\ell}
	P\Big(\Big(\quotient{-1}{v_\ell}\Big)\Big)
	\Big|_{\underline{\zeta}}\\
&=\quotient{n_1!}{(2j_1-n_1)!}\,\partial^{2j_1-n_1}_{v_1}
	\partial^{n_2}_{u_2}\cdots\partial^{n_N}_{u_N}\,
	\prod_{\ell\neq 1}v_\ell^{2j_\ell}\,
	P\Big(-v_1,\quotient{-1}{v_2},\cdots,\quotient{-1}{v_N}\Big)
	\Big|_{\underline{\zeta}}
\end{split}
\end{equation}
Par cons\'equent, la condition cherch\'ee est~: 
$D^{\underline{n}}P(0,\xi_2,\cdots,\xi_N)=0$, pour
$|\underline{n}| +2(j_1-n_1)\leq |\underline{j}|- k -1$. 
La restriction $\xi_\ell\neq 0$, $\ell\neq 1$, n'est pas importante car
l'espace est invariant par translation, cette remarque s'appliquant aussi
\`a la condition trouv\'ee. 
Ainsi, les calculs pr\'ec\'edents conduisent \`a la description suivante
de l'espace des \'etats de CS
\begin{equation}
\label{CSinfty}
\chsispin=\Big\{P\in\esprepspin^{\,{\rm SU}_2}\ \big|\
	D^{\underline{n}}P(0,\xi_2,\cdots,\xi_N)=0, \ |\underline{n}|
	+2(j_1-n_1)
	\leq |\underline{j}|- k -1\Big\}
\end{equation}
si $\xi_1=\infty$. 
De ces r\'esultats, on peut extraire $\chsispin=\big\{0\big\}$
si un des spins est $> k/2$. On peut toujours supposer que
$j_1>k/2$ et $\xi_1=\infty$. Comme $P$ est un polyn\^ome homog\`ene de
degr\'e $|\un j|$, les d\'eriv\'ees $D^{|\un n|}P$ sont nulles
pour $|\un n|> |\un j|$. D'autre part, si $n_1=2j_1$ et 
$|\un n|\leq |\un j|$, on a d'apr\`es l'\'equation~\eqref{CSinfty}
$D^{\underline{n}}P(0,\xi_2,\cdots,\xi_N)=0$. Par
cons\'equent $\partial^{2j_1}_{u_1}P$ est identiquement
nulle en $u_1=0$. Vu l'\'equation~\eqref{babaaurhum}, cela entra\^\i ne
que $P$ est nulle en $u_1=0$. Comme $P$ est invariant par translation
il suit $P\equiv 0$. Ainsi, on a la {\bol r\`egle d'exclusion}
suivante~: $\chsispin=\{0\}$ si un des
spins est $> k/2$, \ie si une des repr\'esentations
$R_\ell$ n'est pas int\'egrable~; on verra 
que les fonctions de Green sont automatiquement nulles (cf.
\'equation~\eqref{sesquiCS}). Ce r\'esultat s'accorde avec 
la construction de l'espace des \'etats de la section 1.{\bf 8}.

De ce qui pr\'ec\`ede, on d\'eduit les espaces explicites pour deux et trois points
d'insertion. Les espaces $\chsispin$ ne doivent pas d\'ependre 
des points d'insertion \`a cause de l'invariance
de M\"obius. Pour deux points d'insertion, on a
$\esprepspin^{\,{\rm SU}_2}=\C(u_1-u_2)^{k}$, si $j_1=j_2=k/2$,
et $\{0\}$ sinon. Ainsi
\[
\CW(\C P^1;\xi_1,\xi_2;j_1,j_2)=\begin{cases}
		       \esprepspin^{\,{\rm SU}_2}, & \mbox{si $j_1=j_2=k/2$},\\
		       \big\{0\big\}, & \text{sinon}.
		\end{cases}
\]
Pour trois points d'insertion, l'espace des tenseurs
invariants est $\C\big\{j_1\,j_2\,j_3\big\}$
si $j_{12},j_{13},j_{23}$ sont des entiers non-n\'egatifs et z\'ero
sinon. On a not\'e $\big\{j_1\,j_2\,j_3\big\}$ 
les polyn\^omes de Clebsch-Gordan~:
$\big\{j_1\,j_2\,j_3\big\}=
	(u_1-u_2)^{j_{12}}\,(u_1-u_3)^{j_{13}}\,(u_2-u_3)^{j_{23}}$,
o\`u $j_{12}=j_1+j_2-j_3$, $j_{13}=j_1-j_2+j_3$ et $j_{23}=-j_1+j_2+j_3$.
Il suit
\[
\CW(\C P^1;\xi_1,\xi_2,\xi_3;j_1,j_2,j_3)=\begin{cases}
		       \esprepspin^{\,{\rm SU}_2}, & \mbox{si $j_1+j_2+j_3\leq k$},\\
		       \big\{0\big\}, & \text{sinon}.
		\end{cases}
\]

Dans la section 2.{\bf 1.2}, on a remarqu\'e que la racine la plus grande
d\'efinit un plongement de $\mathfrak{sl}_2$ dans $\liegc$ et de 
${\rm SU}_2$ dans $G$. On peut alors d\'ecomposer $V_\lambda$ 
via l'action de ${\rm SU}_2$ (cf. la section 1.{\bf 5})
\[
V_{{\lambda_G}}\cong \bigoplus_{j} M^{\lambda_G}
	_{j}\otimes V_{j}.
\]
Un tenseur invariant pour $G$ l'est en particulier pour
${\rm SU}_2$, donc
\[
\esprep^{\,G}\subset \bigoplus_{{\un j}}
	\big(\mathop{\otimes}\limits_\ell
	 M^{\lambda_\ell}_{j_\ell}\big)
	\otimes \esprepspin^{\,{\rm SU}_2}.
\]
Pour trouver quels sont les \'etats de CS parmi
ces tenseurs, il suffit de recommencer l'\'etude pr\'ec\'edente
en regardant la famille \`a un param\`etre $t\mapsto A^{01}_t$
o\`u les champs sont vus comme prenant leurs valeurs dans $\liegc$.
Il suit
\begin{equation}
\label{depthrule}
\CW(\C P^1;\un\xi;\un R)=
	\esprep^{\,G}\cap
	\Big(\bigoplus_{{\un j}}
	\big(\mathop{\otimes}\limits_\ell
	 M^{\lambda_\ell}_{j_\ell}\big)
	\otimes\CW(\C P^1;\un\xi;\un j)\Big).
\end{equation}
Ainsi, chaque composante de $\Psi$ doit correspondre \`a un \'etat de CS
pour ${\rm SU}_2$~\cite{gepner1,gepner2}. \`A l'aide du cas ${\rm SU}_2$,
on obtient la m\^eme r\`egle d'exclusion~: l'espace des \'etats de CS
est r\'eduit \`a z\'ero si une des repr\'esentations 
n'est pas int\'egrable.

\medskip
\section{\'Etats de Chern-Simons elliptiques}

En genre un, on peut utiliser $E_\tau=\C/(\Z+\tau\Z)$ pour repr\'esenter
$\Sigma$. Il n'y a pas de strates de codimension $1$, donc il suffit
de regarder la strate semi-stable $\CA^{01}_{ss}$.
L'analyse de la strate semi-stable requiert une stratification plus
fine de $\CA^{01}_{ss}$. Les d\'etails peuvent \^etre trouv\'es
dans~\cite{gaw94:genus1,gaw89:coset}.
Dans chaque orbite, il existe une connexion plate~\cite{gunning:vb}, \ie
il existe $h\in\CG^\C$ telle que $\wih A={}^hA$ soit plate.
Remarquons qu'il n'y a pas unicit\'e de $\wih A$.
Regardons le transport parall\`ele $\gamma$ de $\wih A$~:
\[
\gamma(\widex) =\expordonne^{\int_{0}^\widex \wih A}
\]
qui est une application du rev\^etement universel de $\Sigma$ dans $\Gc$. 
On peut toujours conjuguer l'holonomie, cette op\'eration
donnant un fibr\'e isomorphe au fibr\'e initial.
Le groupe fondamental \'etant engendr\'e par deux \'el\'ements commutants,
l'holonomie est donn\'ee par une paire de matrices 
commutantes $(\gamma_1, \gamma_2)$ telles que
$\gamma(z+1)=\gamma(z)\,\gamma_1$ et $\gamma(z+\tau)
=\gamma(z)\,\gamma_2$.
Les champs $A^{01}$ sont donc de la forme $(\gamma h)^{-1}\de(\gamma h)$,
cette repr\'esentation \'etant unique modulo la multiplication \`a gauche 
de $\gamma h$ par un \'el\'ement constant de $\CG^\C$.

Si on peut conjuguer
simultan\'ement les deux matrices $\gamma_i$ \`a un \'el\'ement de $T\dC$,
on peut se ramener \`a $\gamma_1=\ee^{-2\pi i \Phi}$ et 
$\gamma_2=\ee^{-2\pi i \Theta}$, avec $\Phi, \Theta\in\lietc$. 
On utilise la libert\'e laiss\'ee dans
le choix de $\wih A$ pour prendre $\gamma'\gamma$ avec $\gamma'=\ee^{2\pi i\Phi z}$.
L'holonomie est alors donn\'ee par $\gamma_1=1$ et $\gamma_2=
\ee^{-2\pi i u}$, avec $u=\Theta-\tau \Phi\in\lietc$.
Pour le champ plat, on utilise 
\[
\wih A_u=A^{10}_u+A^{01}_u=\quotient{\pi u}{\imtau}(dz-d\overline{z}).
\]
Introduisons les applications
\[
h_v=\ee^{\pi(v\bar z-\bar v z)/\imtau}.
\]
Les $h_v$ sont dans $\CG^\C$ si elles sont monovalu\'ees, ce qui
n'arrive que pour $v\in Q^\vee+\tau Q^\vee$.
En tous les cas, on a ${}^{h_v^{-1}}A^{01}_u=A^{01}_{u+v}$.
Si $u\not\in P^\vee+\tau P^\vee$, les seules transformations 
de jauge reliant 
les $A^{01}_u$ sont $h=wh_v$ o\`u $v\in Q^\vee+\tau Q^\vee$ 
et $w\in W$, le groupe de Weyl. L'ensemble de ces champs 
forment une strate $\CA^{01}_0$ ouverte dense dans
$\CA^{01}_{ss}$ et donc dans $\CA^{01}$. 
Cette section mise \`a part, on s'int\'eresse surtout
\`a la strate $\CA^{01}_0$. Notons que $A^{01}_u$ 
s'\'ecrit aussi $\gamma_u^{-1}\de\gamma_u$
avec $\gamma_u$ la fonction multivalu\'ee, donn\'ee par
\begin{equation}
\label{gammau}
\gamma_u=\ee^{-\pi u(z-\overline{z})/\imtau},\qquad
\gamma_u(z+1)=\gamma_u(z),
	\qquad\gamma_u(z+\tau)=\ee^{-2\pi i u}\,\gamma_u(z).
\end{equation}
Presque tous les champs de jauge sont donc de la forme
\[
A^{01}={}^{h^{-1}}A^{01}_u=(\gamma_uh)^{-1}\de(\gamma_uh)
\]
avec $u_\alpha=\alpha(u)\not\in \Z+\tau \Z$, quel que soit $\alpha\in\Delta$.
On v\'erifie que la seule ambigu\"\i t\'e restante r\'eside dans la multiplication
\`a gauche par une transformation de jauge constante \`a valeurs dans le sous-groupe
de Cartan $T\dC$.  Si $u\in P^\vee+\tau P^\vee$, on obtient des
strates disjointes de codimension $>1$.
Si les matrices de l'holonomie ne sont pas conjugables
simultan\'ement \`a $T^\C$, les strates sont
de codimension $\geq 1$. 

\`A tout \'etat
de CS, on associe une application holomorphe $\gamma:\lietc
\rightarrow\esprep$ d\'efinie par
\[
\Bgamma(u)=\ee^{-\pi k\,|u|^2/(2\imtau)}\otimes_\ell \big(
	\ee^{-\pi(z_\ell-\overline{z}_\ell)u/\imtau}\big)
	\sous{\hspace{-0.1cm}R_\ell}\Psi(A^{01}_u)
\]
o\`u $|u|^2=\tr\,u^2$.
Au niveau de $\Bgamma$, les conditions assurant que $\Psi$ est 
un \'etat de CS sont
\label{CSgenreun}
\begin{alignat}{2}
\label{CSone}
\Bgamma(u+q^\vee) & =  \Bgamma(u), & & \quad\text{si}\ q^\vee\in Q^\vee,
	\tag{\ref{CSone}.a}\\
\Bgamma(u+\tau q^\vee) & =  \ee^{-\pi i k\,\tr\,q^\vee(\tau q^\vee
	+2u)}\mathop{\otimes}_\ell\Big(\ee^{-2\pi i\,z_\ell\, q^\vee}\Big)
	\sous{\hspace{-0.1cm}R_\ell}\Bgamma(u),
	 & & \quad\text{si}\ q^\vee\in Q^\vee,\tag{\ref{CSone}.b}\\
0 &= \sum_\ell h_\ell\,\Bgamma(u),
	& & \quad\text{si}\ h\in\liet\tag{\ref{CSone}.c},\\
\Bgamma(wuw^{-1})&=\mathop{\otimes}_\ell w\sous{R_\ell}\Bgamma(u), & & \quad
	\text{si}\ w\in N(T)\tag{\ref{CSone}.d},
\end{alignat}
\addtocounter{equation}{1}
et
\begin{equation}
\Big(\sum_\ell\ee^{2\pi is\, z_\ell}(e_\alpha)\sous{R_\ell}\Big)^p
	\Bgamma(u+tp^\vee) = \CO(t^p),
\end{equation}
pour tout $\alpha\in\Delta$, $p^\vee\in Q^\vee$ avec $\alpha(p^\vee)=1$, $u$
tel que $u_\alpha=m+\tau s$ pour $m,s\in\Z$, $p=1,2,\cdots$ et $t\mapsto 0$. 
Les \'equations~(\ref{CSone}.a,b,d) sont la traduction de l'action
de $W\rtimes (Q^\vee+\tau Q^\vee)$. L'\'equation~(\ref{CSone}.c)
refl\`ete l'ambigu\"\i t\'e de notre param\'etrisation. Elle dit aussi
que $\gamma$ est dans $V_{\underline{\lambda}}^T$. La derni\`ere
\'equation assure la possibilit\'e d'\'etendre $\gamma$ 
aux strates de codimension $1$.
Pour $G={\rm SU}_2$, elles sont explicitement donn\'ees dans 
l'article~\cite[\'eq. (4.10)]{gaw94:genus1}.

Sans points d'insertion, les \'etats de CS sont donc
obtenus \`a partir des applications $\gamma$ holomorphes satisfaisant
les \'equations ~(\ref{CSone}.a-b). Une base de solutions est fournie
par les caract\`eres $\chi^k_\lambda(u,\tau)$ de l'alg\`ebre
de Kac-Moody affine $\widehat{L\lieg}\eC$. En particulier,
la dimension de $\CW(E_\tau,\emptyset,\emptyset)$ est \'egale
au nombre de poids dominants int\'egrables de niveau $k$.

\medskip
\section{\'Etats de Chern-Simons en genre sup\'erieur}

La situation en genre $g\geq 2$ est nettement plus difficile.
On ne regarde que le cas $G={\rm SU}_2$ sans points d'insertion.
On s'int\'eresse donc aux fibr\'es vectoriels holomorphes $E$ de rang $2$ et
de d\'eterminant trivial.
L'\'etude n\'ecessite une certaine familiarit\'e avec le langage
des extensions de fibr\'es~\cite{atiyah:extensions,nar-ram}, voir aussi
\cite{bertram0,thaddeus}.

Soient $E'$ et $E''$ deux fibr\'es vectoriels holomorphes au-dessus de $\Sigma$. Une {\bol extension}
de $E''$ par $E'$ est un fibr\'e vectoriel holomorphe $E$ au-dessus de $\Sigma$ tel que la suite
$0\rightarrow E'\rightarrow E\rightarrow E''\rightarrow 0$ soit exacte. Deux extensions de $E''$
par $E'$ sont {\bol \'equivalentes} s'il existe une application $g:E\rightarrow F$ telle que le
diagramme suivant
\begin{equation*}
\begin{matrix}
0 & \rightarrow & E' & \rightarrow & E & \rightarrow & E'' & \rightarrow & 0\\
  &             & \Vert &         & \downarrow & & \Vert & & \\
0 & \rightarrow & E' & \rightarrow & F & \rightarrow & E'' & \rightarrow & 0\\
\end{matrix}
\end{equation*}
soit commutatif. En vertu du lemme des cinq~\cite{rotman}, 
$g$ est n\'ecessairement un isomorphisme.
On dit qu'une s\'equence $0\rightarrow E'\stackrel{i}{\rightarrow} E
\stackrel{p}{\rightarrow} E''\rightarrow 0$ est {\bol triviale} 
(\<<split\>>) si l'une des trois conditions suivantes est remplie  ~: 
(1) $p$ admet un inverse \`a droite, (2) $i$ admet
un inverse \`a gauche, (3) la s\'equence est \'equivalente \`a 
$0\rightarrow E'\rightarrow E'\oplus E''\rightarrow E''\rightarrow 0$. 
Si $\delta$ est l'application qui envoie $H^0(\text{Hom}(E'',E''))$ dans 
$H^1(\text{Hom}(E'',E'))$, l'image de l'identit\'e par $\delta$ est appel\'ee 
la classe d'extension, not\'ee $\delta(E)$. On obtient alors une correspondance
bijective entre l'ensemble des classes d'\'equivalence d'extensions de $E''$ 
par $E'$ et $H^1(\text{Hom}(E'',E'))$. En particulier, si $\ell$ est un 
fibr\'e en droites, $E'=\ell^{-1}$, $E''=\ell$, alors les extensions sont 
\'enum\'er\'ees par $H^1(\ell^{-2})$. D'apr\`es le th\'eor\`eme
de Riemann-Roch, 
\[
\dim H^1(\ell^{-2})=g-1+2\,\deg \ell
\]
si $\deg\ell>0$.

On fixe une bonne fois pour toutes un fibr\'e en droites holomorphe $L$ de
degr\'e $g$. On suppose que $L(-x)^2\not\cong K$. 
D'apr\`es le lemme 5.5 de~\cite{nar-ram}, si $E$ est un fibr\'e
vectoriel holomorphe de rang $2$ et de d\'eterminant trivial, il existe
$x\in\Sigma$ tel que $H^0({\rm Hom}(L(-x)^{-1},E))\neq 0$. Notons $\phi$
un tel homomorphisme. Soient $x_1,\cdots,x_r$ les z\'eros de $\phi$, compt\'es
avec leurs multiplicit\'e, alors $\phi$ induit un plongement de
$L(-x-x_1-\cdots-x_r)^{-1}$ dans $E$. Notons $X_r=x+x_1+\cdots+x_r$.
Tout en dualisant, $E$ devient une extension de $L(-X_r)$ par son dual $L(-X_r)^{-1}$. On a
$\deg L(-X_r)=g-1-r$. Si on suppose en plus que $E$ est stable,
n\'ecessairement $0\leq r<g-1$.

Soit $\CL_r$ une famille $(L(-X_r))$ de fibr\'es 
holomorphes en droites. C'est un fibr\'e holomorphe en droites 
au-dessus de $\Sigma^{r+1}\times\Sigma$~(\footnote{$\Sigma^{r+1}$ est
le produit cart\'esien sym\'etris\'e. On note $pr_1$ la projection
sur le premier facteur pour le fibr\'e $\CL_r$.}),
dont la restriction \`a $pr_1^{-1}(\{x,x_1,\cdots,x_r\})$
est isomorphe \`a $L(-X_r)$, \cad
\[
\CL_r\big|\sous{\{x,x_1,\cdots,x_r\}\times\Sigma}\cong L(-X_r).
\]
Il n'y a pas unicit\'e de la famille.
On peut prendre $pr_1^*(M)\,\CL_r$ pour tout fibr\'e holomorphe en droites
$M$ au-dessus de $\Sigma^{r+1}$. D'une certaine mani\`ere, on souhaite 
construire une famille de $H^1(L(-X_r))$. 
L'objet math\'ematique qui r\'esout le probl\`eme est l'{\bol
image directe}.
Posons $W_r=R^1pr_{1*}(\CL_r^{-2})$,
la premi\`ere image directe de $\CL_r^{-2}$ par $pr_1$. C'est un
fibr\'e vectoriel holomorphe au-dessus de $\Sigma^{r+1}$
de fibre $H^1(L(-X_r)^{-2})$ au-dessus de $x,x_1,\cdots,x_r$~:
\[
R^1pr_{1*}(\CL_r^{-2})\big|\sous{\{x,x_1,\cdots,x_r\}}
=H^1(\CL^{-2}_r\big|\sous{
	\{x,x_1,\cdots,x_r\}\times\Sigma})
	\cong H^1(L(-X_r)^{-2})
\]
qui est de dimension $3g-3-2r$.
Par $\PP W_r$, on entend
le m\^eme fibr\'e mais avec $\PP H^1(L(-x)^{-2})$. La
dimension totale de $\PP W_r$ est $3g-3-r$.
Pour $G={\rm SU}_2$, la dimension de l'espace des modules
est $3g-3$. G\'en\'eriquement, $r=0$, \ie $E$ est  
(\`a isomorphisme pr\`es) une extension de $L(-x)$ par $L(-x)^{-1}$.
On sait alors (loc. cit.) que le sous-espace de $\PP W_0$ constitu\'e des points
correspondant \`a des fibr\'es stables est un rev\^etement 
ramifi\'e \`a $2g$ feuillets d'un sous-espace dense de $\CN_s$. 
On verra deux constructions pour $\CL_0$. L'une est due
\`a Bertram~\cite{bertram0}, voir aussi~\cite{thaddeus},
l'autre est due \`a Gaw\c{e}dzki~\cite{gaw95:higher1}. Bien entendu, c'est
la deuxi\`eme qui est importante pour nous puisqu'elle associe aux points de
$\PP W_0$ un champ de jauge, \cad une d\'ecoupe $s:\PP W_0\mapsto \CA^{01}$,
rencontrant $2g$ fois une orbite g\'en\'erique.

Fixons un point $x_0$ de $\Sigma$ et une base normalis\'ee
de formes holomorphes (cf. section {\bf 3.5}). On note $L_0=L(-x_0)$.
D\'efinissons une $(0,1)$-forme $a_\widex$ par
\[
a_\widex\equiv\pi\,\Big(\int_{x_0}^\widex\omega\Big)\,
	\imtau^{-1}\,\overline{\omega}.
\]
Par construction, $a_\widex$ d\'epend du rel\`evement $\widex$ de $x$ \`a
$\widetilde{\Sigma}$. Notons $L_\widex$ le fibr\'e en droites topologiquement
\'equivalent \`a $L_0$ muni de la structure holomorphe donn\'ee
par l'op\'erateur $\de_{L_\widex}\equiv\de+a_\widex$. Les fibr\'es
$L_\widex$ pour des rel\`evements du m\^eme point de $\Sigma$
sont tous isomorphes \`a $L(-x)$~(\footnote{
Le fibr\'e $\CO(x_o-x)$ est isomorphe au fibr\'e trivial 
$\Sigma\times\C^2$ muni
de la structure holomorphe $\de+a_\widex$.}). 
On consid\`ere ensuite le fibr\'e
$\wi \Sigma\times L_0$ au-dessus de $\wi \Sigma\times \Sigma$
muni de la structure holomorphe induite par l'op\'erateur
$\bar\delta+\de$, o\`u $\bar\delta$ est la d\'eriv\'ee 
dans la direction triviale
de $\wi \Sigma$. On note $\wi\CL_0$ le fibr\'e pr\'ec\'edent 
obtenu en tordant
la structure holomorphe par $\bar\delta+\de+a_\widex$. On a
\[
\wi\CL_0\big|\sous{\{\wi x\}\times\Sigma}= L_\widex\cong L(-x).
\]
Pour obtenir un
fibr\'e au-dessus de $\Sigma\times\Sigma$, on rel\`eve l'action
du groupe fondamental \`a $\wi\CL_0$. Soit $p\in\pi_1\equiv\pi_1(\Sigma)$.
\[
(\widex,\ell_y)\longmapsto(p\widex,c_p(y)^{-1}\ell_y)
\]
o\`u $\ell_y$ est un \'el\'ement de la fibre $L_\widex$ au-dessus
de $y\in\Sigma$ et $c_p$ est la fonction
monovalu\'ee sur $\Sigma$ donn\'ee par
\[
c_p(y)=\ee^{2\pi i\,{\rm Im}\,\Big[\left(\mathop{\int}\limits_p
	\omega\right)\,\imtau^{-1}\,
	\left(\mathop{\int}\limits_{x_0}^y\overline{\omega}\right)\Big]}.
\]
L'action pr\'eserve la structure holomorphe, \ie si $s$ est une section holomorphe
de $\wi\CL_0$ au-dessus de $(\widex,y)$, $c_p(y)^{-1}\,s(p^{-1}\widex,y)$
l'est aussi. En quotientant $\wi\CL_0$ par l'action de $\pi_1$, on obtient
un fibr\'e holomorphe en droites $\CL_0$ au-dessus de $\Sigma\times \Sigma$,
r\'ealisant ainsi explicitement la famille $(L(-x))$.

Soit $\wi W_0$ la premi\`ere image directe de $\wi\CL_0^{-2}$ par la premi\`ere
projection. On peut \'egalement r\'ealiser $\wi W_0$ par
une construction \<<Dolbeaut\>>, en consid\'erant le quotient
du fibr\'e trivial $\wi \Sigma\times\CE^{0,1}(L_0^{-2})$
par le sous-fibr\'e dont la fibre au-dessus de $\{\widex\}\times\Sigma$ est 
$\de_{L_\widex^{-2}}\Gamma(L_0^{-2})$, o\`u $\de_{L_\widex^{-2}}
=\de-2a_\widex$ --- c'est juste l'isomorphisme
entre $H^1(L_\widex^{-2})$ et $H^{0,1}_{{\rm Dol}}(L_\widex^{-2})$.
On obtient $W_0$ en quotientant $\wi W_0$ par l'action du groupe fondamental.

Pour obtenir la d\'ecoupe $s$, on introduit un isomorphisme $\CC^\infty$ entre
$L^{-1}_0\oplus L_0$ et le fibr\'e trivial $\Sigma\times \C^2$~:
\[
U:L^{-1}_0\oplus L_0\rightarrow \Sigma\times \C^2.
\]
On note $E$ le fibr\'e topologiquement \'equivalent \`a $L^{-1}_0\oplus L_0$
muni de la structure holomorphe induite par 
\[
\de+B^{01}_{\widex,b},\qquad B^{01}_{\widex,b}=
	\begin{pmatrix}
	-a_\widex & b\\
	0 & a_\widex
	\end{pmatrix}
\]
o\`u $b\in\CE^{0,1}(L_0^{-2})$. Le fibr\'e $E$ est une extension de 
$L_\widex$ par $L_\widex^{-1}$. L'op\'erateur
$\de+A^{01}_{\widex,b}$, avec
\[
A^{01}_{\widex,b}=UB^{01}_{\widex,b}U^{-1}+U\de U^{-1},
\]
d\'efinit la structure holomorphe sur le fibr\'e trivial
induite via l'isomorphisme $U$ par la structure holomorphe de $E$.
Soit $c$ une constante non-nulle ou $c=c_p$. Pour $v\in\Gamma(L_0^{-2})$, on
pose 
\[
g_{c,v}=\begin{pmatrix}
	c^{-1} & cv\\
	0 & c
	\end{pmatrix}.
\]
Cette derni\`ere est une section $\CC^\infty$ de ${\rm Aut}(L_0^{-1}\oplus L_0)$.
La transformation de jauge $B^{01}_{\widex,b}\mapsto {}^{g_{c,v}^{-1}}
B^{01}_{\widex,b}$ qui en suit pr\'eserve la forme du champ tout 
en changeant ses composantes par
\[
a_\widex\mapsto a_\widex+c^{-1}\de c,\qquad b\mapsto c^2(b+(\de-2a_\widex)v).
\]
Quant aux champs $A^{01}_{\widex,b}$ ils sont reli\'es par 
la transformation de jauge 
\begin{equation}
\label{jauge}
h_{c,v}=Ug_{c,v}U^{-1}\in\CG^\C.
\end{equation}
En particulier, pour $c=c_p$, $a_\widex$ devient $a_{p\widex}$. 
Avec $c$ constante non-nulle, on ne change pas
$a_\widex$ et la classe de $b$ dans $\PP H^1(L_\widex^{-2})$. 
La classe de $b$ d\'ecrit les fibr\'es $E$ de rang $2$ et de d\'eterminant
trivial, extensions de $L_\widex$, associ\'es \`a l'orbite de $A^{01}_{\widex,b}$.
En choisissant $\widex$ dans un domaine fondamental de $\wi \Sigma$ et un
repr\'esentant $b$ dans chaque classe de $\PP H^1(L_\widex^{-2})$, 
on obtient la d\'ecoupe $s:\PP W_0\rightarrow \CA^{01}$.

Choisissons une structure hermitienne sur $L_0$. Elle induit
une structure hermitienne sur $L^{-1}_0\oplus L_0$ et une connexion 
m\'etrique $\nabla$.  On peut supposer
que l'isomorphisme $U$ transporte la m\'etrique sur $L_0^{-1}\oplus
L_0$ vers la structure hermitienne standard sur $\Sigma\times\C^2$. 
La connexion m\'etrique sur $\Sigma\times\C^2$ induite par $U$ est alors
\[
U\nabla U^{-1}=\nabla+U\nabla^{10}U^{-1}+U\de U^{-1}
	=\nabla+A_0
\]
o\`u $\de$ est l'op\'erateur anti-holomorphe pour $L_0^{-1}\oplus L_0$. 
La connexion $A_0$ est unitaire. 

On associe \`a tout \'etat de CS une application
holomorphe $\psi$
\[
\psi(\widex,b)\equiv\ee^{\frac{ik}{2\pi}\mathop{\int}\limits_\Sigma
	\tr\,A^{10}_0\wedge \Axb}\,\Psi(\Axb)
\]
avec $\widex\in\wi\Sigma$ et $b\in\CE^{01}(L_0^{-2})$. 
Seule la normalisation de $\psi$ d\'epend du choix de la structure hermitienne
sur $L_0$ et du choix de $U$~\cite[App. F]{gaw95:higher1}. La fonction
$\Psi$ est enti\`erement d\'etermin\'ee par $\psi$, puisque les orbites
des $A^{01}_{\widex,b}$ forment un sous-espace dense de $\CA^{01}$.
Par transformation de jauge sur $A^{01}_{\widex,b}$, on obtient des contraintes
sur $\psi$. Prenant $h_{c,v}$ comme dans l'\'equation~\eqref{jauge}, 
successivement avec $h_v=h_{1,v}$, $h_c=h_{c,0}$, $c\in\C^*$, et $h_{c_p}=h_{c_p,0}$, on 
trouve
\begin{align*}
\psi(\widex,b+(\de-2 a_\widex)v) & = \ee^{k\,S(h_v,A^{10}_0+\Axb)}
	\,\psi(\widex, b),\\
\psi(\widex,c^2b) & = \ee^{k\,S(h_c,A^{10}_0+\Axb)}\,\psi(\widex, b),\\
\psi(\widex,c_p^2b) & = \ee^{k\,S(h_{c_p},A^{10}_0+\Axb)}\,\psi(\widex, b).
\end{align*}
En calculant explicitement les facteurs, on trouve que les
fonctions $\psi$ doivent se transformer comme
\begin{align}
\label{CSg}
\taga{CSg}
\psi(\widex,\lambda b+(\de-2a_\widex)v) & = \lambda^{k(g-1)}\psi(\widex,b),\\
\tagb{CSg}
\psi(p\widex,c_p^2b) & = \mu(p,\widex)\,\nu(c_p)^k\,\psi(\widex,b)
\nombre
\end{align}
o\`u, si $F_0$ est la forme de courbure de la connexion m\'etrique 
sur $L_0$, $a_i,b_j$ est une base symplectique canonique,
$\imtau$ est la partie imaginaire de la matrice des p\'eriodes normalis\'ee,  
$W_a$ est l'holonomie de la connexion m\'etrique sur $L_0$ suivant
le cycle $a$, on a pos\'e
\begin{align*}
\nu(c) & = \ee^{\frac{i}{2\pi}\int_\Sigma
	F_0\,{\rm ln}\,c}
	\prod_{i=1}^g
	\big(
	W_{a_j}^{-\frac{i}{2\pi}\mathop{\int}_{b_j}c^{-1}dc}
	W_{b_j}^{\frac{i}{2\pi}\mathop{\int}_{a_j}c^{-1}dc}
	\big),\\
\mu(p,\widex) & = 
	\ee^{\pi\big(\int_p\bar\omega\big)
	\imtau^{-1}\big(\int_p\omega\big)
	+2\pi\big(\int_p\bar\omega\big)
	\imtau^{-1}\big(\mathop{\int}\limits
	_{x_0}^\widex\omega\big)}.
\end{align*}
G\'eom\'etriquement, les propri\'et\'es~(\ref{CSg}.a-b) disent
que $\psi$ est une section holomorphe de la $k$-i\`eme puissance
d'un fibr\'e holomorphe en droites ${\rm DET}$ au-dessus de $\PP W_0$~:
\[
{\rm DET}=\varpi^*\big(L^2K\CO(-x_0)^{2(g-1)}\big)\,{\rm Hf}(W_0){}^{1-g}
\]
o\`u $\varpi$ est la projection associ\'ee au fibr\'e $\PP W_0$ et ${\rm Hf}(W_0)$
est le fibr\'e de Hopf au-dessus de $\PP W_0$~(\footnote{
Si $V$ est un espace vectoriel sur $\C$ de dimension finie. La puissance
$n$-i\`eme du fibr\'e (tautologique) de Hopf
est $V^*\times\C$ modulo la relation d'\'equivalence
$(v,z)\sim(\lambda v,\lambda^{-n}z)$ pour $\lambda\in\C^*$. On obtient donc
un fibr\'e au-dessus de $\PP V=V^*/\C^*$.}). Le diagramme suivant \<<r\'esume\>>
le contenu de la section.
\[
\begin{CD}
	{\rm DET}^k\\
	@A{\psi}AA\\
	\PP W_0 @>s>> \CA^{01} \\
	@V{\varpi}VV\\
	\Sigma\\
\end{CD}
\]
\noindent On a not\'e ${\rm DET}$
ce fibr\'e car il est isomorphe au quotient par $\pi_1$ du
fibr\'e d\'eterminant de la famille d'op\'erateurs $\de+B^{01}_{\widex,b}$
sur $L_0^{-1}\oplus L_0$. On a donc construit un plongement
$\Psi\mapsto \psi$ de $\CW(\Sigma,\emptyset,\emptyset)$ dans $H^0({\rm DET}^k)$.
Ce dernier espace \'etant de dimension finie, l'espace des \'etats de CS 
l'est aussi. Pour le cas avec points d'insertion, on se reportera \`a 
l'article~\cite{gaw95:higher3}.

Comme on l'a d\'ej\`a dit il n'y a pas unicit\'e de $\CL_0$ et $W_0$. 
Dans les articles~\cite{bertram0,thaddeus}, les auteurs proposent 
une autre r\'ealisation, avec laquelle on 
peut identifier l'espace des \'etats de CS. Leur famille
est plus naturelle, mais moins explicite. On pose
\[
\CL_0'=pr_2^*(L)(-\Delta)
\]
o\`u $\Delta$ est la diagonale. Clairement, $\CL_0'$ est une famille
$(L(-x))$. Le passage entre $\CL_0$ et $\CL_0'$ est explicit\'e par
\[
\CL_0'\cong pr_1^*\big(\CO(-x_0)\big)\, \CL_0.
\]
Ceci permet~\cite{thaddeus}
de trouver la dimension de $\CW(\Sigma,\emptyset,\emptyset)$, leur r\'esultat
s'accordant avec celui de Verlinde. Pour plus de d\'etails,
on consultera avec avantage~\cite[section 4]{gaw95:higher1}.

\medskip
\section{Formule de Verlinde}

La formule de Verlinde~\cite{verlinde} 
donne la dimension de l'espace des \'etats de
CS $\chsi$. 
Bien que ce dernier d\'epende de la structure complexe
de $\Sigma$ et des points d'insertion, sa dimension 
$N(g)\sous{\un R}$ ne devrait pas en d\'ependre (la d\'ependance de niveau $k$
est sous-entendue dans la notation).
L'affirmation en question est : l'ensemble des \'etats de CS
devrait former un fibr\'e vectoriel
holomorphe $\ChSi$ au-dessus de l'espace des modules
$\CM_{g,N}$ des surfaces de Riemann de genre $g$ avec $N$ points marqu\'es.
Avec le langage utilis\'e dans la section {\bf 7}, on d\'efinit un fibr\'e
vectoriel holomorphe $\CL$ au-dessus de l'espace des modules 
des $G^\C$-fibres holomorphes et des surfaces de Riemann
avec des points. Les sections holomorphes de $\CL$ sur la sous-vari\'et\'e \`a
surface de Riemann et points fixes donnent les 
\'etats de CS. Le {\bol fibr\'e
de Friedan-Shenker} est alors le fibr\'e
$\ChSi=R^0pr\sous{\CM_{g,N}}\CL$ au-dessus de $\CM_{g,N}$
--- cf. chapitre 4 pour une plus ample discussion.
Avec notre d\'efinition de l'espace des \'etats de CS,
pour $G={\rm SU}_2$, ceci est d\'emontr\'e rigoureusement dans les 
articles~\cite{gaw91:zero} pour le genre z\'ero
et~\cite{gaw94:genus1} pour le genre un. En fait,
les r\'esultats de~\cite{gaw91:zero} entra\^\i nent
par r\'ecurrence~\cite{ramadas} la formule de Verlinde
en genre quelconque, toujours pour le groupe ${\rm SU}_2$. 
D\'ej\`a $N(g)\sous{\un R}$ ne d\'epend pas de l'ordre dans $\un R$, et
ajouter ou enlever la repr\'esentation triviale ne modifie pas
la dimension. On utilise
$N\sous{R,R'}=\delta\sous{\bar R,R'}=N^{^{R,R'}}$ pour monter
ou descendre les indices de $N$. On note $N\sous{\un R}$ 
la dimension en genre z\'ero.

Pour $\Sigma=\C P^1$, on a vu que l'espace des \'etats de CS est 
un sous-espace de $\bibendum^G$ donn\'e par l'\'equation~\eqref{depthrule}.
Dans la limite $k\mapsto\infty$, c'est tout $\bibendum^G$.
Notons ${}^\infty N\sous{\un R}$ la dimension 
de $\bibendum^G$. On peut calculer
${}^\infty N\sous{\un R}$ \`a partir de la th\'eorie des repr\'esentations de $G$.
L'ingr\'edient principal est l'anneau des repr\'esentations introduit
dans la section 1.{\bf 1.2} et la formule d'int\'egration de Weyl.
On trouve
\[
{}^\infty N\sous{\un R}=\big|\liet/(2\pi Q^\vee)\big|^{-1}
	\mathop{\int}\limits_{\liet/2\pi Q^\vee/W}
	\prod_\ell \chi\sous{\lambda_\ell}(u)\,
	\big|\Pi(u)\big|^2\,d^{2r}u
\] 
o\`u $\chi\sous{\lambda_\ell}$ est le caract\`ere de la repr\'esentation $R_\ell$,
dans $\un R$, de plus haut poids $\lambda_\ell$.

Pour obtenir la formule de Verlinde, \cad comprendre ce qui se passe
avec $k$ fini, on doit faire appel au {\bol principe de factorisation}
d\^u \`a Friedan et Shenker~\cite{friedan}. On commence par \'etendre
le fibr\'e $\ChSi$ \`a la compactification $\bar\CM_{g,N}$ de l'espace
des modules~\cite{deligne-mumford,knudsen}. Cette compactification autorise
\`a prendre des surfaces singuli\`eres en des points doubles (ordinaires). Pour obtenir
un point double, on  peut sur une surface lisse
r\'eduire \`a un point (double) un cycle de deux mani\`eres diff\'erentes~:

\begin{figure}[h]
\makebox[5cm][l]{\hspace{1.5cm}
\includegraphics{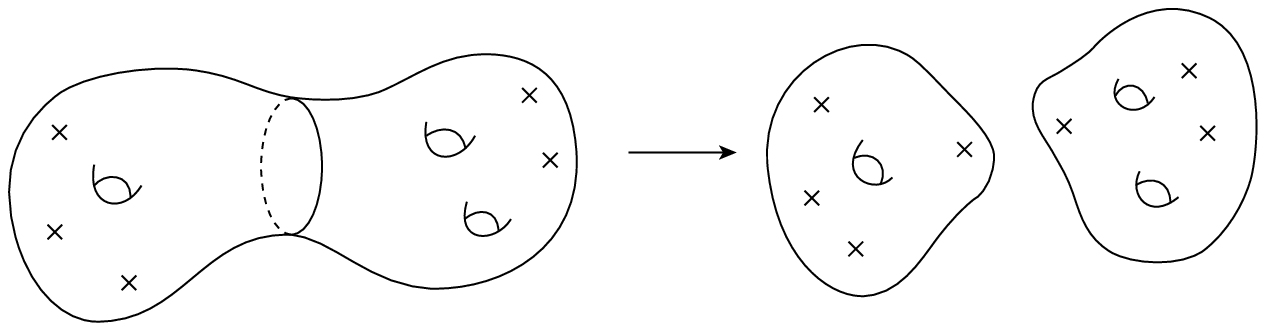}}
\caption{Compactification de l'espace des modules I}
\end{figure}
\newpage
\begin{figure}[h]
\makebox[5cm][l]{\hspace{1.5cm}
\includegraphics{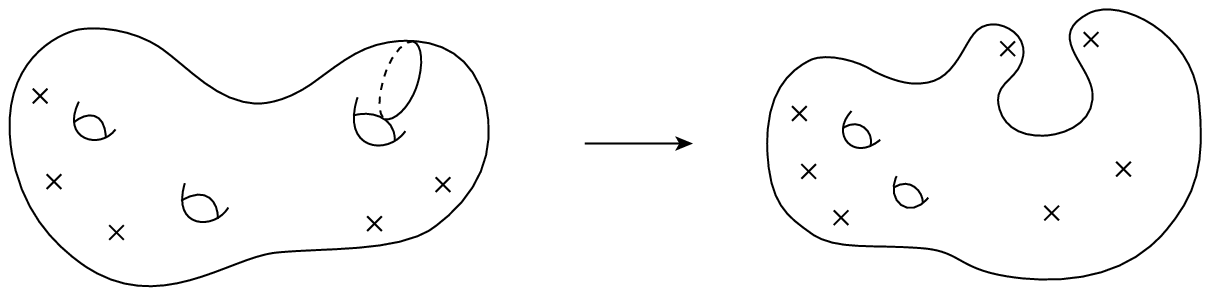}}
\caption{Compactification de l'espace des modules II}
\end{figure}

En enlevant les points doubles, dans le premier cas, on obtient
deux surfaces $\Sigma'$ et $\Sigma''$ de genre $g_1$ et $g_2$ avec $g_1+g_2=g$
et marqu\'ees respectivement par $(\un\xi',\xi_0)$ et $(\un\xi'',\xi_\infty)$ avec
$(\un\xi',\un\xi'')=\un\xi$, dans le second cas, on obtient une surface
$\Sigma'$ de genre $g-1$ avec deux points marqu\'es en plus $\xi_0,\xi_\infty$.
Le principe de factorisation affirme que
\[
\CW(\Sigma,\un\xi,\un R)
	\cong\bigoplus_{R\atop
	\mbox{\tiny int\'egrable}}
	\CW\big(\Sigma',(\un\xi',\xi_0),(\un R',R)\big)
	\,\otimes\,
	\CW\big(\Sigma'',(\un\xi'',\xi_\infty),(\un R'',\bar R)\big),
\]
o\`u bien s\^ur $(\un R',\un R'')=\un R$, et
\[
\CW(\Sigma, \un \xi,\un R)\cong
	\bigoplus_{R\atop\mbox{\tiny int\'egrable}}
	\CW\big(\Sigma',(\un\xi,\xi_0,\xi_\infty),
	(\un R, R,\bar R)\big).
\]
Ce principe a \'et\'e montr\'e, 
dans le contexte pr\'esent, en genre z\'ero
et un (loc. cit.) pour certains cas. En admettant sa validit\'e,
on a les relations de r\'ecurrence suivantes
\begin{align*}
N(g)\sous{\un R} & = \sum_{r\atop
	g_1+g_2=g}
	 N(g_1)\sous{(\un R',R)}\,N(g_2)^R{}\sous{\un R''},\\
N(g)\sous{\un R} & = \sum_R
	N(g-1)^R{}\sous{(\un R',R)}.
\end{align*}
Pour trouver $N\sous{\un R}$, on n'utilise plus l'anneau des
repr\'esentations, mais sa contrepartie~: l'{\bol anneau de fusion}.
Sans entrer dans les d\'etails, l'anneau ${\bf K}(G)$ a
pour constantes de structure les entiers ${}^\infty N$ pour
trois repr\'esentations et l'anneau de fusion utilise plut\^ot
les entiers $N$. 

Si $R$ est une repr\'esentation int\'egrable de niveau $k$, on pose
$\wih R=2\pi (\lambda_R+\rho)/(k+g^\vee)$. On montre alors que
\[
N\sous{\un R}=\big|P^\vee/(k+g^\vee)Q^\vee\big|^{-1}
	\sum_{R\atop\mbox{\tiny int\'egrable}}
	\prod_\ell \chi\sous{\lambda_\ell}(\wih R)\,
	\big|\Pi(\wih R)\big|^2.
\]
En genre sup\'erieur, on d\'eduit 
la formule de Verlinde par r\'ecurrence et
\[
N(g)\sous{\un R}=\sum_{\un R'\atop
	\mbox{\tiny int\'egrables}}
	N^{^{\un R'}}{}\sous{(\un R',\un R)}
\]
o\`u $\un R'$ est une s\'equence de $g$ repr\'esentations int\'egrables.
Il suit
\[
N(g)\sous{\un R}=\big|P^\vee/(k+g^\vee)Q^\vee\big|^{-1+g}
	 \sum_{R\atop\mbox{\tiny int\'egrable}}
	\prod_\ell \chi\sous{\lambda_\ell}(\wih R)\,
	\big|\Pi(\wih R)\big|^{2(1-g)}.
\]

\newpage

La quantit\'e d'articles sur le sujet 
est proportionnelle \`a l'int\'er\^et qu'il a
suscit\'e~:~\cite{degio,mooreetseiberg} pour les th\'eories
conformes rationnelles,~\cite{beauville,gepner1} pour
l'anneau de fusion, par exemple~\cite{faltings} 
pour un traitement rigoureux
du point de vue de la g\'eom\'etrie alg\'ebrique,~\cite{witten:jones} 
pour la th\'eorie de CS et~\cite{tsuchiya} pour
une preuve du principe de factorisation. 
En physique, la formule de Verlinde est \'ecrite
souvent un peu diff\'eremment. \'Ecrivons-la pour $G={\rm SU}_2$. 
On introduit la matrice $S$ donnant la transformation modulaire 
$\tau\mapsto-1/\tau$ des caract\`eres de Kac-Moody affines~:
\[
S^j{}_{j'}=\sqrt{\quotient{2}{k+2}}\,
{\rm sin}\,\quotient{\pi(2j+1)(2j'+1)}{k+2}.
\]
On a alors la formule \`a trois points
\[
N^{j_1}{}_{j_2,j_3}=\sum_{2j=0}^k S^{j}{}_{j_1}
S^{j}{}_{j_2}S^{j}{}_{j_3}
	\big/S^j{}_0,
\]
forme sous laquelle la formule de Verlinde apparut pour
la premi\`ere fois~\cite{verlinde}. Par r\'ecurrence, on trouve
\[
N(g)_{j_1,\cdots,j_N}=\sum_{2j=0}^k
	\big(S^j{}_0)^{2(1-g)}\,\prod_\ell\big(S^j{}_{j_\ell}\big/
	S^j{}_0\big).
\]
Sans points d'insertion, on trouve
\[
\dim\CW(\Sigma,\emptyset,\emptyset)=N(g)
	=\big(\quotient{k+2}{2}\big)^{g-1}
	\sum_{2j=0}^k
	\big({\rm sin}\, \quotient{\pi(2j+1)}{k+2}\big)^{2-2g},
\]
le r\'esultat confirm\'e dans~\cite{thaddeus} par des consid\'erations 
sur les espaces de modules $\CN$. 

\medskip
\section{Factorisation holomorphe}

On note $\underline{\xi}$ une s\'equence de $N$ points sur $\Sigma$ et 
$\underline{R}$ une s\'equence de $N$ repr\'esentations 
irr\'eductibles de $G$.
Chaque espace de repr\'esentation $V_{\lambda_\ell}$ peut \^etre muni d'un produit 
scalaire hermitien canonique, ainsi $\esprep=\otimes_\ell V_{\lambda_\ell}$ 
h\'erite d'un produit scalaire not\'e  $\lsca.,.\rsca$.
On d\'efinit un {\bol produit scalaire (formel) sur l'espace
des \'etats de CS}~ $\chsi$~:
\begin{equation}
\label{bargman}
\Blsca\Psi,\Psi'\Brsca=\mathop{\int}\limits_{\CA}\lsca\Psi(A^{01}),
	\Psi'(A^{01})\rsca\,\exponen\,DA
\end{equation}
o\`u $DA$ est une mesure de Lebesgue formelle sur l'espace 
$\CA$ des connexions unitaires.

On a vu qu'une propri\'et\'e importante de l'espace des \'etats de CS
est qu'il est de dimension finie~; ce r\'esultat a des cons\'equences 
tr\`es importantes pour le mod\`ele de WZNW. Formellement, les fonctions 
de Green modifi\'ees sont analytiques en $A^{10}$ et en $A^{01}$. 
En plus, d'apr\`es l'\'equation~\eqref{platitudeW}, 
pour tout $\Bv\in\esprep$,
$\widegamma(A)\,\Bv$ v\'erifie la contrainte 
requise pour qu'une fonctionnelle 
de $A^{01}$~--- \`a $A^{10}$ fix\'e~--- soit un \'etat de CS. 
Soit $(\Psi_{p})$ une base de l'espace $\chsi$. 
La constatation pr\'ec\'edente 
nous permet d'affirmer que les fonctions de Green 
modifi\'ees se d\'ecomposent comme suit
\[
\widegamma(A)=\sum_{p}\Upsilon_{p}(A^{10})\otimes\Psi_{p}(A^{01})
\in\overline{V}_{\hspace{-0.06cm}\un \lambda}
\otimes V_{\un \lambda}\cong
\mathop{\bigotimes}\limits_\ell{\rm End}(V_{\lambda_\ell})
\]
o\`u $\Upsilon_{p}(A^{10})\in\overline{V}_{\hspace{-0.06cm}\un\lambda}
\cong V_{\overline{\underline{\lambda}}}$.

Au cours du pr\'ec\'edent chapitre, on a vu que les fonctions de Green 
sont invariantes par une transformation PCT. Si on ne tient 
pas compte du changement de structure complexe, on a donc
\[
\widegamma(A)^\dagger=\widegamma(-A^\dagger).
\]
En cons\'equence, gr\^ace \`a l'ind\'ependance lin\'eaire des $\Psi_{q}$,
les $\overline{\Upsilon_{p}(-(A^{10})^\dagger)}$ sont aussi des \'etats 
de CS. Il existe donc une matrice hermitienne $h$ telle que
\begin{equation}
\label{sesquiCS}
\widegamma(A)=\sum_{p,q}h^{pq}\,\,\overline{\Psi_{q}(-(A^{10})^\dagger)}
\otimes\Psi_{p}(A^{01})\,.
\end{equation}
Il est important de noter que le choix de la base est fait 
pour $\underline{\xi}$ et $\J$ fix\'es. L'\'equation~\eqref{sesquiCS}
est la {\bol factorisation holomorphe} de la d\'ependance de
$\widegamma(A)$ pour le champ de jauge $A$.

Pour fixer la matrice $h$, on peut proc\'eder par manipulations 
fonctionnelles~\cite{witten:holo}. Soit $B$ un champ de jauge unitaire 
auxiliare. On s'int\'eresse \`a la quantit\'e suivante~:
\[
\int\widegamma(B^{10}+A^{01})\,\widegamma(A^{10}+B^{01})\,
\exponenB\,DB.
\]
Commen\c{c}ons par remplacer les fonctions de Green par des 
int\'egrales fonctionnelles~:
\begin{equation*}
\int\mathop{\otimes}\limits_\ell (g_{1}g_{2})(\xi_\ell )\sous{R_\ell }\,
	\ee^{-k\,S(g_{1},B^{10}+A^{01})
	-k\,S(g_{2},A^{10}+B^{01})}\,\exponenB Dg_{1}\,Dg_{2}\,DB
\end{equation*}
avec
\begin{equation*}
\begin{split}
\ee^{-k\,S(g_{1},B^{10}+A^{01})}&\,\ee^{-k\,S(g_{2},A^{10}+B^{01})}\,
	\exponenB\\
=&\,\ee^{-k\,S(g_{1})-k\,S(g_{2})}\,\ee^{
	-\frac{ik}{2\pi}\,\int\tr\,
	\left\lbrace g_{1}\partial g_{1}^{-1}\wedge\, A^{01}+
	A^{10}\wedge\,g_{2}^{-1}\de g_{2}\right\rbrace}\\
&\,\ee^{-\frac{ik}{2\pi}\,\int\tr\,\left\lbrace
	B^{10}\wedge\,g_{1}^{-1}\de g_{1}
	+g_{1}B^{10}g_{1}^{-1}\wedge\, A^{01}
	+g_{2}\partial g_{2}^{-1}\wedge\,B^{01}
	+g_{2}A^{10}g_{2}^{-1}\wedge\,B^{01}
	-B^{10}\wedge\,B^{01}\right\rbrace}.
\end{split}
\end{equation*}
L'int\'egrale sur $B$ est gaussienne. Elle est donc calculable. On obtient
\begin{equation*}
\begin{split}
\int\mathop{\otimes}\limits_\ell (g_{1}g_{2})(\xi_\ell )\sous{R_\ell }&\,
	\ee^{-k\,S(g_{1})-k\,S(g_{2})}\,\ee^{
	-\frac{ik}{2\pi}\,\int\tr\,
	\left\lbrace g_{1}\partial g_{1}^{-1}\wedge\, A^{01}+
	A^{10}\wedge\,g_{2}^{-1}\de g_{2}\right\rbrace}\\
&\,\ee^{-\frac{ik}{2\pi}\,\int\tr\,
	(g_{2}\partial g_{2}^{-1}+g_{2}A^{10}g_{2}^{-1})\wedge\,
	(g_{1}^{-1}\partial g_{1}+g_{1}^{-1}A^{01}g_{1})}
	Dg_{1}\,Dg_{2}\\
=&\,\int\mathop{\otimes}\limits_\ell (g_{1}g_{2})(\xi_\ell )\sous{R_\ell }\,
	\ee^{-k\,S(g_{1}g_{2},A)}\,\exponen Dg_{1}Dg_{2}.
\end{split}
\end{equation*}
Pour obtenir la derni\`ere \'egalit\'e, on utilise l'invariance de la mesure 
de Haar (formelle). On trouve ainsi la fonction de Green modifi\'ee 
$\widegamma(A)$.

Reprenons l'int\'egrale de d\'epart, mais rempla\c{c}ons cette fois 
les fonctions de Green par leurs expressions en fonction d'une base 
des \'etats de CS~:
\begin{equation*}
\begin{split}
\sum_{p,q,r,s}h^{pq}h^{rs}\,\overline{\Psi_{s}(-(A^{10})^\dagger)}\,
	&\left(\int\lsca\Psi_{q}(B^{01}),\Psi_{r}(B^{01}\rsca\,
	\exponenB DB\right)\,\otimes\Psi_{p}(A^{01})\\
&=\sum_{p,q,r,s}h^{pq}H_{qr}h^{rs}\,\overline{\Psi_{s}(-(A^{10})^\dagger)} 
	\otimes\Psi_p(A^{01})
\end{split}
\end{equation*}
o\`u $H_{pq}\equiv\Blsca\Psi_{p},\Psi_{q}\Brsca$.
La comparaison des deux r\'esultats nous donne $hHh=h$.
Si on suppose que la matrice $h$ est 
inversible, on obtient l'\'equation
\begin{equation}
\label{inverse}
h^{pq}=(H^{-1})^{pq}\,.
\end{equation}

On peut r\'e\'ecrire ce r\'esultat ind\'ependamment
du choix de la base $(\Phi_p)$.
Soit $e(A^{01})$ l'appli\-cation d'\'evaluation ~:
$\chsi\ni\Psi\longmapsto\Psi(A^{01})\in\esprep$.
On peut identifier $e(A^{01})$ 
avec un \'el\'ement de $\chsi^*\otimes\esprep$, 
not\'e $\boldsymbol{\gamma}(\underline{\xi},\underline{R},A^{01})$.
On appellera $\boldsymbol{\gamma}$ la {\bol fonction de Green chirale}. 
\'Etant donn\'e une base de $\chsi$, on a 
\[
\boldsymbol{\gamma}(\underline{\xi},\underline{R},A^{01})=\sum_p\Psi^{p*}
	\otimes\Psi_{p}(A^{01})
\]
o\`u $\Psi^{p*}$ est une base de $\chsi^*$.
On obtient alors~:
\[
\widegamma(A)=\bigl\langle
	\boldsymbol{\gamma}(\underline{\xi},\underline{R},
	-(A^{10})^\dagger)
	,\,\boldsymbol{\gamma}
	(\underline{\xi},\underline{R},A^{01})
	\bigr\rangle
\]
o\`u $\langle.,.\rangle$ est le produit scalaire sur $\chsi^*$, induit 
par le produit scalaire sur $\chsi$ d\'efini par~\eqref{bargman} et
on a $h^{pq}=\langle\Psi^{q*},\Psi^{p*}\rangle$.

\medskip
\section{Mod\`ele de WZNW hyperbolique}

D'apr\`es ce qui pr\'ec\`ede, il suffit de calculer le produit scalaire des \'etats
de CS pour obtenir les fonctions de Green du mod\`ele de WZNW, \cad
\[
\scal\Psi\scal^2=\int\sca\Psi(B^{01})\sca^2\,
	\ee^{-\frac{ik}{2\pi}\,\int\tr\,(B^{01})^\dagger\wedge B^{01}}
	\,DB.
\]
On commence par \'etudier la situation en genre 
$\geq 2$~\cite{gaw95:higher3,gaw95:higher2,gaw95:higher1}. 
Les genres z\'ero~\cite{gaw91:scalar,gaw89:quadrature,gaw89:coset}
et un~\cite{gaw97:unitarity,gaw89:coset} seront trait\'es s\'epar\'ement.
On effectue le changement de variables
\[
B^{01}\ =\ \hAL{h^{-1}}(n)
\]
o\`u $n\mapsto A^{01}(n)$ est la d\'ecoupe de l'espace des 
champs de jauge \'evoqu\'ee
auparavant, qui choisit localement un champ de jauge dans chaque orbite, 
et $n$ est un param\`etre sur l'espace des modules $\CN_s$.
Pour $G={\rm SU}_2$, c'est l'application $(\widex,b)\mapsto
A^{01}_{\widex,b}$ construite dans la section {\bf 7}.
Pour simplifier un peu les notations, on utilise $A^{01}$ pour $A^{01}(n)$. 
La fonction $h$ est unique. Au niveau infinit\'esimal, cela signifie que
$\partialA$ n'a pas de modes z\'ero.
Une telle manipulation restera formelle mais sera particuli\`erement soign\'ee.
On esp\`ere ainsi arriver \`a des int\'egrales gaussiennes 
(toujours de dimension infinie)
qu'on sait alors parfaitement d\'efinir. On d\'ecompose l'int\'egrale en une 
int\'egrale sur $\CG^\C$ et une int\'egrale sur l'espace des modules 
$\CA^{01}/\CG^\C$. Ainsi,
\[
\scal\Psi\scal^2=\int \sca \Psi(\hAL{h^{-1}})\sca^2\,
	\ee^{-\frac{ik}{2\pi}\,\int\tr\,
	(\hAL{h^{-1}})^\dagger\wedge\hAL{h^{-1}}}
	j(h,n)\,Dh\,\prod_\alpha d^2n^\alpha
\]
o\`u $Dh$ est la mesure de Haar formelle sur $\Gc$ et $j(h,n)$ 
est le Jacobien du
changement de variables. Pour une variation infinit\'esimale de 
$h$ et de $n$, on a
\[
\delta B^{01}=\delta (\hAL{h^{-1}}(n))={}^{h^{-1}}\partialA(h^{-1}\delta h)+h^{-1}\,(
	\partial_{n^\alpha}A^{01}\,\delta n^\alpha)\,h
\]
o\`u ${}^{h^{-1}}\partialA\equiv {\rm Ad}_{h^{-1}}\partialA{\rm Ad}_h$. On munit l'espace
$\CA^{01}$ de la norme naturelle $L^2$, \cad
\[
||A^{01}||^2=i\int_\Sigma \tr\, (A^{01})^\dagger\wedge A^{01}.
\]
Gr\^ace au produit scalaire $\langle\ |\ \rangle$
induit par cette norme, on peut d\'ecomposer $\CA^{01}$ en une somme d'espaces
orthogonaux~: $\CA^{01}=F\oplus F^\bot$, $F$ \'etant ${\rm Im}\,({}^{h^{-1}}\partialA)$.
Soit $\phi$ une $(1,0)$-forme \`a valeurs dans $\liegc$ telle que
\[
\int_\Sigma\tr\, \phi\wedge {}^{h^{-1}}\partialA\Lambda=0
\]
pour toute fonction $\Lambda$ lisse sur $\Sigma$ prenant ses valeurs dans $\liegc$.
Apr\`es int\'egration par parties, 
on constate que c'est compl\`etement \'equivalent \`a
dire que $\phi$ est un mode z\'ero de ${}^{h^{-1}}\overline{D}$, o\`u $\overline{D}$ est
l'op\'erateur $\overline{\partial}+[A^{01},.]_+$ --- $[.,.]_+$ est l'anticommutateur~---
agissant sur les $(1,0)$-formes sur
$\Sigma$ \`a valeurs dans $\lieg^\C$ pour donner une $2$-forme sur $\Sigma$
\`a valeurs dans $\liegc$.  On a encore ${}^{h^{-1}}\overline{D}={\rm Ad}_{h^{-1}}
\overline{D}\,{\rm Ad}_h$. De l\`a, on voit qu'une base de $F^\bot$ est donn\'ee par
les vecteurs $(h^{-1}\phi_\alpha\, h)^\dagger$, pour $(\phi_\alpha)$ une base du noyau
de $\overline{D}$. Les $\phi_\alpha$ d\'ependent du param\`etre $n$, comme $\overline{D}$.
On peut assimiler ceux-ci \`a des \'el\'ements de l'espace cotangent \`a l'espace
des modules. On suppose que les $\phi_\alpha$ varient
holomorphiquement avec $n$.
La d\'eriv\'ee holomorphe du changement de variables est de la forme
\begin{equation*}
\frac{\delta (\hAL{h^{-1}}(n))}{\delta\,(h,n)}
	=\begin{pmatrix}
	{}^{h^{-1}}\partialA & \cdots\\
	0 & (h^{-1}\,\partial_{n^\alpha}A^{01}\,h)^\bot
	\end{pmatrix}
	\ 
	\begin{matrix}
	\leftarrow & h^{-1}\delta h\\
	\leftarrow & \delta n^\alpha
	\end{matrix}
\end{equation*}
o\`u la partie \<<$\bot$\>> est la projection de $h^{-1}\,\partial_{n^\alpha}A^{01}\,h$
sur $F^\bot$, \ie
\[
(h^{-1}\,\partial_{n^\alpha}A^{01}\,h)^\bot=
	\sum_{\gamma,\beta} (h^{-1}\phi_\gamma\, h)^\dagger\ (\Delta^{-1})_{\gamma\beta}\ 
	\big\langle (h^{-1}\phi_\beta\, h)^\dagger\,\big|\,h^{-1}\,\partial_{n^\alpha}A^{01}\,h
	\big\rangle
\]
o\`u $\Delta_{\gamma\beta}=\big\langle (h^{-1}\phi_\gamma\, h)^\dagger\,\big|\,
(h^{-1}\phi_\beta\, h)^\dagger\big\rangle$. On obtient sans probl\`eme
\begin{align*}
(h^{-1}\,\partial_{n^\alpha}A^{01}\,h)^\bot=&\,
	 i\sum_{\gamma,\beta} (h^{-1}\phi_\gamma\, h)^\dagger\ (\Delta^{-1})_{\gamma\beta}\ 
	\int_\Sigma\tr\, \phi_\beta\wedge\partial_{n^\alpha}A^{01},
	\\
\Delta(hh^\dagger,n)_{\gamma\beta}=&\,i\int_\Sigma \tr\, (hh^\dagger)^{-1}\phi_\gamma
	\wedge hh^\dagger\,\phi_\beta^\dagger.
\end{align*}
Le Jacobien du changement de variables est donc
\begin{align*}
j(h,n)&=\left|\frac{\partial(\hAL{h^{-1}}(n))}{\partial(h,n)}\right|^2
	={\rm det}\,\big(({}^{h^{-1}}\partialA)^\dagger\,{}^{h^{-1}}\partialA\big)\ 
	{\rm det}\,\big((h^{-1}\,\partial_{n^\alpha}A^{01}\,h)^{\bot\dagger}\,
	(h^{-1}\,\partial_{n^\alpha}A^{01}\,h)^\bot\big)\\
&={\rm det}\,\big(({}^{h^{-1}}\partialA)^\dagger\,{}^{h^{-1}}\partialA\big)\ 
	{\rm det}\,\Delta(hh^\dagger,n)^{-1}\ 
	\Big|{\rm det}\, \Big(\int_\Sigma\tr\,\phi_\beta\wedge\partial_{n^\alpha}A^{01}
	\Big)\Big|^2.
\end{align*}
Dans ce produit, le d\'eterminant ${\rm det}\,\big(({}^{h^{-1}}\partialA)^\dagger
\,{}^{h^{-1}}\partialA\big)$ doit \^etre r\'egularis\'e, 
par exemple on peut utiliser une r\'egularisation z\'eta~\cite{raysinger}.
On extrait la d\'ependance en $h$ \`a l'aide de l'anomalie chirale~(\footnote{
Cette formule m\'eriterait \`a elle seule une longue discussion. 
Il s'agit de calculer l'anomalie
chirale \`a partir du th\'eor\`eme de Riemann-Roch-Grothendieck-Quillen
pour le fibr\'e d\'eterminant ${\rm DET}\,\partialA$ de la famille
d'op\'erateurs $\partialA$ muni de la m\'etrique de Quillen. 
On trouve une explication d\'etaill\'ee dans~\cite{gaw:book} (cf. aussi~\cite{alvarez:higher,
bost:bourbaki,polyakov}).})~:
\[
{\rm det}'\,\big(({}^{h^{-1}}\partialA)^\dagger\,{}^{h^{-1}}\partialA\big)\
	{\rm det}\,\Delta(hh^\dagger,n)^{-1}
	=\ee^{2g^\vee\,S(hh^\dagger,A)}
	\,{\rm det}'\,\big(\partialA^\dagger\,\partialA\big)\
	{\rm det}\,\Delta(1,n)^{-1}
\]
o\`u $A$ est le champ de jauge complet $-(A^{01})^\dagger+A^{01}$.
On note $\Delta(n)$ la matrice $\Delta(1,n)$. D'autre part, de l'invariance globale
des \'etats de CS on tire
\[
\sca\Psi(\hAL{h^{-1}})\sca^2\,
	\ee^{\frac{ik}{2\pi}\,\int\tr\,(\hAL{h^{-1}})^\dagger\wedge\hAL{h^{-1}}}
	=\lsca\Psi(A^{01}),
	\mathop{\otimes}\limits_\ell (hh^\dagger)(\xi_\ell )^{-1}\sous{R_\ell }\,\Psi(A^{01})
	\rsca\,
	\ee^{-\frac{ik}{2\pi}\,\int\tr\,(A^{01})^\dagger\wedge A^{01}+k\,S(hh^\dagger,A)}.
\]
En fin de compte, apr\`es le changement de variables $B^{01}\mapsto {}^{h^{-1}}A^{01}(n)$,
le produit scalaire est 
\begin{equation}
\begin{split}
\scal\Psi\scal^2=\int&
	\lsca\Psi(A^{01}),
	\mathop{\otimes}\limits_\ell (hh^\dagger)(\xi_\ell )^{-1}\sous{R_\ell }\,\Psi(A^{01})
	\rsca\,
	\ee^{-\frac{ik}{2\pi}\,\int\tr\,(A^{01})^\dagger\wedge A^{01}}
	\,\ee^{(k+2g^\vee)\,S(hh^\dagger,A)}\\
&{\rm det}'\,\big(\partialA^\dagger\,\partialA\big)\
	{\rm det}\,\Delta(n)^{-1}\,
	\Big|{\rm det}\, \Big(\int_\Sigma\tr\,\phi_\beta\wedge\partial_{n^\alpha}A^{01}
	\Big)\Big|^2\,
	D(hh^\dagger)\,\prod_\alpha d^2n^\alpha.
\end{split}
\end{equation}
Dans l'int\'egrande $h$ n'intervient que sous la forme $hh^\dagger$. 
On peut donc r\'eduire 
l'int\'egration sur $h$ en une int\'egration sur $hh^\dagger$ qui prend 
ses valeurs dans l'espace sym\'etrique $\Gc/G$ non-compact. 
La mesure $D(hh^\dagger)$ est donc le produit local des mesures 
$\Gc$-invariantes sur $\Gc/G$. Ce n'est pas \'etonnant 
car au d\'ebut, on int\`egre sur les champs de jauge unitaires. 
Ceci ne fait que refl\'eter
l'invariance initiale du produit scalaire. 
Il appara\^\i t que le produit scalaire est exprim\'e
par une int\'egrale fonctionnelle 
sur un mod\`ele de WZNW pour le groupe $\Gc/G$,
de niveau $-(k+2g^\vee)$, et par la fonction de partition des fant\^omes. 
Si $G={\rm SU}_2$, le quotient
${\rm SL}_2/{\rm SU}_2$ est l'espace hyperbolique tridimensionnel $H^+_3$.
Pour cette raison, on appelle {\bol mod\`ele de WZNW 
hyperbolique} le mod\`ele bas\'e
sur $\Gc/G$.

\subsection*{Genre z\'ero}

Sur la sph\`ere de Riemann, l'orbite du champ nul est dense, donc g\'en\'eriquement
on peut trouver $h\in\CG^\C$ tel que 
\[
A^{01}=h^{-1}\de h.
\]
Pour une variation infinit\'esimale de $h$, on a
\[
\delta A^{01}=\partialh (h^{-1}\delta h).
\]
La param\'etrisation en question n'est pas univoque. Il reste la libert\'e de multiplier 
\`a gauche $h$ par un champ constant $h_0\in\Gc$. Une autre mani\`ere de dire est~: 
l'op\'erateur $\partialh$ a des modes z\'ero. En effet, le noyau de 
$\partialh={\rm Ad}_{h^{-1}}\,\de\,{\rm Ad}_h$ est 
engendr\'e par les vecteurs $h^{-1}t^a\,h$. Pour r\'esoudre l'ambigu\"\i t\'e
de la param\'etrisation, on multiplie l'int\'egrale du produit scalaire par
$1=\int\delta(h(\xi_0))\,dh(\xi_0)$, pour un point $\xi_0$ de $\C P^1$. 
\[
\scal\Psi\scal^2=\int
	\sca\Psi(h^{-1}\de h)\sca^2\ 
	\ee^{-\frac{ik}{2\pi}\,\int\tr\,
	(h^{-1}\de h)^\dagger\wedge h^{-1}\de h}
	\,\delta(h(\xi_0))\,
	j(h)\,Dh.
\]
Le Jacobien du changement de variables est~\cite[App. C]{gaw89:coset} 
\[
j(h)\ =\ {\rm const.}\,\,{\rm det}'\big(\partialh^\dagger\partialh\big)
	\,\Big/\,{\rm det}_{a,b}\Big(\int_\Sigma \tr\, 
	hh^\dagger\,t^a\,(hh^\dagger)^{-1}t^b\,d^2z\Big).
\]
Le symbole \<<const.\>> d\'esigne une constante tampon ne d\'ependant 
que de $G$ qu'on peut absorber dans la forme volume.
Gr\^ace \`a l'anomalie chirale, on peut extraire la d\'ependance en $h$. Ainsi,
\[
j(h)\ =\ {\rm const.}\,\, \ee^{2g^\vee\,S(hh^\dagger)}\,
	\Big(\quotient{{\rm det}'\,(-\Delta)}{{\rm aire}}
	\Big)^{-\dim G}.
\]
Apr\`es le changement de variables, le produit scalaire en g\'eom\'etrie
sph\'erique est donc
\begin{equation}
\begin{split}
\scal\Psi\scal^2\ =\ 
	{\rm const.}\,\,\Big(\quotient{{\rm det}'\,(-\Delta)}{{\rm aire}}
	\Big)^{-\dim G}\int
	\lsca\Psi&(0),
	\mathop{\otimes}\limits_\ell (hh^\dagger)
	(\xi_\ell )^{-1}\sous{R_\ell }\,\Psi(0)
	\rsca\\
&\ee^{(k+2g^\vee)\,S(hh^\dagger)}\,
	\delta(hh^\dagger(\xi_0))\,D(hh^\dagger).
\end{split}
\end{equation}

\subsection*{Genre un}

En genre un, presque tous les champs de jauge sont dans l'orbite de 
$A^{01}_u=\pi u d\overline{z}/\imtau$, $u$ \'etant dans un domaine 
fondamental de $\lietc\setminus 
P^\vee+\tau P^\vee$ pour l'action de $W\rtimes(
Q^\vee+\tau Q^\vee)$. Il y a encore 
une ambigu\"\i t\'e r\'esidant dans la multiplication
\`a gauche de $h$ par un champ constant $h_0$ dans le sous-groupe 
de Cartan $T^\C$. Encore une fois, cela se voit sur l'op\'erateur
$\partialAu$ qui a des modes z\'ero. On comprend a posteriori pourquoi
on doit traiter les genres $0$ et $1$ s\'epar\'ement.
Pour r\'esoudre l'ambigu\"\i t\'e, on utilise la
{\bol d\'ecomposition d'Iwasawa} de $\Gc$~: pour tout $h\in\Gc$, il existe
une seule mani\`ere d'\'ecrire $h$ sous la forme
\[
h=\ee^{\sum_{\alpha>0}v_\alpha e_\alpha}\,\ee^{\varphi/2}\,g
\]
avec $g$ dans le groupe compact $G$, $gg^\dagger=1$, 
$v_\alpha\in\C$ et $\varphi\in\liet$. On pose $n=\ee^v$, $v=\sum v_\alpha e_\alpha$ et
$b=n\,\ee^{\varphi/2}$.
On contr\^ole la libert\'e de $h(\xi_0)$, en fixant $\varphi(\xi_0)$. 
Le produit scalaire est 
\begin{equation*}
\begin{split}
\scal\Psi\scal^2=\int
	\lsca\Psi(A^{01}_u),
	\mathop{\otimes}\limits_\ell (hh^\dagger)(\xi_\ell )^{-1}\sous{R_\ell }&\,\Psi(A^{01}_u)
	\rsca\,
	\ee^{-\frac{ik}{2\pi}\,\int\tr\,(A^{01}_u)^\dagger\wedge A^{01}_u}\\
&\ee^{k\,S(hh^\dagger,A_u)}\,j(h,u)\,
	\delta(\varphi(\xi_0))\,
	D(hh^\dagger)\,
	d^{2r}\!u
\end{split}
\end{equation*}
o\`u le Jacobien est (apr\`es l'anomalie chirale pour extraire la d\'ependance en $h$)
\[
j(h,u)={\rm const.}\,\imtau^{-2r}\,
	\ee^{2g^\vee\,S(hh^\dagger,A_u)}\,
	{\rm det}'\,\big(\partialAu^\dagger\,\partialAu\big).
\]
On montre en plus~\cite{gaw89:coset} que
\[
{\rm det}'\,\big(\partialAu^\dagger\,\partialAu\big)=
	{\rm const.}\, \imtau^{2r}\,\ee^{\pi g^\vee|u-u^\dagger|^2/\imtau}
	\,\big|\Pi(u,\tau)\big|^4
\]
o\`u $\Pi$ est le d\'enominateur de Weyl-Kac. On am\`ene tous ces r\'esultats
dans l'int\'egrale et
\begin{equation}
\begin{split}
\scal\Psi\scal^2={\rm const.}\int
	\lsca\Psi&(A^{01}_u),
	\mathop{\otimes}\limits_\ell (hh^\dagger)(\xi_\ell )^{-1}\sous{R_\ell }\,\Psi(A^{01}_u)
	\rsca\,
	\ee^{-\frac{ik}{2\pi}\,\int\tr\,(A^{01}_u)^\dagger\wedge A^{01}_u}\\
&\ee^{(k+2g^\vee)\,S(hh^\dagger,A_u)}\,
	\ee^{\pi g^\vee|u-u^\dagger|^2/\imtau}
	\,\big|\Pi(u,\tau)\big|^4\,
	\delta(\varphi(\xi_0))\,
	D(hh^\dagger)\,
	d^{2r}\!u.
\end{split}
\end{equation}

\medskip
\section{Repr\'esentation en gaz de Coulomb}

Pour le moment, rien ne nous permet d'affirmer que le changement de variables
conduit \`a des int\'egrables plus faciles \`a traiter. Bien s\^ur, c'est le cas. 
En param\'etrisant $h$ par l'interm\'ediaire de la d\'ecomposition
d'Iwasawa, on tombe sur des int\'egrales gaussiennes. Pour le cas le
plus simple $G={\rm SU}_2$, on l'imagine tr\`es facilement vu la forme de
l'action
\[
S(hh^\dagger)=\quotient{i}{2\pi}\int\partial\varphi\wedge\de\varphi
	-\quotient{i}{2\pi}\int\ee^{-2\varphi}\,
	\da\bar v\wedge\de v
\]
pour
\[
h=\begin{pmatrix}
	1 & v \\
	0 & 1 
	\end{pmatrix}
	\begin{pmatrix}
	\ee^{\varphi/2} & 0\\
	0 & \ee^{-\varphi/2} \\
	\end{pmatrix}
	g
\]
avec $v\in\C$, $\varphi\in\R$ et $g\in {\rm SU}_2$. Maintenant, on continue
l'\'etude en groupe g\'en\'eral uniquement pour les genres $0$ et $1$.
On n'aborde pas le dernier cas connu~: 
genre $\geq 2$, $G={\rm SU}_2$. Au passage, 
c'est le seul cas o\`u on a construit une d\'ecoupe explicite de l'espace 
des modules en genre sup\'erieur. 
Les r\'ef\'erences les plus abouties sur le calcul du produit scalaire
sont~\cite{gaw91:scalar} pour le genre z\'ero,~\cite{gaw97:unitarity}
pour le genre un --- les deux articles traitant le groupe quelconque
avec points d'insertion --- et~\cite{gaw95:higher3,gaw95:higher1}
pour le genre sup\'erieur, dans ce dernier cas uniquement
pour ${\rm SU}_2$.
Dans les nouvelles coordonn\'ees, la mesure 
invariante $D(hh^\dagger)$ est
\[
D(bb^\dagger)=\prod_{j=1}^rd\varphi^j
\prod_{\alpha>0}d^2\big(\ee^{-\varphi_\alpha/2}v_\alpha\big)
\]
o\`u $\varphi^j=\tr\,\varphi h^j$ et $\varphi_\alpha=\alpha(\varphi)$. 

En genre z\'ero, l'action de WZNW dans les coordonn\'ees d'Iwasawa devient~:
\[
S(hh^\dagger)=-\quotient{i}{4\pi}\int\tr\,\da\varphi\wedge\de\varphi
	-\quotient{i}{2\pi}\int\tr\,\ee^\varphi\,(n^{-1}
	\de n)^\dagger\,\ee^{-\varphi}\wedge n^{-1}\de n
\]
avec, en plus, $(hh^\dagger)^{-1}=(n\,\ee^\varphi n^\dagger)^{-1}$.
Pour le moment, on a donc
\begin{equation}
\begin{split}
\scal\Psi\scal^2&=
	{\rm const.}\,\Big(\quotient{{\rm det}'\,(-\Delta)}{{\rm aire}}
	\Big)^{-\dim G}\int
	\lsca\Psi(0),
	\mathop{\otimes}\limits_\ell (n\ee^\varphi n^\dagger)
	(\xi_\ell )^{-1}\sous{R_\ell }\,\Psi(0)\rsca\\
&\ee^{-\frac{i(k+2g^\vee)}{4\pi}\int\tr\,\partial\varphi\wedge\de\varphi}\ 
\ee^{-\frac{i(k+2g^\vee)}{2\pi}\int\tr\,\ee^{\varphi}\,
	(n^{-1}\de n)^\dagger\ee^{-\varphi}\wedge n^{-1}\de n}\ 
	\delta(hh^\dagger(\xi_0))\,D(hh^\dagger).
\end{split}
\end{equation}

En genre un, le calcul est pratiquement identique.
On montre~\cite[App. B]{gaw89:coset} qu'il est toujours possible 
de d\'efinir une action de WZNW pour des champs tordus. 
La formule de Polyakov-Wiegmann jaug\'ee reste valable. 
Ainsi, bien que $\gamma_u$ soit multivalu\'ee 
(cf. \'equation~\eqref{gammau}), on peut quand m\^eme \'ecrire
\[
S(hh^\dagger, A_u)=S(\gamma_ubb^\dagger\gamma_u^\dagger)
-S(\gamma_u\gamma_u^\dagger).
\]
Pour calculer ces deux actions, on introduit les fonctions
interm\'ediaires
\begin{gather*}
\begin{aligned}
n_u&=\gamma_u\,n\,\gamma_u^{-1},\\
\varphi_u&=\varphi+\chi_u+\chi_u^\dagger,
\end{aligned}
\qquad
\begin{aligned}
n_u(z+\tau)&=\ee^{-2\pi iu}\,n_u(z)\,\ee^{2\pi iu},\\
\ee^{\varphi_u(z+\tau)}&=\ee^{-2\pi iu}\,
\ee^{\varphi_u(z)}\,\ee^{2\pi iu^\dagger}
\end{aligned}
\end{gather*}
o\`u  $\chi_u=-\pi u(z-\overline{z})/\imtau$.
Dans ces coordonn\'ees, on a $\gamma_ubb^\dagger\gamma_u^\dagger
=n_u\,\ee^{\varphi_u}\,n_u^\dagger$. La formule 
de Polyakov-Wiegmann~(2.\ref{PW})
reste valable pour des champs tordus uniquement 
dans la direction $\tau$ par
\[
g_1(z+\tau)=a\,g_1(z)\,b,\qquad g_2(z+\tau)=b^{-1}\,g_2(z)\,c.
\]
Par cons\'equent, 
\[
S(\gamma_ubb^\dagger\gamma_u^\dagger)=
	S(n_u)+S(n_u^\dagger)+S(\ee^{\varphi_u})
	-\Gamma(n_u,\ee^{\varphi_u}\,n_u^\dagger)
	-\Gamma(\ee^{\varphi_u},n_u^\dagger).
\]
Le dernier terme est nul car $\ee^{\varphi_u}\bot n_u^\dagger$ 
pour la forme de Killing. On montre (loc. cit.) 
que $S(n_u)=S(n_u^\dagger)=0$ et
\[
S(\ee^{\varphi_u})=-\quotient{i}{4\pi}\int_\Sigma\tr\,
\partial\varphi_u\wedge\de\varphi_u.
\]
On trouve alors que, dans les coordonn\'ees
d'Iwasawa, l'action de WZNW est donn\'ee par
\[
S(hh^\dagger, A_u)=-\quotient{i}{4\pi}\int_\Sigma\tr\,
\partial\varphi\wedge\de\varphi
	-\quotient{i}{2\pi}\int_\Sigma\tr\,
	\ee^{\varphi_u}\,(n_u^{-1}\de n_u)^\dagger\ee^{-\varphi_u}
	\wedge n_u^{-1}\de n_u.
\]
Pour la mesure invariante, comme on a
\[
n_u=\ee^{\sum_{\alpha>0}v'_\alpha\,e_\alpha}\qquad\mbox{o\`u}\qquad
	v'_\alpha=\ee^{-\pi(z-\bar z)u_\alpha/\imtau}\ v_\alpha,
\]
il suit
\[
D(hh^\dagger)=\prod_{j=1}^rd\varphi^j\ \prod_{\alpha>0\atop z}d^2
	\big(\ee^{-\alpha(\varphi_u(z))/2}v'_\alpha(z)\big).
\]
On change ensuite la fonction
$\Psi(A^{01}_u)$ en la fonction holomorphe 
$\Bgamma:\lietc\rightarrow \esprep$~:
\[
\lsca\Psi(A^{01}_u),
\mathop{\otimes}\limits_\ell (hh^\dagger)
(\xi_\ell )^{-1}\sous{R_\ell }\,\Psi(A^{01}_u)\rsca
=\ee^{\,\pi k\,(|u|^2+|u^\dagger|^2)/(2\imtau)}
	\,\lsca\Bgamma(u),
	\mathop{\otimes}\limits_\ell(n_u\ee^{\varphi_u}
	n_u^\dagger)(\xi_\ell)^{-1}\sous{R_\ell}
	\Bgamma(u)\rsca.
\]
\`A la fin, on trouve
\begin{align*}
\scal\Psi\scal^2=\,{\rm const.}\,\int&
	\lsca\Bgamma(u),
	\mathop{\otimes}\limits_\ell(
	n_u\ee^{\varphi_u}n_u^\dagger)(\xi_\ell)^{-1}\sous{R_\ell}
	\Bgamma(u)\rsca\ 
	\ee^{-\frac{i(k+2g^\vee)}{4\pi}\int\tr\,
	\partial\varphi\wedge\de\varphi}\\
&\ee^{-\frac{i(k+2g^\vee)}{2\pi}\int\tr\,\ee^{\varphi_u}\,
	(n_u^{-1}\de n_u)^\dagger\ee^{-\varphi_u}\wedge n_u^{-1}\de n_u}\ 
	\ee^{\frac{\pi(k+2g^\vee)}{2\imtau}\,|u-u^\dagger|^2}\\
&\big|\Pi(u,\tau)|^4\ \delta(\varphi(\xi_0))\  
	\prod_{j=1}^rd\varphi^j\ \prod_{\alpha>0\atop z}d^2
	\big(\ee^{-\alpha(\varphi_u(z))/2}v'_\alpha(z)\big)\,d^{2r}\!u .
\end{align*}

Pour rendre les int\'egrales quadratiques, on utilise
de nouvelles variables $(\eta_\alpha)_{\alpha>0}$ d\'efinies
par
\begin{equation}
\label{inversion}
n^{-1}\de n=\sum_{\alpha>0}\de\eta_\alpha\,e_\alpha.
\end{equation}
Le changement de variables $(v_\alpha)\mapsto (\eta_\alpha)$, pour le genre
z\'ero, ou $(v'_\alpha)\mapsto(\eta_\alpha)$, pour le genre un,
est clairement expliqu\'e dans les articles originaux. 
On utilise alors un \<<truc\>> qui permet d'inverser 
la relation~\eqref{inversion}. 
Ainsi, $n_u$ est obtenu en fonction des $\eta_\alpha$ par
l'interm\'ediaire d'int\'egrales en $y_s$ sur les
fonctions de Green de l'op\'erateur $\de$, \'eventuellement
tordu.  On int\`egre alors sur les $\eta_\alpha$. 
On appelle l'int\'egrale r\'esiduelle sur les $\varphi^j$ 
une {\bol repr\'esentation en gaz de Coulomb}. 
On utilise cette terminologie \`a cause de la ressemblance 
avec la repr\'esentation en int\'egrale fonctionnelle
des poids de Boltzmann pour un gaz de Coulomb. Dans ce langage,
on a des charges externes aux points $\xi_\ell$, des charges d'\'ecran
aux points $y_s$, et une charge \`a l'infini en $\xi_0$.
Apr\`es int\'egration sur les $\varphi^j$, on tombe
enfin sur des int\'egrales de dimension finie.

Si $\lambda_1,\cdots,\lambda_N$ est la s\'equence de plus hauts poids,
on note $\langle \lambda|=\otimes_\ell\langle\lambda_\ell|$, 
$\un\alpha$ une s\'equence $(\alpha_{1,1},\cdots,\alpha_{1,K_1},\alpha_{2,1},\cdots\cdots,
\alpha_{N,K_N})\equiv(\alpha_1,\cdots,\alpha_K)$ de
$K=\sum_\ell K_\ell$ racines simples telles que
\[
\sum_{s=1}^K\alpha_s=\sum_{\ell=1}^N\lambda_\ell
\]
et $\un y=(y_{1,1},\cdots,y_{1,K_1},\alpha_{2,1},\cdots\cdots,
y_{N,K_N})\equiv(y_1,\cdots,y_K)$ une s\'equence de $K$ points sur $\Sigma$.
Deux s\'equences $\un\alpha$ de racines simples ne diff\`erent
que par une permutation $\sigma\in\mathfrak{S}_K$. On introduit
la constante $\kappa=k+g^\vee$.

En genre z\'ero, le produit scalaire est 
\[
\scal \Psi\scal^2={\rm const.}\,
\int\left\vert\,\ee^{-\frac{1}{\kappa}\,S^{(0)}(\un{z},\un{y})}
\langle\, G^{(0)}(\un{z},\un{y}),\Psi(0)\,\rangle\,\right\vert^2\ 
\prod\limits_{s=1}^Kd^2y_s
\]
avec
\[
S^{(0)}(\un{z},\un{y})=
	\sum\limits_{\ell<\ell'}\tr(\lambda_\ell\lambda_{\ell'})\,
	\ln(z_\ell-z_{\ell'})-\sum\limits_{\ell,s}\tr(\lambda_\ell\alpha_s)\,
	\ln(z_\ell-y_s)+\sum\limits_{s<s'}
	\tr(\alpha_s\alpha_{s'})\, \ln(y_s-y_{s'})
\]
mais aussi
\[
G^{(0)}(\un{z},\un{y})=\sum\limits_{\un K} \sum\limits_{\sigma\in \mathfrak{S}_K}F_{\un{K}}
(\un{z},\sigma\un{y})
\langle\lambda|\mathop{\otimes}\limits_\ell(e_{(\sigma\alpha)_{\ell,1}}
\cdots e_{(\sigma\alpha)_{\ell,K_\ell}})_\ell,
\]
prenant ses valeurs dans l'espace dual \`a $V_{\un\lambda}$, et
\[
F_{\un{K}}(\un{z},\un{y})=
{_1\over^{\pi^{K}}}\prod\limits_\ell
{_1\over^{z_\ell-y_{\ell,1}}}
{_1\over^{y_{\ell,1}-y_{\ell,2}}}\cdots
{_1\over^{y_{\ell,K_\ell-1}-y_{\ell,K_\ell}}}.
\]

En genre un, on obtient une expression tout \`a fait semblable
\[
\scal \Psi\scal^2={\rm const.}\,\tau_2^{-r/2}
	\int\ee^{\frac{\pi \kappa}{2\tau_2}\,|w-\bar w|^2}\,\left\vert\,
	\ee^{-\frac{1}{\kappa}\,S^{(1)}(\tau,\un{z},\un{y})}
	\langle\, G^{(1)}(\tau,u,\un{z},\un{y}),\theta(u)\,\rangle\,
	\right\vert^2\ 
	d^{2r}u\prod\limits_{s=1}^Kd^2y_s
\]
avec  $\theta(u)=\Pi(u,\tau)\,\gamma(u)$, $\Pi$ \'etant
le d\'enominateur de Weyl-Kac, et
\[
w\equiv u+{_1\over^{\kappa}}\sum\limits_{\ell=1}^Nz_\ell\lambda_\ell
-{_1\over^{\kappa}}\sum\limits_{s=1}^Ky_s\alpha_s.
\]
La fonction $S^{(1)}$ est holomorphe et  multivalu\'ee en $\tau$, $z_\ell$ et $y_s$,
la fonction $G^{(1)}$ est holomorphe multivalu\'ee et prend ses
valeurs dans l'espace dual \`a $V_{\un\lambda}$. Explicitement, on a
\[
S^{(1)}(\tau,\un{z},\un{y})=
	\sum\limits_{\ell<\ell'}\tr(\lambda_\ell\lambda_{\ell'})\,
        \widetilde{\vartheta}_1
	(z_\ell-z_{\ell'})-\sum\limits_{\ell,s}\tr(\lambda_\ell\alpha_s)\,
        \widetilde{\vartheta}_1
	(z_\ell-y_s)+\sum\limits_{s<s'}
        \tr(\alpha_s\alpha_{s'})\, 
	\widetilde{\vartheta}_1
	(y_s-y_{s'})
\]
avec  $\widetilde{\vartheta}_1(z)\equiv\vartheta_1(z)/\vartheta'_1(0)$,
\[
G^{(1)}(\tau,u,\un{z},\un{y})=
	\sum\limits_{\un K} \sum\limits_{\sigma\in \mathfrak{S}_K}
	F_{\un{K},\sigma\un\alpha}
	(\tau,u,\un{z},\sigma\un{y})
	\langle\lambda|\mathop\otimes\limits_\ell(e_{(\sigma\alpha)_{\ell,1}}
	\cdots e_{(\sigma\alpha)_{\ell,K_\ell}})_\ell
\]
o\`u 
\begin{align*}
F_{\un{K},\un{\alpha}}(\tau,u,\un{z},\un{y})=\quotient{1}{\pi^K}
	\prod\limits_\ell\, &
	P_{\alpha_{\ell,1}(u)+\cdots+\alpha_{\ell,K_\ell}(u)}
	(z_\ell-y_{\ell,1})\,
	P_{\alpha_{\ell,2}(u)+\cdots+\alpha_{\ell,K_\ell}(u)}
	(y_{\ell,1}-y_{\ell,2})\\
&\times\,\cdots P_{\alpha_{\ell,K_\ell}(u)}(y_{\ell,K_\ell-1}-y_{\ell,K_\ell})
\end{align*}
o\`u $P_x(y)$ est la fonction d\'efinie dans la section {\bf 4.4}.
Les fonctions sont multivalu\'ees s\'epar\'ement, mais mises 
ensemble, on obtient un int\'egrand monovalu\'e.

Le probl\`eme de la convergence de ces int\'egrales reste
ouvert. Il a \'et\'e montr\'e
dans un certain nombre de cas~\cite{gaw96:elliptic,gaw89:quadrature}
et il a \'et\'e conjectur\'e pour le cas g\'en\'eral~\cite{gaw89:construc}
que le produit scalaire converge si, et seulement si,
on le calcule sur des \'etats de CS qui satisfont les r\`egles de
fusion.



\chapter[Syst\`emes de Hitchin. Connexion de KZB]%
	{Syst\`emes de Hitchin\\ Connexion de Knizhnik-Zamolodchikov-Bernard}

\medskip
\section{G\'eom\'etrie symplectique}

Comme r\'ef\'erences sur le sujet, j'utilise
les ouvrages~\cite{abraham,mcduff}
et l'article~\cite{cartier:int}.

\subsection{Structure symplectique}

Une {\bol vari\'et\'e de Poisson} est une vari\'et\'e $M$
\'eventuellement de dimension infinie~--- par commodit\'e, on 
utilise la cat\'egorie r\'eelle mais l'expos\'e reste valable dans 
la cat\'egorie complexe~--- munie d'un {\bol crochet de Poisson}, \cad d'une 
application bilin\'eaire antisym\'etrique $\{.,.\}$ sur l'espace 
$\CC^\infty(M)$ v\'erifiant la r\`egle de Leibniz et l'identit\'e de Jacobi~:

	{\bf 1)} $\{\varphi_1,\varphi_2\}=-\{\varphi_2,\varphi_1\}$
		\quad(antisym\'etrie)~;

	{\bf 2)} $\{\varphi,\varphi_1\varphi_2\}=\{\varphi,\varphi_1\}\,
		\varphi_2+\varphi_1\{\varphi,\varphi_2\}$
		\quad(r\`egle de Leibniz)~;

	{\bf 3)} $\{\varphi_1,\{\varphi_2,\varphi_3\}\}
		+\{\varphi_2,\{\varphi_3,\varphi_1\}\}+
		\{\varphi_3,\{\varphi_1,\varphi_2\}\}=0$
		\quad(identit\'e de Jacobi).

\noindent 
Le {\bol tenseur de Poisson} $J\in\Lambda^2 TM$ est d\'efini
par $\{\varphi,\psi\}=J(d\varphi\wedge d\psi)$. 
On consid\`ere aussi l'application $\widetilde{J}:T^*M\rightarrow TM$ 
telle que $\langle \widetilde{J}(d\varphi), d\psi\rangle=\{\varphi,\psi\}$
avec des notations \'evidentes. 
Deux fonctions $\varphi,\psi\in\CC^\infty(M)$ sont dites {\bol en involution}
si leur crochet de Poisson est nul. Une {\bol feuille symplectique}
est une sous-vari\'et\'e maximale de $M$ sur laquelle toute fonction
$\varphi$ en involution avec tous les $\psi\in\CC^\infty(M)$
est constante.

Une {\bol vari\'et\'e symplectique} $M$ est une vari\'et\'e~---
de dimension paire si $M$ est de dimension finie~---
 munie d'une $2$-forme $\omega$
ferm\'ee ($d\omega=0$) et non-d\'eg\'en\'er\'ee, 
appel\'ee {\bol forme symplectique}. On dit que $\omega$ est
non-d\'eg\'en\'er\'ee si l'application $\lambda:TM\rightarrow T^*M$
d\'efinie par $\omega(v,w)=\langle w,\lambda(v)\rangle$ est inversible. 
On note $\widetilde{J}$ son inverse. 
Une vari\'et\'e symplectique est toujours une vari\'et\'e de Poisson avec
\[
\{\varphi,\psi\}\,\equiv\,\langle \widetilde{J}(d\varphi), d\psi\rangle
        =\omega(\widetilde{J}(d\psi),\widetilde{J}(d\varphi)).
\]
Une feuille symplectique d'une vari\'et\'e de Poisson $M$
poss\`ede une structure symplectique naturelle telle que son
crochet de Poisson co\"{\i}ncide avec la restriction du crochet
de Poisson de $M$. 
Si $\varphi\in\CC^\infty(M)$, son champ de vecteurs hamiltonien
est $v_\varphi=\widetilde{J}(d\varphi)$, 
soit $d\varphi=i(v_\varphi)\,\omega$. 
Plus g\'en\'eralement, on dit qu'un {\bol champ de vecteurs} $v$ est
{\bol hamiltonien} (resp. {\bol canonique})
si $i(v)\,\omega$ est exacte (resp. ferm\'ee). Lorsque $v$ est hamiltonien,
une fonction $H_{v}\in\CC^\infty(M)$, telle que $dH_{v}
=i(v)\,\omega$, est appel\'ee un {\bol Hamiltonien} associ\'e \`a $v$.
Soit $\mathfrak{X}(M)$ l'alg\`ebre de Lie des champs de vecteurs sur $M$,
munie du crochet de Lie $[.,.]$ d\'efini par le commutateur.
L'application $\boldsymbol{v}:\CC^\infty(M)\rightarrow \mathfrak{X}(M)$ qui
envoie une fonction sur son champ de vecteurs hamiltonien est
un homomorphisme d'alg\`ebres de Lie.

\subsection{Dual d'une alg\`ebre de Lie}

Soit $G$ un groupe de Lie et
$\lieg$ son alg\`ebre de Lie identifi\'ee avec $T_eG$. 
L'action du groupe $G$ sur lui-m\^eme ($h\mapsto ghg^{-1}$) induit
une action naturelle de $G$ sur $\lieg$, l'action adjointe $\Ad$, et, par
transposition, l'action coadjointe $\Ad^*$ sur $\lieg^*$. Les 
g\'en\'erateurs infinit\'esimaux (cf. p. \pageref{geneinfini})
de ces deux actions sont l'action adjointe $\ad$ de $\lieg$ sur 
$\lieg$ et l'action coadjointe $\ad^*$ de $\lieg$ sur $\lieg^*$.
Afin de rendre les choses plus explicites, posons $L_g:h\mapsto gh$ la 
translation \`a gauche par $g$ et $R_g:h\mapsto hg$ la translation 
\`a droite par $g$. Si $X,Y\in\lieg$, $\xi\in\lieg^*$,
\begin{gather*}
\begin{aligned}
\Ad_gX & =T_e(R_{g^{-1}}L_g)\,X,\\
\langle Y,\Ad^*_g\,\xi\rangle &=\langle \Ad_{g^{-1}}\,Y, \xi\rangle,
\end{aligned}
\qquad
\begin{aligned}
\ad_XY & =[X,Y],\\
\langle Y,\ad^*_X\,\xi\rangle & =-\langle[X,Y],\xi\rangle.
\end{aligned}
\end{gather*}
Pour simplifier, on note
\[
\Ad_gX=gXg^{-1},\qquad
	\Ad^*_g\,\xi= g\xi g^{-1},\qquad
	\ad^*_X\,\xi=[X,\xi].
\]

Les {\bol orbites coadjointes} $\CO$ poss\`edent une structure naturelle de
vari\'et\'e symplectique. Soit $\xi$ pris dans l'orbite $\CO$.
On note $G_\xi$ le groupe d'isotropie de $\xi$ et
$\lieg_\xi$ son alg\`ebre de Lie~:
\[
G_\xi=\{g\in G\ \big|\ g\xi g^{-1}=\xi\}\qquad {\rm et} \qquad
\qquad \lieg_\xi=\{X\in\lieg\ \big|\ [X,\xi]=0\}.
\]
L'orbite  $\CO$ est naturellement isomorphe au quotient
$G/G_\xi$, d'o\`u $T_\xi\CO\cong \lieg/\lieg_\xi$. La forme
bili\-n\'eaire antisym\'etrique sur $\lieg$
\[
\omega_\xi(X,Y)=\langle[X,Y],\xi\rangle=-\langle Y,[X,\xi]\rangle
\]
descend bien sur $\lieg/\lieg_\xi$ car $\omega_\xi(X,.)=0$ 
\'equivaut \`a $X\in\lieg_\xi$. L'application $\xi\mapsto\omega_\xi$ d\'efinit
une $2$-forme sur $\CO$. Utilisant l'identit\'e de Jacobi, on
v\'erifie ensuite que $\omega$ est ferm\'ee. 

Le dual d'une alg\`ebre
de Lie est un exemple de vari\'et\'e qui n'est pas symplectique mais
qui admet une structure de Poisson. Soient $\varphi,\psi\in\CC^\infty(\lieg^*)$
et $\xi\in\lieg^*$, le {\bol crochet de Lie-Poisson} est 
\[
\{\varphi,\psi\}_{\mbox{\tiny\rm L.P.}}(\xi)=
	\langle\, [d\varphi(\xi),d\psi(\xi)],\xi\,\rangle
\]
o\`u on a identifi\'e une diff\'erentielle $d\varphi$ en un point $\xi$ avec
une fonction lin\'eaire sur $\lieg^*$, \ie un \'el\'ement de $\lieg^{**}
\cong\lieg$. Si $(t^a)$ est une base de $\lieg$, $[t^a,t^b]
=if^{abc}\,t^c$, les fonctions
$\xi^a=\langle t^a,.\rangle$ sur $\lieg^*$ forment un syst\`eme
de coordonn\'ees sur $\lieg^*$. On obtient
\[
\{\varphi,\psi\}_{\mbox{\tiny\rm L.P.}}=
	\sum_{a,b,c} if^{abc}\,\frac{\partial\varphi}
	{\partial\xi^a}\,\frac{\partial\psi}{\partial\xi^b}\,\xi^c.
\]
Les feuilles symplectiques de $(\lieg^*,\{.,.\}
_{\mbox{\tiny\rm L.P.}})$ sont les orbites coadjointes.

\section{Moments}

Soit $(M,\omega)$ une vari\'et\'e symplectique, $\{.,.\}$ le crochet de Poisson
associ\'e, $G$ un groupe de Lie d'alg\`ebre de Lie $\lieg$. 
Le groupe $G$ agit sur $M$ par $\Phi:
G\times M\ni (g,x)\mapsto \Phi(g,x)=\Phi_g(x)\in M$. On suppose que 
l'{\bol action} est {\bol symplectique}, \cad que $\Phi_g$ est un 
symplectomorphisme, \cad $\Phi_g^*\omega=\omega$
pour tout $g\in G$. L'action infinit\'esimale
induit un homomorphisme d'alg\`ebres de Lie $\theta:
\lieg\rightarrow \mathfrak{X}(M)$ associant
\`a tout $X\in\lieg$ le champ de vecteurs
\[
\theta(X)\,=\,-\,\quotient{d}{dt}\Big|\sous{t=0}\Phi_{{\rm exp}\,tX},
\]
appel\'e {\bol
g\'en\'erateur infinit\'esimal} de l'action\label{geneinfini}.
On v\'erifie ais\'ement que $(\Phi_g)_*\,\theta(X)=\theta({\rm Ad}_gX)$
et $\theta([X,Y])=[\theta(X),\theta(Y)]$. Notons que l'image 
de $\theta$ est form\'ee de champs de vecteurs symplectiques.

On suppose en plus que l'{\bol action} est {\bol hamiltonienne} \cad qu'il
existe une application
$\widetilde{\mu}:\lieg\rightarrow\CC^\infty(M)$
factorisant $\theta$ par l'interm\'ediaire de $\boldsymbol{v}$~:
\begin{equation}
\label{factorisation}
\theta\,:\,\lieg\stackrel{\widetilde{\mu}}{\longrightarrow}
        \CC^\infty(M)\stackrel{\boldsymbol{v}}{\longrightarrow}
        \mathfrak{X}(M),
\end{equation}
soit $d\widetilde{\mu}(X)=i(\theta(X))\,\omega$. En clair, on suppose
que l'image de $\theta$ est form\'ee de champs de vecteurs 
hamiltoniens. Si $\widetilde{\nu}$ factorise $\theta$ 
suivant~\eqref{factorisation}, il est clair que toute autre application
factorisant $\theta$ est de la forme $\widetilde{\mu}=
\widetilde{\nu}+\sigma$, o\`u 
$\sigma:\lieg\rightarrow \R$ est une application lin\'eaire. 
On adjoint \`a $\widetilde{\mu}$ l'application
$\mu:M\rightarrow\lieg^*$ d\'efinie par
\begin{equation}
\label{moment}
\langle X,\mu(x)\rangle\equiv\widetilde{\mu}(X)(x).
\end{equation}
L'application $\mu$ est donc d\'etermin\'ee \`a une 
constante $\sigma\in\lieg^*$ pr\`es. Lorsque l'action est hamiltonienne,
on dit que $\mu$ est un {\bol moment pour l'action}.

On a remarqu\'e que $\theta$ 
et $\boldsymbol{v}$ sont des homomorphismes d'alg\`ebres, il est donc naturel 
de trouver sous quelles contraintes on peut factoriser $\theta$ par un
homomorphisme d'alg\`ebres $\widetilde{\mu}$. Si c'est possible, on dit 
que l'{\bol action} de $G$ est {\bol poissonnienne}.
Notons que $\widetilde{\mu}$ est un homomorphisme d'alg\`ebres si,
et seulement si,
l'application $\mu:M\rightarrow\lieg^*$ est un morphisme de Poisson, \cad
$\{\varphi\circ \mu,\psi\circ \mu\}=\{\varphi,\psi\}_{\mbox{\tiny\rm L.P.}}
\circ \mu$.  Fixons un $\widetilde{\mu}$ factorisant $\theta$. On a
\[
v_{\widetilde{\mu}([X,Y])}=\theta([X,Y])=[\theta(X),\theta(Y)]
	=[v_{\widetilde{\mu}(X)},v_{\widetilde{\mu}(Y)}]=v_{
	\{\widetilde{\mu}(X),\widetilde{\mu}(Y)\}}.
\]
Il existe donc une application $\chi:\lieg\times\lieg\rightarrow \R$ telle que
\[
\{\widetilde{\mu}(X),\widetilde{\mu}(Y)\}=\widetilde{\mu}([X,Y])
        +\chi(X,Y).
\]
\`A partir de l'identit\'e de Jacobi
pour $\{.,.\}$ et de~: $\{\widetilde{\mu}([X,Y]),\widetilde{\mu}(Z)\}
=\{\{\widetilde{\mu}(X),\widetilde{\mu}(Y)\},\widetilde{\mu}(Z)\}$, on d\'eduit
\[
\chi([X,Y],Z)+\chi([Y,Z],X)+\chi([Z,X],Y)=0,
\]
\cad $\chi$ est un $2$-cocycle de l'alg\`ebre de Lie $\lieg$.
Si on change $\mu$ en $\mu+\sigma$ alors $\chi(X,Y)$ est
remplac\'e par $\chi(X,Y)-\sigma([X,Y])$. L'action de $G$ permet donc de fixer
$\chi$ \`a un cobord pr\`es, \ie une action hamiltonienne d\'etermine
une classe $[\chi]$ dans $H^2(\lieg,\R)$. En effet, un cobord est
un cocycle de la forme $\chi(X,Y)=\sigma([X,Y])$, 
o\`u $\sigma:\lieg\rightarrow\R$ est une application lin\'eaire.
Clairement, l'action est poissonnienne si, et seulement si, $[\chi]=0$, \ie
$\chi$ est un certain cobord $\sigma$. L'application $\widetilde{\mu}
+\sigma$ r\'ealise l'homomorphisme d'alg\`ebres d\'esir\'e. Cet homomorphisme
n'est pas unique. La libert\'e est cette fois donn\'ee par
une application lin\'eaire $\sigma:\lieg\rightarrow\R$ telle que
$\sigma([X,Y])=0$. Ainsi, une action poissonnienne d\'etermine une classe 
dans $H^1(\lieg,\R)$. De ce qui pr\'ec\`ede, on constate que si
$H^1(\lieg,\R)=H^2(\lieg,\R)=0$, condition r\'ealis\'ee si $\lieg$ est 
semi-simple, alors l'action est poissonnienne et un moment pour l'action 
est unique. 

On se propose de reformuler la notion de moment.
Soient $M$ et $N$ deux vari\'et\'es quelconques. Soient $\Phi$ et $\Psi$
deux actions de $G$ sur respectivement $M$ et $N$. Soit $f:M\rightarrow N$
une application $\CC^\infty$. On dit que $f$ est {\bol \'equivariante} pour ces
deux actions de $G$ si, pour tout $g\in G$, on a $f\circ\Phi_g
=\Psi_g\circ f$. On dit que l'action est (infinit\'esimalement) 
{\bol $\epsilon$-\'equivariante} si $Tf\circ \theta_\Phi=\theta_\Psi\circ f$, 
o\`u $\theta_\Phi$ et $\theta_\Psi$ sont les g\'en\'erateurs infinit\'esimaux
des actions $\Phi$ et $\Psi$. On se replace dans le cadre d'une action
hamiltonienne factorisant $\theta$ par un certain $\widetilde{\mu}$.
On va montrer le r\'esultat suivant
\[
\mbox{\rm \'equivariance de } \mu\quad
	\Longrightarrow\quad \mbox{\rm $\epsilon$-\'equivariance de }  \mu
	\quad\Longleftrightarrow
	\quad \mbox{ \rm  action poissonnienne.}
\]
La premi\`ere implication est g\'en\'erale. Dans le contexte qui nous
int\'eresse, l'\'equivariance de $\mu$ dit~: $\mu\circ\Phi_g={\rm Ad}_g^*\circ
\mu$ et l'$\epsilon$-\'equivariance dit~: $T\mu\circ\theta(X)={\rm ad}_X^*
\circ\mu$.
Pour obtenir la partie $\Longleftrightarrow$, on commence par d\'eriver
l'\'equation~\eqref{moment}~: $d\widetilde{\mu}(X). v
=\langle X,T\mu\circ v\rangle$. On en d\'eduit facilement
\[
\langle Y,T\mu\circ\theta(X)\rangle=-\{\widetilde{\mu}(X),\widetilde{\mu}(Y)\}
	\qquad = \qquad -\widetilde{\mu}([X,Y])=\langle Y,[X,\mu]\rangle,
\]
d'o\`u l'\'equivalence annonc\'ee. 
En r\'esum\'e, \'etant donn\'ee une action hamiltonienne, pour montrer
qu'elle est poissonnienne, il suffit de trouver une application 
$\widetilde{\mu}$ factorisant $\theta$ telle que $\mu$ est \'equivariante.

Par la suite, on consid\`ere l'action du groupe $\CG^\C$ 
des transformations de jauge qui est de dimension infinie. 
On est alors surtout int\'eress\'e par une application 
moment \'equivariante.

\subsection{R\'eduction symplectique}

Soient $G$ un groupe de Lie, $\Phi$ une action hamiltonienne de $G$ sur
une vari\'et\'e symplectique $(M,\omega)$. On se donne $\mu:M\rightarrow
\lieg^*$ un moment pour l'action, suppos\'e \'equivariant. 
Le quotient $\CP_\xi\equiv\mu^{-1}(\xi)/G_\xi=\mu^{-1}(\CO)/G$,
appel\'e {\bol espace des phases r\'eduit}, est correctement d\'efini.
Dans le cas particulier $\xi=0$, $\mu^{-1}(0)/G$ est appel\'e le 
{\bol quotient de Marsden-Weinstein}, parfois not\'e $M/\!\!/G$.
On suppose que l'action de $G$ et le moment sont tels que $\CP_\xi$ est
une vari\'et\'e.
L'espace des phases r\'eduit supporte une unique structure symplectique
$\omega_{\rm red}$ telle que 
\begin{equation}
\label{symred}
\pi^*\omega_{{\rm red}}=i^*\omega
\end{equation}
o\`u $\pi:\mu^{-1}(\xi)\rightarrow\CP_\xi$ est la projection canonique
et $i:\mu^{-1}(\xi)\rightarrow M$ est l'injection naturelle.
Soit $m\in \mu^{-1}(\xi)$. Si $v$ appartient \`a $T_m\mu^{-1}(\xi)$, on note
$\widetilde{v}$ son image par $T\pi$. La formule~\eqref{symred} dit seulement
$\omega_{{\rm red}}(\widetilde{v},\widetilde{w})=\omega(v,w)$, pour
$v,w\in T_m\mu^{-1}(\xi)$. Comme $\pi$
et $T\pi$ sont surjectives, l'unicit\'e de $\omega_{{\rm red}}$
 est imm\'ediate. Si $\xi$ et $\xi'$ sont dans la m\^{e}me orbite
coadjointe $\CO$, alors de fa\c{c}on naturelle
$\CP_\xi\cong\CP_{\xi'}\cong \mu^{-1}(\CO)/G\equiv \CP_{\CO}$
et cet isomorphisme pr\'eserve
la structure symplectique.

\subsection{Espaces cotangents}

On a d\'ej\`a vu que tout espace cotangent $M=T^*N$ \`a une vari\'et\'e $N$
admet une structure symplectique canonique. Notons
$\pi$ la projection canonique $\pi:M\rightarrow N$ et $T\pi$
son application tangente $T\pi:TM\rightarrow TN$.
Soient $x\in N$ et $\tau\in T^*_x N$. Si $v\in T_\tau M$, on pose
$\alpha_0(v)\,=\,\langle T\pi (v),\tau\rangle$.
La $2$-forme $\omega_0=d\alpha_0$ sur $M$ est automatiquement ferm\'ee. 
$\omega_0$ n'est pas
d\'eg\'en\'er\'ee. Suivant le {\bol th\'eor\`eme de Darboux}, toute
vari\'et\'e symplectique de dimension finie est localement symplectomorphe 
\`a un fibr\'e cotangent muni de sa structure symplectique canonique.

Soit $G$ un groupe de Lie agissant sur $N$ via un diff\'eomorphisme $\Phi$. 
L'application $\Phi_{g^{-1}}$ se rel\`eve en un symplectomorphisme
$\Psi_g\equiv T^*\Phi_{g^{-1}}:M\rightarrow M$.
C'est un fait beaucoup plus g\'en\'eral puisque
 tout diff\'eomorphisme $f:N\rightarrow N$
se rel\`eve en un  symplectomorphisme de $(M,\omega_0)$, mieux~:
$(T^*f)^*\alpha_0=\alpha_0$. 
On a donc une action symplectique de $G$ sur $M$. 
Soient $x\in N$ et $\tau\in T^*_x N$, 
d'o\`u $\Psi_g(\tau)\in T^*_{\Phi_g(x)}N$.
En fait, l'action $\Psi$ est hamiltonienne et l'application 
$\mu:M\rightarrow\lieg^*$ d\'efinie par
\begin{equation}
\label{momentcot}
\langle X,\mu(\tau)\rangle=-\alpha_0(\theta_\Psi(X))_\tau
\tag{\ref{momentcot}.a}
\end{equation}
est un moment \'equivariant pour l'action. En effet, comme $\Psi$ pr\'eserve la 
$1$-forme $\alpha_0$ et $\Psi_{{\rm exp}\,tX}$ est un flot pour 
$\theta_\Psi(X)$, on a $\pounds_{\theta_\Psi(X)}\,\alpha_0=0$. 
D'apr\`es la formule de Cartan, 
$d\widetilde{\mu}(X)=-i(\theta_\Psi(X))\,d\alpha_0=i(\theta_\Psi(X))\,\omega_0$.
Ainsi $v_{\widetilde{\mu}(X)}=\theta_\Psi(X)$ et l'action
est hamiltonienne. 
Comme $\Psi_g=T^*\Phi_{g^{-1}}$, on a $\pi\circ\Psi_g=\Phi_g\circ\pi$ 
et $T\pi\circ\theta_\Psi=\theta_\Phi\circ\pi$. Il suit
\begin{equation}
\widetilde{\mu}(X)(\tau)=-\langle T\pi\circ\theta_\Psi(X),\tau\rangle
	=-\langle \theta_\Phi(X)\circ \pi(\tau),\tau\rangle
	=-\langle \theta_\Phi(X)(x),\tau\rangle.\tag{\ref{momentcot}.b}
\addtocounter{equation}{1}
\end{equation}
L'\'equivariance de $\mu$, \cad $\mu\circ\Psi_g=g\mu g^{-1}$, s'\'ecrit 
aussi $\widetilde{\mu}(X)(\Psi_g(\tau))=\widetilde{\mu}(g^{-1}Xg)(\tau)$.
Or
\begin{align*}
\widetilde{\mu}(g^{-1}X g)(\tau)&=-\langle \theta_\Phi(g^{-1}Xg)(x),\tau\rangle
	=-\langle \Phi_g^*\theta_\Phi(X)(x),\tau\rangle\\
&=-\langle (T\Phi_{g^{-1}})\circ\theta_\Phi(X)\circ\Phi_g(x),\tau\rangle
	=-\langle \theta_\Phi(X)\circ\Phi_g(x),
        (T^*\Phi_{g^{-1}})(\tau)\rangle\\
&=\widetilde{\mu}(X)(\Psi_g(\tau)),
\end{align*}
donc $\mu$ est \'equivariante. En conclusion, toute action d'un groupe 
sur une vari\'et\'e
se rel\`eve en une action hamiltonienne sur le fibr\'e cotangent, un
moment pour l'action --- \'equivariant par construction ---
\'etant donn\'e par l'\'equation~(\ref{momentcot}.a)
ou l'\'equation~(\ref{momentcot}.b). 

On peut regarder la r\'eduction symplectique de $M$. Soit $\xi\in\lieg^*$.
Le quotient $\mu^{-1}(\xi)/G_\xi$ admet une structure symplectique.
En plus, on montre que 
\[
\mu^{-1}(\xi)/G_\xi\,\cong \,T^*(N/G_\xi)
\]
si, et seulement si, $\lieg_\xi=\lieg$.

\subsection{Syst\`emes int\'egrables}

Suivant le contexte, on peut d\'efinir de plusieurs fa\c{c}ons
un syst\`eme int\'egrable~: syst\`eme alg\'ebrique compl\`etement 
int\'egrable~\cite{DonaM}, int\'egrabilit\'e par quadrature, etc... 
L'int\'egrabilit\'e
qui nous int\'eresse est l'int\'egrabilit\'e par quadrature.
Les syst\`emes int\'egrables constituent le parfait exemple de l'interaction
entre physique~\cite{babelon} et math\'ematiques~\cite[vol. 4 
et 16]{dynamical}. 
Soit $(\CM,\omega)$ une vari\'et\'e symplectique de dimension $2n$. On dit
qu'un syst\`eme --- ayant pour espace des phases $\CM$ --- est {\bol
int\'egrable} si
on peut r\'esoudre ses \'equations du mouvement par {\bol quadrature}, \cad
par un nombre fini d'op\'erations alg\'ebriques et de calculs int\'egrals.
Liouville a montr\'e qu'il suffit de trouver $n$ fonctions ind\'ependantes
(dont le Hamiltonien) en involution. L'\'enonc\'e pr\'ecis, d\^u
\`a Arnold~\cite{arnold}, est le suivant~:

\begin{thm}{~(Liouville-Arnold)}%
Soient $n$ fonctions lisses $f_1,\cdots,f_n$ sur $\CM$, 
en involution. On suppose que les fonctions $f_i$ sont ind\'ependantes
sur les surfaces de niveau  
$\CM_\lambda=\{x\in \CM\ \big|\ 
	f_i(x)=\lambda_i,\ i=1,\cdots, n\}$, \ie
$df_1\wedge\cdots\wedge df_n\neq 0$ en tout point de $\CM_\lambda$.
Si $\CM_\lambda$ est compacte et connexe alors

	{\bf 1)} $\CM_\lambda$ est une vari\'et\'e diff\'eomorphe \`a un tore de dimension $n$,
		le {\bol tore de Liouville} $\mathbb{T}^n$~;

	{\bf 2)} dans un voisinage de $\CM_\lambda$,
	on peut introduire des coordonn\'ees $I_1,\cdots,I_n,\varphi_1,
	\cdots,\varphi_n$, $0\leq\varphi_i\leq 2\pi$, appel\'ees
	{\bol variables d'action-angle}, telles que
	la forme symplectique soit de la forme $\omega=\sum_{i=1}^n
	dI_i\wedge d\varphi_i$. Les $\varphi_i$ sont des coordonn\'ees
	sur $\mathbb{T}^n$. Les $I_i$ sont des coordonn\'ees dans la direction
	transverse \`a $\mathbb{T}^n$, ne d\'ependant que des $f_i$.
	On choisit pour le Hamiltonien $H$ une des $f_i$ par exemple $f_1$. 
	Dans les coordonn\'ees $(I_i,\varphi_i)$, les \'equations du mouvement
	sont de la forme~:
	\[
		\dot{I}_i=0,\qquad
		\dot{\varphi}_i=\{H,I_i\}=\omega_i(I_1,\cdots,I_n)~;
	\]

	{\bf 3)} un syst\`eme int\'egrable peut \^etre int\'egr\'e 
	par quadrature.

\end{thm}

Un concept particuli\`erement performant dans l'analyse des syst\`emes
int\'egrables est d\^u \`a Lax.
Une {\bol paire de Lax} est form\'ee de deux fonctions $L$ et $M$ sur
$\CM$ \`a valeurs dans une alg\`ebre de Lie $\lieg$ telle que la dynamique
du syst\`eme se r\'eduit \`a l'\'equation diff\'erentielle suivante
\[
\frac{dL}{dt}=[L,M]
\]
o\`u le crochet $[.,.]$ est le crochet de $\lieg$. 
On montre facilement qu'un syst\`eme
int\'egrable au sens de Liouville poss\`ede toujours une paire de Lax.
Notons que, m\^eme si elle existe, une paire de Lax n'est pas unique ---
l'alg\`ebre $\lieg$ peut m\^eme changer.

Les puissances de $L$ satisfont \'egalement l'\'equation diff\'erentielle
pr\'ec\'edente. Ainsi, si $I$ est une fonction $\Ad$-invariante sur $\lieg$,
les fonctions $I(L^n)$ sont des int\'egrales du mouvement, \ie sont en 
involution avec le Hamiltonien du syst\`eme. 
Pour obtenir un syst\`eme int\'egrable, on doit trouver suffisamment de 
fonctions ind\'ependantes en involution. On commence par supposer que $L$ 
d\'epend d'un {\bol param\`etre spectral} $\lambda$. Les quantit\'es 
int\'eressantes sont alors les coefficients dans le 
d\'eveloppement en $\lambda$ des $\tr\,L^n(\lambda)$. On suppose que celles-ci
sont ind\'ependantes. Il faut en plus que les int\'egrales soient
en involution, \ie
\[
\{\tr\,L^n(\lambda),\tr\,L^m(\mu)\}=0.
\]
La seule \'equation de Lax ne garantit pas cette propri\'et\'e. Par contre,
on montre~\cite{babvial} que la propri\'et\'e
d'involution des valeurs propres d'une matrice de Lax $L$ est \'equivalente
\`a l'existence d'une fonction sur l'espace des phases \`a valeurs dans
$\lieg\otimes \lieg$, not\'ee $r$, telle que
\[
\{L_1(\lambda),\, L_2(\mu)\}=[r_{12}(\lambda,\mu),L_1(\lambda)]-
	[r_{21}(\lambda,\mu),L_2(\mu)].
\]
On a utilis\'e les notations fix\'ees p.~\pageref{tenseurr}. En plus,
$r_{12}(\lambda,\mu)=r_{ab}(\lambda,\mu)\,t^a\otimes t^b$
et $r_{21}$ est obtenue en \'echangeant les espaces $1$ et $2$.
L\`a encore la matrice $r$ n'est pas unique. 
Ainsi, les $H_n(\lambda)=\tr\,L^n(\lambda)$ sont en involution. On obtient
facilement que
\[
\{H_n(\lambda),L(\mu)\}=[L(\mu),M_n(\lambda,\mu)]
\]
o\`u $M_n(\lambda,\mu)=n\,\tr_1\,L(\lambda)^{n-1}_1\,r_{21}(\lambda,\mu)$.
Le cas le plus simple est celui pour lequel la matrice 
$r$ est constante et v\'erifie
l'{\bol \'equation de Yang-Baxter classique} (EYBC)~:
\[
[r_{12},r_{13}]+[r_{12},r_{23}]+[r_{32},r_{13}]=0.
\]
Il existe bien des g\'en\'eralisations de cette \'equation~:
EYBC modifi\'ee, EYBC dynamique avec une matrice $r$ dynamique, 
EYBQuantique avec une matrice $R$ quantique, etc...

\medskip
\section{Syst\`emes de Hitchin}

La construction de Hitchin commence par la r\'eduction symplectique de
$T^*\CA^{01}$ (de dimension infinie) par le groupe des transformations
de jauge (de dimension infinie) pour obtenir un espace symplectique de
dimension finie, l'espace cotangent \`a $\CN_s$. On regarde l'espace cotangent
$T^*\CA^{01}$ comme l'ensemble 
des paires $(A^{01},\varphi^{10})$, o\`u $A^{01}\in\CA^{01}$ et
$\varphi^{10}$ est une $(1,0)$-forme sur $\Sigma$ \`a valeurs dans $\liegc$,
appel\'ee {\bol champ de Higgs}. Il existe une $1$-forme canonique $\alpha$
sur $T^*\CA^{01}$ donn\'ee par
\[
\langle \delta A^{01}, \delta\varphi^{10}\,|\,\alpha\rangle =
\int_\Sigma \tr\,\varphi^{10}\wedge\delta A^{01}
\]
o\`u $\delta A^{01}$ repr\'esente un vecteur tangent \`a $\CA^{01}$ en 
$A^{01}$ et $\delta\varphi^{10}$ est un vecteur tangent en $\varphi^{10}$.
L'espace cotangent $T^*\CA^{01}$ muni de la $2$-forme $\Omega\equiv d\alpha$ 
est un espace symplectique de dimension infinie. Concr\`etement, on a 
\[
\langle\delta A^{01}_1,\delta\varphi^{10}_1,
        \delta A^{01}_2,\delta\varphi^{10}_2\,|\,\Omega\rangle
=\int_\Sigma \tr\, (\delta\varphi^{10}_1\wedge\delta A^{01}_2
        -\delta\varphi^{10}_2\wedge\delta A^{01}_1).
\]
On peut relever l'action $\Phi_h:A^{01}\mapsto {}^hA^{01}$ de $\CG^\C$ 
sur $\CA^{01}$ en une action hamiltonienne $\Psi_h$ ($=T^*\Phi_{h^{-1}}$)
sur l'espace cotangent $T^*\CA^{01}$ par
$(A^{01},\varphi^{10})\mapsto ({}^hA^{01},{}^h\varphi^{10})$
o\`u ${}^h\varphi^{10}=h\,\varphi^{10}h^{-1}$~(\footnote{
En effet, pour tout $v=h\,\delta A^{01} h^{-1}$, vecteur tangent 
\`a $\CA^{01}$ en ${}^hA^{01}$, on doit avoir
$\langle v,\Psi_h(A^{01},\varphi^{10})\rangle=
\langle T\Phi_{h^{-1}}(v), (A^{01},\varphi^{10})\rangle$.
Ainsi, comme $T\Phi_{h^{-1}}(\delta A^{01})=h^{-1}\delta A^{01} h$, on a 
$\langle h\,\delta A^{01} h^{-1}\,,\,({}^hA^{01},{}^h\varphi^{10})\rangle =
\langle \delta A^{01}\,,\,(A^{01},\varphi^{10})\rangle$. On trouve
imm\'ediatement ${}^h\varphi^{10}=h\,\varphi^{10}h^{-1}$.}). 
Soit $\lambda$ un \'el\'ement de l'alg\`ebre de Lie de $\CG^\C$.
Le g\'en\'erateur infinit\'esimal de l'action $\Phi$ est
$\theta(\lambda)(A^{01})=\de\lambda+A^{01}\lambda-\lambda A^{01}$.
Par cons\'equent, un moment $\mu:T^*\CA^{01}\rightarrow
(\mbox{\rm Lie}\,\CG^\C)^*$ est donn\'e par
\[
\langle \lambda\,,\,\mu(A^{01},\varphi^{10})\rangle
=-\langle \theta(\lambda)(A^{01})\,,\,(A^{01},\varphi^{10})\rangle.
\]
Ainsi, on a
\[
\langle \lambda\,,\,\mu(A^{01},\varphi^{10})\rangle=
	-\int_\Sigma \tr\,\varphi^{10}\wedge 
	(\de\lambda+A^{01}\lambda-\lambda A^{01})
\]
Apr\`es int\'egration par parties, on constate qu'un moment 
pour l'action de $\CG^\C$ sur $T^*\CA^{01}$ est 
\[
\mu(A^{01},\varphi^{10})\,= -[\de \varphi^{10}+A^{01}\wedge\varphi^{10}
                +\varphi^{10}\wedge A^{01}].
\]
Cette derni\`ere prend ses valeurs dans l'espace
des $2$-formes sur $\Sigma$ \`a valeurs
dans $\lieg^\C$, \ie dans le dual de l'alg\`ebre de Lie de $\CG^\C$. Par 
construction, le moment est \'equivariant vis \`a vis de $\CG^\C$. 
Par r\'eduction symplectique,
la $2$-forme $\Omega$ descend en une forme symplectique $\omega$ sur 
$\CP\equiv\mu^{-1}(0)/\CG^\C$. Comme un champ de Higgs tel que 
$\mu(A^{01},\varphi^{10})=0$ induit une forme lin\'eaire sur $\CA^{01}$
nulle sur les vecteurs de $\CA^{01}$ tangents \`a l'orbite de $\CG^\C$, on
a $\CP\cong T^*\CN$\label{zeromodehit}. Pour obtenir un espace des
phases lisse, on doit se restreindre \`a la strate stable de $\CA^{01}$, 
op\'eration toujours sous-entendue m\^eme si cela n'appara\^\i t 
pas explicitement dans les notations. 

On peut g\'en\'eraliser cette construction. Soient $\underline{\xi}$ une 
s\'equence de $N$ points distincts sur $\Sigma$ et $(\CO_\ell)$ une
s\'equence d'orbites coadjointes dans la sous-alg\`ebre de Cartan 
$\lietc$. On pose
\[
\CO=\Big\{\sum_\ell\lambda_\ell\,\delta_{\xi_\ell}\ \Big|\ 
	\lambda_\ell\in\CO_\ell\Big\}
\]
o\`u $\delta_{\xi_\ell}$ est la mesure de Dirac en $\xi_\ell$.
On peut r\'eduire symplectiquement $T^*\CA^{01}$ suivant
\[
\CP_{\CO}\,=\,\mu^{-1}(\{\CO\})\big/\CG^\C\,\cong\,
        \mu^{-1}\Big(\sum_\ell\lambda_\ell\,\delta_{\xi_\ell}\Big)
        \bigg/\CG^\C_{\underline{\xi},\underline{\lambda}}
\]
o\`u $\CG^\C_{\underline{\xi},\underline{\lambda}}$ est le sous-groupe
de $\CG^\C$ fixant $\sum_\ell\lambda_\ell\,\delta_{\xi_\ell}$ dans l'action 
coadjointe de $\CG^\C$. Il existe une g\'en\'eralisation des notions de stabilit\'e
et semi-stabilit\'e \`a des paires de Higgs $(A^{01},\varphi^{10})$~(\footnote{
La g\'en\'eralisation du syst\`eme de Hitchin au cas avec
points d'insertion est pr\'esent\'ee dans~\cite{beauville3,EnR,Mark,Nekr}.
G\'eom\'etriquement, une paire de Higgs tordue par $L$, fibr\'e
en droites de degr\'e positif, est form\'ee d'un fibr\'e
$E$ et d'un homomorphisme $\varphi:E\rightarrow E\otimes L$. Un sous-fibr\'e de
Higgs est une paire de Higgs $(F,\psi)$ telle que $F$ est un sous-fibr\'e
$\varphi$-invariant de $E$ et $\psi$ est la restriction de $\varphi$. La paire 
$(E,\varphi)$ est stable (resp. semi-stable) si pour tout
sous-fibr\'e de Higgs $\mu(F)<\mu(E)$ (resp. $\mu(F)\leq\mu(E)$).
Il existe alors un espace des modules des fibr\'es de Higgs tordus par $L$
lisse, en ne consid\'erant que la partie stable, avec
une compactification semi-stable.}).
En ne consid\'erant que des paires de Higgs stables, on obtient
un espace $\CP_\CO$ lisse admettant une compactification (semi-stable).
La forme $\Omega$ descend sur la partie lisse de 
$\CP_\CO$ pour donner une forme symplectique $\omega$.

Soit $p$ un polyn\^ome homog\`ene sur $\lieg^\C$, $\Gc$-invariant, de 
degr\'e $d_p$. Par exemple, pour $G=\Su$, on a un seul choix 
$p_2(X)=\tr\,X^2$. L'application 
$T^*\CA^{01}\rightarrow \Gamma(K^{d_p})$
qui envoie un point $(A^{01},\varphi^{10})\in T^*\CA^{01}$ 
en la diff\'erentielle $p(\varphi^{10})$ de degr\'e $d_p$ sur $\Sigma$,
est constante sur les $\CG^\C$-orbites comme $p$ est $\Gc$-invariant. 
Si $\mu(A^{01},\varphi^{10})=0$, alors $p(\varphi^{10})$
est holomorphe. On obtient ainsi une {\bol application de Hitchin}
\[
h_p:\CP\rightarrow H^0(K^{d_p})\,.
\]
Les composantes de $h_p$ sont en involution sur $\CP$ --- elles sont d\'ej\`a en
involution sur $T^*\CA^{01}$ puisqu'elles ne d\'ependent que de $\varphi^{10}$. 
En prenant un syst\`eme complet de polyn\^omes invariants,
Hitchin a montr\'e qu'on obtient un syst\`eme complet de Hamiltoniens
en involution sur $\CP$, appel\'e {\bol syst\`eme de Hitchin}~\cite{hitchin}.
Pour les groupes matriciels, les valeurs des Hamiltoniens en un point de $\CP$
sont cod\'ees dans la {\bol courbe spectrale} $\CS$. La courbe spectrale
est l'ensemble des $\lambda\in K$ tels que 
\[
\det(\lambda-\varphi^{10})=0.
\]
La courbe spectrale est un rev\^etement ramifi\'e de $\Sigma$.
Par exemple, pour $G^\C={\rm GL}_r\C$,
\[
\det(\lambda-\varphi^{10})=\sum_{i=1}^r
	\lambda^{r-i}\,\tr\,\Lambda^i\varphi^{10}
\]
et $\CS$ est un rev\^etement ramifi\'e \`a $r$ feuillets
de genre $g(\CS)=\dim\CN_s$ ($=r^2(g-1)+1$ si $g\geq 2$). Les espaces propres $\ell_\lambda$
forment un fibr\'e en droites holomorphe $\ell$ au-dessus de $\CS$. Le fibr\'e
$\ell$ est un point dans une vari\'et\'e jacobienne de $\CS$ sur laquelle
les Hamiltoniens induisent des flots lin\'eaires. En clair, 
la courbe spectrale contient l'information qui permet de trouver 
les variables d'action et la vari\'et\'e jacobienne contient celle qui permet 
de trouver les variables d'angle. Pour les autres groupes $G^\C$,
les tores de Liouville sont des sous-vari\'et\'es des vari\'et\'es
jacobiennes des courbes spectrales.

Avec des points d'insertion sur $\Sigma$, on obtient de m\^eme
une application de Hitchin
\[
h_p:\CP_\CO\mapsto H^0\big(K^{d_p}\big(d_p\,\sumini\,\xi_\ell\big)\big)
\]
prenant ses valeurs dans l'espace des $d_p$-formes m\'eromorphes
avec \'eventuellement des p\^oles d'ordre $\leq d_p$ en $\xi_\ell$.
On obtient encore un syst\`eme int\'egrable sur $\CP_\CO$.

\medskip
\section{Espace des modules des surfaces de Riemann}

L'{\bol espace des modules} $\moduleR$ est l'ensemble des classes 
d'isomorphismes d'objets $(\Sigma,S)$ o\`u $\Sigma$ est une surface 
de Riemann (compacte, connexe) de genre $g$ et $S$ est une s\'equence ordonn\'ee
de $N$ points $\xi_{1},\cdots,\xi_{N}$ deux \`a deux distincts sur 
$\Sigma$.

Soit $\pi_{g,N}$ le groupe libre engendr\'e par les g\'en\'erateurs 
$a_1,\cdots,a_g,b_1,\cdots,b_g,c_1,\cdots,c_N$ et la relation
\[
\big(\prod_{i=1}^ga_ib_ia_i^{-1}b_i^{-1}\big)c_1\cdots c_N=1.
\]
Une surface de Riemann  marqu\'ee de genre $g$ est une classe 
d'\'equivalence de triplets $(\Sigma,\alpha,S)$ o\`u
$\alpha:\pi_{g,N}\rightarrow\pi_1(\Sigma\setminus S)$ est un isomorphisme 
tel que $\alpha(c_i)$ est proprement homotope \`a un lacet orient\'e 
positivement autour de $\xi_{i}$ et si $N=0$ tel que la forme 
d'intersection $(\alpha(a_i),\alpha(b_i))=+1$. Deux triplets 
$(\Sigma,\alpha,S)$ et $(\Sigma',\alpha',S')$ sont 
\'equivalents s'il existe un isomorphisme $\iota:\Sigma\rightarrow\Sigma'$
tel que $\iota(\xi_i)=\xi'_i$ et $\iota_*\alpha$ diff\`ere de $\alpha'$
par un automorphisme int\'erieur, o\`u $\iota_*$ est l'application induite 
par $\iota$ sur les groupes fondamentaux. L'{\bol espace de Teichm\"uller}
$\CT_{g,N}$ est l'ensemble de toutes les surfaces de Riemann marqu\'ees 
de genre $g$.

Le {\bol groupe modulaire} $\Gamma_{g,N}$ est l'ensemble des automorphismes 
$\sigma$ de $\pi_{g,N}$ tels que $\sigma(c_i)$ est conjugu\'e \`a $c_i$ et, si 
$N=0$, $\sigma$ pr\'eserve l'orientation modulo les automorphismes 
int\'erieurs. Le groupe $\Gamma_{g,N}$ est un sous-groupe discret des 
automorphismes de $\CT_{g,N}$ agissant sur $\CT_{g,N}$ via 
$(\Sigma,\alpha,S)\mapsto (\Sigma,\alpha\circ\sigma,S)$. L'espace des 
modules $\moduleR$ est alors le quotient de $\CT_{g,N}$ par le groupe 
modulaire et a pour dimension $3g-3+N$ tant que $2g-2+N>0$. Les cas 
exotiques sont $(g,N)=(0,0),(0,1),(0,2),(1,0)$~:

--- en genre z\'ero, les automorphismes de $\C P^1$ sont les transformations 
	de M\"obius. On peut donc envoyer trois points sur $0,1,\infty$. 
	L'espace des modules est donc r\'eduit \`a un point pour $N\leq 3$. 
	Par contre, $\CM_{0,4}=\C P^1\setminus\{0,1,\infty\}$ et 
	$\CM_{0,N}\cong(\C P^1\setminus\{0,1,\infty\})^{N-3}\setminus 
	\{\text{diagonales}\}$, si $N\geq 5$~;

--- les rappels sur les courbes elliptiques montrent que
	$\CM_{1,0}$ est \'egal \`a $\poincare/{\rm PSL}_2\Z$. 
	L'appli\-cation $j$ \'etablit un isomorphisme entre $\CM_{1,0}$ et 
	le plan complexe $\C$. Il est facile de compactifier l'espace des 
	modules. On utilise une compactification \`a la Alexandroff de 
	$\CM_{1,0}$, alors $\overline{\CM}_{1,0}$ est la sph\`ere de 
	Riemann. Trivialement, $\CM_{1,1}=\CM_{1,0}$.

Enfin, signalons que l'espace $\CM_{g,N}$ poss\`ede une compactification 
naturelle~: l'{\bol espace des modules des surfaces de Riemann stables} 
$\overline{\CM}_{g,N}$, due \`a Knudsen-Deligne-Mumford. Cette pr\'esentation
inspir\'ee par celle de Mumford~\cite{mumford:jac} n'est pas tr\`es adapt\'ee
\`a notre travail.

Soit $\Sigma$ une vari\'et\'e bidimensionnelle (compacte, connexe, orient\'ee,
sans points marqu\'es) de genre $g$. Soient $\mathbb{MET}$ l'ensemble des 
m\'etriques Riemanniennes sur $\Sigma$, $\mathbb{WEYL}$ le groupe 
des transformations de Weyl --- \cad les application $\ee^\sigma$ o\`u 
$\sigma\in\CC^\infty(\Sigma,\R)$~--- $\text{Diff}_+$ l'ensemble des 
diff\'eomorphismes de $\Sigma$ sur lui-m\^eme pr\'eservant l'orientation et
$\text{Diff}_{+,0}$ la composante connexe de l'identit\'e
dans $\text{Diff}_+$ --- c'est un sous-groupe normal de $\text{Diff}_+$. 
Soit $\mathbb{WEYL}\rtimes\text{Diff}_+$ le produit semi-direct
o\`u la loi de groupe est 
\[
(f,\ee^\sigma)\,(f',\ee^{\sigma'})=(f\circ f',\ee^{\sigma'}\,(\ee^\sigma\circ f')).
\]
Ce groupe agit sur $\mathbb{MET}$ par~: $\gamma\mapsto \ee^\sigma\,(f^*\gamma)$.
Le groupe de Weyl agissant librement sur l'espace des m\'etriques, 
l'espace quotient $\mathbb{CONF}=\mathbb{MET}/\mathbb{WEYL}$ 
n'a pas de points singuliers. D'apr\`es le chapitre 2, 
c'est l'ensemble des classes conformes de m\'etriques 
sur $\Sigma$, mais aussi l'ensemble $\mathbb{COMPL}$ des structures complexes 
sur $\Sigma$. Pour ce dernier,
deux structures complexes $\J$ et $\J'$ sont \'equivalentes si,
 et seulement si,
il existe un \'el\'ement $f$ de $\text{Diff}_+$ tel que $f:(\Sigma,\J')
\rightarrow(\Sigma,\J)$ soit un automorphisme pour les structures complexes
sous-jacentes. On voit facilement que si $\gamma$ et $\gamma'$ sont des
m\'etriques compatibles avec respectivement $\J$ et $\J'$ alors $f^*\gamma$
et $\gamma'$ sont dans la m\^eme classe conforme.
Clairement, les quotients $\mathbb{CONF}/
\text{Diff}_+$ et $\mathbb{COMPL}/\text{Diff}_+$ sont identiques, c'est
l'{\bol espace des modules}
\[
\CM\sous{\Sigma}=\mathbb{MET}/\mathbb{WEYL}\rtimes\text{Diff}_{+}.
\]
L'espace des modules n'est pas une vari\'et\'e complexe. 
On dit que $\CM\sous{\Sigma}$ est une \<<orbifold\>>.
En effet, l'existence d'isom\'etries implique que
l'action de $\text{Diff}_+$ sur
$\mathbb{MET}$ n'est pas libre. Sur $\mathbb{CONF}$ ou $\mathbb{COMPL}$, 
c'est l'existence
d'automorphismes conformes  qui fait que l'action n'est pas libre --- un
automorphisme conforme est un diff\'eomorphisme $f$ de $\Sigma$ tel que
$f^*\gamma$ et $\gamma$ soient dans la m\^eme classe conforme. Par contre,
l'{\bol espace de Teichm\"uller} 
\begin{equation}
\label{teich}
\CT\sous{\Sigma}=\mathbb{MET}/\mathbb{WEYL}\rtimes\text{Diff}_{+,0}
\end{equation}
est une vari\'et\'e complexe d\`es que $g\geq 2$.
Dans ce cas, il n'y a qu'un nombre fini d'automor\-phismes conformes
($\leq 84(g-1)$) et aucun n'est dans $\text{Diff}_{+,0}$.
Pour pallier le petit handicap sur le genre, on introduit l'espace 
$\mathbb{GAUSS}$ des m\'etriques de courbure gaussienne constante~:
\begin{equation*}
\mathbb{GAUSS}\equiv\begin{cases}
        \{\gamma\in\mathbb{MET}
        \ \big|\ K_{\gamma}=+1\}, & \text{si $g=0$},\\
        \{\gamma\in\mathbb{MET}\ \big|\ K_{\gamma}=0\ \text{et}\
        \text{vol}\,\Sigma=1\},
                        & \text{si $g=1$},\\
        \{\gamma\in\mathbb{MET}\ \big|\ K_{\gamma}=-1\}, &
        \text{si $g\geq 2$}.
        \end{cases}
\end{equation*}
En genre un, on introduit la condition suppl\'ementaire car, la caract\'eristique
d'Euler \'etant nulle, on ne dispose pas de normalisation sur le volume.
L'espace des m\'etriques $\mathbb{MET}$ est un 
$\mathbb{WEYL}$\,-fibr\'e principal au-dessus de l'espace contractile 
$\mathbb{CONF}$. Ce fibr\'e est donc trivial. Par exemple, 
une trivialisation associe \`a une classe conforme une m\'etrique de courbure 
gaussienne \'egale \`a $1$ (resp. $0$, resp. $-1$) en genre $0$ (resp. $1$, 
resp. $\geq 2$). L'existence est assur\'ee par le th\'eor\`eme 
d'uniformisation de Riemann. Par contre on a unicit\'e uniquement
en genre $\geq 2$ et l'espace $\mathbb{CONF}$ se confond donc avec l'espace 
$\mathbb{GAUSS}$. Pour obtenir une construction \<<agr\'eable\>> 
en n'importe quel genre de l'espace de Teichm\"uller, on peut poser
$\CT\sous{\Sigma}=\mathbb{GAUSS}/\text{Diff}_{+,0}$. 
En quotientant l'espace de Teichm\"uller par le 
{\bol groupe modulaire de Teichm\"uller} $\Gamma\sous{\Sigma}=\text{Diff}_+/
\text{Diff}_{+,0}$, on retombe sur l'espace des modules, \ie 
$\CM\sous{\Sigma}=\CT\sous{\Sigma}/\Gamma\sous{\Sigma}$.

Pour d\'ecrire l'espace tangent \`a l'espace des modules $\CM_{g,0}$,
on utilise la repr\'esentation $\mathbb{COMPL}/\text{Diff}_{+,0}$.
Soit $\J$ une structure presque complexe sur $\Sigma$ et 
$z$ un syst\`eme de coordonn\'ees complexes pour $\J$, \cad
$\J=-i\,\partial_z\otimes dz+i\,\partialbarz\otimes d\overline{z}$.
Faisons varier la structure complexe $\J\mapsto\J'=\J+\delta\J$, 
o\`u $\delta\J=\deltanu\,\partial_z\otimes dz+
\overline{\deltanu}\,\partialbarz\otimes d\overline{z}+
\deltamu\,\partial_z\otimes d\overline{z}
+\overline{\deltamu}\,\partialbarz\otimes dz$. 
On a utilis\'e le fait que $\J$ est vraiment
la complexification d'un automorphisme du fibr\'e tangent r\'eel. 
Comme $\J^2=-1$, on a aussi 
$\J\,\delta\J+\delta\J\,\J=0$, ce qui donne $\deltanu=0$.
Notons $\delta\mu=\deltamu\,\partial_z\otimes d\bar z$, donc
$\J'=\J+\delta\mu+\overline{\delta\mu}$.
Si $z'=z+\delta z$ est un 
syst\`eme de coordonn\'ees complexes pour $\J'$, \ie
$\J'=-i\,\partial_{z'}\otimes d{z'}+i\,\partial_{\hspace{0.02cm}
\overline{z}'}\otimes d\overline{z}'$, on obtient facilement~: 
$\delta\beltrami=-2i\,\partialbarz(\delta z)$. 
Il faut encore d\'eterminer les variations provenant d'une reparam\'etrisation. 
L'alg\`ebre de Lie de $\text{Diff}_{+,0}$ est l'alg\`ebre des champs de
vecteurs sur $\Sigma$. Par cons\'equent, les variations dues \`a l'action
de $\text{Diff}_{+,0}$ sont les
\'el\'ements de la forme $\delta\mu=\overline{\partial}
(\delta v)$, o\`u $\delta v$ est un $(1,0)$-champ de vecteur $\CC^\infty$.
Si on utilise le langage de la cohomologie de Dolbeaut, on a montr\'e que
$T\sous{\J}\CM_{g,0}\cong H_{{\rm Dol}}^{0,1}(K^{-1})
\cong H^1(K^{-1})$. Par dualit\'e de Serre, 
l'espace cotangent est isomorphe \`a $H^0(K^2)$, \ie 
les formes quadratiques holomorphes. D'apr\`es la 
formule de Riemann-Roch,
\[
h^0(K^2)-h^1(K^2)=\text{deg}\,K^2+1-g.
\]
Or, $\deg K^2=4g-4>2g-2$ si $g\geq 2$, donc $h^1(K^2)=0$. Par cons\'equent, 
$\dim \CM_{g,0}=3g-3$ en genre $\geq 2$.

Si on utilise $\mathbb{CONF}/\text{Diff}_{+,0}$, on raisonne comme suit.
On choisit une m\'etrique $\gamma$ et $\mathbf{x}=(x,y)$
un syst\`eme de coordonn\'ees isothermes
pour $\gamma$, \ie $\gamma=\ee^{\sigma}\,|dz|^2$ avec $z=x+iy$.
Dans un autre syst\`eme de coordonn\'ees, $\gamma+\delta\gamma=\ee^{\sigma+\delta \sigma}
\,|dz+\delta\mu\, d\overline{z}|^2$.
Il suit imm\'ediatement que
$\delta\mu=\delta\gamma_{\overline{z}\,\overline{z}}/2\gamma_{z\overline{z}}$.
Un vecteur tangent \`a $\mathbb{CONF}$ est donc un \'el\'ement de
$\Lambda^{01}(K^{-1})$. Il faut encore d\'eterminer les variations
dues \`a l'action de $\text{Diff}_{+,0}$.
La variation de la m\'etrique due \`a l'action infinit\'esimale de
$\text{Diff}_{+,0}$ est
$\gamma+\delta\gamma=\gamma(\mathbf{x}+\delta\mathbf{x})_{\mu\nu}
\,d(x+\delta x)^\mu d(y+\delta y)^\nu$, d'o\`u
$\delta\gamma=\nabla_\mu \delta v_\nu+\nabla_\nu \delta v_\mu$
avec $\delta v_\mu=\gamma_{\mu\nu}\,\delta x^\nu$ et $\nabla_\mu$
la connexion de Levi-Cevita de $\gamma$. On a juste montr\'e
que $\delta \gamma=\pounds_{\delta v}\gamma$, ce qui est imm\'ediat quand on
remarque que l'alg\`ebre de Lie de $\text{Diff}_{+,0}$ est l'alg\`ebre
des champs de vecteurs sur $\Sigma$. On isole le terme de trace non-nulle
qui peut \^etre absorb\'e par transformation de Weyl, ainsi
$\delta\gamma={\rm div}(\delta v)\,\gamma +P\,(\delta v)$,
o\`u $(P X)_{\mu\nu}=\nabla_\mu X_\nu+\nabla_\nu X_\mu- ({\rm div}\,X)
\gamma_{\mu\nu}$. Ainsi, $P$ envoie les champs de vecteurs
sur les vecteurs tangents \`a l'orbite de $\text{Diff}_{+,0}$ \`a $\gamma$.
La variation de structure complexe correspondante est $\delta\mu=(P\,
\delta v)_{\overline{z}\,\overline{z}}/2\gamma_{z\overline{z}}=
\partialbarz \delta v^z$, si $\delta v=\delta v^z\partial_z
+\delta v^{\overline{z}}\partialbarz$. L'espace tangent \`a l'espace des
modules est donc $\Lambda^{01}(K^{-1})/\overline{\partial}\Gamma(K^{-1})$,
\cad $H^1(K^{-1})$.
Au passage, on voit que le noyau de $P$ est form\'e des champs de vecteurs
qui engendrent les automorphismes conformes pour $\gamma$,
appel\'es champs de Killing conformes. On voit facilement que
${\rm Ker}\,P=H^0(K^{-1})$. En genre $0$, $h^0(K^{-1})=3$, il
y a donc $6$ vecteurs de Killing conformes (r\'eels), qui engendrent
$\Slc$. En genre $1$, $h^0(K^{-1})=1$. Si $g\geq 2$, $h^0(K^{-1})=0$, il
n'y a donc pas de champs de Killing conformes.

On conserve les hypoth\`eses pr\'ec\'edentes, mais on suppose en plus
donn\'ee une s\'equence $S=\un\xi$ de $N$ points sur la surface avec  $N>0$.
Soit $\text{Diff}_+^{S}$ le sous-groupe des diff\'eomorphismes de $\Sigma$
pr\'eservant l'orientation et laissant $S$ invariante point par point.
Soit $\text{Diff}_{+,0}^{S}$ la composante connexe de l'identit\'e dans
$\text{Diff}_{+,0}^{S}$. Le groupe modulaire est $\Gamma\sous{\Sigma}^S=
\text{Diff}_{+}^S/\text{Diff}_{+,0}^S$, l'espace des modules est le
quotient $\CM\sous{\Sigma}^S=\mathbb{CONF}/\text{Diff}_{+}^S$ et
l'espace de Teichm\"uller est
$\CT\sous{\Sigma}^S=\mathbb{CONF}/\text{Diff}_{+,0}^S$.
On a toujours l'identification $\CM\sous{\Sigma}^S=\CT^S\sous{\Sigma}/
\Gamma\sous{\Sigma}^S$. Les espaces tangents et cotangents \`a
l'espace des modules sont
\[
T\CM_{g,N}=H^1\big(K^{-1}\big(-\sumini\,\xi_\ell\big)\big),
	\qquad T^*\CM_{g,N}=H^0\big(K^2\big(\sumini\,\xi_\ell\big)\big).
\]
Le th\'eor\`eme de Riemann-Roch donne la dimension
de $\CM_{g,N}$ pour $2g-2+N>0$~: $\dim \CM_{g,N}=3g-3+N$.

On utilise aussi un troisi\`eme espace des modules~: l'espace des modules $\CM_{g,N}^\star$
des surfaces de Riemann de genre $g$, \'etant
donn\'es  $N$ points distincts $\xi_1,\cdots,\xi_N$ sur $\Sigma$
et, en chacun de ces points, un $1$-jet de param\`etre holomorphe local, centr\'e  en
$\xi_\ell$. On identifie $\CM^\star_{g,N}$ avec l'espace des structures complexes
sur $\Sigma$ et des suites de $N$ \'el\'ements du fibr\'e
cotangent $T^*\Sigma$ se projetant sur $\Sigma$ en des points diff\'erents,
le tout modulo l'action des diff\'eomorphismes pr\'eservant l'orientation. 
Si $(z,\bar z)$ sont les coordonn\'ees pour un $\J$, un \'el\'ement du fibr\'e
cotangent est $(\xi_\ell,\alpha\,dz(\xi_\ell)+\bar\alpha\,d\bar z(\xi_\ell))$ et
$\alpha\,dz(\xi_\ell)$ est le $1$-jet de param\`etre local en $\xi_\ell$.
Un vecteur tangent \`a $\CM_{g,N}^\star$ est un \'el\'ement $\delta\J$ et
$N$ vecteurs $Y_i$ tangents \`a $T^*\Sigma$, cette fois modulo les vecteurs
provenant de l'action des champs de vecteurs $\delta v=\delta v^z\,\partial_z
+\delta v^{\bar z}\,\partialbarz$
sur $\Sigma$. On a vu que, localement, $\delta\J=\delta\mu+\bar{\delta\mu}$ avec $\delta\mu
^z_{\bar z}=-2i\,\partialbarz(\delta z)$, $\delta z$ \'etant d\'etermin\'e
modulo les $\delta v^z$ holomorphes. On peut choisir $\delta\mu$ nulle au 
voisinage des $\xi_\ell$, puis utiliser la libert\'e holomorphe pour annuler
les $Y_i$.  L'action du groupe des diff\'eomorphismes revient
\`a supprimer les $\delta\mu^z_{\bar z}$ de la forme $\partialbarz\delta v^z$
o\`u $\delta v^z$ est un $(1,0)$-champ de vecteurs. On peut toujours choisir 
$\delta v^z$  telle que $\delta v^z\big/(z-z_\ell)^2$ soit lisse au voisinage
de $\xi_\ell$ et, ensuite, choisir
$\delta\mu$ telle que $\delta\mu^z_{\bar z}\big/(z-z_\ell)^2$ soit lisse
au voisinage de $\xi_\ell$. Finalement, 
\[
T\CM_{g,N}^\star=H^{0,1}_{\rm Dol}\big(K^{-1}\big(-2\,\sumini\,\xi_\ell
	\big)\big)\cong 
	H^1\big(K^{-1}\big(-2\,\sumini\,\xi_\ell\big)\big),
        \qquad T^*\CM_{g,N}^\star=H^0\big(K^2\big(2\,\sumini\,\xi_\ell\big)\big).
\]
En particulier, si $g-1+N>0$, on trouve $\dim \CM^\star_{g,N}=3g-3+2N$.

\medskip
\section{Connexion de KZB}

\`A partir de l'espace des \'etats de Chern-Simons $\chsi$
on construit un fibr\'e vectoriel $\ChSi$ au-dessus de l'espace des 
modules $\moduleR$ des surfaces de Riemann de genre $g$ marqu\'ees 
par $N$ points. Ici, il s'agit d'analyser le comportement d'un \'etat de 
Chern-Simons quand on fait varier la structure complexe de $\Sigma$ ainsi
que les points. Formellement, la fibre au-dessus du point 
$(\J,\underline{\xi})\in\moduleR$ est l'espace $\chsi$\,, \ie
\[
\ChSi\left|_{\J,\underline{\xi}}\right.=\chsi.
\]
Une section de $\ChSi$ est une application
\[
\boldsymbol{\Psi}:\moduleR\ni(\J,\underline{\xi})\rightarrow
	\boldsymbol{\Psi}(\J,\underline{\xi},.)\in\chsi\,.
\]
Un \'etat de Chern-Simons sera donc une fonctionnelle
$\Bpsi(\J,\underline{\xi},.)$ holomorphe de $A^{01}$, v\'erifiant l'identit\'e
de Ward (chirale)~:
\begin{equation}
\label{CSmoduli}
{_\delta\over^{\delta A^{10}}}\Bpsi=0\quad\text{et}\quad
	\Bpsi(\J,\underline{\xi},A^{01})=\ee^{-k\,S(h,A^{01})}
	\,\mathop{\otimes}\limits_\ell 
	h(\xi_\ell )\sous{R_\ell }\,\Bpsi(\J,\underline{\xi},\hAR{h^{-1}}).
\end{equation}
Jusqu'\`a maintenant, on a \'ecrit $\Bpsi$ comme une fonctionnelle de $A^{01}$.
Dans la plupart des cas, on garde cette notation.
N\'eanmoins, le fait m\^eme de consid\'erer $A^{01}$ sous-entend 
qu'on utilise la structure complexe. Quand on fait varier $\J$, on doit
imp\'erativement r\'etablir la d\'ependance en $A=A^{10}+A^{01}$
o\`u $A^{10}=-(A^{01})^\dagger$.

Dans la section suivante, on construit une structure holomorphe sur
$\ChSi$, \ie un op\'erateur $\overline{\con}$ comme d\'efini au chapitre 3.
D'autre part, le produit scalaire de Bargman sur $\chsi$ induit 
une structure hermitienne naturelle sur $\ChSi$. 
Le fibr\'e $\ChSi$ devient (formellement) un fibr\'e vectoriel 
holomorphe hermitien.

Si $E$ est un fibr\'e vectoriel holomorphe, donc
d\'ej\`a muni d'une structure holomorphe, on dit qu'une connexion
est {\bol compatible avec une structure holomorphe} $\conbar$ si
$\nabla^{01}= \conbar$.
Une {\bol structure hermitienne} sur $E$ est la donn\'ee d'un produit hermitien
$\Blsca\,,\,\Brsca_x$ sur toute fibre $E_x$ de $E$, tel que pour tout ouvert
$U$ de $X$ et pour toutes sections $\xi$, $\eta$ de $E$ au-dessus
de $U$, l'application $U\ni x\mapsto \Blsca\xi(x),\eta(x)\Brsca_x\in\C$ est
$\CC^\infty$. Tout fibr\'e vectoriel admet une structure hermitienne.
 On dit qu'une connexion est {\bol compatible avec une structure
hermitienne} sur $E$ si $d\Blsca\xi,\eta\Brsca=\Blsca\nabla\xi,
\eta\Brsca+\Blsca\xi,\nabla\eta\Brsca$, $\forall\xi,\eta\in\Gamma(E)$.

Soit $E$ un fibr\'e vectoriel holomorphe hermitien.
Il existe une unique connexion $\nabla$ sur $E$ compatible
avec les structures hermitienne et holomorphe, appel\'ee
{\bol connexion m\'etrique}.  La courbure de la connexion m\'etrique
est une $(1,1)$-forme \`a valeurs dans ${\rm End}\,E$.

La {\bol connexion de Knizhnik-Zamolodchikov-Bernard},
not\'ee $\Bnabla$, est la connexion m\'e\-trique sur $\ChSi$.
Ainsi, la connexion de KZB est l'unique connexion compatible 
avec la structure complexe~:
\begin{equation}
\KZmubar=\con_{\overline{\delta\mu}},\qquad
\KZzbar=\con_{\overline{z}_\ell }
\end{equation}
et avec la structure hermitienne~:
$\partial\Blsca\Bpsi',\Bpsi\Brsca=
\Blsca\conbar\Bpsi',\Bpsi\Brsca+\Blsca\Bpsi',\Bnabla^{10}\Bpsi\Brsca$
si $\Bpsi$ et $\Bpsi'$ appartiennent \`a $\Gamma(\ChSi)$ (on dira aussi 
que la connexion de KZB est {\bol unitaire}).
Cette derni\`ere condition va nous permettre d'extraire $\KZz$ et $\KZmu$. 
La fa\c{c}on dont on d\'erive la connexion est parall\`ele \`a
la d\'emarche de Axelrod, Della Pietra et Witten~\cite{axelrod}. La diff\'erence majeure
r\'eside dans l'utilisation de la g\'eom\'etrie complexe, comme cela a \'et\'e 
propos\'e dans~\cite{gaw89:construc}. On cons\-truit
la connexion de KZB par des manipulations sur les int\'egrales fonctionnelles.
Ensuite, on utilise une d\'ecoupe de l'espace des modules
$s:\CN\mapsto \CA^{01}$. Ainsi, si on conna\^\i t une d\'ecoupe
explicite, on devrait \^etre en mesure de d\'ecrire compl\`etement
la connexion de KZB. On fera cela explicitement en genres z\'ero et un.
On utilise
le produit scalaire des \'etats de Chern-Simons dont, en g\'en\'eral, on ne 
peut pas d\'emontrer la convergence. N\'eanmoins, notre construction
conduit \`a des expressions ne posant quant \`a elles aucun probl\`eme
de convergence. Une fois qu'on a obtenu le produit scalaire des
\'etats de Chern-Simons et la connexion de KZB, la question 
la plus importante est~: {\bol la connexion de KZB est-elle unitaire
pour une structure hermitienne sur $\ChSi$ ?} Notre construction 
est bien bas\'ee sur l'unitarit\'e de la connexion mais uniquement
au niveau formel. Si on admet que le produit scalaire converge pour
les \'etats de Chern-Simons, la r\'eponse est oui~: en genre un pour $G={\rm SU}_2$
avec un point d'insertion~\cite{gaw96:elliptic} --- dans ce cas on montre
aussi que le produit scalaire converge --- et pour la situation
g\'en\'erale~\cite{gaw97:unitarity}.  
La condition d'unitarit\'e de la connexion r\'ev\`ele d'int\'eressantes relations
avec l'{\bol Ansatz de Bethe}~\cite{babujian,BabF,gaw96:elliptic,
gaw97:unitarity,gaw91:scalar,ReshV}.

La connexion en question apparut pour la premi\`ere fois dans un article
de Knizhnik et Zamolodchikov~\cite{kz} pour le genre z\'ero (cf. 
section 2.{\bf 4.8}), puis elle
fut obtenue en genre sup\'erieur par 
Bernard~\cite{bernard:kzbanyg,bernard:kzb}.
Comme pour l'espace des \'etats de Chern-Simons, il existe
d'autres constructions de la connexion de KZB~\cite{felder:houches,
hitchin:flat}.
En genre un, notre connaissance de la connexion de KZB a \'et\'e
consid\'erablement am\'elior\'ee par les travaux~\cite{EtinK,gaw94:genus1,
felder:integral,felder:elliptic}. En genre $2$, van Geemen et de Jong 
ont trouv\'e une formule explicite pour la connexion~\cite{VGDJ}.
Leur approche est un peu diff\'erente puisqu'ils \'etudient la
connexion de Hitchin~\cite{hitchin:flat}. Cette derni\`ere devrait,
au moins pour ce qui nous int\'eresse, co\"{\i}ncider avec la
connexion de KZB. Hitchin consid\`ere un fibr\'e vectoriel
au-dessus de l'espace de Teichm\"uller dont
les fibres sont les sections globales d'une puissance du
fibr\'e d\'eterminant au-dessus de $\CN_s$. La connexion de Hitchin
est une connexion projectivement plate sur ce fibr\'e vectoriel.
En genre $\geq 2$, il n'existe pas
de description compl\`ete de la connexion de KZB.
D'autres versions de la connexion de KZB existent comme 
la \<<connexion\>> d\'eform\'ee $q$KZB~\cite{felder:ICM}.

\subsection{Structure holomorphe sur {$\ChSi$}}

Pour simplifier, on sous-entend la d\'ependance en $\J$ et $\underline{\xi}$
de $\Bpsi$. Commen\c{c}ons par rappeler quelques propri\'et\'es 
\'el\'ementaires de la d\'erivation fonctionnelle. Par d\'efinition, 
\begin{align*}
\delta\Bpsi(A)=\int_\Sigma\tr\,&{_{\delta\Bpsi}\over^{\delta A}}\wedge\delta A
	=\int_\Sigma\tr\,{_{\delta\Bpsi}\over^{\delta A^{01}}}
	\wedge\delta A^{01}+\int_\Sigma\tr\,
	{_{\delta\Bpsi}\over^{\delta A^{10}}}\wedge\delta A^{10}\\
	&=\int_\Sigma{_{\delta\Bpsi}\over^{\delta A\sur{a}_{z}}}
		\,\delta A\sur{a}_{z}\,d^2z
	+\int_\Sigma{_{\delta\Bpsi}\over^{\delta A\sur{a}_{\overline{z}}}}
	\,\delta A\sur{a}_{\overline{z}}\,d^2z.
\end{align*}
Les identit\'es 
\begin{align*}
{_{\delta\Bpsi}\over^{\delta A^{10}}}&=
	\left({_{\delta\Bpsi}\over^{\delta A}}\right)^{01}=
	-i\,{_{\delta\Bpsi}\over^{\delta A_{z}}}\,d\overline{z}=
	-i\,{_{\delta\Bpsi}\over^{\delta A\sur{a}_{z}}}
	\,t^a\,d\overline{z},\\
{_{\delta\Bpsi}\over^{\delta A^{01}}}&=
	\left({_{\delta\Bpsi}\over^{\delta A}}\right)^{10}=
	i\,{_{\delta\Bpsi}\over^{\delta A_{\overline{z}}}}\,dz=
	i\,{_{\delta\Bpsi}\over^{\delta A\sur{a}_{\overline{z}}}}\,t^a\,dz
\end{align*}
clarifient le passage entre ces trois d\'efinitions.
On utilise $d_{\delta\mu}$ (resp. $d_{\overline{\delta\mu}}$) pour
la d\'eriv\'ee holomorphe (resp. anti-holomorphe)
dans la direction $\J$ \`a $\underline{\xi}$ et $A$ fix\'es, et
$\partial_{z_\ell }$ (resp. $\partialzbar$) pour
la d\'eriv\'ee holomorphe (resp. anti-holomorphe)
dans la direction $\xi_\ell $.
Comme la d\'ecomposition $A=A^{10}+A^{01}$ d\'epend du choix 
d'une structure complexe,
la d\'eriv\'ee fonctionnelle par rapport \`a $A^{10}$ ne commute pas
avec les d\'eriv\'ees dans la direction $\J$. Effectivement, on a
\begin{align}
\label{commutateurs}
\left[d_{\delta\mu},{_\delta\over^{\delta A^{10}}}\right]\Bpsi&=
	-{_i\over^2}\,{_{\delta\Bpsi}\over^{\delta A^{01}}}\,\delta\mu,
	\tag{\ref{commutateurs}.a}\\
\left[d_{\overline{\delta\mu}},{_\delta\over^{\delta A^{10}}}\right]\Bpsi&=
	-{_i\over^2}\,{_{\delta\Bpsi}\over^{\delta A^{10}}}\,
	\overline{\delta\mu}\tag{\ref{commutateurs}.b}.
\addtocounter{equation}{1}
\end{align}
Montrons par exemple la premi\`ere de ces deux \'egalit\'es~:
\begin{equation*}
\begin{split}
\left[d_{\delta\mu},{_\delta\over^{\delta A^{10}}}\right]\Bpsi&=
	\left[d_{\delta\mu},{_\delta\over^{\delta A}}\,
	{_{1-i\J}\over^2}\right]\Bpsi=d_{\delta\mu}\,\left({_{\delta\Bpsi}
	\over^{\delta A}}\,{_{1-i\J}\over^2}\right)-{_\delta\over^{\delta A}}
	\,(d_{\delta\mu}\Bpsi)\,{_{1-i\J}\over^2}\\
&={_{\delta\Bpsi}\over^{\delta A}}\,\left(-{_{i}\over^2}\delta\mu\right)
	=-{_i\over^2}\,{_{\delta\Bpsi}\over^{\delta A^{01}}}\,\delta\mu.
\end{split}
\end{equation*}

Se donner une structure holomorphe sur $\ChSi$ revient \`a d\'efinir
un op\'erateur $\conbar:\Gamma(\ChSi)\rightarrow\Lambda^{01}(\ChSi)$.
Soit $\Bpsi\in\Gamma(\ChSi)$, on pose 
\begin{align*}
\con_{\overline{\delta\mu}}\,\Bpsi &=d_{\overline{\delta\mu}}\,\Bpsi+
	{_k\over^{8\pi}}\int_\Sigma\tr\,A^{01}\wedge(A^{01}
	\,\overline{\delta\mu})\,\Bpsi,\\
\con_{\overline{z}_\ell }\Bpsi &=\partialzbar\Bpsi
	+(A_{\overline{z}_\ell })_\ell \,\Bpsi
\end{align*}
o\`u $(A_{\overline{z}_\ell })_\ell \equiv 
	A\sur{a}_{\overline{z}_\ell }(\xi_\ell )\,t^a_\ell$ 
agit sur le $\ell$-i\`eme facteur de $\esprep$.
Pour que cette d\'efinition ait un sens, il faut que 
l'op\'erateur $\conbar$ envoie une section de $\ChSi$ en une forme
de $\Lambda^{01}(\ChSi)$. Soit $\Bpsi\in\Gamma(\ChSi)$.
Commen\c{c}ons par nous occuper de 
$\con_{\overline{\delta\mu}}\,\Bpsi$. La d\'ependance en $A^{10}$ 
est vite r\'egl\'ee puisque
\begin{equation*}
\begin{split}
{_\delta\over^{\delta A^{10}}}(\con_{\overline{\delta\mu}}\,\Bpsi)&=
	-\left[d_{\overline{\delta\mu}},
	{_\delta\over^{\delta A^{10}}}\right]\Bpsi
	+d_{\overline{\delta\mu}}\,\left({_{\delta\Bpsi}\over^{\delta A^{10}}}
	\right)+{_\delta\over^{\delta A^{10}}}\,
	\Big({_k\over^{8\pi}}\int_\Sigma\tr\,A^{01}
	\wedge(A^{01}\,\overline{\delta\mu})\Big)\\
&={_i\over^2}\,{_{\delta\Bpsi}\over^{\delta A^{10}}}\,\overline{\delta\mu}=0.
\end{split}
\end{equation*}
Si on utilise l'\'equation d\'efinissant $\conbar$ et 
l'identit\'e de Ward pour l'\'etat $\Bpsi$,
\[
\con_{\overline{\delta\mu}}\,\Bpsi(A^{01})=\Big(d_{\overline{\delta\mu}}+
	{_k\over^{8\pi}}\,\int_\Sigma\tr\,A^{01}\wedge(A^{01}\,
	\overline{\delta\mu})\Big)\,\Big(\ee^{-k\,S(h,A^{01})}\,
	\mathop{\otimes}\limits_\ell h(\xi_\ell )\sous{R_\ell }\,
	\Bpsi(\hAR{h^{-1}})\Big).
\]
Sans entrer dans les d\'etails, disons qu'on a besoin
de deux ingr\'edients pour mener \`a bien les calculs. 
La d\'eriv\'ee de l'action par rapport \`a $\overline{\delta\mu}$ est 
\begin{equation*}
\begin{split}
d_{\overline{\delta\mu}}\,(k\,S(h,A^{01}))=&\,
	d_{\overline{\delta\mu}}\Big(-{_{ik}\over^{4\pi}}
	\int_\Sigma\tr\,(h^{-1}dh)\,\posiJ\wedge(h^{-1}dh)\,\negaJ\\
	&\phantom{d_{\overline{\delta\mu}}}\quad+{_{ik}\over^{2\pi}}
	\int_\Sigma\tr\,(h\,dh^{-1})\,\posiJ\wedge
	A\,\negaJ\Big)\\
=&\,{_k\over^{8\pi}}\int_\Sigma\tr\,(h^{-1}\de h)\,\overline{\delta\mu}\wedge
	h^{-1}\de h-{_k\over^{4\pi}}\int_\Sigma\tr\,
	(h\de h^{-1})\,\overline{\delta\mu}\wedge A^{01}.
\end{split}
\end{equation*}
Afin de calculer les d\'eriv\'ees par rapport \`a la structure complexe de
$\Bpsi(\hAR{h^{-1}})$, 
on doit r\'etablir la d\'ependance en $A=A^{10}+A^{01}$. Ainsi, 
\[
\Bpsi(\hAR{h^{-1}})=\Bpsi(\hAL{h^\dagger}+\hAR{h^{-1}})=
\Bpsi(\hA{h^\dagger}\,\posiJ+\hA{h^{-1}}\,\negaJ)
\]
et
\[
d_{\overline{\delta\mu}}\,\Bpsi(\hAR{h^{-1}})=(d_{\overline{\delta\mu}}
	\,\Bpsi)(\hAR{h^{-1}})+\int_\Sigma\tr\,
	{_{\delta\Bpsi}\over^{\delta A^{10}}}\wedge(\hA{h^\dagger}\,
	{_i\over^2}\,\overline{\delta\mu})
	+\int_\Sigma\tr\,{_{\delta\Bpsi}\over^{\delta A^{01}}}
	\wedge(\hA{h^{-1}}\,{_{-i}\over^2}\,\overline{\delta\mu}).
\]
Les deux int\'egrales sont nulles donc $d_{\overline{\delta\mu}}\,
\Bpsi(\hAR{h^{-1}})=(d_{\overline{\delta\mu}}\,\Bpsi)(\hAR{h^{-1}})$.
Si on regroupe tous ces r\'esultats, on trouve l'identit\'e de Ward
pour $\con_{\overline{\delta\mu}}\,\Bpsi$~:
\[
(\con_{\overline{\delta\mu}}\,\Bpsi)(A^{01})=\ee^{-k\,S(h,A^{01})}\,
	\mathop{\otimes}\limits_\ell h(\xi_\ell )\sous{R_\ell }\,
	(\con_{\overline{\delta\mu}}\,\Bpsi)(\hAR{h^{-1}}).
\]
L'holomorphie de $\con_{\overline{z}_\ell }\Bpsi$ est imm\'ediate. 
Pour obtenir l'identit\'e de Ward, on raisonne exactement comme avec
$\con_{\overline{\delta\mu}}\,\Bpsi$~:
\begin{equation*}
\begin{split}
\con_{\overline{z}_\ell }\Bpsi(A^{01})=&\,
	\ee^{-k\,S(h,A^{01})}\,
	\Bigl(\,\mathop{\otimes}\limits_{\ell'<\ell}
	h(\xi_{\ell'})\sous{R_{\ell'}}\Bigr)
	\otimes h(\xi_\ell)\sous{R_\ell}\,
	(h^{-1}\partialzbar h)(\xi_\ell )_\ell\otimes
	\Bigl(\,\mathop{\otimes}\limits_{\ell'>\ell}
	h(\xi_{\ell'})\sous{R_{\ell'}}\Bigr)\,\Bpsi(\hAR{h^{-1}})\\
&+\ee^{-k\,S(h,A^{01})}\,\mathop{\otimes}\limits_{\ell'}
	h(\xi_{\ell'})\sous{R_{\ell'}}\,
	(\partialzbar\Bpsi)(\hAR{h^{-1}})
	+\ee^{-k\,S(h,A^{01})}\,(A_{\overline{z}_\ell })_\ell \,
	\mathop{\otimes}\limits_{\ell'}h(\xi_{\ell'})\sous{R_{\ell'}}
	\,\Bpsi(\hAR{h^{-1}})\\
=&\,\ee^{-k\,S(h,A^{01})}\,\mathop{\otimes}\limits_{\ell'}
	h(\xi_{\ell'})\sous{R_{\ell'}}\,
	(\con_{\overline{z}_\ell }\Bpsi)(\hAR{h^{-1}}).
\end{split}
\end{equation*} 

En r\'esum\'e, les composantes $\KZmubar$ et $\KZzbar$ de la connexion
de KZB sont 
\begin{align}
\label{KZantiholo}
\KZmubar &=d_{\overline{\delta\mu}}+
        {_k\over^{8\pi}}\int_\Sigma\tr\,A^{01}\wedge(A^{01}
        \,\overline{\delta\mu}),\tag{\ref{KZantiholo}.a}\\
\KZzbar &=\partialzbar
        +(A_{\overline{z}_\ell })_\ell.
	\tag{\ref{KZantiholo}.b}
\addtocounter{equation}{1}
\end{align}

\subsection{Unitarit\'e. Calcul formel de $\Bnabla$}

Encore une fois, rappelons que les calculs sont effectu\'es sur un plan formel.
En particulier, on ne pr\^ete pas attention aux probl\`emes de convergence.
Pour trouver $\KZmu$, on utilise l'unitarit\'e de la connexion~:
$\Blsca\Bpsi',\KZmu\Bpsi\Brsca=d_{\delta\mu}\Blsca\Bpsi',\Bpsi\Brsca
-\Blsca\KZmubar\,\Bpsi',\Bpsi\Brsca$. Le principe est simple.
On transforme le c\^ot\'e droit de 
l'\'egalit\'e afin d'obtenir le produit scalaire de $\Bpsi'$ avec
un \'etat de Chern-Simons. Par d\'erivation sous le signe int\'egral du 
premier terme, on a
\begin{multline*}
d_{\delta\mu}\int\lsca\Bpsi'(A),\Bpsi(A)\rsca\,\exponen\,DA\\
=\int\Big(\lsca(d_{\overline{\delta\mu}}\,\Bpsi')(A),\Bpsi(A)\rsca
	+\lsca\Bpsi'(A),(d_{\delta\mu}\Bpsi)(A)\rsca\\
+{_k\over^{4\pi}}\int_\Sigma\tr\,A^{10}\wedge (A^{10}\delta\mu)\ 
	\lsca\Bpsi'(A),\Bpsi(A)\rsca
\Big)\,\exponen\,DA.
\end{multline*}
Si on utilise ce r\'esultat et l'\'equation~(\ref{KZantiholo}.a), on obtient
\begin{equation*}
\begin{split}
\Blsca\Bpsi',\KZmu\Bpsi\Brsca
	=&\,\int\Big(\lsca\Bpsi'(A),(d_{\delta\mu}\Bpsi)(A)\rsca\\
&+{_k\over^{8\pi}}\int_\Sigma\tr\,A^{10}\wedge (A^{10}\delta\mu)
	\ \lsca\Bpsi'(A),\Bpsi(A)\rsca\Big)\,\exponen\,DA.
\end{split}
\end{equation*} 
Ensuite, on int\`egre par parties (en $A$) le second terme. Pour cela, 
il suffit de remarquer que
\[
\Big(\int_\Sigma\tr\,A^{10}\wedge(A^{10}\delta\mu)\Big)\,\exponen
	=-{_{4\pi i}\over^k}\,\int_\Sigma\tr\,A_{z}
	\,{_{\delta}\over^{\delta A_{\overline{z}}}}
	\Big(\exponen\Big)\,\delta\mu^{z}_{\overline{z}}\,d^2z.
\]
Ici, il est techniquement plus ais\'e de travailler en coordonn\'ees~:
\[
\Blsca\Bpsi',\KZmu\Bpsi\Brsca=\int\lsca\Bpsi'(A),\Big(d_{\delta\mu}
	+{_{i}\over^2}\,\int_\Sigma\tr\,A_{z}\,
	{_{\delta}\over^{\delta A_{\overline{z}}}}
	\,\delta\mu\sur{z}_{\overline{z}}\,d^2z\Big)
	\Bpsi(A)\rsca\,\exponen\,DA.
\]
On ne peut pas conclure car $\big(d_{\delta\mu}+\frac{i}{2}\,
\int\tr\,A_{z}\,\frac{\delta}{\delta A_{\overline{z}}}
\,\delta\mu^{z}_{\overline{z}}\,d^2z\big)\Bpsi$ n'est pas 
un \'etat de Chern-Simons. Par exemple, il n'y a pas holomorphie. 
On ne s'est pas tromp\'e de m\'ethode, on s'est juste arr\^et\'e trop t\^ot.
D'apr\`es l'\'equation~(\ref{commutateurs}.a), c'est
plut\^ot $\big(d_{\delta\mu}+i\,
\int\tr\,A_{z}\,\frac{\delta}{\delta A_{\overline{z}}}
\,\delta\mu^{z}_{\overline{z}}\,d^2z\big)\Bpsi$ qui est holomorphe.
On laisse momentan\'ement ce terme de c\^ot\'e et on 
int\`egre par parties une seconde fois le terme r\'esiduel~:
\begin{equation*}
\begin{split}
\Blsca\Bpsi',\KZmu\Bpsi\Brsca=\int\lsca\Bpsi'(A),&\Big(d_{\delta\mu}
	+i\,\int_\Sigma\tr\,A_{z}\,
	\deltaz\,\delta\mu\sur{z}_{\overline{z}}\,d^2z\\
&+{_{i\pi}\over^k}\int_\Sigma\tr\,
	\deltaz\,\deltaz\,\delta\mu\sur{z}_{\overline{z}}\,d^2z\Big)
	\Bpsi(A)\rsca\,\exponen\,DA.
\end{split}
\end{equation*}
On obtient donc une expression (formelle) pour la composante $\KZmu$ de
la connexion,
\[
\KZmu\Bpsi=\Big(d_{\delta\mu}+i\int_\Sigma\tr\,A_{z}\,
	\deltaz\,\deltamu\,d^2z+{_{i\pi}\over^k}\int_\Sigma\tr\,
	\deltaz\,\deltaz\,\deltamu\,d^2z\Big)\Bpsi.
\]
Le r\'esultat fait intervenir le Laplacien sur l'espace des champs de jauge,
par essence singulier. Pour que la 
d\'emonstration soit tout \`a fait compl\`ete, on doit montrer que le 
c\^ot\'e gauche est bien un \'etat de Chern-Simons~--- c'est l'objet 
de la fin de cette section. Parfois, il est plus pratique 
d'utiliser la connexion sous la forme suivante
\begin{equation}
\label{KZZ}
\KZmu\Bpsi=\Big(d_{\delta\mu}+{_i\over^2}\int_\Sigma\tr\,A^{10}\wedge
	\Big({_\delta\over^{\delta A^{01}}}\,\delta\mu\Big)
	+{_{\pi}\over^{2k}}\int_\Sigma\tr\,{_\delta\over^{\delta A^{01}}}\wedge
	\Big({_\delta\over^{\delta A^{01}}}\,\delta\mu\Big)\Big)\Bpsi.
\end{equation}

Pour calculer $\KZz$, on proc\`ede de la m\^eme fa\c{c}on. 
Suivant l'unitarit\'e de la connexion, on a $\Blsca\Bpsi',\KZz\Bpsi\Brsca=
\partial_{z_\ell}\Blsca \Bpsi',\Bpsi\Brsca-\Blsca\KZzbar\,\Bpsi',\Bpsi\Brsca$.
Le calcul est nettement plus simple. Il repose sur 
l'\'equation~(\ref{KZantiholo}.b) et 
$\lsca(A_{\overline{z}_\ell })_\ell \Bpsi'(A),\Bpsi(A)\rsca
=-\lsca\Bpsi'(A),(A_{z_\ell })_\ell \Bpsi(A)\rsca$.
On obtient
\[
\Blsca\Bpsi',\KZz\Bpsi\Brsca=
	\int\lsca\Bpsi'(A),(\partial_{z_\ell}+(A_{z_\ell })_\ell )
	\Bpsi(A)\rsca\,\exponen\,DA.
\]
Comme pr\'ec\'edemment, on int\`egre par parties. 
\`A la fin, on trouve
\[
\KZz\Bpsi=\Big(\partial_{z_\ell}-{_{2\pi}\over^k}\,t^a_\ell \,
	{_\delta\over^{\delta A\sur{a}_{\overline{z}_\ell }}}\Big)\Bpsi.
\]
Cette composante de la connexion est \'egalement singuli\`ere.
On expliquera dans la prochaine section comment traiter les termes divergents.

Montrons la propri\'et\'e laiss\'ee en suspens~:
si $\Bpsi\in\Gamma(\ChSi)$ alors aussi $\KZmu\Bpsi,~\KZz\Bpsi\in
\Gamma(\ChSi)$.
L'holomorphie ne pose aucun probl\`eme.
L'obtention des identit\'es de Ward pour $\KZmu\Bpsi$ est 
relativement longue, mais elle suit le m\^eme chemin que pour 
$\KZmubar\,\Bpsi$. En cons\'equence, on ne donne que les points
saillants de la preuve. Pour commencer, 
\begin{equation*}
\begin{split}
\KZmu\Bpsi(A^{01})=&
	\Big(
	d_{\delta\mu}+{_i\over^2}\int_\Sigma\tr\,A^{10}\wedge
	\Bigl({_\delta\over^{\delta A^{01}}}\,\delta\mu\Bigr)
	+{_{\pi}\over^{2k}}\int_\Sigma\tr\,{_\delta\over^{\delta A^{01}}}\wedge
	\Bigl({_\delta\over^{\delta A^{01}}}\,\delta\mu\Bigr)
	\Big)\\       
	&\Big(\ee^{-k\,S(h,A^{01})}\,\mathop{\otimes}\limits_\ell 
	h(\xi_\ell )\sous{R_\ell }\,\Bpsi(\hAR{h^{-1}})\Big).
\end{split}
\end{equation*}
Faisant agir $\KZmu$ sur $\ee^{-k\,S(h,A^{01})}\,\otimes_\ell  
h(\xi_\ell )\sous{R_\ell }$, on arrive \`a
\begin{equation*}
\begin{split}
\KZmu\Bpsi(A^{01})=
	\ee^{-k\,S(h,A^{01})}\,\mathop{\otimes}\limits_\ell 
	h(\xi_\ell )\sous{R_\ell }\, & \Big\lbrace
	d_{\delta\mu}+{_i\over^2}\int_\Sigma\tr\,(A^{10}-h\partial h^{-1})
	\wedge \Big({_\delta\over^{\delta A^{01}}}\,\delta\mu\Big)\\
&\ +{_{\pi}\over^{2k}}\int_\Sigma\tr\,{_\delta\over^{\delta A^{01}}}
	\wedge \Big({_\delta\over^{\delta A^{01}}}\,\delta\mu\Big)
	\Big\rbrace\,\Bpsi(\hAR{h^{-1}}).
\end{split}
\end{equation*}
Si on revient \`a la d\'efinition de la d\'eriv\'ee fonctionnelle, 
on montre facilement que
\[
\quotient{\delta}{\delta A^{01}}\,\Bpsi(\hAR{h^{-1}})=
	h\,{_{\delta\Bpsi}\over^{\delta A^{01}}}(\hAR{h^{-1}})\,h^{-1}.
\]
On v\'erifie alors que
\[
{_i\over^2}\int_\Sigma\tr\,(A^{10}-h\partial h^{-1})\wedge
	\Big({_\delta\over^{\delta A^{01}}}\,\delta\mu\Big)\,
	\Bpsi(\hAR{h^{-1}})={_i\over^2}\int_\Sigma\tr\,\hAL{h^\dagger}
	\wedge\Big({_{\delta\Bpsi}\over^{\delta A^{01}}}\,\delta\mu\Big).
\]
Pour obtenir la d\'eriv\'ee par rapport \`a $\delta\mu$, on 
r\'etablit la d\'ependance compl\`ete en $A$ et
\[
d_{\delta\mu}\Bpsi(\hAL{h^\dagger}+\hAR{h^{-1}})=(d_{\delta\mu}\Bpsi)
	(\hAR{h^{-1}})+{_i\over^2}\int_\Sigma\tr\,(\hAR{h^{-1}}-
	\hAL{h^\dagger})\wedge\left({_{\delta\Bpsi}\over^{\delta A^{01}}}\,
	\delta\mu\right).
\]
\`A la fin, les termes se compensent bien pour donner l'identit\'e de Ward~:
\[
(\KZmu\Psi)(A^{01})=\ee^{-k\,S(h,A^{01})}\,\mathop{\otimes}\limits_\ell 
	h(\xi_\ell )\sous{R_\ell }\,(\KZmu\Psi)(\hAR{h^{-1}}).
\]
Il va sans dire qu'on doit proc\'eder de m\^eme avec $\KZz\Bpsi$.

\subsection{R\'egularisation}

Nous venons de construire la connexion de KZB par des manipulations 
fonctionnelles. Comme nous avons oubli\'e (volontairement) les probl\`emes 
de convergences pos\'es par l'int\'egrale fonctionnelle, on a trouv\'e des 
termes singuliers. N\'eanmoins, on a obtenu au chapitre 2 
des comportements \`a courtes distances qui permettent de 
convenablement r\'egulariser la connexion de KZB.
Les termes qui nous emb\^etent correspondent \`a l'insertion
d'un ou deux courants, or 
\begin{equation}
\label{opsingulier}
\begin{split}
{_\delta\over^{\delta A\sur{a}_{\overline{z}}}}\,\Bpsi&=
	\left(-{_1\over^\pi}\,{_{t^a_\ell }\over^{z-z_\ell }}
	+\cdots\right)\,\Bpsi,\\
{_\delta\over^{\delta A\sur{a}_{\overline{z}}}}\,
	{_\delta\over^{\delta A\sur{b}_{\overline{w}}}}\,\Bpsi&
	=\left({_k\over^{2\pi^2}}\,{_{\delta\sur{ab}}\over^{(z-w)^2}}
	-{_i\over^\pi}\,{_{f^{abc}}\over^{z-w}}+\cdots\right)\,\Bpsi,
\end{split}
\end{equation}
si $A$ est un champ de jauge nul autour des points d'insertion soit $z$, 
$w$ et $z_\ell $. En particulier, 
\[
\tr\,\Bigl({_\delta\over^{\delta A_{\overline{z}}}}\,
	{_\delta\over^{\delta A_{\overline{w}}}}\Bigr)\,\Bpsi
	=\left({_k\over^{4\pi^2}}\,{_{\dim G}\over^{(z-w)^2}}+
	\cdots\right)\,\Bpsi.
\]

Pour obtenir une expression valable pour n'importe quel champ de jauge, on 
utilise le transport parall\`ele le long du segment $[z,w]$~:
$\boldsymbol{e}_{z,w}=\expordonner^{\,\int_z^wA^{01}}$.
Rappelons que l'exponentielle ordonn\'ee $f(t)=
\expordonner^{\,\int_0^t\rho(\sigma)\,d\sigma}$ est la solution de 
l'\'equation diff\'erentielle $f'(t)=f(t)\,\rho(t)$, $t\in[0,1]$, 
avec la condition initiale $f(0)=1$.  Soit $h\in\CG^\C$ telle 
que $\hAR{h^{-1}}=0$ autour de $z$, \cad $A^{01}=h\de h^{-1}$ au voisinage 
de $z$. Un calcul classique conduit \`a
\[
h(z)^{-1}\,\boldsymbol{e}_{z,w}\,h(w)=
	\expordonner^{\,\int_z^{w}h^{-1}\partial h}
\]
si $w$ est au voisinage de $z$. On en d\'eduit le d\'eveloppement 
limit\'e suivant
\begin{equation}
\begin{split}
h(z)^{-1}\,\boldsymbol{e}_{z,w}\,h(w)=\text{exp}\Bigl\lbrace
	&(w-z)\,(h^{-1}\partial_z h)(z)+{_1\over^2}\,(w-z)^2\,
	\partial_z(h^{-1}\partial_zh)(z)\\
&+{_1\over^2}\,(w-z)\,(\overline{w}-\overline{z})\,
	\partialbarz(h^{-1}\partial_zh)(z)+\text{o}(|w-z|^2)
	\Bigr\rbrace.
\end{split}
\label{developpement}
\end{equation}
Dans la suite, on note $\boldsymbol{e}'_{z,w}$ le d\'eveloppement limit\'e, 
\cad $\boldsymbol{e}_{z,w}=h(z)\,\boldsymbol{e}'_{z,w}\,h(w)^{-1}$.
On obtient alors les d\'eveloppements \`a courtes distances en champ de jauge 
arbitraire 
\begin{equation}
\begin{split}
\tr\,\left(t^a\,\text{Ad}_{\boldsymbol{e}_{z_\ell ,z}}\,
	{_\delta\over^{\delta A_{\overline{z}}}}\right)\,\Bpsi&=
	\left({_1\over^{2\pi}}\,{_{t^a_\ell }\over^{z-z_\ell }}+\cdots\right)\,\Bpsi,\\
\tr\,\left(\text{Ad}_{\boldsymbol{e}_{z,w}}
	\Bigl({_\delta\over^{\delta A_{\overline{w}}}}\Bigr)\,
	{_\delta\over^{\delta A_{\overline{z}}}}\right)\,\Bpsi
	&=\left({_k\over^{4\pi^2}}\,{_{\dim G}\over^{(z-w)^2}}+
	\cdots\right)\,\Bpsi.
\end{split}
\end{equation}
Il est maintenant facile de d\'efinir la r\'egularisation des 
op\'erateurs~\eqref{opsingulier}~:
\begin{equation}
\begin{split}
\singulier\quotient{\delta}{\delta A\sur{a}_{\overline{z}_\ell}}\singulier
	&\equiv \lim_{z\rightarrow z_\ell }\left(
	2\,\tr\,\left(t^a\,\text{Ad}_{\boldsymbol{e}_{z_\ell ,z}}\,
	{_\delta\over^{\delta A_{\overline{z}}}}\right)
	-{_1\over^\pi}\,{_1\over^{z-z_\ell }}\,t^a_\ell \right),\\
\tr\,\singulier\deltaz\,\deltaz\singulier 	
	&\equiv \lim_{w\rightarrow z}\left(
	\tr\,\text{Ad}_{\boldsymbol{e}_{z,w}}
	\Bigl({_\delta\over^{\delta A_{\overline{w}}}}\Bigr)\,
	{_\delta\over^{\delta A_{\overline{z}}}}-
	{_k\over^{4\pi^2}}\,{_{\dim G}\over^{(z-w)^2}}\right).
\end{split}
\end{equation}
Ces deux op\'erateurs sont finis --- dans la derni\`ere \'equation
on suppose que $z\neq z_\ell$. 
La renormalisation du produit scalaire
des \'etats de Chern-Simons introduit une d\'ependance dans la
m\'etrique. Par cons\'equent, la connexion d\'epend aussi de 
la m\'etrique. On utilise une m\'etrique localement plate, \cad
compatible avec la structure complexe et plate autour des points d'insertion
et du support de $\delta \mu^z_{\bar z}$.
Dans un changement de structure complexe infinit\'esimal,  on suppose
que la m\'etrique change par $\delta\gamma=\frac{i}{2}\,\delta\mu^z_{\bar z}\,d\bar z^2
-\frac{i}{2}\,\bar{\delta\mu^z_{\bar z}}\,d z^2$. Ainsi,
\begin{equation}
\label{belleamie}
\delta\gamma_{z\bar z}=0,\qquad\delta\gamma^{zz}=\quotient{2}{i}\,\delta\mu^z_{\bar z}.
\end{equation}
Pour obtenir les bonnes expressions
pour la connexion, on doit changer $k$~: $k\mapsto \kappa=k+g^\vee$.
Si on ne translate pas $k$, on n'a plus une connexion.
La connexion de KZB r\'egularis\'ee 
\begin{align}
\label{KZren}
\KZmu&=d_{\delta\mu}+i\int_\Sigma\tr\,A_{z}\,
	{_\delta\over^{\delta A_{\overline{z}}}}\,\deltamu\,d^2z
	+{_{i\pi}\over^{\kappa}}
	\int_\Sigma\tr\,\singulier
	\deltaz\,\deltaz\singulier\deltamu\,d^2z,
	\tag{\ref{KZren}.a}\\
\KZz&=\partial_{z_\ell}-{_{2\pi}\over^{\kappa}}\,t^a_\ell \,
	\singulier{_\delta\over^{\delta A\sur{a}_{\overline{z}_\ell }}}
	\singulier
	\tag{\ref{KZren}.b}
\addtocounter{equation}{1}
\end{align}
pr\'eserve bien l'espace des sections de $\ChSi$.
Dans la premi\`ere \'equation, les op\'erateurs
$\frac{\delta}{\delta A_{\overline{z}}}$ et $\tr\,\singulier
\frac{\delta}{\delta A_{\overline{z}}}\,
\frac{\delta}{\delta A_{\overline{z}}}\singulier$ sont singuliers quand on fait
tendre $z$ vers $z_\ell$. On commence donc par int\'egrer sur $\Sigma$ priv\'e
de \<<petits\>> voisinages autour des points d'insertion, qu'on fait ensuite
tendre vers z\'ero.

\subsection{Construction de Sugawara}

Au chapitre 2, on a utilis\'e la construction de Sugawara sans d\'emonstration.
Ce qui manquait alors \'etait la connexion de KZB.
D'apr\`es la factorisation holomorphe des fonctions de Green modifi\'ees,
\begin{equation}
\label{holfac0}
\widegamma(0)=\sum_{p,q}h^{pq}\,\Psi_{p}(0)\otimes
	\overline{\Psi_{q}(0)}
\end{equation}
o\`u $\Psi_p$ est une base de l'espace des \'etats de Chern-Simons et
$(h^{pq})$ est la matrice inverse de $(\Psi_p,\Psi_q)$.
On choisit la base de telle sorte que $\Bnabla\Psi_p=0$ 
en $(\J,\underline{\xi})\in\moduleR$. En d\'erivant la
derni\`ere \'equation dans la direction holomorphe en $\J$,
on trouve facilement~(\footnote{
Au chapitre 2, on avait remarqu\'e que le d\'eveloppement \`a courtes
distances~\eqref{ward12} nous \'echappait. Pour l'obtenir, il faudrait reprendre
ce qui suit en champ non-nul.})
\begin{eqnarray*}
d_{\delta\mu}\widegamma(0)&=&\sum_{p,q}h^{pq}\,d_{\delta\mu}\Psi_{p}(0)\otimes
        \overline{\Psi_{q}(0)}\\
&=& -{_{i\pi}\over^{\kappa}}\,{\widetilde{Z}}\,
        \int_\Sigma\tr\,\singulier
        \deltaz\,\deltaz\singulier\,\widegamma(0)\,\deltamu\,d^2z.
\end{eqnarray*}
D'apr\`es l'\'equation~\eqref{belleamie}~:
$\delta\mu^z_{\bar z}=\frac{i}{2}\,\delta\gamma^{zz}$. Ceci et
les d\'efinitions du tenseur d'\'energie-impulsion et des courants nous
am\`enent directement \`a la construction de Sugawara~:
\begin{equation}
\label{SUGA}
\langle\,T_{zz}\,\grl\,\rangle=
{_2\over^{\kappa}}\,\langle\,\mathop{\lim}\limits_{w\rightarrow z}%
\left(\tr\,J_w\,J_z-{_{k\,\dim G}\over^{4\,(w-z)^2}}\right)
\,\grl\,\rangle.
\end{equation}

Un autre point laiss\'e de c\^ot\'e \'etait~: les champs $g(\xi_\ell)\sous{R_\ell}$
sont des champs primaires de poids conformes $(\Delta_\ell,\Delta_\ell)$.
Pour d\'emontrer cette assertion, il suffit de
d\'emontrer les d\'eveloppements~(2.\ref{ward9}.a-b).
D\'erivons l'\'equation~\eqref{holfac0} dans la direction holomorphe
en $\xi_\ell$~:
\[
\partial_{z_\ell}\widegamma(0)=\quotient{2\pi}{\kappa}\,
	t^a_\ell\,\singulier
	\quotient{\delta}{\delta A\sur{a}_{\overline{z}_\ell}}\singulier
	\widegamma(0)
\]
d'o\`u 
\begin{equation}
\label{COUR}
\partial_{z_\ell}\langle\,\grm\,\rangle=-\quotient{2}{\kappa}\,\,t^a_\ell\,
	\langle\,\mathop{\lim}\limits_{z\rightarrow z_\ell}
	\Big(J^a_z+\quotient{1}{z-z_\ell}\,t^a_\ell\Big)\grm\,\rangle.
\end{equation}
Les \'equations~(\ref{SUGA}-\ref{COUR}) sont les versions quantiques de 
$T_{zz}=\frac{2}{k}\,\tr\,\big(J_z\big)^2$
et $J_z=-\frac{k}{2}\,\partial_zg\,g^{-1}$ (cf. p.~\pageref{SUGACOUR}).
Reprenons l'\'equation~(2.\ref{ward7bis}), sym\'etris\'ee en $z$ et $w$, 
apr\`es sommation en $a=b$,
\begin{equation*}
\begin{split}
\langle\,J^{a}_{w}&\,J^{a}_{z}%
\,\grm\,\rangle\,=\,{_{k\,{\dim G}/2}\over^{(z-w)^2}}%
\,\langle\,\grm\,\rangle+\quotient{1}{z-z_\ell}\,\quotient{1}{w-z_\ell }\,
	t^a_\ell t^a_\ell\,\langle\,\grm\,\rangle\\
&-\sum_\ell{_{t^a_\ell }\over^{z-z_\ell }}\,%
\langle\,\big(J^{a}_{w}+\quotient{1}{w-z_\ell}\big)\,\grm\,\rangle
-\sum_\ell{_{t^a_\ell }\over^{w-z_\ell }}\,%
\langle\,\big(J^{a}_{z}+\quotient{1}{z-z_\ell}\big)\,\grm\,\rangle+\cdots
\end{split}
\end{equation*}
On prend la limite quand $w$ tend vers $z$. Il appara\^\i t
le tenseur d'\'energie d'impulsion~:
\[
\langle\,T_{zz}\,\grm\,\rangle
=\quotient{1}{(z-z_\ell)^2}\,\quotient{C_\ell}{\kappa}\,
	\langle\,\grm\,\rangle
	-\quotient{1}{z-z_\ell}\,\quotient{2}{\kappa}\,
	\langle\,\big(J^a_z+\quotient{1}{z-z_\ell}\,t^a_\ell\big)\grm\,\rangle
	+\cdots
\]
o\`u $C_\ell$ est le Casimir quadratique $t^a_\ell t^a_\ell$ dans la
repr\'esentation $R_\ell$. Cette \'equation et l'\'egalit\'e~\eqref{COUR}
donnent le d\'eveloppement~(2.\ref{ward9}.a)
\[
T(z)\,g(\xi_\ell )\sous{R_\ell }=
        {_{\Delta_\ell }\over^{(z-z_\ell )^2}}\,
        g(\xi_\ell )\sous{R_\ell }+
        {_1\over^{z-z_\ell }}\,\partial_{z_\ell }
        g(\xi_\ell )\sous{R_\ell }+\cdots
\]
On a donc d\'emontrer que $g(\xi_\ell)\sous{R_\ell}$ est un champ
primaire de poids conformes $(\Delta_\ell,\Delta_\ell)$ avec
$\Delta_\ell=C_\ell/\kappa$. Ainsi, la
connexion de KZB est la source g\'eom\'etrique du tenseur
d'\'energie-impulsion et des champs primaires $g(\xi_\ell)\sous{R_\ell}$
du mod\`ele de WZNW.

\subsection{La connexion en m\'etrique g\'en\'erale}

Jusqu'\`a maintenant, on s'est int\'eress\'e \`a la connexion de KZB
dans une m\'etrique localement plate et satisfaisant les \'equations
~\eqref{belleamie}. Il est possible d'extraire
des r\'esultats pr\'ec\'edents la connexion en m\'etrique g\'en\'erale --- cf.
aussi~\cite{eguchi} pour les th\'eories conformes en m\'etrique
g\'en\'erale.
Lorsqu'on change la m\'etrique par transformation de Weyl, le produit
scalaire est multipli\'e par (cf. \'eq. (2.\ref{weyl1}) et (2.\ref{weyl2}))
\[
\ee^{-\frac{ic}{24\pi}\,S_L(\sigma)}\,
\prod_\ell\ee^{\Delta_\ell\sigma(\xi_\ell)}
\]
o\`u $c$ est la charge centrale du mod\`ele de WZNW, \cad
$c=k\,\dim G/\kappa$. On doit alors rajouter
\begin{equation}
\label{DER}
d_{\delta\mu}\big(
	-\quotient{ic}{24\pi}\,S_L(\sigma)+\sum_\ell\Delta_\ell\,\sigma(\xi_\ell)
	\big)
\end{equation}
\`a $\KZmu$ et $\Delta_\ell\,\partial_{z_\ell}\sigma$ \`a $\KZz$.
La forme la plus g\'en\'erale de la connexion de KZB est donc
\begin{align}
\label{KZMET}
\KZmu&=d_{\delta\mu}+i\int_\Sigma\tr\,A_{z}\,
        {_\delta\over^{\delta A_{\overline{z}}}}\,\deltamu\,d^2z
        +{_{i\pi}\over^{\kappa}}
        \int_\Sigma\tr\,\singulier
        \deltaz\,\deltaz\singulier\deltamu\,d^2z,
        \non\\
&\quad -\quotient{ic}{24\pi}\,\int_\Sigma\Big(
	\delta(\ln \gamma_{z\bar z})R_\gamma+t_{zz}\,\delta\mu^z_{\bar z}\,d^2z\Big)
	+\sum_\ell\Delta_\ell\,\delta(\ln \gamma_{z\bar z})(\xi_\ell),
	\taga{KZMET}\\
\KZz&=\partial_{z_\ell}-{_{2\pi}\over^{\kappa}}\,t^a_\ell \,
        \singulier{_\delta\over^{\delta A\sur{a}_{\overline{z}_\ell }}}
        \singulier+\Delta_\ell\,\partial_{z_\ell}(\ln \gamma_{z\bar z})(\xi_\ell).
	\tagb{KZMET}
\addtocounter{equation}{1}
\end{align}
Dans ces \'equations, $\gamma_{z\bar z}$ est une m\'etrique riemannienne
$\gamma=2\,\gamma_{z\bar z}dzd\bar z$ sur $\Sigma$ pour une
certaine structure complexe et changeant avec celle-ci suivant
\[
\delta\gamma=\gamma_{z\bar z}\,\Big(
	\quotient{i}{2}\,\delta\mu^z_{\bar z}\,d\bar z^2+
	2\,\delta(\ln \gamma_{z\bar z})\,dzd\bar z
	-\quotient{i}{2}\,\bar{\delta\mu^z_{\bar z}}\,d z^2\Big).
\]
La courbure de $\gamma$ est $R_\gamma=\de\da\ln\gamma_{z\bar z}$ et 
$t_{zz}=\partial_z^2\ln\gamma_{z\bar z}-\frac{1}{2}\,(\partial_z\ln\gamma_{z\bar z})^2$.
Les termes compl\'ementaires dans l'\'equation~(\refa{KZMET}) sont
\'egaux ensemble \`a la d\'eriv\'ee~\eqref{DER} en $\sigma=\ln \gamma_{z\bar z}$.  
Le second terme correspond \`a un changement de classe conforme de m\'etrique
et les autres au changement dans la classe conforme par $\delta\sigma
=\delta(\ln\gamma_{z\bar z})$.

La contrepartie g\'eom\'etrie des d\'eveloppements \`a courtes distances
~(2.\ref{ward11}) ou leurs \'equivalents alg\'ebriques~(2.\ref{vir})
est~: {\bol la connexion de KZB est projectivement plate}, \cad
\[
\Bnabla^2=\quotient{c}{2}\,\Bnabla_{\rm Quillen}^2
	-\sum_\ell\Delta_\ell\,R_\gamma(\xi_\ell).
\]
Ici, $\Bnabla_{\rm Quillen}^2$ est la courbure de la connexion (m\'etrique) de Quillen sur
le fibr\'e d\'eterminant de la famille d'op\'erateurs
$\de$ agissant sur les fonctions sur $\Sigma$.

\subsection{Quantification des syst\`emes de Hitchin}

Si $p$ est le polyn\^ome quadratique $p_2=\tr$, alors l'application
de Hitchin $h_{p_2}$
prend ses valeurs dans $H^0(K^2)$ --- dans le cas sans points d'insertion.
On a vu que c'est aussi l'espace cotangent \`a l'espace des modules
$\CM_{g}$. On peut coupler une classe $[\delta\mu]$ de diff\'erentielles
de Beltrami dans $H^1(K^{-1})=T\CM_g$ avec une diff\'erentielle 
quadratique $\rho$ dans $H^0(K^2)$ par
\[
\int_\Sigma\rho\,\delta\mu.
\]
Les diff\'erentielles de la forme 
$\partial_z\otimes\de(\delta v^z)$ se couplent
avec un $\rho$ pour donner z\'ero, donc on a une bonne d\'efinition
sur les classes $[\delta\mu]$. En particulier, pour $\rho=h_{p_2}$,
on peut d\'efinir un nouveau Hamiltonien par
\begin{equation}
\label{bel}
h_{\delta\mu}=\int_\Sigma h_{p_2}\,\delta\mu.
\end{equation}
Si on admet des points d'insertion, le Hamiltonien de Hitchin est 
une forme quadratique dans $H^0(K^2(2\sum_\ell\xi_\ell))
=T^*\CM_{g,N}^\star$, \cad
m\'eromorphe avec \'eventuellement des
p\^oles d'ordre $\leq 2$ en $\xi_\ell$. On peut encore coupler $h_{p_2}$ 
avec une classe $[\delta\mu]\in H^1(K^{-1}(-2\sum_\ell\xi_\ell)=
T\CM_{g,N}^\star$ de formes de Beltrami
par l'\'equation~\eqref{bel}. Les Hamiltoniens $h_{\delta\mu}$, pour
diff\'erentes $\delta\mu$, sont en involution sur $\CP_\CO$ --- sur $\CP$ s'il
n'y a pas de points d'insertion.

L'espace des phases $\CP\cong T^*\CN_s$ peut 
\^etre quantifi\'e (g\'eom\'etriquement).
L'espace des sections holomorphes de la $k$-i\`eme puissance du 
fibr\'e d\'eterminant $\CL$ au-dessus de $\CN_{ss}$ est notre 
espace des \'etats
quantiques. Ce n'est rien d'autre que l'espace des \'etats de Chern-Simons 
$\CW(\Sigma,\emptyset,\emptyset)$.
La connexion de KZB $\KZmu$ donn\'ee par
l'\'equation~(\ref{KZZ}) est un op\'erateur du second
ordre de symbole principal proportionnel au Hamiltonien $h_{\delta\mu}$
--- on remplace $\frac{\delta}{\delta A^{01}}$ 
par $\frac{2k}{\pi i}\,\varphi^{10}$. 
On peut donc consid\'erer la connexion $\KZmu$ comme la quantification
de $\frac{2k}{\pi}\,h_{\delta\mu}$ reliant deux $H^0(\CL^k)$
pour des structures complexes qui diff\`erent d'un $\delta\mu$.
Par le changement d'\'echelle $\delta\mu\rightarrow \kappa\,\delta\mu$
dans $\KZmu$, on obtient dans la limite $\kappa\mapsto 0$ des op\'erateurs
agissant dans un seul $H^0(\CL^{-g^\vee})$, \cad \`a structure complexe
fixe. 

On peut aussi quantifier l'espace des phases $\CP_\CO$.
\`A l'orbite coadjointe $\CO_\ell$ au point $\xi_\ell$
on associe l'espace de repr\'esentation $V_{\lambda_\ell}$ de $G$. L'espace
des \'etats quantiques est l'espace des \'etats de Chern-Simons $\chsi$. 
La connexion $\KZmu$ est toujours la quantification du Hamiltonien 
$\frac{2k}{\pi}\, h_{\delta\mu}$. Cette fa\c{c}on de proc\'eder permet de
traiter en m\^eme temps les deux parties de la connexion, \cad
$\KZz$ et $\KZmu$.

\medskip
\section{Formules basses}

Au chapitre 3, on a construit une d\'ecoupe $s:\CN\ni n\mapsto A^{01}(n)
\in\CA^{01}$ de l'espace des modules. On se propose de 
r\'e\'ecrire les op\'erateurs apparaissant dans la 
connexion de KZB en fonction des d\'eriv\'ees de $\Bpsi$
par rapport \`a $n$. L'\'etape suivante consistera \`a utiliser
des d\'ecoupes explicites.

\subsection{Premi\`ere variation}

Calculons la variation de $\Bpsi(\hAR{h^{-1}})$ par rapport \`a
$\delta\,(\hAR{h^{-1}})=h^{-1}\,\bigl(\partialA(\delta h\,h^{-1})+
\delta A^{01}\bigl)\,h$ --- on sous-entend le point $z$ o\`u le 
champ $A^{01}$ est \'evalu\'e. L'op\'erateur 
$\partialA=\de+[A^{01},.]$ agit sur les fonctions $\CC^\infty$ sur 
$\Sigma$ \`a valeurs dans $\liegc$ pour donner une $(0,1)$-forme sur 
$\Sigma$ \`a valeurs dans $\liegc$.  On utilise deux m\'ethodes distinctes 
afin de trouver le premier ordre de $\delta\bigl(\Bpsi(\hAR{h^{-1}})\bigr)$, 
ensuite on compare les r\'esultats. D'une part, d'apr\`es l'identit\'e de 
Ward sur $\Bpsi$, c'est
\begin{equation*}
\begin{split}
\delta\bigl(\ee^{k\,S(h,A^{01})}&\,\hinv\,\Bpsi(A^{01})\bigr)
	=\expoS\,\hinv\,(\delta\Bpsi)(A^{01})-\sum_m(h^{-1}\delta h)
	(\xi_m)_m\,\Bpsi(\hAR{h^{-1}})\\
&\,+{_{ik}\over^{2\pi}}\int_\Sigma\tr\,
	\left(h\,\partial(\hAR{h^{-1}})\,h^{-1}\delta h\,h^{-1}
	\,+h\,\partial h^{-1}\wedge\delta A^{01}\right)\,\Bpsi(\hAR{h^{-1}}).\\
\end{split}
\end{equation*}
D'autre part, d'apr\`es la d\'efinition de la d\'erivation fonctionnelle, c'est
aussi
\begin{equation*}
\begin{split}
\int_\Sigma\tr\,\deltapsi(\hAR{h^{-1}})\wedge\delta\,(\hAR{h^{-1}})
	=&\,\int_\Sigma\tr\,\overline{D}\,\left(h\,\deltapsi
	(\hAR{h^{-1}})\,h^{-1}\right)\,
	\delta h \, h^{-1}\\
&\,+\int_\Sigma\tr\,h\,\deltapsi(\hAR{h^{-1}})\,h^{-1}\wedge \delta A^{01}
\end{split}
\end{equation*}
o\`u $\Dbar=\de+[A^{01},.]_+$ --- $[.,.]_+$ est l'anticommutateur~--- 
c'est l'op\'erateur agissant sur les $(1,0)$-formes sur
$\Sigma$ \`a valeurs dans $\lieg^\C$ pour donner une $2$-forme sur $\Sigma$ 
\`a valeurs dans $\liegc$. Cet op\'erateur intervient quand on int\`egre 
par parties un terme en $\partialA$. Le noyau de $\Dbar$ est isomorphe
\`a l'espace cotangent en $A^{01}$ \`a $\CN$ et son conoyau est vide.
On note $(\phi_\alpha)$ une base de $\ker\Dbar$. 
Soit enfin l'op\'erateur inverse $\overline{D}^{-1}:\Lambda^2(\Sigma,\liegc)
\rightarrow\Lambda^{1,0}(\Sigma,\liegc)$ d\'efini par 
$\Dbar\,\Dbar^{-1}\varpi=\varpi$ et
\begin{equation}
\label{modulinit}
\int_\Sigma\tr\,(\Dbar^{-1}\varpi)\wedge\partial_{n^\alpha}A^{01}=0,\qquad
			\alpha=1,\cdots,\dim\CN.
\end{equation}
Comparons les r\'esultats pr\'ec\'edents~: avec $\delta h\,h^{-1}$ 
arbitraire et 
$\delta A^{01}$ nulle, on obtient
\begin{equation}
\label{preums1}
\begin{split}
\overline{D}\,\Big(h\,\deltapsi(\hAR{h^{-1}})\,h^{-1}\Big)
	=&\,{_{ik}\over^{2\pi}}\,h\,\partial(\hAR{h^{-1}})\,h^{-1}
	\,\Bpsi(\hAR{h^{-1}})\\
&\,-2\sum_mh\,t^a\,h^{-1}\,\delta\sur{(2)}(z-z_m)\,d^2z\,t^a_m\,
	\Bpsi(\hAR{h^{-1}}),
\end{split}
\end{equation}
--- sauf mention contraire $h$ et $A^{01}$ sont pris en $z$ --- tandis qu'avec $\delta h\,h^{-1}$ 
nulle et $\delta A^{01}$ arbitraire,
\begin{equation}
	\label{preums2}
	\begin{split}
	\int_\Sigma\tr\,h\,\deltapsi(\hAR{h^{-1}})\,h^{-1}\wedge&
	\,\partialpha A^{01}=
	\expoS\,\hinv\,(\partialpha\Bpsi)(A^{01})\\
&\,+{_{ik}\over^{2\pi}}\,\Big(\int_\Sigma\tr\,
	h\,\partial h^{-1}\wedge\partialpha A^{01}\Big)\,\Bpsi(\hAR{h^{-1}}).
	\end{split}
\end{equation}
Appliquons $\Dbar^{-1}$ \`a l'\'equation~\eqref{preums1}~:
\begin{equation*}
\begin{split}
h\,\deltapsi(\hAR{h^{-1}})\,h^{-1}
	=&\,{_{ik}\over^{2\pi}}\,\Dbar^{-1}\Big(h\,\partial(\hAR{h^{-1}})\,
	h^{-1}\Big)\,\Bpsi(\hAR{h^{-1}})\\
&\,-2\sum_m\Dbar^{-1}\bigl(h\,t^a\,h^{-1}\,
	\delta\sur{(2)}(z-z_m)\,d^2z\bigr)\,t^a_m\,
	\Bpsi(\hAR{h^{-1}})+c^\alpha\,\phi_\alpha
\end{split}
\end{equation*}
o\`u $c^\alpha$ sont des nombres complexes. Afin de d\'eterminer ces derniers, 
on introduit l'\'egalit\'e pr\'ec\'edente dans l'\'equation~\eqref{preums2}~:
\begin{equation*}
\begin{split}
c^\alpha\int_\Sigma\tr\,\phi_\alpha\wedge\partial_{n^\beta}A^{01}
	=&\,\expoS\,\hinv\,(\partial_{n^\beta}\Bpsi)(A^{01})\\
	&\,+{_{ik}\over^{2\pi}}\int_\Sigma\tr\,(h\partial h^{-1})\wedge
	\partial_{n^\beta}A^{01}\ \Bpsi(\hAR{h^{-1}}).
\end{split}
\end{equation*}
Soit $\Omega$ l'inverse de la matrice d'\'el\'ements
$\int\tr\,\phi_\alpha\wedge\partial_{n^\beta}A^{01}$. 
On trouve $c^\alpha$ en multipliant la derni\`ere \'egalit\'e
\`a droite par $\Omega^{\beta\alpha}$. Ensuite, on r\'einjecte $c^\alpha$ 
\begin{equation*}
\begin{split}
\deltapsi&(\hAR{h^{-1}})
	=h^{-1}\,\Biggl\lbrace
	{_{ik}\over^{2\pi}}\,\Dbar^{-1}\Big(h\,\partial(\hAR{h^{-1}})\,h^{-1}
	\Big)
	\,\Bpsi(\hAR{h^{-1}})\\
&\,-2\sum_m\Dbar^{-1}\bigl(h\,t^a\,h^{-1}\,
	\delta\sur{(2)}(z-z_m)\,d^2z\bigr)\,t^a_m\,
        \Bpsi(\hAR{h^{-1}})\\
&+\left(\expoS\,\hinv\,(\partial_{n^\beta}\Bpsi)(A^{01})
	+{_{ik}\over^{2\pi}}\int_\Sigma\tr\,(h\partial h^{-1})\wedge
	\partial_{n^\beta}A^{01}\ \Bpsi(\hAR{h^{-1}})\right)
	\,\Omega^{\beta\alpha}\phi_\alpha\Biggr\rbrace\,h,
\end{split}
\end{equation*}
\cad la premi\`ere variation de $\Bpsi$ en $\hAR{h^{-1}}$. En $h=1$, on 
obtient la d\'eriv\'ee fonctionnelle de $\Bpsi$ en $A^{01}$~:
\begin{equation}
\label{premvar}
{_{\delta\Bpsi}\over^{\delta A^{01}}}
	={_{ik}\over^{2\pi}}\,\Dbar^{-1}(\partial A^{01})\,\Bpsi-
	2\sum_m\Dbar^{-1}\bigl(t^a\,\delta\sur{(2)}(z-z_m)\,d^2z\bigr)
	\,t^a_m\,\Bpsi
	+\Omega^{\beta\alpha}\phi_\alpha\,\partial_{n^\beta}\,\Bpsi.
\end{equation}

\subsection{Seconde variation}

\'Etudions la variation de l'avant-derni\`ere \'equation par rapport 
\`a $\delta(\hAR{h^{-1}})$ en $w\neq z$ pour $h=1$, $\delta h\neq 0$ 
et $\delta A^{01}=0$, \cad $\delta(\hAR{h^{-1}})=\partialA(\delta h)(w)$. 
On introduit la notation tensorielle afin de distinguer les deux copies de
$\liegc$ entrant dans le calcul, ainsi que la trace $\tr_1$ dans le premier 
espace. Pour commencer, on a
\[
\delta\bigl(\Bpsi(\hAR{h^{-1}})\bigr)
	=\quotient{ik}{2\pi}\,\int_\Sigma\tr\,\delta h\,\partial A^{01}\ 
	\Bpsi(A^{01})-\sum_m\delta h(\xi_m)_m\ \Bpsi(A^{01}).
\]
Ensuite, comme pour obtenir la premi\`ere variation, on calcule de deux
fa\c{c}ons la quantit\'e $\delta\Big(\frac{\delta\Bpsi(\hAR{h^{-1}})}{\delta A^{01}(z)}\Big)$.
\begin{equation}
\label{sec1}
\begin{split}
\int_\Sigma&\tr_1\,\deltaW\otimes\deltaZpsi(A^{01})
	\wedge\partialA(\delta h)(w) 
	=-\left[\delta h,\deltaZpsi\right]\\
&+{_{ik}\over^{2\pi}}\,\overline{D}^{-1}\bigl(
	[\delta h,\partial A^{01}]+\partial\,\partialA(\delta h)\bigr)\,\Bpsi
	-2\sum_m\Dbar^{-1}\bigl([\delta h,t^a]\,\delta\sur{(2)}(z-z_m)
	\,d^2z\bigr)\,t^a_m\,\Bpsi\\
&-\quotient{ik}{2\pi}\Big(\int_\Sigma\tr\,\partial\delta h\wedge
	\partial_{n^\beta}A^{01}\Big)\ \Omega^{\beta\alpha}\phi_\alpha\ \Bpsi
	+\left\lbrace {_{ik}\over^{2\pi}}\int_\Sigma\tr\,\delta h\,
	\partial A^{01}-\sum_m\delta h(\xi_m)_m\right\rbrace\\
&\times\,\left\lbrace {_{ik}\over^{2\pi}}\,\overline{D}^{-1}(\partial A^{01})
	-2\sum_{m'}\Dbar^{-1}\bigl(t^a\,\delta\sur{(2)}(z-z_{m'})\,d^2z\bigr)
	\,t^a_{m'}+\Omega^{\beta\alpha}\phi_\alpha\,\partial_{n^\beta}
	\right\rbrace\Bpsi\\
=&-\left[\delta h,\deltaZpsi\right]+\left\lbrace{_{ik}\over^{2\pi}}
	\int_\Sigma\tr\,\delta h\,\partial A^{01}-\sum_m\delta h(\xi_m)_m
	\right\rbrace\deltaZpsi\\
&-{_{ik}\over^{2\pi}}\left\lbrace\overline{D}^{-1}
	\overline{D}(\partial\delta h)+\Big(\int_\Sigma\tr\,\partial\delta 
	h\wedge\partial_{n^\beta} A^{01}\Big)
	\Omega^{\beta\alpha}\phi_\alpha\right\rbrace\Bpsi\\
&-2\sum_m\Dbar^{-1}\bigl([\delta h,t^a]\,\delta\sur{(2)}(z-z_m)
	\,d^2z\bigr)\,t^a_m\,\Bpsi,
\end{split}
\end{equation}
apr\`es avoir utilis\'e l'\'equation~\eqref{premvar} ainsi que l'\'egalit\'e 
$[\delta h,\partial A^{01}]+\partial\partialA(\delta h)
=-\overline{D}(\partial\delta h)$. Notons \'egalement que la d\'efinition de 
$\overline{D}^{-1}$ entra\^\i ne que
\begin{equation}
\label{rork}
\phi=\overline{D}^{-1}\overline{D}\phi
	+\Big(\int_\Sigma\tr\,\phi\wedge\partial_{n^\beta} A^{01}\Big)
	\Omega^{\beta\alpha}\phi_\alpha
\end{equation}
pour tout $\phi\in\Lambda^{1,0}(\Sigma,\liegc)$. 
En particulier, cette \'egalit\'e avec $\phi=\partial\delta h$ permet de 
simplifier encore l'\'equation~\eqref{sec1}~:
\begin{equation*}
\begin{split}
\int_\Sigma\tr_1\,\deltaW&\otimes\deltaZpsi(A^{01})
	\wedge\partialA(\delta h)(w)\\ 
	=&-\left[\delta h,\deltaZpsi\right]+\left\lbrace
	{_{ik}\over^{2\pi}}\int_\Sigma\tr\,\delta h\,\partial A^{01}
	-\sum_m\delta h(\xi_m)_m\right\rbrace\deltaZpsi\\
&-\left\lbrace\quotient{ik}{2\pi}\,\partial\delta h
	+2\sum_m\Dbar^{-1}\bigl([\delta h,t^a]\,\delta\sur{(2)}
	(z-z_m)\,d^2z\bigr)\,t^a_m \right\rbrace\,\Bpsi.
\end{split}
\end{equation*}
La variation $\delta h$ \'etant arbitraire, on peut l'\'eliminer. Par 
exemple, montrons comment transformer le dernier terme \`a cet effet. 
Soit $\Green$ 
le noyau de $\overline{D}^{-1}$, \cad l'application \`a valeurs dans 
$\text{End}\,\liegc$ telle que
\[
\overline{D}^{-1}\varpi(z)=\Big(\int_\Sigma\Green(z,v)
	\left(\varpi(v)\right)\Big)dz
\]
pour $\varpi\in\Lambda^{2}(\Sigma,\liegc)$. Le dernier terme devient
$-2\sum_m\Green(z,z_m)\bigl([\delta h(z_m),t^a]\bigr)\,dz\,t^a_m\,\Bpsi$,
soit apr\`es transformation
\[
4i\,f^{abc}\sum_m\int_\Sigma\tr_1\,
	\delta\sur{(2)}(w-z_m)\,d^2 w\, t^b\otimes
	\Green(z,z_m)(t^c)\,dz\ t^a_m\,\Bpsi\ \delta h(w).
\]
Des consid\'erations similaires nous conduisent \`a
\begin{equation}
\label{sec2}
\begin{split}
\overline{D}_w\deltaW&\otimes\deltaZpsi=
	-2\,\delta\sur{(2)}(w-z)\,t^a\,d^2w\otimes\left[t^a,\deltaZpsi\right]
	+{_{ik}\over^{2\pi}}\,\partial A^{01}(w)\otimes\deltaZpsi\\
&-2\sum_m\delta\sur{(2)}(w-z_m)\,t^a\,d^2w\otimes t^a_m\,\deltaZpsi
	-{_{ik}\over^{\pi}}\,\partial_z\delta(w-z)\,t^a\,d^2w\otimes 
	t^a\,dz\,\Bpsi\\
&+4i\,f^{abc}\sum_m\delta\sur{(2)}(w-z_m)\,t^a\,d^2w\otimes
	\Green(z,z_m)(t^b)\,dz\ t^c_m\,\Bpsi
\end{split}
\end{equation}
o\`u $\Dbar_w$ est l'op\'erateur $\Dbar$ correspondant \`a l'insertion de 
$A^{01}$ en $w$. Appliquons l'op\'erateur $\overline{D}^{-1}_w$ \`a 
l'\'equation~\eqref{sec2}.
\begin{equation*}
\begin{split}
\deltaW\otimes&\deltaZpsi=
	-2\Big(\int_\Sigma\Green(w,u)\left(t^a\right)\,\delta\sur{(2)}(u-z)
	\,d^2u\Big)dw\otimes\left[t^a,\deltaZpsi\right]\\
&+{_{ik}\over^{2\pi}}\Big(\int_\Sigma\Green(w,u)\
	\left(\partial A^{01}(u)\right)\Big)dw\otimes\deltaZpsi\\
&-2\sum_m\Big(\int_\Sigma\Green(w,u)(t^a)\,\delta\sur{(2)}(u-z_m)\,d^2u\Big)
	dw\otimes t^a_m\,\deltaZpsi\\
&-{_{ik}\over^{\pi}}\Big(\int_\Sigma\Green(w,u)\left(t^a\right)\,
	\partial_z\delta\sur{(2)}(u-z)\,d^2u\Big)dw\otimes t^a\,dz\,\Bpsi\\
&+4i\,f^{abc}\sum_m\Big(\int_\Sigma\Green(w,u)(t^a)\,
	\delta\sur{(2)}(u-z_m)\,d^2u\Big)
	dw\otimes\Green(z,z_m)(t^b)\,dz\ t^c_m\,\Bpsi\\
& -\phi_\rho(w)\otimes c^\rho(z)\,dz
\end{split}
\end{equation*}
o\`u les $c^\rho$ sont des fonctions \`a valeurs dans $\liegc$ \`a 
d\'eterminer. La seconde variation est donc donn\'ee par l'\'equation
\begin{equation}
\label{deuxvar}
\begin{split}
\deltaw&\otimes\deltazpsi=
	2i\,\Green(w,z)\left(t^a\right)\otimes\left[t^a,\deltazpsi\right]
	+{_{k}\over^{2\pi}}\left(\int_\Sigma\Green(w,.)
	\left(\partial A^{01}\right)\right)\otimes\deltazpsi\\
&+2i\sum_m\Green(w,z_m)(t^a)\otimes t^a_m\,\deltazpsi
	+{_{ik}\over^{\pi}}\,\partial_z\Green(w,z)\left(t^a\right)\,
	\otimes t^a\,\Bpsi\\
&-4i\,f^{abc}\sum_m\Green(w,z_m)(t^a)\otimes\Green(z,z_m)(t^b)\,t^c_m\,\Bpsi
	+\phi_{\rho,w}\otimes c^\rho(z)
\end{split}
\end{equation}
o\`u $\phi_\rho(w)=\phi_{\rho,w}\,dw$, $\phi_{\rho,w}$ \`a valeurs 
dans $\liegc$. Pour conclure, il suffit de d\'eterminer les $c^\rho$. 
On organise le calcul en deux \'etapes. D\'ej\`a, la d\'eriv\'ee suivante 
fait  appara\^\i tre $c^\rho$ en fonction d'une d\'eriv\'ee fonctionnelle~: 
\begin{equation*}
\label{pre}
\partial_{n^\gamma}\left(\deltazpsi\right)=\int_\Sigma\tr_1\,
	\deltaW\otimes\deltazpsi
	\wedge\partial_{n^\gamma}A^{01}(w)
	=i\,c^\rho(z)\int_\Sigma\tr\,\phi_\rho(w)\wedge
	\partial_{n^\gamma}A^{01}(w)~;
\end{equation*}
en effet, \`a droite dans l'\'equation~\eqref{deuxvar}, les autres
termes appartiennent \`a l'image de $\overline{D}_w^{-1}$ et 
donnent z\'ero quand on les int\`egre contre $\partial_{n^\gamma}A^{01}(w)$,
vu l'\'equation~\eqref{modulinit}. 
Il suit
\begin{equation}
\label{cestlui}
c^\rho(z)=-i\,\partial_{n^\gamma}\left(\deltazpsi\right)\Omega^{\gamma\rho}.
\end{equation}
Ensuite, d'apr\`es l'\'equation~\eqref{premvar}, on a
\begin{equation*}
\begin{split}
\partial_{n^\gamma}\left(\deltaZpsi\right)=&
	{_{ik}\over^{2\pi}}\,\partialn\Dbar^{-1}(\partial A^{01})\,\Bpsi
	+{_{ik}\over^{2\pi}}\,\Dbar^{-1}(\partialn\partial A^{01})\,\Bpsi
	+{_{ik}\over^{2\pi}}\,\Dbar^{-1}(\partial A^{01})\,\partialn\Bpsi\\
&+\partialn\Omega^{\beta\alpha}\phi_\alpha\,\partial_{n^\beta}\,\Bpsi
	+\Omega^{\beta\alpha}\,\partialn\phi_\alpha\,\partial_{n^\beta}\,\Bpsi	
	+\Omega^{\beta\alpha}\phi_\alpha\,\partialn\partial_{n^\beta}\Bpsi\\
&-2\sum_m\partial_{n^\gamma}\Dbar^{-1}\bigl(t^a\,\delta\sur{(2)}
	(z-z_m)\,d^2z\bigr)\,t^a_m\,\Bpsi\\
&-2\sum_m\Dbar^{-1}\bigl(t^a\,\delta\sur{(2)}
        (z-z_m)\,d^2z\bigr)\,t^a_m\,\partial_{n^\gamma}\Bpsi.
\end{split}
\end{equation*}
Le c\^ot\'e droit de cette \'equation contient trois termes qu'on peut expliciter compl\`etement.
D'apr\`es l'\'equation~\eqref{rork}, on a
\[
(\partial_{n^\gamma}\overline{D}^{-1})\,\varpi=
	\overline{D}^{-1}\overline{D}
	(\partial_{n^\gamma}\overline{D}^{-1})\,\varpi
	+\Big(\int_\Sigma\tr\,(\partialn\Dbar^{-1})\,
	\varpi\wedge\partial_{n^\nu}A^{01}
	\Big)\Omega^{\nu\alpha}\phi_\alpha
\]
pour $\varpi=\partial A^{01}\in\Lambda^{2}(\Sigma,\liegc)$. Or, 
apr\`es d\'erivation par rapport 
\`a $n^\gamma$ des \'equations d\'efinissant 
l'op\'erateur $\overline{D}^{-1}$, on trouve
\begin{gather*}
\left[\partial_{n^\gamma}A^{01},\overline{D}^{-1}
	\varpi\right]_++\overline{D}(\partial_{n^\gamma}
	\overline{D}^{-1})\,\varpi=0,\\
\int_\Sigma\tr\,(\partial_{n^\gamma}\overline{D}^{-1})\,\varpi\wedge
	\partial_{n^\nu}
	A^{01}+\int_\Sigma\tr\,\overline{D}^{-1}
	\varpi\wedge\partial_{n^\gamma}
	\partial_{n^\nu}A^{01}=0,
\end{gather*}
par cons\'equent
\[
(\partial_{n^\gamma}\overline{D}^{-1})\,\varpi=
	-\overline{D}^{-1}\left([\Dbar^{-1}\varpi,\partialn A^{01}]_+\right)
	-\Big(\int_\Sigma\tr\,\Dbar^{-1}
	\varpi\wedge\partialn\partial_{n^\nu}A^{01}
	\Big)\Omega^{\nu\alpha}\phi_\alpha.
\]
Ensuite, $\Omega$ est l'inverse de la matrice d'\'el\'ements
$\int\tr\,\phi_\alpha\wedge\partial_{n^\beta}A^{01}$ donc
\[
\partialn \Omega^{\beta\alpha}=-\Omega^{\beta\mu}\int_\Sigma
	\tr\,\left(\partialn\phi_\mu\wedge\partial_{n^\nu}A^{01}+
	\phi_\mu\wedge\partialn\partial_{n^\nu}A^{01}
	\right)\Omega^{\nu\alpha}.
\]
Enfin, $\phi_\alpha$ \'etant un mode z\'ero de $\overline{D}$, 
on a $\Dbar \phi_\alpha=0$.
On d\'erive cette derni\`ere \'equation par rapport \`a 
$n^\gamma$ puis on utilise
l'\'equation~\eqref{rork}~; on trouve
\[
\partialn\phi_\alpha=-\Dbar^{-1}\left([\phi_\alpha,\partialn A^{01}]_+\right)
	+\Big(\int_\Sigma\tr\,\partialn\phi_\alpha\wedge\partial_{n^\nu}A^{01}
	\Big)\Omega^{\nu\mu}\phi_\mu.
\]
En fin de compte, on a montr\'e que
\begin{equation*}
\begin{split}
\partialn\left({_{\delta\Bpsi}\over^{\delta A^{01}}}\right)=&\,
	{_{ik}\over^{2\pi}}\left\lbrace-\overline{D}^{-1}
	\left([\Dbar^{-1}(\partial A^{01}),\partialn A^{01}]_+\right)
	-\Big(\int_\Sigma\tr\,\Dbar^{-1}(\partial A^{01})\wedge
	\partialn\partial_{n^\nu}A^{01}
	\Big)\Omega^{\nu\alpha}\phi_\alpha\right\rbrace\Bpsi\\
&+{_{ik}\over^{2\pi}}\,\Dbar^{-1}(\partialn\partial A^{01})\,\Bpsi
	+{_{ik}\over^{2\pi}}\,\Dbar^{-1}(\partial A^{01})\,\partialn\Bpsi\\
&-\Omega^{\beta\mu}\left\lbrace\int_\Sigma\tr\,
	\left(\partialn\phi_\mu\wedge\partial_{n^\nu}A^{01}+
	\phi_\mu\wedge\partialn\partial_{n^\nu}A^{01}
	\right)\right\rbrace\Omega^{\nu\alpha}\phi_\alpha\,
	\partial_{n^\beta}\,\Bpsi\\
&+\Omega^{\beta\mu}\left\lbrace-\Dbar^{-1}
	\left([\phi_\mu,\partialn A^{01}]_+\right)
	+\Big(\int_\Sigma\tr\,\partialn\phi_\mu\wedge\partial_{n^\nu}A^{01}
	\Big)\Omega^{\nu\alpha}\phi_\alpha\right\rbrace
	\partial_{n^\beta}\,\Bpsi\\
&+\Omega^{\beta\alpha}\phi_\alpha\,\partialn\partial_{n^\beta}\Bpsi
	+2\sum_m\Biggl\lbrace
	 \Dbar^{-1}\bigl([\Dbar^{-1}(t^a\,\delta\sur{(2)}(z-z_m)\,d^2z),
			\partial_{n^\gamma}A^{01}]_+
	\bigr)\\
&+\left(\int_\Sigma\tr\,\Dbar^{-1}(t^a\,\delta\sur{(2)}(z-z_m)\,d^2z)
	\wedge\partial_{n^\gamma}\partial_{n^\nu}A^{01}\right)
	\Omega^{\nu\alpha}\,\phi_\alpha\Biggr\rbrace t^a_m\,\Bpsi.
\end{split}
\end{equation*}
\`A nouveau, on fait intervenir la d\'eriv\'ee fonctionnelle de $\Bpsi$~:
\begin{equation*}
\begin{split}
\partialn\left({_{\delta\Bpsi}\over^{\delta A^{01}}}\right)=
	&\, -\Dbar^{-1}\Big(\left[{_{\delta\Bpsi}\over^{\delta A^{01}}},
	\partialn A^{01}\right]_+\Big)-\Big(\int_\Sigma\tr\,\deltapsi
	\wedge\partialn\partial_{n^\nu}A^{01}\Big)\Omega^{\nu\alpha}
	\phi_\alpha\\
&+{_{\delta}\over^{\delta A^{01}}}(\partialn\Bpsi)+
	{_{ik}\over^{2\pi}}\,\Dbar^{-1}(\partialn\partial A^{01})\Bpsi\\
=&\,-\int_\Sigma\Green(z,u)\Big(\left[{_{\delta\Bpsi}\over^{\delta A^{01}(u)}}
	,\partialn A^{01}(u)\right]_+\Big)dz
	-\Big(\int_\Sigma\tr\,\deltapsi\wedge\partialn
	\partial_{n^\nu}A^{01}\Big)\Omega^{\nu\alpha}\phi_\alpha\\
&+{_{\delta}\over^{\delta A^{01}}}\,(\partialn\Bpsi)+
	{_{ik}\over^{2\pi}}\,\Big(\int_\Sigma\Green(z,u)\left(
	\partialn\partial A^{01}(u)\right)\Big)dz\,\Bpsi.
\end{split}
\end{equation*}
De ces \'equations, on tire une expression pour $c^\rho$.
On injecte alors cette derni\`ere dans l'\'equa\-tion~\eqref{deuxvar} 
pour arriver \`a
\begin{equation*}
\begin{split}
\deltaw&\otimes\deltazpsi=
	2i\,\Green(w,z)\left(t^a\right)\otimes\left[t^a,\deltazpsi\right]
	+{_{k}\over^{2\pi}}\Big(\int_\Sigma\Green(w,.)\left(
	\partial A^{01}\right)\Big)\otimes\deltazpsi\\
&+2i\sum_m\Green(w,z_m)(t^a)\otimes t^a_m\,\deltazpsi
	+{_{ik}\over^{\pi}}\,\partial_z\Green(w,z)\left(t^a\right)
	\,\otimes t^a\,\Bpsi\\
&-4i\,f^{abc}\sum_m\Green(w,z_m)(t^a)\otimes\Green(z,z_m)(t^b)\,t^c_m\,\Bpsi
	+\phi_{\rho,w}\otimes\Biggl\lbrace\int_\Sigma\Green(z,.)
	\Big(\left[{_{\delta\Bpsi}\over^{\delta A^{01}}},
	\partialn A^{01}\right]_+\Big)\\
&+\Big(\int_\Sigma\tr\,\deltapsi\wedge\partialn\partial_{n^\nu}A^{01}\Big)
	\Omega^{\nu\alpha}\phi_{\alpha,z}
	-i{_{\delta}\over^{\delta A_{\overline{z}}}}\,(\partialn\Bpsi)
	-{_{ik}\over^{2\pi}}\,\Big(\int_\Sigma\Green(z,.)
	\left(\partialn\partial A^{01}\right)\Big)
	\Bpsi\Biggr\rbrace\Omega^{\gamma\rho}.
\end{split}
\end{equation*}
Rappelons enfin la formule donnant la premi\`ere d\'eriv\'ee de $\Bpsi$~:
\begin{equation}
\label{premiere}
\deltazpsi
	=-i\Omega^{\mu\rho}\phi_{\rho,z}\,\partial_{n^\mu}\,\Bpsi
	+{_{k}\over^{2\pi}}\,\int_\Sigma\Green(z,.)\left(\partial A^{01}\right)
	\,\Bpsi+2i\sum_m\Green(z,z_m)(t^a)\,t^a_m\,\Bpsi.
\end{equation}

\subsection{R\'egularisation}

On a d\'ej\`a analys\'e le comportement singulier de ces d\'eriv\'ees 
quand $w$ tend vers $z$ ou $z$ tend vers $z_\ell$. On se propose 
d'appliquer le m\^eme processus de r\'egularisation aux termes divergents 
intervenant dans le membre droit de ces formules. Clairement, il suffit
de se concentrer sur la fonction de Green de $\Dbar$ et sur sa 
d\'eriv\'ee par rapport \`a la seconde variable. Soit $\varpi\in\Lambda^2
(\Sigma,\liegc)$. Par construction, on a
\[
\left(\partialbarw+\left[A_{\overline{w}},.\right]\right)\,
	\int_\Sigma\Green(w,.)(\varpi)\,d\overline{w}\wedge dw=\varpi(w).
\]
Soit $h\in\CG^\C$ une transformation de jauge telle que 
${}^{h^{-1}}A^{01}=0$ au voisinage de $z$. \`A cette condition, on a 
$\partialbarw+\left[A_{\overline{w}},.\right]=
\text{Ad}\sous{h(w)}\,\partialbarw\,\text{Ad}\sous{h(w)^{-1}}$. Notons $\id$ 
l'application identit\'e de $\text{End}\,\liegc$. Il suit
$\partialbarw\,\text{Ad}\sous{h(w)^{-1}}\Green(w,z)=\frac{1}{2i}\,
\text{Ad}\sous{h(z)^{-1}}\,\delta\sur{(2)}(w-z)\,\id$, pour $w$ au 
voisinage de $z$. On sait aussi que $\pi\,\delta\sur{(2)}(w-z)=
\partialbarw\,(w-z)^{-1}$, d'o\`u
\[
\Green(w,z)=\quotient{1}{2\pi i}\,\text{Ad}\sous{h(w)}\,
	\text{Ad}\sous{h(z)^{-1}}\quotient{\id}{w-z}+\cdots
\]
et les pointill\'es symbolisent toujours des termes r\'eguliers en 
$w$ au voisinage de $z$. Apr\`es d\'erivation, 
\[
\partial_z\Green(w,z)=\quotient{1}{2\pi i}\,\text{Ad}\sous{h(w)}\,
	\text{Ad}\sous{h(z)^{-1}}\,\quotient{\id}{(w-z)^2}+
	\quotient{1}{2\pi i}\,\text{Ad}\sous{h(w)}\,
	\text{Ad}\sous{h(z)^{-1}}\,\text{ad}\sous{(h\partial_zh^{-1})(z)}\,
	\quotient{\id}{w-z}+\cdots
\]
Utilisons le transport parall\`ele $\boldsymbol{e}_{z,w}$ pour 
r\'e\'ecrire ces deux d\'eveloppements limit\'es sans faire appel \`a $h$~:
\begin{equation}
\label{regulapartrans}
\begin{split}
	\text{Ad}_{\boldsymbol{e}_{z,w}}\,\Green(w,z)&=\quotient{1}{2\pi i}
	\,\quotient{\id}{w-z}+\widetilde{\Green}(w,z),\\
\text{Ad}_{\boldsymbol{e}_{z,w}}\,\partial_z\Green(w,z)&=\quotient{1}{2\pi i}\,
	\quotient{\id}{(w-z)^2}-\quotient{1}{4\pi i}\,
	\quotient{\overline{w}-\overline{z}}{w-z}\,
	\left[\partial_zA_{\overline{z}}(z),.\right]+
	\widetilde{\partial_z\Green}(w,z)
\end{split}
\end{equation}
o\`u $\widetilde{\Green}$ et $\widetilde{\partial_z\Green}$ sont des 
fonctions r\'eguli\`eres en $w$ au voisinage de $z$. Montrons par exemple 
la seconde \'egalit\'e. L'ingr\'edient principal de la d\'emonstration est 
l'\'equation~\eqref{developpement}~: $\boldsymbol{e}_{z,w}=h(z)\,
\boldsymbol{e}'_{z,w}\,h(w)^{-1}$. Ainsi,
\begin{equation*}
\begin{split}
\text{Ad}_{\boldsymbol{e}_{z,w}}\,\partial_z\Green(w,z)=&\,
	\quotient{1}{2\pi i}\,\text{Ad}\sous{h(z)}\,
	\text{Ad}_{\boldsymbol{e}'_{z,w}}\,\text{Ad}\sous{h(z)^{-1}}\,
	\quotient{\id}{(w-z)^2}\\
&+\quotient{1}{2\pi i}\,\text{Ad}\sous{h(z)}\,
	\text{Ad}_{\boldsymbol{e}'_{z,w}}\,
	\text{Ad}\sous{h(z)^{-1}}\,\text{ad}\sous{(h\partial_zh^{-1})(z)}\,
	\quotient{\id}{w-z}+\cdots.
\end{split}
\end{equation*}
Or $\boldsymbol{e}'_{z,w}=\ee^\chi$, donc $\text{Ad}_{\boldsymbol{e}'_{z,w}}=
\ee^{\text{ad}_\chi}=1+\text{ad}_\chi+\frac{1}{2}\,\text{ad}_\chi^2+\cdots$. 
Ainsi,
\begin{equation*}
\begin{split}
\text{Ad}_{\boldsymbol{e}_{z,w}}\,\partial_z\Green(w,z)=&\,
	\quotient{1}{2\pi i}\,\quotient{\id}{(w-z)^2}
	+\quotient{1}{2\pi i}\,\text{Ad}\sous{h(z)}\,
	\text{ad}\sous{(h^{-1}\partial_zh)(z)}\,
	\text{Ad}\sous{h(z)^{-1}}\,\quotient{\id}{w-z}\\
&+\quotient{1}{4\pi i}\,
	\text{Ad}\sous{h(z)}\,\text{ad}\sous{\partialbarz
	(h^{-1}\partial_zh)(z)}
	\,\text{Ad}\sous{h(z)^{-1}}\,\quotient{\overline{w}-
	\overline{z}}{w-z}\,\id+\quotient{1}{2\pi i}\,
	\text{ad}\sous{(h\partial_zh^{-1})(z)}\,
	\quotient{\id}{w-z}+\cdots.
\end{split}
\end{equation*}
Or $\text{Ad}\sous{h(z)}\,\text{ad}\sous{(h^{-1}\partial_zh)(z)}\,
\text{Ad}\sous{h(z)^{-1}}=\text{ad}\sous{\text{Ad}\sous{h(z)}\,
(h^{-1}\partial_zh)(z)}=-\text{ad}\sous{(h\partial_zh^{-1})(z)}$, 
donc le deuxi\`eme terme compense le dernier. On traite de la m\^eme 
fa\c{c}on le troisi\`eme terme, la conclusion est alors imm\'ediate.
Pour la premi\`ere d\'eriv\'ee de $\Bpsi$, on s'int\'eresse \`a la 
quantit\'e suivante
\begin{equation*}
\begin{split}
\tr\,\Bigl(t^a\,&\text{Ad}_{\boldsymbol{e}_{z_\ell,z}}\,\deltazpsi\Bigr)
	=-i\,\Omega^{\mu\rho}\,\tr\,
	\left(t^a\,\text{Ad}_{\boldsymbol{e}_{z_\ell,z}}\,\phi_{\rho,z}\right)
		\,\partial_{n^\mu}\,\Bpsi
	+\quotient{k}{2\pi}\,\tr\,t^a\,\text{Ad}_{\boldsymbol{e}_{z_\ell,z}}\,
		\int_\Sigma\Green(z,.)(\partial A^{01})\ \Bpsi\\
	&+2i\sum_{m\neq \ell}\tr\,t^a\,\text{Ad}_{\boldsymbol{e}_{z_\ell,z}}\,
		\Green(z,z_m)(t^b)\,t^b_m\,\Bpsi
	+2i\,\tr\,t^a\,\left(\quotient{1}{2\pi i}\,\quotient{t^b}{z-z_\ell}
		+\widetilde{\Green}(z,z_\ell)(t^b)\right)\,t^b_\ell\,\Bpsi.
\end{split}
\end{equation*}
Le seul terme singulier restant est compens\'e par la r\'egularisation 
propos\'ee.  En fin de compte,
\begin{equation}
\label{premierederivee}
\tag{\ref{premierederivee}.a}
\begin{split}
\singulier\quotient{\delta\Bpsi}{\delta A\sur{a}_{\overline{z}_\ell}}
	\singulier=&\,-2i\,\Omega^{\mu\rho}\,\tr\,(t^a\,\phi_{\rho,z_\ell})\,
	\partial_{n^\mu}\,\Bpsi +\quotient{k}{\pi}\,\tr\,t^a\,
	\int_\Sigma\Green(z_\ell,.)(\partial A^{01})\ \Bpsi\\
	&+4i\sum_{m\neq \ell}\tr\,t^a\,\Green(z_\ell,z_m)(t^b)\,t^b_m\,\Bpsi
	+4i\,\tr\,t^a\,\widetilde{\Green}(z_\ell,z_\ell)(t^b)\,t^b_\ell\,\Bpsi.
\end{split}
\end{equation}
Ensuite, la r\'egularisation du Laplacien donne 
\begin{equation*}
\begin{split}
\tr\,&\singulier\deltaz\,\deltazpsi
	\singulier=2i\,\tr\,\widetilde{\Green}(z,z)
	\left(t^a\right)\left[t^a,\deltazpsi\right]
	+{_{k}\over^{2\pi}}\,\tr\,\Big(\int_\Sigma\Green(z,.)
	\left(\partial A^{01}\right)\Big)\deltazpsi\\
&+2i\sum_m\tr\,\Green(z,z_m)(t^a)\, t^a_m\,\deltazpsi
	+{_{ik}\over^{\pi}}\,\tr\,\widetilde{\partial_z\Green}(z,z)
	\left(t^a\right)\,t^a\,\Bpsi\\
&-4i\,f^{abc}\sum_m\tr\,\Green(z,z_m)(t^a)\,\Green(z,z_m)(t^b)\,t^c_m\,\Bpsi
	+\tr\,\phi_{\rho,z}\Biggl\lbrace\int_\Sigma\Green(z,.)
	\Big(\left[{_{\delta\Bpsi}\over^{\delta A^{01}}},
	\partialn A^{01}\right]_+\Big)\\
&+\Big(\int_\Sigma\tr\,\deltapsi\wedge\partialn\partial_{n^\nu}A^{01}\Big)
	\Omega^{\nu\alpha}\phi_{\alpha,z}
	-i{_{\delta}\over^{\delta A_{\overline{z}}}}\,(\partialn\Bpsi)
	-{_{ik}\over^{2\pi}}\,\Big(\int_\Sigma\Green(z,.)
	\left(\partialn\partial A^{01}\right)\Big)
	\Bpsi\Biggr\rbrace\Omega^{\gamma\rho}.
\end{split}
\addtocounter{equation}{1}
\end{equation*}
Finalement, pour obtenir une 
formule ne comportant que des d\'eriv\'ees de $\Bpsi$ 
par rapport aux param\`etres $n^\nu$, on combine la derni\`ere \'equation avec 
l'\'equation~\eqref{premvar}~:
\begin{equation}
\tag{\ref{premierederivee}.b}
\tr\singulier\deltaz\,\deltazpsi\singulier=
	-\Omega^{\alpha\gamma}\Omega^{\nu\rho}\,
	\tr\,(\phi_{\gamma,z}\phi_{\rho,z})\,
	\partial_{n^\alpha}\partial_{n^\nu}\Bpsi
	+\Xi^\alpha(z)\,\partial_{n^\alpha}\Bpsi+\Upsilon(z)\,\Bpsi
\end{equation}
avec
\begin{equation*}
\begin{split}
\Xi^\alpha(z)=&\,2\,\Omega^{\alpha\gamma}\,\tr\,
	\widetilde{\Green}(z,z)\left(t^a\right)\,
	\left[t^a,\phi_{\gamma,z}\right]-{_{ik}\over^{\pi}}\,
	\Omega^{\alpha\gamma}\,\tr\,\phi_{\gamma,z}\int_\Sigma\Green(z,.)
	\left(\partial A^{01}\right)\\
&+\Omega^{\alpha\gamma}\,\Omega^{\nu\rho}\,\tr\,
	\phi_{\rho,z}\int_\Sigma\Green(z,.)
	\left(\left[\phi_\gamma,\partial_{n^\nu}A^{01}\right]_+\right)\\
&+\Omega^{\alpha\gamma}\,\Omega^{\nu\rho}\,\Omega^{\epsilon\mu}\,
	\Big(\int_\Sigma\tr\,\phi_\gamma\wedge\partial_{n^\nu}
	\partial_{n^\epsilon}A^{01}\Big)\tr\,\phi_{\rho,z}\,\phi_{\mu,z}\\
&+4\,\Omega^{\alpha\gamma}\sum_m\tr\,\phi_{\gamma,z}\,
	\Green(z,z_m)(t^a)\,t^a_m
\end{split}
\end{equation*}
et 
\begin{equation*}
\begin{split}
\Upsilon(z)=&\,{_{ik}\over^{\pi}}\,\tr\,
	\widetilde{\partial_z\Green}(z,z)\left(t^a\right)\,t^a
	+{_{ik}\over^{\pi}}\,\tr\,\widetilde{\Green}(z,z)\left(t^a\right)\,
	\left[t^a,\int_\Sigma\Green(z,.)\left(\partial A^{01}\right)\right]\\
&+{_{k^2}\over^{4\pi^2}}\,\tr\,\Big(\int_\Sigma\Green(z,.)
	\left(\partial A^{01}\right)\Big)^2
	-{_{ik}\over^{2\pi}}\,\Omega^{\alpha\gamma}\,\tr\,\phi_{\gamma,z}
	\int_\Sigma\Green(z,.)\left(\partial\partial_{n^\alpha}A^{01}\right)\\
&+{_{ik}\over^{2\pi}}\,\Omega^{\alpha\gamma}\,\tr\,\phi_{\gamma,z}\int_\Sigma
	\Green(z,.)\left([\Dbar^{-1}(\partial A^{01}),
	\partial_{n^\alpha}A^{01}]_+\right)\\
&+{_{ik}\over^{2\pi}}\,\Omega^{\alpha\gamma}\,\Omega^{\nu\rho}
	\tr\,\phi_{\gamma,z}\,\phi_{\rho,z}
	\int_\Sigma\tr\,\Dbar^{-1}(\partial A^{01})\wedge
	\partial_{n^\alpha}\partial_{n^\nu}A^{01}\\
&-4\sum_m\tr\,\widetilde{\Green}(z,z)(t^a)\,[t^a,\Green(z,z_m)(t^b)]\,t^b_m
        +\quotient{2ik}{\pi}\sum_m\tr\,\Green(z,z_m)(t^a)
        \int_\Sigma\Green(z,.)(\partial A^{01})\,t^a_m
\end{split}
\end{equation*}
\begin{equation*}
\begin{split}
\phantom{\Upsilon(z)}
&-4\sum_{m,m'}\tr\,\Green(z,z_m)(t^a)\,\Green(z,z_{m'})
	(t^b)\,t^a_m\,t^b_{m'}\\
&-2\,\Omega^{\alpha\gamma}\sum_m\tr\,\phi_{\gamma,z}\int_\Sigma\Green(z,w)
	\left([\Green(w,z_m)(t^a)\,dw,
	\partial_{n^\alpha}A^{01}(w)]_+\right)\,t^a_m\\
&-2\,\Omega^{\alpha\gamma}\,\Omega^{\nu\rho}\,\tr\,\phi_{\gamma,z}\,
	\phi_{\rho,z}\sum_m\int_\Sigma\Green(u,z_m)(t^a)\,du\wedge
	\partial_{n^\alpha}\partial_{n^\nu}A^{01}(u)\,t^a_m.
\end{split}
\end{equation*}
En r\'esum\'e, les formules utiles pour le calcul de la connexion de KZB sont 
celles donn\'ees par les \'equations~\eqref{premiere}, \eqref{regulapartrans}, 
(\ref{premierederivee}.a-b).

\medskip
\section{Connexion de KZB sph\'erique}

En genre z\'ero, l'orbite du champ nul est dense dans $\CA^{01}$, \ie presque tous
les champs sont de la forme $A^{01}=h^{-1}\de h$, avec $h\in\CG^\C$ 
fix\'ee \`a une constante pr\`es~: $h\mapsto h_0h$. 
L'espace des \'etats de Chern-Simons est un sous-fibr\'e du fibr\'e trivial
de fibre $\esprep^{\,G}$.
On sait aussi qu'il n'existe qu'une structure complexe, 
donc seules $\KZz$ et $\KZzbar$ ont un sens.
Sur la sph\`ere, la fonction de Green de $\Dbar$ est
$\Green(z,w)=\frac{1}{2\pi i}\,\frac{1}{z-w}$.
Comme en plus le transport parall\`ele est trivial, on a 
$\widetilde{\Green}\equiv 0$ d'o\`u
\[
\singulier\quotient{\delta\Bpsi}{\delta A\sur{a}_{\overline{z}_\ell}}
	\singulier =\quotient{1}{\pi}\sum_{m\neq \ell}
	\frac{1}{z_\ell-z_m}\,t^a_m\,\Bpsi.
\]
On retrouve bien la connexion de Knizhnik-Zamolodchikov d\'ej\`a
aper\c{c}ue au chapitre 2
\[
\KZz\Bpsi=\partial_{z_\ell}\Bpsi-\quotient{1}{\kappa}\,H^{(0)}_\ell\,\Bpsi,\qquad
\KZzbar\Bpsi=\partialzbar\Bpsi.
\]
o\`u $H^{(0)}_\ell$ est le Hamiltonien de Gaudin~:
\begin{equation}
\label{GAUDIN}
H^{(0)}_\ell(\un\xi)=2
        \sum_{m\neq\ell}\frac{t^a_\ell\,t^a_m}{z_\ell-z_m}.
\end{equation}
L'exposant $(0)$ est l\`a pour signaler qu'on travaille en genre z\'ero.
La platitude de la connexion est une cons\'equence de
l'\'equation $[H_\ell^{(0)},H_{\ell'}^{(0)}]=0$, \'equation
\'equivalente \`a l'EYBC pour $t^a_1\,t^a_2\big/z$:
\[
\Big[\frac{t^a_n\,t^a_{m}}{z_n-z_{m}}\,,\,
        \frac{t^a_{m}\,t^a_{\ell}}{z_{m}-z_{\ell}}\Big]+
        \mbox{permutations cycliques}\,=\,0\, .
\]

\medskip
\section{Syst\`emes de Hitchin sph\'eriques}

Pour identifier le syst\`eme de Hitchin, on doit r\'esoudre l'\'equation
$\mu(A^{01},\varphi^{10})=\sum_\ell\lambda_\ell\,\delta_{z_\ell}$, \cad
\[
\de({}^h\varphi^{10})=\sum_\ell \nu_\ell \delta_{z_\ell }
\]
o\`u $\nu_\ell =h(z_\ell)\lambda_\ell\,h(z_\ell )^{-1}$ appartient \`a l'orbite coadjointe
de $\lambda_\ell $ dans $\liegc$~: $\CO_\ell =\{h_0\lambda_\ell\,h_0^{-1}\,|
\,h_0\in \Gc\}$. Cette \'equation admet des solutions si, et seulement
si, l'int\'egrale sur $\Sigma$ du terme de droite est nulle, \ie
si la somme des r\'esidus est nulle~:  $\sum_\ell \nu_\ell =0$. On a alors
\[
{}^h\varphi^{10}(z)=\quotient{1}{2\pi i}\sum_\ell \frac{\nu_\ell }{z-z_\ell }\,dz.
\]
Soit $\Gc_\ell $ le sous-groupe de $\Gc$ fixant $\lambda_\ell $ pour l'action
adjointe. Il suit
\[
\CG^\C_{\underline{z},\underline{\lambda}}=
\Big\{h\in\CG^\C\,\Big|\, h(z_\ell )\in G_\ell ^\C\,,\ \forall \ell\Big\}.
\]
L'espace des phases r\'eduit est 
\[
\CP^{(0)}_{\CO}
        \,\cong\,\Big\{\underline{\nu}
	\in\mathop{\times}_{\ell=1}^N\CO_\ell \,\Big|\,
         \sum_\ell \nu_\ell =0\Big\}\bigg/\Gc.
\]
Chaque orbite coadjointe peut \^etre munie d'une structure symplectique naturelle
donnant le crochet de Poisson $\{\nu^a,\nu^b\}=if^{abc}\nu^c$ si $\nu=\nu^a t^a$
et $(t^a)$ est la base orthogonale de $\liegc$. La structure symplectique
induite sur $\CP_\CO^{(0)}$ co\"{\i}ncide avec celle qu'on obtient apr\`es
r\'eduction symplectique de $\times_\ell \CO_\ell $ par l'action
diagonale de $\Gc$. 

Comme $p$ est un polyn\^ome $\Gc$-invariant, on a
$p(\varphi^{10})=p({}^h\varphi^{10})$. L'application de Hitchin est donc
\[
h_{p}(\underline{\nu})(z)=p\Big(\quotient{1}{2\pi i}\sum_\ell \frac{\nu_\ell }
        {z-z_\ell }\Big)\,(dz)^{d_p},
\]
soit encore
\[
h_{p}(\underline{\nu})(z)=\sum_{n_1,\cdots,n_{d_p}}
	\frac{p(\nu_{n_1},\cdots,\nu_{n_{d_p}})}
	{(z-z_1)\cdots(z-z_{d_p})}\,\Big(\quotient{dz}{2\pi i}\Big)^{d_p}.
\]
En particulier, si $p=p_2$, $h_{p_2}$ prend ses valeurs dans l'espace
des formes quadratiques m\'eromorphes et
\[
h_{p_2}(\underline{\nu})(z)=\quotient{1}{2}\sum_\ell 
 	\bigg(\frac{1}{(z-z_\ell )^2}\,\delta_\ell
        +\frac{1}{z-z_\ell}\,h^{(0)}_\ell 
        \bigg)\Big(\quotient{dz}{2\pi i}\Big)^2
\]
o\`u, \`a une constante multiplicative pr\`es, $h^{(0)}_\ell $ est le r\'esidu en 
$z=z_\ell $, \cad
\[
h^{(0)}_\ell (\un\nu)=2\sum_{m\neq \ell}\frac{\nu^a_\ell \nu^a_{m}}{z_\ell -z_{m}}
\qquad\mbox{et}\qquad \delta_\ell =\nu^a_\ell \nu^a_\ell .
\]

Soit $\delta\mu$ une diff\'erentielle de Beltrami lisse \`a l'infini
et se comportant comme $\CO((z-z_\ell )^2)$ autour des points d'insertion.
Sur la sph\`ere de Riemann, 
$\delta\mu=\de(\delta v)$ avec $\delta v=\delta v^z\,
\partial_z$ un champ de vecteur lisse sur $\C P^1$. Ce dernier
est compl\`etement d\'etermin\'e modulo
les sections holomorphes de $K^{-1}$. Sur la 
sph\`ere de Riemann, $h^0(K^{-1})=3$~;
les \'el\'ements de $H^0(K^{-1})$ engendrent ${\rm SL}_2$~:
$(a+bz+cz^2)\partial_z$. La nouvelle coordonn\'ee complexe sur $\C P^1$ est
donn\'ee par $z'=z+\delta v^z$. La coordonn\'ee du point $\xi_\ell$ change
donc de $\delta z_\ell=\delta v^z(\xi_\ell)$ et le $1$-jet en $\xi_\ell$
change de $\delta\chi_\ell=\chi_\ell\,\da_z(\delta v^z)(\xi_\ell)$.
On peut alors calculer $h_{\delta\mu}$ en scindant l'int\'egration en
une int\'egrale sur de petites boules autour des points d'insertion et 
une int\'egrale sur le reste de $\Sigma$~:
\begin{align*}
h_{\delta\mu}&=\quotient{1}{2}\sum_\ell\int_\Sigma
	\bigg(\frac{1}{(z-z_\ell )^2}\,\delta_\ell
        +\frac{1}{z-z_\ell}\,h^{(0)}_\ell 
        \bigg)\Big(\quotient{dz}{2\pi i}\Big)^2
        \ \de(\delta v)\\
        &={_1\over^{4\pi i}}\sum_\ell
        \left(\delta_\ell\,\chi_\ell^{-1}\delta\chi_\ell
        \,+\, h^{(0)}_\ell\,\delta z_\ell\right).
\label{hmu0}
\end{align*}

On sait que la connexion de KZ est donn\'ee par
\begin{align*}
\KZmu\Bpsi &=
\sum\limits_\ell\Big(\delta\chi_\ell\,\big(\da_{\chi_\ell}-\chi_\ell^{-1}\Delta_\ell\big)
\,+\,\delta z_\ell\,\big(\da_{z_\ell}-{_1\over^{\kappa}} H^{(0)}_\ell\big)
\Big)\Bpsi\, ,\label{KZ01}\\
\KZmubar\Bpsi &=
\sum\limits_\ell\Big({\delta\bar\chi}_\ell\,\da_{\bar\chi_\ell}
\,+\,{\delta\bar z}_\ell\,\da_{\bar z_\ell}
\Big)\Bpsi
\label{KZ02}
\end{align*}
o\`u 
\[
H^{(0)}_\ell(\un\xi)=2
        \sum_{m\neq\ell}\frac{t^a_\ell\,t^a_m}{z_\ell-z_m}\qquad
	\mbox{et}\qquad
\Delta_\ell=C_\ell/\kappa.
\]
Mise \`a part la translation $k\mapsto\kappa=k+g^\vee$, on d\'eduit
$\Delta_\ell$ et $\frac{1}{\kappa}H_\ell^{(0)}$ de
$\frac{2k}{\pi}\,\frac{1}{4\pi i}\,\delta_\ell$ et 
$\frac{2k}{\pi}\,\frac{1}{4\pi i}\,h_\ell^{(0)}$
par la quantification g\'eom\'etrique des orbites coadjointes,
\cad le remplacement des $\nu^a_\ell$ par les op\'erateurs 
$\frac{\sqrt{2i}\pi}{k}\,t^a_\ell$. 
Le crochet de Poisson devient $\frac{k}{\sqrt{2i}\pi}$ 
fois le commutateur --- $\frac{1}{k}$ joue le r\^ole
de la constante de Planck.

\medskip
\section{Connexion de KZB elliptique}

On calcule la connexion de KZB en genre un, \cad d\'efinie
sur une courbe elliptique $\Sigma=E_\tau=\C\,/\,(\Z+\tau\Z)$.
On sait que l'orbite de $A^{01}_u=\pi u\,
\imtau^{-1}\,d\overline{z}$, $u\in\lietc\setminus P^\vee+\tau P^\vee$, 
est dense dans $\CA^{01}$. On utilise la base 
$\baseB$ form\'ee des
$e_\alpha$, $\alpha\in\Delta$, et des $h^j$, $j=1,\cdots, r$, base orthonormale
de la sous-alg\`ebre de Cartan $\lietc$. Remarquons que les formules 
donnant la connexion sont obtenues pour une base orthogonale $\baseA=(t^a)$,
$\tr\, t^at^b=\delta^{ab}/2$. Une partie du jeu consiste
\`a jongler entre $\baseA$ et $\baseB$.

L'op\'erateur $\Dbar=\de+[A^{01}_u,.]_+$ agit sur les $(1,0)$-formes 
$\phi$ sur $\Sigma$ \`a valeurs dans $\liegc$. 
On d\'ecompose $\phi$ sur la base $\baseB$, 
$\phi=\phi^\alpha\,e_\alpha+\phi^j\,h^j$, ainsi
$\Dbar\phi=\de\phi^j\,h^j+\de_\alpha\, \phi^\alpha\,e_\alpha$,
o\`u $\de_\alpha=\de+(\pi u_\alpha/\imtau)\,d\overline{z}$.
Si $\phi$ est un mode z\'ero de $\Dbar$, $\phi^j$ est constante 
en tant que
fonction anti-analytique et $(\Z+\tau\Z)\,$-p\'eriodiques. Ensuite, 
l'\'egalit\'e
\[
\de_\alpha=\ee^{\pi u_\alpha(z-\overline{z})/\imtau}\,
	\de\ \ee^{-\pi u_\alpha(z-\overline{z})/\imtau}
\]
permet de d\'eduire que 
$\phi^{\alpha}=0$, d\`es que $u\not\in P^\vee+\tau P^\vee$, ce qu'on a 
suppos\'e au d\'ebut. Ainsi, le noyau de $\Dbar$ est 
engendr\'e par les $\phi^j_{0}=h^j\,dz$. 
Maintenant, on s'int\'eresse \`a la fonction
de Green $\Green$ de $\Dbar^{-1}$. 
D\'ecomposons la fonction de Green sur $\baseB$~:
$\Green(w,z)=\green_\alpha(w,z)\,e_\alpha+\green_j(w,z)\,h^j$. 
Chaque composante est une fonction 
$(\Z+\tau\Z)\,$-p\'eriodiques \`a valeurs dans $\C$. Seules 
$\green_j(w,z)(h^j)$, $\green_{\alpha}(w,z)(e_\alpha)$
ne sont pas nulles --- on les note simplement $G_j, G_\alpha$~---
elles satisfont les \'equations suivantes
\begin{gather}
\label{greencond}
\partialbarw\,G_j(w,z)=
	\quotient{1}{2i}\,\delta\sur{(2)}(w-z)+{\rm cte},
	\qquad \int_\Sigma G_j(w,z)\,d^2w=0,\\
\big(\partial_{\overline{w}}+\quotient{\pi u_\alpha}{\imtau}\big)\,G_\alpha(w,z)=
	\quotient{1}{2i}\,\delta\sur{(2)}(w-z).\non
\end{gather}
Une mani\`ere \'economique d'obtenir $G_\alpha$ utilise la fraction 
$P$ construite \`a partir 
de la fonction th\^eta de Jacobi $\vartheta_1$~: 
\[
G_\alpha(w,z)=\quotient{1}{2\pi i}
	\,\ee^{\pi u_\alpha (w-\overline{w}-z+\overline{z})/\imtau}
		\,\noyau_{u_\alpha}(w-z).
\]
La fonction de Green de $\de$ sur $E_\tau$ n'est pas unique. Par contre,
la deuxi\`eme condition de~\eqref{greencond}
assure l'unicit\'e de $G_j$. Le r\'esultat est  
\[
G_j(w,z)=\quotient{1}{2\pi i}\,\rho(w-z)+
	\quotient{w-\overline{w}-z+\overline{z}}{2i\imtau}.
\]
Celui-ci ne d\'ependant pas de $j$, on le note $H$.
Cette fonction v\'erifie la premi\`ere condition de~\eqref{greencond} 
puisque
$\partial_{\overline{w}}\,H(w,z)=\frac{1}{2i}\,\delta\sur{(2)}(w-z)
-\frac{1}{2i\imtau}$. Remarquons que le deuxi\`eme crit\`ere 
impos\'e par~\eqref{greencond} est aussi
rempli car $\frac{1}{2\pi i}\,\rho(y)+\frac{1}{2i\imtau}\,
(y-\overline{y})$ est la d\'eriv\'ee totale de la fonction monovalu\'ee~:
$\frac{1}{2\pi i}\,\ln|\vartheta_1(y)|^2+
\frac{1}{4i\imtau}\,(y-\overline{y})^2$.
Pour r\'egulariser les fonctions de Green on utilise les 
\'equations~\eqref{regulapartrans} pour le transport parall\`ele
de $A^{01}_u$~:
\begin{equation*}
\begin{split}
\widetilde{\Green}(w,z)&=\ee^{\nu u_\alpha}\green_\alpha(w,z)\,e_\alpha
			+\green_{j}(w,z)\,h^j
			-\quotient{1}{2\pi i}\,\quotient{\id}{w-z}\\
\widetilde{\partial_z\Green}(w,z)&=
		\ee^{\nu u_\alpha}\partial_z\green_\alpha(w,z)\,e_\alpha
		+\partial_z\green_{j}(w,z)\,h^j
		-\quotient{1}{2\pi i}\,\quotient{\id}{(w-z)^2}
\end{split}
\end{equation*}	
o\`u $\nu=\pi\,(\bar w-\bar z)/\imtau$.
On note $\widetilde{G}_\alpha$ la partie r\'eguli\`ere de $G_\alpha$
et ainsi de suite avec les autres fonctions. On trouve sans probl\`eme~:
\begin{gather*}
\widetilde H=0,\qquad\widetilde{G}_\alpha=
	-\quotient{iu_\alpha}{2\imtau}+\quotient{1}{2\pi i}\,\rho(u_\alpha),
	\qquad\widetilde{\partial_z H}=
	\quotient{1}{\pi i}\,\eta_1-\quotient{1}{2i\imtau},\\
\widetilde{\partial_zG}_\alpha=\quotient{\pi iu_\alpha^2}{4\imtau^2}
			+\quotient{iu_\alpha}{2\imtau}\,\rho(u_\alpha)
			-\quotient{1}{4\pi i}\,
	\frac{\vartheta_1''}{\vartheta_1}(u_\alpha)
			-\quotient{1}{2\pi i}\,\eta_1~;
\end{gather*}
pour des raisons triviales, on a supprim\'e la d\'ependance en $z$.

Maintenant, on poss\`ede tous les ingr\'edients n\'ecessaires au calcul 
de la connexion. 
Reprenons l'\'equation~(\ref{premierederivee}.a)~:
\begin{equation*}
\begin{split}
\singulier\quotient{\delta\Bpsi}{\delta A\sur{a}_{\overline{z}_\ell}}
        \singulier
        =&\,-2i\,\Omega^{j\ell}\,\tr\,(t^a\,\phi^\ell_{0,z_\ell})\,
	\partial_{u^j}\,\Bpsi
        +4i\sum_{m\neq \ell}\tr\,t^a\,\Green(z_\ell,z_m)(t^b)\,t^b_m\,\Bpsi\\
&+4i\,\tr\,t^a\,\widetilde{\Green}(z_\ell,z_\ell)(t^b)\,t^b_\ell\,\Bpsi
\end{split}
\end{equation*}
o\`u $u=u^jh^j$. On v\'erifie simplement que $\Omega=\frac{i}{2\pi}\,1$.
Cela dit, dans ces formules, on utilise une base orthogonale, 
en l'occurrence $\baseA$. Or, on a calcul\'e les fonctions de Green 
par rapport \`a $\baseB$. 
On doit donc d'abord effectuer le changement ad\'equat. 
On n'a pas besoin de conna\^\i tre explicitement les matrices de passage entre 
$\baseA$ et $\baseB$. Si $e_\alpha=P^{\alpha a}t^a$, $h^j=P^{ja}t^a$ et
inversement $t^a=P_{a\alpha}e_\alpha+P_{aj}h^j$, on montre que
\begin{equation}
\label{changebase}
\tr\,(t^ah^j)\,t^a=P_{aj}t^a=\quotient{1}{2}\,h^j,\qquad
\tr\,(t^ae_\alpha)\,t^a=\tr(e_\alpha e_{-\alpha})\,
P_{a,-\alpha}t^a=\quotient{1}{2}\,e_\alpha.
\end{equation}
\`A l'aide de ces formules, on trouve 
\begin{equation*}
\begin{split}
-\quotient{2\pi}{\kappa}\,t^a_\ell\,&\singulier
	\quotient{\delta\Bpsi}{\delta A\sur{a}
	_{\overline{z}_\ell}}\singulier
	=\,-\quotient{1}{\kappa}\,\sum_{j=1}^r(h^j)_\ell\,\partial_{u^j}\Bpsi
	-\quotient{2\pi i}{\kappa}\sum_{m\neq \ell}
	\Bigl\lbrace\sum_{\alpha}G_\alpha(z_\ell,z_m)\,
	(e_\alpha)_\ell\,(e_{-\alpha})_m\\
&+\sum_{j=1}^rH(z_\ell,z_m)\,(h^j)_\ell\,(h^j)_m\Bigr\rbrace\,\Bpsi
	-\quotient{2\pi i}{\kappa}
	\sum_{\alpha>0}\widetilde{G}_\alpha\,(X_\alpha)_\ell
	\,\Bpsi.
\end{split}
\end{equation*}
Au chapitre 3, on a repr\'esent\'e les \'etats de Chern-Simons par des 
applications holomorphes $\Bgamma:\lietc\rightarrow \esprep$, avec
\[
\Bpsi(A^{01}_u)=
	\ee^{\pi k|u|^2/(2\imtau)}\,\otimes_\ell
	\left(\ee^{\pi(z_\ell-\overline{z}_\ell)u
	/\imtau}\right)\sous{\hspace{-0.08cm}R_\ell}\,\Bgamma(u)
\]
et $\Bgamma$ satisfait les conditions de la p.~\pageref{CSgenreun}.
La connexion de KZB est consid\'erablement plus simple quand on regarde son action sur $\Bgamma$.
En fait, on utilise m\^eme la description en termes de fonctions th\^eta~: 
$\theta(u)=\Pi(u,\tau)\,\Bgamma(u)$. On note $\delta(\tau,u,\underline{\xi})$ 
la fonction qui r\'ealise le changement de fonction $\Bpsi(A^{01}_u)
=\delta(\tau,u,\underline{\xi})\,
\theta(u)$. Le r\'esultat est~:
\[
\KZz\theta\equiv\delta^{-1}\KZz\Bpsi=
	\partial_{z_\ell}\theta-\quotient{1}{\kappa}
	\,H^{(1)}_\ell\,\theta
\]
o\`u $H^{(1)}_\ell$ est une g\'en\'eralisation du Hamiltonien de Gaudin~:
\[
H^{(1)}_\ell(\tau,\underline{\xi})=\sum_{j=1}^r(h^j)_\ell\,\partial_{u^j}
	+\sum_{m\neq\ell}\Big\lbrace
	\sum_\alpha\noyau_{u_\alpha}
	(z_\ell-z_m)\,(e_\alpha)_\ell\,(e_{-\alpha})_m
	+\sum_{j=1}^r\rho(z_\ell-z_m)\,(h^j)_\ell\,(h^j)_m\Big\rbrace.
\]

Le lecteur d\'esireux de refaire le chemin  menant \`a $\KZz\theta$
pourra regarder attentivement ce qui suit~; il y trouvera tous les
ingr\'edients n\'ecessaires pour mener \`a bien les calculs.
Dans la partie $\KZmu$ de la connexion, entre autres, intervient
la premi\`ere d\'eriv\'ee de $\Bpsi$ donn\'ee par
l'\'equation~\eqref{premiere}, qui dans le cas en question se lit
\[
\deltazpsi
        =\quotient{1}{2\pi}\,h^j\,\partial_{u^j}\,\Bpsi
        +i\sum_m\left\lbrace G_\alpha(z,z_m)\,e_\alpha\,(e_{-\alpha})_m
	+G_j(z,z_m)\,h^j\,(h^j)_m\right\rbrace\Bpsi.
\]
Pour une surface elliptique $E_\tau$, la variation de la structure 
complexe est repr\'esent\'ee par
\[
\delta\J=\quotient{\delta\tau}{\imtau}\,\partial_z\otimes d\overline{z}
	+\quotient{\delta\overline{\tau}}{\imtau}\,
	\partialbarz\otimes dz.
\]
En particulier, $\deltamu=\delta\tau/\imtau$ est constante.
Comme de plus $A^{10}_u=-(A^{01}_u)^\dagger=
-\pi u^\dagger/\imtau dz$, $u^\dagger
=\overline{u}^j\,h^j\in\lietc$, 
la premi\`ere d\'eriv\'ee contribue dans $\KZmu$ par
\[
i\int_\Sigma\tr\,A_{z}\,
        \deltazpsi\,\deltamu\,d^2z
	=-\quotient{i}{2\imtau}\,
	\delta\tau\,\sum_{j=1}^r\overline{u}^j\,\partial_{u^j}\Bpsi.
\]
D'autre part, le Laplacien sur l'espace des champs de jauge  
est donn\'e par l'\'equation~(\ref{premierederivee}.b)~:
\[
\tr\,\singulier\deltaz\,\deltazpsi\singulier=\quotient{1}{4\pi^2}
	\,\Delta_u\Bpsi
        +\Xi^j(z)\,\partial_{u^j}\Bpsi+\Upsilon(z)\,\Bpsi,
\]
o\`u $\Delta_u=\sum_j(\partial_{u^j})^2$ et
\begin{equation*}
\begin{split}
\Xi^j(z)=&\,2\,\Omega^{jk}\,\tr\,\widetilde{\Green}(z,z)\left(t^a\right)
        \,\left[t^a,\phi^k_{0,z}\right]
+\Omega^{jk}\Omega^{\ell m}\,\tr\,\phi^m_{0,z}\int_\Sigma\Green(z,.)
        \Big(\left[\phi^k_0,\partial_{u^\ell}A^{01}\right]_+\Big)\\
        &+4\,\Omega^{jk}\sum_m\tr\,\phi^k_{0,z}\,\Green(z,z_m)(t^a)\,t^a_m,\\
=&\,\quotient{i}{2\pi}\,\sum_{\alpha}\alpha^j\,\widetilde{G}_\alpha
        +\quotient{i}{\pi}\sum_\ell H(z,z_\ell)\,(h^j)_\ell,
\end{split}
\end{equation*}
o\`u $\alpha^j=\alpha(h^j)$ --- \`a ne pas confondre
avec la racine simple $\alpha_j$ --- ainsi que
\begin{equation*}
\begin{split}
\Upsilon(z)=&\,{_{ik}\over^{\pi}}\,\tr\,\widetilde{\partial_z\Green}
	(z,z)\left(t^a\right)\,t^a
        -4\sum_m\tr\,\widetilde{\Green}(z,z)(t^a)\,
	[t^a,\Green(z,z_m)(t^b)]\,t^b_m\\
        &-4\sum_{m,m'}\tr\,\Green(z,z_m)(t^a)
	\,\Green(z,z_{m'})(t^b)\,t^a_m\,t^b_{m'}\\
&-2\,\Omega^{jk}\sum_m\tr\,\phi^k_{0,z}\int_\Sigma\Green(z,w)
        \left([\Green(w,z_m)(t^a)\,dw,\partial_{u^j}A^{01}(w)]_+\right)\,t^a_m,\\
=&\,\quotient{ik}{2\pi}\,\Big\lbrace
        \sum_\alpha\widetilde{\partial_zG}_\alpha
        +\sum_{j=1}^r\widetilde{\partial_zH}\Big\rbrace
        -\sum_{\alpha}\widetilde{G}_\alpha\sum_\ell
        H(z,z_\ell)\,(X_\alpha)_\ell\\
&-\sum_{\ell,m}\Big\lbrace
        \sum_\alpha G_{-\alpha}(z,z_\ell)\,G_{\alpha}(z,z_m)\,
        (e_\alpha)_\ell(e_{-\alpha})_m
        +\sum_{j=1}^rH(z,z_\ell)\,H(z,z_m)\,(h^j)_\ell(h^j)_m
        \Big\rbrace.
\end{split}
\end{equation*}
Bien que relativement long, le calcul utilise uniquement
les \'equations~\eqref{changebase}.
La contribution du Laplacien dans $\KZmu$ est alors
\begin{equation*}
\begin{split}
{_{i\pi}\over^{\kappa}}
        \int_\Sigma\tr\,\singulier
        \deltaz\,\deltazpsi\singulier\deltamu\,d^2z=
	\quotient{1}{\kappa}\,\delta\tau\,
	\Big\lbrace&
	\quotient{i}{4\pi}\,\Delta_u\Bpsi-\sum_{\alpha>0}\alpha^j
	\widetilde{G}_\alpha\,\partial_{u^j}\Bpsi\\
&-\quotient{k}{2}\Big(
	\sum_\alpha \widetilde{\partial_zG}_\alpha+\sum_{j=1}^r
	\widetilde{\partial_zH}\Big)\,\Bpsi
	+i\pi\,\CI\Big\rbrace.
\end{split}
\end{equation*}
On a isol\'e la partie $\CI$  comportant des int\'egrales non-triviales~:
\begin{equation*}
\begin{split}
-\imtau\,\CI=\sum_{\ell,m}\Bigl\lbrace&
        \sum_\alpha\int_\Sigma G_{-\alpha}(z,z_\ell)\,G_\alpha(z,z_m)\,d^2z
        \,(e_\alpha)_\ell\,(e_{-\alpha})_m\\
&+\sum_{j=1}^r\int_\Sigma H(z,z_\ell)\,H(z,z_m)\,d^2z\,(h^j)_\ell\,(h^j)_m
        \Bigr\rbrace.
\end{split}
\end{equation*}
On calculera ces int\'egrales apr\`es le changement de fonction $\Bpsi\mapsto
\theta$. Pour le moment, utilisant la forme explicite des fonctions de Green,
on peut \'ecrire 
\[
\KZmu\Bpsi=d_{\delta\mu}\Bpsi
	+\delta\tau\,\Big(\quotient{1}{\kappa}\,\quotient{i}{4\pi}
	\,\Delta_u\Bpsi+\sum_{j=1}^rX^j\,\partial_{u^j}\Bpsi+X\,\Bpsi\Big)
\]
avec 
\begin{align*}
X^j=&\,\quotient{i}{2\imtau}\,(u^j-\overline{u}^j)
        +\quotient{1}{\kappa}\,\quotient{1}{2\pi i}\,\varphi^j,\\
X=&\,-\quotient{k}{\kappa}\,\Big\lbrace
	\sum_{\alpha>0}\Big(\quotient{\pi iu_\alpha^2}{4\imtau^2}
                        +\quotient{iu_\alpha}{2\imtau}\,\rho(u_\alpha)
                        -\quotient{1}{4\pi i}\,
        \frac{\vartheta_1''}{\vartheta_1}(u_\alpha)
                        -\quotient{1}{2\pi i}\,\eta_1\Big)\\
&\,+\sum_{j=1}^r
	\Big(\quotient{1}{2\pi i}\,\eta_1-\quotient{1}{4i\imtau}\Big)
	\Big\rbrace+\quotient{i\pi}{\kappa}\,\CI.
\end{align*}
Il est commode d'introduire les deux variables
\[
\varphi^j=\quotient{\pi ku^j}{\imtau}-\sum_{\alpha>0}\alpha^j\rho(u_\alpha)
        \quad   \mbox{et} \quad
        \sigma^j=\sum_\ell(z_\ell-\overline{z}_\ell)\,(h^j)_\ell
\]
car elles interviennent tout le temps, par exemple avec
$\delta^{-1}\partial_{u^j}\delta = \varphi^j+\frac{\pi}{\imtau}\,\sigma^j$.
Comme pour $\KZz\Bpsi$, on effectue le changement de fonction 
$\Bpsi\mapsto\theta$. Apr\`es un peu de calculs, on obtient
\[
\delta^{-1}\,\KZmu\Bpsi=
	\delta^{-1}\,d_{\delta\mu}(\delta\,\theta)
	+\delta\tau\,\Big(\quotient{1}{\kappa}\,\quotient{i}{4\pi}
        \,\Delta_u\Bpsi+\sum_{j=1}^r Y^j\,\partial_{u^j}\theta+Y\,\theta\Big)
\]
avec
\begin{align*}
Y^j=&\,\quotient{1}{\kappa}\,\quotient{i}{2\imtau}\,\sigma^j+
	\quotient{i}{2\imtau}\,(u^j-\overline{u}^j),\\
Y=&\,\quotient{i}{2\imtau}\,\sum_{j=1}^r(u^j-\overline{u}^j)\,
	\delta^{-1}\partial_{u^j}\delta
	-\quotient{\pi ik}{4\imtau^2}\,|u|^2
	+\quotient{1}{\kappa}\,\quotient{i\pi}{4\imtau^2}\sum_{j=1}^r(
	\sigma^j)^2
	+\quotient{i\pi}{\kappa}\,\delta^{-1}\CI\,\delta\\
&\,+\quotient{1}{\kappa}\quotient{1}{4\pi i}\Big(
        \sum_{\alpha>0}|\alpha|^2\rho'(u_\alpha)
        +\sum_{\alpha,\beta>0}\tr\,\alpha\beta\,
        \rho(u_\alpha)\,\rho(u_\beta)\Big)\\
&\,+\quotient{k}{\kappa}\,\quotient{1}{4\pi i}\Big(
	(d-3r)\,\eta_1+\sum_{\alpha>0}\frac{\vartheta''_1}{\vartheta_1}
	(u_\alpha)\Big)
\end{align*}
o\`u $d$ est la dimension de $\lieg$ et $r$ son rang.
\`A premi\`ere vue, les deux derniers termes sont surprenants. Toutefois 
d'apr\`es la section {\bf 4.4} ce ne sont que des formes
d\'eguis\'ees du d\'enominateur de Weyl-Kac. Ainsi $Y$ vaut
\[
Y=\quotient{i}{2\imtau}\,\sum_{j=1}^r(u^j-\overline{u}^j)\,
        \delta^{-1}\partial_{u^j}\delta
        -\quotient{\pi ik}{4\imtau^2}\,|u|^2
        +\quotient{1}{\kappa}\,\quotient{i\pi}{4\imtau^2}\sum_{j=1}^r(
        \sigma^j)^2+\Pi^{-1}\partial_\tau\Pi
        +\quotient{i\pi}{\kappa}\,\delta^{-1}\CI\,\delta.
\]

Il est plus astucieux de 
ne pas chercher \`a calculer imm\'ediatement les int\'egrales dans
$\delta^{-1}\CI\,\delta$ et
d'attendre d'avoir fini de d\'ebroussailler la zone. On en vient maintenant 
\`a un point cl\'e du calcul. \`A la fin on d\'esire exprimer 
$\KZtau\theta\equiv\delta^{-1}\KZtau\Bpsi$ en fonction des d\'eriv\'ees 
partielles $\partial_\tau$, $\partial_{u^j}$, 
$\partialbaruj$, $\partial_{z_\ell}$ 
et $\partialzbar$ de $\theta$. Il faut donc conna\^\i tre le passage de 
$\KZmu$ \`a $\KZtau$ ainsi que celui de $d_{\delta\mu}$ \`a $\partial_\tau$, 
... Na\"\i vement, on pourrait croire que cette transformation est 
simplement $d_{\delta\mu}=\delta\tau\,\partial_\tau$. En fait, 
la d\'erivation de $\delta\mu$ agit comme une d\'eriv\'ee totale sur 
la structure complexe alors que les d\'eriv\'ees partielles en $u$, ... 
supposent un choix pr\'ealable de structure complexe. En effet,
le choix d'une d\'ecomposition en parties analytiques et anti-analytiques du 
champ et des coordonn\'ees des points d'insertion pr\'esuppose une structure 
complexe. En principe, on a 
\[
d_{\delta\mu}=\delta\tau\,\partial_\tau+
\sum_j\delta u^j\,\partial_{u^j}+\sum_j\delta\overline{u}^j\,\partialbaruj+
\sum_\ell\,
(\delta z_\ell\,\partial_{z_\ell}+\delta\overline{z}_\ell\,\partialzbar).
\]
Par exemple, pour trouver les variations de $u^j$ et $\overline{u}^j$, 
il suffit 
de trouver quelles $\delta u^j$ et $\delta\overline{u}^j$ sont
telles que la variation du champ complet $A_u$ par rapport \`a 
celles-ci compense la 
variation du champ par rapport \`a la structure complexe 
($\delta\overline{\tau}=0$),
\cad $\delta_uA_u+\delta_\tau A_u=0$. On finit par trouver
\begin{gather*}
\delta\tau\KZtau=\KZmu-\delta\tau\sum_\ell
	\quotient{z_\ell-\overline{z}_\ell}{2i\imtau}\,\KZz,\\
d_{\delta\mu}=\delta\tau\Big(\partial_\tau+\sum_j\quotient{u^j-\overline{u}^j}
	{2i\imtau}\,\partial_{u^j}+\sum_\ell
	\quotient{z_\ell-\overline{z}_\ell}{2i\imtau}\,\partial_{z_\ell}
	\Big).
\end{gather*}
Quand on effectue ces quelques changements, presque tous les termes
disparaissent et
\begin{align*}
\KZtau\theta=&\,\partial_\tau\theta+\quotient{1}{\kappa}\,\quotient{i}{4\pi}\,
	\Delta_u\theta+\quotient{1}{\kappa}\,
	\quotient{i\pi}{4\imtau^2}\,\sum_j(\sigma^j)^2
	\,\theta
	+\quotient{i\pi}{\kappa}\,\delta^{-1}\CI\,\delta\,\theta\\
&\,+\quotient{1}{\kappa}\,\quotient{1}{2i\imtau}
	\sum_\ell(z_\ell-\overline{z}_\ell)
	\sum_{m\neq \ell}
	\Big\lbrace\sum_\alpha\noyau_{u_\alpha}
	(z_\ell-z_m)\,(e_\alpha)_\ell\,(e_{-\alpha})_m\\
&\,+\sum_{j=1}^r\rho(z_\ell-z_m)\,(h^j)_\ell\,(h^j)_m\Big\rbrace\,\theta.
\end{align*}

Il ne reste plus qu'\`a calculer $\delta^{-1}\CI\,\delta$. D\'ej\`a,
l'action adjointe de $\delta$ sur $\CI$ n'intervient que 
sur $(e_\alpha)_\ell\,(e_{-\alpha})_m$ et supprime les exponentielles
dans le produit $G_{-\alpha}(z,z_\ell)\,G_\alpha(z,z_m)$.
Par cons\'equent,
\[
\delta^{-1}\CI\,\delta=
\sum_{\ell,m}\Big\lbrace\quotient{1}{4\pi^2\imtau}\sum_\alpha
	\Delta_\alpha(z_\ell,z_m)\,(e_\alpha)_\ell\,(e_{-\alpha})_m
	-\quotient{1}{\imtau}\sum_{j=1}^r\Delta(z_\ell,z_m)\,(h^j)_\ell\,(h^j)_m\Big\rbrace
\]
o\`u 
\begin{align*}
\Delta_\alpha(z_\ell,z_m)=&\,\int_\Sigma \noyau_{-u_\alpha}(z-z_\ell)\,
	\noyau_{u_\alpha}(z-z_m)\,d^2z,\\
\Delta(z_\ell,z_m)=&\, \int_\Sigma H(z,z_\ell)\,H(z,z_m)\,d^2z.
\end{align*}
Ces quantit\'es ne d\'ependent que de la diff\'erence 
$y=z_\ell-z_m$. En cherchant 
leur d\'eriv\'ee par rapport \`a $\overline{y}$, on d\'eduit ais\'ement que, 
si $y\neq 0$, 
\begin{align*}
\Delta(y)=&\,
	-\quotient{1}{8\imtau}\,(y-\overline{y})^2-\quotient{1}{4\pi}\,
	(y-\overline{y})\,\rho(y)-\quotient{\imtau}{8\pi^2}\,
	\frac{\vartheta_1''}{\vartheta_1}(y)+C,\\
\Delta_\alpha(y)=&\,
	\pi\,(y-\overline{y})\,\noyau_{u_\alpha}(y)-
	\imtau\,\partial_x\noyau_{u_\alpha}(y).
\end{align*}
La constante $C$ est fix\'ee par la condition $\int_\Sigma \Delta(y)\,
d^2y=0$. On peut d\'eduire de ces r\'esultats les int\'egrales aux points 
co\"{\i}ncidants en prenant la limite $y\rightarrow 0$ sur la droite r\'eelle
\[
\Delta(0)=\quotient{3\imtau}{4\pi^2}\,\eta_1+C,\qquad
\Delta_\alpha(0)=\imtau\,\rho'(u_\alpha).
\]
On peut peut-\^etre calculer explicitement la constante $C$ mais,
comme elle dispara\^\i t d'elle-m\^eme dans le calcul, ce n'est pas
important. La fin ne pose aucun probl\`eme. Il suffit 
de scinder $\delta^{-1}\CI\,\delta$ en deux sommes, l'une portant sur toutes 
les configurations et l'autre \'evitant les points co\"{\i}ncidants. La 
premi\`ere somme contient les trois termes contenant $y-\overline{y}$ --- 
\ie les termes nuls par passage \`a la limite $y\rightarrow 0$, $y\in\R$ --- 
et la seconde le prolongement analytique des trois autres. 
Presque tous les termes se compensent et $\KZtau\theta=\partial_\tau\theta-
\frac{1}{\kappa}\,H^{(1)}_0\,\theta$ o\`u
\[
H^{(1)}_0(\tau,\underline{\xi})=-\quotient{i}{4\pi}\,\Delta_u-\quotient{i}{4\pi}
\sum_{\ell,m}
	\Big\lbrace\sum_\alpha\partial_x\noyau_{u_\alpha}(z_\ell-z_m)\,
	(e_\alpha)_\ell\,(e_{-\alpha})_m+
	\quotient{1}{2}\sum_{j=1}^r\frac{\vartheta''_1}{\vartheta_1}
	(z_\ell-z_m)\,(h^j)_\ell\,(h^j)_m\Big\rbrace.
\]

Enfin, il reste le calcul de $\conbar$. On trouve sans difficult\'e
$\con_{\overline{z}_\ell}\theta=\partialzbar\theta$ et
$\con_{\overline{\tau}}\,\theta=\partialbartau\theta
	+\sum_j\frac{u^j-\overline{u}^j}{2i\imtau}\,\partialbaruj\theta$.
En plus, $\Bgamma$ est une fonction holomorphe donc de d\'eriv\'ee 
$\partialbaruj\Bgamma$ nulle. 
En r\'esum\'e, la connexion de KZB exprim\'ee dans son action sur les 
fonctions th\^eta est
\begin{gather*}
\KZzbar= \partialzbar,\qquad\KZtaubar=\partialbartau,\\
\KZz=\partial_{z_\ell}-\quotient{1}{\kappa}\,H^{(1)}_\ell,
\qquad\KZtau=\partial_\tau-\quotient{1}{\kappa}\,H^{(1)}_0
\end{gather*}
o\`u
\begin{equation*}
\begin{split}
H^{(1)}_\ell(\tau,\underline{\xi})=&\sum_{j=1}^r(h^j)_\ell\,\partial_{u^j}
        +\sum_{m\neq\ell}\Big\lbrace
        \sum_\alpha\noyau_{u_\alpha}
        (z_\ell-z_m)\,(e_\alpha)_\ell\,(e_{-\alpha})_m
        +\sum_{j=1}^r\rho(z_\ell-z_m)\,(h^j)_\ell\,(h^j)_m\Big\rbrace,\\
H^{(1)}_0(\tau,\underline{\xi})=&-\quotient{i}{4\pi}\,\Delta_u-\quotient{i}{4\pi}
\sum_{\ell,m}
        \Big\lbrace\sum_\alpha\partial_x\noyau_{u_\alpha}(z_\ell-z_m)\,
        (e_\alpha)_\ell\,(e_{-\alpha})_m+
        \quotient{1}{2}\sum_{j=1}^r\frac{\vartheta''_1}{\vartheta_1}
        (z_\ell-z_m)\,(h^j)_\ell\,(h^j)_m\Big\rbrace.
\end{split}
\end{equation*}
La platitude de la connexion est assur\'ee par les relations~:
$[H_\ell^{(1)},H^{(1)}_{\ell'}]=0$, pour $\ell,\ell'=0,1,\cdots,N$,
\'equations \'equivalentes \`a l'EYBC dynamique~\cite{felder:elliptic}.
Le cas le plus simple est $G={\rm SU}_2$ avec l'insertion
d'un spin $j$ en $z$. La seule partie non-triviale est 
\[
\KZtau=\partial_\tau+\quotient{i}{2\pi\kappa}\,\partial_v^2
	-\quotient{ij(j+1)}{2\pi\kappa}\big(\wp(v)+2\,\eta_1\big)
\]
o\`u $\kappa=k+2$, $\wp$ est la fonction de Weierstra\ss\ et $u=vh/\sqrt{2}$.
On a utilis\'e l'\'egalit\'e $\partial_x\noyau_{v}(0)=\rho'(v)
=-\wp(v)-2\eta_1$. Il appara\^\i t l'op\'erateur de Lam\'e
$\partial_v^2+c\,\wp(v)$ --- pour le groupe $G={\rm SU}_n$ c'est 
l'op\'erateur de Calogero-Sutherland elliptique (g\'en\'eralis\'e).

\medskip
\section{Syst\`emes de Hitchin elliptiques}

On reprend les notations de la section 3.{\bf 6}. Les champs sont param\'etris\'es
par $u\in\lietc\setminus (P^\vee+\tau P^\vee)$ et $h\in\CG^\C$~:
$A^{01}={}^{h^{-1}}\hspace{-0.15cm}A^{01}_u=(\gamma_uh)^{-1}\de(\gamma_uh)$.
Pour \'eviter les ambigu\"\i t\'es dans la param\'etrisation, on doit
identifier les paires suivantes
\[
(u,h)\,\sim\,(wuw^{-1},\, wh)\,\sim\,(u+q^\vee,\,
h^{-1}_{q^\vee}\,h)\,
\sim
        \,(u+\tau q^\vee,\, h^{-1}_{\tau q^\vee}\,h),
\]
avec  $q^\vee$ dans le r\'eseau des coracines $Q^\vee$ et $w$ 
dans le groupe de Weyl $W$. Comme en genre z\'ero, on doit r\'esoudre
l'\'equation
\begin{equation}
\label{genreun-H}
\de\,({}^{\gamma_uh}\varphi^{01})\,=\,\sum_\ell\nu_\ell\,\delta_{z_\ell}
\end{equation}
o\`u $\nu_\ell=(\gamma_uh)(z_\ell)\,\lambda_\ell\,(\gamma_uh)(z_\ell)^{-1}$.
On d\'ecompose $\nu_\ell$ sur la base $\baseB$~: 
$\nu_\ell=\sum_\alpha\nu_\ell^{-\alpha}\,e_\alpha
+\nu^0_\ell$ o\`u $\nu^0_\ell=\nu_\ell^j\,h^j\in{\lietc}$.
On peut r\'esoudre l'\'equation~\eqref{genreun-H} \`a condition
que $\sum_\ell\nu^0_\ell=0$. Dans ce cas
\[
{}^{\gamma_u h}\varphi^{01}(z)\ =\ \Big(\varphi_0
        \,+\,\sum_\ell\big(\sum_\alpha P_{u_\alpha}(z-z_\ell)\,
        \nu_\ell^{-\alpha}e_\alpha
        +\rho(z-z_\ell)\,\nu^0_\ell\,\big)\Big)\frac{_{dz}}{^{2\pi i}}
\]
o\`u $\varphi_0=\varphi^j_0\,h^j$ est une constante arbitraire.
La r\'eduction symplectique donne
\[
\CP^{(1)}_{\CO}\,\cong\,
        \Big\{(u,\varphi_0,\underline{\nu})\in T^*{\lietc}\times
        (\mathop{\times}\limits_{\ell=1}^N\CO_\ell)\,\Big|\,
	\sum_\ell\nu^0_\ell=0\Big\}
        \bigg/ W{\rtimes} (Q^\vee+\tau Q^\vee)
\]
o\`u l'action du groupe $W{\rtimes} (Q^\vee+\tau Q^\vee)$ identifie les paires
\begin{eqnarray*}
(u,\varphi_0,\underline{\nu}) & \sim &
        (wuw^{-1}, w\varphi_0w^{-1}, w
        \underline{\nu}w^{-1})\,\sim\,
        \left(u+q^\vee,\varphi_0,
        (h^{-1}_{q^\vee}(z_\ell)\nu_\ell h_{q^\vee}(z_\ell))\right)\\
        & \sim &
        \left(u+\tau q^\vee,\varphi_0,
        (h^{-1}_{\tau q^\vee}(z_\ell)\,\nu_\ell\,h_{\tau q^\vee}(z_\ell))
        \right).
\end{eqnarray*}
La r\'eduction symplectique de $\CP^{(1)}_{\CO}$ est aussi celle de
$T^*{\lietc}\times(\times_\ell\CO_\ell)$ par
le groupe  $W{\rtimes} (Q^\vee+\tau Q^\vee)$. Maintenant
on d\'eduit ais\'ement l'application de Hitchin pour $p=p_2$
\begin{equation*}
\begin{split}
h_{p_2}(u,\varphi_0,\underline{\nu})(z)&=
	p_2 \Big(\varphi_0
        \,+\,\sum_\ell\big(\sum_\alpha P_{u_\alpha}(z-z_\ell)\,
        \nu_\ell^{-\alpha}e_\alpha
        +\rho(z-z_\ell)\,\nu^0_\ell\,\big)\Big)
	\left(\frac{_{dz}}{^{2\pi i}}\right)^2\\
&=\Big(-\quotient{1}{2}\,\sum_\ell\rho'(z-z_\ell)\,\delta_\ell
        +2\sum_\ell\rho(z-z_\ell)\,h^{(1)}_\ell+4\pi i\,h^{(1)}_0
        \Big)\left(\frac{_{dz}}{^{2\pi i}}\right)^2
\end{split}
\end{equation*}
o\`u, comme en genre z\'ero, $\delta_\ell=\nu^a_\ell\nu^a_\ell$ et
\begin{align*}
h^{(1)}_\ell(u,\varphi_0,\underline{\nu})&=\sum_{j=1}^r \nu^j_\ell\varphi_0^j+
	\sum_{m\neq \ell}\Big\{
        \sum_\alpha P_{u_\alpha}(z_\ell-z_m)\,\nu_\ell^\alpha\nu_m^{-\alpha}
        \,+\,\sum_{j=1}^r\rho(z_\ell-z_m)\,\nu^j_\ell\nu^j_m\Big\},\\
h^{(1)}_0(u,\varphi_0,\underline{\nu})&=
	-\quotient{i}{4\pi}\sum_{j=1}^r\varphi_0^j\varphi_0^j
        -\quotient{i}{4\pi}\sum_{\ell,m}\Big\{
        \sum_\alpha\partial_xP_{u_\alpha}(z_\ell-z_m)
        \nu^\alpha_\ell\nu^{-\alpha}_m
        +\frac{_1}{^2}
        \sum_{j=1}^r \frac{\vartheta_1''}{\vartheta_1}(z_\ell-z_m)
        \nu^j_\ell\nu_m^j\Big\}.
\end{align*}

Soit $\delta\mu=\delta\mu_{\bar z}^z\,\partial_z\otimes d\bar z$
une diff\'erentielle de Beltrami se comportant comme
$\CO((z-z_\ell)^2)$ au voisinage des points d'insertion.
Pour la nouvelle structure complexe, la coordonn\'ee complexe
est $z'=z+\frac{z-\bar z}{2i\tau_2}\,\delta\tau+{\delta v}^z$
avec $\partialbarz z'=\delta\mu_{\bar z}^z$.
On exige que $\delta v^z(z+1)=\delta v^z(z+\tau)
=\delta v^z(z)$. La variation $\delta\tau$ est d\'etermin\'ee
par la condition que l'int\'egrale de $\delta\mu_{\bar z}^z$ 
sur $\Sigma$ est \'egale \`a $\frac{i}{2\tau_2}\,\delta\tau$.
La variation ${\delta v}^z$ est unique \`a une constante
multiplicative pr\`es. On a $z'(z+1)=z'(z)+1$ tandis
que $z'(z+\tau)=z'+\tau'$ pour $\tau'=\tau+\delta\tau$. 
La nouvelle courbe elliptique est donc isomorphe \`a
$E_{\tau'}$. Les coordonn\'ees des
points d'insertion changent en $z'_\ell=z_\ell+\delta z_\ell$ 
et les $1$-jets de param\`etre local changent en 
$\chi'_\ell=\chi_\ell+\delta\chi_\ell$, avec
\[
\delta z_\ell=\quotient{z_\ell-\bar z_\ell}{2i\tau_2}
	\,\delta\tau+{\delta v}^z(\xi_\ell),\qquad
	\chi_\ell^{-1}\delta\chi_\ell =
	\quotient{\delta\tau}{2i\tau_2}+\da_z\delta v^z(\xi_\ell).
\]
En int\'egrant $h_{p_2}$ contre $\delta\mu$ sur $\Sigma$ priv\'ee
de petites boules autour des points d'insertion et apr\`es une
int\'egration par parties, on trouve
\[
h_{\delta\mu}\,=\,\int_\Sigma h_{p_2}\delta\mu
        ={_1\over^{2\pi i}}\sum_\ell
        \left(\quotient{1}{2}\,\delta_\ell\,\chi_\ell^{-1}\delta\chi_\ell
        +2\,h^{(1)}_\ell\,\delta z_\ell+2\,h^{(1)}_0\,\delta\tau\right).
\]
La connexion de KZB en genre un est donn\'ee par
\begin{align*}
\KZmu&=
\sum\limits_\ell\Big(\delta\chi_\ell\,\big(\da_{\chi_\ell}-\chi_\ell^{-1}\Delta_\ell\big)
+\delta z_\ell\,\big(\da_{z_\ell}-\quotient{1}{\kappa}\, H^{(1)}_\ell\,\big)
+\delta\tau\,\big(\da_{\tau}-\quotient{1}{\kappa} \,H^{(1)}_0\,\big)
\Big),\\
\KZmubar&=
\sum\limits_\ell\Big(\delta\bar\chi_\ell\,\da_{\bar\chi_\ell}
+\delta \bar z_\ell\,\da_{\bar z_\ell}+\delta\bar\tau\,
\da_{\bar\tau}\Big)
\end{align*}
o\`u $\Delta_\ell$ est le poids conforme en $\xi_\ell$. Mis
\`a part la translation $k\mapsto \kappa$, on d\'eduit
$\Delta_\ell$, $\frac{1}{\kappa}\,H^{(1)}_\ell$ et 
$\frac{1}{\kappa}\,H^{(1)}_0$ de
$\frac{2k}{\pi}\,\frac{1}{4\pi i}\,\delta_\ell$, 
$\frac{2k}{\pi}\,\frac{1}{\pi i}\,h^{(1)}_\ell$ et
$\frac{2k}{\pi}\,\frac{1}{\pi i}\,h^{(1)}_0$ par la quantification
g\'eom\'etrique des orbites coadjointes, \cad  
\[
\nu^a_\ell\mapsto \quotient{\pi\sqrt{2i}}{k}\,t^a_\ell,
	\qquad \varphi_o^j\mapsto 
	\quotient{\pi\sqrt{2i}}{2k}\,\partial_{u^j},\qquad
	\nu^\alpha_\ell\mapsto
	\quotient{\pi\sqrt{2i}}{2k}\,(e_\alpha)_\ell,
	\qquad \nu^j_\ell\mapsto
	\quotient{\pi\sqrt{2i}}{2k}\,(h^j)_\ell.
\]
Les diff\'erences dans les facteurs multiplicatifs viennent de celles
dans les normalisations des bases de $\liegc$.



\chapter[Syst\`emes de Hitchin (articles)]{Syst\`emes de Hitchin\\(articles)}

\protect\hspace{\parindent}%
Dans ce chapitre, on reproduit les deux articles \'ecrits avec 
K.~Gaw\c{e}dzki~:

\begin{description}

\item[1997] {\bol Self-duality of the ${\rm SL}_2$ Hitchin integrable
        system at genus two}, pr\'epublication IHES/P/97/80 et
	solv-int/9710025, accept\'e pour publication dans Communications
        in Mathematical Physics.

\item[1998] {\bol Hitchin systems at low genera},
        pr\'epublication IHES/P/98/21 et hep-th/9803101.

\end{description}

\noindent Certaines parties du deuxi\`eme article sont aussi pr\'esent\'ees dans le texte
en fran\c{c}ais.

\medskip
\section*{R\'esum\'e}

\addcontentsline{toc}{section}{\protect\numberline{}R\'esum\'e}

On se place exclusivement en genre deux. Le groupe $G$ est ${\rm SU}_2$.
Une surface de Riemann de genre deux est toujours hyperelliptique, 
\ie $\Sigma$ est la normalisation de la courbe alg\'ebrique plane 
d'\'equation
\[
y^2=(x-a_1)\cdots(x-a_6).
\]
Dor\'enavant, on ne marque plus la diff\'erence 
entre $\Sigma$ et la courbe hyperelliptique $C$.
La courbe $C$ est le rev\^etement ramifi\'e $C\ni(x,y)
\mapsto x\in \C P^1$ du plan projectif, \`a deux feuillets, 
au-dessus des points $a_1,\cdots, a_6$. On suppose
que les $a_\ell$ sont tous finis. Les formes holomorphes et les formes quadratiques
holomorphes sont de la forme
\[
\omega=(a+bx)\,\frac{dx}{y},\qquad
	\beta= (a+bx+cx^2)\,\Big(\frac{dx}{y}\Big)^2.
\]
On note $(\omega^a)$ la base de $H^0(K)$.
La Jacobienne de $C$ est ${\rm Jac}\,C=\C^2/(\Z^2+\tau\Z^2)$, o\`u 
$\tau$ est la matrice des p\'eriodes normalis\'ee.
Il y a $16$ structures spin $\delta=\delta'+\tau\delta''$ --- \cad
des fibr\'es en droites holomorphes $L$ de degr\'e $1$ tels que $L^2=K$
--- avec $\delta',\delta''\in\frac{1}{2}\,\Z^2/\Z^2$. Les structures spin 
impaires $\delta_\ell=\delta_\ell'+\tau\delta''_\ell$
sont au nombre de $6$ et sont \'enum\'er\'ees par les points de
Weierstra\ss\  $a_\ell$, \cad les z\'eros de leurs sections holomorphes.
Les structures paires sont au nombre de $10$, ce sont celles
qui n'admettent pas de section holomorphe. 

En genre deux, l'espace des fonctions th\^eta de degr\'e deux $\Theta_2
\equiv{\bf Th}_2$ est de dimension $4$. Une base de $\Theta_2$ est donn\'ee par
\[
\theta_{e}(z)=\sum_{n\in\Z^g}\ee^{2\pi i\,(n+e).\tau(n+e)
        +4\pi i\,(n+e).z},
\]
o\`u $e\in\frac{1}{2}\,\Z^2/\Z^2$. La fonction
\begin{equation}
\label{double}
\vartheta(u+v)\,\vartheta(u-v)=\sum\limits_e\theta_e(u)\,\theta_e(v)
\end{equation}
est une fonction th\^eta de degr\'e deux en $u$ mais aussi en $v$.
L'application $v\mapsto\vartheta(\cdot+v)\,\vartheta(\cdot-v)$
d\'etermine un plongement de la {\bol surface de Kummer} ${\rm Jac}\,C\big/\Z^2$
sur la quartique $\CK$ dans l'espace projectif tridimensionnel
$\PP\Theta_2$ --- $\Z^2$ est l'involution $L\mapsto L^{-1}K$ ou
$v\mapsto -v$. La fonction th\^eta~\eqref{double} d\'etermine
une forme quadratique non-d\'eg\'en\'er\'ee sur l'espace $\Theta_2^*$
dual \`a $\Theta_2$. On peut alors identifier $\Theta_2^*$ et 
$\Theta_2$ en envoyant $\phi\in\Theta_2^*$ vers
$\iota (\phi)\in\Theta_2$ par
\[
\iota(\phi)(u)=\langle\,\vartheta(u+\cdot)\,\vartheta(u-\cdot)\,,
\,\phi\,\rangle.
\]
Cette identification \'echange la base $(\theta_e)$ de $\Theta_2$ et la
base duale $(\theta^*_e)$ de $\Theta_2^*$, mais aussi
la surface de Kummer $\CK$ avec sa forme duale $\CK^*\subset
\PP\Theta^*_2$. La surface de Kummer duale est compos\'ee
des formes lin\'eaires proportionnelles aux formes 
d'\'evaluation $\phi_u$~:
\[
\langle\,\theta\,,\,\phi_u\,\rangle=\theta(u).
\]
Le groupe des structures de spin $(\Z\big/2\Z)^4$
agit sur ${\bf Th}_k$, pour $k$ pair, par les endomorphismes
$[\delta]$, $\delta=(\delta',\delta'')$, d\'efinis par
\[
\big([\delta]\,\theta\big) (u)=\ee^{\pi i k\, \delta''\cdot\tau \delta''+
	2\pi i k\, \delta''\cdot u}\,\theta(u+\delta'+\tau \delta'').
\]
Si $k$ n'est pas divisible par $4$, l'action est seulement projective~:
$[\delta_1]\,[\delta_2]=(-1)^{4\,\delta'_1\cdot\delta_2''}\, [\delta_1+\delta_2]$~;
cette action se rel\`eve au groupe de Heisenberg. Pour $k=2$, 
\begin{equation}
\label{heisenberG}
[\delta]\,\theta_e=(-1)^{4\,\delta'\cdot e}\,\theta_{e+\delta''}.
\end{equation}
L'action $[\delta]$ pr\'eserve la surface de Kummer $\CK$ et l'action
transpos\'ee $[\delta]^t$ pr\'eserve la surface de Kummer duale $\CK^*$.

En genre deux, l'espace des modules des fibr\'es vectoriels stables
de rang deux et de d\'eterminant trivial est canoniquement
isomorphe \`a $\PP\Theta_2\setminus\CK$~\cite{nar-ram}. L'isomorphisme associe
\`a un ${\rm SL}_2$-fibr\'e $E$ la fonction th\^eta de degr\'e deux
s'annulant aux points $u\in\C^2$ correspondant aux duaux des sous-fibr\'es 
en droites de $E$. Contrairement au cas g\'en\'eral,
la compactification semi-stable est lisse et
\[
\CN_{ss}\cong \PP\Theta_2.
\]
Les points dans la surface de Kummer repr\'esentent donc les 
classes de Seshadri de fibr\'es semi-stables non-stables. L'espace
des phases du syst\`eme de Hitchin pour $G={\rm SU}_2$,
sans points d'insertion, est donc
\[
T^*\CN_{ss}\cong T^*\PP\Theta_2\cong
	\big\{\,(\theta,\phi)\in\Theta_2\times\Theta_2^*\,\big|\,
	\theta\neq 0,\ \langle\theta,\phi\rangle=0\big\}
	\big/\C^*,
\]
o\`u $t\in\C^*$ identifie les paires par
$(\theta,\phi)\sim(t\,\theta,t^{-1}\,\phi)$. En tant qu'espace symplectique,
c'est la r\'eduction symplectique de $T^*\big(\Theta_2\setminus\{0\}\big)$
par l'action de $\C^*$. En utilisant les bases $(\theta_e)$ et $(\theta_e^*)$,
on a les d\'ecompositions
\[
\theta=\sum_eq_e\,\theta_e,\qquad\phi=\sum_ep_e\,\theta_e^*.
\]
On peut alors repr\'esenter $T^*\CN_{ss}$ par l'espace des paires
$(q,p)\in\C^4\times\C^4$, $q\neq0$, $q\cdot p=0$,
avec l'identification $(q,p)\sim(t\,q,t^{-1}\,p)$, muni de
la forme symplectique $dp\wedge dq$.

On montre que l'application de Hitchin $h_{p_2}:T^*\PP\Theta_2
\rightarrow H^0(K^2)$ a une forme particuli\`erement simple~:
\begin{equation}
\label{hitchin2}
h_{p_2}=\quotient{1}{2}
	\sum_{\ell,\ell'=1\atop
	{\ell\neq \ell'}}^6 \frac{r_{\ell \ell'}}{(x-a_\ell)(x-a_{\ell'})}\,\left(
	\quotient{dx}{2\pi i}\right)^2
	=\sum_{\ell =1}^6
	\frac{h_\ell^{(2)}}{x-a_\ell}\,\left(
        \quotient{dx}{2\pi i}\right)^2
\end{equation}
o\`u 
\[
r_{\ell \ell'}(\theta,\phi)=\quotient{1}{16}\,
	\langle\,[\delta_\ell]\,\theta\,,\,[\delta_{\ell'}]^t\,\phi\,\rangle
	\,\langle\,[\delta_{\ell'}]\,\theta\,,\,[\delta_\ell]^t\,\phi\,\rangle
\]
o\`u $\delta_\ell,\delta_{\ell'}$ sont les structures spin impaires
aux points $a_\ell$ et $a_{\ell'}$ et
\[
h^{(2)}_\ell(\un a,\theta,\phi)=\sum_{\ell'\neq\ell}^6
	\frac{r_{\ell \ell'}}{a_\ell-a_{\ell'}}.
\]
Gr\^ace \`a l'\'equation~\eqref{heisenberG}, on peut r\'ecrire les
$r_{\ell \ell'}$ en fonction de $q$ et $p$. On obtient des polyn\^omes
homog\`enes de degr\'e $2$ en $q_e$ et $p_e$. L'\'egalit\'e
$\sum_{\ell'\neq\ell}r_{\ell \ell'}=0$, pour tout $\ell$, garantie
que $h_{p_2}$ est une forme quadratique holomorphe. Des identit\'es
similaires montrent que les $h^{(2)}_\ell$ ne sont pas tous ind\'ependants. 
Comme on sait que le syst\`eme est int\'egrable, il y 
a bien entendu $3$ Hamiltoniens ind\'ependants. 

La d\'emonstration du r\'esultat~\eqref{hitchin2} se fait en quatre
\'etapes. Les deux premi\`eres sont le fait de van Geemen 
et Previato~\cite{VGP}. Ceux-ci montrent que pour tout $\theta\neq 0$ et
pour tout $u\in\C^2$ tel que $\theta(u)=0$, on a
\[
h_{p_2}(\theta,\phi_u)=-\quotient{1}{16\pi^2}\,\big(\partial_{u^a}
	\theta(u)\,\omega^a\big)^2.
\]
Cette \'equation donne le polyn\^ome quadratique $h_{p_2}(\theta,.)$
sur la quartique $\CK^*_\theta=\CK^*\cap\PP\theta^\bot$ dans le sous-espace
projectif de $\Theta_2^*$ perpendiculaire \`a $\theta$.
En principe cela fixe compl\`etement l'application de Hitchin. 
Ensuite, ils observent que, pour chaque point de Weierstra\ss, la conique
\[
\CC_\ell=\big\{\C^*\phi\in\PP\theta^\bot\ \big|\ 
	h_{p_2}(\theta,\phi)\big|_{a_\ell}=0\big\}
\]
est la r\'eunion de deux bitangentes \`a $\CK^*_{\theta}$.
Les \'equations des bitangentes \`a la surface de Kummer sont
connues depuis bien longtemps. On peut
alors obtenir l'application de Hitchin modulo un facteur
multiplicatif ne d\'ependant que de $\theta$. Les derni\`eres
\'etapes sont celles d\'ecrites dans l'article reproduit
ci-apr\`es. On montre que l'application de Hitchin poss\`ede
une propri\'et\'e d'{\bol auto-dualit\'e}~:
\[
h_{p_2}(\iota(\phi),\iota^{-1}(\theta))=h_{p_2}(\theta,\phi),
\]
\ie $h_{p_2}(p,q)=h_{p_2}(q,p)$. Cette propri\'et\'e n'est pas
du tout \'evidente quand on regarde l'espace des modules.
Maintenant, on sait que le facteur multiplicatif est une constante.
Celle-ci est fix\'ee par un calcul assez triste de $h_{p_2}$
en des points sp\'eciaux de $\CK\times\CK^*$.

La propri\'et\'e d'involution des Hamiltoniens de Hitchin $h^{(2)}_\ell$ est
\'equivalente aux \'equations de type Yang-Baxter
\[
\Big\{\frac{r_{\ell\ell'}}{a_\ell-a_{\ell'}},
	\frac{r_{\ell'\ell''}}{a_{\ell'}-a_{\ell''}}\Big\}
	+\mbox{per. cycl.}=0,\qquad
\Big\{\frac{r_{\ell\ell'}}{a_\ell-a_{\ell'}},
        \frac{r_{nn'}}{a_{n}-a_{n'}}\Big\}
	=0
\]
avec $\{\ell,\ell'\}\cap\{n,n'\}=\emptyset$. Les r\'esultats de~\cite{VGDJ}
sur la connexion de KZB --- avec les m\^emes hypoth\`eses de d\'epart ---
permet de voir que les Hamiltoniens $h^{(2)}_\ell$ sont une version modifi\'ee
du {\bol syst\`eme de Neumann}~\cite{AT,neumann} dont on peut
trouver la trace dans la Jacobienne des courbes hyperelliptiques~\cite{mumford2}.
L'espace des phases $T^*\PP^3$ peut \^etre identifi\'e avec l'orbite
coadjointe $\CO$ du groupe complexe ${\rm SL}_4$ compos\'ee des matrices
de rang $1$ sans trace $|q\rangle\langle p|$. En utilisant le rev\^etement double
${\rm SL}_4\rightarrow {\rm SO}_6$ et l'identification $\mathfrak{sl}_4
\cong\mathfrak{so}_6$, l'orbite $\CO$ devient l'orbite compos\'ee
des matrices antisym\'etriques $J=(J_{\ell\ell'})$ de rang $2$ et de carr\'e
nul~: $J^2=0$. Ces matrices sont de la forme $J_{\ell\ell'}=
Q_\ell P_{\ell'}-P_\ell\, Q_{\ell'}$ avec $P,Q$ vecteurs de $\C^6$ engendrant
un sous-espace isotrope~: $Q^2=Q\cdot P=P^2=0$. La forme symplectique est
$dP\wedge dQ$. Les fonctions $J_{\ell\ell'}$ sur $\CO$ ont pour crochets
de Poisson
\begin{equation}
\label{sosix}
\begin{cases}
\big\{J_{\ell\ell'},J_{\ell'\ell''}\big\}=-J_{\ell\ell''}
	&\mbox{si $\ell,\ell',\ell''$ sont diff\'erents},\\
\big\{J_{\ell\ell'},J_{nn'}\big\} =0
	&\mbox{si $\ell,\ell',n,n'$ sont diff\'erents}.
\end{cases}
\end{equation}
Les formules donnant le passage de $p,q$ \`a $P,Q$ sont explicit\'ees
dans~\cite{gawtran}. On trouve
\begin{equation}
\label{hamiHIT}
r_{\ell\ell'}=-\quotient{1}{4}\,J_{\ell\ell'}^2,\qquad
h_\ell^{(2)}(\un a)=-\quotient{1}{4}\sum_{m\neq\ell}
	\frac{J^2_{\ell m}}{a_\ell-a_m}.
\end{equation}
L'involution des Hamiltoniens $h^{(2)}_\ell$ est une cons\'equence directe 
des relations alg\'ebriques dans $\mathfrak{so}_6$. Le syst\`eme
de Neumann bien que tr\`es proche est diff\'erent du 
syst\`eme de Hitchin. Il prend en compte
l'orbite coadjointe de ${\rm SO}_N$ form\'ee des matrices antisym\'etriques 
de rang $2$ et de carr\'e non-nul. Contrairement \`a nos orbites,
les orbites y ont des formes r\'eelles non-triviales.

On peut adapter la m\'ethode de Lax sur le syst\`eme de Neumann~\cite{AT}
pour traiter notre syst\`eme de Hitchin. La matrice de Lax
$L(\lambda)=(L(\lambda)_{\ell\ell'})$ est donn\'ee par
\[
L(\lambda)_{\ell\ell'}=\lambda\,J_{\ell\ell'}+a_\ell\,\delta_{\ell\ell'}.
\]
Les crochets de Poisson prennent la forme
\[
\big\{L(\lambda)_1,\, L(\mu)\big\}_2
	=\big[L(\lambda)_1,r^-(\lambda,\mu)\big]
	-\big[L(\mu)_2,r^+(\lambda,\mu)]
\]
o\`u les matrices $r$ sont 
\[
r^\pm(\lambda,\mu)=\quotient{\lambda\mu}{\lambda+\mu}\,C\pm
	\quotient{\lambda\mu}{\lambda-\mu}\,T,
\]
avec $T(v_\ell\otimes v_{\ell'})=v_{\ell'}\otimes v_\ell$
et $C(v_\ell\otimes v_{\ell'})=\delta_{\ell\ell'}\sum_kv_k\otimes v_k$
pour une base orthonormale de $\C^6$, et
$r^\pm$ satisfont l'EYBC. La connaissance d'une matrice de Lax permet
de faire beaucoup de choses. Dans~\cite{gawtran}, on utilise
celle-ci pour trouver les variables d'angles. La courbe
spectrale $\CS$ y est une courbe hyperelliptique de genre $3$. Les solutions
du probl\`eme aux vecteurs propres pour la matrice de Lax,
sont cod\'ees dans un sous-fibr\'e holomorhe en droites $L$  de degr\'e $4$
du fibr\'e de rang $2$ $W=\C^2\otimes\CO(4\infty_1
+4\infty_2)$ au-dessus de $\CS$ --- $\infty_1$ et $\infty_2$ sont
les points au-dessus de l'infini dans la courbe hyperelliptique
$\CS$. En plongeant $L$ dans la Jacobienne $J^4(\CS)$ de $\CS$ par
l'application d'Abel, on d\'eduit que les Hamiltoniens $h^{(2)}_\ell$
engendrent un flot constant sur $J^4(\CS)$. Ainsi, les {\bol variables 
d'angle} sont les coordonn\'ees sur la Jacobienne de $\CS$ et
les {\bol variables d'action} sont les Hamiltoniens $h^{(2)}_\ell$.
Dans le langage originel de Hitchin, la courbe spectrale
est la courbe $\CS'$ d'\'equation
\[
y^2=(x-a_1)\cdots(x-a_6),\qquad
\sigma^2=-4\sum_{\ell,\ell'\atop
	\ell\neq\ell'}^6
	J^2_{\ell\ell'}\,\prod_{m\neq\ell,\ell'}(x-a_m)
\]
et de genre $5$. C'est un rev\^etement double de $\Sigma$. La construction
g\'en\'erale donnerait pour les variables d'angle des points
dans une vari\'et\'e Prym tridimensionnelle, dans la Jacobienne
de dimension $5$ des fibr\'es en droites de degr\'e $-2$ sur $\CS'$.

Comme en genre z\'ero et un, on peut coupler $h_{p_2}$ avec 
une diff\'erentielle de Beltrami. Le changement de structure
complexe $\delta\mu$ bouge les points de ramification $a_\ell$.
Soit $\omega'{}^a=\omega^a+\delta\omega^a+\bar\delta\omega^a$
repr\'esentant la nouvelle base de $H^0(K)$ avec $\delta\omega^a$
une $(0,1)$-forme et $\bar\delta\omega^a$ une $(1,0)$-forme.
Pour que $\omega'{}^a$ soit bien une $(1,0)$-forme
pour la nouvelle structure complexe et pour qu'elle
soit bien ferm\'ee, on doit n\'ecessairement avoir
\[
\bar\delta\omega^a=\omega^a\delta\mu,\qquad
	\de\delta\omega^a=-\partial(\omega^a\delta\mu).
\]
L'\'equation sur $\delta\omega^a$ a toujours des solutions. Les solutions sont 
obtenues modulo les formes ab\'eliennes. Pour fixer cette libert\'e
on demande que $\int_{a_\alpha}\omega'{}^\beta=\delta^{\alpha\beta}$.
On peut r\'ealiser $\Sigma$ comme un rev\^etement de $\C P^1$ par
l'application $\Sigma\ni \xi\mapsto x(\xi)= \omega^2(\xi)\big/\omega^1(\xi)$.
Le nouveau rev\^etement est
\[
x'(\xi)=\frac{\omega'{}^2}{\omega'{}^1}(\xi)
	=x(\xi)+\frac{\delta\omega^2}{\omega^1}(\xi)
	-x(\xi)\,\frac{\delta\omega^1}{\omega^1}(\xi).
\]
On obtient les points de ramification $\zeta'_\ell$ de $x'$ en 
cherchant les solutions de l'\'equation $\partial'x'(\zeta_\ell')=0$.
En utilisant $\zeta'_\ell=\zeta_\ell+\delta \zeta_\ell$, on obtient
\[
\partial^2x(\zeta_\ell)\,\delta \zeta_\ell+
	\partial\Big(\frac{\delta\omega^2}{\omega^1}\Big)(\zeta_\ell)
	-\partial\Big(x\,\frac{\delta\omega^1}{\omega^1}\Big)(\zeta_\ell)=0.
\]
Comme $\de x(\zeta_\ell)=0$ la forme 
quadratique $\partial^2x(\zeta_\ell)$ est bien d\'efinie. Les
points de ramifications sont isol\'es et $\partial^2x(\zeta_\ell)\neq 0$.
On peut donc r\'esoudre l'\'equation pr\'ec\'edente pour $\delta\zeta_\ell$.
Les points de ramification $\zeta'_\ell$ sont envoy\'es
sur
\[
a'_\ell=x(\zeta_\ell+\delta\zeta_\ell)
	=a_\ell+\frac{\delta\omega^2}{\omega^1}(\zeta_\ell)
	-a_\ell\,\frac{\delta\omega^1}{\omega^1}(\zeta_\ell).
\]
Il suit 
\[
\delta a _\ell=a'_\ell-a_\ell=\frac{\delta\omega^2}{\omega^1}(\zeta_\ell)
        -a_\ell\,\frac{\delta\omega^1}{\omega^1}(\zeta_\ell).
\]
On est en mesure de coupler $h_{p_2}$ avec une diff\'erentielle
de Beltrami $\delta\mu$. D\'ej\`a, de ce qui pr\'ec\`ede, on trouve
\begin{align*}
\de\delta x=\de(x'-x)=\frac{\de\delta\omega^2}{\omega^1}
        &-x\,\frac{\de\delta\omega^1}{\omega^1}
	=-\frac{\da(\omega^2\,\delta\mu)}{\omega^1}
	+x\,\frac{\da(\omega^1\,\delta\mu)}{\omega^1}\\
	&=-\frac{\de(x\,\omega^1\,\delta\mu)}{\omega^1}+
	x\,\frac{\da(\omega^1\,\delta\mu)}{\omega^1}
	=dx\,\delta\mu.
\end{align*}
On int\`egre alors $h_{p_2}$ contre $\delta\mu$ sur $\Sigma$ priv\'e
de petites boules autour des points de ramification et de
deux points \`a l'infini, on trouve
\[
h_{\delta\mu}=\int_\Sigma h_{p_2}\,\delta\mu=
	\quotient{1}{\pi i}\sum_{\ell=1}^6h^{(2)}_\ell\,\delta a_\ell.
\]
La diff\'erence dans le facteur multiplicatif provient du rev\^etement
double.

Le fibr\'e d\'eterminant $\CL$ au-dessus de l'espace des modules $\CN_{ss}\cong
\PP\Theta_2$ s'identifie \`a l'espace dual au fibr\'e tautologique 
au-dessus de $\PP\Theta_2$. Ainsi $H^0(\CL^k)$ est l'espace
des polyn\^omes homog\`enes $\psi$ de degr\'e $k$ au-dessus de $\Theta_2$.
L'alg\`ebre $\mathfrak{so}_6$ agit sur $\Theta_2$ par les op\'erateurs
diff\'erentiels, toujours not\'es $J_{\ell\ell'}$, satisfaisant les
relations de commutations~\eqref{sosix} o\`u crochets de Poisson sont
remplac\'es par commutateurs. Dans le langage de $(p,q)$,
les op\'erateurs sont obtenus apr\`es le remplacement des $p_\ell=p_{\delta_\ell}$ 
dans $J_{\ell\ell'}$ par $-\partial_{a_\ell}$ --- on ne l'a pas \'ecrit
mais on peut obtenir les \'el\'ements de $J$ en fonction des $p_\ell$~\cite{gawtran}.
La  connexion de KZB a \'et\'e obtenue indirectement par van Geemen
et de Jong~\cite{VGDJ}. On trouve qu'\`a une $1$-forme pr\`es ---
van Geemen et de Jong d\'efinissent seulement la classe projective de connexions ---
\begin{align*}
\KZmu\,\psi &=\sum_{\ell=1}^6\delta a_\ell\big(\partial_{a_\ell}
	-\quotient{1}{\kappa}\,H^{(2)}_\ell\big)\psi,\\
\KZmubar\,\psi &= \sum_\ell\delta\bar a_\ell\,\partial_{\bar a_\ell}\psi
\end{align*}
o\`u
\[
H^{(2)}_\ell(\un a)=-\quotient{1}{2}\sum_{m\neq\ell}
	\frac{J_{\ell m}^2}{a_\ell-a_m}.
\]
On peut alors retrouver $\frac{1}{\kappa}\,H^{(2)}_\ell$ en quantifiant
le Hamiltonien $\frac{2k}{\pi}\,\frac{1}{\pi i}\,
h^{(2)}_\ell$ par le remplacement $J_{\ell m}\mapsto\frac{\sqrt{i}\pi}{k}\,J_{\ell m}$.
Les Hamiltoniens quantiques sont des op\'erateurs diff\'erentiels
d'ordre $2$ sur $\Theta_2\cong\C^4$. La connexion $\Bnabla$
est projectivement plate.

\newpage


\def\tilde{\widetilde}
\def\hat{\widehat}

\renewcommand{\theequation}{\arabic{section}.\arabic{equation}}

\renewcommand{\bar}{\overline}
\newcommand{\be}{\begin{eqnarray}}
\newcommand{\en}{\end{eqnarray}\vs 0.5 cm}
\newcommand{\no}{\noindent}
\newcommand{\vs}{\vskip}
\newcommand{\hs}{\hspace}
\newcommand{\p}{\partial}

\renewcommand{\NC}{{\mathbb{C}}}
\renewcommand{\NZ}{{\mathbb{Z}}}
\renewcommand{\NP}{{\mathbb{P}}}
\newcommand{\Ng}{{\mathfrak{g}}}
\newcommand{\Ker}{{\rm Ker}}
\newcommand{\llambda}{\mbox{\boldmath $\lambda$}}
\newcommand{\aalpha}{\mbox{\boldmath $\alpha$}}
\newcommand{\xx}{\mbox{\boldmath $x$}}
\newcommand{\xxi}{\mbox{\boldmath $\xi$}}
\newcommand{\kk}{\mbox{\boldmath $k$}}
\newcommand{\Lie}{\hbox{Lie}}
\newcommand{\w}{{\rm w}}
\newcommand{\s}{\hspace{0.05cm}}
\newcommand{\m}{\hspace{0.025cm}}
\newcommand{\ch}{{\rm ch}}
\newcommand{\ra}{{\rightarrow}}
\newcommand{\mt}{{\mapsto}}
\newcommand{\hf}{{_1\over^2}}
\newcommand{\Ssl}{{\mathfrak{sl}}}
\newcommand{\Sso}{{\mathfrak{so}}}
\newcommand{\Di}{{\slash\hs{-0.21cm}\partial}}
\newcommand{\bx}{{\bf x}}
\newcommand{\SL}{{\rm SL}}
\newcommand{\SO}{{\rm SO}}

\HRule

\addcontentsline{toc}{section}{\protect\numberline{}{\bf Self-duality of
        the ${\rm SL}_2$ Hitchin integrable system at genus 2}}

\begin{center}
{\Large \bf Self-duality of the ${\rm SL}_2$ Hitchin integrable\\
        system at genus 2}\\

\vskip 1cm

{\sc Krzysztof Gaw\c{e}dzki}\\
I.H.E.S., C.N.R.S., F-91440 Bures-sur-Yvette, France\\

\vskip 0.5 cm

{\sc Pascal Tran-Ngoc-Bich}\\
Universit\'e de Paris-Sud, F-91405 Orsay, France

\vskip 1cm

\begin{quote}
{\bf Abstract.} We revisit the Hitchin integrable system
\cite{hitchin}\cite{VGP} whose phase space is the bundle cotangent 
to the moduli space $\mathcal{N}$ of holomorphic ${\rm SL}_2$-bundles 
over a smooth complex curve of 
genus 2. As shown in \cite{nar-ram}, $\m\CN$ 
may be identified with the 3-dimensional projective space of theta 
functions of the 2$^{\rm nd}$ order, i.e. $\CN\cong \NP^{3}$. 
We prove that the Hitchin system on $T^*\CN\cong T^*\NP^{3}$ 
possesses a remarkable symmetry: it is invariant under 
the interchange of positions and momenta. This property allows 
to complete the work of van Geemen-Previato \cite{VGP} which,
basing on the classical results on geometry of the Kummer quartic 
surfaces, specified the explicit form of the Hamiltonians
of the Hitchin system. The resulting integrable system resembles 
the classic Neumann systems which are also self-dual. 
Its quantization produces a commuting family of differential operators 
of the 2$^{\rm nd}$ order acting on homogeneous polynomials in four 
complex variables. As recently shown by van Geemen-de Jong 
\cite{VGDJ}, these operators realize 
the Knizhnik-Zamolodchikov-Bernard-Hitchin
connection for group ${\rm SU}_2$ and genus 2 curves.
\end{quote}
\end{center}

\medskip
\section{Introduction}

In \cite{hitchin}, Nigel Hitchin has discovered an interesting 
family of classical integrable models related to modular
geometry of holomorphic vector bundles or to 2-dimensional
gauge fields. The input data for Hitchin's construction
are a complex Lie group $G$ and a complex  curve $\Sigma$ 
of genus $\gamma$. The configuration space of the integrable 
system is the moduli space $\CN$ of (semi)stable holomorphic 
$G$-bundles over $\Sigma$. This is a finite-dimensional 
complex variety and Hitchin's construction is done in the holomorphic
category. It exhibits a complete family of Poisson-commuting 
Hamiltonians on the (complex) phase space $T^*\CN$.  
The Hitchin Hamiltonians have open subsets of abelian varieties 
as generic level sets on which they induce additive
flows \cite{hitchin}. More recently, Hitchin's construction 
was extended to the case of singular or punctured curves 
\cite{Mark}\cite{Nekr}\cite{EnR} providing a unified construction 
of a vast family of classical integrable systems. For $\Sigma=\NC P^1$ 
with punctures, one obtains this way the so called Gaudin chains
and for $G={\rm SL}_N$ and $\Sigma$ of genus 1 with one puncture,
the elliptic Calogero-Sutherland models which
found an unexpected application in the 
supersymmetric 4-dimensional gauge theories \cite{DonaW}.

In Section 2 of the present paper we briefly
recall the basic idea of Hitchin's construction. The main  
aim of this contribution is to treat in detail the
case of $G={\rm SL}_2$ and $\Sigma$ of genus 2 (no punctures).
The genus 2 curves are hyperelliptic, i.e.\s\s given
by the equation
\qq
\zeta^2\s=\s\prod\limits_{s=1}^6(\lambda-\lambda_s)
\label{hellp}
\qqq
where $\lambda_s$ are 6 different complex numbers.
The semistable moduli space $\CN$ has a particularly simple form 
for genus 2, \cite{nar-ram}: it is the
projectivized space of theta functions of the 2$^{\rm nd}$ 
order:
\qq
\CN\ =\ \NP H^0(L_{\Theta}^2)
\label{be}
\qqq
where $L_\Theta$ is the theta-bundle over the Jacobian $J^1$
of (the isomorphism classes of) degree 
$\gamma-1=1$ line bundles\footnote{we use
the multiplicative notation for the tensor product of line 
bundles} $l$ over $\Sigma$. \s${\rm dim}_\C H^0(L_\Theta^2)=4$
so that $\CN\cong\NP^3$. This picture of $\CN$ is related 
to the realization of ${\rm SL}_2$-bundles as extensions 
of degree 1 line bundles. We review some 
of the results in this direction in Section 3 using 
a less sophisticated language than that of the original
work \cite{nar-ram}. The relation between the extensions 
and the theta functions is lifted to the level of the cotangent 
bundle $T^*\CN$ in Section 4. The language of extensions 
proves suitable for a direct description of the Hitchin Hamiltonians 
on $T^*\CN$. The main aim is, however, to present the Hitchin system 
as an explicit 3-dimensional family of integrable systems 
on $T^*\NP^3$, parametrized by the moduli of the curve.
This was first attempted, and almost achieved, in reference 
\cite{VGP}.

Let us recall that the Hitchin Hamiltonians 
are components of the map
\qq
\CH\m:\ T^*\CN\ \longrightarrow\ H^0(K^2)
\label{Hm}
\qqq
with values in the (holomorphic) quadratic differentials
($K$ denotes the canonical bundle of $\Sigma$).
Due to relation (\ref{be}), the map $\CH$ may be viewed as 
a $H^0(K^2)$-valued function of pairs $(\theta,\phi)$ where 
$\theta\in H^0(L_{\Theta}^2)$ and $\phi$ from the dual space
$H^0(L_\Theta^2)^*$ are s.t. ${\langle\m}\theta,
\m\phi{\m\rangle}=0$. Fix a holomorphic trivialization 
of $L_\Theta$ around $l\in J^1$ and denote by $\phi_l$ 
the linear form that computes the value of the theta function 
at $l$. As was observed in \cite{VGP}, 
\qq
\CH(\theta,\m\phi_l)\ =\ -\m{_1\over^{16\pi^2}}
\s(d\theta(l))^2
\label{onK}
\qqq
(with appropriate normalizations). In the above   
formula, $\theta$ is viewed as a function
on $J^1$ and $d\theta(l)$ as an element of $H^0(K)$.
Since $\theta(l)=0$, the equation is consistent with 
changes of the trivialization of $L_\Theta$.

The map $J^1\ni l\mapsto\phi_l$ induces
an embedding of the Kummer surface $J^1/\NZ_2$
with $l$ and $l^{-1}K$ identified into a quartic
$\CK^*$ in $\NP H^0(L_\Theta^2)^*$. The Kummer quartic
is a carrier of a rich but classical structure, a subject 
of an intensive study of the nineteenth century geometers, 
see \cite{Kummer} and also the last chapter of \cite{griffiths}. 
The reference \cite{VGP} used the relation (\ref{onK}) and 
a mixture of the classical results and of more modern
algebraic geometry to recover an explicit form of the components
of the Hitchin map $\CH$ up to a multiplication by a function on 
the configuration space. The authors of \cite{VGP} checked
that the simplest way to fix this ambiguity leads 
to Poisson-commuting functions but they fell short of showing 
that the latter coincide with the ones of the Hitchin 
construction.

Among the aims of the present paper is to fill the 
gap left in \cite{VGP}. We observe that the proposal 
of \cite{VGP} has a remarkable {\bf self-duality} property: 
it is invariant under the interchange of the positions 
and momenta in $T^*\NP^3$. We show that the Hitchin 
construction leads to a system with the same symmetry. This limits
the ambiguity left by the analysis of \cite{VGP} to a multiplication
of the components of $\CH$ by constants. A direct check based on
Eq.\s\s(\ref{onK}) fixes the normalizations and results 
in a formula for the Hitchin map which uses the hyperelliptic 
description (\ref{hellp}) of the curve. Namely,
\qq 
\CH\ =\ -\m{_1\over^{128\m\pi^2}}\s\sum\limits_{1\leq s\not=t\leq 6}
{r_{st}\over(\lambda-\lambda_s)(\lambda-\lambda_t)}\,(d\lambda)^2
\label{GR}
\qqq
where $r_{st}$ are explicit polynomials in $(\theta,\phi)$
given, upon representation of $(\theta,\phi)$ by
pairs $(q,p)\in\NC^4\times\NC^4$, by Eqs.\s\s(\ref{rs})
below. The above expression for $\CH$ has a similar form as that  
for the Hitchin map on the Riemann sphere with 
6 insertion points $\lambda_s$, see e.g. Sect. 4 of \cite{gaw97:unitarity}, 
except for the structure of the terms $r_{st}$. This is not 
an accident but is connected to the reduction of conformal field 
theory on genus 2 surfaces to an orbifold theory in genus 0
\cite{Knizh}\cite{Zamo}. We plan to return to this relation 
in a future publication.

Let us discuss in more details how we establish the 
self-duality of the Hitchin Hamiltonians. The main tool 
here is an explicit expression for the values of the 
Hitchin map off the Kummer quartic $\CK^*$ which
we obtain in Section 5. Our formula 
for $\CH(\theta,\phi)$ requires a choice 
of a pair of perpendicular 2-dimensional
subspaces $(\Pi,\Pi^\perp)$ where $\theta\in\Pi\subset
H^0(L_\Theta^2)$ and $\phi\in\Pi^\perp\subset H^0(L_\Theta^2)^*$
(there is a complex line of such choices). 
The plane $\Pi^\perp$ corresponds to a line $\NP\Pi^\perp$ 
in $\NP H^0(L_\Theta^2)^*$ which intersects the Kummer quartic 
$\CK^*$ in four points $\NC^*\phi_{l_j},\ j=1,2,3,4,$ (counting
with multiplicity). Whereas the analysis of \cite{VGP}
was mainly concerned with the geometry of bitangents
to $\CK^*$ with two pairs of coincident $\phi_{l_j}$'s,
we concentrate on the generic situation with $\phi_{l_j}$'s 
different. Then any two of them, say $\NC^*\phi_{l_1}$ 
and $\NC^*\phi_{l_2}$, span $\Pi^\perp$. $\Pi$
is composed of the 2$^{\rm nd}$ order theta functions vanishing 
at $l_1$ and $l_2$. In particular, 
\qq
\phi\s=\s a_1\m\phi_{l_1}\m+\m a_2\m\phi_{l_2}\quad
{\rm and}\quad\theta(l_1)\s=\s0\s=\s\theta(l_2)\s.
\label{ss0}
\qqq
Let $x_1+x_2$ and $x_3+x_4$ be the divisors of $l_1l_2$ and
of $l_1l_2^{-1}K$, respectively, where $x_i$ are four 
points\footnote{the other two lines of intersection 
of $\NP\Pi^\perp$ with $\CK^*$ correspond 
to $l_3$ and $l_4$ with $l_1l_3=\CO(x_1+x_3)$,
$l_1l_3^{-1}K=\CO(x_2+x_4)$, $l_1l_4=\CO(x_1+x_4)$,
$l_1l_4^{-1}K=\CO(x_2+x_3)$} in $\Sigma$.
If $l_1^2\not=K$, which holds in a general situation,
then the quadratic differential $\CH(\theta,\m\phi)$
is determined by its values at $x_i$ which, as we show
in Section 5, are
\qq
\CH(\theta,\m\phi)(x_i)\ = \ 
-\m {_1\over^{16\pi^2}}\s\left(a_1\s d\theta(l_1)\pm a_2\s 
d\theta(l_2)\right)^2(x_i)\s.
\label{fr}
\qqq
Sign plus is taken for $x_1$ and $x_2$
and sign minus for $x_3$ and $x_4$. Note that for $\phi=\phi_l$ 
with $\theta(l)=0$ the above equation reproduces
the result (\ref{onK}).

As we recall at the end of Section 3, there exists an almost natural 
linear isomorphism $\iota$ between $H^0(L_\Theta^2)^*$ 
and $H^0(L_\Theta^2)$. What follows is independent of the remaining 
ambiguity in the choice of $\iota$. The identity 
$\langle\m\theta,\phi\m\rangle=\langle\m
\iota(\phi),\m\iota^{-1}(\theta)\m\rangle$ implies that
if $(\theta,\phi)$ is a perpendicular pair then so
is $(\theta',\m\phi')$ where $\theta'=\iota(\phi)$ and
$\phi'=\iota^{-1}(\theta)$.  Thus $\iota$ 
interchanges the positions and momenta in $T^*\CN$. 
We may take $(\Pi',\m{\Pi'}^\perp) 
=(\iota(\Pi^\perp),\m\iota^{-1}(\Pi))$ as a pair of perpendicular 
subspaces containing $(\theta',\phi')$. The line 
$\NP{\Pi'}^\perp$ meets $\CK^*$ in four points 
$\NC^*\phi_{l'_j}$. Equivalently, 
$\NC^*\iota(\phi_{l'_j})$ are the points of intersection of $\NP\Pi$ 
with the Kummer quartic $\CK=\iota(\CK^*)\subset\NP H^0(L_\Theta^2)$. 
In general situation, ${\Pi'}^\perp$ is spanned by any pair 
of $\phi_{l'_j}\m$'s so that 
\qq
\phi'\s=\s a'_1\m\phi_{l'_1}\m+\m a'_2\m\phi_{l'_2}\quad{\rm and}
\quad\theta'(l'_1)\s=\s0\s=\s\theta'(l'_2)
\label{ss2}
\qqq
which is the dual version of relations (\ref{ss0}).
Equivalently,
\qq
\theta\s=\s a'_1\s\iota(\phi_{l'_1})\s+\s a'_2\s\iota(\phi_{l'_2})
\quad{\rm and}\quad{\langle\m}\iota(\phi_{l'_1}),
\s\phi{\m\rangle}\s=\s0\s=\s
{\langle\m}\iota(\phi_{l'_2}),\s\phi{\m\rangle}\s.
\label{ss1}
\qqq
Let $y_i$ be the points associated 
to $l'_j$ the same way as the points $x_i$ were associated 
to $l_j$.  \s$l'_j$ may be chosen so that $y_i$ and $x_i$ coincide 
modulo the natural involution of $\Sigma$ fixing the six 
Weierstrass points. Formula (\ref{fr}) implies then that
\qq
\CH(\theta',\m\phi')(y_i)\ = \ 
-\m {_1\over^{16\pi^2}}\s\left(a'_1\s d\theta'(l'_1)\pm a'_2\s 
d\theta'(l'_2)\right)^2(y_i)\s.
\label{frd}
\qqq
Points $y_i$ in Eq.\s\s(\ref{frd}) may be replaced by $x_i$ 
since the quadratic differentials are equal at point $x$ 
if and only if they are equal at the image of $x$ by the
involution of $\Sigma$. A direct calculation of 
the coefficients $a_1,\s a_2$ and $a'_1,\s a'_2$ appearing 
on the right hand sides of Eqs.\s\s(\ref{fr})
and (\ref{frd}) shows then that both expressions coincide,
establishing the self-duality of $\CH$. The verification 
of this equality is the subject of Section 6.

In Section 7, we recall the main result of reference \cite{VGP}
and show how the self-duality may be used to complete
the analysis performed there and to obtain the explicit form 
(\ref{GR}) of the Hitchin map. We briefly discuss the relation
of that form to the classical Yang-Baxter equation.

An appropriate quantization of Hitchin Hamiltonians leads to operators 
acting on holomorphic sections of powers of the determinant line bundle 
over $\CN$ and defining the Knizhnik-Zamolodchikov-Bernard-Hitchin
\cite{kz}\cite{bernard:kzb}\cite{bernard:kzbanyg}\cite{hitchin:flat} connection.
In our case, the sections of the powers
of the determinant bundle are simply homogeneous
polynomials on $H^0(L_\Theta^2)$. It is easy to quantize
the Hamiltonians corresponding to the components of 
the Hitchin map (\ref{GR}) in such a way that one obtains
an explicit family of commuting 2$^{\rm nd}$ order differential
operators acting on such polynomials. The corresponding connection 
coincides with the explicit form of the (projective) KZBH connection
worked out recently\footnote{we thank B. van Geemen for attracting
our attention to ref. \cite{VGDJ} and for pointing out that this work
may be used to fix indirectly the precise form of the Hitchin map}
in \cite{VGDJ}. 

The quantization of the genus 2 Hitchin system
is briefly discussed in Conclusions, where we also mention other possible
directions for further research. Four appendices which close the paper 
contain some of more technical material.

We would like to end the presentation of our paper
by expressing some regrets. We apologize to Ernst Eduard Kummer 
and other nineteenth century giants for our insufficient 
knowledge of their classic work. The apologies
are also due to few contemporary algebraic geometers who could
be interested in the present work for an analytic character
of our arguments. To the specialist in integrability
we apologize for the yet incomplete analysis of the integrable 
system studied here and, finally, we apologize to ourselves 
for not having finished this work 2 years ago.

\setcounter{equation}{0}
\medskip
\section{Hitchin's construction}

Let us assume, for simplicity, that the complex Lie group $G$ is simple,
connected and simply connected. We shall denote by $\Ng$
its Lie algebra. The complex curve $\Sigma$ will be 
assumed smooth, compact and connected. Topologically,
all $G$-bundles on $\Sigma$ are trivial and the complex 
structures in the trivial bundle may be described by giving
operators $\de+A$ where $A$ are
smooth $\Ng$-valued 0,1-forms on $\Sigma$ \cite{atiyahbott}.
Let $\CA$ denote the space of such forms 
(i.e. of chiral gauge fields). The group $\CG$ of 
local (chiral) gauge transformations composed of smooth maps 
$h$ from $\Sigma$ to $G$ acts on operators $\de+A$ 
by conjugation and on the gauge fields $A$ by
\qq
A\ \longmapsto\ {}^h\hspace{-0.14cm}A\s\equiv\s 
h\s A\s h^{-1}\s+\s h\s \de h^{-1}\s.
\nonumber
\qqq
Two holomorphic $G$-bundles are equivalent iff the
corresponding gauge fields are in the same orbit of $\CG$.
Hence the space of orbits $\CA/\CG$ coincides with
the (moduli) space of inequivalent holomorphic $G$-bundles.
It may be supplied with a structure of a variety provided one
gets rid of bad orbits. This may be achieved by limiting 
the considerations to (semi)stable 
bundles, i.e. such that the vector bundle associated 
with the adjoint representations of $G$ contains only 
holomorphic subbundles with negative (non-positive) 
first Chern number. For $\gamma>1$, the moduli space 
$\CN_s\equiv\CA_s/\CG$ 
of stable $G$-bundles is a smooth complex variety with 
a natural compactification to a variety $\CN_{ss}$, 
the (Seshadri-)moduli space of semistable bundles \cite{nar-ram}.

The complex cotangent bundle $T^*\CN_s$ may be obtained
from the infinite-dimensional bundle $T^*\CA_s$ by 
the symplectic reduction. $T^*\CA_s$ may be realized as
the space of pairs $(A,\Phi)$ where $\Phi$
is a (possibly distributional) $\Ng$-valued 1,0-form on $\Sigma$,
$A\in\CA_s$ and the duality with the vectors 
$\delta A$ tangent to $\CA$ is given by
\qq
\int_{_\Sigma} \tr\ \Phi\wedge\delta A
\nonumber
\qqq
with $tr$ standing for the Killing form.
The action of the local gauge group $\CG$ on $\CA_s$
lifts to a symplectic action on $T^*\CA_s$ by
\qq
\Phi\ \longmapsto\ {}^h\hs{-0.05cm}\Phi\s\equiv\s 
h\s\Phi\s h^{-1}\ .
\nonumber
\qqq
The moment map $\mu$ for the action of $\CG$ on $T^*\CN_s$
is
\qq
\mu(A,\Phi)\ =\ \de\Phi+A\wedge\Phi
+\Phi\wedge A\ \equiv\ \de_{_{A}}\Phi\ .
\nonumber
\qqq
Note that it takes values in $\Ng$-valued 2-forms on $\Sigma$.
These may be naturally viewed as elements of the space 
dual to the Lie algebra of $\CG$. The symplectic reduction
of $T^*\CA_s$ realizes $T^*\CN_s$ as the
space of $\CG$-orbits in the zero level of $\mu$:
\qq
T^*\CN_s\ \cong\ \mu^{-1}(\{0\})\s/\s\CG\ .
\nonumber
\qqq

For a homogeneous $G$-invariant polynomial $P$ on $\Ng$ of
degree $d_P$, the gauge invariant expression \s$P(\Phi)$ 
defines a section of the bundle $K^{d_P}$ of $d_P$-differentials 
on $\Sigma$. If $\Phi$ is in the zero level of $\mu$
then \s$P(\Phi)$ is also holomorphic. Hence the map 
\s$\Phi\s\mapsto\s P(\Phi)\s$ induces a map 
\qq
\CH_P\s:\ T^*\CN_s\ \longrightarrow\ H^0(K^{d_P})
\nonumber
\qqq
into the finite dimensional vector space of holomorphic 
differentials of degree $d_P$ on $\Sigma$. The components
of such vector-valued Hamiltonians clearly Poisson-commute 
since upstairs (on $T^*\CA_s$) they depend only 
on the momentum variables $\Phi$. By a beautiful
argument, Hitchin showed \cite{hitchin} that taking 
all polynomials $P$ one obtains a complete system 
of Hamiltonians in involution and that the 
collection of maps $\CH_P$ defines in generic 
points a foliation of $T^*\CN_s$ into (open subsets of)
abelian varieties.

Let us briefly sketch Hitchin's argument for $G={\rm SL}_2$.
There is only one (up to normalization) non-trivial 
invariant polynomial $P_2$ on $\Ssl_2$ given by, say, 
half of the Killing
form. $\CH\equiv\CH_{P_2}$ maps into the space of quadratic 
differentials. A non-trivial holomorphic quadratic differential 
$\rho$ determines a (spectral) curve $\Sigma'\subset K$ given 
by the equation
\qq
\xi^2=\rho(\pi(\xi))
\label{1}
\qqq
where \s$\xi\in K$ and $\pi$ is the projection of $K$
on $\Sigma$. The map $\xi\mapsto-\xi$ gives an involution 
$\sigma$ of $\Sigma'$. Restriction of $\pi$ to $\Sigma'$ 
is a 2-fold covering of $\Sigma$ ramified over $4(\gamma-1)$       
points fixed by $\sigma$, the zeros of $\rho$. 
$\Sigma'$ has genus $\gamma'=4\gamma-3$. \m If \m$\rho\m=\m 
\frac{1}{2}\s \tr\s\s(\Phi)^2\m$ then 
relation (\ref{1}) coincides with the eigen-value equation
\qq
{\rm det}\s(\Phi-\s\xi\cdot I)\s=\s 0
\nonumber
\qqq
for the Lax matrix $\Phi$.
Let for each $0\not=\xi\in\Sigma'$, $l_\xi$ denote
the corresponding eigen-subspace of $\Phi$. By continuity,
$l_\xi$ extend to vanishing $\xi$ in $\Sigma'$ and
\s$\cup_{_{\xi}}l_{\xi}$ forms a line subbundle $l$ 
of $\Sigma'\times\NC^2$. In fact, $l$ is a holomorphic subbundle
with respect to the complex structure defined on
$\Sigma'\times\NC^2$ by $\de+A\circ\pi$. The degree
of $l$ is $-2(\gamma-1)$. Besides,
\qq
l\s(\sigma^*l)\s=\s\pi^*K^{-1}\s.
\label{K}
\qqq    
Conversely, given $\Sigma'$ and a holomorphic 
line bundle $l$ of degree $-2(\gamma-1)$ on it satisfying 
(\ref{K}), we may recover a rank 2 holomorphic bundle $E$ 
of trivial determinant over $\Sigma$ as a pushdown of $l$ 
to $\Sigma$. Thus for $0\not=\xi\in\Sigma'$, $E_{\pi(\xi)}
=l_\xi\oplus l_{-\xi}$. $E$ corresponds to a unique
holomorphic ${\rm SL}_2$-bundle which, if stable (what happens 
on an open subset of $l$'s) defines a point in the moduli 
space $\CN_s$. A holomorphic 1,0-form 
with values in the traceless endomorphisms of $E$
acting as multiplication by $\pm\xi$ on 
$l_{\pm\xi}\subset E_{\pi(\xi)}$ defines then
a unique covector of $T^*\CN_s$. Thus $\Sigma'$ encodes
the values of the quadratic Hitchin Hamiltonian $\CH$
(i.e.\s\s of the action variables) whereas the line bundles 
$l$ satisfying relation (\ref{K}) form the abelian (Prym) 
variety (of the angle variables) describing the level set 
of $\CH$.

\setcounter{equation}{0}
\medskip
\section{${\rm SL}_2$ moduli space at genus 2}

We shall present briefly the description
of the moduli space $\CN_s$ for $G={\rm SL}_2$
and $\gamma=2$ which was worked out in \cite{nar-ram}.

Let us start by recalling some basic facts about 
theta functions. We shall use a coordinate rather
than an abstract language. The space of degree $\gamma-1$ 
holomorphic line bundles 
forms a Jacobian torus $J^{\gamma-1}$ of complex dimension
$\gamma$. Fixing a marking (a symplectic homology basis
$(A_a,B_b),\ {a},{b}=1,\dots,\gamma$),
we may identify $J^{\gamma-1}$ with $\NC^\gamma/(\NZ^\gamma
+\tau\NZ^\gamma)$. $\tau\equiv(\tau^{{a}{b}})$ 
is the period matrix,
i.e. $\tau^{{a}{b}}=\int_{B_{{b}}}\omega^{a}$
where $\omega^{a}$ are the basic holomorphic
forms on $\Sigma$ normalized so that $\int_{A_{a}}
\omega^{b}=\delta^{{a}{b}}$. The point $0\in\NC^\gamma$
corresponds in $J^{\gamma-1}$ to a (marking dependent) spin 
structure $S_0$, i.e. a degree 1 bundle such that 
$S_0^2=K$. $u\in\NC^\gamma$ describes the line bundle 
$V(u)S_0$ where $V(u)$ is the flat line bundle 
with the twists $\ee^{2\pi i u^{b}}$ along the $B_{b}$ 
cycles. The set of degree 1 bundles $l$ with non-trivial
holomorphic sections forms a divisor $\Theta$ of a holomorphic
line bundle $L_\Theta$ over $J^{\gamma-1}$. Holomorphic
sections of the $k$-th power ($k>0$) of $L_\Theta$ 
are called theta function of order $k$. With the 
use of a marking, they may be represented by holomorphic 
functions $u\mapsto\theta(u)$ on $\NC^2$ satisfying
\qq
\theta(u+p+\tau q)\s=\s\ee^{-\pi i k\s q\cdot\tau q-2\pi i
k\s q\cdot u}\ \theta(u)
\qqq
for $p,q\in\NZ^\gamma$. The functions
\qq
\theta_{k,e}(u)\s=\s\sum\limits_{n\in{\NZ}^\gamma}
\ee^{\s\pi i k\s
(n+e/k)\cdot\tau(n+e/k)\s+\s 2\pi i k\s(n+e/k)\cdot u}
\qqq
where $e\in\NZ^\gamma/k\NZ^\gamma$ form a basis of
the theta functions of order $k$. Hence ${\rm dim}\s H^0(L_\Theta^k)
=k^\gamma.$ In particular, the Riemann theta function
$\theta_{1,0}(u)\equiv\vartheta(u)$ represents
the unique (up to normalization) non-trivial holomorphic
section of $L_\Theta$. It vanishes on the set
$$\{\sum\limits_{i=1}^{\gamma-1}\smallint_{x_0}^{x_i}\omega-
\Delta\ \vert\ x_1\in\Sigma,\dots,x_{\gamma-1}\in\Sigma\}$$ 
representing the divisor $\Theta$. Here $\Delta\in\NC^\gamma$ 
denotes the ($x_0$-dependent) vector of Riemann constants.
All theta functions of order 1 and 2 are even functions
of $u$.

For $\gamma=2$, the divisor $\Theta$ is formed by the bundles
$\CO(x)$ with divisors $x\in\Sigma$. \m$\CO(x)=
V(\int_{x_0}^x\omega-\Delta)S_0$. The pullback 
of the theta bundle $L_\Theta$ by means of the map
$x\mapsto\CO(x)$ is equivalent to the canonical bundle $K$.
The equivalence assigns 1,0-forms to functions representing
sections of the pullback of $L_\Theta\m$:
\qq
\epsilon^{{a}{b}}\m\da_{b}\vartheta(
\smallint_{x_0}^x\omega-\Delta)\quad\mapsto\quad
\omega^a(x)\s.
\label{ide}
\qqq
This is consistent since vanishing 
of $\vartheta(\int\limits_{x_0}^x\omega-\Delta)$ implies that
\qq
\da_{a}\vartheta(\smallint_{x_0}^x\omega-\Delta)\ \omega^{a}(x)=0\ .
\nonumber
\qqq
Hence any multivalued function on $\Sigma$ picking up a factor 
$\ee^{-\pi i\tau^{{a}{a}}-2\pi i(\int_{x_0}^x\omega^{a}
-\Delta^{a})}$ when $x$ goes around the $B_{a}$
cycle and univalued around the $A_{a}$ cycles may be 
identified with a 1,0-form on $\Sigma$.

As already suggested by the discussion at the end of 
Sect.\s\s 2, for the ${\rm SL}_2$ group it is more convenient to
use the language of holomorphic vector bundles (of rank 2 and
trivial determinant) than to work with principal
${\rm SL}_2$-bundles. Of course the first ones are
just associated to the second ones by the fundamental
representation of $\SL_2$. Any stable rank 2
bundle $E$ with trivial determinant is an extension 
of a degree 1 line bundle $l$
(\cite{nar-ram}, Lemmas 5.5 and 5.8), i.e. it appears
in an exact sequence of holomorphic vector bundles 
\qq
0\m\longrightarrow\m l^{-1}\smash{\mathop{\longrightarrow}^\sigma}
\m E\m \smash{\mathop{\longrightarrow}^\varpi}
\s l\m \longrightarrow\m 0\ .
\label{ES}
\qqq
The inequivalent extensions (\ref{ES}) are classified
by the cohomology classes in $H^1(l^{-2})$. 
This may be seen as follows.
Taking a section of $\varpi$, \m i.e.\s\s a smooth bundle 
homomorphism $s:\s l\rightarrow E$ such that $\varpi\circ s=id_l$,
we infer that $\varpi\s\de s=0$ and hence that $\de s=\sigma\s b$
for $b$ a 0,1-form with values in ${\rm Hom}(l,l^{-1})=l^{-2}$,
\m i.e. $b\in\wedge^{01}(l^{-2})$. $b$ is determined up to 
$\de\varphi$ where $\varphi$ is a smooth section of $l^{-2}$, 
i.e. $\varphi\in\Gamma(l^{-2})$. The class $[b]$ 
in $\wedge^{01}(l^{-2})/\Gamma(l^{-2})\s\cong\s
H^1(l^{-2})$ determines the extension (\ref{ES}) up
to equivalence. Each $b$ corresponds to an extension:
one may simply take $E$ equal to $l^{-1}\oplus l$
with the $\de$-operator given by $\de_{_{l^{-1}\oplus\m l}}+
\s\begin{pmatrix}
0&b\\0&0\end{pmatrix}$. \m Proportional $[b]$
correspond to equivalent bundles $E$. If $E$ is a stable
bundle then the extension (\ref{ES}) is necessarily
nontrivial, i.e. $[b]\not=0$.

Let $C_E$ denote the set of degree 1 line 
bundles $l$ s.t. \s$H^0(l\otimes E)\not=0\s$
(equivalently, s.t. $E$ is an extension of $l$).
This is a complex 1-dimensional variety. It was shown
in \cite{nar-ram} that $C_E$ characterizes the bundle
$E$ up to isomorphism and that there exists a
theta function $\theta$ of the 2$^{\rm nd}$ order which
vanishes exactly on $C_E$. The assignment
\s$E\s\mapsto\s\NC^*\theta\s$ gives an injective map 
\qq
m:\s\CN_s\s\longrightarrow\s\NP H^0(L_\Theta^2)\ .
\label{map}
\qqq
Let $V(u_1)S_0\equiv l_{u_1}\in C_E$. $E$ may be
realized as an extension of $l_{u_1}$ which is characterized
by $[b]\in H^1(l_{u_1}^{-2})$. Then one may take
\qq
\theta(u)\ = \int_{_\Sigma}\s K(x;u_1,u)
\wedge b(x)\ .
\label{thetaE}
\qqq
where
\qq
K(x;u_1,u)\ =\ \vartheta(\smallint_{x_0}^x\omega
-u_1-u-\Delta)\ \vartheta(\smallint_{x_0}^x
\omega-u_1+u-\Delta)\cr
\cdot\left(\epsilon^{{a}{b}}\s\da_{b}\vartheta
(\smallint_{x_0}^x\omega-\Delta)\right)^{-1}
\omega^{a}(x)
\label{Kk}
\qqq 
(it does not depend on the choice of $a=1,2$). \m
Let us explain the above formulae. \s$K(x;u_1,u)\m$, 
\m in its dependence
on \s$x\m$, \m is a multivalued holomorphic 1,0-form.
More exactly, the function
\qq
x\ \mapsto\ \vartheta(\smallint_{x_0}^x\omega-u_1-u-\Delta
)
\label{s_2}
\qqq
is multivalued around the $B_{a}$-cycles picking up the factor
$$\ee^{-\pi i\tau^{{a}{a}}\s-\s2\pi i(\int_{x_0}^x
\omega^{a}-u_1^{a}-u^{a}-\Delta^{a})}$$
when \s$x\s$ goes around \s$B_{a}\s$ so that it describes
an element \s$s_2\in H^0(l_{u_1} l_{u})\s$
(non-vanishing if \s$u_1+u\s\not\in\s\NZ^2+\tau\NZ^2\m$).
\m Similarly,
$$x\ \mapsto\ \vartheta(\smallint_{x_0}^x\omega
-u_1+u-\Delta)\ 
\left(\epsilon^{{a}{b}}\m\da_{b}\vartheta
(\smallint_{x_0}^x\omega-\Delta)\right)^{-1}
\omega^{a}(x)$$
picks up the factor $$\ee^{\s2\pi i(u_1^{a}-u^{a})}$$
when $x$ goes around $B_{a}$ and describes a 
holomorphic 1,0-form $\chi$ with values in $l_{u_1}l_u^{-1}\s$
(non-vanishing if \s$u_1-u\s\not\in\s\NZ^2+\tau\NZ^2\s$). 
\s The product \s$s_2\chi=K(\m\cdot\s;u_1,u)\s$ 
is a holomorphic 1,0-form with values in \s$l_{u_1}^{\m2}\s$ 
and it may be paired with \s$b\in\wedge^{01}(l_{u_1}^{-2})\s$ via
the integral over \s$x\s$ on the r.h.s. of Eq.\s\s(\ref{thetaE}).
The integral is independent of the choice of the representative $b$ 
of the cohomology class $[b]$. In its dependence on $u$,
\m$K(x;u_1,u)$ is a theta 
function of the 2$^{\rm nd}$ order and so is $\theta(u)$. 
In Appendix 1 we check explicitly that $\theta$ given
by Eq.\s\s(\ref{thetaE}) possesses the required property.

The product of the two shifted Riemann 
theta functions \s$\vartheta(u'-u)\s
\vartheta(u'+u)\s$ is a theta function
of the 2$^{\rm nd}$ order both in $u'$ and in $u$
(and it is invariant under the interchange 
$u'\leftrightarrow u$). Let $\iota$ denote the (marking 
dependent) linear isomorphism between the spaces 
$H^0(L_\Theta^2)^*$ and $H^0(L_\Theta^2)$ defined by
\qq
\iota(\phi)(u)\s=\s{\langle\m}\vartheta(\s\cdot\s-u)
\s\m\vartheta(\s\cdot\s+u)\s\m,
\m\s\s\phi{\m\rangle} 
\label{lini}
\qqq
An easy calculation shows that
\qq
\vartheta(u'-u)\s\s\vartheta(u'+u)
\ =\ \sum\limits_{e}\theta_{2,e}(u')\s\s\theta_{2,e}(u)\s.
\qqq
Hence $\iota$ interchanges the basis $(\theta_{2,e})$ 
of $H^0(L_\Theta^2)$ with the dual basis $(\theta_{2,e}^{\s*})$
of $H^0(L_\Theta^2)^*$. Denote by $\phi_{u}$ the linear form 
on $H^0(L_\Theta^2)$ that computes the value of the theta 
function at point $u\in\NC^2$. The Kummer quartic 
$\CK^*\subset H^0(L_\Theta^2)^*$, \s$\CK^*
=\{\m\NC^*\phi_{u'}\s\vert\s u'\in\NC^2\m\}\m$ 
is mapped by the isomorphism $\iota$ into a quartic 
\s$\CK\subset H^0(L_\Theta^2)\s$ of theta functions  
proportional to
\qq
u\ \mapsto\ \vartheta(u'-u)\s\s\vartheta(u'+u)
\nonumber
\qqq
for some $u'\in\NC^2$.

One may define a projective action of $(\NZ/2\NZ)^4$ 
on $H^0(L_\Theta^2)$ by assigning to an element 
$(e,e')\in(\NZ/2\NZ)^4$, with $e,e'=(0,0),\s(1,0),
\s(0,1)$ or $(1,1)$, a linear transformation $U_{e,e'}$ s.t.
\qq
(U_{e,e'}\theta)(u)\ =\ \ee^{\m{1\over2}\m\pi i\s
e'\cdot\m\tau\m e'\s+\s 2\pi i\s e'\cdot\m u}\s\s
\theta(u+\hf(e+\tau e'))\s.
\label{disa}
\qqq
The relation \s$U_{e_1,e_1'}\s U_{e_2,e_2'}\ =\ 
(-1)^{e_1\cdot e_2'}\s U_{e_1+e_2,\m e_1'+e_2'}\m$
holds so that $U$ lifts to the Heisenberg group.
In the action on the basic theta functions,
\qq
U_{e_1,e_1'}\m\theta_{2,e}\ =\ (-1)^{e_1\cdot\m e} 
\s\s\theta_{2,\m e+e_1'}\s.
\label{acone}
\qqq
The marking-dependence of the isomorphism
$\iota$ of Eq.\s\s(\ref{lini}) is given by the action 
of $(\NZ/2\NZ)^4$. It is easy to check that this 
action preserves \s$\CK$ and that the transposed action 
of $(\NZ/2\NZ)^4$ preserves $\CK^*$. The $(\NZ/2\NZ)^4$ 
symmetry of the Kummer quartics allows to find easily 
their defining equation, see Appendix 3.

It was shown in \cite{nar-ram} that the image of $\CN_s$ 
under the map (\ref{map}) contains all non-zero
theta functions of the 2$^{\rm nd}$ order except the
ones in the the Kummer quartic $\CK$. 
The latter correspond, however, to the (Seshadri equivalence 
classes of) semistable but not stable bundles so that the map 
$m$ extends to an isomorphism between $\CN_{ss}$ and 
$\NP H^0(L_\Theta^2)$ showing that $\CN_{ss}$ is a smooth 
projective variety.

\setcounter{equation}{0}
\medskip
\section{Cotangent bundle}

Let us describe the cotangent space of
$\CN_s$ at point $E$. The covectors tangent
to $\CN_s$ at $E$ may be identified with holomorphic
1,0-forms $\Psi$ with values in the bundle of traceless
endomorphisms of $E$. We may assume that $E$ is an 
extension of a line bundle $l$ of degree 1 realized as
$l^{-1}\oplus l$ with $\de_{_E}=\de_{_{l^{-1}\oplus\m l}}
\hs{-0.08cm}+B$ where \s$B=\begin{pmatrix}0&b\\0&0\end{pmatrix}\m$. 
\s Then 
\qq
\Psi\ =\ \begin{pmatrix}-\mu&\nu\\\eta&\mu\end{pmatrix}
\label{Lax}
\qqq
where $\mu\in\wedge^{10}$, $\nu\in\wedge^{10}(l^{-2})$,
$\eta\in\wedge^{10}(l^2)$ and 
\qq
\de_{_{l^2}}\eta=0\s,\hspace{1cm}\de\mu=-\eta\wedge b\s,
\hspace{1cm}\de_{_{l^{-2}}}\nu=2\mu\wedge b\ .
\label{rel}
\qqq
It is easy to relate the above description of covectors
tangent to $\CN_s$ to the one of Sect.\s\s2.
Let $\CU:\s l^{-1}\oplus l\rightarrow\Sigma\times\NC^2$ be
a smooth isomorphism of rank 2 bundles with trivial determinant.
Then $\CU\s\de_{_E}\CU^{-1}=\de+A$ for a certain $\Ssl_2$-valued 0,1-form
$A$ and $\Phi=\CU\Psi\CU^{-1}$ 
satisfies $\de_{_{A}}\Phi=0$. The $\CG$ orbit 
of $(A,\Phi)$ is independent of the choice of $\CU$
and the quadratic Hitchin Hamiltonian takes value 
${1\over 2}\s\tr\s(\Phi)^2$ on it. The latter expression 
is clearly equal to $\frac{1}{2}\s\tr\s(\Psi)^2
=\s\mu^2+\eta\s\nu$ which, as easily follows from relations 
(\ref{rel}), defines a holomorphic quadratic differential. Hence
\qq
\CH(E,\Psi)\s=\s \mu^2+\eta\s\nu\ .
\label{hit}
\qqq
We would like to express the latter using the theta function
description of $T^*\CN_{ss}=T^*\NP H^0(L_{\Theta}^2)$
where the covectors tangent to $\CN_{ss}$ at $\NC^*\theta$ 
are represented by linear forms $\phi$ on $H^0(L_{\Theta}^2)$ 
\m s.t. ${\langle\m}\theta,\m\phi{\m\rangle}=0\m$.

Let $l=l_{u_1}\in C_E$, \m i.e. $\theta(u_1)=0$
for the theta function corresponding to $E$.
We shall assume that $l^2\not=K$ i.e.
that \s$2u_1\s\not\in\s\NZ^2+\tau\NZ^2$. 
\m An infinitesimal variation $\delta E$ of the bundle $E$ 
in $\CN_s$ may be achieved by changing 
\s$\de_{_E}=\de_{_{l^{-1}\oplus\m l}}+B\s$ with \s$B=
\begin{pmatrix}0&b\\0&0\end{pmatrix}\s$ to 
\qq
\de_{_{l^{-1}\oplus\m l}}+\begin{pmatrix}\pi\m\delta 
u_1({\rm Im}\s\tau)^{-1}\bar\omega
&b+\delta b\\0&-\pi\m\delta u_1({\rm Im}\s\tau)^{-1}\bar\omega
\end{pmatrix}
\ \equiv\ \de_{_E}+\s\delta B
\label{DB}
\qqq
(all other variations of $\de_{_E}$ may be obtained from (\ref{DB}) 
by infinitesimal gauge transformations). Clearly
\qq
{\langle\m}\delta E\s,\s\Psi{\m\rangle}\ =\ \int_{_\Sigma}
\tr\s\s\Psi\wedge\delta B\ =\ -\s 2\pi
\delta u_1({\rm Im}\s\tau)^{-1}\int_{_\Sigma}\mu\wedge\bar
\omega\s+\s\int_{_\Sigma}\eta\wedge\delta b\ .
\label{25}
\qqq
Note that the line bundle $l_{u_1}$ with the $\de$-operator 
changed to $\de_{_{l_{u_1}}}\hspace{-0.2cm}
-\pi\m\delta u_1({\rm Im}\s\tau)^{-1}
\bar\omega$ is equivalent to $l_{u_1+\delta u_1}\equiv l'$ and the
equivalence is established by multiplication by the
multivalued function
$x\mapsto\ee^{\s 2\pi i\s\delta u_1({\rm Im}\s\tau)^{-1}
\int_{x_0}^x{\rm Im}\s\omega}$. Hence $l^{-1}\oplus l$
with the $\de$-operator given by Eq.\s\s(\ref{DB}) is equivalent to 
${l'}^{-1}\oplus l'$ with the $\de$-operator $\de_{_{{l'}^{-1}\oplus
\m l'}}+\s\begin{pmatrix}0&{b+\delta'b}\\0&0\end{pmatrix}\s$ where
\s$\delta'b(x)\ =\ \delta b\s -\s 4\pi i\s\delta u_1({\rm Im}
\s\tau)^{-1}(\smallint\limits_{x_0}^x{\rm Im}\s\omega)\s b(x)\m.$
\s The last bundle corresponds by the relation (\ref{thetaE})
to the theta function  
\qq
(\theta+\delta\theta)(u)\ =\ \int_{_\Sigma}
K(x;\m u_1+\delta u_1,\m u)\wedge (b(x)+\delta'b(x))\ .
\nonumber
\qqq
Hence $\delta E$ is represented by the variation 
\qq
\delta\theta(u)\ =\ 
-\ 2\pi\m\delta u_1^{a}\s ({\rm Im}\s\tau)^{-1}_{\s{a}{b}}
\int_{_\Sigma}L^b(x;u_1,u)\wedge b(x)\s
+\s\int_{_\Sigma}K(x;u_1,u)\wedge
\delta b(x)
\label{nd}
\qqq
of the theta function, where
\qq
L^{a}(x;u_1,u)\ =\ K(x;u_1,u)\s
\smallint_{x_0}^x(\omega^{a}-\bar\omega^{a})
\s-\s{_1\over^{2\pi}}{\rm Im}\s\tau^{{a}{b}}
\s\da_{u_1^{b}}K(x;u_1,u)\ .
\label{gal}
\qqq
Note that as functions of $x$, \s$L^a(x;u_1,u)$
are 1,0-forms with values in $l_{u_1}^2$ 
(as are $K(x;u_1,u)$). They are not holomorphic:
\qq
\de_x\m L^{a}(x;u_1,u)\ 
=\ K(x;u_1,u)\wedge\bar\omega^{a}(x)\s.
\nonumber
\qqq
As functions of $u$, \m$L^{a}(x;u_1,u)$ are theta 
functions of the 2$^{\rm nd}$ order.

We would like to find an explicit form of the Lax matrix $\Psi$ 
representing the linear form $\phi$ on $H^0(L_{\Theta}^2)$
s.t. ${\langle\m}\theta,\m\phi{\m\rangle}=0$. We shall
achieve this goal partially, finding the entries $\eta$
and $\mu$ of the matrix (\ref{Lax}). The correspondence 
between $\Psi$ and $\phi$ is determined by the equality
\qq
{\langle\m}\delta E\s,\s\Psi{\m\rangle}\ 
=\ {\langle\m}\delta\theta\s,\s\phi{\m\rangle}
\nonumber
\qqq
Since the left hand side is given by Eq.\s\s(\ref{25})
and $\delta\theta$ by Eq.\s\s(\ref{nd}), we obtain
\qq
&&\hs{1cm}-\s 2\pi\m
\delta u_1({\rm Im}\s\tau)^{-1}\int_{_\Sigma}\mu\wedge\bar
\omega\s+\s\int_{_\Sigma}\eta\wedge\delta b\cr
&&=\ 
-\m2\pi\m\delta u_1^{a}\s ({\rm Im}\s\tau)^{-1}_{\s{a}{b}}
\int_{_\Sigma}\hspace{-0.2cm}{\langle\m}L^b(x;u_1,\s\cdot\s)\m,
\s\phi{\m\rangle}\wedge\m b(x)\s
+\s\int_{_\Sigma}\hspace{-0.2cm}
{\langle\m}K(x;u_1,\s\cdot\s)\m,\s\phi{\m\rangle}\wedge\m\delta b(x)\s.
\hspace{0.5cm}
\label{equll}
\qqq
Taking $\delta u_1=0$ we infer that
\qq
\eta(x)\ =\ {\langle\m}K(x;u_1,\s\cdot\s)\s,\s\phi{\m\rangle}
\label{eta}
\qqq
is the lower left entry of the matrix $\Psi$ corresponding
to the linear form $\phi$. 

It is easy to find the entry $\mu$ of $\Psi$ 
representing the linear form $\phi_{u_1}$ (recall
that $\phi_{u_1}$ computes the value of a theta 
function in $H^0(L_\Theta^2)$ at point $u_1$). 
Since $K(x;u_1,u_1)=0\m$, it follows from Eq.\s\s(\ref{eta})
that $\eta=0$ in this case. Eq.\s\s(\ref{equll}) reduces 
then to
\qq
&&-\s 2\pi\m\delta u_1({\rm Im}\s\tau)^{-1}
\int_{_\Sigma}\mu\wedge\bar
\omega\s=\s\delta u_1^{a}\int_{_\Sigma}
\da_{u_1^{a}} K(x;u_1,u_1)\wedge b(x)\s\cr
&&=\s -\s\delta u_1^{a}\int_{_\Sigma}
\da_{u^{a}} K(x;u_1,u_1)\wedge b(x)\s
=\s-\s\delta u_1^{a}\s\s\da_{a}
\theta(u_1)\s.
\nonumber
\qqq
This fixes $\mu$ uniquely:
\qq
\mu\ =\ {_i\over^{4\pi}}\s\da_{a}\theta(u_1)\s\s
\omega^{a}\ .
\label{mu}
\qqq
Let us check that there exists $\nu\in\wedge^{10}(l_{u_1}^{-2})$
such that the last equation of (\ref{rel}) holds. For this
it is necessary and sufficient that 
\qq
\int_{_\Sigma}{\kappa}\m\s\mu\wedge b\ =\ 0
\label{cons}
\qqq
for a non-zero holomorphic section ${\kappa}$ of \m$l_{u_1}^2=
V(2u_1)K$
(\m${\rm dim}\s H^0(l_{u_1}^2)=1$ if \s$2u_1\s\not\in\s\NZ^2+\tau\NZ^2$).
\m But such a section may be represented by the function
\qq
x\ \mapsto \ \vartheta(\smallint_{x_0}^x\omega-2u_1-\Delta)
\nonumber
\qqq 
so that, recalling the definition (\ref{Kk}), we obtain
\qq
\int_{_\Sigma}{\kappa}\m\s\omega^{a}\wedge b\ =\ 
\int_{_\Sigma}
\epsilon^{{a}{b}}\s\da_{u^{b}} K(x;u_1,u_1)\wedge b(x)
\ =\ \epsilon^{{a}{b}}\s\s\da_{b}\theta(u_1)\ .
\label{kob}
\qqq
Hence the relation (\ref{cons}) follows for $\mu$ given by
Eq.\s\s(\ref{mu}). The 1,0-form $\nu$ satisfying the last relation 
of (\ref{rel}) is now unique since $H^0(l_{u_1}^{-2}K)=\{0\}$.

We would like to find the entry $\mu$ 
of $\Psi$ corresponding to more general linear forms 
$\phi$ s.t. ${\langle\m}\theta,
\m\phi{\m\rangle}=0$. Recall that $\theta$
with $\theta(u_1)=0$ may be given by formula (\ref{thetaE}) 
with $b\in\wedge^{0,1}(l_{u_1}^{-2})$. Note that any
2$^{\rm nd}$-order theta function \s$\delta\theta$
vanishing at $u_1$ and not in the Kummer quartic $\CK$
may be written as
\qq
\delta\theta(u)\ =\ \int_{_\Sigma}K(x;u_1,u)\wedge
\delta b(x)
\label{dth}
\qqq
with $\delta b\in\wedge^{01}(l_{u_1}^{-2})$ since 
it corresponds to an extension of $l_{u_1}$. 
The space of $\delta\theta$ vanishing at $u_1$
is 3-dimensional, as well as the space
$H^1(l_{u_1}^{-2})$ of classes $[\delta b]$ 
and the assumption that $\delta\theta\not\in\CK$ 
is obviously superfluous. Set for a linear form $\psi$ on 
$H^0(L_{\Theta}^2)$,
\qq
\eta_\psi(x)\s=\ {\langle\m}K(x;u_1,\s\cdot\s)\m,\s\psi{\m\rangle}\ .
\label{etab}
\qqq
$\eta_\psi$ defines a holomorphic 1,0-form with values
in $l_{u_1}^2$. We have
\qq
{\langle\m}\delta\theta\m,\m\psi{\m\rangle}\ =\s\int_{_\Sigma}
\eta_\psi\wedge\delta b
\label{dta}
\qqq
for \s$\delta\theta\s$ given by Eq.\s\s(\ref{dth}). By 
dimensional count, the map $\psi\mapsto\eta_\psi$ is onto 
$H^0(l_{u_1}^2K)$ with the 1-dimensional kernel spanned 
by $\phi_{u_1}$. Specifying Eq.\s\s(\ref{dta}) to $\delta\theta
\propto\theta$, we obtain the relation
\qq
{\langle\m}\theta\m,\m\psi{\m\rangle}\ =\s\int_{_\Sigma}\eta_\psi
\wedge b
\label{dtE}
\qqq
which determines the class $[b]\in H^1(l_{u_1}^{-2})$ in terms
of $\theta$. On the other hand, taking $\psi=\phi$ 
in Eq.\s\s(\ref{etab}), we infer that $\eta=0\m$ if and only 
if $\phi$ is proportional to $\phi_{u_1}$, 
the case studied before. 

If $\eta_\phi\not=0$ then $\mu$ depends on the choice
of the representative $b$ in the class $[b]\in 
H^1(l_{u_1}^{-2})$ characterizing $E$ as the extension
of $l_{u_1}$. Under the transformation $b\mapsto b+\de\varphi$
where $\varphi$ is a section of $l_{u_1}^{-2}$, 
\qq
\eta\mapsto\eta\m,\ \quad\mu\mapsto\mu+\varphi\m\eta\m,\ \quad
\nu\mapsto\nu-2\m \varphi\m\mu-\varphi^2\eta\s.
\nonumber
\qqq
The pairing of the theta functions $L^a(x;u_1,\s\cdot\s)$  
of Eq.\s\s(\ref{gal}) with the linear form $\phi$
gives two 1,0-forms with values in $l_{u_1}^2$:
\qq
\chi^{a}(x)\ =\ {\langle\m}L^{a}(x;\m\cdot\s,u_1)\m,
\s\phi{\m\rangle}\quad\ \ 
{\rm s.t.}\quad\ \de\chi^{a}\s
=\s\eta\wedge\bar\omega^{a}\s.
\label{cih}
\qqq
Specifying the equality (\ref{equll}) to the case
with $\delta b=0$, we infer the relation
\qq
\int_{_\Sigma}\mu\wedge\bar\omega^a\ =\ \int_{_\Sigma}\chi^a
\wedge b
\label{cfm}
\qqq
which, together with the equation 
\qq
\de\mu=-\eta\wedge b
\label{cfm1}
\qqq
determines $\mu$ completely. In Appendix 2, we show 
that $\mu$ fixed this way satisfies the relation 
$\int_{_\Sigma}{\kappa}\s\mu\wedge b=0$ and hence defines 
a unique 1,0-form $\nu$ with values in 
$l_{u_1}^{-2}$ s.t. $\de\nu=2\m\mu\wedge b$.

\setcounter{equation}{0}
\medskip
\section{Hitchin Hamiltonians}

From the relation (\ref{hit}) and the explicit form of
$\Psi$ corresponding to $\phi_{u_1}$ ($\eta$ vanishing, $\mu$
given by Eq.\s\s(\ref{mu})\m), 
one obtains
\qq
\CH(\m\theta\m,\s a_1\m\phi_{u_1})\ =\ -\m{_1
\over^{16\pi^2}}\s a_1^2\s\s
(\s \da_{a}\theta(u_1)\s\s\omega^{a}\s)^2\s.
\label{first}
\qqq
The right hand side is a quadratic differential.
Eq.\s\s(\ref{first}), whose projective
version was first obtained in \cite{VGP}, 
is consistent with the rescaling $\theta\mapsto 
t\s\theta$ and $\phi\mapsto t^{-1}\phi$ for $t\in\NC^*$.
It describes the value of the Hitchin map
$\CH$ on the special covectors, namely those 
represented by the pairs $(\theta\m,\s\phi)$ 
s.t. $\NC^*\phi$ is in the intersection $\CK^*_E$
of the Kummer quartic $\CK^*$ with the plane 
${\langle\m}\theta,\m\phi{\m\rangle}=0$.
The linear span of $\CK^*_E$ gives the whole cotangent
space $T^*_E\CN_{ss}$. Indeed, any theta function of
the 2$^{\rm nd}$ order $\delta\theta$ which vanishes 
on $C_E$ has to be proportional to $\theta$ and defines 
a zero vector in $T_E\CN_{ss}$.
$\CK^*_E$ is itself a quartic. Hence the restriction 
of the quadratic polynomial $\CH$ to six lines in $\CK^*_E$ 
in a general position determines $\CH$ completely.

It is possible to find a more explicit description
of the values of $\CH$ away from $\CK^*_E$ and this is
the main aim of the rest of the present section.
Suppose then that the entry $\eta$ in $\Psi$
does not vanish. Let $x_i$, $i=1,\dots,4$, be its four zeros.
We shall assume that $\eta$ cannot be written
as $\kappa\m\omega$ for $\kappa\in H^0(l_{u_1}^2)$ and 
$\omega\in H^0(K)$. This is true for 
generic $\phi$. In this case, \s$\eta\m=\m 
a_2\m\eta_{\phi_{u_2}}\s$ for some $a_2\in\NC^*$ 
and for $u_2$ satisfying
\qq
u_1+u_2\s=\s\smallint_{x_0}^{x_1}\omega
+\smallint_{x_0}^{x_2}\omega-2\Delta\ \quad{\rm and}\ \quad
u_1-u_2\s=\s \smallint_{x_0}^{x_3}\omega
+\smallint_{x_0}^{x_4}\omega-2\Delta\s,
\label{upm}
\qqq
$u_1\pm u_2\s\not\in\s\NZ+\tau\NZ$.
\s Indeed, \m$\eta_{\phi_{u_2}}(x)\m$ 
is a holomorphic section of $l_{u_1}^2K$
represented by the multivalued function
\s$\vartheta(\smallint_{x_0}^x
\omega-u_1-u_2-\Delta)\s\m\vartheta(\smallint_{x_0}^x\omega
-u_1+u_2-\Delta)\s$ vanishing exactly
at $x_i$ and such a section is unique up to normalization.
We infer that in the action on the theta functions
of Eq.\s\s(\ref{dth}), the linear 
forms $\phi$ and $a_2\m\phi_{u_2}$
coincide. Since Eq.\s\s(\ref{dth}) gives all theta functions
vanishing at $u_1$, it follows that 
\qq
\phi\s =\s a_1\s\phi_{u_1}\s+\s a_2\s\phi_{u_2}
\label{alpha}
\qqq
for some $a_1\in\NC$. Let us stress that, to fix normalizations, 
$u_1$ and $u_2$ should be viewed as elements of $\NC^2$ 
with $x_i$ in relations (\ref{upm}) belonging to the covering space 
$\tilde\Sigma$ of $\Sigma$. The relation 
${\langle\m}\theta\m,\m\phi{\m\rangle}\s=0$ 
implies that $\theta(u_2)=0$.

Summarizing, we have shown that a generic pair $(\theta,\phi)$ 
s.t. ${\langle\m}\theta,\m\phi{\m\rangle}=0$ 
may be obtained by first choosing $u_1$ and $u_2$ 
s.t. $2u_1,\m 2u_2,\m u_1\pm u_2\s\not\in\s\NZ+\tau\NZ$
\m and then taking $\theta$ from the 2-dimensional space of theta 
functions vanishing at $u_1$ and $u_2$ and $\phi$ from 
the orthogonal subspace. 
The zeros $x_i$ of $\eta$ are determined from Eqs.\s\s(\ref{upm})
(as the zeros of $\vartheta(\smallint_{x_0}^x
\omega-u_1\pm u_2-\Delta)$). For simplicity, we shall 
assume that they are distinct (this is true for generic $\phi$).
Then the differentials $\da\eta(x_i)\in(l_{u_1}^2K^2)_{x_i}$
do not vanish.

A quadratic differential $\rho\in H^0(K^2)$
is determined by its values at four points $x_i$ which form 
a divisor of $l_{u_1}^2K\not=K^2$. Since ${\rm dim}\m H^0(K^2)=3$,
there is one linear relation satisfied by all $\rho(x_i)\m$:
\qq
\sum\limits_{i=1}^4\rho(x_i)\s{\kappa}(x_i)\s\da\eta(x_i)^{-1}\s=\s 0
\nonumber
\qqq
for $0\not={\kappa}\in H^0(l_{u_1}^2)$. \m It expresses the fact that 
the sum of residues of the meromorphic 1,0-form $\rho\m{\kappa}
\m\eta^{-1}$ has to vanish. For $\rho=\CH(\theta,\phi)=\mu^2+\eta
\nu$,
\qq
\rho(x_i)\s=\s\mu(x_i)^2
\nonumber
\qqq
so that it is enough to know $\mu(x_i)$ in order to determine
\m$\CH(\theta,\phi)$. \m Note that although
the 1,0-form $\mu$ depends on the choice of the representative $b$
of the class $[b]\in H^1(l_{u_1}^{-2})$ defined by Eq.\s\s(\ref{dtE}),
the values $\mu(x_i)$ are invariant since under $b\mapsto b+\de\varphi$
the 1,0-form $\mu$ changes to $\mu+\varphi\m\eta$.

It remains to find $\mu(x_i)$. Consider the meromorphic function
$\eta_\psi\eta^{-1}$. Viewed as a distribution, 
$\de(\eta_\psi\eta^{-1})$ is supported at the poles 
of $\eta_\psi\eta^{-1}$ and
\qq
\int_{_\Sigma}\mu\wedge\de(\eta_\psi\eta^{-1})\s=\s
-2\pi i\sum\limits_{i=1}^4\mu(x_i)\m\eta_\psi(x_i)\s\da\eta(x_i)^{-1}
\nonumber
\qqq
for any (smooth) 1,0-form $\mu$. \m In particular, for $\mu$ 
satisfying Eq.\s\s(\ref{cfm1}) we obtain
\qq
\sum\limits_{i=1}^4\mu(x_i)\s\eta_\psi(x_i)\s\da\eta(x_i)^{-1}
\s=\s{_1\over^{2\pi i}}\int_{_\Sigma}\eta_\psi\wedge b\s
=\s{_1\over^{2\pi i}}\s\m{\langle\m}\theta,\m\psi{\m\rangle}\s.
\label{21}
\qqq
Recall that $\eta_\psi$ run through the three-dimensional
space $H^0(l_{u_1}^2K)$. If $\eta_\psi(x_i)=0$ for all $i$
then $\eta_\psi$ has to be proportional to 
$\eta=a_2\m\eta_{\phi_{u_2}}$. Hence vectors $(\eta_\psi(x_i))$ 
form a 2-dimensional subspace in $\mathop{\oplus}
\limits_i(l_{u_1}^2K)_{x_i}$ and equations (\ref{21})
determine vector $(\mu(x_i))\in\mathop{\oplus}\limits_i K_{x_i}$ 
up to a 2-dimensional ambiguity spanned by $(\omega^a(x_i))$
(indeed, as the residues of the meromorphic 1,0-form
$\eta_\psi\eta^{-1}\omega^a$, the numbers 
$\omega^a(x_i)\m\eta_\psi(x_i)\s\da\eta(x_i)^{-1}$ sum to zero).
It is clearly enough to take for $\psi$ in Eq.\s\s(\ref{21})
any two linear forms independent of $\phi_{u_1}$ and $\phi_{u_2}$.
In the generic situation, we may choose the forms 
$\da_a\phi_{u_1}$ defined by 
\qq
{\langle\m}\theta\m,\s\da_a\phi_{u_1}{\m\rangle}\s=\s\da_{a}\theta(u_1)\s.
\nonumber
\qqq
Denoting the corresponding 1,0-forms $\eta_\psi$ by $\eta'_a$,
we obtain 2 relations for $\mu(x_i)\m$:
\qq
\sum\limits_{i=1}^4\mu(x_i)\s\eta'_a(x_i)\s\da\eta(x_i)^{-1}
\s=\s{_1\over^{2\pi i}}\s\da_a\theta(u_1)\s.
\label{22}
\qqq
Alternatively, we may choose for $\psi$ the linear forms 
$\da_a\phi_{u_2}$ corresponding to 1,0-forms $\eta''_a$. 
This gives the relations
\qq
\sum\limits_{i=1}^4\mu(x_i)\s\eta''_a(x_i)\s\da\eta(x_i)^{-1}
\s=\s{_1\over^{2\pi i}}\s\da_a\theta(u_2)\s.
\label{24}
\qqq
$\eta''_a$ must be linearly dependent from $\eta'_a$ and $\eta$
(in the generic situation):
\qq
\eta''_a\ =\ D_a^b\s\eta'_b\s+\s\eta
\label{lind}
\qqq
leading via Eqs.\s\s(\ref{22}) and (\ref{24}) to the relation 
\qq
\da_a\theta(u_2)\ =\ D_a^b\s\s\da_b\theta(u_1)\s. 
\nonumber
\qqq

We need 2 more equations to determine $\mu(x_i)$. They may be 
obtained from Eqs.\s\s(\ref{cfm}) fixing the holomorphic 
contributions to $\mu$. Indeed, using the 2$^{\rm nd}$ equation 
in (\ref{cih}), and Eq.\s\s(\ref{cfm1}) we infer that
\qq
\int_{_\Sigma}\mu\wedge\bar\omega^a\s=\s\int_{_\Sigma}(\mu\m\eta^{-1})
\s\eta\wedge\bar\omega^a\s=\s\int_{_\Sigma}(\mu\m\eta^{-1})\s\de\chi^a
\s=\s\int_{_\Sigma}\chi^a\wedge\de(\mu\m\eta^{-1})\cr
=\s\int_{_\Sigma}\chi^a\wedge b\s-\s2\pi i\sum\limits_{i=1}^4
\mu(x_i)\s\chi^a(x_i)\s\da\eta(x_i)^{-1}
\label{tbr}
\qqq
so that Eq.\s\s(\ref{cfm}) implies that
\qq
\sum\limits_{i=1}^4
\mu(x_i)\s\chi^a(x_i)\s\da\eta(x_i)^{-1}\s=\s0\s.
\label{mie}
\qqq
These are the two missing equations. To see this, 
repeat the calculation (\ref{tbr}) for $\mu$
replaced by $\omega^b$. This gives the relation
\qq
{_1\over^\pi}\s{\rm Im}\m\tau^{ab}\s=\s\sum\limits_{i=1}^4
\omega^b(x_i)\s\chi^a(x_i)\s\da\eta(x_i)^{-1}\s.
\nonumber
\qqq
Suppose now that \m$d_a\m\chi^a(x_i)+e\s\eta_\psi(x_i)=0$
\s for $i=1,\dots,4$. It follows that
\qq
0\s=\s\sum\limits_{i=1}^4
\omega^b(x_i)\left(d_a\chi^a(x_i)
+e\m\eta_\psi(x_i)\right)\da\eta(x_i)^{-1}
\s=\s {_1\over^\pi}\s{\rm Im}\m\tau^{ab}\s d_a
\nonumber
\qqq
so that $d_a=0$. Hence the vectors $(\chi^a(x_i))$ span
a 2-dimensional subspace of $\mathop{\oplus}\limits_iK_{x_i}$
transversal to the 2-dimensional subspace spanned
by the vectors $(\eta_\psi(x_i))$ and the linear equations
(\ref{21}) and (\ref{mie}) determine $\mu(x_i)$ completely.

It is enough to consider the case $\phi=\phi_{u_2}$. 
\m Indeed, the shift \s$\phi\s\mapsto\s\phi\m+\m a_1
\m\phi_{u_1}\s$ results in the change
\qq
\mu\ \ \ \mapsto\ \ \ \mu\s+\s{_{i}\over^{4\pi}}\s a_1\s
\s\da_a\theta(u_1)\m\s\omega^a\s,
\nonumber
\qqq
see Eq.\s\s(\ref{mu}). Identifying 1,0-forms with 
multivalued functions by the relation (\ref{ide}) 
and setting \m$\chi_a=2\pi\m({\rm Im}\m\tau)^{-1}_{ab}
\chi^b$, \s${w}_i=\smallint_{x_0}^{x_i}\omega\m-\m\Delta$,
\s$G_1=G_{12}=-G_2$ and $G_3=G_{34}=-G_4$ where
\qq
G_{ij}\ =\ {\rm det}\begin{pmatrix}\da_1\vartheta({w}_i)&
\da_1\vartheta({w}_j)\\\da_2\vartheta({w}_i)&
\da_2\vartheta({w}_j)\end{pmatrix}\s,
\nonumber
\qqq
we obtain
\qq
&&\da\eta(x_1)\s=\s G_1\ 
\vartheta({w}_1-{w}_3-{w}_4)\s,
\quad\quad\quad
\ \chi_a(x_1)\s =\s -\da_a\vartheta({w}_2)\ 
\vartheta({w}_1-{w}_3-{w}_4)\s,\cr
&&\da\eta(x_2)\s =\s G_2\ \vartheta({w}_2
-{w}_3-{w}_4)\s,
\quad\quad\quad
\ \chi_a(x_2)\s =\s -\da_a\vartheta({w}_1)\ 
\vartheta({w}_2-{w}_3-{w}_4)\s,\cr
&&\da\eta(x_3)\s =\s G_3\ 
\vartheta({w}_3-{w}_1-{w}_2)\s,
\quad\quad\quad
\ \chi_a(x_3)\s =\s -\da_a\vartheta({w}_4)\ 
\vartheta({w}_3-{w}_1-{w}_2)\s,\cr
&&\da\eta(x_4)\s =\s G_4\ 
\vartheta({w}_4-{w}_1-{w}_2)\s,
\quad\quad\quad
\ \chi_a(x_4)\s =\s -\da_a\vartheta({w}_3)\ 
\vartheta({w}_4-{w}_1-{w}_2)\s,\cr\cr
&&\eta'_a(x_1)\s =\s \da_a\vartheta({w}_1)\ 
\vartheta({w}_2+{w}_3+{w}_4)\s,\quad
\eta''_a(x_1)\s =\s\ \ \da_a\vartheta({w}_2)\ 
\vartheta({w}_1-{w}_3-{w}_4)\s,\cr
&&\eta'_a(x_2)\s =\s \da_a\vartheta({w}_2)\ 
\vartheta({w}_1+{w}_3+{w}_4)\s,\quad
\eta''_a(x_2)\s =\s\ \ \da_a\vartheta({w}_1)\ 
\vartheta({w}_2-{w}_3-{w}_4)\s,\cr
&&\eta'_a(x_3)\s =\s \da_a\vartheta({w}_3)\ 
\vartheta({w}_1+{w}_2+{w}_4)\s,\quad
\eta''_a(x_3)\s =\s -\da_a\vartheta({w}_4)\ 
\vartheta({w}_3-{w}_1-{w}_2)\s,\cr
&&\eta'_a(x_4)\s =\s \da_a\vartheta({w}_4)\ 
\vartheta({w}_1+{w}_2+{w}_3)\s,
\quad\eta''_a(x_4)\s =\s -\da_a\vartheta({w}_3)\ 
\vartheta({w}_4-{w}_1-{w}_2)\s.
\nonumber
\qqq
Given these values, it is easy to find the explicit
form of the matrix $(D^b_a)$ appearing 
in the relation between the derivatives
of $\da_a\theta$ at $u_1$ and $u_2$
by specifying Eq.\s\s(\ref{lind})
to two of the points $x_i$. One form of these relations is
\qq
&&\da_2\vartheta({w}_3)\s\da_1\theta(u_2)
\s-\s\da_1\vartheta({w}_3)\s\da_2\theta(u_2)\s\m\cr\cr
&&\hspace{2cm}=\s-\m{_{\vartheta({w}_3-{w}_1-{w}_2)}
\over^{\vartheta({w}_1+{w}_2+{w}_4)}}
\s\s(\da_2\vartheta({w}_4)\s\da_1\theta(u_1)
\s-\s\da_1\vartheta({w}_4)\s\da_2\theta(u_1))\s,\cr\cr
&&\da_2\vartheta({w}_4)\s\da_1\theta(u_2)
\s-\s\da_1\vartheta({w}_4)\s\da_2\theta(u_2)\s\cr\cr
&&\hspace{2cm}=\s-\m{_{\vartheta({w}_4-{w}_1-{w}_2)}
\over^{\vartheta({w}_1+{w}_2+{w}_3)}}
\s\s(\da_2\vartheta({w}_3)\s\da_1\theta(u_1)
\s-\s\da_1\vartheta({w}_3)\s\da_2\theta(u_1))\s.
\nonumber
\qqq
Let us denote
\s$\tilde\mu(x_i)\m=\m\mu(x_i)/G_i\m$. 
\m Eqs.\s\s(\ref{mie}) have the general solution
\qq
(\tilde\mu(x_1),\dots,\tilde\mu(x_4))\ =\ 
g_1\s\m(G_{34},0,G_{23},-G_{24})\s+\s
g_2\s\m(0,G_{34},G_{13},-G_{14})
\nonumber
\qqq
and Eqs.\s\s(\ref{24}) fix the values of $g_1$ and $g_2$ to 
\qq
&&g_1\ =\ -\m{\da_2\vartheta({w}_1)\s\da_1\theta(u_2)
\m-\m\da_1\vartheta({w}_1)\s\da_2\theta(u_2)\over
{4\pi i\s G_{12}\m G_{34}}}\cr\cr
&&g_2\ =\ \s\s\m{\da_2\vartheta({w}_2)\s\da_1\theta(u_2)
\m-\m\da_1\vartheta({w}_2)\s\da_2\theta(u_2)\over{4\pi i\s 
G_{12}\m G_{34}}}\s.
\nonumber
\qqq
This leads to the following simple result:
\qq
\mu(x_i)\s=\s\pm\s{_i\over^{4\pi}}\s\m(\m\da_2\vartheta({w}_i)
\s\da_1\theta(u_2)\s-\s\da_1\vartheta({w}_i)\s\da_2
\theta(u_2)\m)
\label{u1u2}
\qqq
or, in a more abstract notation from the introduction,
\qq
\mu(x_i)\s=\s\pm\s{_i\over^{4\pi}}\s\m d\theta(l_{u_2})
\nonumber
\qqq
with the plus sign for $i=1,2$ and the minus one for $i=3,4$.

Since the Hitchin Hamiltonian is quadratic in $\phi$
and its values on $\phi_{u_1}$ and $\phi_{u_2}$
are given by Eq.\s\s(\ref{first}), it follows that
\qq
&&\CH(\m\theta\s,\m\s a_1\s\phi_{u_1}\m+\m a_2
\s\phi_{u_2})\cr\cr
&&\hspace{1cm}=\ a_1^2\s\CH(\m\theta\m,\s\phi_{u_1})\s
+\s a_2^2\s\CH(\theta\m,\s\phi_{u_2})
\s+\s2\m a_1a_2\s(\m c_1\m(\omega^1)^2\s
+\s c_2\m\omega^1\omega^2
\s+\s c_3\m(\omega^2)^2\m)\s.\hspace{0.8cm}
\nonumber
\qqq
The mixed term may be found from the linear equations
\qq
&&{_i\over^{4\pi}}\s(\da_2\vartheta({w}_i)\s\da_1\theta(u_1)-
\da_1\vartheta({w}_i)\s\da_2\theta(u_1))
\s\s\m\tilde\mu(x_i)\s\s G_i\cr\cr
&&\hspace{1.5cm}=\ c_1\s\da_2\vartheta
({w}_i)\s\da_2\vartheta({w}_i)
\s-\s c_2\s\da_2\vartheta({w}_i)\s\da_1\vartheta({w}_i)
\s+\s c_3\s\da_1\vartheta({w}_i)\s\da_1\vartheta({w}_i)\s.
\nonumber
\qqq
Their explicit solution leads to the expression
\qq
&&\hspace{-0.6cm}\CH(\m\theta\s,\m\s a_1\s\phi_{u_1}\m
+\m a_2\s\phi_{u_2})
\ =\ -\m{_1\over^{16\pi^2}}\s\s(\m a_1
\s\da_a\theta(u_1)\s\omega^a
\s+\s a_2\s\da_a\theta(u_2)\s\omega^a\m)^2\s\cr
&&+\s\s{_{a_1\m a_2}\over^{4\pi^2\s G_{13}\m G_{23}}}
\s\s(\da_2\vartheta({w}_3)\s\da_1\theta(u_1)
-\da_1\vartheta({w}_3)\s\da_2\theta(u_1))\s\label{ost}\cr
&&\ \ \ \cdot\ (\da_2\vartheta({w}_3)\s\da_1\theta(u_2)
-\da_1\vartheta({w}_3)\s\da_2\theta(u_2))
\ \da_a\vartheta({w}_1)\s\m\da_b
\vartheta({w}_2)\s\m\omega^a\omega^b\s.
\qqq
The second term on the right hand side hand side
is a quadratic differential that vanishes at $x_1$ and $x_2$
and is equal to \s$\frac{a_1\m a_2}{4\pi^2}\s\s
\da_a\theta(u_1)\s\m\da_b\theta(u_2)\s\omega^a\omega^b\s$
at $x_3$ and $x_4$ so that
\qq
\CH(\theta,\m\phi)(x_i)\ =\ 
-\m{_1\over^{16\pi^2}}\s\s(\m a_1
\s\da_a\theta(u_1)\s\omega^a(x_i)
\s\pm\s a_2\s\da_a\theta(u_2)\s\omega^a(x_i)\m)^2\s
\label{frr}
\qqq
where sign plus should be taken for $x_1$ and $x_2$ and
sign minus for $x_3$ and $x_4$. This is the result
(\ref{fr}) described in Introduction.

\setcounter{equation}{0}
\medskip
\section{Self-duality}

We would like to compare the values of the Hitchin Hamiltonians
on the dual pairs $(\theta,\phi)$ and $(\theta',\m\phi')$ where
$\theta'=\iota(\phi)$ and $\phi'=\iota^{-1}(\theta)$
with $\iota$ defined by Eq.\s\s(\ref{lini}).
Recall that, given $u_1$ s.t. $\theta(u_1)=0$, we associated to 
the linear form \m$\phi\m$ a 1,0-form $\eta$
by Eq.\s\s(\ref{eta}).  Viewed as a holomorphic 
section of \s$l_{u_1}^2K$, 
\qq
\eta(x)\ =\ {\langle\m}\vartheta(\smallint_{x_0}^x\omega
-u_1-\s\cdot\s-\Delta)\ \vartheta(\smallint_{x_0}^x
\omega-u_1+\s\cdot\s-\Delta)\s\s,\ \phi{\m\rangle}\s.
\nonumber
\qqq
Let us denote
\qq
u_i'\s=\s\smallint_{x_0}^{x_i}\omega\m-\m u_1-\Delta\s.
\label{uip}
\qqq
The vanishing of $\eta(x_i)$ implies then that
the linear form $\phi$ annihilates the theta functions
\qq
u\ \ \mapsto\ \ \vartheta(u'_i-u)\s\m\vartheta
(u'_i+u)\s=\iota(\phi_{u'_i})(u)
\label{4theta}
\qqq
and also, if we rewrite $\eta(x_i)$ as $\iota(\phi)(u'_i)$,  
that $\theta'(u'_i)=0$. Since $\phi=
a_1\m\phi_{u_1}+a_2\m\phi_{u_2}$ and $\phi_{u_1}$ annihilates 
the theta functions (\ref{4theta}) as well, it follows that  
they belong to $\Pi$. Hence $\NC^*\iota(\phi_{u'_i})$ are 
the 4 points of intersection of the line $\NP\Pi$ with the Kummer 
quartic $\CK$. Equivalently, $\NC^*\phi_{u'_i}$ are the points
of intersection of $\NP\m{\Pi'}^\perp$ with $\CK^*$. 
In the generic situation, any pair of theta functions 
$\phi_{u'_i}$ spans ${\Pi'}^\perp$ and since 
$\phi'\in{\Pi'}^\perp$, we may write
\qq
\phi'\ =\ a'_1\s\phi_{v_1}\s+\s a'_2\s\phi_{v_2}
\label{aap}
\qqq
or, equivalently,
\qq
\theta\ =\ a'_1\s\iota(\phi_{v_1})\s
+\s a'_2\s\iota(\phi_{v_2})\s.
\label{aapp}
\qqq

The involution $l\mapsto l^{-1}K$ of the Jacobian $J^1$ 
lifts to $\NC^2$ to the flip of sign of $u$. By restriction 
to the bundles $\CO(x)$, it induces the involution $x\mapsto x'$ 
of $\Sigma$ which leaves 6 Weierstrass points invariant.
The latter involution lifts to an involution
(without fixed points) of the covering space
$\tilde\Sigma$ determined by the equation
\qq
\smallint_{x_0}^{x}\omega\s-\s\Delta\ 
=\ -\smallint_{x_0}^{x'}\omega\s+\s\Delta\s.
\label{invo}
\qqq
Definitions (\ref{uip}) together with Eqs.\s\s(\ref{upm})
give the relations
\qq
u'_1-u'_2\s=\s\smallint_{x_0}^{x_1}\omega
-\smallint_{x_0}^{x_2}\omega\ \quad{\rm and}\ \quad
u'_1+u'_2\s=\s -\smallint_{x_0}^{x_3}\omega
-\smallint_{x_0}^{x_4}\omega+2\Delta
\nonumber
\qqq
holding in $\NC^2$, with $x_i\in\tilde\Sigma$.
They may be rewritten as
\qq
u'_1-u'_2\s=\s\smallint_{x_0}^{x_1}\omega
+\smallint_{x_0}^{x'_2}\omega\s-\s2\Delta\ \quad{\rm and}\ \quad
u'_1+u'_2\s=\s \smallint_{x_0}^{x'_3}\omega
+\smallint_{x_0}^{x'_4}\omega-2\Delta\s,
\label{upmq}
\qqq
which, upon the flip of the sign of $u'_2$ leaving 
$\phi_{u'_2}$ unchanged, provides the dual version
of relations (\ref{upm}) corresponding to points 
$x_1,x'_2,x'_3,x'_4\in\tilde\Sigma$.
Applying the previous result (\ref{frr}) and
using the possibility to exchange a point with
its image under the involution of $\Sigma$ in the argument
of a quadratic differential, we infer that
\qq
\CH(\theta',\m\phi')(x_i)\ =\ 
-\m{_1\over^{16\pi^2}}\s\s(\m a'_1
\s\da_a\theta'(u'_1)\s\omega^a(x_i)
\s\mp\s a_2\s\da_a\theta'(u'_2)\s\omega^a(x_i)\m)^2\s.
\label{frrr}
\qqq
The sign minus should be taken for $x_1$ and $x_2$ and sign plus 
for $x_3$ and $x_4$. The exchange of signs in comparison
with Eq.\s\s(\ref{frr}) is due to the flip $u'_2\mapsto-u'_2$.

In order to compare expressions (\ref{frr}) and (\ref{frrr})
we shall calculate the coefficients $a_{1,2}$ and $a'_{1,2}$
of the linear combinations (\ref{alpha}) and (\ref{aap}).
Note that the definition $\theta'=\iota(\phi)$ implies that
\qq
\theta'(\smallint_{x_0}^x\omega-u_1-\Delta)\ =\ a_2\s\s\s
\vartheta(\smallint_{x_0}^x\omega-u_1-u_2-\Delta)
\s\s\vartheta(\smallint_{x_0}^x\omega-u_1+u_2-
\Delta)\s.
\nonumber
\qqq
Taking the derivative over $x$ at $x_1$, we obtain
\qq
\da_a\theta'(u'_1)\m\s\omega^a(x_1)\ = 
\ -\m a_2\s\s\s\vartheta(w_1-w_3-w_4)\s\s\da_a
\vartheta(w_2)\s\m\omega^a(x_1)
\nonumber
\qqq
where we employed Eqs.\s\s(\ref{upm}) and the abbreviated
notations $w_i=\smallint_{x_0}^{x_i}-\Delta$. Hence
\qq
a_2\ =\ -\m{_{\da_a\theta'(u'_1)\m\s\omega^a(x_1)}\over
^{\vartheta(w_1-w_3-w_4)\s\s\da_a
\vartheta(w_2)\s\m\omega^a(x_1)}}\s.
\label{a22}
\qqq
Similarly,
\qq
\theta'(\smallint_{x_0}^x\omega-u_2-\Delta)\ =\ a_1\s\s\s
\vartheta(\smallint_{x_0}^x\omega-u_1-u_2-\Delta)
\s\s\vartheta(\smallint_{x_0}^x\omega+u_1-u_2-
\Delta)\s.
\nonumber
\qqq
Taking the derivative at $x=x_1$ and noting that 
\m$w_1-u_2=-u'_2\m$, \m we infer that
\qq
a_1\ =\ {_{\da_a\theta'(u'_2)\m\s\omega^a(x_1)}\over
^{\vartheta(w_1+w_3+w_4)\s\s\da_a
\vartheta(w_2)\s\m\omega^a(x_1)}}\s.
\label{a11}
\qqq
To calculate $a'_{1,2}$, we note that Eq.\s\s(\ref{aapp})
implies that
\qq
\theta(\smallint_{x_0}^x\omega-v_1-\Delta)\ =\ a'_2\s\s\s\vartheta(
\smallint_{x_0}^x\omega-u'_1-u'_2-\Delta)\s\s
\vartheta(\smallint_{x_0}^x\omega-u'_1+u'_2-\Delta)\s.
\nonumber
\qqq
Upon derivation at $x=x_1$ and with the use of relations 
(\ref{upmq}) and (\ref{invo}), this gives
\qq
a'_2\ =\ -\m{_{\da_a\theta(u_1)\m\s\omega^a(x_1)}\over
^{\vartheta(w_1+w_3+w_4)\s\s\da_a
\vartheta(w_2)\s\m\omega^a(x_1)}}\s.
\label{a22p}
\qqq
Finally, since
\qq
\theta(\smallint_{x_0}^x\omega+v_2-\Delta)\ =\ a'_1\s\s\vartheta(
\smallint_{x_0}^x\omega-u'_1+u'_2-\Delta)\s\s
\vartheta(\smallint_{x_0}^x\omega+u'_1+u'_2-\Delta)\s.
\nonumber
\qqq
and \s$w_1+u'_2=u_2$ we infer that
\qq
a'_1\ =\ -\m{_{\da_a\theta(u_2)\m\s\omega^a(x_1)}\over
^{\vartheta(w_1-w_3-w_4)\s\s\da_a
\vartheta(w_2)\s\m\omega^a(x_1)}}\s.
\label{a11p}
\qqq
Substitution of expressions (\ref{a11}),(\ref{a22}),(\ref{a11p})
and (\ref{a22p}) shows equality of the right hand sides of
Eqs.\s\s(\ref{frr}) and (\ref{frrr}) for $x_i=x_1$. Since
there is a full symmetry between points $x_i$ (hidden in
our arbitrary choices of the order and the signs 
of $u_j\m$'s and $u'_j\m$'s), the self-duality  
\qq
\CH(\theta,\m\phi)\ =\ \CH(\theta',\m\phi')
\label{seld}
\qqq
follows.

\setcounter{equation}{0}
\medskip
\section{van Geemen-Previato's result and beyond}

The genus 2 curves are hyperelliptic.
The map \s$H^0(K)\ni\omega\s\mapsto\s\omega(x)\s$
defines an element of $\NP H^0(K)^*$ and varying $x\in\Sigma$
one obtains a realization of $\Sigma$ as a ramified double cover  
\m$\NP H^0(K)^*\cong\NP^1\m$. One may use the 1,0-forms 
$\omega^a\in H^0(K)$ to define the homogeneous coordinates 
on $\NP H^0(K)^*$. Then  
\qq
\lambda(x)\ =\ {_{\omega^2(x)}\over^{\omega^1(x)}}\ =\ 
-\m{_{\da_1\vartheta(\smallint_{x_0}^x\omega-\Delta)}
\over^{\da_2\vartheta(\smallint_{x_0}^x\omega-\Delta)}}
\label{inho}
\qqq
becomes the inhomogeneous coordinate of the image in $\NP^1$ 
of the point $x\in\Sigma$. \m If $x'$ is the image
of $x$ under the involution \s$\CO(x)\mapsto \CO(-x)K=\CO(x')\m$,
\s i.e.\s\s\m if
\qq
\smallint_{x_0}^x\omega+\smallint_{x_0}^{x'}\omega-2\Delta\ \in\ 
\NZ+\tau\NZ\quad\quad{\rm then}\quad\quad
\lambda(x)\s=\s\lambda(x')\s.
\nonumber
\qqq
Hence the involution $x\mapsto x'$ permutes the sheets of the 
covering $\Sigma\mapsto\NP^1$ ramified over the 6 Weierstrass 
points $x_s$, $s=1,\dots,6$, fixed by the involution. $\CO(x_s)$ 
is an odd spin structure. i.e. 
\qq
\smallint_{x_0}^{x_s}\omega-\Delta\ =\ E_s\ \ 
{\rm mod}\s(\NZ^2+\tau\NZ^2)
\nonumber
\qqq
and
\qq
\lambda_s\equiv\lambda(x_s)\s=\s
-\m{_{\da_1\vartheta(E_s)}
\over^{\da_2\vartheta(E_s)}}
\label{wp}
\qqq
where $E_s=\frac{1}{2}\,(e_s+\tau\m e_s')$ with $e_s,\m e_s'=(1,0),\s(0,1)$ 
or $(1,1)$ such that $e_s\cdot\m e_s'$ is odd. The possibilities 
are:
\qq
&&e_1=(1,0),\ e_1'=(1,0);\quad e_2=(1,1),\ e_2'=(1,0);
\quad e_3=(0,1),\ e_3'=(0,1);\cr\label{ordr}\cr
&&e_4=(1,1),\ e_4'=(0,1);\quad e_5=(0,1),\ e_5'=(1,1);
\quad e_6=(1,0),\ e_6'=(1,1).
\qqq
and we shall number the Weierstrass points (in a marking-dependent
way) in agreement with this list. $\Sigma$ may be identified 
with the hyperelliptic curve given by the equation 
\qq
\zeta^2\ =\ \prod\limits_{s=1}^6(\lambda-\lambda_s)
\label{hell}
\qqq
with the involution mapping $(\lambda,\zeta)$ to $(\lambda,-\zeta)$.
The expressions
\qq
\omega^1=C\s{{d\lambda}\over\zeta}\quad\ \ {\rm and}\quad\ \ 
\omega^2=C\s{{\lambda\m d\lambda}\over\zeta}\s,
\label{newf} 
\qqq
where $C$ is a constant, give the basis of holomorphic 1,0-forms 
of $\Sigma$ (the right hand sides vanish exactly where
the left hand sides do).

Let us recall the main result of \cite{VGP} based on 
the analysis of the formula (\ref{first}) for the Hitchin 
Hamiltonians on the Kummer quartic $\CK^*$. It will be 
convenient to identify the pairs $(\theta,\phi)$ 
s.t. ${\langle\m}\theta,\phi{\m\rangle}=0$ 
with pairs $(q,p)\in\NC^4\times\NC^4$ s.t. $q\cdot p=0$
by the relations
\qq
&&\theta\ =\ q_1\s\theta_{2,(0,0)}\s+\s q_2\s\theta_{2,(1,0)}
\s+\s q_3\s\theta_{2,(0,1)}\s+\s q_4\s\theta_{2,(1,1)}\s,\cr
&&\phi\ =\ p_1\s\theta_{2,(0,0)}^{\s*}\s+\s p_2\s\theta_{2,(1,0)}
^{\s*}\s+\s p_3\s\theta_{2,(0,1)}^{\s*}\s+\s 
p_4\s\theta_{2,(1,1)}^{\s*}\s.
\nonumber
\qqq
The symplectic form of $T^*\NP^3$ is the standard
$dp\wedge dq$ and the isomorphism $\iota$ interchanges $p$
and $q$. By examining the values of the quadratic differentials 
given by $\CH$ at the Weierstrass points $x_s$, van Geemen and 
Previato showed that
\qq
\CZ_s(q)\ =\ \{\s p\s\s|\s\s q\cdot p=0,\ 
\CH(q,p)(x_s)=0\s\}
\nonumber
\qqq
is a union of a pair of bitangents to $\CK^*$.
Then classical results giving the equations
for bitangents to the Kummer surface permitted the authors
of \cite{VGP} to write an almost explicit formula for $\CH(x_s)$ 
in the form
\qq
\CH(q,p)(x_s)\ =\ h_s\s\sum\limits_{t\not= s}{{r_{st}(q,p)}
\over{\lambda_s-\lambda_t}}
\label{VGP}
\qqq
where $r_{st}=r_{ts}$ are homogeneous polynomials, 
\qq
&&r_{12}(q,p)\ =\ \ \ \m(q_1p_1+q_2p_2-q_3p_3-q_4p_4)^2\s,\cr
&&r_{13}(q,p)\ =\ \ \ \m(q_1p_4-q_2p_3-q_3p_2+q_4p_1)^2\s,\cr
&&r_{14}(q,p)\ =\ -(q_1p_4+q_2p_3-q_3p_2-q_4p_1)^2\s,\cr
&&r_{15}(q,p)\ =\ -(q_1p_3-q_2p_4-q_3p_1+q_4p_2)^2\s,\cr
&&r_{16}(q,p)\ =\ \ \ \m(q_1p_3+q_2p_4+q_3p_1+q_4p_2)^2\s,\cr
&&r_{23}(q,p)\ =\ -(q_1p_4-q_2p_3+q_3p_2-q_4p_1)^2\s,\cr
&&r_{24}(q,p)\ =\ \ \ \m(q_1p_4+q_2p_3+q_3p_2+q_4p_1)^2\s,\cr
&&r_{25}(q,p)\ =\ \ \ \m(q_1p_3-q_2p_4+q_3p_1-q_4p_2)^2\s,\label{rs}\cr
&&r_{26}(q,p)\ =\ -(q_1p_3+q_2p_4-q_3p_1-q_4p_2)^2\s,\cr
&&r_{34}(q,p)\ =\ \ \ \m(q_1p_1-q_2p_2+q_3p_3-q_4p_4)^2\s,\cr
&&r_{35}(q,p)\ =\ \ \ \m(q_1p_2+q_2p_1+q_3p_4+q_4p_3)^2\s,\cr
&&r_{36}(q,p)\ =\ -(q_1p_2-q_2p_1-q_3p_4+q_4p_3)^2\s,\cr
&&r_{45}(q,p)\ =\ -(q_1p_2-q_2p_1+q_3p_4-q_4p_3)^2\s,\cr
&&r_{46}(q,p)\ =\ \ \ \m(q_1p_2+q_2p_1-q_3p_4-q_4p_3)^2\s,\cr
&&r_{56}(q,p)\ =\ \ \ \m(q_1p_1-q_2p_2-q_3p_3+q_4p_4)^2\s
\qqq
and $h_s\in K^2_{x_s}$ could still depend on $q$.
In the original language of pairs $(\theta,\phi)$,
and of the $(\NZ/2\NZ)^4$-action (\ref{acone})
on $H^0(L_\Theta^2)$ one has
\qq
r_{st}(\theta,\phi)\ =\ \ {\langle\m}U_{e_s,e'_s}U_{e_t,e'_t}\m\theta\m,
\s\phi{\m\rangle}\s{\langle\m}U_{e_t,e'_t}U_{e_s,e'_s}
\m\theta\m,\s\phi{\m\rangle}
\nonumber
\qqq
with $e_s,\m e'_s$ from the list (\ref{ordr}).
The polynomials $r_{st}$ are self-dual:
\qq
r_{st}(q,p)\ =\ r_{st}(p,q)
\qqq
and the self-duality of $\CH$ proven in the present paper forces 
coefficients $h_s$ in Eq.\s\s(\ref{VGP}) to be $q$-independent 
filling partially the gap left in \cite{VGP}. 
An easy but important identity is
\qq
\sum\limits_{t\not=s}r_{st}(q,p)\ =\ (q\cdot p)^2\ =\ 0
\label{sum}
\qqq
for any fixed $s$. It implies that the Hamiltonians
(\ref{VGP}) are preserved up to normalization
by the isomorphisms of the hyperelliptic surfaces induced
by the fractional action \s$\lambda\s\mapsto\s\lambda'=
{a\lambda+b\over c\lambda+d}\s$ of $\SL_2$ on $\NP^1$.

We would still like to fix the values of the constants
$h_s$ in Eqs.\s\s(\ref{VGP}). We claim that they are such 
that the Hitchin map is given by Eq.\s\s(\ref{GR}), i.e. that
\qq
\CH(q,p)\ \ =\ \ -\m{_1\over^{128\m\pi^2}}\s\m\sum\limits_{s,
t=1,\dots,6,\atop s\s\not=\s t}{r_{st}(q,p)\over
(\lambda-\lambda_s)(\lambda-\lambda_t)}\s\m(d\lambda)^2\s.
\label{glr}
\qqq
First note that the above formula is consistent with the
$\SL_2$ transformations. Indeed, relations
(\ref{sum}) imply that
\qq
\sum\limits_{s\s\not=\s t}{r_{st}\over
(\lambda'-\lambda'_s)(\lambda'-\lambda'_t)}\s\m(d\lambda')^2
\ =\ \sum\limits_{s\s\not=\s t}{r_{st}\over
(\lambda-\lambda_s)(\lambda-\lambda_t)}\s\m(d\lambda)^2
\nonumber
\qqq
for $\lambda'={a\lambda+b\over c\lambda+d}$. 
Taking, in particular, $\lambda'=\lambda^{-1}$ one verifies
that the quadratic differentials (\ref{glr}) are
regular at infinity. They are also regular at the branching
points since ${d\lambda\over\sqrt{\lambda-\lambda_s}}$
is a local holomorphic differential around $x_s$. 
Hence the r.h.s. of Eq.\s\s(\ref{glr}) is indeed
a (holomorphic) quadratic differential. Thus 
Eq.\s\s(\ref{glr}) is equivalent to relations (\ref{VGP})
with \s$h_s= {(d\lambda)^2\over(\lambda-\lambda_s)}
\vert_{_{x_s}}\m$, \s modulo an overall normalization. 
To prove Eq.\s\s(\ref{glr}) we shall verify it 
at a point of the phase space for which $\CH(q,p)(x_s)\not=0$
for $s\not=1$. This will fix $h_s$ for $s\not=1$ and hence
all of them (two quadratic differentials equal at points
$x_s$ with $s\not=1$ have to coincide).

Consider a pair $(\theta,\phi_{u_1})$ lying in the product
$\CK\times\CK^*$ of the Kummer quartics with
\qq
\theta(u)\ =\ \ee^{{1\over 2}\m\pi i\s e'_1\cdot\tau\m e'_1\m
+\m 2\pi i\s e'_1\cdot u_1}\ \vartheta(u_1+E_1-u)
\ \vartheta(u_1+E_1+u)\hspace{1.5cm}\cr\cr
=\s\sum\limits_e\s(U_{e_1,e'_1}\theta_{2,e})(u_1)\ 
\theta_{2,e}(u)
\label{121}
\qqq
for $e_1=e'_1=(1,0)$. Note that ${\langle\m}\theta,\phi_{u_1}
\m\rangle=0$.
Eq.\s\s(\ref{first}) together with the relations (\ref{newf}) 
and the equation
\qq
\da_a\theta(u_1)\ =\ -\m\ee^{{1\over 2}\m\pi i\s e_1'\cdot\m\tau 
\m e'_1\s+\s 2\pi i\s e'_1\cdot\m u_1}\s\s\m\da_a\vartheta(E_1)
\ \vartheta(2u_1+E_1)
\nonumber
\qqq
results in the identity
\qq
\CH(\theta,\phi_{u_1})\ =\  
-\m {_{C^2}\over^{16\pi^2}}\s\s\ee^{\m\pi i\s e'_1\cdot\m\tau\m e'_1
\s+\s 4\pi i\s e'_1\cdot\m u_1}\s\s(\da_2\vartheta(E_1))^2\s\s 
\vartheta(2u_1+E_1)^2\s\s(\lambda-\lambda_1)^2
\s\s{_{(d\lambda)^2}\over^{\zeta^2}}
\hspace{0.4cm}
\label{123}
\qqq
where $C$ is the constant appearing in Eq.\s\s(\ref{newf}).
Note that $\CH(\theta,\phi_{u_1})\not=0$ as long as
\s$\vartheta(2u_1+E_1)\not=0$. It follows 
that $\CH(\theta,\phi_{u_1})$ is a quadratic
differential proportional to \s$(\lambda-\lambda_1)^2  
\s{(d\lambda)^2\over\zeta^2}$ which has the $4^{\m\rm th}$ 
order zero at $x_1$. The latter property characterizes it
uniquely up to normalization.

It is not difficult to check that Eq.\s\s(\ref{glr}) gives 
a quadratic differential with the same property. Indeed,
in the language of $q\m$'s and $p\m$'s, the linear form 
\m$\phi_{u_1}$ corresponds to a vector $p\in\NC^4$ 
and $\theta$ to $q=(p_2,-p_1,p_4,-p_3)$. A straightforward
verification shows that $r_{1\m t}(q,p)=0$ for all $t\not=1$.
This implies that the quadratic differential given
by Eq.\s\s(\ref{glr}) vanishes to the second order
at $x_1$. The condition that it vanishes to the fourth 
order is
\qq
\sum\limits_{s\s\not=\s t,\atop s,t\s\not=\s 1}
r_{st}((p_2,-p_1,p_4,-p_3),\s p\m)\ 
\prod\limits_{v\s\not=\s1,s,t}
(\lambda_1-\lambda_v)\ =\ 0\s.
\nonumber
\qqq
A direct calculation shows that this is exactly the equation
(\ref{kuma}) of the Kummer quartic with the coefficients
(\ref{kumb}) so that it holds for
$p$ corresponding to $\phi_{u_1}$. This establishes
proportionality between the Hitchin map and
the right hand side of Eq.\s\s(\ref{glr}) with 
a coefficient that may be still curve-dependent. 

Fixing the overall normalization of the Hitchin map
is more involved. We shall calculate the value
of the quadratic differential on the right hand side 
of Eq.\s\s(\ref{123}) at $\lambda=\lambda_2$ and 
compare it to the value given by Eq.\s\s(\ref{glr}). 
Since this is somewhat technical, we defer the argument
to Appendix 4.

The system with Hamiltonians (\ref{VGP}) bears some similarity 
to the classic Neumann systems\footnote{we thank M. Olshanetsky 
for attracting our attention to this fact}, also anchored
in modular geometry \cite{mumford2}\cite{AT}. The Hamiltonians
of a Neumann system have the form
\qq
\CH_s\ =\ \sum\limits_{1\leq t\not= s\leq n}{J_{st}^2
\over{\lambda_s-\lambda_t}}
\label{Neum}
\qqq
where $J_{st}\s=\s q_sp_t-q_tp_s$ are the functions on $T^*\NC^n$
generating the infinitesimal action of the complex group $\SO_n$:
\qq
&&\hbox to 5cm{$\{J_{st},J_{tv}\}\s=\s -J_{sv}$\hfill}{\rm for\ \ }
s,t,v\ \ {\rm different},\cr
&&\label{son}\cr
&&\hbox to 5cm{$\{J_{st},J_{vw}\}\s=\s0$\hfill}{\rm for\ \ }
s,t,v,w\ \ {\rm different}.
\qqq
The fact that the Hamiltonians (\ref{VGP}) (with constant $h_s$)
Poisson commute reduces, as is well known, to the identities
\qq
&&\hbox to 5cm{$\{r_{st}+r_{sv}\m,\s r_{tv}\}\s=\s 0$\hfill}
{\rm and\  cyclic\ permutations\ thereof}\s,\cr
\label{cyb1}\cr
&&\hbox to 5cm{$\{r_{st}\m,\s r_{vw}\}\s=\s0$\hfill}
{\rm for}\quad\quad\{s,t\}\cap\{v,w\}=\emptyset\s.
\qqq
If we set $r_{st}=J_{st}^{\m2}$ for the Neumann system, then 
Eqs.\s\s(\ref{cyb1}) follow from the relations (\ref{son}). 
It appears that the same algebra stands behind 
the fact\footnote{this is the classical version 
of the observation of \cite{VGDJ}}
that $r_{st}$ given by Eq.\s\s(\ref{rs}) verify 
(\ref{cyb1}). The phase space \s$T^*\CN_{ss}\s\cong\s
\{\s(q,p)\s\m\vert\s\m q\cdot p=0\s\}/\NC^*\m$, \m where
$\NC^*$ acts by $(q,p)\mapsto(tq,\m t^{-1}p)$, may be 
identified with the coadjoint orbit of the group $\SL_4$ 
composed of the traceless complex 4$\times$4 matrices 
$\vert p\rangle\langle q\vert$ of rank 1. Using the isomorphism 
of the complex Lie algebras $\Ssl_4\cong \Sso_6$, we obtain 
the functions $J_{st}=-J_{ts}$ on this $\SL_4$-orbit  
which generate the action of $\Sso_6$ and have the Poisson
brackets given by (\ref{son}). A straightforward check shows
that, for $r_{st}$ of Eq.\s\s(\ref{rs}),
\qq
r_{st}\ =\ -\m4\m J_{st}^{\m 2}
\qqq
so that Eq.\s\s(\ref{cyb1}) follows from the $\Sso_6$-algebra
(\ref{son}). 

Upon the introduction of the rational 
functions \s${r_{st}\over\lambda}\m$,
Eqs.\s\s(\ref{cyb1}) take the form 
\qq
&&\{{_{r_{st}}\over^{\lambda_s-\lambda_t}}\m,\s
{_{r_{sv}}\over^{\lambda_s-\lambda_v}}\}\ +\ \{
{_{r_{st}}\over^{\lambda_s-\lambda_t}}\m,\s
{_{r_{tv}}\over^{\lambda_t-\lambda_v}}\}
\ +\ \{{_{r_{sv}}\over^{\lambda_s-\lambda_v}}\m,
\s{_{r_{tv}}\over^{\lambda_t-\lambda_v}}\} 
\ =\ 0\s,\cr
\label{cyb2}\cr
&&\{{_{r_{st}}\over^{\lambda_s-\lambda_t}}\m,\s
{_{r_{vw}}\over^{\lambda_v-\lambda_w}}\}\ =\ 0\ 
\quad\quad\quad {\rm for}\quad\ \{s,t\}
\cap\{v,w\}=\emptyset\s.
\qqq
The first of these identities is, essentially, the classical 
Yang-Baxter equation. Note, however, that $r_{st}$, 
unlike in the Gaudin and Neumann systems, is not an element 
of a product of two copies of a Poisson algebra 
of functions: there is no sign of an explicit product 
structure, or of a reduction thereof, in our phase space. 
The important question is whether $r_{st}$ come from 
a rational solution of the CYBE. The conformal field theory 
work \cite{Knizh}\cite{Zamo} suggests that the answer may 
be positive, at least in some sense. 

The knowledge of the explicit form of the quadratic differentials
$\CH(q,p)$ allows to write the explicit equations for
the genus 5 spectral curve of the $\SL_2$ Hitchin system at genus 2,
see Eq.\s\s(\ref{1}). They take the form
\qq
\zeta^2\ =\ \prod\limits_{s=1}^6(\lambda-\lambda_s)\s,\quad\ \ 
\xi^2\ =\ \sum\limits_{s\s\not=\s t}r_{st}(q,p)
\prod\limits_{v\s\not=\s s,t}(\lambda-\lambda_v)\s.
\label{specc}
\qqq
The involution of the spectral curve flips the sign of $\xi$.
To extract explicit formulae for the angle variables describing 
the point on the Prym variety of the spectral curve, we would
need, however, a more explicit knowledge of the entire
Lax matrix $\Psi$.

\setcounter{equation}{0}
\medskip
\section{Conclusions}

The main result of the present paper is the proof
of self-duality of the Hitchin Hamiltonians 
on the cotangent bundle to the moduli space 
of the holomorphic $\SL_2$ bundles on a genus 2 complex 
curve. The result was based on an expression for 
the Hitchin Hamiltonians off the Kummer quartic on
which the values of the Hamiltonians were determined
in \cite{VGP}. Using the self-duality,
we were able to complete the analysis of \cite{VGP}
and to obtain the explicit formula (\ref{GR})
for the Hitchin map (\ref{Hm}) giving the action
variables of the integrable system. The explicit 
formula for the angle variables remains still to be found. 
An interesting open problem is an extension of the present 
work to the case with insertion points.

Another important problem related to Hitchin's construction 
is the quantization of the corresponding integrable systems. 
For the $\SL_2$ case such a quantization is essentially
provided by the Knizhnik-Zamolodchikov-Bernard-Hitchin
connection \cite{kz}\cite{bernard:kzb}\cite{bernard:kzbanyg} which describes 
the variation of conformal blocks of the ${\rm SU}_2$ WZW conformal
field theory under the change of the complex structure 
of the curve. The (partition function) conformal blocks 
are holomorphic sections of the $k^{\m\rm th}$-power 
of the determinant line bundle over the moduli space $\CN_{ss}$ 
($k$ is the level of the WZW theory). In our case, they are simply 
$k^{\m\rm th}$-order homogeneous polynomials on $H^0(L_\theta^2)$.
It is easy to quantize the Hitchin Hamiltonians
\qq
H_s\ =\ \sum\limits_{t\s\not=\s s}{r_{st}\over\lambda_s-\lambda_t}\s.
\nonumber
\qqq
If one keeps the original formulae (\ref{rs}) for $r_{st}$ 
in which $p_i$ stands now for ${1\over i}\s\da_{q_i}$, the
relations (\ref{cyb1}) or (\ref{cyb2}) still hold after
the replacement of the Poisson brackets by the commutators. 
One obtains this way the commuting operators $H_s$ mapping 
the space of homogeneous, degree $k$ polynomials 
in variables $q$ into itself. Note, however, that now
\qq
\sum\limits_{t\s\not=\s s}r_{st}\ =\ -\m k(k+4)
\nonumber
\qqq
for each fixed $s$ so that the quantization changes the conformal 
properties of the Hamiltonians. A direct construction of 
the projective version of the KZBH connection for group ${\rm SU}_2$ 
and genus 2 has been recently given in ref.\s\s\cite{VGDJ} 
by following Hitchin's approach \cite{hitchin:flat}. It is consistent
with the above {\it ad hoc} quantization of the classical
Hitchin Hamiltonians.

The integral formulae for the conformal blocks 
\cite{BabF,ReshV,EtinK} or, equivalently, 
the integral formulae for the scalar product of
the conformal blocks \cite{gaw97:unitarity} have been used
at genus 0 and 1 to extract the Bethe Ansatz
eigen-vectors and eigen-values of the quantized 
version of the quadratic Hitchin Hamiltonians.
The Bethe-Ansatz type diagonalization of the 
quantization of the genus 2 Hitchin Hamiltonians
is among the issues that will have to be examined.

Finally, as we stressed in the text, the relations
between the conformal WZW field theory on a genus 2
surface and an orbifold theory in genus 0 requires 
further study.

\setcounter{section}{1}
\renewcommand{\theequation}{A\arabic{section}.\arabic{equation}}

\setcounter{equation}{0}
\medskip
\section*{Appendix 1}

Let us check that $\theta$ given by Eq.\s\s(\ref{thetaE})
vanishes if and only if $$H^0(l_u\otimes E)\s
=\s\{\s(s_1,s_2)\ |\ s_2\in H^0(l_u 
l_{u_1}),\ \de_{_{l_u^{-1}l_{u_1}}}
\hspace{-0.1cm}s_1+s_2\s b=0\s\}\s\ \not=\ \s0\s.$$
For \s$u-u_1\in\NZ^2+\tau\NZ^2\s$ the 1$^{\rm st}$
theta function on the r.h.s. of Eq.\s\s(\ref{Kk}) vanishes
but  \s$l_u=l_{u_1}\s$ and $l_{u_1}\in C_E\m$. Assume
now that \s$u-u_1\s\not\in\s\NZ^2+\tau\NZ^2\m$.
Then \s${\rm dim}\ H^0(l_u^{-1}l_{u_1}K)=1\s$
with a non-zero \s$\chi\in H^0(l_u^{-1}l_{u_1}K)$. 
The necessary and sufficient condition for the solvability 
of the equation \s$\de_{_{l_ul_{u_1}^{-1}}}s_1+s_2\m b=0\s$ 
for a given \s$s_2\in H^0(l_ul_{u_1})\s$ is 
\qq
\int_{_\Sigma}\chi\m s_2\s b=0\s.
\label{soco}
\qqq
If \s$u+u_1\in \NZ^2+\tau\NZ^2\s$ then
\s$l_ul_{u_1}=K\s$ and \s${\rm dim}\s H^0(l_ul_{u_1})=2\s$ 
so that there always is a non-zero solution but
also \s$\theta(u)=0\s$ in this case due to the vanishing of the 
$2^{\rm nd}$ theta function on the r.h.s. of Eq.\s\s(\ref{Kk}).
Finally, if \s$u\pm u_1\s\not\in\s\NZ^2+\tau\NZ^2\s$ then
\s$s_2\in H^0(l_ul_{u_1})\s$ has to be proportional
to the element defined by (\ref{s_2}) and the condition
(\ref{soco}) coincides with the equation \s$\theta(u)=0\m$.

\addtocounter{section}{1}
\setcounter{equation}{0}
\medskip
\section*{Appendix 2}

Let us show that the 1,0-form $\mu$ satisfying relations 
(\ref{cfm}) and (\ref{cfm1}) automatically fulfills the condition
\qq
\int_{_\Sigma}{\kappa}\s\mu\wedge b\s=\s0\s.
\label{orth}
\qqq
Among the infinitesimal gauge field variations $\delta B$
given by Eq.\s\s(\ref{DB}) there are ones which are equivalent 
to infinitesimal gauge transformations:
\qq
\delta B\s=\s\de\Lambda\s+\s[B,\Lambda]\s.
\nonumber
\qqq
Explicitly, for \s$\Lambda=\begin{pmatrix}{-\sigma}&\varphi\\\kappa
&\sigma\end{pmatrix}\s$ with $\sigma$ a function, $\varphi$ a section 
of $l_{u_1}^{-2}$ and $\kappa$ a section of $l_{u_1}^2$, 
this requires that
\qq
\de\kappa\s=\s0\s,\quad\quad
\pi\s\delta u_1\m({\rm Im}\m\tau)^{-1}\bar\omega\s=\s
-\de\sigma+\kappa\m b\s,\quad\quad\delta b\s=\s\de\varphi
+2\m\sigma\m b\s.
\label{three}
\qqq
Such variations may only change the normalization
of the theta function $\theta$. Integrating the second of 
the above relations against forms $\omega^a$ and using 
Eq.\s\s(\ref{kob}) we find that 
\qq
\delta u_1^a\s=\s-{_1\over^{2\pi i}}\s\epsilon^{ab}
\da_{b}\theta(u_1)
\label{du0}
\qqq
for the proper normalization of ${\kappa}$. For such $\delta u_1$
the first term on the right hand side 
of Eq.\s\s(\ref{nd}) gives a theta
function vanishing at $u=u_1$ and may be compensated by the
second term. The 3$^{\rm rd}$ equation of (\ref{three})
gives the compensating $\delta b\in\wedge^{01}(l_{u_1}^{-2})$.
Pairing Eq.\s\s(\ref{nd}) with the above $\delta u_1$ 
and $\delta b$ with the linear form $\phi$, we obtain 
the identity
\qq
{_1\over^i}\s\epsilon^{ab}
\da_{b}\theta(u_1)\s({\rm Im}\m\tau)_{ac}^{-1}
\int_{_\Sigma}\chi^c\wedge b\s+\s2\int_{_\Sigma}
\sigma\s\eta\wedge b\ =\ 0\ .
\label{cri}
\qqq
On the other hand,
\qq
\int_{_\Sigma}{\kappa}\s\mu\wedge b\s=\s
\int_{_\Sigma}\mu\wedge\de\sigma\s-\s{_1\over^{2i}}
\s\epsilon^{ab}\da_b\theta(u_1)\s({\rm Im}\m)^{-1}_{ac}
\int_{_\Sigma}\mu\wedge\bar\omega^c\s\cr
=\s-\int_{_\Sigma}\sigma\s\eta\wedge b\s-\s
{_1\over^{2i}}\s\epsilon^{ab}\da_b\theta(u_1)\s({\rm Im}
\m)^{-1}_{ac}\int_{_\Sigma}\chi^c\wedge b\ =\ 0
\nonumber
\qqq
where we have subsequently used the 2$^{\rm nd}$ equation
in (\ref{three}) with $\delta u_1$ given by Eq.\s\s(\ref{du0}), 
the relation $\de\mu=-\eta\wedge b$ and Eq.\s\s(\ref{cfm})
fixing $\mu$ and, finally, the identity (\ref{cri}).

\addtocounter{section}{1}
\setcounter{equation}{0}
\medskip
\section*{Appendix 3}

It is not difficult to see that there exist a non-zero 
element $P\in S^4 H^0(L_\Theta^2)$, a homogeneous polynomial 
of degree 4 on $H^0(L_\Theta^2)^*$, s.t. 
\qq
P(\phi_{u'})\s=\s0
\nonumber
\qqq
for all $u'\in\NC^2$. Indeed, ${\rm dim}\s S^4H^0(L_\Theta^2)
=({7\atop3})=35$ but the map $u'\mapsto P(\phi_{u'})$ 
defines an even theta function of order 8 and 
${\rm dim}\s H^0_{\rm even}(L_\Theta^8)=34$. 
$P$ is a quartic expression in $\theta_{2,e}(u')$
which vanishes for all $u'$. It has to be preserved
by the $(\NZ/2\NZ)^4$-action (\ref{acone}) and
hence it must be of the form
\qq
P&=&c_1\s(\theta_{2,(0,0)}^{\s4}+\s\theta_{2,(1,0)}^{\s4}
+\s\theta_{2,(0,1)}^{\s4}+\s\theta_{2,(1,1)}^{\s4})\cr\cr
&+&c_2\s(\theta_{2,(0,0)}^{\s2}\s\m\theta_{2,(1,0)}^{\s2}+
\s\theta_{2,(0,1)}^{\s2}\s\m\theta_{2,(1,1)}^{\s2})\cr\cr
&+&c_3\s(\theta_{2,(0,0)}^{\s2}\s\m\theta_{2,(0,1)}^{\s2}+
\s\theta_{2,(1,0)}^{\s2}\s\m\theta_{2,(1,1)}^{\s2})\cr\cr
&+&c_4\s(\theta_{2,(0,0)}^{\s2}\s\m\theta_{2,(1,1)}^{\s2}+
\s\theta_{2,(1,0)}^{\s2}\s\m\theta_{2,(0,1)}^{\s2})\cr\cr
&+&c_5\s\theta_{2,(0,0)}\s\m\theta_{2,(1,0)}\s\m\theta_{2,(0,1)}
\s\m\theta_{2,(1,1)}\s.
\nonumber
\qqq
It is not difficult to calculate the values of coefficients
$c_i$. Denoting $\alpha\equiv\theta_{2,(0,0)}(0)$,
\s$\beta\equiv\theta_{2,(1,0)}(0)$, 
\s$\gamma\equiv\theta_{2,(0,1)}(0)\s$
and \s$\delta\equiv\theta_{2,(1,1)}(0)$,
one has
\qq
c_1&=&\ \ \m(\alpha^2\beta^2-\gamma^2\delta^2)
(\alpha^2\gamma^2-\beta^2\delta^2)(\alpha^2\delta^2
-\beta^2\gamma^2)\s,\cr\cr
c_2&=&-(\alpha^4+\beta^4-\gamma^4-\delta^4)
(\alpha^2\gamma^2-\beta^2\delta^2)(\alpha^2\delta^2
-\beta^2\gamma^2)\s,\cr\label{em}\cr
c_3&=&-(\alpha^4-\beta^4+\gamma^4-\delta^4)
(\alpha^2\beta^2-\gamma^2\delta^2)(\alpha^2\delta^2
-\beta^2\gamma^2)\s,\cr\cr
c_4&=&-(\alpha^4-\beta^4-\gamma^4+\delta^4)
(\alpha^2\beta^2-\gamma^2\delta^2)(\alpha^2\gamma^2
-\beta^2\delta^2)\s,\cr\cr
c_5&=&\ \ \m2\m \alpha\beta\gamma\delta\m
[(\alpha^4-\beta^4+\gamma^4-\delta^4)^2-\m 
4(\alpha^2\gamma^2-\beta^2\delta^2)^2]\s.
\qqq
If we use the basis dual to $(\theta_{2,e})$ to
identify $\phi\in H^0(L_\Theta^2)^*$ with a vector
$p=(p_1,p_2,p_3,p_4)\in\NC^4$, the equation
of the Kummer quartic $\CK^*$ becomes
\qq
c_1\s(p_1^4+p_2^4+p_3^4+p_4^4)\s+\s c_2\s
(p_1^2\m p_2^2\s+\s p_3^2\m p_4^2)
\s+\s c_3\s(p_1^2\m p_3^2\s+\s p_2^2\m p_4^2)\s\m\cr
\label{kuma}\cr
+\s c_4\s(p_1^2\m p_4^2\s+\s p_2^2\m p_3^2)
\s+\s c_5\s p_1\m p_2\m p_3\m p_4\ =\ 0\s.
\qqq
Similarly, identifying $\theta\in H^0(L_\Theta^2)$ with
$q=(q_1,q_2,q_3,q_4)\in\NC^4$ with the help of the basis
$(\theta_{2,e})$, the same equation with $p$ replaced by
$q$ defines the Kummer quartic $\CK$, compare \cite{Kummer}, 
page 81.

We shall also need another well known presentation of the above
equation using the inhomogeneous coordinates of the Weierstrass 
points $\lambda_s$ given by Eq.\s\s(\ref{wp}). It is usually
obtained by beautiful geometric considerations about quadratic
line complexes, see \cite{griffiths}. It may be also obtained
analytically by observing that the multivalued functions
\qq
x\ \mapsto\ \theta_{2,e}(\smallint_{x_0}^x\omega-\Delta)
\nonumber
\qqq
transform like bilinears in \s$\da_a\vartheta(
\smallint_{x_0}^x\omega-\Delta)\m$, \s i.e.\s\s that they
represent quadratic differentials. It follows that
\qq
\sum\limits_e\theta_{2,e}(E_s)\s\s\theta_{2,e}(
\smallint_{x_0}^x\omega-\Delta)\ =\ \vartheta(E_s+
\smallint_{x_0}^x\omega-\Delta)\s\s\vartheta(E_s-
\smallint_{x_0}^x\omega+\Delta)\s\cr
=\ D_s\s\s\left(\da_1\vartheta(E_s')\s\m\da_2\vartheta(
\smallint_{x_0}^x\omega-\Delta)\s-\s
\da_2\vartheta(E_s')\s\m\da_1\vartheta(
\smallint_{x_0}^x\omega-\Delta)\right)\s\label{rlti}\cr
\ \cdot\s\left(\da_1\vartheta(E_s'')\s\m\da_2\vartheta(
\smallint_{x_0}^x\omega-\Delta)\s-\s
\da_2\vartheta(E_s'')\s\m\da_1\vartheta(
\smallint_{x_0}^x\omega-\Delta)\right)\s
\qqq
where $E_s=\frac{1}{2}\,(e_s+\tau e'_s)\s$ is an odd characteristics
from the list (\ref{ordr}) and $E_s'\m,\s E_s''$ are the
two other ones s.t. $E_s+E_s'=E_s''\s\m{\rm mod}\m(\NZ^2+\tau\NZ^2)$.
The odd characteristics $E_s,\m E_s',\m E_s''$ are
either a permutation of $E_1,\m E_4,\m E_5$ or 
a permutation of $E_2,\m E_3,\m E_6$.
The relations (\ref{rlti}) hold since both sides represent
a quadratic differential with double zeros at the Weierstrass
points corresponding to $E_s'$ and $E_s''$. One may obtain
expressions for the coefficients $D_s$ by the de l'Hospital
rule applied twice at those points. Specifying then
$\smallint_{x_0}^x\omega-\Delta$ to $E_s$ or to 3 remaining
odd characteristics one obtains relations for
quadratic combinations of $\theta_{2,e}(0)$ of the form
\s$\pm\alpha^2\pm\beta^2\pm\gamma^2\pm\delta^2\s$
with 2 plus and 2 minus signs as well as for
\s$\alpha\beta\pm\gamma\delta\m$, 
\s$\alpha\gamma\pm\beta\delta\m$
and \s$\alpha\delta\pm\beta\gamma\m$. 
These relations may be used to compute the ratios 
of the coefficients $c_i$ (\ref{em})
which become functions of $\lambda_s$ only.
One obtains this way an alternative expression  
for the coefficients $c_i$ 
\qq
c_1&=&(\lambda_1-\lambda_2)(\lambda_3-\lambda_4)
(\lambda_5-\lambda_6)\s,\cr\cr
c_2&=&2(\lambda_1-\lambda_2)
((\lambda_3-\lambda_5)(\lambda_4-\lambda_6)
+(\lambda_3-\lambda_6)(\lambda_4-\lambda_5))\s,\cr\cr
c_3&=&-2(\lambda_3-\lambda_4)((\lambda_1-\lambda_5)
(\lambda_2-\lambda_6)
+(\lambda_1-\lambda_6)(\lambda_2-\lambda_5))\s,
\label{kumb}\cr\cr
c_4&=&2(\lambda_5-\lambda_6)((\lambda_1-\lambda_3)
(\lambda_2-\lambda_4)
+(\lambda_1-\lambda_4)(\lambda_2-\lambda_3))\s,\cr\cr
c_5&=&-2(\lambda_1-\lambda_3)((\lambda_4-\lambda_5)
(\lambda_2-\lambda_6)
+(\lambda_4-\lambda_6)(\lambda_2-\lambda_5))\cr
&&-2(\lambda_1-\lambda_4)((\lambda_3-\lambda_5)
(\lambda_2-\lambda_6)
+(\lambda_3-\lambda_6)(\lambda_2-\lambda_5))\cr
&&-2(\lambda_1-\lambda_5)((\lambda_2-\lambda_4)
(\lambda_3-\lambda_6)
+(\lambda_2-\lambda_3)(\lambda_4-\lambda_6))\cr
&&-2(\lambda_1-\lambda_6)((\lambda_2-\lambda_4)
(\lambda_3-\lambda_5)
+(\lambda_2-\lambda_3)(\lambda_4-\lambda_5))\s.
\qqq
equivalent to the previous one up to normalization.
Note that the $\SL_2$ transformations 
\s$\lambda_s\mapsto{a\lambda_s+b\over c\lambda_s+d}\s$ 
preserve the form the quartic equation.
The virtue of the analytic approach is that it
also provides useful expressions for the non-homogeneous 
ratios like e.g.
\qq
{\alpha\beta+\gamma\delta\over\alpha^2\gamma^2-\beta^2\delta^2}
\ =\ -\s{\ee^{-\m{1\over 2}\m \pi i\s(1,0)\cdot\tau\m(1,0)}\over
2\s\m C^2\s\s(\da_2\vartheta(E_1))^2}\ {(\lambda_2-\lambda_5)
(\lambda_2-\lambda_6)(\lambda_3-\lambda_4)\over\lambda_1-\lambda_2}\s.
\label{tbus}
\qqq
$C^2$ is given by the equations
\qq
C^2\s=\s\hf\s{_{(\da_1\vartheta)^3\s\da_2^3\vartheta
\s-\s 3\s(\da_1\vartheta)^2\s\da_2\vartheta
\s\da_1\da_2^2\vartheta
\s+\s 3\s\da_1\vartheta\s(\da_2\vartheta)^2
\s\da_1^2\da_2\vartheta
\s-\s(\da_2\vartheta)^3
\s\da_1^3\vartheta}\over^{(\da_2\vartheta)^4}}\bigg\vert_{E_s}
\ \prod\limits_{t\s\not=\s s}(\lambda_s-\lambda_t)\hspace{0.4cm}
\nonumber
\qqq
holding for any fixed $s$. It is not difficult to see by 
differentiating twice Eq.\s\s(\ref{inho}) at $x=x_s$ 
that $C$ is the same constant that appears in Eq.\s\s(\ref{newf}). 
The expression (\ref{tbus}) is used below to fix the 
normalization of the Hitchin map.

\addtocounter{section}{1}
\setcounter{equation}{0}
\medskip
\section*{Appendix 4}

We shall show here that the overall normalization of the Hitchin map
is as in Eq.\s\s(\ref{glr}). Since 
\qq
&&\ee^{\m\pi i\s e_1'\cdot\tau\m e_1'\s+\s 4\pi i\s e_1'
\cdot u_1}\s\s\vartheta(2u_1+E_1)^2\s\cr\cr
&&=\s-\m\ee^{\m\pi i\s e_1'\cdot\tau\m e_1'}
\s\s\vartheta(2u_1+E_1)\s\s\vartheta(2u_1-E_1)
=\s-\m\ee^{\m\pi i\s e_1'\cdot\m\tau\m e_1'}\s\sum\limits_e
\theta_{2,e}(E_1)\s\s\theta_{2,e}(2u_1)\s\cr
&&=\s-\m\ee^{\m{1\over 2}\m\pi i\s(1,0)\cdot\tau(1,0)}
\s\sum\limits_e(-1)^{(1,0)\cdot e}\s\s\theta_{2,e+(1,0)}(0)
\s\s\theta_{2,e}(2u_1)\s,
\nonumber
\qqq
the coefficient of ${(d\lambda)^2\over\zeta^2}$ 
on the right hand side of Eq.\s\s(\ref{123}) takes 
at $\lambda=\lambda_2$ the value
\qq
\m{_{C^2}\over^{16\pi^2}}\s\s\ee^{\m{1\over 2}\m\pi i\s(1,0)
\cdot\tau(1,0)}\ (\da_2\vartheta(E_1))^2\ 
(\lambda_1-\lambda_2)^2\s\s(\beta\m\theta_{2,(0,0)}(2u_1)
\s-\s\alpha\m\theta_{2,(1,0)}(2u_1)\cr
+\s\delta\m\theta_{2,(0,1)}(2u_1)\s
-\s\gamma\m\theta_{2,(1,1)}(2u_1))
\label{ha}
\qqq
in the notations of Appendix 3.
This coefficient should coincide with the one obtained from the 
right hand side of Eq.\s\s(\ref{glr}) which is equal to
\qq
-\m{_1\over^{64\pi^2}}\sum\limits_{t\s\not=\s2}r_{2\m t}(q,p)\s
\prod\limits_{v\s\not=\s2,t}(\lambda_2-\lambda_v)
\label{ho}
\qqq
calculated at $(q,p)$ corresponding to $(\theta,\phi_{u_1})$
with $\theta$ given by Eq.\s\s(\ref{121}). The respective
values of $r_{st}$ are:
\qq
r_{1\m t}&=& 0\s,\cr\cr
r_{23}&=&2\s(-\m\alpha\gamma^2
\m\theta_{2,(0,0)}(2u_1)\s-\s\beta\delta^2\s\theta_{2,(1,0)}(2u_1)
\s-\s\gamma\alpha^2\s\theta_{2,(0,1)}(2u_1)\cr
&&-\s\delta\beta^2\s\theta_{2,(1,1)}(2u_1)
\s-\s\beta\gamma\delta\s\theta_{2,(0,0)}(2u_1)
\s-\s\alpha\gamma\delta\s\theta_{2,(1,0)}(2u_1)\cr
&&-\s\alpha\beta\delta\s\theta_{2,(0,1)}(2u_1)
\s-\s\alpha\beta\gamma\s\theta_{2,(1,1)}(2u_1)\m)\s,\cr\cr
r_{24}&=&2\s(\m\alpha\gamma^2
\s\theta_{2,(0,0)}(2u_1)\s+\s\beta\delta^2\s\theta_{2,(1,0)}(2u_1)
\s+\s\gamma\alpha^2\s\theta_{2,(0,1)}(2u_1)\cr
&&+\s\delta\beta^2\s\theta_{2,(1,1)}(2u_1)
\s-\s\beta\gamma\delta\s\theta_{2,(0,0)}(2u_1)
\s-\s\alpha\gamma\delta\s\theta_{2,(1,0)}(2u_1)\cr
&&-\s\alpha\beta\delta\s\theta_{2,(0,1)}(2u_1)
\s-\s\alpha\beta\gamma\s\theta_{2,(1,1)}(2u_1)\m)\s,\label{00}\cr\cr
r_{25}&=&2\s(\m\alpha\delta^2
\s\theta_{2,(0,0)}(2u_1)\s+\s\beta\gamma^2\s\theta_{2,(1,0)}(2u_1)
\s+\s\gamma\beta^2\s\theta_{2,(0,1)}(2u_1)\cr
&&+\s\delta\alpha^2\s\theta_{2,(1,1)}(2u_1)
\s+\s\beta\gamma\delta\s\theta_{2,(0,0)}(2u_1)
\s+\s\alpha\gamma\delta\s\theta_{2,(1,0)}(2u_1)\cr
&&+\s\alpha\beta\delta\s\theta_{2,(0,1)}(2u_1)
\s+\s\alpha\beta\gamma\s\theta_{2,(1,1)}(2u_1)\m)\s,\cr\cr
r_{26}&=&2\s(-\m\alpha\delta^2
\s\theta_{2,(0,0)}(2u_1)\s-\s\beta\gamma^2\s\theta_{2,(1,0)}(2u_1)
\s-\s\gamma\beta^2\s\theta_{2,(0,1)}(2u_1)\cr
&&-\s\delta\alpha^2\s\theta_{2,(1,1)}(2u_1)
\s+\s\beta\gamma\delta\s\theta_{2,(0,0)}(2u_1)
\s+\s\alpha\gamma\delta\s\theta_{2,(1,0)}(2u_1)\cr
&&+\s\alpha\beta\delta\s\theta_{2,(0,1)}(2u_1)
\s+\s\alpha\beta\gamma\s\theta_{2,(1,1)}(2u_1)\m)\s.
\nonumber
\qqq
Multiplying the coefficients at subsequent
$\theta_{2,e}(2u_1)$ in expression (\ref{ha}) 
by $\alpha,\s-\beta,\s\gamma$ and $-\delta$, respectively, 
and summing them up we obtain
\qq
\m{_{C^2}\over^{8\pi^2}}\s\s\ee^{\m{1\over 2}\m\pi i\s 
(1,0)\cdot\tau(1,0)} 
\ (\da_2\vartheta(E_1))^2\ 
(\lambda_1-\lambda_2)^2\s\s(\alpha\beta+\gamma\delta)\s.
\nonumber
\qqq
A similar operation on expression (\ref{ho}) gives
\qq
-\m{_1\over^{16\pi^2}}\s\s(\lambda_1-\lambda_2)
(\lambda_2-\lambda_5)
(\lambda_2-\lambda_6)(\lambda_3-\lambda_4)
\s(\alpha^2\gamma^2-\beta^2\delta^2)\s.
\nonumber
\qqq
The equality of the two expressions follows from 
Eq.\s\s(\ref{tbus}). This verifies the correctness
of the overall normalization of the Hitchin map in 
Eq.\s\s(\ref{glr}).



\newpage
\addcontentsline{toc}{section}{\protect\numberline{}{\bf
	 Hitchin systems at low genera}}
\setcounter{section}{0}
\newcommand{\LieG}{\mathfrak{g}}
\newcommand{\LieT}{\mathfrak{t}}
\newcommand{\groupesl}{\mathfrak{sl}}
\newcommand{\groupeso}{\mathfrak{so}}

\renewcommand{\theequation}{\arabic{section}.\arabic{equation}}

\HRule
\begin{center}
{\Large \bf
	Hitchin systems at low genera}\\
\vskip 1cm

{\sc Krzysztof Gaw\c{e}dzki}\\
I.H.E.S., C.N.R.S., F-91440 Bures-sur-Yvette, France\\

\vskip 0.5 cm

{\sc Pascal Tran-Ngoc-Bich}\\
Universit\'e de Paris-Sud, F-91405 Orsay, France

\vskip 1cm

\begin{quote}
{\bf Abstract.} The paper gives a quick account of 
the simplest cases of the Hitchin integrable systems
and of the Knizhnik-Zamolodchikov-Bernard connection
at genus $0$, $1$ and $2$. In particular,
we construct the action-angle variables of the
genus $2$ Hitchin system with group ${\rm SL}_2$ by exploiting 
its relation to the classical Neumann integrable systems.
\end{quote}
\end{center}

\setcounter{equation}{0}
\medskip
\section{Hitchin systems}

As was realized by Hitchin in \cite{hitchin},
a large family of integrable systems may be obtained
by a symplectic reduction of a chiral 2-dimensional
gauge theory. Let $\Sigma$ denote a closed Riemann surface 
of genus $g$ and let $G$ be a complex Lie group which
we shall assume simple, connected and simply connected. 
We shall denote by $\CA$ the space of $\LieG$-valued
0,1-gauge fields\footnote{one may work in a fixed
smoothness class and use the Sobolev norms to define 
topology in $\CA$} $A=A_{\overline{z}}\,d\overline{z}$
on $\Sigma$. Hitchin's construction \cite{hitchin} associates 
to $\Sigma$ and $G$ an integrable system obtained by a symplectic 
reduction of the infinite-dimensional complex symplectic 
manifold $T^*\CA$ of pairs $(A,\Phi)$ where $\Phi=\Phi_{z}\, dz$ 
is a $\LieG$-valued 1,0-Higgs field. The holomorphic symplectic 
form on $T^*\CA$ is
\qq
\int_\Sigma\tr\,\, \delta\Phi\s\s\delta A
\qqq
where $\tr$ stands for the Killing form on $\LieG$ normalized
so that $\tr\,\phi^2=2$ for the long roots $\phi$.
The local gauge transformations $h\in{\CG}\equiv {\rm Map}(\Sigma,G)$
act on $T^*\CA$ by 
\qq
A\longmapsto {}^hA\,\equiv\, hAh^{-1}+h\de h^{-1},\ \ \quad
\Phi\longmapsto {}^h\Phi\,\equiv\, h\Phi h^{-1}
\qqq
preserving the symplectic form. The corresponding
moment map $\,\mu:\, T^*\CA\longrightarrow {\rm Lie}(\CG)^*\cong$
$\wedge^2(\Sigma)\otimes \LieG\,$ takes the form
\qq
\mu(A,\Phi)\ =\ \de\Phi+A\Phi+\Phi A.
\qqq
The symplectic reduction gives the reduced phase space
\qq
\CP\ =\ {\mu}^{-1}(\{0\})\,/\,\CG
\label{CP}
\qqq
with the symplectic structure induced from that of $T^*\CA$.
$\CP$ may be identified with the complex cotangent 
bundle $T^*\CN$ to the orbit space $\CN=\CA/\CG$
and $\CN$, in turn, with the moduli space of holomorphic 
$G$-bundles on $\Sigma$. More precisely,
care should be taken to avoid non-generic bad orbits
in order to obtain tractable orbit spaces. This may be done 
by considering only gauge fields $A$ leading to stable 
$G$-bundles forming smooth moduli space $\CN_s$
or those leading to semi-stable bundles giving rise
to a, generally singular, compactification $\CN_{ss}$
of $\CN_s$. In what follows we shall be somewhat cavalier 
about such details. 

The Hitchin system has $\CP$ as its phase space.
Its Hamiltonians are obtained the following way. 
Let $p$ be a homogeneous ${\rm Ad}$-invariant polynomial
on $\LieG$ of degree $d_p$. Then
\qq
h_p(A,\Phi)\ =\ p(\Phi)=p(\Phi_z)(dz)^{d_p}
\qqq
defines a $d_p$-differential on $\Sigma$ which
is holomorphic if $\mu(A,\Phi)=0$. Since 
$h_p$ is constant on the orbits of $\CG$,
it descends to the reduced phase space:
\qq
h_p\,:\,\CP\longrightarrow H^{0}(K^{d_{p}})\m.
\qqq
Here $K$ stands for the canonical bundle 
(of covectors $\propto dz$) and $H^0(K^{d_p})$
is the (finite-dimensional) vector space of the holomorphic
$d_p$-differentials on $\Sigma$. The (components of)
$h_p$ Poisson-commute (they Poisson-commute already 
as functions on $T^*\CA$ since they depend only 
on the "momenta" $\Phi$). The point of Hitchin's
construction is that, by taking a complete 
system of polynomials $p$, one obtains on $\CP$
a complete system of Hamiltonians in involution. 
For the matrix groups, the values of Hamiltoniens 
$h_p$ at a point of $\CP$ may be encoded 
in the spectral curve $\CC$ obtained by solving 
the characteristic equation
\qq
\det(\Phi-\xi)\,=\,0
\label{det}
\qqq
for $\xi\in K$. The spectral curve of the eigenvalues $\xi$ 
is a ramified cover of $\Sigma$. The corresponding eigenspaces 
of $\Phi$ form then a holomorphic line bundle over $\CC$
belonging to a subspace of the Jacobian of $\CC$
on which the Hamiltonians $h_p$ induce linear flows.

For the quadratic polynomial $p_2=\frac{1}{2}\tr$, 
the map $h_{p_2}$ takes values in the space
of holomorphic quadratic differentials $H^0(K^2)$. This is
the space cotangent to the moduli space of complex curves $\Sigma$.
Variations of the complex structure of $\Sigma$ are described 
by Beltrami differentials $\delta\mu=\delta\mu_{\bar z}^z\,
\partial_z d\bar z$ such that $z'=z+\delta z$ with
$\da_{\bar z}\delta z=\delta\mu_{\bar z}^z$ gives new 
complex coordinates. The Beltrami differentials $\delta\mu$ may 
be paired with quadratic differentials $\beta$ by 
\qq
(\beta,\delta\mu)\ \mapsto\ \int_{\Sigma}\beta\,\delta\mu. 
\label{pair}
\qqq
The differentials $\delta\mu=\de(\delta\xi)$, where $\delta\xi$ 
is a vector field on $\Sigma$, describe variations of the complex 
structure due to diffeomorphisms of $\Sigma$ and they pair 
to zero with $\beta$. The quotient space $H^1(K^{-1})$
of differentials $\delta\mu$ modulo $\de(\delta\xi)$
is the tangent space to the moduli space of curves $\Sigma$ 
and $H^0(K^2)$ is its dual. The pairing (\ref{pair}) defines 
then for each $[\delta\mu]\in H^1(K^{-1})$
a Hamiltonian 
\qq
h_{\delta\mu}\ \equiv\ \int_\Sigma h_{p_2}\m\delta\mu\m.
\label{hmu}
\qqq
The Hamiltonians $h_{\delta\mu}$ Poisson-commute 
for different $\delta\mu$.

Hitchin's construction possesses a natural generalization
\cite{Mark}\cite{Nekr}\cite{EnR}\cite{gaw97:unitarity}. Let 
$x_n\in\Sigma$ be a finite family of distinct points in $\Sigma$
and $\CO_n$ a family of (co)adjoint orbits in $\LieG^*\cong \LieG$. 
\qq
\CO\ =\ \{\,\sum\limits_n\lambda_n\delta_{x_n}\ \vert\ \lambda_x
\in{\CO}_n\,\},
\qqq
where $\delta_x$ stands for the Dirac delta measure at $x$, 
forms a coadjoint orbit of the group $\CG$ of local gauge 
transformations. In the symplectic reduction we may replace 
definition (\ref{CP}) with
\qq
\CP_{_{\CO}}\ =\ \mu^{-1}(\{\CO\})\,/\,\CG\ \cong\ 
\mu^{-1}(\sum\lambda_n\delta_{x_n})\,/\,
\CG_{\underline{\lambda},\underline{x}}
\label{CPG}
\qqq
where $\CG_{\underline{\lambda},\underline{x}}$ is the 
subgroup of $\CG$ fixing $\sum\lambda_n\delta_{x_n}$.
Upon restriction to properly defined stable  
pairs $(A,\Phi)$, $\CP_{_{\CO}}$ gives a smooth space 
with a semi-stable compactification \cite{Mark}.
Its second representation in (\ref{CPG}) allows 
to equip $\CP_{_{\CO}}$ with the symplectic structure 
inherited from $T^*\CA$. The reduced Hamiltonians $h_p$ 
take now values in $H^0(K^{d_p}(d_p\sum x_n))$, i.e. define 
meromorphic $d_p$-differentials with possible poles at $x_n$
of order $\leq d_p$ and they still define an integrable
system on $\CP_{_{\CO}}$. In particular, $h_{p_2}$ takes values
in $H^0(K^2(2\sum x_n))$ which is dual to $H^1(K^{-1}(-2\sum x_n))$,
the tangent space to the moduli space of curves $\Sigma$ with
fixed punctures $x_n$ and first jets at $x_n$ of holomorphic 
local parameters $z_n$, $z_n(x_n)=0$. The corresponding Beltrami 
differentials $\delta\mu$ behave like $\CO(z_n^2)$ around $x_n$ 
and they are taken modulo $\de(\delta\xi)$ where the vector fields 
$\xi$ are also $\CO(z_n^2)$ around $x_n$ (such vector fields
do not change the first jets at $x_n$ of the local parameters 
$z_n$). \m$\delta\mu$ may still be coupled to quadratic 
differentials $\beta\in H^0(K^2(2\sum x_n))$ by (\ref{pair}) 
and Eq.\s\s(\ref{hmu}) defines for $[\delta\mu]\in 
H^1(K^{-1}(-2\sum x_n))$ Hamiltonians on $\CP_{_{\CO}}$ 
that are in involution.

\setcounter{equation}{0}
\medskip
\section{Knizhnik-Zamolodchikov-Bernard connection}

The phase space $\CP\cong T^*\CN$ may be 
(geometrically) quantized by considering the space $H^0({\CL}^k)$ 
of holomorphic sections of the $k^{\rm th}$ power of the determinant 
line bundle ${\CL}$ over $\CN$ (more exactly, over its semi-stable 
version $\CN_{ss}$) as the space of quantum states. Such sections
are given by holomorphic functions $\psi$ on $\CA$ satisfying 
the Ward identity
\qq
\label{CS}
 \psi(A)=\ee^{-k\,S(h,A)}\,\psi({}^{h^{-1}}\hspace{-0.15cm}A)
\qqq
for $h\in\CG$ and with $S(h,A)$ standing for the action of the 
gauged Wess-Zumino-Novikov-Witten (WZNW) model.
The identity (\ref{CS}) expresses the gauge invariance 
on the quantum level. 
The vector spaces $H^0({\CL}^k)$ arise naturally in the context
of the WZNW model and of the Chern-Simons theory \cite{witten:jones}.
They are finite-dimensional and their dimension is given 
by the Verlinde formula \cite{verlinde}. Put together for different
complex structures of $\Sigma$, they form a holomorphic vector 
bundle $\CW$ over the moduli space of complex curves. 
In the language of functions $\psi$, the $\de$-operator 
of this bundle is given by
\qq
\de_{\overline{\delta\mu}}\psi\,=\,
\left(d_{\overline{\delta\mu}}\,+\,{_k\over^{4\pi i}}
\int_{\Sigma}\tr\,\m (A\m\overline{\delta\mu})\, A\right)\psi
\qqq
where $d_{\overline{\delta\mu}}$ differentiates $\psi$
viewed as a function of the unitary gauge field 
$B=-A^*+A=-A_{\bar z}^*dz+A_{\bar z}d\bar z$ 
(functions of $B$ are naturally identified 
for different complex structures on $\Sigma$). The bundle
$\CW$ may be equipped with a projectively flat connection 
$\nabla^{\rm KZB}$ \cite{witten:jones}\cite{hitchin:flat} which may be traced 
back to the works of Knizhnik-Zamolodchikov \cite{kz} 
and Bernard \cite{bernard:kzb}\cite{bernard:kzbanyg}. In the present 
description of $\CW$, the KZB connection takes the form 
\cite{gaw89:construc}
\qq
&&\nabla^{\rm KZB}_{\delta\mu}\psi\,=\,\left(d_{\delta\mu}\,-\,
	\int_\Sigma\tr\,A^*\m(
	\frac{_\delta}{^{\delta A}}\,\delta\mu)
	\,-\,\frac{_{\pi i}}{^\kappa}
	\int_\Sigma\tr\ {}^\bullet_\bullet\,
	\frac{_\delta}{^{\delta A}}
	(\frac{_\delta}{^{\delta A}}\,\delta\mu)
	\m{}^\bullet_\bullet\right)\psi\, ,\label{KZB}\\
	\cr
	&&\nabla^{\rm KZB}_{\overline{\delta\mu}}\psi\ =\ 
	\de_{\overline{\delta\mu}}\,\psi
	\qqq
where $\kappa=k+g^\vee$ with $g^\vee$ denoting the dual Coxeter 
number of $G$. The symbol ${}^\bullet_\bullet\ {}^\bullet_\bullet$  
indicates that one should remove the singularity at the coinciding 
points of $\frac{\delta}{\delta A(x)}\frac{\delta}{\delta A(y)}\psi$
before setting $x=y$. How this is precisely done depends on some 
choices (e.g. of a projective connection or a metric on each 
$\Sigma$) but the choices lead to connections differing
by addition of a scalar form. 

The second order operator on the right hand side of 
Eq.\s\s(\ref{KZB}) has the principal symbol (obtained by 
replacement of $\frac{\delta}{\delta A}$ by
$\frac{k}{2\pi i}\Phi$) proportional to the Hitchin Hamiltonian 
$h_{\delta\mu}$. The KZB connection $\nabla^{\rm KZB}_{\delta\mu}$
may be considered a quantization of $\frac{k}{2\pi i}h_{\delta\mu}$ 
which, instead of acting in a fixed space $H^0({\CL}^k)$ relates 
two such spaces for the complex structures differing 
by $\delta\mu$\cite{hitchin:flat}. Note that if we rescale 
$\delta\mu\mapsto\kappa\delta\mu$,
we should obtain from the KZB connection 
in the limit $\kappa\to0$ operators acting in the space 
$H^0({\CL}^{-g^\vee})$ corresponding to a fixed complex structure. 
This space becomes non-trivial if we admit singular sections
of $\CL$ or work with higher cohomologies of $\CL^{-g^\vee}$. 
It also admits a quantization of the non-quadratic Hitchin 
Hamiltonians \cite{BeilDr}\cite{Frenkl}. For $k\not=-g^\vee$
we may also obtain from Eq.\s\s(\ref{KZB}) operators
in a single space if we chose a local trivialization 
of the bundle $\CW$ (or of a bundle $\CW'\supset\CW$).

The above quantization extends to the case of the phase space
$\CP_{_{\CO}}$ if the coadjoint orbits $\CO_n$ associated
to points $x_n\in\Sigma$ correspond to irreducible holomorphic 
representations of $G$ in vector spaces $V_n$ (i.e. to irreducible 
unitary representations of the compact form of $G$). 
The quantum states are now represented by holomorphic maps 
on $\CA$ with values in $V=\mathop{\otimes}\limits_nV_n$ 
satisfying the Ward identities
\qq
\label{CS1}
 \psi(A)=\ee^{-k\,S(h,A)}\,\mathop{\otimes}\limits_nh(x_n)
 \,\psi({}^{h^{-1}}\hspace{-0.15cm}A)
\qqq
for $h\in\CG$ generalizing Eq.\s\s(\ref{CS}). The spaces 
of solutions are still finite-dimensional and form a holomorphic
vector bundle over the moduli space of punctured curves with first 
jets of local parameters at the punctures. The complex structure 
and the KZB connection are given by the same formulae with
Beltrami differentials $\delta\mu$ restricted to behave 
like $\CO(z_n^2)$ at the punctures. Since $\frac{\delta}
{\delta A(x)}\psi=\CO(z_n^{-1})$ and $\tr\, {}^\bullet_\bullet
\frac{\delta}{{\delta A(x)}}\frac{\delta}{{\delta A(x)}}
{}^\bullet_\bullet\m\psi=\CO(z_n^{-2})$\m\ around $x_n$,
there is no problem of convergence of the integrals over $\Sigma$. 
Again, $\nabla^{\rm KZB}_{\delta\mu}$ may be viewed as the quantization
of the Hitchin Hamiltonian $\frac{k}{2\pi i}h_{\delta\mu}$ and 
all the above remarks apply.

\setcounter{equation}{0}
\medskip
\section{Genus zero}

Up to diffeomorphisms, there is only one 
Riemann surface of genus zero: the Riemann sphere $\NP^1
=\NC\cup\{\infty\}$. On $\NP^1$, the gauge orbit of the zero gauge 
field is open and dense in $\CA$, i.e. the generic gauge field 
takes the form
\qq
A=h^{-1}\de h
\qqq
where $h\in\CG$ is determined up to left multiplication by
a constant $g\in G$. The equation $\mu(A,\Phi)=
\sum_n\lambda_n\delta_{z_n}$, with $\lambda_n$ belonging
to the (co)adjoint orbit $\CO_n$ associated to the puncture 
$z_n$, becomes
\qq
\de({}^h\Phi)=\sum_n\nu_n\delta_{z_n}
\qqq
where $\nu_n=h(z_n)\lambda_n h(z_n)^{-1}\in\CO_n$. 
This equation has a (unique) solution 
\qq
{}^h\Phi(z)\,=\,\sum_n\frac{\nu_n}{z-z_n}\,\frac{_{dz}}{^{2\pi i}}
\qqq
if and only if the sum of residues is zero, i.e.\s\s if 
$\sum_n\nu_n=0$. We obtain then for $\CP_{_{\CO}}$ defined 
by Eq.\s\s(\ref{CPG}):
\qq
       \CP_{_{\CO}}\,\cong\,
       \Big\{\underline{\nu}\in\mathop{\times}\limits_n\CO_n
       \,\Big|\,\sum_n\nu_n=0\Big\}\bigg/ {G}\,.
\qqq
The (co)adjoint orbits carry a natural symplectic structure
leading to the Poisson bracket $\{\nu^a,\nu^b\}\,
=\,i\m f^{abc}\m\nu^b$ for $\nu^a=\tr\, t^a\nu$ where $t^a$ 
are the generators of $\LieG$ s.t. $\tr\, t^at^b=\delta^{ab}$ 
and $[t^a,t^b]=if^{abc}t^c$.
It is easy to check that the complex symplectic structure 
on $\CP_{_{\CO}}$ coincides with the one obtained by 
the symplectic reduction of $\times_n\CO_n$ with respect 
to the diagonal action of ${G}$. 

The Hamiltonians $h_p$ are  
\qq
h_{p}(z)=p\Big(\sum_n\frac{\nu_n}
	{z-z_n}\Big)\,(\frac{_{dz}}{^{2\pi i}})^{d_p}.
\qqq
In the special case $p_2=\frac{1}{2}\tr$, 
\qq
h_{p_2}\,=\,{_1\over^2}\sum_{n,m}
	\frac{\nu_n^a\nu_m^a}{(z-z_n)(z-z_m)}
	\,(\frac{_{dz}}{^{2\pi i}})^2\,=\,
	\sum_n
	\left(\frac{1}{(z-z_n)^2}\,\delta_n
	+\frac{1}{z-z_n}\,h_n
	\right)(\frac{_{dz}}{^{2\pi i}})^2
\label{h21}
\qqq
where 
\qq
\delta_n={_1\over^2}\m\nu^a_n\nu^a_n\,,\qquad
h_n=\sum_{m\neq n}\frac{\nu^a_n\nu^a_{m}}{z_n-z_{m}}\,.
\qqq
Let $\delta\mu$ be a Beltrami differential regular at infinity
and behaving like $\CO((z-z_n)^2)$ around the insertions.
Necessarily, $\delta\mu=\de(\delta\xi)$ for a regular vector
field $\delta\xi=\delta\xi^z\da_z$ on $\NP^1$. $\delta\xi$
is determined up to infinitesimal M\"{o}bius transformations
$(a+bz+cz^2)\da_z$. We may take $z'=z+\delta\xi^z$ as the new 
complex coordinate on $\NP^1$ with the modified complex structure. 
The modification is then equivalent to the shift $\delta z_n=\delta
\xi^z(z_n)$ of the insertion points and the shift 
$\delta\chi_n=\chi_n\m\da_z(\delta\xi_n^z)(z_n)$ of the first jet 
of the local parameter at the punctures parametrized 
by the $\da_z$-derivative $\chi_n$ of the parameter at $z_n$. 
An easy calculation involving cutting out small balls
around the insertions and integration by parts shows that
\qq
h_{\delta\mu}\ =\ \sum_n\int_\Sigma
	\left(\frac{1}{(z-z_n)^2}\,\delta_n
	+\frac{1}{z-z_n}\,h_n
	\right)(\frac{_{dz}}{^{2\pi i}})^2\,\,\de(\delta\xi)\cr
	=\ {_1\over^{2\pi i}}\sum_n
	\left(\delta_n\m\,\chi_n^{-1}\delta\chi_n
	\,+\, h_n\,\delta z_n\right).
\label{hmu012}
\qqq

The quantum states $\psi$ at genus zero may be
labeled by their values $\psi(0)$ at $A=0$
which belong to the subspace $V^G$ of the $G$-invariant 
tensors in $V\equiv\mathop{\otimes}_n V_n$. Indeed, 
$\psi$ is determined by its values on the dense $\CG$-orbit 
of $A=0$. Hence the bundle $\CW$ is a subbundle of the trivial 
bundle with the fiber $V^G$. The KZ(B) connection reduces 
in this case \cite{kz} to the formula
\qq
&&\nabla^{\rm KZ}_{\delta\mu}\psi(0)\ =\ 
\sum\limits_n\left(\delta\chi_n(\da_{\chi_n}-\chi_n^{-1}\Delta_n)
\,+\,\delta z_n(\da_{z_n}-{_1\over^{\kappa}}\m H_n)
\right)\psi(0)\, ,\label{KZ012}\\
&&\nabla^{\rm KZ}_{\overline{\delta\mu}}\psi(0)\ =\ 
\sum\limits_n\left(\overline{\delta\chi}_n\da_{\bar\chi_n}
\,+\,\overline{\delta z}_n\da_{\bar z_n}
\right)\psi(0)
\label{KZ022}
\qqq
where 
\qq
\Delta_n={_1\over^{2\kappa}}\m t^a_nt^a_n\, ,\quad\quad
H_n\,=\,\sum\limits_{m\not=n}\frac{t_n^at_m^a}{z_n-z_m}\s
\label{H0} 
\qqq
with $t^a_n$ denoting the action of the generator $t^a$ in
the factor $V_n$ of $V$. $\Delta_n$ is a number, the 
{\it conformal weight} assigned in the WZNW theory
to the irreducible representation of $G$ in $V_n$ \cite{kz}.
Note that, modulo the shift $k\mapsto\kappa=k+g^\vee$,
$\Delta_n$ and $\frac{1}{\kappa}H_n$ may be obtained from 
$\frac{k}{(2\pi i)^2}\delta_n$ 
and $\frac{1}{(2\pi i)^2}h_n$, respectively, by the (geometric) 
quantization of the coadjoint orbits which replaces the functions 
$\nu^a_n$ by the operators $\frac{2\pi}{ki}t^a_n$ so that 
the Poisson bracket turns into $\frac{ki}{2\pi}$ times the commutator 
($\frac{2\pi}{k}$ plays the role of the Planck constant).
The flatness of the connection $\nabla^{{\rm KZ}}$,
\m$(\nabla^{{\rm KZ}})^2=0$, follows
from the equation $[H_n,H_m]=0$, equivalent to the classical 
Yang-Baxter equation (CYBE) for $\frac{t^a_1t^a_2}{z}$:
\qq
\Big[\frac{t^a_n\,t^a_{m}}{z_n-z_{m}}\,,\,
	\frac{t^a_{m}\,t^a_{p}}{z_{m}-z_{p}}\Big]+
	{\rm cyclic\ permutations}\,=\,0\, .
	\label{CYBE}
\qqq

\setcounter{equation}{0}
\medskip
\section{Genus one}

At genus one, every Riemann surface is isomorphic 
to an elliptic curve $E_\tau\equiv\NC/(\NZ+\tau\NZ)$ 
where $\tau$ is a complex number of positive imaginary part $\imtau$. 
Denote by $\Delta$ ($\Delta_+$) the set of (positive) roots of $\LieG$,
by $e_\alpha$ the step generators attached to the roots $\alpha$ 
and let $(\eta^j)$ be an orthonormal basis of the Cartan algebra $\LieT$. 
We set $u_\alpha=\tr\,u\alpha$ and $u^j=\tr\,u\eta^j$ for $u\in \LieT$.
We shall need some elliptic functions: the Jacobi theta function
\qq
\vartheta_1(z)=-i\sum\limits_{\ell=-\infty}^\infty
(-1)^\ell\,\ee^{\pi i\tau
(\ell+\frac{1}{2})^2+\pi i z\,(2\ell+1)}\, ,
\qqq
the Green function $P_x$ of the twisted $\de$-operator
\qq
P_x(z)=\frac{\vartheta_1'(0)\,\vartheta_1(x+z)}{\vartheta_1(x)\,
\vartheta_1(z)},
\qqq
with the properties $P_x(z+1)=P_x(z)$, $P_x(z+\tau)=\ee^{-2\pi i x}
P_x(z)$ and $P_x(z)=\frac{1}{z}+\CO(1)$ around $z=0$,
the function 
\qq
\rho=\vartheta_1'/\vartheta_1
\qqq
s.t. $\rho(z+1)=\rho(z)$, $\rho(z+\tau)=\rho(z)-2\pi i$, 
$\rho(z)=\frac{1}{z}+\CO(1)$ around $z=0$
and, finally, the Weyl-Kac denominator
\qq
\Pi(u)=\ee^{2\pi i\tau d/24}
	\prod_{\alpha\in\Delta_+}\,(\ee^{\pi iu_\alpha}-
	\ee^{-\pi iu_\alpha})\,
	\prod\limits_{\ell=1}^\infty\Big[(1-\ee^{2\pi 
	i\ell\tau})^{r}
	\,\prod_{\alpha\in\Delta}(1-\ee^{2\pi 
	i\ell\tau}\ee^{2\pi iu_\alpha})
	\Big]
\qqq
where $d$ denotes the dimension and $r$ the rank of $G$.

On $E_\tau$ a generic gauge field is in the orbit of $A_u=
\pi u\,d\bar{z}/\imtau$, for $u\in{\LieT}$, i.e.
\qq
A={}^{h^{-1}}\hspace{-0.15cm}A^{01}_u=(h_uh)^{-1}\de(h_uh)
\qqq
where $h_u=\ee^{\pi(u\bar{z}-\bar{u}z)/\imtau}$. Consequently, 
the gauge fields may be parametrized by $u$ and $h$. To avoid 
ambiguities, we have to identify the pairs as follows
\qq
(u,\m h)\,\sim\,(wuw^{-1},\, wh)\,\sim\,(u+q^\vee,\,
h^{-1}_{q^\vee}\,h)\,
\sim
	\,(u+\tau q^\vee,\, h^{-1}_{\tau q^\vee}\,h),
\qqq
for $q^\vee$ in the coroot lattice $Q^\vee$ and $w$ in the 
the normalizer $N$ of ${\LieT}$ in $G$. Similarly to 
the genus zero case, we have to solve the equation
\qq
\de\,({}^{h_uh}\Phi)\,=\,\sum_n\nu_n\delta_{z_n}
\qqq
where $\nu_n=(h_uh)(z_n)\,\lambda_n\,(h_uh)(z_n)^{-1}$. 
Decomposing $\nu_n=\sum_\alpha\nu_n^{-\alpha}e_\alpha
+\nu^0_n$ with $\nu^0_n=\nu_n^j\eta^j\in{\LieT}$, we can solve 
the above equation if and only if $\sum_n\nu^0_n=0$. 
In that case,
\qq
{}^{h_u h}\Phi(z)\ =\ \left(\varphi_0
	\,+\,\sum_n\Big(\sum_\alpha P_{u_\alpha}(z-z_n)\,
	\nu_n^{-\alpha}e_\alpha
	+\rho(z-z_n)\,\nu^0_n\Big)\right)\frac{_{dz}}{^{2\pi i}}
\qqq
for an arbitrary constant $\varphi_0=\varphi^j_0\m\eta^j$. 
Performing the symplectic reduction, we find
\qq
\CP_{_{\CO}}\,\simeq\,
	\Big\{\m(u,\varphi_0,\underline{\nu})\in T^*{\LieT}\times
	(\mathop{\times}\limits_n\CO_n)\,\Big|\,\sum_n\nu^0_n=0\Big\}
	\bigg/ N{\rtimes} (Q^\vee+\tau Q^\vee)
\qqq
where the action of $N{\rtimes} (Q^\vee+\tau Q^\vee)$ implements
the identifications
\begin{eqnarray*}
(u\m,\s\varphi_0\m,\s\underline{\nu}) & \sim & 
	(wuw^{-1}\m,\s w\varphi_0w^{-1}\m,\s w
	\underline{\nu}w^{-1})\,\sim\,
	\left(u+q^\vee,\s\varphi_0\m,\s
	(h^{-1}_{q^\vee}(z_n)\m\nu_n\m h_{q^\vee}(z_n))\right)\\
	& \sim &
	\left(u+\tau q^\vee\m,\s\varphi_0\m,\s
	(h^{-1}_{\tau q^\vee}(z_n)\,\nu_n\,h_{\tau q^\vee}(z_n))
	\right).
\end{eqnarray*}
The symplectic structure of $\CP_{_{\CO}}$ is that of the reduction
of \m$T^*{\LieT}\times(\mathop{\times}\limits_n\CO_n)$ by
the group $N{\rtimes} (Q^\vee+\tau Q^\vee)$.
Now it is easy to write down the Hitchin Hamiltonians. Let us 
us do it for $p_2=\frac{1}{2}\tr$. A straightforward computation 
identifying the pole terms leads to
\qq
h_{p_2}\ =\ \Big\{
	-\sum_n\rho'(z-z_n)\,\delta_n
	\,+\,\sum_n\rho(z-z_n)\,h_n\,+\,h_0
	\Big\}\Big(\frac{_{dz}}{^{2\pi i}}\Big)^2
\qqq
where, as before, $\delta_n=\frac{1}{2}\nu^a_n\nu^a_n$ and  
\qq
&&h_0={_1\over^2}\sum_{j=1}^r\varphi_0^j\varphi_0^j
	+{_1\over^2}\sum_{m,n}\Big\{
	\sum_\alpha\partial_xP_{u_\alpha}(z_n-z_m)
	\m\nu^\alpha_n\nu^{-\alpha}_m
	+\frac{_1}{^2}
	\sum_{j=1}^r \frac{\vartheta_1''}{\vartheta_1}(z_n-z_m)
	\m\nu^j_n\nu_m^j\Big\},\hspace{1cm}\\
&&h_n=\sum_{j=1}^r \nu^j_n\varphi_0^j\,+\,\sum_{m\neq n}\Big(
	\sum_\alpha P_{u_\alpha}(z_n-z_m)\,\nu_n^\alpha\nu_m^{-\alpha}
	\,+\,\sum_{j=1}^r\rho(z_n-z_m)\,\nu^j_n\nu^j_m\Big).
\qqq
Note the similarity to the genus 0 case (\ref{h21}).

Let $\delta\mu=\delta\mu_{\bar z}^z\m\da_zd\bar z$ 
be a Beltrami differential on $E_\tau$
behaving like $\CO((z-z_n)^2)$ around the insertions.
The modified complex structure corresponds to the
complex coordinate $z'=z+\frac{z-\bar z}{2i\tau_2}
\m\delta\tau\m+\m{\delta\xi}^z$ 
s.t. $\da_{\bar z}z'=\delta\mu_{\bar z}^z$. 
We require that $\delta\xi^z(z+1)=\delta\xi^z(z+\tau)
=\delta\xi^z(z)$. $\delta\tau$ is determined from the condition
that the integral of $\delta\mu_{\bar z}^z$ over
$E_\tau$ is equal to that of $\frac{i}{2\tau_2}\delta\tau$.
${\delta\xi}^z$ is unique up to an additive constant.
Note that $z'(z+1)=z'(z)+1$ whereas $z'(z+\tau)=z'+\tau'$
where $\tau'=\tau+\delta\tau$. Hence the deformed
curve is isomorphic to $E_{\tau'}$ with the punctures
moved to $z'_n=z_n+\delta z_n$ and the first jets
of local parameters changed to $\chi'_n=\chi_n+\delta\chi_n$
with
\qq
\delta z_n={z_n-\bar z_n\over2i\tau_2}\m\delta\tau\m
+\m{\delta\xi}^z(z_n)\,,\quad\quad\chi_n^{-1}\delta\chi_n
={\delta\tau\over2i\tau_2}\m+\m\da_z\delta\xi^z(z_n)\,.
\qqq
Again by a straightforward calculation substituting
$\delta\mu_{\bar z}^z=\da_{\bar z}z'$, cutting out
small balls around points $z_n$ and integrating by parts,
we obtain
\qq
h_{\delta\mu}\,=\,\int_{E_\tau}h_{p_2}\m\delta\mu\ 
	=\ {_1\over^{2\pi i}}\sum_n
	\left(\delta_n\,\chi_n^{-1}\delta\chi_n
	\,+\, h_n\,\delta z_n\,+\,{_1\over^{2\pi i}}\, 
	h_0\,\delta\tau\right).
\label{hmu1}
\qqq

The quantum states $\psi$ at genus one may be characterized
by giving holomorphic functions $\tilde\psi(u)$ on $\LieT$ 
with values in $V^T$, the subspace of $T$-invariant tensors
in the product $V$ of the representation spaces,
\qq
\tilde\psi(u)\,=\,\Pi(u)\,\ee^{-\pi k\,\tr\,u^2/(2\imtau)}
\mathop{\otimes}\limits_n \Big(\ee^{-\pi(z_n-\bar{z}_n)
u/\imtau}\Big)_{n}\psi(A_u)\,.
\label{st1}
\qqq
The KZB connection takes the form \cite{EtinK,gaw94:genus1,felder:elliptic,felder:integral,gaw97:unitarity}
\qq
&&\nabla^{\rm KZB}_{\delta\mu}\tilde\psi\,=\, 
\sum\limits_n\left(\delta\chi_n(\da_{\chi_n}-\chi_n^{-1}\Delta_n)
\,+\,\delta z_n(\da_{z_n}-{_1\over^{\kappa}}\m H_n)
\,+\,\delta\tau(\da_{\tau}-{_1\over^{2\pi i\kappa}}\m H_0)
\right)\tilde\psi\,,\hspace{1cm}\label{KZB1}\\
&&\nabla^{\rm KZB}_{\overline{\delta\mu}}\tilde\psi\,=\, 
\sum\limits_n\left(\overline{\delta\chi}_n\da_{\bar\chi_n}
\,+\,\overline{\delta z}_n\da_{\bar z_n}\,+\,\overline{\delta\tau}
\m\da_{\bar\tau}\right)\tilde\psi
\label{KZB2}
\qqq
where $\Delta_n$ is as before and the operators
$\frac{1}{k}H_0$ and $\frac{1}{k}H_n$ are obtained 
from the Hamiltonians $\frac{k}{(2\pi i)^2}h_0$ and 
$\frac{k}{(2\pi i)^2}h_n$ by the replacement
\qq
\varphi_0^j\mapsto {_{2\pi}\over^{ki}}\partial_{u^j},\qquad
	\mu^\alpha_n\mapsto {_{2\pi}\over^{ki}}e_{\alpha\m n},
	\qquad \mu^j_n\mapsto {_{2\pi}\over^{ki}}\eta^j_n\,,
\qqq
i.e. by the geometric quantization.
The resulting Hamiltonians $H_0$ and $H_n$ act on general 
meromorphic functions on $\LieT$ with values in $V^T$.
The flatness of the KZB connection is ensured by their 
commutation: $[H_n,H_m]=0$, for $n,m=0,1,\dots$, following from
the so called dynamical CYBE~\cite{felder:elliptic}.

\setcounter{equation}{0}
\medskip
\section{Genus two}

\subsection{Curve and its Jacobian}

Let $\Sigma$ be a curve of genus 2. Choosing a marking,
i.e.\s\s a symplectic homology basis $(A^a,B^a)$, $a=1,2,$ on $\Sigma$,
we may fix the corresponding basis $(\omega^a)$ of the
holomorphic 1,0-forms (abelian differentials)
s.t. $\int_{A^a}\omega^b=\delta^{ab}$. 
The $B^a$-periods of the abelian differentials give rise 
to the symmetric period matrix $\tau$, $\tau^{ab}=
\int_{B^a}\omega^b$, with a positive imaginary part. 
The map
\qq
\Sigma\ni x\,\mapsto\,z(x)={\omega^2(x)\over\omega^1(x)}
\label{doco}
\qqq
realizes $\Sigma$ as a double covering of $\NP^1$
ramified over six Weierstrass points $x_n$ or as
a hyperelliptic curve given by the equation
\qq
y^2=\prod_{n=1}^6(z-z_n)\,.
\label{curve}
\qqq
The coordinates $z_n=z(x_n)$ are assumed to be finite. 
This may be always achieved by an appropriate choice 
of the marking. Curve $\Sigma$ may be viewed as composed 
of the points $(y,z)$ and of two points at infinity. 
The covering of $\NP^1$ is $(y,z)\mapsto z$. In this 
representation, the holomorphic 1,0-forms and the 
holomorphic quadratic differentials on $\Sigma$ 
have the form
\qq
\omega=(a+bz)dz/y\,,\qquad \beta=(a+bz+cz^2)(dz)^2/y^2\,,
\label{qdiff}
\qqq
respectively. In particular,
\qq
\omega^1\ \propto\ dz/y\,,\qquad\omega^2
\ \propto\ z\m dz/y\,
\qqq
with the same proportionality constant.

The Jacobian $J^1$ of the degree 1 holomorphic line bundles 
may be represented as $\NC^2/(\NZ^2+\tau\NZ^2)$. 
In particular, the spin structures $L$, $L^2=K$, correspond
to points $e+\tau e'$ with $e\in\frac{1}{2}\NZ^2/\NZ^2$.
There are 6 odd spin structures $e_n+\tau e'_n$ 
labeled by the Weierstrass points $z_n$, the zeros
of the holomorphic sections they admit, and 10 
even spin structures without holomorphic sections.

\subsection{Theta functions}

The degree 1 bundles with holomorphic sections form 
the theta-divisor in $J^1$. The holomorphic sections of the 
$k^{\rm th}$-power of the corresponding theta bundle
over $J^1$ may be represented by the holomorphic theta
functions of degree $k$ on $\NC^2$ defined by the relations
\qq
\ee^{\m\pi i k\s n\cdot\tau n\m+\m
2\pi i k\s n\cdot u}\,\theta(u+m+\tau n)\,=\,\theta(u)
\label{thetaa}
\qqq
for $m,n\in\NZ^2$. They form the space $\Theta_{k}$ of dimension $k^2$. 
For $k=1$ there is a single (up to normalization) 
theta function 
\qq
\vartheta(u)\,=\,\sum\limits_{n\in\NZ^2}\ee^{\m\pi i\s n\cdot\tau n
+2\pi i\s n\cdot u}\,,
\qqq
the Riemann theta function.
For $k=2$ there are four independent $\theta$-functions.
One can take them as
\qq
\theta_{e}(u)\,=\,\sum\limits_{n\in\NZ^2}\ee^{\m2\pi i\s (n+e)\cdot
\tau(n+e)\m+\m4\pi i\s (n+e)\cdot u}
\label{base}
\qqq
for $e\in\frac{1}{2}\NZ^2/\NZ^2$. 
\qq
\vartheta(u+v)\m\vartheta(u-v)\s=\s\sum\limits_e\theta_e(u)\m
\theta_e(v)
\label{thth}
\qqq
is a second order theta function in both $u$ and $v$.
The map \s$v\,\mapsto\,\vartheta(\s\cdot\s+v)\m\vartheta
(\s\cdot\s-v)\s$ determines an embedding of the Kummer 
surface $J^1/\NZ_2\cong \NC^2/(\NZ^2+\tau\NZ^2)/\NZ_2$ 
onto a quartic surface $\CK$ in the 3-dimensional projective 
space $\NP\Theta_2\s$ ($\NZ_2$ maps the degree 1 line bundles 
$L$ into $L^{-1}K$ or $v\in\NC^2$ into $-v$).

The double theta function (\ref{thth}) determines a non-degenerate
symmetric quadratic form on the space $\Theta_2^*$ dual to $\Theta_2$.
It permits to identify $\Theta_2^*$ with $\Theta_2$ by
sending $\phi\in\Theta_2^*$ to $\iota(\phi)\in\Theta_2$ defined by
\qq
\iota(\phi)(u)\,=\,\langle\m\vartheta(u+\s\cdot\m\s)\m
\vartheta(u-\s\cdot\m\s)\m,\s\phi\m\rangle\s.
\label{iota}
\qqq
The identification exchanges the basis $(\theta_e)$ of $\Theta_2$ 
with the dual basis $(\theta_e^*)$ and the Kummer quartic $\CK$
with its dual version $\CK^*\subset\NP\Theta_2^*$. $\CK^*$ 
is composed of linear forms proportional to the evaluation
forms $\phi_u$ defined by
\qq
\langle\theta,\m\phi_u\rangle\,=\,\theta(u)\,.
\qqq

The group $(\frac{1}{2}\NZ/\NZ)^4$ of spin structures acts 
on $\Theta_k$ for even $k$ by endomorphisms $U_{e,e'}$ 
defined by 
\qq
(U_{e,e'}\m\theta)(u)\,=\,\ee^{\m\pi i k\s e'\cdot\tau e'\m+\m
2\pi i k\s e'\cdot u}\,\theta(u+e+\tau e')\,.
\qqq
For $k$ not divisible by $4$ this action is only projective:
$U_{e,e'}U_{f,f'}=(-1)^{4\s e\cdot f'}\, U_{e+f,\m e'+f'}$ and
it lifts to a Heisenberg group. For $k=2$,
\qq
U_{e,e'}\m\theta_{e''}=(-1)^{4\s e\cdot e''}\,\theta_{e'+e''}\,.
\label{act1}
\qqq
The action of \s$U_{e,e'}$ preserves the Kummer quartic $\CK\subset
\NP\Theta_2$ and the action of the transposed endomorphisms 
$U_{e,e'}^{\ t}$ preserves $\CK^*$.

\subsection{Moduli space of ${\rm SL}_2$-bundles}

In the fundamental paper~\cite{nar-ram}, Narasimhan 
and Ramanan proved that the  moduli space $\CN_{s}$ of the stable 
${\rm SL}_2$ holomorphic bundles is canonically isomorphic to $\NP\Theta_2
\setminus\CK$. The isomorphism associates to an ${\rm SL}_2$-bundle $E$
the second order theta function $\theta$ vanishing at the points 
$u\in\NC^2$ 
corresponding to the duals of the line-subbundles of the rank 2 
bundle associated to $E$. In other words, if $A$ is the gauge
field whose $\CG$-orbit corresponds to $E$ and if $L_u$ denotes 
the degree 1 line bundle corresponding 
to $u\in\NC^2/(\NZ^2+\tau\NZ^2)$ then
the theta function $\theta$ associated to $E$ vanishes at $u$
if and only if there exists a pair $s=(s_1,s_2)$ composed
of sections of $L_u$ s.t. $(\de+A)s=0$. The semistable 
compactification of the moduli space $\CN_{s}$ is
\qq
\label{thetaiso}
\CN_{ss}\ \cong\ \NP\Theta_2
\qqq
and, exceptionally, it is smooth. The points of the Kummer quartic 
$\CK$ represent (classes of) the semistable but not stable bundles. 
Hence for $G={\rm SL}_2$ the phase space of the Hitchin system on the genus 
2 curve with no insertions is 
\qq
T^*\CN_{ss}\ \cong\ T^*\NP\Theta_2\ \cong
\ \{\s(\theta,\phi)\in\Theta_2\times\Theta_2^{\m*}
\s\s\s|\s\s\s \theta\not=0,\ \langle\theta,\phi\rangle=0\s\}
\Big/\NC^\times
\qqq
with the action of $t\in\NC^\times$ given by $(\theta,\phi)
\mapsto(t\m\theta,\m t^{-1}\phi)$. As a symplectic space,
it is the symplectic reduction of $T^*(\Theta_2\setminus\{0\})$
by the action of $\NC^\times$. Using
the bases $(\theta_e)$ and $(\theta_e^*)$ to decompose
\qq
\theta=\sum q_e\theta_e\,,\qquad\phi=\sum p_e\theta_e^*\,,
\label{decomplo}
\qqq
we may represent $T^*\CN_{ss}$ as the space of pairs
$(q,p)\in\NC^4\times\NC^4$, $q\not=0$, $q\cdot p=0$, 
with the identification $(q,p)\sim(t\m q,\m t^{-1}p)$ 
and the symplectic form induced from $dp\cdot dq$.

\subsection{Hitchin map for $G={\rm SL}_2$}

The Hitchin map $h_{p_2}\m:\m T^*\NP\Theta_2\rightarrow H^0(K^2)$
appears to take a particularly simple form resembling
the genus 0 formula (\ref{h21}):
\qq
h_{p_2}\ =\ {_1\over^{2}}\sum\limits_{n,m=1\atop n\not=m}^6
{r_{nm}\over(z-z_n)(z-z_m)}\s({_{dz}\over^{2\pi i}})^2
\ =\ \sum\limits_{n=1}^6{h_n\over z-z_n}\s
({_{dz}\over^{2\pi i}})^2\hspace{0.4cm}
\label{h22}
\qqq
where
\qq
r_{nm}(\theta,\m\phi)\ =\ {_1\over^{16}}\s
\langle\m U_{e_n,e'_n}\theta\m,\s
U_{e_m,e'_m}^{\ t}\phi\m\rangle\s
\langle\m U_{e_m,e'_m}\theta\m,\s 
U_{e_n,e'_n}^{\ t}\phi\m\rangle\s
\label{rnm}
\qqq
(the last two factors on the right hand side coincide modulo sign)
and
\qq
h_n\ =\ \sum\limits_{m\not=n}^6{r_{nm}\over{z_n-z_m}}\,.
\label{h2n}
\qqq
With the help of Eq.\s\s(\ref{act1}), $r_{nm}$'s may be
rewritten in the language of $q$'s and $p$'s as explicit 
homogeneous polynomials of order 2 in $q_e$ and in $p_e$,
see below. The identity
\qq
\sum\limits_{m\not=n} r_{nm}\,=\,0
\label{srnm}
\qqq
holding for each $n$ guarantees that 
\s$\prod\limits_{n=1}^6(z-z_n)\sum\limits_{n,m=1\atop n\not=m}^6
\frac{r_{nm}}{(z-z_n)(z-z_m)}\s$ is a quadratic polynomial in $z$.
It follows that the right hand side of Eq.\s\s(\ref{h22}) 
determines a quadratic differential on $\Sigma$ of the general
form given by Eq.\s\s(\ref{qdiff}). 

The equality (\ref{h22}) is not immediate. It was established 
in four steps. We shall only enumerate them here. The two first 
crucial steps were performed in \cite{VGP}. It was shown 
there that for any $\theta\not=0$ and $u\in\NC^2$ s.t.  
$\theta(u)=0$,
\qq
h_{p_2}(\theta,\phi_u)\ =\ -{_1\over^{16\pi^2}}\s(\da_{u^a}\theta(u)
\s\omega^a)^2\,.
\label{onKum}
\qqq
The above equation describes the quadratic polynomial 
$h_{p_2}(\theta,\s\cdot\s)$
(with values in $H^0(K^2))$ on the quartic $\CK^*_\theta
=\CK^*\cap\NP\theta^\perp$ in the projectivized subspace 
of $\Theta_2^*$ perpendicular to $\theta$. In principal, 
it determines $h_{p_2}$ completely. It was 
observed then that the above formula implies that for each 
Weierstrass point $z_n$ the conic 
\qq
\CC_n\ =\ \{\m\NC^\times\phi\s\in\s\NP\theta^\perp\ 
\vert\ h_{p_2}(\theta,\phi)\vert_{z_n}=0\m\}
\label{conic}
\qqq
is in fact the union of two bitangents to $\CK_\theta^*$. 
The explicit equations for the bitangents to the Kummer quartics 
known since about a century permitted then to establish 
Eq.\s\s(\ref{h22}) up to multiplication by a $\theta$-dependent
factor \cite{VGP}. The other steps in the proof of the
formula (\ref{h22}) were taken in \cite{GaTr} were it was 
established that the Hitchin map $h_{p_2}$ possesses
the important self-duality property:
\qq
h_{p_2}(\iota(\phi),\m\iota^{-1}(\theta)\m)\ =\ h_{p_2}
(\theta,\m\phi)
\label{sdu}
\qqq
or that $h_{p_2}(q,p)=h_{p_2}(p,q)$ in the language of (\ref{decomplo}).
This property, far from obvious in the original formulation
of the Hitchin system, restricted the ambiguity on the right
hand side of Eq.\s\s(\ref{h22}) to a (possibly curve-dependent) 
constant factor. The latter was fixed in \cite{GaTr} by a
tedious calculation of $h_{p_2}$ at special points of
$\CK\times\CK^*$.

\subsection{Deformations of complex structure}

For the hyperelliptic curve, 
the variations of the complex structure 
described by the Beltrami differentials $\delta\mu$
on $\Sigma$ change the images $z_n$ 
of the ramification points of the covering of $\NP^1$. 
Let us find these changes. Let
\qq
{\omega'}^a\,=\,\omega^a\s+\s\delta\omega^a+\bar\delta\omega^a
\qqq
denote a deformed basis $({\omega'}^a)$ of the abelian 
differentials with $\delta\omega^a$ of 1,0-type and $\bar\delta
\omega^a$ of 0,1-type in the original complex structure. 
$\delta\omega^a$ and $\bar\delta\omega^a$ have to satisfy
the relations
\qq
\bar\delta\omega^a\,=\,\omega^a\m\delta\mu\,,\qquad
\de(\delta\omega^a)\,=\,-\,\da(\omega^a\m\delta\mu)
\label{sp0}
\qqq
stating, respectively, that ${\omega'}^a$ is of the 1,0-type
in the deformed structure and that it is a closed form.
The equation for $\delta\omega^a$ always has solutions.
They are defined modulo abelian differentials and the ambiguity
may be fixed by demanding that $\int_{A^a}{\omega'}^b=\delta^{ab}$.
The deformed covering map onto $\NP^1$ is then
\qq
z'(x)\,=\,{{\omega'}^2(x)\over{\omega'}^1(x)}\,=\,
z(x)\,+\,{\delta\omega^2\over\omega^1}(x)\,-\,
z(x)\,{\delta\omega^1\over\omega^1}(x)\,.
\label{zpx}
\qqq
The ramification points $x'_n$ of the map $z'$ are determined by
solving the equation
\qq
\da' z'(x'_n)\,=\,0\,
\qqq
which, upon rewriting $x'_n=x_n+\delta x_n$ becomes
\qq
(\da)^2z(x_n)\s\delta x_n\,+\,\da\m({\delta\omega^2\over\omega^1})(x_n)
-\da\m( z\m{\delta\omega^1\over\omega^1})(x_n)\ =\ 0\,.
\qqq
Since $\de\m z(x_n)=0$, the quadratic differential $\m(\da)^2z(x_n)$ 
is well defined. Besides, since the ramification points are
isolated, it does not vanish so that one may solve the above 
equations for $\delta x_n$. The ramification points $x'_n$ are 
mapped to
\qq
z'_n\,=\, z'(x_n+\delta x_n)\,=\, z_n\,+\,{\delta\omega^2\over\omega^1}
(x_n)\,-\, z_n\m{\delta\omega^1\over\omega^1}(x_n)\s.
\qqq
We infer that
\qq
\delta z_n\equiv z'_n-z_n\,=\,{\delta\omega^2\over\omega^1}
(x_n)\,-\, z_n\m{\delta\omega^1\over\omega^1}(x_n)
\label{dzpn}
\qqq
are the variations of $z_n$ corresponding to the Beltrami
differential $\delta\mu$.

We may now find the values of the Hitchin Hamiltonians 
$\, h_{\delta\mu}\,=\,\int_\Sigma h_{p_2}\s\delta\mu\,$
related to the Beltrami differentials.
Note that, by virtue of Eqs.\s\s(\ref{zpx}) and (\ref{sp0}),
\qq
\de(\delta z)\,\equiv\,
\de\m(z'-z)\,=\,\,\de\m(\delta\omega^2)\m/\m\omega^1\,-\,
z\,\de\m(\delta\omega^1)\m/\m\omega^1\s=\s
-\m\da\m(\omega^2\m\delta\mu)\m/\m\omega^1\,\,\cr\cr
+\,z\,\da\m(\omega^1\m\delta\mu)\m/\m\omega^1
=\,-\m\da\m(z\s\omega^1\m\delta\mu)\m/\m\omega^1
\,+\,z\s\da\m(\omega^1\m\delta\mu)\m/\m\omega^1
\ =\ (dz)\,\delta\mu\,.
\label{dzpn1}
\qqq
A straightforward integration by parts over the region
in $\Sigma$ without small balls around the Weierstrass
points $x_n$ and around 2 points at infinity gives now:
\qq
h_{\delta\mu}\ =\ \int_\Sigma\sum\limits_{n=1}^6
{h_n\over z-z_n}\s ({_{dz}\over^{2\pi i}})^2\m\delta\mu\ =\ 
({_1\over^{2\pi i}})^2\int_\Sigma\sum\limits_{n=1}^6
{h_n\over z-z_n}\s dz\,\de(\delta z)\cr
\ =\ {_1\over^{\pi i}}\sum\limits_{n=1}^6 h_n\,\delta z_n\,.
\hspace{0.5cm}
\label{h2dm}
\qqq
The comparison with Eq.\s\s(\ref{hmu012}) shows an additional 
factor 2 which comes from the double covering.

\subsection{Relation to Neumann systems}

The Poisson-commutation of the Hamiltonians 
$h_n$ (of which any 3 give independent action variables
of our integrable system) is equivalent to the CYB-like 
equations
\qq
\{{r_{nm}\over z_n-z_m}\m,\,{r_{mp}\over z_m-z_p}\}\,
+\,{\rm cycl.}\,=\,0\,,\qquad\{{r_{nm}\over z_n-z_m}\m,\,
{r_{pq}\over z_p-z_q}\}\,=\,0
\label{CYBL}
\qqq
for $\{n,m\}\cap\{p,q\}=\emptyset$. The above relations may
may be directly checked, as noticed in \cite{VGP}. The more 
recent observations of the paper \cite{VGDJ} 
on the Knizhnik-Zamolodchikov-Bernard connection 
in the same setup (see below) permit to identify 
the integrable system with the Hamiltonians $h_n$ 
of Eq.\s\s(\ref{h2n}): it is a modified version of the classical 
genus 2 Neumann systems \cite{neumann,mumford2,AT} whose original 
version is also rooted in the modular geometry of hyperelliptic 
curves. This goes as follows. 

The phase space $T^*\NP^3$ (without
the zero section) may be identified with the coadjoint orbit 
$\CO_1$ of the complex group ${\rm SL}_4$ composed of the traceless 
rank 1 matrices $\vert q\rangle \langle p\vert$. The action 
of ${\rm SL}_4$ in $\wedge^2\NC^4$ preserves
the quadratic form induced by the exterior product 
on $\wedge^2\NC^4$ and the identification 
$\wedge^4\NC^4\cong\NC^1$. It leads to the double covering
${\rm SL}_4\rightarrow {\rm SO}_6$ if we choose in $\wedge^2\NC^4$ 
the Pl\"{u}cker basis turning the quadratic form into
the sum of squares. The inverse relation is the complexified 
version of the twistor calculus. Upon the identification of 
$\groupesl_4$ with $\groupeso_6$, the coadjoint orbit $\CO_1$ becomes 
the one composed of (complex) rank 2 antisymmetric matrices
$J=(J_{nm})$ of square zero: $J^2=0$. Such matrices
are of the form $J_{nm}=Q_nP_m-P_nQ_m$ with vectors $P,Q\in\NC^6$ 
spanning an isotropic subspace, i.e. with $Q^2=Q\cdot P=P^2=0$.
Given $(q,p)$ with $q\cdot p=0$, in order to find 
the corresponding pair $(Q,P)$, it is enough to complete vector 
$q$ to a basis $(q,f_1,f_2,f_3)$ of $\NC^4$ s.t. 
$f_1\cdot p=f_2\cdot p=0$ and $f_3\cdot p=1$ and to set
\qq
Q=q\wedge f_1\,,\quad R=f_2\wedge f_3\,,\quad P=q\wedge 
f_2\m/\m Q\cdot R
\label{PQ} 
\qqq
in the language of $\wedge^2\NC^4\cong\NC^6$. 
In the Pl\"{u}cker coordinates, we may take
\qq
&&\hbox to 4cm{$Q_1\s=\s-(q_1p_3+q_4p_2)\,,$\hfill}\quad
\hbox to 4cm{$Q_2\s=\s i(q_1p_3-q_4p_2)\,,$\hfill}\quad
\hbox to 4cm{$Q_3\s=\s-i(q_1p_2+q_4p_3)\,,$\hfill}\cr\cr
&&\hbox to 4cm{$Q_4\s=\s q_1p_2-q_4p_3\,,$\hfill}\quad
\hbox to 4cm{$Q_5\s=\s q_2p_2+q_3p_3\,,$\hfill}\quad
\hbox to 4cm{$Q_6\s=\s\, i(q_2p_2+q_3p_3)\,,$\hfill}\cr\cr
&&\hbox to 4cm{$P_1\s=\s-{1\over2}\,{{q_2p_4+q_3p_1}
\over{q_2p_2+q_3p_3}}\,,$\hfill}\quad
\hbox to 4cm{$P_2\s=\s{i\over2}\,{{q_2p_4-q_3p_1}
\over{q_2p_2+q_3p_3}}\,,$\hfill}\quad
\hbox to 4cm{$P_3\s=\s{i\over2}\,{{q_3p_4+q_2p_1}
\over{q_2p_2+q_3p_3}}\,,$\hfill}\cr\cr
&&\hbox to 4cm{$P_4\s=\s-{1\over2}{{q_3p_4-q_2p_1}
\over{q_2p_2+q_3p_3}}\,,$\hfill}\quad
\hbox to 4cm{$P_5\s=\s{1\over2}\,,$\hfill}\quad
\hbox to 4cm{$P_6\s=\s-{i\over2}$\hfill}
\nonumber
\qqq
where $q_1\equiv q_{(0,0)}$, \s$q_2\equiv q_{({1\over 2},0)}$,
\s$q_3\equiv q_{(0,{1\over 2})}$, \s$q_4\equiv q_{({1\over 2},
{1\over 2})}$ and similarly for $p$'s.
In terms of $(Q,P)$, the symplectic form is $dP\cdot dQ$.
The functions $J_{nm}$ on $\CO_1$ have the Poisson bracket
\qq
&&\hbox to 6cm{$\{J_{nm},\m J_{mp}\}\ =\ -J_{np}$\hfill}
{\rm for}\ \ n,m,p\ \ \ \ \ {\rm different},\label{soa1}\\
&&\hbox to 6cm{$\{J_{nm},\m J_{pq}\}\ =\ 0$\hfill}{\rm for}
\ \ n,m,p,q\ \ {\rm different}\label{soa2}
\qqq
A straightforward check shows now that
\qq
&&J_{12}\ =\ \ \ \m{_i\over^2}(q_1p_1+q_2p_2-q_3p_3-q_4p_4)\s,\quad
J_{13}\ =\ -{_i\over^2}(q_1p_4-q_2p_3-q_3p_2-q_4p_1)\s,\hspace{1.06cm}\cr
&&J_{14}\ =\ \ \ \m{_1\over^2}(q_1p_4+q_2p_3-q_3p_2-q_4p_1)\s,\quad
J_{15}\ =\ -{_1\over^2}(q_1p_3-q_2p_4-q_3p_1+q_4p_2)\s,\hspace{1.06cm}\cr
&&J_{16}\ =\ \ \ \m{_i\over^2}(q_1p_3+q_2p_4+q_3p_1+q_4p_2)\s,\quad
J_{23}\ =\ -{_1\over^2}(q_1p_4-q_2p_3+q_3p_2-q_4p_1)\s,\hspace{1.06cm}\cr
&&J_{24}\ =\ -{_i\over^2}(q_1p_4+q_2p_3+q_3p_2+q_4p_1)\s,\quad
J_{25}\ =\ \ \ \m{_i\over^2}(q_1p_3-q_2p_4+q_3p_1-q_4p_2)\s,\hspace{1.06cm}
\label{Js}\\
&&J_{26}\ =\ \ \ \m{_1\over^2}(q_1p_3+q_2p_4-q_3p_1-q_4p_2)\s,\quad
J_{34}\ =\ -{_i\over^2}(q_1p_1-q_2p_2+q_3p_3-q_4p_4)\s,\hspace{1.06cm}\cr
&&J_{35}\ =\ -{_i\over^2}(q_1p_2+q_2p_1+q_3p_4+q_4p_3)\s,\quad
J_{36}\ =\ -{_1\over^2}(q_1p_2-q_2p_1-q_3p_4+q_4p_3)\s,\hspace{1.06cm}\cr
&&J_{45}\ =\ \ \ \m{_1\over^2}(q_1p_2-q_2p_1+q_3p_4-q_4p_3)\s,\quad
J_{46}\ =\ -{_i\over^2}(q_1p_2+q_2p_1-q_3p_4-q_4p_3)\s,\hspace{1.06cm}\cr
&&J_{56}\ =\ \ \ \m{_i\over^2}(q_1p_1-q_2p_2-q_3p_3+q_4p_4)\nonumber
\qqq
and that
\qq
r_{nm}\ =\ -{_1\over^4}\,(J_{nm})^2\s.
\label{jsq}
\qqq
The equations (\ref{CYBL}) assuring that the Hamiltonians
\s$h_n=-{1\over 4}\sum\limits_{m\not= n}\frac{J_{nm}^{\s2}}{z_n-z_m}$ 
Poisson-commute follow directly from the 
$\groupeso_6$ algebra (\ref{soa1}), (\ref{soa2}). The original
Neumann systems are very similar but involve the
coadjoint orbits of ${\rm SO}_{N}$ composed from rank 2 antisymmetric
matrices of square $\not=0$ \cite{AT}. Such orbits, contrary
to the one we consider, have nontrivial standard real forms.

\subsection{Lax matrix approach}

Although the change of the orbit modifies dimensional 
counts and many details, the methods used in the analysis 
of the Neumann systems, in particular the Lax method developed
in \cite{AT}, generalize with minor variations to
our system and permit to find explicitly the angle variables
of the genus 2 Hitchin system. The Lax matrix may be taken as
$L(\zeta)=(L_{nm} (\zeta))$ with
\qq
L_{nm}(\zeta)\ =\ \zeta\, J_{nm}\,+\, z_n\m\delta_{nm}\,.
\label{lax}
\qqq
As in \cite{AT}, the Poisson brackets (\ref{soa1}), (\ref{soa2}) 
may be rewritten in the matrix form as
\qq
\{L(\zeta)\otimes 1\m,\s 1\otimes L(\zeta')\}
\ =\ [L(\zeta)\otimes 1\m,\, r^-(\zeta,\zeta')]
\ -\ [1\otimes L(\zeta')\m,\, r^+(\zeta,\zeta')]
\label{r1r2}
\qqq                                
where the $r$-matrices
\qq
r^{\pm}(\zeta,\zeta')\,=\,{_{\zeta\m\zeta'}\over^{\zeta+\zeta'}}\, 
C\,\pm\,{_{\zeta\m\zeta'}\over^{\zeta-\zeta'}}\, T 
\label{rpm}
\qqq
with $C_{mn,qp}=\delta_{mq}\delta_{np}$ and 
$T_{mn,qp}=\delta_{mp}\delta_{nq}$ \m satisfy the CYBE.
The above form of the Poisson bracket implies immediately that
\qq
\{\m\tr\, L(\zeta)^\ell\m,\m\,\tr\, L(\zeta')^{\ell'}\m\}\ =\ 0
\qqq
for all $\zeta,\m\zeta'$. Since
\qq
{_{d^2}\over^{d^2\zeta}}\,\tr\,L(0)^\ell\ =\ 
2\m\ell\sum\limits_{n,m=1\atop n\not=m}^6z_n^{\s\ell-1}
{J_{nm}^{\s2}\over z_n-z_m}\,,
\qqq
the Hamiltonians 
$h_n=-{1\over 4}\sum\limits_{m\not= n}{J_{nm}^{\s2}\over z_n-z_m}$
may be expressed as combinations of the quantities 
$\tr\, L(\zeta)^\ell$. It is not difficult to see that the converse
is also true.

More generally, Eq.\s\s(\ref{rpm})
implies that
\qq
\{\m\tr\, L(\zeta)^\ell\m,\, L(\zeta')\m\}
\ =\ [\m M_\ell(\zeta,\zeta')\m,\, L(\zeta')\m]
\label{M1}
\qqq
with
\qq
M_\ell(\zeta,\zeta')\ =\ \ell\,{_{\zeta\m\zeta'}
\over^{\zeta-\zeta'}}\, L(\zeta)^{\ell-1}\,+\,
\ell\,{_{\zeta\m\zeta'}\over^{\zeta+\zeta'}}\, 
L(-\zeta)^{\ell-1}\,.
\label{M2}
\qqq
It follows that the commuting time evolutions of the Lax 
matrix $L(\zeta)$ generated by the Hamiltonians $h_n$,
\qq
\delta_nL(\zeta)\,
=\,\{\m h_n\m,\, L(\zeta)\m\}\,\delta t_n\,,
\label{dtn}
\qqq
are isospectral. In other words, the spectral curve $\CS$ given 
by the characteristic equation
\qq
\det\m(L(\zeta)-z)\ =\ 0
\label{che}
\qqq
is left invariant by the dynamics generated by any of the
Hamiltonians $h_n$. An easy calculation using the fact that
the matrix $J$ has rank 2 gives
\qq
\det\m(L(\zeta)-z)\ =\ \prod\limits_{n=1}^6(z-z_n)\Big(
1\,+\,{_1\over^2}\m\zeta^2\sum\limits_{n,m=1\atop n\not=m}^6
{J_{nm}^{\s 2}\over(z-z_n)(z-z_m)}\Big)\,.
\label{che1}
\qqq
Upon the substitution $\sigma\m=\m {1\over i\m\zeta}\m
\prod_n(z-z_n)$,
the characteristic equation (\ref{che}) becomes
\qq
\sigma^2\ =\ {_1\over^2}\prod\limits_{n=1}^6(z-z_n)^2
\sum\limits_{n,m=1\atop n\not=m}^6{J_{nm}^{\s 2}
\over(z-z_n)(z-z_m)}\ \equiv\ P(z)\,.
\label{che2}
\qqq
Since $P(z)$ is a polynomial in $z$ of order
8 (see the remark after Eq.\s\s(\ref{srnm})),
this is the equation of a hyperelliptic curve $\CS$
of genus 3 composed of pairs $(\sigma,z)$ and 2 points
$p^\pm_\infty$ corresponding to $z=\infty$. We shall
consider only such points of the phase space that
$\CS$ is smooth. The 1,0-forms 
\qq
\Omega^b\,=\,z^b\, dz\m/\m\sigma
\label{Omb}
\qqq
with $b=0,1,2$ form a basis of the abelian differentials 
on $\CS$.

We shall search for the eigenvectors $X=(X_n)$ of the 
Lax matrix.
This will allow to adapt the arguments described in great detail
in Sect.\s\s4 of \cite{mumford2} to the present case.
The eigenvector equations
\qq
\zeta\m J_{nm}\m X_m\ =\ \zeta\m Q_n\,(P\cdot X)
-\zeta\m P_n\,(Q\cdot X)\ =\ (z-z_n)\m X_n
\label{evec}
\qqq
imply that
\qq
(z-z_n)\m X_n\ =\ a\m Q_n\,+\, b\m P_n\s.
\qqq
Upon multiplication by $\sigma\m=\m\frac{1}{i}
\m\zeta^{-1}\prod_n(z-z_n)$,
Eq.\s\s(\ref{evec}) becomes a system of 2 linear equations
for $a$ and $b$:
\qq
&&(\sigma+V)\s a\m\ +\ \m i\m W\s b\ =\ 0\,,\cr
&&-\m i\m U\s a\,\s+\,\s(\sigma-V)\s b\ =\ 0
\label{s22}
\qqq
where
\qq
&&U(z)\ =\ \ \prod\limits_{n=1}^6(z-z_n)\sum\limits_{n=1}^6
{Q_n^{\s2}\over z-z_n}\s,\cr
&&V(z)\ =\ i\m\prod\limits_{n=1}^6(z-z_n)\sum\limits_{n=1}^6
{Q_nP_n\over z-z_n}\s,\cr
&&W(z)\ =\ \ \prod\limits_{n=1}^6(z-z_n)\sum\limits_{n=1}^6
{P_n^{\s2}\over z-z_n}
\label{UVW}
\qqq
are $4^{\rm th}$-order polynomials in $z$. The non-trivial
solution exists if 
\qq
\sigma^2\ =\ U(z)W(z)+V(z)^2\ =\ P(z)
\qqq
where the last equality follows by a straightforward check.
The system (\ref{s22}) of linear equations defines a holomorphic
line subbundle $L$ of the rank 2 bundle $W=\NC^2\otimes
\CO(4p^+_\infty+4p^-_\infty)$ over the hyperelliptic 
curve $\CS$ (the coefficients behave as $z^4$
at infinity). As solutions of (\ref{s22}) we may take 
for example
\qq
a\,=\,\sigma-V(z)\,,\quad b\,=\, i\, U(z)\,\ \qquad{\rm or}
\ \qquad a\,=\,-\m i\, W(z)\,,\quad b\,=\,\sigma+V(z)\s.
\label{sol}
\qqq
Since $a$ and $b$ are proportional to $z^4$ at infinity,
they define holomorphic sections of $L\subset W$
regular at $p^\pm_\infty$. They vanish at four points 
\qq
p'_\alpha=(V(z'_\alpha),\m z'_\alpha)\qquad\qquad{\rm or}\qquad
\qquad p''_\alpha=(-\m V(z''_\alpha),\m z''_\alpha)\,,
\qqq
respectively, where $z'_\alpha$ are the roots of $U$ 
and $z''_\alpha$ are those of $W$. Hence the degree 
of the line bundle $L$ is equal to 4. $H^0(L)$ has dimension
2 and is spanned by the two solutions (\ref{sol}).

\subsection{Angle variables}

The knowledge of the bundle $L$ may be encoded in the
image $w$ of $L$ in the Jacobian $J^4(\CS)$ under the Abel map:
\qq
w\ =\ \sum\limits_{\alpha=1}^4
\int\limits_{p_0}^{p'_\alpha}
\Omega\ =\ 
\sum\limits_{\alpha=1}^4
\int\limits_{p_0}^{p''_\alpha}
\Omega
\label{angl}
\qqq
were $\Omega=(\Omega^b)$ is the vector of the abelian 
differentials (\ref{Omb}) on $\CS$ and $p_0$ is a fixed point 
of $\CS$. Under the infinitesimal
time evolution (\ref{dtn}) inducing the changes $\delta_nU$,
$\delta_nV$, $\delta_nW$, of the polynomials $U,V,W$,
the image of the Abel map changes by
\qq
\delta_nw^b\ =\ \sum\limits_{\alpha=1}^4{{z'_\alpha}^b\ 
\delta_n z'_\alpha\over V(z'_\alpha)}
\ =\ -\sum\limits_{\alpha=1}^4{{z''_\alpha}^b\  
\delta_n z''_\alpha\over V(z''_\alpha)}\s.
\qqq
The variations of the zeros of $U$ are
\qq
\delta_n z'_\alpha\,=\,-\,{\delta_n U(z'_\alpha)\over
U'(z'_\alpha)}
\qqq
and similarly for $\delta_nz''_\alpha$.
A direct calculation gives
\qq
\delta_n U(z)\,=\,\{\m h_n\m,\, U(z)\m\}\,\delta t_n\ 
=\ 4\, i\,\prod\limits_{m\not=n}(z_n-z_m)^{-1}\,{V(z_n)\m 
U(z)\m-\m U(z_n)\m V(z)\over z-z_n}\,\delta t_n\,,\hspace{0.7cm}
\qqq
see \cite{mumford2}, page 3.69. Hence
\qq
\delta_n w^b\ =\ 4\, i\prod\limits_{m\not=n}(z_n-z_m)^{-1}
\ U(z_n)\,\delta t_n\ \sum\limits_{\alpha=1}^4{{z'_\alpha}^b
\over(z'_\alpha-z_n)\,\m U'(z'_\alpha)}\,.
\qqq
The vanishing of the sum of residues of the meromorphic
form ${z^b\,\m dz\over (z-z_n)\, U(z)}$ implies that the last 
sum is equal to $-\m z_n^b\, U(z_n)^{-1}$ so that
\qq
\delta w^b\ =\ \{\m w^b\m,\, h_n\m\}\,\delta t_n
\ =\ {_4\over^i}\m\prod\limits_{m\not=n}(z_n-z_m)^{-1}
\,\m z_n^b\,\m\delta t_n
\qqq
which does not depend on the phase-space variables.
We infer that the Hamiltonians $h_n$ generate constant flows
on the complex torus $J^4(\CS)\cong\NC^3/\Lambda$ where 
$\Lambda$ is the lattice of periods of $\Omega$. 
Modulo a constant linear transformation, the coordinates 
$w^b$, $b=0,1,2,$ provide together with 3 of the Hamiltonians 
$h_n$ a Darboux coordinate system for $T^*\NP^3$. We have thus 
found the angle variables of the Hitchin system (as the angles
of $J^4(\CS)$).

It should be stressed that the above approach based 
on the Lax matrix is simpler then the one obtained by
following the general procedure for the Hitchin systems. 
In particular, Eq.\s\s(\ref{det}) giving the spectral curve 
$\CC$ in the general approach is, as shown in \cite{GaTr},   
\qq
\xi^2\ =\ -\, 4\prod\limits_{n=1}^6(z-z_n)
\sum\limits_{n,m=1\atop n\not=m}^6{J_{nm}^{\s 2}
\over(z-z_n)(z-z_m)}
\qqq
to which one has to add the equation (\ref{curve}) 
of the original curve $\Sigma$ of genus 2. $\CC$ is 
of genus 5 and it is a ramified cover of both $\Sigma$ 
(by forgetting $\xi$) and $\CS$ (by setting 
$\sigma={i\over 2\sqrt{2}}\m\xi\m y$). While the 
general construction would give the angle variables
as those of a 3-dimensional Prym variety 
in the 5-dimensional Jacobian of degree $-2$ line bundles
on $\CC$, the Lax approach gave them as the angles
of the degree 4 Jacobian of $\CS$.

\subsection{KZB connection}

The determinant bundle $\CL$ over the moduli space 
$\CN_{ss}\cong\NP\Theta_2$ of the holomorphic ${\rm SL}_2$-bundles 
over the genus 2 curve coincides with the dual of the tautological 
bundle on $\NP\Theta_2$ so that $H^0(\CL^k)$ is the space 
of homogeneous polynomials $\Psi$ of degree $k$
on the space $\Theta_2$. The Lie algebra $\groupeso_{6}$ acts in
the space $\Theta_2$ by the first order differential 
operators, still denoted by $J_{nm}$, satisfying the commutation 
relations (\ref{soa1}), (\ref{soa2}) with the Poisson bracket replaced
by the commutator. In the $(p,q)$-language they are obtained
by replacing $p_n$'s in the expressions (\ref{Js}) for $J_{nm}$ by
$-\da_{x_n}$. The KZB connection for the case in question
has been work out in \cite{VGDJ}. It takes the form 
(up to a scalar 1-form)
\qq
&&\nabla^{\rm KZB}_{\delta\mu}\,\Psi\ =\ 
\sum\limits_{n=1}^6\delta z_n\,(\da_{z_n}
-{_1\over^{\kappa}}\m H_n)
\,\Psi\, ,\label{KZ21}\\
&&\nabla^{\rm KZB}_{\overline{\delta\mu}}\,\Psi\ =\ 
\sum\limits_n\overline{\delta z}_n\,\da_{\bar z_n}
\,\Psi
\label{KZ22}
\qqq
where 
\qq
H_n\ =\ -\m{_1\over^2}\sum\limits_{m\not=n}^6{J_{nm}^{\s 2}
\over z_n-z_m}
\qqq
so that ${1\over k} H_n$ is a quantization of ${2k\over
(2\pi i)^2}\m h_n$ obtained by the replacement $J_{nm}\mapsto 
{2\pi\over k\m i}\m J_{nm}$ in the classical expression 
for $h_n$. The quantum Hamiltonians $H_n,\ n=1,\dots,6$, 
are commuting second order differential operators on 
$\Theta_2\cong\NC^4$.

\setcounter{equation}{0}
\medskip
\section{Conclusions}

We have described above in explicit terms the
Hitchin integrable systems and the Knizhnik-
Zamolodchikov-Bernard connection
in the genus 0, 1 and 2 geometries, the last case only for $G={\rm SL}_2$
and with no punctures. The main original contribution of the paper
is the construction of the angle variables of the genus 2 system.  
It is a modification of a similar construction, based on the use 
of a Lax matrix, for the classical Neumann system. 
The diagonalization of the quantized Hamiltonians $H_n$
entering the genus 2 KZB connection for $G={\rm SL}_2$ as well
as the identification of the genus 2 Hitchin systems 
with punctures and for different groups remain open problems.


\small

\bibliographystyle{amsplain}
\bibliography{art,livre}


\end{document}